\newif\ifdraft\draftfalse  % set true to show comments
\newif\ifanon\anonfalse    % set true to suppress names, etc.
\newif\iffull\fullfalse   % set true for long version
\newif\ifappendices\appendicestrue

\PassOptionsToPackage{usenames,dvipsnames,svgnames,table}{xcolor}
\documentclass[sigplan,acmsmall]{acmart}

\usepackage[usenames,dvipsnames,svgnames,table]{xcolor}
\usepackage{amsmath}
\usepackage{mathtools}
\usepackage{bussproofs}
\usepackage{amsthm}
\usepackage{csvsimple}
\usepackage{thmtools,thm-restate}
\usepackage{changepage}
\usepackage{booktabs}
\usepackage{amssymb}
\usepackage[inline]{enumitem}
\usepackage{multirow,bigdelim}
\usepackage{multicol}
\usepackage{siunitx}
\usepackage{listings}
\usepackage{sansmath}
\usepackage{url}
\usepackage{flushend}
\usepackage{microtype}
\usepackage[utf8]{inputenc}
\usepackage{mathpartir}
\usepackage{empheq}
\usepackage{array}
\usepackage{pgfplots}
\usepackage{stmaryrd}
\usepackage{courier}
\usepackage{qtree}
\usepackage[normalem]{ulem}
\usepackage{relsize}
\usepackage{tikz}
\usepackage{algorithm}
\usepackage[noend]{algpseudocode}
\usepackage{graphicx}
\usepackage{subcaption}
\usepackage{textcomp}
\usepackage{tabularx}
\usepackage{stackengine}
\usepackage{caption}
\usepackage{wrapfig}
\usepackage{remreset}

\renewcommand\footnotetextcopyrightpermission[1]{}

\usetikzlibrary{
  er,
  matrix,
  shapes,
  arrows,
  positioning,
  fit,
  calc,
  pgfplots.groupplots,
  arrows.meta
}
\tikzset{>={Latex}}

\definecolor{dkblue}{rgb}{0,0.1,0.5}
\definecolor{dkgreen}{rgb}{0,0.6,0}
\definecolor{dkred}{rgb}{0.6,0,0}
\definecolor{dkpurple}{rgb}{0.7,0,0.4}
\definecolor{olive}{rgb}{0.4, 0.4, 0.0}
\definecolor{teal}{rgb}{0.0,0.5,0.5}
\definecolor{orange}{rgb}{0.9,0.6,0.2}
\definecolor{lightyellow}{RGB}{255, 255, 179}
\definecolor{lightgreen}{RGB}{170, 255, 220}
\definecolor{teal}{RGB}{141,211,199}
\definecolor{darkbrown}{RGB}{121,37,0}

\makeatletter
\def\renewtheorem#1{%
  \expandafter\let\csname#1\endcsname\relax
  \expandafter\let\csname c@#1\endcsname\relax
  \gdef\renewtheorem@envname{#1}
  \renewtheorem@secpar
}
\def\renewtheorem@secpar{\@ifnextchar[{\renewtheorem@numberedlike}{\renewtheorem@nonumberedlike}}
\def\renewtheorem@numberedlike[#1]#2{\newtheorem{\renewtheorem@envname}[#1]{#2}}
\def\renewtheorem@nonumberedlike#1{  
\def\renewtheorem@caption{#1}
\edef\renewtheorem@nowithin{\noexpand\newtheorem{\renewtheorem@envname}{\renewtheorem@caption}}
\renewtheorem@thirdpar
}
\def\renewtheorem@thirdpar{\@ifnextchar[{\renewtheorem@within}{\renewtheorem@nowithin}}
\def\renewtheorem@within[#1]{\renewtheorem@nowithin[#1]}
\makeatother

\pgfplotsset{
    every non boxed x axis/.style={} 
}

\newcommand{\FINISH}[3]{\ifdraft\textcolor{#1}{[#2: #3]}\fi}
\newcommand{\bcp}[1]{\FINISH{dkred}{B}{#1}}

\theoremstyle{definition}
\renewtheorem{theorem}{Theorem}
\renewtheorem{lemma}{Lemma}
\renewtheorem{corollary}{Corollary}
\renewtheorem{definition}{Definition}
\newtheorem{property}{Property}
\theoremstyle{plain}
\theoremstyle{remark}

\theoremstyle{remark}
\newtheorem{case}{Case}
\makeatletter
\@addtoreset{subcase}{case}
\@addtoreset{case}{lemma}
\@addtoreset{case}{theorem}
\@addtoreset{case}{corollary}
\@addtoreset{case}{definition}
\@addtoreset{case}{definition}
\@removefromreset{theorem}{section}
\makeatother

\algnewcommand\algorithmicswitch{\textbf{switch}}
\algnewcommand\algorithmicmatch{\textbf{match}}
\algnewcommand\algorithmiccase{\textbf{case}}
\algnewcommand\algorithmicwith{\textbf{with}}
\algnewcommand\algorithmicforeach{\textbf{foreach}}
\algnewcommand\Assert[1]{\State \algorithmicassert(#1)}%
\algdef{SE}[SWITCH]{Switch}{EndSwitch}[1]{\algorithmicmatch\ #1\ \algorithmicwith}{\algorithmicend\ \algorithmicswitch}%
\algdef{SE}[CASE]{Case}{EndCase}[1]{$|~$ #1 $\rightarrow$}{\algorithmicend\ \algorithmiccase}%
\algdef{SE}[FOREACH]{ForEach}{EndForEach}[2]{\algorithmicforeach\ #1 $\in$ #2}{\algorithmicend\ \algorithmicforeach}%
\algdef{SE}[CaseTwo]{CaseTwo}{EndCaseTwo}[2]{$|~$ #1 $\rightarrow$ #2}{\algorithmicend\ \algorithmiccase}%
\algtext*{EndSwitch}%
\algtext*{EndCase}%
\algtext*{EndCaseTwo}%
\algtext*{EndSecondCase}%
\algtext*{EndForEach}%

\newcommand{\CF}[1]{{\small \texttt{#1}}}         % Code Font

\newcommand{\PCF}[1]{\textproc{#1}}
\newcommand{\VarCF}[1]{{\color{darkbrown} \CF{#1}}}
\newcommand{\StringCF}[1]{\CF{\textcolor{blue}{#1}}}
\newcommand{\Regex}{\ensuremath{\mathit{S}}}         % Regular Expression

\newcommand{\BooleanAnd}{\ensuremath{~\wedge~}}
\newcommand{\BooleanOr}{\ensuremath{\vee}}
\newcommand{\BooleanImplies}{\ensuremath{\Rightarrow}}
\newcommand{\Rewrite}{\ensuremath{\rightarrow}}
\newcommand{\RewriteAtom}{\ensuremath{\Rewrite_\Atom}}
\newcommand{\RewriteDNF}{\ensuremath{\Rewrite_\DNFRegex}}
\newcommand{\ConcatDNF}{\ensuremath{\odot}}
\newcommand{\ConcatDNFOf}[2]{\ensuremath{#1\ConcatDNF#2}}

\newcommand{\ConcatSequence}{\ensuremath{\odot_{\Sequence}}}
\newcommand{\ConcatSequenceOf}[2]{\ensuremath{#1\ConcatSequence#2}}
\newcommand{\ConcatPermutation}{\ensuremath{\odot}}
\newcommand{\ConcatPermutationOf}[2]{\ensuremath{#1\ConcatPermutation#2}}
\newcommand{\SwapPermutation}{\ensuremath{\circledS}}
\newcommand{\SwapPermutationOf}[2]{\ensuremath{#1\SwapPermutation#2}}
\newcommand{\DistributePermutation}{\ensuremath{\otimes}}
\newcommand{\DistributePermutationOf}[2]{\ensuremath{#1\DistributePermutation#2}}
\newcommand{\DistributeSwapPermutation}{\ensuremath{\otimes^{\mathit{s}}}}
\newcommand{\DistributeSwapPermutationOf}[2]{\ensuremath{#1\DistributeSwapPermutation#2}}
\newcommand{\ConcatSequenceLens}{\ensuremath{\odot_{\SequenceLens}}}
\newcommand{\ConcatSequenceLensOf}[2]{\ensuremath{#1\ConcatSequenceLens#2}}
\newcommand{\ConcatDNFLens}{\ensuremath{\odot}}
\newcommand{\ConcatDNFLensOf}[2]{\ensuremath{#1\ConcatDNFLens#2}}
\newcommand{\SwapSequenceLens}{\ensuremath{\circledS_{\SequenceLens}}}
\newcommand{\SwapSequenceLensOf}[2]{\ensuremath{#1\SwapSequenceLens#2}}
\newcommand{\SwapDNFLens}{\ensuremath{\circledS}}
\newcommand{\SwapDNFLensOf}[2]{\ensuremath{#1\SwapDNFLens#2}}

\newcommand{\OrDNF}{\ensuremath{\oplus}}
\newcommand{\OrDNFOf}[2]{\ensuremath{#1\OrDNF#2}}
\newcommand{\OrDNFLens}{\ensuremath{\oplus}}
\newcommand{\OrDNFLensOf}[2]{\ensuremath{#1\OrDNFLens#2}}
\newcommand{\PutRight}{\ensuremath{\mathit{put}}}
\newcommand{\PutLeft}{\ensuremath{\mathit{get}}}

\newcommand{\RegexAlt}{\ensuremath{\mathit{T}}}         % Regular Expression
\newcommand{\Or}{\ensuremath{~|~}}
\newcommand{\RegexOr}[2]{\ensuremath{#1\Or#2}}

\newcommand{\RegexConcat}[2]{\ensuremath{#1\cdot#2}}
\newcommand{\EmptyString}{\ensuremath{\epsilon}}

\newcommand{\Concat}{\ensuremath{\cdot}}
\newcommand{\Union}{\ensuremath{\cup}}
\newcommand{\Intersect}{\ensuremath{\cap}}
\newcommand{\BigUnion}{\ensuremath{\bigcup}}

\newcommand{\SemanticsOf}[1]{\ensuremath{[ \! [#1] \! ]}}
\newcommand{\SetOf}[1]{\ensuremath{\{#1\}}}
\newcommand{\UnambigItOf}[1]{\ensuremath{#1^{*!}}}
\newcommand{\UnambigConcat}{\ensuremath{\Concat^!}}
\newcommand{\SequenceUnambigConcatOf}[1]{\ensuremath{\UnambigConcat(#1)}}
\newcommand{\UnambigConcatOf}[2]{\ensuremath{#1 \UnambigConcat #2}}
\newcommand{\UnambigOrOf}[2]{\ensuremath{\LanguageOf{#1} \cap \LanguageOf{#2} = \emptyset}}
\newcommand{\Atom}{\ensuremath{\mathit{A}}}          % Atoms
\newcommand{\AtomAlt}{\ensuremath{\mathit{B}}}
\newcommand{\AtomType}{\ensuremath{\mathit{Atom}}}

\newcommand{\Sequence}{\ensuremath{\mathit{SQ}}}
\newcommand{\SequenceType}{\ensuremath{\mathit{Sequence}}}

\newcommand{\SequenceAlt}{\ensuremath{\mathit{TQ}}}
\newcommand{\DNFRegex}{\ensuremath{\mathit{DS}}}         % Regular Expression
\newcommand{\DNFRegexAlt}{\ensuremath{\mathit{DT}}}    %Alt Regex
\newcommand{\DNFRegexType}{\ensuremath{\mathit{DNF}}}

\newcommand{\String}{\ensuremath{\mathit{s}}}        % String
\newcommand{\StringAlt}{\ensuremath{\mathit{t}}}        % StringAlt
\newcommand{\Language}{\ensuremath{L}}
\newcommand{\LanguageOf}[1]{\ensuremath{\mathcal{L}(#1)}}

\newcommand{\RangeIncInc}[2]{\ensuremath{[#1,#2]}}

\newcommand{\Lens}{\ensuremath{\mathit{l}}}
\newcommand{\AtomLens}{\ensuremath{\mathit{al}}}
\newcommand{\IterateAtomType}{\textit{Iterate}}

\newcommand{\SequenceLens}{\ensuremath{\mathit{sql}}}
\newcommand{\SequenceLensType}{\ensuremath{\mathit{SequenceLens}}}
\newcommand{\DNFLens}{\ensuremath{\mathit{dl}}}
\newcommand{\DNFLensType}{\ensuremath{\mathit{DNFLens}}}
\newcommand{\AtomLensType}{\ensuremath{\mathit{AtomLens}}}

\newcommand{\PermutationSetOf}[1]{\ensuremath{S_{#1}}}
\newcommand{\Permutation}{\ensuremath{\sigma}}

\newcommand{\Star}{\ensuremath{^*}}
\newcommand{\StarOf}[1]{\ensuremath{{#1}\Star}}
\newcommand{\ConstLens}{\ensuremath{\mathit{const}}}
\newcommand{\ConstLensOf}[2]{\ensuremath{\ConstLens(#1,#2)}}
\newcommand{\ConcatLens}{\ensuremath{\mathit{concat}}}
\newcommand{\ConcatLensOf}[2]{\ensuremath{\ConcatLens(#1,#2)}}
\newcommand{\ConcatLensShortOf}[2]{\ensuremath{\mathit{c}(#1,#2)}}
\newcommand{\SwapLens}{\ensuremath{\mathit{swap}}}
\newcommand{\SwapLensOf}[2]{\ensuremath{\SwapLens(#1,#2)}}
\newcommand{\SwapLensShortOf}[2]{\ensuremath{\mathit{s}(#1,#2)}}
\newcommand{\OrLens}{\ensuremath{\mathit{or}}}
\newcommand{\OrLensOf}[2]{\ensuremath{\OrLens(#1,#2)}}
\newcommand{\IdentityLens}{\ensuremath{\mathit{id}}}
\newcommand{\IdentityLensOf}[1]{\ensuremath{\IdentityLens_{#1}}}
\newcommand{\IdentityLensShortT}{\ensuremath{\mathit{id}}}
\newcommand{\IdentityLensShortOf}[1]{\ensuremath{\IdentityLensShortT_{#1}}}
\newcommand{\IterateLens}{\ensuremath{\mathit{iterate}}}
\newcommand{\IterateLensOf}[1]{\ensuremath{\mathit{\IterateLens(#1)}}}
\newcommand{\Identity}{\ensuremath{\mathit{id}}}
\newcommand{\Compose}{\ensuremath{\circ}}
\newcommand{\ComposeLensOf}[2]{\ensuremath{#1\mathrel{;}#2}}

\newcommand{\GBar}{\ensuremath{~|~}}

\newcommand{\GEq}{\ensuremath{::=~}}

\newcommand{\Nats}{\ensuremath{\mathbb{N}}}

\newcommand{\InverseOf}[1]{\ensuremath{#1^{-1}}}

\newcommand{\OfType}{\ensuremath{:}}
\newcommand{\OfRewritelessType}{\ensuremath{\,\,\tilde{\OfType}\,\,}}
\newcommand{\MapsBetweenTypeOf}[2]{\ensuremath{#1 \Leftrightarrow #2}}
\newcommand{\ArrowTypeOf}[2]{\ensuremath{#1 \rightarrow #2}}

\newcommand{\ToDNFRegex}{\ensuremath{\Downarrow}}
\newcommand{\ToDNFRegexOf}[1]{\ensuremath{\ToDNFRegex\mkern-4mu #1}}
\newcommand{\ToRegex}{\ensuremath{\Uparrow}}
\newcommand{\ToRegexOf}[1]{\ensuremath{\ToRegex\mkern-4mu #1}}

\newcommand{\SuchThat}{\ensuremath{~|~}}
\newcommand{\That}{\ensuremath{~.~}}

\newcommand{\Alphabet}{\ensuremath{\Sigma}}
\newcommand{\Character}{\ensuremath{c}}

\newcommand{\SequenceLeft}{\ensuremath{[}}
\newcommand{\SequenceRight}{\ensuremath{]}}
\newcommand{\SequenceOf}[1]{\ensuremath{\SequenceLeft#1\SequenceRight}}
\newcommand{\SeqSep}{\ensuremath{\mkern-1mu\Concat\mkern-1mu}}
\newcommand{\DNFLeft}{\ensuremath{\langle}}
\newcommand{\DNFRight}{\ensuremath{\rangle}}
\newcommand{\DNFOf}[1]{\ensuremath{\DNFLeft#1\DNFRight}}
\newcommand{\DNFSep}{\ensuremath{\Or}}
\newcommand{\SequenceLensLeft}{\ensuremath{[}}
\newcommand{\SequenceLensRight}{\ensuremath{]}}
\newcommand{\SequenceLensOf}[1]{\ensuremath{\SequenceLensLeft#1\SequenceLensRight}}
\newcommand{\SeqLSep}{\ensuremath{\mkern-1mu\Concat\mkern-1mu}}
\newcommand{\DNFLensLeft}{\ensuremath{\langle}}
\newcommand{\DNFLensRight}{\ensuremath{\rangle}}
\newcommand{\DNFLensOf}[1]{\ensuremath{\DNFLensLeft#1\DNFLensRight}}
\newcommand{\DNFLSep}{\ensuremath{\Or}}

\newcommand{\ConstantLensRule}{\textsc{Constant Lens}}

\newcommand{\IterateLensRule}{\textsc{Iterate Lens}}
\newcommand{\ConcatLensRule}{\textsc{Concat Lens}}
\newcommand{\SwapLensRule}{\textsc{Swap Lens}}
\newcommand{\OrLensRule}{\textsc{Or Lens}}
\newcommand{\ComposeLensRule}{\textsc{Compose Lens}}
\newcommand{\RewriteRegexLensRule}{\textsc{Rewrite Regex Lens}}

\newcommand{\AtomUnrollstarLeftRule}{\textsc{Atom Unrollstar\SubLeft}}
\newcommand{\AtomUnrollstarRightRule}{\textsc{Atom Unrollstar\SubRight}}
\newcommand{\ParallelAtomStructuralRewriteRule}{\textsc{Parallel Atom Structural Rewrite}}
\newcommand{\ParallelSwapAtomStructuralRewriteRule}{\textsc{Parallel Swap Atom Structural Rewrite}}
\newcommand{\AtomStructuralRewriteRule}{\textsc{Atom Structural Rewrite}}
\newcommand{\DNFStructuralRewriteRule}{\textsc{DNF Structural Rewrite}}
\newcommand{\ParallelDNFStructuralRewriteRule}{\textsc{Parallel DNF Structural Rewrite}}
\newcommand{\ParallelSwapDNFStructuralRewriteRule}{\textsc{Parallel Swap DNF Structural Rewrite}}
\newcommand{\IdentityRewriteRule}{\textsc{Identity Rewrite}}
\newcommand{\DNFReorderRule}{\textsc{DNF Reorder}}

\newcommand{\SequenceLensRule}{\textsc{Sequence Lens}}
\newcommand{\AtomLensRule}{\textsc{Atom Lens}}
\newcommand{\DNFLensRule}{\textsc{DNF Lens}}
\newcommand{\RewriteDNFRegexLensRule}{\textsc{Rewrite DNF Regex Lens}}

\newcommand{\SubLeft}{\textsubscript{L}}
\newcommand{\SubRight}{\textsubscript{R}}

\newcommand{\Set}{\ensuremath{\mathit{S}}}

\newcommand{\OrIdentityRule}{\textit{+ Ident}}
\newcommand{\EmptyProjectionRightRule}{\textit{0 Proj\SubRight{}}}
\newcommand{\EmptyProjectionLeftRule}{\textit{0 Proj\SubLeft{}}}
\newcommand{\ConcatAssocRule}{\textit{\Concat{} Assoc}}
\newcommand{\OrAssociativityRule}{\textit{\Or{} Assoc}}
\newcommand{\OrCommutativityRule}{\textit{\Or{} Comm}}
\newcommand{\DistributivityLeftRule}{\textit{Dist\SubRight{}}}
\newcommand{\DistributivityRightRule}{\textit{Dist\SubLeft{}}}
\newcommand{\ConcatIdentityLeftRule}{\textit{\Concat{} Ident\SubLeft{}}}
\newcommand{\ConcatIdentityRightRule}{\textit{\Concat{} Ident\SubRight{}}}
\newcommand{\SumstarRule}{\textit{Sumstar}}
\newcommand{\ProductstarRule}{\textit{Prodstar}}
\newcommand{\UnrollstarLeftRule}{\textit{Unrollstar\SubLeft{}}}
\newcommand{\UnrollstarRightRule}{\textit{Unrollstar\SubRight{}}}
\newcommand{\StarstarRule}{\textit{Starstar}}
\newcommand{\DicyclicityRule}{\textit{Dicyc}}
\newcommand{\Derivation}{\ensuremath{\mathcal{D}}}

\newcolumntype{L}{>{$}l<{$}}
\newcolumntype{R}{>{$}r<{$}}

\renewcommand{\subsubsection}[1]{\paragraph{{#1}}}

\newcommand{\Examples}{\ensuremath{\mathit{exs}}}

\newcommand{\ParallelRewrite}{\ensuremath{\,\mathrlap{\to}\,{\scriptstyle\parallel}\,\,\,}}
\newcommand{\ParallelRewriteAtom}{\ensuremath{\ParallelRewrite_{\Atom}}}
\newcommand{\ParallelRewriteSwap}{\ensuremath{\ParallelRewrite^{\mathit{swap}}}}
\newcommand{\ParallelRewriteSwapAtom}{\ensuremath{\ParallelRewrite^{\mathit{swap}}_{\Atom}}}

\newcommand{\Property}{\ensuremath{\mathit{p}}}
\newcommand{\Propagator}{\ensuremath{\mathit{q}}}

\newcommand{\DiamondProperty}{\ensuremath{\mathit{confluent}}}
\newcommand{\DiamondPropertyWithPropertyOf}[1]{\ensuremath{\DiamondProperty_{#1}}}
\newcommand{\IsConfluentWithPropertyOf}[2]
    {\ensuremath{\DiamondPropertyWithPropertyOf{#2}(#1)}}
\newcommand{\BisimilarProperty}{\ensuremath{\mathit{bisimilar}}}
\newcommand{\BisimilarPropertyWithPropertyOf}[1]{\ensuremath{\BisimilarProperty_{#1}}}
\newcommand{\IsBisimilarWithPropertyOf}[2]
    {\ensuremath{\BisimilarPropertyWithPropertyOf{#2}(#1)}}

\newcommand{\Sep}{\ensuremath{\$}}

\newcommand{\Cross}{\ensuremath{\times}}

\newcommand{\Sorting}{\ensuremath{\mathit{sorting}}}
\newcommand{\SortingOf}[2]{\ensuremath{\Sorting(#1,#2)}}

\newcommand{\Sort}{\ensuremath{\mathit{sort}}}
\newcommand{\SortOf}[2]{\ensuremath{\Sort(#1,#2)}}

\newcommand{\ListType}{\ensuremath{\mathit{List}}}
\newcommand{\ListTypeOf}[1]{\ensuremath{#1\,\ListType}}
\newcommand{\ListLeft}{\ensuremath{[}}
\newcommand{\ListRight}{\ensuremath{]}}
\newcommand{\ListOf}[1]{\ensuremath{\ListLeft #1 \ListRight}}

\newcommand{\SequenceLeq}{\ensuremath{\leq_{Seq}}}
\newcommand{\AtomLeq}{\ensuremath{\leq_{Atom}}}
\newcommand{\ILSLeq}{\ensuremath{\leq_{\mathit{intlistset}}}}
\newcommand{\ExampledDNFLeq}{\ensuremath{\leq_{DNF}^{\Examples}}}
\newcommand{\ExampledSequenceLeq}{\ensuremath{\leq_{Seq}^{\Examples}}}
\newcommand{\ExampledAtomLeq}{\ensuremath{\leq_{Atom}^{\Examples}}}

\newcommand{\NormalizedStarOf}[1]{\ensuremath{\NormalizedStarOf{#1}_n}}

\newcommand{\DNFLensHasSemanticsOf}[1]{\ensuremath{\xLeftrightarrow{#1}}}
\newcommand{\SatisfiesDNFLensHasSemanticsOf}[3]{\ensuremath{#2\DNFLensHasSemanticsOf{#1}#3}}
\newcommand{\SatisfiesIdentitySemantics}[2]
  {\ensuremath{\SatisfiesDNFLensHasSemanticsOf{\Identity}{#1}{#2}}}
\newcommand{\EquivalenceOf}[1]{\equiv_{#1}}

\newcommand{\SSREquiv}{\ensuremath{\equiv^s}}

\newcommand{\ReflexivityRule}{\textsc{Reflexivity}}
\newcommand{\BaseRule}{\textsc{Base}}
\newcommand{\SymmetryRule}{\textsc{Symmetry}}
\newcommand{\TransitivityRule}{\textsc{Transitivity}}

\newcommand{\BaseRegexType}{\textit{Base}}
\newcommand{\EmptyRegexType}{\textit{Empty}}
\newcommand{\StarRegexType}{\textit{Star}}
\newcommand{\ConcatRegexType}{\textit{Concat}}
\newcommand{\OrRegexType}{\textit{Or}}

\newcommand{\ConstLensType}{\textit{Const}}
\newcommand{\ConcatLensType}{\textit{Concat}}
\newcommand{\IterateLensType}{\textit{Iterate}}
\newcommand{\SwapLensType}{\textit{Swap}}
\newcommand{\OrLensType}{\textit{Or}}
\newcommand{\ComposeLensType}{\textit{Compose}}
\newcommand{\IdentityLensType}{\textit{Identity}}

\newcommand{\StarAtomType}{\textit{Star}}
\newcommand{\MultiConcatSequenceType}{\textit{MultiConcat}}
\newcommand{\MultiOrDNFRegexType}{\textit{MultiOr}}

\newcommand{\AtomToDNF}{\ensuremath{\mathcal{D}}}
\newcommand{\AtomToDNFOf}[1]{\ensuremath{\AtomToDNF(#1)}}
\newcommand{\AtomToDNFLens}{\ensuremath{\mathcal{D}}}
\newcommand{\AtomToDNFLensOf}[1]{\ensuremath{\AtomToDNFLens(#1)}}

\newcommand{\Queue}{\ensuremath{\mathit{Q}}}
\newcommand{\QueueElement}{\ensuremath{\mathit{qe}}}
\newcommand{\QueueElements}{\ensuremath{\QueueElement\mathit{s}}}
\newcommand{\ExpCount}{\ensuremath{\mathit{e}}}
\newcommand{\True}{\ensuremath{\mathit{true}}}
\newcommand{\False}{\ensuremath{\mathit{false}}}

\newcommand{\DictionaryOrderL}{\ensuremath{[}}
\newcommand{\DictionaryOrderR}{\ensuremath{]}}
\newcommand{\DictionaryOrderOf}[1]{\ensuremath{\DictionaryOrderL #1 \DictionaryOrderR}}

\newcommand{\SetOfListOrderL}{\ensuremath{\{}}
\newcommand{\SetOfListOrderR}{\ensuremath{\}}}
\newcommand{\SetOfListOrderOf}[1]{\ensuremath{\SetOfListOrderL #1 \SetOfListOrderR}}

\newcommand{\Int}{\ensuremath{i}}
\newcommand{\UserDef}{\ensuremath{U}}
\newcommand{\UserDefAlt}{\ensuremath{V}}

\newcommand{\Optician}{Optician}
\newcommand{\SynthLens}{\PCF{SynthLens}}
\newcommand{\TypeProp}{\PCF{TypeProp}}
\newcommand{\SynthDNFLens}{\PCF{SynthDNFLens}}
\newcommand{\ToLens}{\ensuremath{\Uparrow}}
\newcommand{\ToLensOf}[1]{\ensuremath{\ToLens{}\mkern-4mu #1}}

\newcommand{\RigidSynth}{\PCF{RigidSynth}}
\newcommand{\RigidSynthInternal}{\PCF{RigidSynthInternal}}
\newcommand{\RigidSynthSequence}{\PCF{RigidSynthSeq}}
\newcommand{\RigidSynthAtom}{\PCF{RigidSynthAtom}}

\newcommand{\CreatePQueue}{\PCF{CreatePQueue}}
\newcommand{\GetTransitiveSet}{\PCF{GetTransitiveSet}}
\newcommand{\GetCurrentSet}{\PCF{GetCurrentSet}}
\newcommand{\Pop}{\PCF{Pop}}
\newcommand{\ExpandOnce}{\PCF{ExpandOnce}}
\newcommand{\ExpandRequired}{\PCF{ExpandRequired}}
\newcommand{\FixProblemElts}{\PCF{FixProblemElts}}
\newcommand{\Expand}{\PCF{Expand}}
\newcommand{\ForceExpand}{\PCF{ForceExpand}}
\newcommand{\Reveal}{\PCF{Reveal}}
\newcommand{\Map}{\PCF{Map}}
\newcommand{\EnqueueMany}{\PCF{EnqueueMany}}
\newcommand{\ReturnVal}[1]{\ensuremath{\Return\,#1}}

\newcommand{\CurrentSet}{\ensuremath{\mathit{CS}}}
\newcommand{\TransitiveSet}{\ensuremath{\mathit{TS}}}

\newcommand{\StringType}{\ensuremath{\mathit{String}}}

\newcommand{\SUBSECTION}[1]{\iffull\subsection{#1}\else\paragraph*{#1}}

\newcommand{\None}{\ensuremath{\mathit{None}}}
\newcommand{\DNFLensOption}{\ensuremath{\DNFLens\mathit{o}}}
\newcommand{\Some}{\ensuremath{\mathit{Some}}}
\newcommand{\SomeOf}[1]{\ensuremath{\Some\,#1}}

\newcommand{\Success}{\ensuremath{\boldsymbol{\color{dkgreen}\checkmark}}}
\newcommand{\Failure}{\ensuremath{\boldsymbol{\color{dkred}\times}}}

\newcommand{\Append}{\ensuremath{+\!\!\!\!+\ }}
\newcommand{\ModernTitle}{\VarCF{modern\_\discretionary{}{}{}title}}
\newcommand{\DNFModernTitle}{\VarCF{dnf\_modern\_title}}
\newcommand{\DNFLegacyTitle}{\VarCF{dnf\_legacy\_title}}
\newcommand{\LegacyTitle}{\VarCF{legacy\_\discretionary{}{}{}title}}
\newcommand{\LegacyTitleP}{\VarCF{legacy\_\discretionary{}{}{}title'}}
\newcommand{\TextChar}{\VarCF{text\_\discretionary{}{}{}char}}

\newcommand{\IntList}{\ensuremath{\mathit{il}}}
\newcommand{\IntListSet}{\ensuremath{\mathit{ils}}}
\newcommand{\StringIntListSet}{\ensuremath{\mathit{sils}}}
\newcommand{\ProjectStrings}{\ensuremath{\mathit{projectstrings}}}
\newcommand{\ProjectStringsOf}[1]{\ensuremath{\mathit{\ProjectStrings(#1)}}}
\newcommand{\ProjectILS}{\ensuremath{\mathit{projectils}}}
\newcommand{\ProjectILSOf}[1]{\ensuremath{\mathit{\ProjectILS(#1)}}}
\newcommand{\Generates}{\ensuremath{\rightsquigarrow}}

\newcommand{\ExampledDNFRegex}{\ensuremath{EDS}}
\newcommand{\ExampledDNFRegexAlt}{\ensuremath{EDT}}

\newcommand{\ExampledSequence}{\ensuremath{ESQ}}
\newcommand{\ExampledSequenceAlt}{\ensuremath{ETQ}}

\newcommand{\ExampledAtom}{\ensuremath{EA}}
\newcommand{\ExampledAtomAlt}{\ensuremath{EB}}

\newcommand{\EmbedExamples}{\ensuremath{\PCF{EmbedExamples}}}
\newcommand{\EmbedExamplesOf}[2]{\ensuremath{\EmbedExamples(#1,#2)}}

\newcommand{\Morpheus}{Morpheus}
\newcommand{\InSynth}{InSynth}

\newcommand{\FullMode}{\textbf{Full}}
\newcommand{\NoCSMode}{\textbf{NoCS}}
\newcommand{\NoFPEMode}{\textbf{NoFPE}}
\newcommand{\NoERMode}{\textbf{NoER}}
\newcommand{\NoUDMode}{\textbf{NoUD}}
\newcommand{\FlashExtractMode}{\textbf{FlashExtract}}
\newcommand{\FlashFillMode}{\textbf{Flash Fill}}
\newcommand{\NaiveMode}{\textbf{Na\"ive}}

\newcommand{\NumMoreThanFlashFill}
{35}
\newcommand{\ExamplesRequiringNonzeroExamples}
{5}
\newcommand{\ExpansionsAllForcedNoLC}
{22}
\newcommand{\BenchmarksCompletedNoLC}
{38}
\newcommand{\ExpansionsNotAllForcedNoLC}
{16}

\newcommand{\ExpansionsPerformedNaiveExpansionSuccess}
{5}
\newcommand{\FlashFillSuccesses}
{3}

\newcommand{\LOC}{3713}

\lstnewenvironment{sflisting}[1][]
  {\lstset{%
    mathescape,
    basicstyle=\small\sffamily,
    aboveskip=5pt,
    belowskip=5pt,
    columns=flexible,
    frame=,
    xleftmargin=1em,#1}\sansmath}
  {}
  \acmMonth{10}
  \acmYear{2017}

\begin{document}

\setlength{\pdfpageheight}{\paperheight}
\setlength{\pdfpagewidth}{\paperwidth}
%\toappear{}

%\conferenceinfo{POPL '16}{January 20--22, 2016, St. Petersburg, FL, USA} 
%\copyrightyear{2016} 
%\copyrightdata{978-1-nnnn-nnnn-n/yy/mm} 

% Uncomment one of the following two, if you are not going for the 
% traditional copyright transfer agreement.

%\exclusivelicense                % ACM gets exclusive license to publish, 
                                  % you retain copyright

%\permissiontopublish             % ACM gets nonexclusive license to publish
                                  % (paid open-access papers, 
                                  % short abstracts)

%\titlebanner{DRAFT---do not distribute}        % These are ignored unless
%\preprintfooter{DRAFT---do not distribute}   % 'preprint' option specified.

\title{Synthesizing Bijective Lenses}

\begin{abstract}
Bidirectional transformations between different data representations
occur frequently in modern software systems.  They appear as serializers and
deserializers, as parsers and pretty printers, as database views and view
updaters, and as a multitude of different kinds of ad hoc data converters. 
Manually building bidirectional transformations---by writing two separate
functions that are intended to be inverses---is tedious and error prone.
A better approach is to use a domain-specific language in
which both directions can be written as a single expression.  However,
these domain-specific languages can be difficult to program in, requiring
programmers to manage fiddly details while working in a complex type system.

We present an alternative approach.  
Instead of coding transformations manually, we synthesize them from
declarative format descriptions and examples.
%This approach will generate invertable pairs of data transformation functions
%while reducing programmer tedium. 
Specifically,
we present \emph{\Optician{}}, a tool for type-directed synthesis of bijective
string transformers.   The inputs to \Optician{} are a pair of ordinary
regular expressions representing two data formats
and a few concrete examples for disambiguation.  The output is a well-typed
program in Boomerang (a bidirectional language based on the
theory of {\em lenses}). %, which behaves as specified by the given examples.
The main technical challenge involves navigating the vast program search
space efficiently enough.  In particular, and unlike most prior work on type-directed 
synthesis, our system operates in the context of a language with a rich equivalence
relation on types (the theory of regular expressions).  Consequently, program
synthesis requires search in two dimensions:  First, our synthesis algorithm must
find a pair of ``syntactically compatible types,'' and second, using the structure
of those types, it must find a type- and example-compliant term.  Our key insight is
that it is possible to
reduce the size of this search space \emph{without losing any computational power}
by defining a new language of lenses designed
specifically for synthesis.  The new language is free from arbitrary function
composition and operates only over types and terms in a new disjunctive normal form.
We prove (1) our new language
is just as powerful as a more natural, compositional, and declarative
language and (2) our synthesis algorithm is sound and complete with respect to the
new language.  We also demonstrate empirically
that our new language changes the synthesis problem from
one that admits intractable solutions to one that admits 
highly efficient solutions, able to synthesize lenses
between complex file formats with great variation in seconds.  
We evaluate \Optician{} on a benchmark suite of 39 examples
that includes both microbenchmarks and realistic examples derived from
other data management systems including Flash Fill, a tool
for synthesizing string transformations in spreadsheets, and Augeas, a tool
for bidirectional processing of Linux system configuration files.
 
%% The main technical challenges \Optician{} addresses are 
%% (1) the combinatorial explosion caused by the complex
%% lens typing rules, and
%% (2) the size and complexity of the regular expressions
%% describing real-world data formats.  We address (1) by shifting the
%% search to a related space where both regular expressions and lenses are
%% constrained to a ``disjunctive normal form.'' We address (2) by 
%% treating named, user-defined regular expressions abstractly for as
%% long as possible. 

%% We evaluate \Optician{} on a benchmark suite of \afm{26} examples
%% that includes both microbenchmarks and realistic examples derived from
%% other data management systems including  
%% FlashFill, a tool
%% for synthesizing string transformations in spreadsheets, and Augeas, a tool
%% for bidirectional processing of Linux system configuration files based on
%% Boomerang.  Our implementation infers lenses for all benchmark programs in
%% under \afm{5} seconds.
\end{abstract}

\ifanon
%\authorinfo{}
%           {}
%           {}
\maketitle
% \vspace*{-6cm}
\else
\author{Anders Miltner}
\affiliation{Princeton University, USA}
\email{amiltner@cs.princeton.edu}

\author{Kathleen Fisher}
\affiliation{Tufts University, USA}
\email{kfisher@eecs.tufts.edu}

\author{Benjamin C. Pierce}
\affiliation{University of Pennsylvania, USA}
\email{bcpierce@cis.upenn.edu}

\author{David Walker}
\affiliation{Princeton University, USA}
\email{dpw@cs.princeton.edu}

\author{Steve Zdancewic}
\affiliation{University of Pennsylvania, USA}
\email{stevez@cis.upenn.edu}

\maketitle
\fi

% \category{D.3.1}
% {Programming Languages}
% {Formal Definitions and Theory}
% [Semantics]
\ifanon\else
\terms{Languages, Theory}

\keywords{Functional Programming, Proof Search, Program Synthesis, Type Theory,
  Bidirectional Programming}
\fi

% begin introduction
\section{Introduction}

Programs that analyze consumer information, performance statistics, transaction
logs, scientific records, and many other kinds of data are essential components
in many software systems.
Oftentimes, the data analyzed comes in \emph{ad hoc} formats, making
tools for reliably parsing, printing, cleaning, and transforming data
increasingly important.
Programmers often need to reliably transform back-and-forth between
formats, not only transforming source data into a target
format but also safely transforming target data back into the source format.
\emph{Lenses}~\cite{Focal2005-long} are back-and-forth
transformations that provide strong guarantees about their round-trip behavior,
guarding against data corruption
while reading, editing, and writing data sources.  

A lens comprises two functions,
\emph{get} and \emph{put}.  The \emph{get} function translates
source data into the target format.  If the target data is updated, the
\emph{put} function translates this edited data back into the
source format.  
A benefit of lens-based languages is that they use a single term
to express both 
\emph{get} and \emph{put}.
Furthermore, well-typed lenses give rise to 
\emph{get} and \emph{put} functions 
guaranteed to satisfy desirable invertibility properties.

Lens-based languages are present in variety of tools and have found mainstream
industrial use.
Boomerang~\cite{boomerang, Matching10} lenses provide
guarantees on transformations between {\em ad hoc} string document formats.
Augeas~\cite{augeas}, a popular tool that reads Linux system configuration
files, uses the \emph{get} part of a lens to transform configuration
files into a canonical tree representation that users can edit
either manually or 
programmatically.  It uses the lens's \emph{put} to merge the edited
results back into the original string format.  Other lens-based languages and
tools include 
GRoundTram~\cite{Hidaka2011GRoundTramAI},
BiFluX~\cite{DBLP:conf/ppdp/PachecoZH14}, 
BiYacc~\cite{DBLP:conf/staf/ZhuK0SH15},
Brul~\cite{DBLP:conf/etaps/ZanLKH16},
BiGUL~\cite{DBLP:conf/pepm/KoZH16}, 
bidirectional variants of 
relational algebra~\cite{BohannonPierceVaughan},
spreadsheet formulas~\cite{DBLP:conf/vl/MacedoPSC14},
graph query languages~\cite{DBLP:conf/icfp/HidakaHIKMN10},
and
XML transformation languages~\cite{DBLP:conf/pepm/LiuHT07}.
Unfortunately, these languages impose fiddly constraints on lenses,
making lens programming slow and tedious.
For example, Boomerang
programmers often must rearrange the order of data items by recursively
using operators that swap adjacent fields.
Furthermore, the Boomerang type checker is very strict, disallowing many programs
because they contain ambiguity about how certain data is transformed.
In short, lens languages provide strong bidirectional guarantees at the cost of
forcing programmers to satisfy finicky type systems.

To make programming with lenses faster and easier, we have
developed \emph{\Optician{}}, a
tool for synthesizing lenses from simple, high-level specifications.
This work continues a recent trend toward streamlining programming tasks
by synthesizing programs in a variety of domain-specific
languages~\cite{flashfill,le-pldi-2014,perelman2014test,morpheus},
many guided by
types~\cite{osera+:pldi15,frankle+:popl16,armando+:pldi16,feser-pldi-2015,morpheus}.  
Specifically, \Optician{} supports the synthesis of \emph{bijective lenses}, a
useful subset of Boomerang. 
As inputs, \Optician{} takes specifications of the source and target formats, plus
a collection of concrete examples of the desired 
transformation.  Format specifications are supplied as ordinary regular
expressions.
Because regular expressions are so widely understood, we anticipate such
inputs will be substantially easier for everyday programmers to work with
than the unfamiliar syntax of lenses.  Moreover, including these format
descriptions communicates a
great deal of information to the synthesis system.  Thus, requiring user input
of regular expressions
makes synthesis robust, 
helps the system scale to large and complex data sources, and 
constrains the search space sufficiently that the user typically needs
to give very few, if any, examples.

Despite the benefits of Boomerang's informative types,
Boomerang is not well-suited to support synthesis directly.
Specifically, Boomerang's types are regular expression
pairs, and each regular expression is equivalent to an infinite
number of other regular expressions.
To synthesize all Boomerang terms, a type-directed synthesizer
must sometimes be able to find, amongst all possible equivalent
regular expressions, the one with the right
syntactic structure to guide the subsequent search for a 
well-typed, example-compatible Boomerang term.  
%% Search for the
%% right Boomerang term is then made even more difficult by the fact that
%% there exist function compositions ($f \circ g$) that cannot be written
%% using some alternative single function $f'$ without composition.
%% The inadmissability of function composition in Boomerang
%% will force any na\"ive, but complete synthesis engine to guess haphazardly, and
%% inefficiently, at intermediate lens types.

To resolve these issues, we introduce a new language of
\emph{Disjunctive Normal Form (DNF) lenses}.
Just as string lenses have pairs of regular expressions as types, DNF lenses
have pairs of \emph{DNF regular expressions} as types.
The typing judgements for DNF lenses limit how equivalences can be
used, 
%and require no composition operator, 
greatly reducing the size
of the search space.
Despite the restrictive syntax and type system of DNF lenses, we prove our
new language is
equivalent to a natural, declarative specification of the 
bijective fragment of Boomerang.

\begin{figure}
  \centering
  \begin{tikzpicture}[auto,node distance=1.5cm]
    \node[text width=1.5cm,minimum height=.6cm,align=center,draw,rectangle] (todnfregex) {\ToDNFRegex{}};
    
    \node[align=right, anchor=east] (regex1) [left = .6cm of todnfregex.north west]{\Regex{}};
    \node[align=right, anchor=east] (regex2) [left = .6cm of todnfregex.south west]{\RegexAlt{}};
    \node[align=right, anchor=east] (exs) [below = .2cm of regex2]{ \Examples{} };
    
    \node[align=center] (dnfregex1) [right = .4cm of todnfregex.north east]{\DNFRegex{}};
    \node[align=center] (dnfregex2) [right = .4cm of todnfregex.south east]{\DNFRegexAlt{}};

    \node[text width=2.3cm,minimum height=.6cm,align=center,draw,rectangle] [right = 1.45cm of todnfregex.east] (synthdnflens) {\SynthDNFLens{}};
    \node[align=center] [above = .7cm of synthdnflens] (optician) {\Optician{}};
    
    \node[align=center] [right = .4cm of synthdnflens] (dnflens) {\DNFLens{}};
    
    \node[text width=1.5cm,minimum height=.6cm,align=center,draw,rectangle] [right = .4cm of dnflens] (tolens) {\ToLens{}};
    
    \node[align=center] [right = .6cm of tolens] (lens) {\Lens{}};

    \path[->] (regex1.east) edge (todnfregex.north west);
    \path[->] (regex2.east) edge (todnfregex.south west);
    
    \path[->] (todnfregex.north east) edge (dnfregex1.west);
    \path[->] (todnfregex.south east) edge (dnfregex2.west);
    
    \path[->] (dnfregex1.east) edge (synthdnflens.north west);
    \path[->] (dnfregex2.east) edge (synthdnflens.south west);
    
    \path[->] (synthdnflens) edge (dnflens);
    
    \path[->] (dnflens) edge (tolens);
    
    \path[->] (tolens) edge (lens);
    \draw[->] ($(exs.east)+(-3pt,0)$) -| node(exsedge) {} (synthdnflens);
    \node[fit={($(todnfregex.west)+(-4pt,0)$) ($(tolens.east)+(4pt,0)$) (exsedge) (dnfregex1) (optician) (dnfregex2)},draw] (surrounding) {};
    % Now place a relation (ID=rel1)
    %\node[text width=2cm,align=center,draw, rectangle] (sketch-gen) [right = .75cm of spec] {\TypeProp{}};
    %\node (below-gen) [below=.5cm of sketch-gen] {};
    %\node[text width=2cm,align=center,draw, rectangle] (sketch-compl)
    %     [right = .25cm of sketch-gen] {\RigidSynth{}};
    %\node (below-compl) [below=.5cm of sketch-compl] {};
    %\node[align=center] (lens) [right = .75cm of sketch-compl] {Lens}; 
    %% Draw an edge between rel1 and node1; rel1 and node2
    %\path[->] (spec) edge node (start-alg) {} (sketch-gen);
    %\path[->] (sketch-gen) edge node(middle) {} (sketch-compl);
    %\path[->] (sketch-compl) edge node[near start](success) {\Success{}} (lens);
    %\draw[<-] (sketch-gen.south) -- +(0,-.5) -| node[above left](failure){\Failure{}} (sketch-compl.south);

    %\node (synth-name) [above=.5cm of middle] {\Optician{}};
    %
    %\node[fit=(sketch-gen) (sketch-compl) (start-alg) (synth-name) (failure) (success) ,draw] (surrounding) {};
  \end{tikzpicture}
  \caption{Schematic Diagram for \Optician{}.  Regular expressions, \Regex{} and
    \RegexAlt{}, and examples, \Examples{}, are given as input.
    First, the function \ToDNFRegex{} converts \Regex{} and \RegexAlt{} into
    their respective DNF forms, \DNFRegex{} and \DNFRegexAlt{}.
    Next, \SynthDNFLens{} synthesizes a DNF lens, \DNFLens{}, from \Regex{},
    \RegexAlt{}, and \Examples{}.
    Finally, \ToLens{} converts \DNFLens{} into \Lens{}, a lens in Boomerang
    that is equivalent to \DNFLens{}.}
  \label{fig:schematic-diagram-synthesis}
\end{figure}
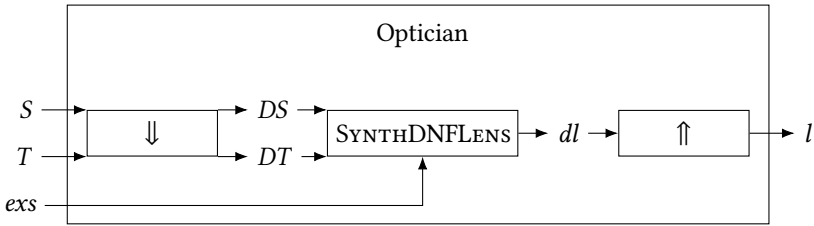

Figure~\ref{fig:schematic-diagram-synthesis} shows a high-level,
schematic diagram for \Optician{}.
First, \Optician{} uses the function \ToDNFRegex{} to convert the input
regular expressions into DNF regular expressions.  Next,
\SynthDNFLens{} performs type-directed synthesis on these DNF regular
expressions and the input examples to synthesize a DNF lens.  Finally, this DNF lens is
converted back into a regular lens with the function \ToLens{}, and returned to the user.

\paragraph*{Contributions.}  \Optician{} makes
bidirectional programming more accessible by obviating the need for
programmers to write lenses by hand.
We begin by briefly reviewing some background on regular expressions and core
lens combinators (\S\ref{preliminaries}).
After we motivate our problem
with an extended, real-world example (\S\ref{overview}),
we offer the following technical contributions:

\begin{itemize}
\item 
  We introduce a new lens language (DNF Lenses) that is suitable for synthesis
  (\S\ref{sec:dnfre} and \S\ref{sec:dnf}).  
  We show how to convert ordinary regular expressions and lenses
  into the corresponding DNF forms, and we prove that DNF lenses are sound
  and complete with respect to the high-level bijective lens syntax.

\item
  We present an efficient, type-directed synthesis algorithm for
  synthesizing lenses (\S\ref{algorithm}).  We prove that if there is a lens
  that satisfies the input specification, this algorithm will return such a lens.

\item
  We evaluate \Optician{}, its optimizations, and
  existing synthesis tools on 39 benchmarks, including
  examples derived from Flash Fill~\cite{gulwani-popl-2014} and the
  Augeas~\cite{augeas} system (\S \ref{evaluation}).  
  We show that our optimizations are critical for synthesizing many of
  the complex lenses in our benchmark suite and that our full
  algorithm succeeds on all benchmarks in under 5
  seconds.  

\item While we are not aware of any other systems for automatically
 synthesizing bijective transformations, we establish a baseline for
 our the effectiveness of our techniques by comparing our synthesis algorithm
 with the one used in Flash Fill~\cite{gulwani-popl-2014}, a well-known
 and influential synthesis system deployed in Microsoft Excel.  Flash Fill only
 synthesizes transformations in one direction, but it was only able to
 complete synthesis of \FlashFillSuccesses{} out of 39 of our benchmarks.  We conjecture
 that the extra information we supply the synthesis system via our 
 regular format descriptions, allows it to scale to significantly more
 complex and varied formats than is possible in 
 current string synthesis systems that do not use this information.
\end{itemize} We close with related work
(\S \ref{sec:related}) and conclusions (\S \ref{sec:conc}).

% end introduction

% begin preliminaries
\section{Preliminaries}
\label{preliminaries}

\paragraph*{Technical Report} Throughout the paper, we will state a number of
theorems.  We have omitted these proofs for space, and have included these
details in the auxiliary technical report.

\SUBSECTION{Regular Expressions}
% \label{sec:regexp}

We use $\Sigma$ to denote the alphabet of individual characters $\Character$;
strings $\String$ and $\StringAlt$ are
elements of $\StarOf{\Sigma}$.  Regular expressions, abbreviated REs, are
used to express \emph{languages}, which are subsets of $\StarOf{\Sigma}$.
REs over $\Sigma$ are:
\begin{center}
  \Regex{},\RegexAlt{} \hspace{1em} \GEq{}
  \hspace{1em} $\String$ \hspace{1em} 
  \GBar{} \hspace{1em} $\emptyset$ \hspace{1em} 
  \GBar{} \hspace{1em} $\StarOf{\Regex}$ \hspace{1em} 
  \GBar{} \hspace{1em} $\RegexConcat{\Regex_1}{\Regex_2}$ \hspace{1em} 
  \GBar{} \hspace{1em} $\RegexOr{\Regex_1}{\Regex_2}$
\end{center}
$\LanguageOf{\Regex}\subseteq \StarOf{\Sigma}$, the language of $\Regex$, is
defined as usual. 

\paragraph*{\iffull Regular Expression \fi Unambiguity}

%These regular expressions are used to inductively define languages.
%\[
%  \begin{array}{lcl}
%    \LanguageOf{\String} &=& \{\String\}\\
%    \LanguageOf{\emptyset} &=& \{\}\\
%    \LanguageOf{\RegexConcat{\Regex_1}{\Regex_2}} &=&
%                                                      \{\StringConcat{\String_1}{\String_2} \SuchThat
%                                                      \String_1\in\LanguageOf{\Regex_1} \BooleanAnd \String_2\in\LanguageOf{\Regex_2}\}\\
%    \LanguageOf{\RegexOr{\Regex_1}{\Regex_2}} &=&
%                                                  \{\String \SuchThat
%                                                  \String\in\LanguageOf{\Regex_1} \BooleanOr \String\in\LanguageOf{\Regex_2}\}\\
%    \LanguageOf{\StarOf{\Regex}} &=&
%                                     \{\String_1\Concat\ldots\Concat\String_n \SuchThat
%                                     n\in\Nats \wedge \String_i\in\LanguageOf{\Regex}\}
%  \end{array}
%\]

The typing derivations of lenses require regular expressions to be written in a
way that parses text unambiguously. $\Regex$ and $\RegexAlt$ are
\textit{unambiguously concatenable}, written
$\Regex\UnambigConcat\RegexAlt$ if, 
for all strings $\String_1,\String_2\in\LanguageOf{\Regex}$ and
$\StringAlt_1,\StringAlt_2\in\LanguageOf{\RegexAlt}$, whenever
$\String_1\Concat\StringAlt_1=\String_2\Concat\StringAlt_2$ it is the case that
$\String_1=\String_2$ and $\StringAlt_1=\StringAlt_2$.
Similarly, $\Regex$ is \textit{unambiguously iterable},
written $\UnambigItOf{\Regex}$ if,
for all $n,m\in\Nats$ and for all strings
$\String_1,\ldots,\String_n,\StringAlt_1,\ldots,\StringAlt_m\in\LanguageOf{\Regex}$,
whenever
$\String_1\Concat\ldots\Concat\String_n=\StringAlt_1\Concat\ldots\Concat\StringAlt_m$
it is the case that $n=m$ and $\String_i=\StringAlt_i$ for all $i$.

A regular expression $\Regex$ is \textit{strongly unambiguous} if one
of the following holds: 
\begin{enumerate*}[label=(\alph*) ]
\item $\Regex = \String$,  or
\item $\LanguageOf{\Regex} = \SetOf{}$, or 
\item $\Regex = \Regex_1 \Concat \Regex_2$ with $\Regex_1 \UnambigConcat
  \Regex_2$, or
\item $\Regex = \Regex_1 \Or \Regex_2$ with $\Regex_1 \Intersect \Regex_2 =
  \emptyset$, 
or
\item $\Regex = \StarOf{(\Regex')}$ with $\UnambigItOf{(\Regex')}$.
\end{enumerate*}
\iffull Moreover, i\else I\fi{}n the recursive cases, $\Regex_1$, $\Regex_2$, and $\Regex'$
must also be strongly unambiguous.

% fig:regex-equivalence-rules

% fig:definitional-equivalence-rules
\paragraph*{\iffull Regular Expression \fi Equivalences}
\Regex{} and \RegexAlt{} are \textit{equivalent}, written $\Regex\equiv\RegexAlt$, if 
$\LanguageOf{\Regex}=\LanguageOf{\RegexAlt}$.
There exists an equational theory for determining whether two regular expressions
are equivalent, presented by Conway \cite{conway}, and proven complete by Krob
\cite{Krob}.\footnote{There are other complete axiomatizations for
  regular expression equivalence, such as Kozen's\cite{kozen-complete} and Salomaa's\cite{salomaa-complete}.
  We focus on Conway's for the sake of specificity, but discuss alternative
  theories in \S\ref{sec:related}.} 
Conway's axioms consist of the semiring axioms (associativity,
commutativity, identities, and distributivity for $|$ and $\cdot$) plus the following 
rules for equivalences involving the Kleene star:
\begin{center}
\begin{tabular}{@{}r@{\hspace{1em}}c@{\hspace{1em}}l@{\hspace{1em}}r@{}}
  \StarOf{(\RegexOr{\Regex{}}{\RegexAlt{}})} & $\equiv$ & \RegexConcat{\StarOf{(\RegexConcat{\StarOf{\Regex{}}}{\RegexAlt{}})}}{\StarOf{\Regex{}}} & \SumstarRule{}\\
  \StarOf{(\RegexConcat{\Regex{}}{\RegexAlt{}})} & $\equiv$ & \RegexOr{\EmptyString{}}{(\RegexConcat{\RegexConcat{\Regex{}}{\StarOf{(\RegexConcat{\RegexAlt{}}{\Regex{}})}}}{\RegexAlt{}})} & \ProductstarRule{} \\
  ${(\Regex{}^*)}^*$ & $\equiv$ & \StarOf{\Regex{}} & \StarstarRule{} \\
  \StarOf{(\RegexOr{\Regex}{\RegexAlt})} & $\equiv$ & $\StarOf{(\RegexConcat{(\RegexOr{\Regex}{\RegexAlt})}{\RegexOr{\RegexAlt}{\RegexConcat{{(\RegexConcat{\Regex}{\StarOf{\RegexAlt}})}^n}{\Regex}}})}\Concat$
                                             $(\EmptyString\Or(\RegexOr{\Regex}{\RegexAlt})\Concat$
                                             $({(\RegexConcat{\Regex}{\StarOf{\RegexAlt}})}^0\Or\ldots\Or{(\RegexConcat{\Regex}{\StarOf{\RegexAlt}})}^n))$ & \DicyclicityRule{}
\end{tabular}
\end{center}

While this equational theory is complete, na\"{i}vely using it in the
context of lens synthesis presents several problems.  
In the context of lens synthesis, we instead use the equational theory
corresponding to the axioms of a \textit{star semiring}~\cite{Droste2009}.
If two regular expressions are equivalent within this equational theory,
they are \emph{star semiring equivalent}, written $\Regex \SSREquiv \RegexAlt$.
The star semiring axioms consist of the semiring axioms plus the following
rules for equivalences involving the Kleene star:
\begin{center}
\begin{tabular}{@{}r@{\hspace{1em}}c@{\hspace{1em}}l@{\hspace{1em}}r@{}}
  \StarOf{\Regex} & $\SSREquiv$ & \EmptyString \Or (\Regex \Concat \StarOf{\Regex}) & \UnrollstarLeftRule{}\\
  \StarOf{\Regex} & $\SSREquiv$ & \EmptyString \Or (\StarOf{\Regex} \Concat \Regex) & \UnrollstarRightRule{}\\
\end{tabular}
\end{center}
In \S\ref{overview}, we provide intuition for why synthesis with full regular
expression equivalence is problematic and justify our choice of using star
semiring equivalence instead.

\SUBSECTION{Bijective Lenses}

% fig:lens-semantics

All bijections between languages are lenses.  We define \emph{bijective lenses} to be
bijections created from the following Boomerang lens combinators, $\Lens$.
\begin{center}
  \begin{tabular}{l@{\ }l@{\ }l@{\ }l}
    % REGEX
     \Lens{} & \GEq{} & & $\ConstLensOf{s_1 \in \StarOf{\Alphabet}}{s_2 \in \StarOf{\Alphabet}}$ \\
            & & \GBar{} & $\IterateLensOf{\Lens}$ \\
            & & \GBar{} & $\ConcatLensOf{\Lens_1}{\Lens_2}$ \\
            & & \GBar{} & $\SwapLensOf{\Lens_1}{\Lens_2}$ \\
            & & \GBar{} & $\OrLensOf{\Lens_1}{\Lens_2}$ \\
            & & \GBar{} & $\ComposeLensOf{\Lens_1}{\Lens_2}$ \\
            & & \GBar{} & $\IdentityLensOf{\Regex}$ 
  \end{tabular}
\end{center}

\noindent The denotation of a lens $\Lens$ is  $\SemanticsOf{\Lens}
\subseteq \StringType \Cross \StringType$.  If $(\String_1,\String_2) \in
\SemanticsOf{\Lens}$, then $\Lens$ \emph{maps between} $\String_1$ and
$\String_2$.
\begin{center}
  \begin{tabular}{@{}rcl}
    $\SemanticsOf{const(\String_1,\String_2)}$ &=& $\SetOf{(\String_1,\String_2)}$\\
    
    $\SemanticsOf{\IterateLensOf{\Lens}}$ &=& $\SetOf{(\String_1\Concat\ldots\Concat\String_n,
                                              \StringAlt_1\Concat\ldots\Concat\StringAlt_n)\SuchThat
                                              n \in \Nats \BooleanAnd
                                              \forall i \in \RangeIncInc{1}{n},
                                              (\String_i,\StringAlt_i)\in\SemanticsOf{\Lens}}$\\
    
    $\SemanticsOf{\ConcatLensOf{\Lens_1}{\Lens_2}}$ &=&
                                                      $\SetOf{(\String_1\Concat\String_2,\StringAlt_1\Concat\StringAlt_2)\SuchThat
                                             (\String_1,\StringAlt_1)\in\SemanticsOf{\Lens_1}\BooleanAnd
                                                 (\String_2,\StringAlt_2)\in\SemanticsOf{\Lens_2}}$\\
    
    $\SemanticsOf{\SwapLensOf{\Lens_1}{\Lens_2}}$ &=&
                                                    $\SetOf{(\String_1\Concat\String_2,\StringAlt_2\Concat\StringAlt_1)\SuchThat
                                             (\String_1,\StringAlt_1)\in\SemanticsOf{\Lens_1}\BooleanAnd
                                                 (\String_2,\StringAlt_2)\in\SemanticsOf{\Lens_2}}$\\
    
    $\SemanticsOf{\OrLensOf{\Lens_1}{\Lens_2}}$ &=&
                                                  $\SetOf{(\String,\StringAlt)
                                                  \SuchThat(\String,\StringAlt)\in\SemanticsOf{\Lens_1}
                                                  \BooleanOr(\String,\StringAlt)\in\SemanticsOf{\Lens_2}}$\\
    
    $\SemanticsOf{\ComposeLensOf{\Lens_1}{\Lens_2}}$ &=&
                                                       $\SetOf{(\String_1,\String_3)\SuchThat\exists\String_2
                                             (\String_1,\String_2)\in\SemanticsOf{\Lens_1}\BooleanAnd
                                                 (\String_2,\String_3)\in\SemanticsOf{\Lens_2}}$\\
    
    $\SemanticsOf{\IdentityLensOf{\Regex}}$ &=& $\SetOf{(\String,\String)
                                              \SuchThat \String\in
                                              \LanguageOf{\Regex}}$
  \end{tabular}
\end{center}

The simplest lens in the combinator language is the constant lens between
strings $\String$, and $\StringAlt$, $\ConstLensOf{\String}{\StringAlt}$.
The lens $\ConstLensOf{\String}{\StringAlt}$, when operated left-to-right,
replaces the string $\String$ with $\StringAlt$, and when operated
right-to-left, replaces string $\StringAlt$ with $\String$.  The identity lens
on a regular expression, $\IdentityLensOf{\Regex}$, operates in both
directions by applying the
identity function to strings in $\LanguageOf{\Regex}$.  The composition
combinator, 
$\ComposeLensOf{\Lens_1}{\Lens_2}$, operates by applying $\Lens_1$ then $\Lens_2$
when operating left to right, and applying $\Lens_2$ then $\Lens_1$ when
operating right to left.

Each of the other lenses manipulates structured data.  For instance,
$\ConcatLensOf{\Lens_1}{\Lens_2}$ operates by applying $\Lens_1$ to the left
portion of a string, and $\Lens_2$ to the right, and concatenating the results.
The combinator
$\SwapLensOf{\Lens_1}{\Lens_2}$ does the same as $\ConcatLensOf{\Lens_1}{\Lens_2}$
but it swaps the results before concatenating.
The combinator $\OrLensOf{\Lens_1}{\Lens_2}$ operates by applying either $\Lens_1$ or $\Lens_2$
to the string.  The combinator $\IterateLensOf{\Lens}$ operates by repeatedly
applying $\Lens$ to subparts of a string.

\paragraph{Lens Typing}

\begin{figure}
  \centering% \footnotesize
  \begin{mathpar}
    \inferrule[]% \ConstantLensRule{}
    {
      \String_1 \in \StarOf{\Sigma}\\
      \String_2 \in \StarOf{\Sigma}
    }
    {
      \ConstLensOf{\String_1}{\String_2} \OfType \String_1 \Leftrightarrow \String_2
    }

    \inferrule[]
    {
      \Lens \OfType \Regex \Leftrightarrow \RegexAlt\\
      \UnambigItOf{\Regex}\\
      \UnambigItOf{\RegexAlt}
    }
    {
      \IterateLensOf{\Lens} \OfType \StarOf{\Regex} \Leftrightarrow \StarOf{\RegexAlt}
    }

    \inferrule[]%\ConcatLensRule{}
    {
      \Lens_1 \OfType \Regex_1 \Leftrightarrow \RegexAlt_1\\\\
      \Lens_2 \OfType \Regex_2 \Leftrightarrow \RegexAlt_2\\\\
      \UnambigConcatOf{\Regex_1}{\Regex_2}\\
      \UnambigConcatOf{\RegexAlt_1}{\RegexAlt_2}
    }
    {
      \ConcatLensOf{\Lens_1}{\Lens_2} \OfType \Regex_1\Regex_2 \Leftrightarrow \RegexAlt_1\RegexAlt_2
    }

    \inferrule[]%\SwapLensRule{}
    {
      \Lens_1 \OfType \Regex_1 \Leftrightarrow \RegexAlt_1\\\\
      \Lens_2 \OfType \Regex_2 \Leftrightarrow \RegexAlt_2\\\\
      \UnambigConcatOf{\Regex_1}{\Regex_2}\\
      \UnambigConcatOf{\RegexAlt_2}{\RegexAlt_1}
    }
    {
      \SwapLensOf{\Lens_1}{\Lens_2} \OfType \Regex_1\Regex_2 \Leftrightarrow \RegexAlt_2\RegexAlt_1
    }
    
    \inferrule[]%\OrLensRule{}
    {
      \Lens_1 \OfType \Regex_1 \Leftrightarrow \RegexAlt_1\\
      \Lens_2 \OfType \Regex_2 \Leftrightarrow \RegexAlt_2\\\\
      \UnambigOrOf{\Regex_1}{\Regex_2}\\
      \UnambigOrOf{\RegexAlt_1}{\RegexAlt_2}
    }
    {
      \OrLensOf{\Lens_1}{\Lens_2} \OfType
      \Regex_1 \Or \Regex_2
      \Leftrightarrow \RegexAlt_1 \Or \RegexAlt_2
    }
    
    \inferrule[]%\ComposeLensRule{}
    {
      \Lens_1 \OfType \Regex_1 \Leftrightarrow \Regex_2\\
      \Lens_2 \OfType \Regex_2 \Leftrightarrow \Regex_3\\
    }
    {
      \ComposeLensOf{\Lens_1}{\Lens_2} \OfType \Regex_1 \Leftrightarrow \Regex_3
    }

    \inferrule[]%\IdentityLensRule{}
    {
      \Regex \text{ is strongly unambiguous}
    }
    {
      \IdentityLensOf{\Regex} \OfType \Regex \Leftrightarrow \Regex
    }

    \inferrule[]%\RewriteRegexLensRule{}
    {
      \Lens \OfType \Regex_1 \Leftrightarrow \Regex_2\\
      \Regex_1 \SSREquiv \Regex_1'\\
      \Regex_2 \SSREquiv \Regex_2'
    }
    {
      \Lens \OfType \Regex_1' \Leftrightarrow \Regex_2'
    }
  \end{mathpar}

  \caption{Lens Typing Rules}
  \label{fig:lens-typing}
\end{figure}

The typing judgement for lenses has the form $\Lens \OfType \Regex
\Leftrightarrow \RegexAlt$, meaning $\Lens$ bijectively maps between
$\LanguageOf{\Regex}$ and $\LanguageOf{\RegexAlt}$.   
Figure~\ref{fig:lens-typing} gives the typing relation.  Many of the
typing derivations require side conditions about
unambiguity.  These side conditions guarantee that the semantics of the language
create a bijective function.  For example, if $\Lens_1 \OfType \Regex_1
\Leftrightarrow \RegexAlt_1$, and $\Lens_2 \OfType \Regex_2 \Leftrightarrow
\RegexAlt_2$, and $\Regex_1$ is not unambiguously concatenable with $\Regex_2$,
then there would exist $\String_1,\String_1'\in\LanguageOf{\Regex_1}$, and
$\String_2,\String_2'\in\LanguageOf{\Regex_2}$ where
$\String_1\Concat\String_2 = \String_1'\Concat\String_2'$, but $\String_1
\neq \String_1'$, and $\String_2 \neq \String_2'$.  The lens
$\ConcatLensOf{\Lens_1}{\Lens_2}$ would no longer necessarily act as a function
when applied from left to right, as
$\Lens_1$ could be applied to both $\String_1$ and to $\String_1'$.
Because any ambiguous RE can be replaced by an equivalent unambiguous one, these
ambiguity constraints do not have an impact on the computational power
of the language.

The typing rule for $\IdentityLensOf{\Regex}$ requires a strongly unambiguous
regular expression.  This unambiguity allows the identity lens to be
derivable from other lenses.\footnote{In practice, we allow regular
expressions that aren't strongly unambiguous to appear in $\IdentityLensOf{\Regex}$, provided that they are expressed as a user defined regular expression.
We elide such user-defined regular expression information
from the theory for the sake of simplicity.}  This requirement does not,
however, reduce expressiveness, as any regular
expression is equivalent to a strongly unambiguous regular
expression~\cite{unambigregex}.

The last rule in Figure~\ref{fig:lens-typing} is a type equivalence rule
that lets the typing rules consider a lens
$\Lens \OfType \Regex_1 \Leftrightarrow \Regex_2$ to have type
$\Regex_1' \Leftrightarrow \Regex_2'$ so long as
$\Regex_1 \SSREquiv \Regex_1$ and
$\Regex_2 \SSREquiv \Regex_2'$.  Notice that this rule uses star semiring equivalence
as opposed to Conway equivalence.  In theory, this reduces the expressiveness of the
type system; in practice, we have not found it restrictive.  We explain and
justify the decision to use star semiring equivalence in the next section.

%% Ideally, this rule would use
%% the full power of the Conway axiomatization $\equiv$, since that
%% theory is complete for regular expression equivalence.  However, for
%% the purposes of our synthesis algorithm we use a finer, but more
%% tractable notion of equivalence $\SSREquiv \subseteq \equiv$.
%% We have not found this restriction to limit
%% the expressivity of lenses in practice.

Finally, it is worthwhile at this point to notice that the problem of
finding a well-typed lens $\Lens$ given a pair of regular
expressions---the lens synthesis problem---would not be difficult if
it were not for lens composition and the type equivalence rules.  When
read bottom up, these two rules apparently require wild guesses at
additional regular expressions to continue driving synthesis
recursively in a type-directed fashion.  In contrast, in the other
rules, the shape of the lens is largely determined by the given types.
The following sections elaborate on this problem and describe our
solution.

% In this subset, the full generality of \ProductstarRule{} is not present,
% \DicyclicityRule{} is not present, \StarstarRule{} is not present,
% and \SumstarRule{} is not present.  \StarstarRule{} is a rule that introduces
% ambiguity, so removing it no longer allows ambiguity to be introduced.
% The removal of \SumstarRule{} no longer allows for different choices of an
% interation to be grouped by when they occur.  \ProductstarRule{} is similar to
% \UnrollstarLeftRule{} and \UnrollstarRightRule, though different parts of a
% regular expression can be unrolled
% on different sides of the star, altering the ordering of the regular expression
% under the star.
% Removing \DicyclicityRule{} removes the ability to split the repetition of the
% star into a quotiented part, and a remainder.

% The primary difference between the regular expression equivalence rules and the
% definitional equivalence rules is that the regular expression equivalence
% rules allow for breaking up parts of the regular expression under a star by
% grouping together different parts of the repetition, whereas the definitional
% equivalence rules breaks parts of a regular expression under a star up only by
% how many times the already defined repetition occurs.

% end preliminaries

% begin overview
\section{Overview}
\label{overview}

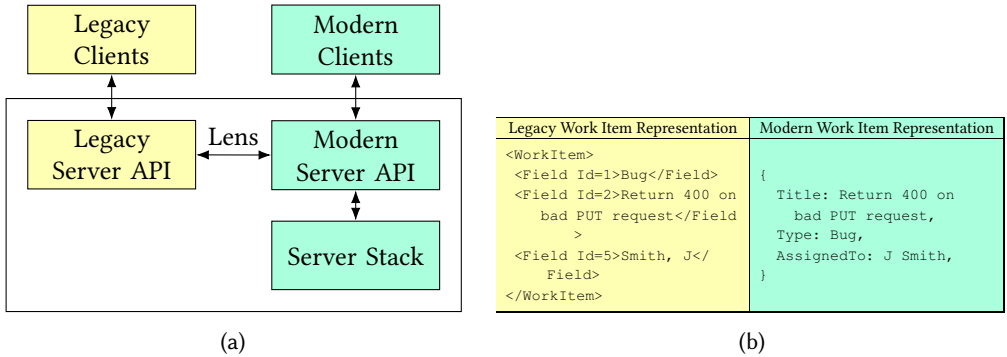
\begin{figure}
  \centering
  \begin{subfigure}[b]{.49\textwidth}
    \centering
    \begin{tikzpicture}[auto,node distance=1.5cm]
      % clients
      \node[align=center,draw,rectangle,minimum width=2cm,minimum height = .9cm, text width=2cm, fill=lightyellow] (old-clients) {Legacy Clients};
      \node[text width=2cm,align=center,draw,rectangle,minimum height = .9cm, text width=2cm, fill=lightgreen] (new-clients) [right = 1cm of old-clients] {Modern Clients};

      % server apis
      \node[text width=2cm,align=center,draw,rectangle,minimum height = .9cm, fill=lightyellow] (old-server-api) [below = .6cm of old-clients] {Legacy Server API};
      \node[text width=2cm,align=center,draw,rectangle,minimum height = .9cm, fill=lightgreen] (new-server-api) [below = .6cm of new-clients] {Modern Server API};

      % server stack
      \node[text width=2cm,align=center,draw,rectangle,minimum height = .9cm, fill=lightgreen] (server-stack) [below = .4cm of new-server-api] {Server Stack};
      
      % Draw an edge between rel1 and node1; rel1 and node2
      \path[<->] (old-clients) edge node {} (old-server-api);
      \path[<->] (new-clients) edge node {} (new-server-api);
      \path[<->] (new-server-api) edge node {} (server-stack);
      \path[<->] (old-server-api) edge node {Lens} (new-server-api);

      % Surrounding box
      \node[fit=(old-server-api) (new-server-api) (server-stack) ,draw, inner sep=.8em] (surrounding) {};
    \end{tikzpicture}
    \caption{}
    \label{subfig:vsts-arch}
  \end{subfigure}
  \begin{subfigure}[b]{.49\textwidth}
    % https://tex.stackexchange.com/questions/168741/showing-two-listings-in-a-table-side-by-side
    \setbox0=\hbox{%
      \begin{minipage}{1.95in}
\begin{lstlisting}[
tabsize=1,
language={xml},
]
<WorkItem>
 <Field Id=1>Bug</Field>
 <Field Id=2>Return 400 on
    bad PUT request</Field>
 <Field Id=5>Smith, J</Field>
</WorkItem>
\end{lstlisting}
      \end{minipage}
    }
    \savestack{\listingA}{\box0}

    \setbox0=\hbox{%
      \begin{minipage}{1.95in}
\begin{lstlisting}[
tabsize=1,
]
{
  Title: Return 400 on
    bad PUT request,
  Type: Bug,
  AssignedTo: J Smith,
}
\end{lstlisting}
      \end{minipage}
    }
    \savestack{\listingB}{\box0}
    \centering
    \resizebox{\columnwidth}{!}{%
      \begin{tabular}{|>{\columncolor{lightyellow}}c|>{\columncolor{lightgreen}}c|}
        \hline
        Legacy Work Item Representation & Modern Work Item Representation \\
        \hline
        \listingA & \listingB \\
        \hline
      \end{tabular}
    }
    \caption{}
    \label{subfig:idealized-representations}
  \end{subfigure}
  \caption{VSTS Architecture Using Lenses.
    In (a), %how do I make it ref correctly, doing a ref makes it appear as 2a
    we show the proposed architecture of VSTS using lenses.  When a legacy
    client requests a work item, the server retrieves the data in a modern format
    through the new APIs, then the lens converts it into a legacy format to
    return to the client.   When a legacy client updates a work item, it
    provides the data in the legacy format to the server.
    The lens then converts this data into the modern format for the new endpoints to
    process.
    Idealized Task representations from legacy and modern web service endpoints
    are given in (b).}
  \label{fig:vsts-with-lenses}
\end{figure}

To highlight the difficulties in synthesizing lenses, we use an extended
example inspired by the evolution of Microsoft's Visual Studio Team
Services (VSTS).  VSTS is 
a collection of web services for
team management -- providing a unified location for source control (e.g. a Git
server), task management (i.e. providing a means to keep track of
TODOs and bugs), and more.
In 2014, to increase third party developer interaction, VSTS
released new web service endpoints~\cite{bharry}.  However, despite VSTS
introducing new, modernized web APIs, they must still maintain the old, legacy
web APIs for continued support of legacy
clients~\cite{team-explorer}.  Instead of maintaining server code for each endpoint,
we envision an architecture that uses a lens to convert resources of
the old form into resources of the new form and vice versa, as shown in
Figure~\ref{subfig:vsts-arch}.  Writing each of these converters by hand is slow
and error prone.  We speed up this process by only requiring users to
input regular expressions and input-output examples.  Furthermore, the generated
lenses are guaranteed 
to map between the provided regular expressions and to act correctly on the
provided examples. 

Consider a ``Work Item,'' a resource that represents a task given to a
team.  Idealized versions of the representations of work items from the new and
old APIs are given in Figure~\ref{subfig:idealized-representations}.  In our
proposed architecture, if an old
client performs an HTTP GET request to receive a work item, the server first
retrieves that work item using the modern API, and then uses the
lens's \PutLeft{} function to convert this task into the legacy format.  Similarly,
if an old client performs an HTTP PUT request to update a work item, the server
first uses the lens's 
\PutRight{} function to convert that data into the
new format, and then inputs the work item in the new representation to the modern APIs.  The two
representations contain the same information, but they are presented
differently.

For simplicity, let's consider only finding the mapping between the ``Title''
field of the task in the legacy and modern formats.
The legacy client accepts inputs of the form 
\begin{lstlisting}
   !legacy_title! = @"<Field Id=2>"@ !text_char!* @"</Field>"@
\end{lstlisting}
while the modern client accepts inputs of the form
\begin{lstlisting}
   !modern_title! = (@"Title:"@ !text_char!* !text_char! @","@)
                 | @""@
\end{lstlisting}
where \TextChar{} is a user-defined data type representing what characters can
be present in a text field (like the title field).
We would like to be able to synthesize $\Lens$, a lens that satisfies the
typing judgement $\Lens \OfType \LegacyTitle{} \Leftrightarrow \ModernTitle$
(i.e.\ $\Lens$ maps between the legacy representation \LegacyTitle{}, and the modern
representation \ModernTitle{}).
Because the modern API omits the title field if it is blank, the lens must
perform different 
actions depending on the number of text characters present,
functionality provided by \OrLens{} lenses.
An \OrLens{} lens applies one of two lenses, depending on which of the lenses'
source types matches the input string.

However, the typing rule for \OrLens{} does not suffice to type check lenses
that map between \LegacyTitle{} and \ModernTitle{}.  While
\CF{\color{darkbrown} modern\_title} is a regular expression with an
outermost disjunction, \CF{\color{darkbrown} legacy\_title}
is a regular expression with an outermost concatenation, so the rule cannot be
directly applied.
We address this problem by allowing conversions between equivalent
regular expression types with the type equivalence rule.
Using this rule, a type-directed synthesis algorithm can convert
\LegacyTitle{} into
\begin{lstlisting}
   !legacy_title'! =   @"<Field Id=2></Field>"@
                  | (@"<Field Id=2>"@ !text_char! !text_char!* @"</Field>"@)
\end{lstlisting}
There exist \OrLens{} lenses between \LegacyTitle{}' and \ModernTitle{}, and the two cases of an empty
and a nonempty number of text characters can be handled separately.  However,
the need to find this equivalent type highlights a significant challenge in
synthesizing bijective lenses.

\paragraph*{Challenge 1: Multi-dimensional Search Space.}
Since regular expression equivalence is decidable, 
it is easy to \emph{check} whether a given lens $\Lens$ with type 
$\Regex_1 \Leftrightarrow \Regex_2$
also has type 
$\Regex_1' \Leftrightarrow \Regex_2'$. 
During synthesis, however, deciding when and how to use type conversion is
difficult because there are infinitely many regular expressions that are
equivalent to the source and target regular expressions.  Does the algorithm
need to consider all of them?  In what order?  To convert from
\CF{\color{darkbrown} legacy\_title} to \CF{\color{darkbrown} legacy\_title'},
the algorithm must first unroll \CF{{\color{darkbrown} text\_char}*} into
\CF{{\color{blue} ""} | {\color{darkbrown} text\_char} {\color{darkbrown}
    text\_char}*}, and then it must
distribute this disjunction on the left and the right.

A related challenge arises from the composition operator,
$\ComposeLensOf{\Lens_1}{\Lens_2}$.  The typing rule for composition
requires that the target type of $l_1$ be the source
type of $l_2$.  To synthesize a composition lens between
$\Regex_1$ and $\Regex_3$, a sound synthesizer must find an intermediate type
$\Regex_2$ and lenses with types $\Regex_1 \Leftrightarrow \Regex_2$ and
$\Regex_2 \Leftrightarrow \Regex_3$.  Searching for the correct
regular expression $\Regex_2$ is again problematic because
the search space is infinite.

%Furthermore, Boomerang has a composition operator to compose multiple
%lenses.Removing the composition operator from Boomerang reduces
%expressivity.  That means, to have a synthesis procedure that can
%synthesize all lenses, the procedure must synthesize composition lenses.

Thus, na\"{i}vely applying type-directed synthesis techniques
involves searching \emph{in three infinite dimensions}.  A complete na\"{i}ve
synthesizer must search for
(1) a \emph{type} consisting of two regular expressions equivalent to the
given ones but with ``similar shapes'' and
(2) a lens \emph{expression} that has the given type and is consistent
with the user's examples.
Furthermore, whenever composition is part of the
expression, na\"{i}ve type-directed synthesis requires a further search for
(3) an {\em intermediate regular expression}.

Our approach to this challenge is to define a new ``DNF syntax'' for types
and lenses that reduces the synthesis search space \emph{in all dimensions}.
In this new language, regular expressions are written in a disjunctive normal
form, where disjunctions are fully distributed over concatenation and where binary
operators are replaced by $n$-ary ones, eliminating associativity rules.
Using DNF regular expressions, when presented with a synthesis problem with type
$(A | B) C \Leftrightarrow A' C' | B' C'$, \Optician{} will first convert this
type into $\DNFOf{\SequenceOf{A \SeqSep C} \DNFSep \SequenceOf{B
    \SeqSep C}} \Leftrightarrow
\DNFOf{\SequenceOf{A' \SeqSep C'} \DNFSep \SequenceOf{B' \SeqSep C'}}$, where
$\DNFOf{\ldots}$ represents $n$-ary disjunction and $\SequenceOf{\ldots}$
represents $n$-ary concatenation.
Like DNF regular expressions, DNF lenses are stratified, with disjunctions
outside of concatenations, and they use $n$-ary operators instead of binary
ones.  Furthermore, DNF lenses do not need a composition operator, eliminating
an entire dimension of search.  This stratification and the lack of
composition creates a very tight relationship between the structure of a well-typed
DNF lens and its DNF regular expression types.

Translating regular expressions into DNF form collapses many equivalent REs
into the same syntactic form.  However, this translation does not fully
normalize regular expressions.
Nor do we want it to: If a synthesizer normalized $\epsilon \Or BB*$ to $B*$, it
would have trouble synthesizing lenses with types like $\EmptyString \Or BB*
\Leftrightarrow \EmptyString \Or CD*$
where the first occurrence of $B$ on the left needs to be transformed into
$C$ while the rest of the $B$s need to be transformed into $D$.
Normalization to DNF eliminates many, but not all, of the regular expression
equivalences that may be needed before a simple, type-directed structural search
can be applied---i.e., DNF regular expressions are only
\emph{pseudo-canonical}.

\begin{figure}
  \centering
  \begin{tikzpicture}[auto,node distance=1.5cm]
    \node[text width=2.3cm,minimum height=.6cm,align=center,draw,rectangle] (typeprop) {\TypeProp{}};
    
    \node[align=right, anchor=east] (dnfregex1) [left = 1cm of typeprop.north west]{\DNFRegex{}};
    \node[align=right, anchor=east] (dnfregex2) [left = 1cm of typeprop.south west]{\DNFRegexAlt{}};
    
    \node[align=right, anchor=east] (exs) [below = .2cm of dnfregex2]{ \Examples{} };
    % Now place a relation (ID=rel1)
    \node (below-gen) [below=.5cm of typeprop] {};
    \node[align=center] (dnfregex'1) [right = .4cm of typeprop.north east]{\DNFRegex{}'};
    \node[align=center] (dnfregex'2) [right = .4cm of typeprop.south east]{\DNFRegexAlt{}'};
    \node[text width=2.3cm,minimum height=.6cm,align=center,draw,rectangle] (rigidsynth) [right = 1.55cm of typeprop] {\RigidSynth{}};
    \node (below-compl) [below=.5cm of rigidsynth] {};
    \node[align=center] (dnflens) [right = 1.1cm of rigidsynth.south east] {\DNFLens{}}; 
    % Draw an edge between rel1 and node1; rel1 and node2
    
    \path[->] (dnfregex1.east) edge (typeprop.north west);
    \path[->] (dnfregex2.east) edge (typeprop.south west);
    \path[->] (typeprop.north east) edge (dnfregex'1);
    \path[->] (typeprop.south east) edge (dnfregex'2);
    \path[->] (dnfregex'1) edge (rigidsynth.north west);
    \path[->] (dnfregex'2) edge (rigidsynth.south west);
    \path[->] (rigidsynth.south east) edge  (dnflens);
    \draw[->] ($(exs.east)+(-3pt,0)$) -| node(exsedge) {} (rigidsynth);
    \draw[->] (rigidsynth.north east) -| node[above right, near start](failure){\Failure{}} +(0,.5) -|  (typeprop.north);

    \node (success) [below = .65cm of failure] {\Success{}};

    \node (synth-name) [above=1cm of dnfregex'1] {\SynthDNFLens{}};
    
    \node[fit={(typeprop) ($(typeprop.west)+(-.5cm,0)$) (exsedge) (synth-name) (failure) (success)} ,draw] (surrounding) {};
  \end{tikzpicture}
  \caption{Schematic Diagram for DNF Lens Synthesis Algorithm.  DNF regular
    expressions, $\DNFRegex$ and $\DNFRegexAlt$, and a set of examples
    $\Examples$ are given as input.  The synthesizer, \TypeProp, uses these
    input DNF regular expressions to propose a pair of equivalent DNF regular
    expressions, $\DNFRegex'$ and $\DNFRegexAlt'$.  The synthesizer
    \RigidSynth{} then attempts to generate a DNF lens, $\DNFLens$, which goes
    between $\DNFRegex'$ and $\DNFRegexAlt'$ and satisfies all the examples in
    $\Examples$.  If \RigidSynth{} is successful, $\DNFLens$
    is returned.  If 
    \RigidSynth{} is unsuccessful, information of the failure is returned to
    \TypeProp{}, which continues proposing candidate DNF regular expressions
    until \RigidSynth{} finds a satisfying DNF lens.}
  \label{fig:2-phased-schematic-diagram}
\end{figure}

Consequently, a type-directed synthesis algorithm must still search through
some equivalent regular expressions.
To handle this search, $\SynthDNFLens{}$ is structured as two
communicating synthesizers, shown in Figure~\ref{fig:2-phased-schematic-diagram}.
The first synthesizer, \TypeProp{}, proposes DNF regular expressions
equivalent to the input DNF regular expressions.
\TypeProp{} uses the axioms of a star semiring to unfold Kleene star
operators in one or both types, to obtain equivalent (but larger) DNF regular
expression types. The second synthesizer, \RigidSynth{}, performs a
syntax-directed search based on the structure of the provided DNF regular
expressions, as well as the input examples. If the second synthesizer finds a
satisfying DNF lens, it returns that lens.  If the second synthesizer fails to
find such a lens, \TypeProp{} learns of that failure,
and proposes new candidate DNF regular expression pairs.

\paragraph*{Star Semiring Equivalence and Rewriting}

One could try to search the space of DNF regular expressions equivalent to
the input regular expressions by turning the Conway axioms into (undirected)
rewrite rules operating on DNF regular expressions
and then trying all possible combinations of rewrites.  Doing so would
be problematic because the Conway axiomatization itself is both highly
nondeterministic and infinitely branching (due to the choice of $n$ in
the dyclicity axiom).

We also want DNF lenses to be closed under composition -- if it is not then we
need to be able to synthesize lenses containing composition operators.  To be
closed under composition, it is sufficient
for the equivalence relation used in the type equivalence rule
to be the equivalence closure of a rewrite system ($\Rewrite$) satisfying
four conditions.
First, if $\Regex \Rewrite \Regex'$, then $\LanguageOf{\Regex} =
\LanguageOf{\Regex'}$.
Second, if $\Regex \Rewrite \Regex'$ and $\Regex$ is strongly unambiguous, then
$\Regex'$ is also strongly unambiguous.
The remaining two properties relate the rewrite rules to the typing derivations
of DNF lenses, when those typing derivations do not use type
equivalence.
To express these properties, 
we use the notation
$\DNFLens \OfRewritelessType \DNFRegex \Leftrightarrow \DNFRegexAlt$
to mean that
if $\DNFLens$ is a DNF lens that
goes between DNF regular expressions $\DNFRegex$ and $\DNFRegexAlt$, then the
typing derivation contains no instances of the type equivalence rule.
Using this notation, we can express the \textit{confluence} property, as
follows:
\begin{definition}[Confluence]
  Whenever $\DNFLens_1 \OfRewritelessType (\ToDNFRegexOf{\Regex_1}) \Leftrightarrow
  (\ToDNFRegexOf{\RegexAlt_1})$,
  if $\Regex_1 \Rewrite
  \Regex_2$ and $\RegexAlt_1 \Rewrite \RegexAlt_2$, there exist regular
  expressions $\Regex_3$, and
  $\RegexAlt_3$ and a DNF lens $\DNFLens_3$, such that:
  \begin{enumerate}
  \item $\Regex_2 \Rewrite \Regex_3$
  \item $\RegexAlt_2 \Rewrite \RegexAlt_3$
  \item $\DNFLens_3 \OfRewritelessType (\ToDNFRegexOf{\Regex_3}) \Leftrightarrow
    (\ToDNFRegexOf{\RegexAlt_3})$
  \item $\SemanticsOf{\DNFLens_3} = \SemanticsOf{\DNFLens_1}$
  \end{enumerate}
\end{definition}
We call the final property \textit{bisimilarity}.  Bisimilarity requires two
symmetric conditions.
\begin{definition}[Bisimilarity]
  Whenever $\DNFLens_1 \OfRewritelessType (\ToDNFRegexOf{\Regex_1}) \Leftrightarrow
  (\ToDNFRegexOf{\RegexAlt_1})$ and 
  $\Regex_1 \Rewrite \Regex_2$, there exist a regular expression $\RegexAlt_2$ and a
  DNF lens $\DNFLens_2$ such that
  \begin{enumerate}
  \item $\RegexAlt_1 \Rewrite \RegexAlt_2$
  \item $\DNFLens_2 \OfRewritelessType (\ToDNFRegexOf{\Regex_2}) \Leftrightarrow
    (\ToDNFRegexOf{\RegexAlt_2})$
  \item $\SemanticsOf{\DNFLens_2} = \SemanticsOf{\DNFLens_1}$
  \end{enumerate}
  To be bisimilar, the symmetric property must also hold for $\RegexAlt_1 \Rewrite
  \RegexAlt_2$.
\end{definition}

Our solution for handling type equivalence is to use $\SSREquiv$, the equivalence relation
generated by the axioms of a star semiring.
This equivalence relation is compatible with our lens synthesis strategy, as
orienting these unrolling rules from left to right presents us with a rewrite
relation that is both confluent and bisimilar, and whose equivalence closure is
$\SSREquiv$.
The star semiring axioms are the 
coarsest subset of regular expression equivalences we could
find that is generated by a rewrite relation and is still confluent and
bisimilar.  We have not been able to prove that this relation is the coarsest
such relation possible, but it is sufficient to cover all the test
cases in our benchmark suite (see \S\ref{evaluation}).  However, it is easy to
show that Conway's axioms (\ProductstarRule{} in particular) are not bisimilar,
which is why we avoid this in our system.

\paragraph*{Challenge 2: Large Types}
DNF lenses are equivalent in expressivity to lenses and the algorithm
\SynthDNFLens{} is quite fast.
Unfortunately, the conversion to DNF incurs an exponential blowup.  In practical
examples, the regular expressions describing complex {\em ad hoc} data
formats may be very large, causing the exponential blowup to have a significant
impact on synthesis time.  The key to addressing this issue is to observe that
users naturally construct large types incrementally, introducing named
abbreviations for major subcomponents.  For example, in the
specification of \LegacyTitle{} and
\ModernTitle{}, the variable \TextChar{} describes which characters
can be present in a title.
To include a large disjunction representing all valid title characters
instead of the concise variable \TextChar{} in the definitions of \LegacyTitle{}
and \ModernTitle{} would be unmaintainable and difficult to read.

Unfortunately, leaving these variables opaque introduces a new dimension of
search.  In addition to searching through the rewrites on regular expressions,
the algorithm must also search through possible \emph{substitutions},
replacements of variables with their definitions.
We designate these two types of equivalences \emph{expansions}, using
``rewrites'' to denote expansions that arise from traversing rewrite rules on
the regular expressions, and
using ``substitutions'' to denote expansions that arise from replacing a
variable with its definition.

Interestingly, \Optician{} can exploit the structure inherent in these named
abbreviations to speed up the search dramatically.  For example, if
\TextChar{} appears just once in both the source and the target types, the system
hypothesizes that the identity lens can be used to convert between
occurrences of \TextChar{}.  On the other hand, if \TextChar{} appears in the
source but not in the target, the system recognizes that, to find a lens,
\TextChar{} must be replaced by its
definition.  In this way, the positions of names
can serve as a guide for applying substitutions and rewrites in the
synthesis algorithm.  By using these named abbreviations, \TypeProp{} guides the 
transformation of regular expression types
during search by deducing when certain expansions must be taken, or when one of
a class of expansions must be taken.
% end overview

% begin formalisms

\section{DNF Regular Expressions}
\label{sec:dnfre}

The first important step in \Optician{} is to convert
regular expression types into \emph{disjunctive normal form} (DNF).
%% A DNF regular expression is, intuitively, a regular expression with
%% all terms fully distributed, and with associativity information removed.
%% Distributing terms and removing associativity information creates a mutually
%% recursive data structure, consisting of the types of
A DNF regular expression, abbreviated DNF RE, is an n-ary disjunction of
sequences, where a sequence alternates between concrete strings and
atoms, and an atom is an iteration of DNF regular expressions.
The grammar below describes the syntax of 
DNF regular expressions (\DNFRegex{},\DNFRegexAlt{}),
sequences (\Sequence{},\SequenceAlt{}), and atoms (\Atom, \AtomAlt)
formally.

\begin{center}
  \begin{tabular}{l@{\ }c@{\ }l@{\ }>{\itshape\/}r}
    % DNF_REGEX
    \Atom{},\AtomAlt{} & \GEq{} & \StarOf{\DNFRegex{}}
% & \StarAtomType{}
\\
    \Sequence{},\SequenceAlt{} & \GEq{} &
                                                       $\SequenceOf{\String_0\SeqSep\Atom_1\SeqSep\ldots\SeqSep\Atom_n\SeqSep\String_n}$ 
%& \MultiConcatSequenceType{} 
\\
    \DNFRegex{},\DNFRegexAlt{} & \GEq{} & $\DNFOf{\Sequence_1\DNFSep\ldots\DNFSep\Sequence_n}$ %& \MultiOrDNFRegexType{} 
  \end{tabular}
\end{center}

Notice that it is straightforward to convert an arbitrary series
of atoms and strings into a sequence:  if there are multiple concrete strings 
between atoms, the strings may be concatenated into a single string.
If there are multiple atoms between concrete strings, the atoms 
may be separated by empty strings, which will sometimes be omitted for
readability.
Notice also that a simple string with no atoms may be represented as
a sequence containing just one concrete string.
In our implementation, names of user-defined regular expressions
are also atoms.  However, we elide such definitions from our theoretical analysis.

Intuitions about DNF regular expressions may be confirmed
by their semantics, which we give by defining the language (set of strings)
that each DNF regular expression denotes:

\begin{trivlist}
  \centering
\item 
%\begin{center}
  \begin{tabular}{@{\ }R@{\ }L}
    \LanguageOf{\StarOf{\DNFRegex}} \ =\  &
                                        \{\String_1\Concat\ldots\Concat\String_n
                                        \SuchThat \forall i \String_i\in\LanguageOf{\DNFRegex}\}\\
    \LanguageOf{\SequenceOf{\String_0\SeqSep\Atom_1\SeqSep\ldots\SeqSep\Atom_n\SeqSep\String_n}}\ =\  & 
%\\
%\multicolumn{2}{L}{\ \ \ \ 
\{\String_0\Concat\StringAlt_1\cdots\StringAlt_n\Concat\String_n \SuchThat \StringAlt_i\in\LanguageOf{\Atom_i}\}
%}
\\
    \LanguageOf{\DNFOf{\Sequence_1\DNFSep\ldots\DNFSep\Sequence_n}}\ =\  &
%\\
%\multicolumn{2}{L}{\ \ \ \ 
\{\String \SuchThat \String \in \LanguageOf{\Sequence_i} \text{\ and $i\in\RangeIncInc{1}{n}$}\}
%}
  \end{tabular}
%\end{center}
\end{trivlist}

A sequence of strings and atoms is \textit{sequence unambiguously
  concatenable},
written $\SequenceUnambigConcatOf{\String_0;\Atom_1;\ldots;\Atom_n;\String_n}$,
if, when $\String_i',\StringAlt_i'\in\LanguageOf{\Atom_i}$ for all $i$, then
$\String_0\String_1'\ldots\String_n'\String_n=
\String_0\StringAlt_1'\ldots\StringAlt_n'\String_n$
implies $\String_i'=\StringAlt_i'$ for all $i$.
A DNF regular expression $\Regex$ is \textit{unambiguously iterable},
written $\UnambigItOf{\Regex}$ if,
for all $n,m\in\Nats$ and for all strings
$\String_1,\ldots,\String_n,\StringAlt_1,\ldots,\StringAlt_m\in\LanguageOf{\Regex}$,
if
$\String_1\Concat\ldots\Concat\String_n=\StringAlt_1\Concat\ldots\Concat\StringAlt_m$
then $n=m$ and $\String_i=\StringAlt_i$ for all $i$.

\paragraph*{Expressivity of DNF Regular Expressions}

% fig:dnf-regex-functions
\begin{figure}
  \raggedright
  $\ConcatSequence{} \OfType{} \ArrowTypeOf{\SequenceType{}}{\ArrowTypeOf{\SequenceType{}}{\SequenceType{}}}$\\
  $\ConcatSequenceOf{[\String_0\SeqSep\Atom_1\SeqSep\ldots\SeqSep\Atom_n\SeqSep\String_n]}{[\StringAlt_0\SeqSep\AtomAlt_1\SeqSep\ldots\SeqSep\AtomAlt_m\SeqSep\StringAlt_m]}=
  [\String_0\SeqSep\Atom_1\SeqSep\ldots\SeqSep\Atom_n\SeqSep\String_n\Concat\StringAlt_0\SeqSep\AtomAlt_1\SeqSep\ldots\SeqSep\AtomAlt_m\SeqSep\StringAlt_m]$\\

  \medskip
  
  $\ConcatDNF{} \OfType{} \ArrowTypeOf{\DNFRegexType{}}{\ArrowTypeOf{\DNFRegexType{}}{\DNFRegexType{}}}$\\
  $\ConcatDNFOf{\DNFOf{\Sequence_1\DNFSep\ldots\DNFSep\Sequence_n}}{\DNFOf{\SequenceAlt_1\DNFSep\ldots\DNFSep\SequenceAlt_m}}=
      \DNFLeft \ConcatSequenceOf{\Sequence_1}{\SequenceAlt_1}\DNFSep \cdots
      \DNFSep \ConcatSequenceOf{\Sequence_1}{\SequenceAlt_m}\DNFSep
      \cdots \DNFSep \ConcatSequenceOf{\Sequence_n}{\SequenceAlt_1}\DNFSep \cdots \DNFSep \ConcatSequenceOf{\Sequence_n}{\SequenceAlt_m} \DNFRight$
  
  \medskip
  
  $\OrDNF{} \OfType{}
  \ArrowTypeOf{\DNFRegexType{}}{\ArrowTypeOf{\DNFRegexType{}}{\DNFRegexType{}}
  }$\\
  $\OrDNFOf{\DNFOf{\Sequence_1\DNFSep\ldots\DNFSep\Sequence_n}}{\DNFOf{\SequenceAlt_1\DNFSep\ldots\DNFSep\SequenceAlt_m}} =
  \DNFOf{\Sequence_1\DNFSep\ldots\DNFSep\Sequence_n\DNFSep\SequenceAlt_1\DNFSep\ldots\DNFSep\SequenceAlt_m}$
  
  \medskip
  
  \AtomToDNF{} \OfType
  \ArrowTypeOf{\AtomType{}}{\DNFRegexType{}}\\
  $\AtomToDNFOf{\Atom} = \DNFOf{\SequenceOf{\EmptyString \SeqSep \Atom \SeqSep
      \EmptyString}}$
  \caption{DNF Regular Expression Functions}
  \label{fig:dnf-regex-functions}
\end{figure}

Any regular expression may be converted into an equivalent DNF regular expression.  
To define the conversion function, we rely on several auxiliary functions defined in 
Figure~\ref{fig:dnf-regex-functions}.  Intuitively, $\DNFRegex{} \ConcatDNF \DNFRegex{}$ concatenates
two DNF regular expressions, producing a well-formed DNF regular expression as a result.
Similarly, $\DNFRegex{} \OrDNF{} \DNFRegex{}$ generates a new DNF regular expression representing the
union of two DNF regular expressions.  Finally, $\AtomToDNF{(\Atom)}$ converts a naked atom into a
well-formed DNF regular expression.
The conversion algorithm itself, written $\ToDNFRegexOf{\Regex}$, is defined below.
\[
  \begin{array}{rcl}
    \ToDNFRegexOf{\String} & = & \DNFOf{\SequenceOf{\String}}\\
    \ToDNFRegexOf{\emptyset} & = & \DNFOf{}\\
    \ToDNFRegexOf{(\StarOf{\Regex})} & = & \AtomToDNFOf{\StarOf{(\ToDNFRegexOf{\Regex})}}\\
    \ToDNFRegexOf{(\RegexConcat{\Regex_1}{\Regex_2})} & = & \ToDNFRegexOf{\Regex_1} \ConcatDNF \ToDNFRegexOf{\Regex_2}\\
    \ToDNFRegexOf{(\RegexOr{\Regex_1}{\Regex_2})} & = & \ToDNFRegexOf{\Regex_1} \OrDNF \ToDNFRegexOf{\Regex_2}\\
  \end{array}
\]
Using $\ToDNFRegex{}$, the definition of \LegacyTitleP{} gets converted into the
DNF regular expression:
\begin{lstlisting}
$\DNFLegacyTitle$ =
  $\DNFLeft$ $\SequenceOf{\StringCF{"<Field Id=2></Field>"}}$
 $\DNFSep\hspace*{.33mm}
  \SequenceOf{\StringCF{"<Field Id=2>"} \SeqSep \TextChar
    \SeqSep \StringCF{""}
    \SeqSep \StarOf{\DNFOf{\SequenceOf{\TextChar}}}
    \SeqSep \StringCF{"<Field Id=2>"}}$ $\DNFRight$
\end{lstlisting}
and the definition of \ModernTitle{} gets converted into the DNF regular
expression:
\begin{lstlisting}
$\DNFModernTitle$ =
  $\DNFLeft$ $ \SequenceOf{\StringCF{"Title:"} \SeqSep \TextChar
    \SeqSep \StringCF{""}
    \SeqSep \StarOf{\DNFOf{\SequenceOf{\TextChar}}}
    \SeqSep \StringCF{","}}$
 $\DNFSep\hspace*{.33mm}
  \SequenceOf{\StringCF{""}}$ $\DNFRight$
\end{lstlisting}
%

%% Intuitively, the definition of $\ToDNFRegex$ states that the DNF regular
%% expression with no sequences
%% corresponds to the empty set.  A single sequence with only a string corresponds
%% to a \BaseRegexType{} regular expression string.
%% The stars of DNF regular expressions correspond to the stars
%% of atoms.  The function $\OrDNF$ corresponds to the regular expression
%% primitive \OrRegexType{}.  The function $\ConcatDNF$ corresponds to the regular
%% expression primitive \ConcatRegexType{}. 
\iffull
\noindent
We formalize the correspondence between regular expressions and
DNF regular expressions via the following theorem.
\fi

\begin{theorem}[$\ToDNFRegex$ Soundness]
  \label{thm:dnfrs}
  For all regular expressions \Regex{},
  $\LanguageOf{\ToDNFRegexOf{\Regex}}=\LanguageOf{\Regex{}}$.
\end{theorem}

%% Furthermore, we can prove that DNF regular expressions are sound with respect to
%% regular expressions,
%% and that \ToDNFRegex{} is surjective, by providing a right inverse, $\ToRegex$.\\
%%   \begin{tabular}{r@{\ }c@{\ }l}
%%     $\ToRegexOf{(\StarOf{\DNFRegex})}$ & = & $\StarOf{(\ToRegexOf{\DNFRegex})}$\\
%%     $\ToRegexOf{\SequenceOf{\String_0}}$ & = & $\String_0$\\
%%     $\ToRegexOf{\SequenceOf{\String_0\SeqSep\Atom_1\SeqSep\ldots
%%     \SeqSep\Atom_n\SeqSep\String_n}} $ & = & $
%%     \ToRegexOf{\SequenceOf{\String_0\SeqSep\Atom_1\SeqSep
%%     \ldots\SeqSep\Atom_{n-1}\SeqSep\String_{n-1}}}$\\
%%                                        & &
%%     $\Concat \ToRegexOf{\Atom_n} \Concat \String_{n+1}$\\
%%     $\ToRegexOf{\DNFOf{}} $ & = & $\emptyset$\\
%%     $\ToRegexOf{\DNFOf{\Sequence_1\DNFSep\ldots\DNFSep\Sequence_n}} $ & = &
%%     $(\ToRegexOf{\DNFOf{\Sequence_1\DNFSep\ldots\DNFSep\Sequence_{n-1}}})
%%     \,\Or\, (\ToRegexOf{\Sequence_n})$
%%   \end{tabular}

%% This function provides a way to find a regular expression which corresponds to a given
%% DNF regular expression.  We formalize this correspondence with the theorem
%% below.

% fig:dnf-regex-rewrites

\paragraph*{DNF Regular Expression Rewrites}

\begin{figure}
  \centering
  \begin{mathpar}
    \centering
    \inferrule[\AtomUnrollstarLeftRule{}]
    {
    }
    {
      \StarOf{\DNFRegex}\RewriteAtom
      \OrDNFOf{\DNFOf{\SequenceOf{\EmptyString}}}{(\ConcatDNFOf{\DNFRegex}{\AtomToDNFOf{\StarOf{\DNFRegex}}})}
    }

    \inferrule[\AtomUnrollstarRightRule{}]
    {
    }
    {
      \StarOf{\DNFRegex}\RewriteAtom
      \OrDNFOf{\DNFOf{\SequenceOf{\EmptyString}}}{(\ConcatDNFOf{\AtomToDNFOf{\StarOf{\DNFRegex}}}{\DNFRegex})}
    }

    \inferrule[\AtomStructuralRewriteRule{}]
    {
      \DNFRegex \Rewrite \DNFRegex'
    }
    {
      \StarOf{\DNFRegex} \RewriteAtom \DNFOf{\SequenceOf{\StarOf{\DNFRegex'}}}
    }

    \inferrule[\DNFStructuralRewriteRule{}]
    {
      \Atom_j \RewriteAtom \DNFRegex
    }
    {
      \DNFOf{\Sequence_1\DNFSep\ldots\DNFSep\Sequence_{i-1}} \OrDNF
      \DNFOf{\SequenceOf{\String_0\SeqSep\Atom_1\SeqSep\ldots\SeqSep\String_{j-1}}}
      \ConcatDNF \AtomToDNFOf{\Atom_j} \ConcatDNF
      \DNFOf{\SequenceOf{\String_j\SeqSep\ldots\SeqSep\Atom_m\SeqSep\String_m}}
      \OrDNF \DNFOf{\Sequence_{i+1}\DNFSep\ldots\DNFSep\Sequence_n}\Rewrite\\
      \DNFOf{\Sequence_1\DNFSep\ldots\DNFSep\Sequence_{i-1}} \OrDNF
      \DNFOf{\SequenceOf{\String_0\SeqSep\Atom_1\SeqSep\ldots\SeqSep\String_{j-1}}}\ConcatDNF\DNFRegex\ConcatDNF\SequenceOf{\String_j\SeqSep\ldots\SeqSep\Atom_m\SeqSep\String_m} \OrDNF
      \DNFOf{\Sequence_{i+1}\DNFSep\ldots\DNFSep\Sequence_n}
    }
    % \inferrule[\ReflexivityRule{}]
    % {
    % }
    % {
    %   \DNFRegex \StarOf{\Rewrite} \DNFRegex
    % }
    % \inferrule[\BaseRule{}]
    % {
    %   \DNFRegex \Rewrite \DNFRegexAlt
    % }
    % {
    %   \DNFRegex \StarOf{\Rewrite} \DNFRegexAlt
    % }
    % \inferrule[\TransitivityRule{}]
    % {
    %   \DNFRegex_1 \StarOf{\Rewrite} \DNFRegex_2\\
    %   \DNFRegex_2 \StarOf{\Rewrite} \DNFRegex_3
    % }
    % {
    %   \DNFRegex_1 \StarOf{\Rewrite} \DNFRegex_3
    % }
  \end{mathpar}
  \caption{DNF Regular Expression Rewrite Rules}
  \label{fig:dnf-regex-rewrites}
\end{figure}

There are many fewer equivalences on DNF regular expressions than there are on
regular expressions, but there
still remain pairs of DNF regular expressions that, while syntactically different,
are semantically identical.  
%% In order to search the space of equivalent DNF 
%% In our synthesis algorithm we need to allow for a rewrite system on DNF regular
%% expressions to search through the equivalences.
%% Using the star equivalences, the synthesis algorithm applies rewrites
%% to the stars in a DNF
%% regular expression for type-directed synthesis to eventually be applied to the
%% rewritten regular expressions.  These rewrites are the ones used in
%% \textbf{Step 2: Traversing Equivalences}.
Figure~\ref{fig:dnf-regex-rewrites} defines a collection of rewrite rules on DNF regular expressions
designed to search the space of equivalent DNF REs.  
%We write $\StarOf{\Rewrite}$ for the reflexive, transitive closure of $\Rewrite{}$.
This directed rewrite system helps limit
the search space more than the non-directional equivalence $\SSREquiv$ relation.
However, because the rewrite rules are confluent, it is just as powerful as the
$\SSREquiv$ relation.
 
%% Intuitively, applications of these rewrite rules correspond to traversing
%% through equivalences of the DNF regular expressions.
%% However, by having rewrite rules, instead of a rule similar to
%% \RewriteRegexLensRule{}, instead of having to try all equivalences, we are able
%% to limit the fan-out and only traverse a few.

Because disjunctive normal form flattens a series of unions or concatenations
into an n-ary sum-of-products,
there is no need for rewriting rules that manage associativity or
distributivity.  Moreover, the lens term language and synthesis algorithm itself manages out-of-order summands, so we also have no need of rewriting
rules to handle commutativity of unions.  Hence, the rewriting system
need only focus on rewrites that involve Kleene star.
The rule \AtomUnrollstarLeftRule{} is a directed rewrite rule
designed to mirror \UnrollstarLeftRule{}.  Intuitively, it unfolds any atom 
$\StarOf{\DNFRegex{}}$ into $\RegexOr{\EmptyString{}}{(\RegexConcat{\DNFRegex{}}{\StarOf{{\DNFRegex{}}}})}$.  However, 
$\RegexOr{\EmptyString{}}{(\RegexConcat{\DNFRegex{}}{\StarOf{{\DNFRegex{}}}})}$ is
not a \emph{DNF} regular expression.  Hence, the rule uses DNF concatenation (\ConcatDNF)
and union (\OrDNF) in place of regular expression concatenation and union
to ensure a DNF regular expression is constructed.  The rule
\AtomUnrollstarRightRule{} mirrors the rule \UnrollstarRightRule{} in 
a similar way.

The rules \AtomStructuralRewriteRule{} and \DNFStructuralRewriteRule{}
make it possible to rewrite terms involving Kleene star that are
nested deep within a DNF regular expression, while ensuring that the
resulting term remains in DNF form.

%% With the atoms rewritten to DNF regular expressions, \DNFStructuralRewriteRule{}
%% provides a way to combine the rewritten atom into the broader regular
%% expression\afm{, through singling out the atom, expanding it, and recombining it into
%% the regular expression}.
%% Similarly, if a DNF regular expression is rewritten to a different
%% DNF regular expression, the rewrite rule \AtomStructuralRewriteRule{} can extend
%% this rewrite to rewriting atoms containing those regular expressions.
%% The symbol $\StarOf{\Rewrite}$ is used to denote the transitive and reflexive
%% closure of $\Rewrite$.  \afm{These correspond closely with the definitional
%% equivalence rules.  Many of the semiring axioms are normalized in the regular
%% expression into this form, and these rewrites correspond to applications of the
%% rewrites associated with \StarRegexType{}.  This does not handle \OrRegexType{}
%% commutativity -- this commutativity is handled in the typing of the DNF lenses,
%% not in the rewrite rules, to make the search space narrower.}

\section{DNF Lenses}
\label{sec:dnf}

The syntax of DNF lenses (\DNFLens{}),
sequence lenses (\SequenceLens{})
and atom lenses (\AtomLens{}) is defined below. DNF lenses and
sequence lenses both contain permutations ($\sigma$) that help
describe how these lenses act on data.
%We can now define a syntax for DNF lenses between DNF regular
%expressions.  
\begin{center}
  \begin{tabular}{@{}r@{\ }c@{}l@{}}
    % REGEX
    \AtomLens{} & \GEq{} & $\IterateLensOf{\DNFLens}$ \\
    \SequenceLens{} & \GEq{} & $(\SequenceLensOf{(\String_0,\StringAlt_0)\SeqLSep\AtomLens_1\SeqLSep\ldots\SeqLSep\AtomLens_n\SeqLSep(\String_n,\StringAlt_n)},\sigma)$ \\
    \DNFLens{} & \GEq{} & $(\DNFLensOf{\SequenceLens_1\DNFLSep\ldots\DNFLSep\SequenceLens_n}, \sigma)$ \\
  \end{tabular}
\end{center}

   %!legacy_title'! = @"<Field Id=2></Field>"@
   %| (@"<Field Id=2>"@ !text_char! !text_char!* @"</Field>"@)
   
A DNF lens consists of a list of sequence lenses and a permutation.
Much like a DNF regular expression is a list of disjuncted sequences, a
DNF lens contains a list of $\OrLens{}$ed sequence lenses.
DNF lenses also contain a permutation that provides
information about 
which sequences are mapped to which by the internal sequence lenses.
As an example, consider a DNF lens that maps between data with type
\DNFLegacyTitle{} and data with type \DNFModernTitle{}.
In such a lens, the permutation $\sigma$ 
indicates whether data matching $\SequenceOf{\StringCF{"<Field Id=2></Field>"}}$
will be translated to
data matching $\SequenceOf{\StringCF{"Title:"} \SeqSep \TextChar
    \SeqSep \StringCF{""}
    \SeqSep \StarOf{\DNFOf{\SequenceOf{\TextChar}}}
    \SeqSep \StringCF{","}}$ or $\SequenceOf{\StringCF{""}}$, and likewise for
the other sequence in \DNFLegacyTitle{}.  In this case, we would use the
permutation that swaps the order, as the first sequence in \DNFLegacyTitle{}
gets mapped to the 
second in \DNFModernTitle{}, and vice-versa.
As we will see in a moment, these permutations make 
it possible to construct a well-typed lens between two DNF regular expressions
regardless of the order in which clauses in a DNF regular expression appear,
thereby eliminating the need to consider equivalence modulo commutativity 
of these clauses.

A sequence lens consists of a list of atom lenses separated by pairs of 
strings, and a permutation.
Intuitively, much like a sequence is a list of concatenated atoms and strings, a
sequence lens is a list of concatenated atom lenses and string pairs.
Sequence lenses also contain a permutation that makes \PutLeft{} and \PutRight{}
reorder data, allowing sequence
lenses to take the job of both $\ConcatLens{}$ and $\SwapLens{}$.
If there are $n$ elements in the series then
the DNF sequence lens divides an input string up into $n$ substrings.
The $i^{th}$ such substring is transformed by the $i^{th}$ element of the series.
More precisely, if that $i^{th}$ element is an atom lens, then the  $i^{th}$
substring is transformed according to that atom lens.  If the $i^{th}$ element
is a pair of strings $(s_1,s_2)$ then that pair of strings
acts like a constant lens: when used from left-to-right, such a lens 
translates string $s_1$ into $s_2$; when used from right-to-left, such a
lens translates string $s_2$ into $s_1$. 
After all of the substrings have been transformed,
the permutation describes how to rearrange the 
substrings transformed by the atom lenses to obtain the final output.
As an example, consider a sequence lens that maps between data with type
$\SequenceOf{\StringCF{"Title:"} \SeqSep \TextChar
  \SeqSep \StringCF{""}
  \SeqSep \StarOf{\DNFOf{\SequenceOf{\TextChar}}}
  \SeqSep \StringCF{","}}$
and data with type 
$\SequenceOf{\StringCF{"<Field Id=2>"} \SeqSep \TextChar
    \SeqSep \StringCF{""}
    \SeqSep \StarOf{\DNFOf{\SequenceOf{\TextChar}}}
    \SeqSep \StringCF{"<Field Id=2>"}}$.  We desire no
  reorderings
  between the atoms $\TextChar$ and $\StarOf{\DNFOf{\SequenceOf{\TextChar}}}$,
  so the permutation associated with this lens would be
  the identity permutation.

An atom lens is an iteration of a DNF lens; its semantics is similar to
the semantics of ordinary iteration lenses.
In our implementation, identity transformations between
user-defined regular expressions are also atom lenses.
However, we elide such definitions from our theoretical analysis.

The semantics of DNF Lenses, sequence lenses and atom lenses 
is defined formally below.

\begin{center}
  \begin{tabular}{@{}r@{\ }c@{}l@{}}
    % REGEX
    $\SemanticsOf{\IterateLensOf{\DNFLens}}$ & \GEq{} & $\SetOf{(\String_1\Concat\ldots\Concat\String_n,
    \StringAlt_1\Concat\ldots\Concat\StringAlt_n)\SuchThat
    n\in\Nats\BooleanAnd(\String_i,\StringAlt_i)\in\SemanticsOf{\DNFLens}}$ \\
    $\SemanticsOf{(\SequenceLensOf{(\String_0,\StringAlt_0)\SeqLSep
      \AtomLens_1\SeqLSep\ldots\SeqLSep\AtomLens_n
      \SeqLSep(\String_n,\StringAlt_n)},\Permutation)}$ & \GEq{} & $\SetOf{
    (\String_0\String_1'\ldots\String_n'\String_n,
    \StringAlt_0\StringAlt_{\Permutation(1)}'\ldots
    \StringAlt_{\Permutation(n)}'\StringAlt_n)\SuchThat
    (\String_i',\StringAlt_i')\in\SemanticsOf{\AtomLens_i}}$ \\
    $\SemanticsOf{(\DNFLensOf{\SequenceLens_1\DNFLSep\ldots\DNFLSep
      \SequenceLens_n},\Permutation)}$ & \GEq{} & $\SetOf{(\String,\StringAlt)\SuchThat
    (\String,\StringAlt)\in\SequenceLens_i\text{ for some $i$}}$ \\
  \end{tabular}
\end{center}

\begin{figure}
  \centering
  \begin{mathpar}
    \inferrule[] %\AtomLensRule{}
    {
      \DNFLens \OfRewritelessType \DNFRegex \Leftrightarrow \DNFRegexAlt\\
      \UnambigItOf{\DNFRegex}\\
      \UnambigItOf{\DNFRegexAlt}
    }
    {
      \IterateLensOf{\DNFLens} \OfRewritelessType \StarOf{\DNFRegex}
      \Leftrightarrow \StarOf{\DNFRegexAlt}
    }

    \inferrule[] %\SequenceLensRule{}
    {
      \AtomLens_1 \OfRewritelessType \Atom_1 \Leftrightarrow \AtomAlt_1\\
      \ldots\\
      \AtomLens_n \OfRewritelessType \Atom_n \Leftrightarrow \AtomAlt_n\\\\
      \sigma \in \PermutationSetOf{n}\\
      \SequenceUnambigConcatOf{\String_0 ; \Atom_1 ; \ldots ; \Atom_n ; \String_n}\\
      \SequenceUnambigConcatOf{\StringAlt_0 ; \AtomAlt_{\sigma(1)} ; \ldots ;
        \AtomAlt_{\sigma(n)} ; \StringAlt_n}
    }
    {
      (\SequenceLensOf{(\String_0,\StringAlt_0)\SeqLSep\AtomLens_1\SeqLSep\ldots\SeqLSep\AtomLens_n\SeqLSep(\String_n,\StringAlt_n)},\sigma) \OfRewritelessType
      \SequenceOf{\String_0\SeqSep\Atom_1\SeqSep\ldots\SeqSep\Atom_n\SeqSep\String_n}\Leftrightarrow
      \SequenceOf{\StringAlt_0\SeqSep\AtomAlt_{\sigma(1)}\SeqSep\ldots\SeqSep\AtomAlt_{\sigma(n)}\SeqSep\StringAlt_n}
    }

    \inferrule[] %\DNFLensRule{}
    {
      \SequenceLens_1 \OfRewritelessType \Sequence_1 \Leftrightarrow \SequenceAlt_1\\
      \ldots\\
      \SequenceLens_n \OfRewritelessType \Sequence_n \Leftrightarrow \SequenceAlt_n\\\\
      \sigma \in \PermutationSetOf{n}\\
      i \neq j \Rightarrow \LanguageOf{\Sequence_{i}} \cap \LanguageOf{\Sequence_{j}}=\emptyset\\
      i \neq j \Rightarrow \LanguageOf{\SequenceAlt_{i}} \cap \LanguageOf{\SequenceAlt_{j}}=\emptyset\\
    }
    {
      (\DNFLensOf{\SequenceLens_1\DNFLSep\ldots\DNFLSep\SequenceLens_n},\sigma) \OfRewritelessType
      \DNFOf{\Sequence_1\DNFSep\ldots\DNFSep\Sequence_n}
      \Leftrightarrow \DNFOf{\SequenceAlt_{\sigma(1)}\DNFSep\ldots\DNFSep\SequenceAlt_{\sigma(n)}}
    }

    \inferrule[] %\RewriteDNFRegexLensRule{}
    {
      \DNFRegex' \StarOf{\Rewrite} \DNFRegex\\
      \DNFRegexAlt' \StarOf{\Rewrite} \DNFRegexAlt \\
      \DNFLens \OfRewritelessType \MapsBetweenTypeOf{\DNFRegex}{\DNFRegexAlt}
    }
    {
      \DNFLens \OfType \MapsBetweenTypeOf{\DNFRegex'}{\DNFRegexAlt'}
    }

  \end{mathpar}
  \caption{DNF Lens Typing}
  \label{fig:dnf-lens-typing}
\end{figure}

\paragraph*{Type Checking}
Figure~\ref{fig:dnf-lens-typing} presents the type checking rules for
DNF lenses.  In order to control where regular expression rewriting
may be used (and thereby reduce search complexity), 
the figure defines two separate typing judgements.
The first judgement has the form 
$\DNFLens \OfRewritelessType \DNFRegex \Leftrightarrow \DNFRegexAlt$.
It implies that the lens $\DNFLens$ is a well-formed
bijective map between the
languages of $\DNFRegex$ and $\DNFRegexAlt$.  This judgement does
not include the rule for rewriting the types of the source or target
data.  The second judgement has the form 
$\DNFLens \OfType \DNFRegex \Leftrightarrow \DNFRegexAlt$.  
It rewrites the source and target types, and then searches for a
DNF lens with the rewritten types.

One of the key differences between these typing judgements and the
judgements for ordinary lenses are the permutations.  For example,
in the rule for typing DNF lenses, the permutation $\sigma$
indicates how to match  
sequence types in the domain 
($\Sequence_1\ldots\Sequence_n$) and the 
range ($\SequenceAlt_{\sigma(1)}\ldots\SequenceAlt_{\sigma(n)}$).
Permutations are used in a similar way in the typing rule for
sequence lenses.

%% There is one rule for each syntactic form
%% Similarly to the language of lenses, the typing of the lenses
%% correspond closely to the syntax for the lenses themselves.
%% The typing of these DNF lenses are defined in Figure~\ref{fig:dnf-lens-typing}.
%% The typing $\DNFLens \OfType \DNFRegex \Leftrightarrow \DNFRegexAlt$ means that
%% $\DNFLens$ maps bijectively between the languages of $\DNFRegex$ and
%% $\DNFRegexAlt$.  The typing $\DNFLens \OfRewritelessType \DNFRegex
%% \Leftrightarrow \DNFRegexAlt$ means that $\DNFLens$ maps bijectively between the
%% languages of $\DNFRegex$ and $\DNFRegexAlt$, and no rewrites were used in the
%% application of the typing.

\iffull
These DNF lenses only express bijections that are already expressible in the
language of lenses, and they can express everything expressible in the
language of lenses.
\fi

\paragraph*{Properties}  While DNF lenses have a restrictive syntax, they
remain as powerful as ordinary bijective lenses.  The following theorems
characterize the relationship between the two languages.

\begin{theorem}[DNF Lens Soundness]
  \label{thm:dnfls}
  If there exists a derivation of $\DNFLens \OfType \MapsBetweenTypeOf{\DNFRegex}{\DNFRegexAlt}$,
  then there exist a lens, $\ToLensOf{\DNFLens}$, and regular expressions, $\Regex$ and 
$\RegexAlt$, such that $\ToLensOf{\DNFLens} \OfType \MapsBetweenTypeOf{\Regex}{\RegexAlt}$ and
  $\ToDNFRegexOf{\Regex}=\DNFRegex$ and
  $\ToDNFRegexOf{\RegexAlt}=\DNFRegexAlt$ and
  $\SemanticsOf{\ToLensOf{\DNFLens}}=\SemanticsOf{\DNFLens}$.
\end{theorem}

\begin{theorem}[DNF Lens Completeness]
  \label{thm:dnflc}
  If there exists a derivation for $\Lens \OfType \Regex \Leftrightarrow
  \RegexAlt$,
  then there exists a DNF lens $\DNFLens$ such that
  $\DNFLens \OfType (\ToDNFRegexOf{\Regex}) \Leftrightarrow (\ToDNFRegexOf{\RegexAlt})$ and $\SemanticsOf{\Lens}=\SemanticsOf{\DNFLens}$.
\end{theorem}

\paragraph*{Discussion} 
DNF lenses are significantly better suited to synthesis than regular
bijective lenses.  First, they contain no composition operator.
Second, the use of equivalence (rewriting) is highly constrained:  Rewriting
may only be used once at the top-most level as opposed to interleaved
between uses of the other rules.  Consequently, a type-directed synthesis
algorithm may be factored into two discrete steps: one step that
searches for an effective pair of regular expressions
and a second step that is directed by the syntax of the regular expression
types that were discovered in the first step.
%% All rules except for \RewriteDNFRegexLensRule{} have a simple inductive
%% algorithm that comes from the type.  Furthermore, even
%% \RewriteDNFRegexLensRule{} can only be applied once at the end of the typing
%% derivation, and it is not computationally intensive to find what rewrites can
%% apply to a given DNF regular expression.
%% Furthermore, there is no composition operator, so
%% we have reduced the dimensions we are searching in from 3 to 2.  Before, given a
%% specification, multiple rules could be applicable, equivalent regular
%% expressions would have to be searched through, and regular expressions for
%% composition would need to be found.
%% For DNF lenses, only the correct rewrites for the regular expressions, and the
%% correct permutations for each rule application, must be found.
% end formalisms

% begin implementation
\section{Algorithm}
\label{algorithm}

\paragraph*{Synthesis Overview}
Algorithm~\ref{alg:synthlens} presents the synthesis procedure.
\SynthLens{} takes the source and target regular
expressions $\Regex$ and $\RegexAlt$, and a list of examples $\Examples$, as
input. First, \SynthLens{} validates the unambiguity of the input regular
expressions, $\Regex$ and $\RegexAlt$, and confirms that they parse the
input/output examples, $\Examples$. Next, the algorithm converts
$\Regex$ and $\RegexAlt$ into DNF regular expressions
$\DNFRegex$ and $\DNFRegexAlt$
using the $\ToDNFRegex$ operator.
It then calls \SynthDNFLens{}
on $\DNFRegex$, $\DNFRegexAlt$, and the examples
to create a DNF lens $\DNFLens$.  
Finally, it uses \ToLens{} to convert $\DNFLens$ to a Boomerang lens.

% finally tweaked for readability with \Beautify{}.

\begin{algorithm}
  \caption{\SynthLens}
  \label{alg:synthlens}
  \begin{algorithmic}[1]
    \Function{SynthDNFLens}{$\DNFRegex,\DNFRegexAlt,\Examples$}
    \State $\Queue \gets \Call{\CreatePQueue}{(\DNFRegex,\DNFRegexAlt),0}$
    \While{$\True$}
    \State $(\QueueElement,\Queue) \gets \Call{\Pop}{\Queue}$
    \State $(\DNFRegex',\DNFRegexAlt',\ExpCount) \gets \QueueElement$
    \State $\DNFLensOption \gets
    \Call{\RigidSynth}{\DNFRegex',\DNFRegexAlt',\Examples}$
    \Switch{\DNFLensOption}
    \CaseTwo {\SomeOf{\DNFLens}}{\ReturnVal{\DNFLens}}
    \EndCaseTwo
    \Case {\None}
    \State $\QueueElements \gets \Call{\Expand}{\DNFRegex,\DNFRegexAlt,\ExpCount}$
    \State $\Queue \gets \Call{\EnqueueMany}{\QueueElements,\Queue}$
    \EndCase
    \EndSwitch
    \EndWhile
    \EndFunction
    
    \Statex
    \Function{SynthLens}{$\Regex,\RegexAlt,\Examples$}
    \State $\Call{Validate}{\Regex,\RegexAlt,\Examples}$
    \State $(\DNFRegex,\DNFRegexAlt) \gets
    (\ToDNFRegexOf{\Regex},\ToDNFRegexOf{\RegexAlt})$
    \State $\DNFLens \gets \Call{SynthDNFLens}{\DNFRegex,\DNFRegexAlt,\Examples}$
    \State $\ReturnVal{\ToLensOf{\DNFLens}}$
    \EndFunction
  \end{algorithmic}
\end{algorithm}

\SynthDNFLens{} starts by creating a priority queue $\Queue$ to manage
the search for a DNF lens.  Each element $\QueueElement$  in the queue
is a tuple of 
the source DNF regular expression $\DNFRegex'$, 
the target DNF regular expression $\DNFRegexAlt'$, 
and a count $\ExpCount$ of the number
of expansions needed to produce this pair of DNF
regular expressions from the originals $\DNFRegex$ and
$\DNFRegexAlt$.  (Recall that an expansion is a 
use of a rewrite rule or the substitution of a user-defined
definition for its name.) The priority of each queue element is $\ExpCount$; DNF regular
expressions that have undergone fewer expansions will get priority.
The algorithm initializes the queue with
$\DNFRegex$ and $\DNFRegexAlt$, which have an expansion count of
zero.  The algorithm then proceeds by iteratively examining the
highest priority element from the queue (this examination corresponds to
\TypeProp{} in Figure~\ref{fig:2-phased-schematic-diagram}), and using the 
function \RigidSynth{} to try to find a rewriteless DNF lens
between the popped source and target DNF regular expressions that satisfy
the examples \Examples{}.  If successful, the algorithm returns the
DNF lens $\DNFLens$.  Otherwise, the function \Expand{} produces
a new set of candidate DNF regular expression pairs that can
be obtained from $\DNFRegex$ and $\DNFRegexAlt$ by
applying various expansions to the source and target DNF regular
expressions.

\paragraph*{\Expand}
\begin{algorithm}
  \caption{\Expand{}}
  \label{alg:synthlens2}
  \begin{multicols}{2}
  \begin{algorithmic}[1]
    \Function{ExpandRequired}{$\DNFRegex,\DNFRegexAlt$,$\ExpCount$}
    \State $\CurrentSet_{\DNFRegex} \gets \Call{\GetCurrentSet}{\DNFRegex}$
    \State $\CurrentSet_{\DNFRegexAlt} \gets \Call{\GetCurrentSet}{\DNFRegexAlt}$
    \State $\TransitiveSet_{\DNFRegex} \gets \Call{\GetTransitiveSet}{\DNFRegex}$
    \State $\TransitiveSet_{\DNFRegexAlt} \gets
    \Call{\GetTransitiveSet}{\DNFRegexAlt}$
    \State $r \gets \False$
    \ForEach{$(\UserDef, \Int)$}{$\CurrentSet_{\DNFRegex} \setminus
      \TransitiveSet_{\DNFRegexAlt}$}
    \State $r \gets \True$
    \State $(\DNFRegex,\ExpCount) \gets \Call{\ForceExpand}{\DNFRegex,\UserDef,\Int,\ExpCount}$
    \EndForEach
    \ForEach{$(\UserDef, \Int)$}{$\CurrentSet_{\DNFRegexAlt} \setminus
      \TransitiveSet_{\DNFRegex}$}
    \State $r \gets \True$
    \State $(\DNFRegexAlt,\ExpCount) \gets \Call{\ForceExpand}{\DNFRegexAlt,\UserDef,\Int,\ExpCount}$
    \EndForEach
    \If {$r$}
    \State $\ReturnVal{\Call{\ExpandRequired}{\DNFRegex,\DNFRegexAlt,\ExpCount}}$
    \EndIf
    \State $\ReturnVal{(\DNFRegex,\DNFRegexAlt,\ExpCount)}$
    \EndFunction

    \Statex
    
    \Function{FixProblemElts}{$\DNFRegex,\DNFRegexAlt$,$\ExpCount$}
    \State $\CurrentSet_{\DNFRegex} \gets \Call{\GetCurrentSet}{\DNFRegex}$
    \State $\CurrentSet_{\DNFRegexAlt} \gets
    \Call{\GetCurrentSet}{\DNFRegexAlt}$
    \State $\QueueElements \gets []$
      \ForEach{$(\UserDef, \Int)$}{$\CurrentSet_{\DNFRegexAlt} \setminus
      \CurrentSet_{\DNFRegex}$}
    \State $\QueueElements \gets \QueueElements \Append \Call{\Reveal}{\DNFRegex,\UserDef,\Int,\ExpCount,\DNFRegexAlt}$
    \EndForEach
    \ForEach{$(\UserDef, \Int)$}{$\CurrentSet_{\DNFRegex} \setminus
      \CurrentSet_{\DNFRegexAlt}$}
    \State $\QueueElements \gets \QueueElements \Append \Call{\Reveal}{\DNFRegexAlt,\UserDef,\Int,\ExpCount,\DNFRegex}$
    \EndForEach
    \State $\ReturnVal{\QueueElements}$
    \EndFunction

    \Statex
    
    \Function{Expand}{$\DNFRegex,\DNFRegexAlt$,$\ExpCount$}
    \State $(\DNFRegex,\DNFRegexAlt,\ExpCount) \gets
    \Call{ExpandRequired}{\DNFRegex,\DNFRegexAlt,\ExpCount}$
    \State $\QueueElements \gets
    \Call{FixProblemElts}{\DNFRegex,\DNFRegexAlt,\ExpCount}$
    \Switch{\QueueElements}
    \CaseTwo {[]}{$\ReturnVal{\Call{\ExpandOnce}{\DNFRegex,\DNFRegexAlt,\ExpCount}}$}
    \EndCaseTwo
    \CaseTwo {\_}{$\ReturnVal{\QueueElements}$}
    \EndCaseTwo
    \EndSwitch
    \EndFunction
  \end{algorithmic}
  \end{multicols}
\end{algorithm}

Intelligent expansion inference is key to the efficiency of \Optician{}.
\Expand{}, shown in Algorithm~\ref{alg:synthlens2}, codifies this inference.
It makes critical use of the locations of user-defined data types, measured 
by their \emph{star depth}, which is the number of nested \Star{}'s the data type occurs beneath.
Star depth locations are useful because we can quickly compute the
current star depths of user-defined data types (with \GetCurrentSet{}) and the
star depths of user-defined data types reachable via expansions
(with \GetTransitiveSet{}).
Furthermore, the star depths of user-defined data types have the following useful property:

\begin{property}
\label{prop:udpairs}
  If $\UserDef$ is present at star depth $\Int$ in $\DNFRegex$ and there
  exists a rewriteless
  DNF lens $\DNFLens$ such that $\DNFLens \OfRewritelessType \DNFRegex
  \Leftrightarrow \DNFRegexAlt$, then $\UserDef$ is also present at star depth
  $\Int$ in $\DNFRegexAlt$.  The symmetric property is true if $\UserDef$ is
  present at star depth $\Int$ in $\DNFRegexAlt$.
\end{property}

Property~\ref{prop:udpairs} means that if there is a rewriteless DNF lens
between two DNF regular expressions, then the same user-defined
data types must be present at the same locations in both of the DNF regular
expressions. We use this
property to determine when certain rules must be applied and to direct the search
to rules that make progress towards this required alignment.

\Expand{} has three major components: \ExpandRequired{}, \FixProblemElts{}, and
\ExpandOnce{}, which we discuss in turn.
\ExpandRequired{} performs expansions that \emph{must} be taken.  In particular,
if a user-defined data type $\UserDef$ at star depth $\Int$ is impossible 
to reach through any number of expansions on the opposite side, then that 
user-defined data type \emph{must} be replaced by its definition at
that depth.
For example, 
consider trying to find a lens between $\DNFOf{\SequenceOf{\LegacyTitle}}$ and
$\DNFOf{\SequenceOf{\ModernTitle}}$.
No matter how many expansions are performed on
\VarCF{modern\_title}, the user-defined type \VarCF{legacy\_title}
will not be exposed 
%at depth 0 (or any other depth) 
because the set of possible reachable pairs of
data types and star depths in \VarCF{modern\_title} is $\{(\ModernTitle,0),
$ $(\TextChar,0),(\TextChar,1)\}$.  Because no number of expansions will reveal
$\VarCF{\LegacyTitle}$ on the right, 
the algorithm must replace $\VarCF{\LegacyTitle}$ with its definition
on the left in order to find a lens.
\ExpandRequired{} continues until it finds all forced expansions.

\ExpandRequired{} finds all the expansions that must be performed, but
it does not
perform any other expansions.  However, there are many situations where it is
possible to infer that one of a set of expansions must be performed
without forcing any individual expansion.
In particular, for any pair of types that have a rewriteless lens
between them, for each (user-defined type, star depth) pair $(\UserDef,\Int)$ on one side, that
same pair must be present on the other side.  \FixProblemElts{}
identifies when there is a $(\UserDef,\Int)$ pair present on only one
side.  After identifying these problem elements, it 
calls 
\Reveal{} to find the expansions that will reveal this element.  
For example,
after $\DNFOf{\SequenceOf{\LegacyTitle{}}}$ has been expanded to
\[\DNFOf{\SequenceOf{\StringCF{"<Field Id=2>"} \SeqSep
    \StarOf{\DNFOf{\SequenceOf{\TextChar}}} \SeqSep
    \StringCF{"</Field>"}}}
\]
\noindent
and $\DNFOf{\SequenceOf{\ModernTitle{}}}$ has been expanded to
\[\DNFOf{\SequenceOf{\StringCF{"Title:"} \SeqSep \TextChar
    \SeqSep \StringCF{""}
    \SeqSep \StarOf{\DNFOf{\SequenceOf{\TextChar}}}
    \SeqSep \StringCF{","}}
  \DNFSep
  \SequenceOf{\StringCF{""}}}
\]
\noindent
we can see that the modern expansion
has an instance of \TextChar{} at depth 0, where the legacy one does not.
For a lens to exist between the two types,
\TextChar{} must be revealed at star depth 0 in the legacy expansion.
Revealing \TextChar{} at depth zero will give back two candidate DNF
regular expressions,
one from an application of
\AtomUnrollstarLeftRule{}, and one from an application of
\AtomUnrollstarRightRule{}.

Together \ExpandRequired{} and \FixProblemElts{} apply many
expansions, but by themselves they are not sufficient.
Typically, when \FixProblemElts{} and \ExpandRequired{} do not find all the
necessary expansions, the input data formats expect large amounts of similar
information.
For example, in trying to synthesize the identity
transformation between \CF{\textcolor{blue}{""} | \textcolor{darkbrown}{U} |
  \textcolor{darkbrown}{UU}(\textcolor{darkbrown}{U}*)} and
\CF{\textcolor{blue}{""} | \textcolor{darkbrown}{U}(\textcolor{darkbrown}{U}*)},
\ExpandRequired{} and \FixProblemElts{} find no forced expansions.
An expansion is necessary, but the set of pairs
($\SetOf{(\VarCF{U},0),(\VarCF{U},1)}$) is the same for both sides.  
When this situation arises, the algorithm uses the \ExpandOnce{}
function to conduct a purely enumerative search, implemented by performing all
single-step expansions.

\paragraph*{\RigidSynth} 
The function \RigidSynth{}, shown in Algorithm~\ref{alg:rigidsynth},
implements the portion of \SynthLens{} that
generates a lens from the types and examples without using any equivalences.
Intuitively, it aligns the structures of the source and target regular
expressions by finding appropriate permutations of nested sequences
and nested atoms, taking into account the information contained in the
examples.  Once it finds an alignment, it generates the
corresponding lens. 

\begin{algorithm}
  \caption{\RigidSynth}
  \label{alg:rigidsynth}
  \begin{algorithmic}[1]
    \Function{\RigidSynthAtom}{$\Atom{},\AtomAlt{},\Examples$}
    \Switch{$(\Atom{}, \AtomAlt)$}
    \Case {(\UserDef, \UserDefAlt)}
    \If {$\UserDef \AtomLeq^\Examples \UserDefAlt \BooleanAnd \UserDefAlt
      \AtomLeq^\Examples \UserDef$}
    \State \ReturnVal{\SomeOf{\IdentityLensOf{\UserDef}}}
    \Else
    \State \ReturnVal{\None}
    \EndIf
    \EndCase
    \Case {(\StarOf{\DNFRegex},\StarOf{\DNFRegexAlt})}
    \Switch{$\RigidSynth(\DNFRegex,\DNFRegexAlt,\Examples)$}
    \CaseTwo{\SomeOf{\DNFLens}}{\ReturnVal{\IterateLensOf{\DNFLens}}}
    \EndCaseTwo
    \CaseTwo {\None}{$\ReturnVal{\None}$}
    \EndCaseTwo
    \EndSwitch
    \EndCase
    \CaseTwo {\_}{\ReturnVal{\None}}
    \EndCaseTwo
    \EndSwitch
    \EndFunction

    \Statex

    \Function{\RigidSynthSequence}{$\Sequence,\SequenceAlt,\Examples$}
    \State $\SequenceOf{\String_0 \SeqSep \Atom_1 \SeqSep \ldots \SeqSep \Atom_n
      \SeqSep \String_n} \gets
    \Sequence$
    \State $\SequenceOf{\StringAlt_0 \SeqSep \AtomAlt_1 \SeqSep \ldots \SeqSep
      \AtomAlt_m \SeqSep \StringAlt_m} \gets
    \SequenceAlt$
    \If {$n \neq m$}
    \State $\ReturnVal{\None}$
    \EndIf
    \State $\sigma_1 \gets \SortingOf{\AtomLeq^\Examples}{\ListOf{\Atom_1
        \SeqSep \ldots \SeqSep \Atom_n}}$
    \State $\sigma_2 \gets \SortingOf{\AtomLeq^\Examples}{\ListOf{\AtomAlt_1
        \SeqSep \ldots \SeqSep \AtomAlt_n}}$
    \State $\sigma \gets \InverseOf{\sigma_1} \Compose \sigma_2$
    \State $\mathit{ABs} \gets
    \Call{Zip}{\ListOf{\Atom_1 \SeqSep \ldots \SeqSep
        \Atom_n},\ListOf{\AtomAlt_{\sigma(1)} \SeqSep \ldots \SeqSep \AtomAlt_{\sigma(n)}}}$
    \State $\mathit{alos} \gets
    \Call{\Map}{\RigidSynthAtom(\Examples),\mathit{ABs}}$
    \Switch{$\Call{AllSome}{\mathit{alos}}$}
    \CaseTwo {\SomeOf{\ListOf{\AtomLens_1 \SeqSep \ldots \SeqSep
          \AtomLens_n}}}{$\ReturnVal{\SomeOf{(\SequenceLensOf{(\String_0,\StringAlt_0)
            \SeqSep \AtomLens_1 \SeqSep \ldots \SeqSep \AtomLens_n \SeqSep (\String_n,\StringAlt_n)},\InverseOf{\sigma})}}$}
    \EndCaseTwo
    \CaseTwo {\None}{$\ReturnVal{\None}$}
    \EndCaseTwo
    \EndSwitch
    \EndFunction

    \Statex

    \Function{\RigidSynth}{$\DNFRegex,\DNFRegexAlt,\Examples$}
    \State $\DNFOf{\Sequence_1 \DNFSep \ldots \DNFSep \Sequence_n} \gets
    \DNFRegex$
    \State $\DNFOf{\SequenceAlt_1 \DNFSep \ldots \DNFSep \SequenceAlt_m} \gets
    \DNFRegexAlt$
    \If {$n \neq m$}
    \State $\ReturnVal{\None}$
    \EndIf
    \State $\sigma_1 \gets
    \SortingOf{\SequenceLeq^\Examples}{\ListOf{\Sequence_1 \DNFSep \ldots
        \DNFSep \Sequence_n}}$
    \State $\sigma_2 \gets
    \SortingOf{\SequenceLeq^\Examples}{\ListOf{\SequenceAlt_1 \DNFSep \ldots
        \DNFSep \SequenceAlt_n}}$
    \State $\sigma \gets \InverseOf{\sigma_1} \Compose \sigma_2$
    \State $\mathit{STQs} \gets
    \Call{Zip}{\ListOf{\Sequence_1 \DNFSep \ldots \DNFSep
        \Sequence_n},\ListOf{\SequenceAlt_{\sigma(1)} \DNFSep \ldots \DNFSep \SequenceAlt_{\sigma(n)}}}$
    \State $\mathit{sqlos} \gets
    \Call{\Map}{\RigidSynthSequence(\Examples),\mathit{STQs}}$
    \Switch{$\Call{AllSome}{\mathit{sqlos}}$}
    \CaseTwo {\SomeOf{\ListOf{\SequenceLens_1 \DNFSep \ldots \DNFSep
          \SequenceLens_n}}}{$\ReturnVal{\SomeOf{(\DNFLensOf{\SequenceLens_1
            \DNFSep \ldots \DNFSep \SequenceLens_n},\InverseOf{\sigma})}}$}
    \EndCaseTwo
    \CaseTwo {\None}{$\ReturnVal{\None}$}
    \EndCaseTwo
    \EndSwitch
    \EndFunction
  \end{algorithmic}
\end{algorithm}

Searching for aligning permutations requires care, as na\"{i}vely
considering all permutations between two DNF regular
expressions $\DNFOf{\Sequence_1 \DNFSep \ldots \DNFSep \Sequence_n}$ and
$\DNFOf{\SequenceAlt_1 \DNFSep \ldots \DNFSep \SequenceAlt_n}$ would require time
proportional to $n!$.  A better approach is to identify elements of
the source and target DNF regular expressions that match and to
leverage that information to create candidate permutations.

\RigidSynth{} performs this identification via orderings on
sequences (\SequenceLeq), and atoms (\AtomLeq).
To determine if one expression is less than the other, the algorithm converts
each expression into a list of its subterms and returns whether the
lexicographic ordering determines the first list less than the second.
These orderings are carefully constructed so that equivalent terms have lenses
between them.  For
example, between two sequences, $\Sequence$ and $\SequenceAlt$, there is a lens
$\SequenceLens \OfRewritelessType \Sequence \Leftrightarrow \SequenceAlt$ if,
and only if, $\Sequence \SequenceLeq \SequenceAlt$ and $\SequenceAlt
\SequenceLeq \Sequence$.  Through these orderings, aligning the components
reduces to merely sorting and zipping lists.  Furthermore, through composing the
permutations required to sort the sequences, the algorithm discovers the
permutation used in the lens.

As an example, consider trying to find a DNF lens between
\begin{lstlisting}
   $\DNFLeft$ $\SequenceOf{\StringCF{"<Field Id=2></Field>"}}$
  $\DNFSep\hspace*{.33mm}
  \SequenceOf{\StringCF{"<Field Id=2>"} \SeqSep \TextChar
    \SeqSep \StringCF{""}
    \SeqSep \StarOf{\DNFOf{\SequenceOf{\TextChar}}}
    \SeqSep \StringCF{"<Field Id=2>"}}$ $\DNFRight$
\end{lstlisting}
and
\begin{lstlisting}
   $\DNFLeft$ $\SequenceOf{\StringCF{"Title:"} \SeqSep \TextChar
    \SeqSep \StringCF{""}
    \SeqSep \StarOf{\DNFOf{\SequenceOf{\TextChar}}}
    \SeqSep \StringCF{","}}$
  $\DNFSep\hspace*{.33mm}
  \SequenceOf{\StringCF{""}}$ $\DNFRight$
\end{lstlisting}
As \RigidSynth{} considers the legacy DNF regular expression, it
orders its two sequences
by maintaining the existing order:  first $\SequenceOf{\StringCF{"<Field
    Id=2></Field>"}}$, then
$\SequenceOf{\StringCF{"<Field Id=2>"} \SeqSep \TextChar
    \SeqSep \StringCF{""}
    \SeqSep \StarOf{\DNFOf{\SequenceOf{\TextChar}}}
    \SeqSep \StringCF{"<Field Id=2>"}}$.
In contrast,
\RigidSynth{} reorders 
the two sequences of the modern DNF regular expression,
making $\SequenceOf{\StringCF{""}}$ first,
and $\SequenceOf{\StringCF{"Title:"} \SeqSep \TextChar
    \SeqSep \StringCF{""}
    \SeqSep \StarOf{\DNFOf{\SequenceOf{\TextChar}}}
    \SeqSep \StringCF{","}}$ second;
the overall permutation is a swap.  
As a result, the two string sequences become aligned, as do the two
complex sequences. 

Then, the algorithm calls \RigidSynthSequence{} on the two aligned sequence pairs.
There are no atoms in both
$\SequenceOf{\StringCF{"<Field Id=2></Field>"}}$ and
$\SequenceOf{\StringCF{""}}$, trivially creating the 
sequence lens:
\begin{lstlisting}
$(\SequenceLensOf{(\StringCF{"<Field
    Id=2></Field>"},\StringCF{""})},\Identity)$
\end{lstlisting}
Next, the sequences
$\SequenceOf{\StringCF{"<Field Id=2>"} \SeqSep \TextChar
  \SeqSep \StringCF{""}
  \SeqSep \StarOf{\DNFOf{\SequenceOf{\TextChar}}}
  \SeqSep \StringCF{"<Field Id=2>"}}$
and
$\SequenceOf{\StringCF{"Title="} \SeqSep
  \VarCF{text\_char} \SeqSep
  \SeqSep \StringCF{""}
  \SeqSep \StarOf{\DNFOf{\SequenceOf{\TextChar}}}
  \SeqSep \StringCF{","}}$
would be sent
to \RigidSynthSequence{}.
In \RigidSynthSequence{}, the atoms would not be reordered, aligning \VarCF{text\_char} with
\VarCF{text\_char}, and $\StarOf{\DNFOf{\SequenceOf{\VarCF{text\_char}}}}$ with
$\StarOf{\DNFOf{\SequenceOf{\VarCF{text\_char}}}}$.
Immediately, \RigidSynthAtom{} finds the
identity transformation on \VarCF{text\_char}, and will recurse to find
$\IterateLensOf{(\DNFLensOf{(\SequenceLensOf{(\StringCF{""},\StringCF{""})
      \SeqSep \IdentityLensOf{\TextChar} \SeqSep (\StringCF{""},\StringCF{""}),
  \Identity})}, \Identity)}$ 
for $\StarOf{\DNFOf{\SequenceOf{\VarCF{text\_char}}}}$.
  Then, these generated atom lenses are combined into a
sequence lens.  Lastly, the two sequence lenses are used with the swapping
permutation to create the final DNF lens.

By incorporating information about how examples are parsed in the
orderings,
\SynthLens{} guarantees not only that there will be a lens between the regular
expressions, but also
that the lens will satisfy the examples.
For example if
$\Sequence \SequenceLeq^{\Examples} \SequenceAlt$ and
$\SequenceAlt \SequenceLeq^{\Examples} \Sequence$ (where
$\SequenceLeq^{\Examples}$ is the ordering incorporating example information)
then there is not only a
sequence lens between $\Sequence$ and $\SequenceAlt$, but there is one that also
satisfies the examples.
Incorporating parse tree information lets the synthesis algorithm differentiate
between previously indistinguishable
subcomponents; a \TextChar{} that parsed only \StringCF{"a"} would become
less than a \TextChar{} that parsed only \StringCF{"b"}.  The details of
these orderings are formalized in
\ifappendices
Section~\ref{alg-correctness}.
\else
the full version of this paper~\cite{?}.
\fi

\paragraph*{Correctness}
We have proven two theorems demonstrating the correctness of our algorithm.

\begin{theorem}[Algorithm Soundness]
  \label{thm:alg-soundness}
  For all lenses $\Lens$, regular expressions $\Regex$ and $\RegexAlt$, and
  examples $\Examples$, 
  if $\Lens = \SynthLens(\Regex,\RegexAlt,\Examples)$, then
  $\Lens \OfType \Regex \Leftrightarrow \RegexAlt$ and for all
  $(\String,\StringAlt)$ in $\Examples$, $(\String,\StringAlt) \in
  \SemanticsOf{\Lens}$.
\end{theorem}

\begin{theorem}[Algorithm Completeness]
  \label{thm:alg-completeness}
  Given regular expressions $\Regex$ and $\RegexAlt$, and a set
  of examples $\Examples$, if there exists a lens $\Lens$ such that
  $\Lens \OfType \Regex \Leftrightarrow \RegexAlt$ and for all
  $(\String,\StringAlt)$ in $\Examples$, $(\String,\StringAlt) \in
  \SemanticsOf{\Lens}$, then $\SynthLens(\Regex,\RegexAlt,\Examples)$ will
  return a lens.
\end{theorem}

Theorem~\ref{thm:alg-soundness} states that when we return a lens, that 
lens will match the specifications.  Theorem~\ref{thm:alg-completeness} states
that if there is a DNF lens that satisfies the specification, then we will
return a lens, but not necessarily the same one.  However, from
Theorem~\ref{thm:alg-soundness}, we know that this lens will match the
specifications.  The proofs of these theorems, and the previous ones,
are
\ifappendices
provided in the appendix.
\else
submitted to POPL 2018 as a part of the anonymous supplementary.
material.
\fi

\paragraph*{Simplification of Generated Lenses}
While our system takes in only partial specifications, there can be multiple
lenses that satisfy the specifications.  To help users
determine if the synthesized lens is correct, \Optician{} transforms the
generated code to make it easily readable.  \Optician{} (1) maximally 
factors the $\ConcatLens$s and $\OrLens$s, (2) turns lenses that perform identity
transformations into identity lenses, and (3) simplifies the regular expressions
the identity lenses take as an argument.  Performing these transformations and
pretty printing the generated lenses make the synthesized lenses easy to
understand.

\paragraph*{Compositional Synthesis}
Most synthesis problems can be divided into subproblems.  For example, if the
format $\Regex_1 \Concat \Regex_2$ must be converted into $\RegexAlt_1 \Concat
\RegexAlt_2$, one might first work on the $\Regex_1 \Leftrightarrow \RegexAlt_1$ and
$\Regex_2 \Leftrightarrow \RegexAlt_2$ subproblems.  After those subproblems have
been solved, the lenses they generate can be combined into a solution for
$\Regex_1 \Concat \Regex_2 \Leftrightarrow \RegexAlt_1 \Concat \RegexAlt_2$.

Our tool allows users to specify multiple synthesis problems in a single file,
and allows the later, more complex problems to use the results generated by
earlier problems.  This tactic allows \Optician{} to scale to problems of just
about any size and complexity with just a bit more user input.  This
compositional interface also provides users greater control over the
synthesized lenses and allows reuse of intermediate synthesized abstractions.
The compositional synthesis engine allows
lenses previously defined manually by the user, and lenses in the Boomerang standard
library to be included in synthesis.

% end implementation

% begin evaluation
\section{Evaluation}
\label{evaluation}

We have implemented \Optician{} in \LOC{} lines
of OCaml code.  We have integrated our synthesis algorithm into Boomerang, so
users can input synthesis tasks in place of lenses.  We have published this code
on GitHub, with a link given in the non-anonymized supplementary material.

We evaluate our synthesis algorithm by applying it to 39{}
benchmark programs.
All evaluations were performed on a 2.5 GHz Intel Core i7 processor with 16 GB
of 1600 MHz DDR3 running macOS Sierra.

\paragraph*{Benchmark Suite Construction}
We constructed our benchmarks by adapting examples from
Augeas~\cite{augeas} and 
Flash Fill~\cite{gulwani-popl-2014} and by handcrafting specific
examples to test various 
features of the algorithm.

Augeas is a configuration editing system for Linux that uses lens
combinators similar to those in Boomerang. However, it transforms
strings on the left to structured trees on the right rather than
transforming strings to strings.
We adapted these Augeas lenses to our setting by converting the
right-hand sides to strings that correspond to serialized versions
of the tree formats.  
Augeas also supports {\em asymmetric lenses}~\cite{Focal2005-long},
which are more general than the bijective lenses \Optician{} can synthesize.
We adapted these examples by adding extra fields to the
target to make the transformations bijective and thus suitable
for our study.
We derived 29 of the benchmark tests by
adapting the first 27 lenses in alphabetical order, as well as the lenses
\CF{aug/xml-firstlevel} and \CF{aug/xml} that were referenced
by the `A' lenses.
Furthermore, the 12 last synthesis problems derived
from Augeas were tested after \Optician{} was
finalized, demonstrating that the optimizations were not
overtuned to perform well on the testing data.

Flash Fill is a system that allows users to specify common string
transformations by example~\cite{gulwani-popl-2014}.  
Many of the examples from Flash Fill are non-bijective because
the user's goal is often to extract information.  We were able to
adapt some examples by adding information to the target so the
resulting transformation was bijective.
We derived three benchmarks from examples in the
Flash Fill paper~\cite{gulwani-popl-2014} that were close to bijections.

Finally, we added custom examples to highlight weaknesses of
our algorithm (\CF{cap-prob} and \CF{2-cap-prob}) 
and to test situations for which we thought the tool would be
particularly useful (\CF{workitem-probs}, \CF{date-probs}, \CF{bib-prob},
and \CF{addr-probs}).   These examples convert between work item formats, date
formats, bibliography formats, and address formats, respectively.

We have both complex and simple synthesis tasks in our benchmark suite.
We generate lenses of sizes between 5 AST
nodes, for simple problems like changing how dates are represented, and
305 AST nodes, for complex tasks like transforming arbitrary XML of depth
3 or less to a dictionary representation.

\begin{figure}
  \includegraphics{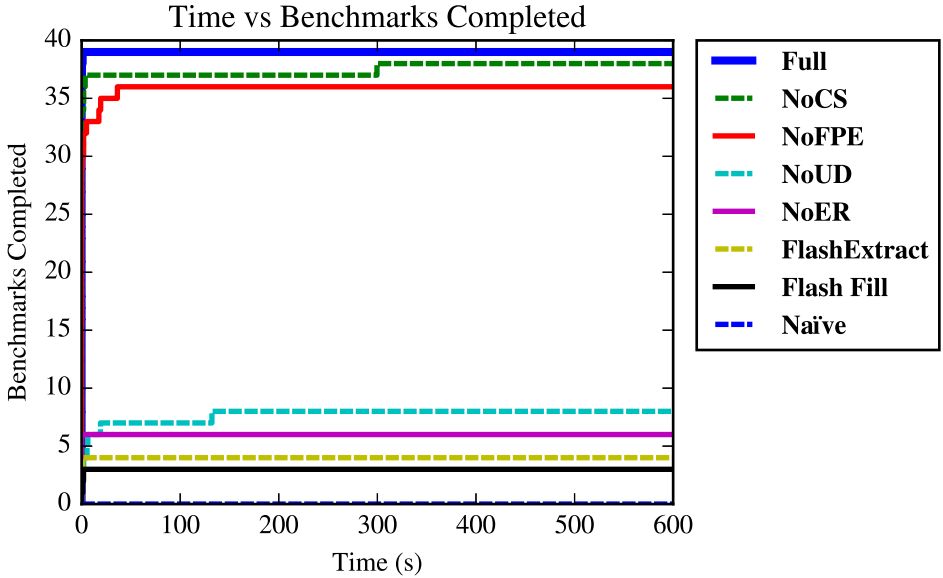}
  \caption{
    Number of benchmarks that can be solved by a given algorithm in a given
    amount of time. \FullMode{} is the full synthesis
    algorithm.
    \NoCSMode{} is
    the synthesis algorithm using all optimizations but without using
    a library of
    existing lenses.  \NoFPEMode{} is the core DNF synthesis algorithm
    augmented with user-defined data types with
    forced expansions performed.  \NoERMode{} is the core synthesis
    augmented with
    user-defined data types.  \NoUDMode{} is the core synthesis algorithm.
    \FlashExtractMode{} is the existing FlashExtract system.  \FlashFillMode{} is
    the existing Flash Fill system.  \NaiveMode{} is na\"{i}ve type-directed 
    synthesis on the bijective lens combinators.  Our synthesis algorithm performs
    better than the na\"{i}ve approach and other string transformation systems,
    and our optimizations speed up the algorithm enough that all tasks become
    solvable.
  }
  \label{fig:synthesis-times}
\end{figure}

\paragraph*{Impact of Optimizations}
We developed a series of optimizations that improve the performance of the
synthesis algorithm dramatically.  To determine the relative importance of these
optimizations, we developed the 5 different modes that run the synthesis
algorithm with various optimizations enabled.  These modes are:

\begin{itemize}
\item[\FullMode{}:] All optimizations are enabled, and compositional synthesis
  is used.
\item[\NoCSMode{}:] Like \FullMode{}, but compositional synthesis is not used.
\item[\NoFPEMode{}:] Like \NoCSMode{}, but \FixProblemElts{} is never called,
  expansions are only forced through \ExpandRequired{} or processed enumeratively
  through \ExpandOnce{}.
\item[\NoERMode{}:]  Like \NoFPEMode{}, but all the expansions taken are generated
  through enumerative search from \ExpandOnce{}.
\item[\NoUDMode{}:]  User-defined data types are no longer kept abstract until
  needed.  All user-defined regular expressions get replaced by their
  definition at the start of synthesis.
\end{itemize}

We ran \Optician{} in each mode over our benchmark suite.
We summarize the results of these tests in
Figure~\ref{fig:synthesis-times}.
\FullMode{} synthesized all 39 benchmarks, \NoCSMode{} synthesized 48
benchmarks, \NoFPEMode{} synthesized 36 benchmarks, \NoERMode{} synthesized 6
benchmarks, \NoUDMode{} synthesized 8 benchmarks, and \NaiveMode{} synthesized 0
benchmarks.  \Optician{}'s optimizations make
synthesis effective against a wide range of complex data formats.

Interestingly, \NoERMode{} performs \emph{worse} than \NoUDMode{}.
Adding in user defined data types introduces the additional search through
substitutions. 
The cost of this additional search outweighs the savings that these data type
abstractions provide.  In particular, because of the large fan-out of possible 
expansions, \NoERMode{} can only synthesize lenses which
require \ExpansionsPerformedNaiveExpansionSuccess{} or fewer expansions.  However, some lenses require over 50
expansions.  Without a way to intelligently traverse
expansions, the need to search through substitutions makes synthesis
unbearably slow.

\begin{figure}
  \includegraphics{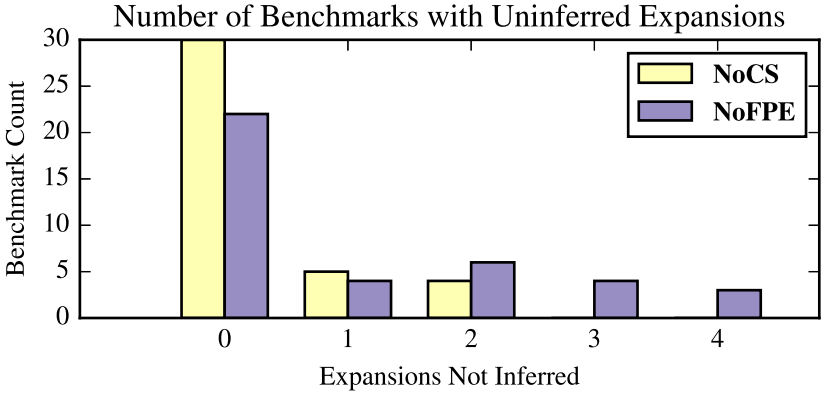}
  \caption{Number of expansions found using enumerative search for tasks
    requiring a given number of
    expansions.  \NoCSMode{} is using the full inference
  algorithm.
    \NoFPEMode{} only counts forced inferences as found by
  the \ExpandRequired{} function.  Both systems are able
    to infer the vast majority of expansions. Full inference only rarely
    requires enumerative search.}
  \label{fig:uninferred-exps}
\end{figure}

In \NoFPEMode{}, we can determine that many expansions are forced, so 
an enumerative search is often unnecessary.
Figure~\ref{fig:uninferred-exps} shows that in a majority of examples, all the
expansions can be identified as required, minimizing the impact of the
large fan-out.  While unable to infer every expansion for all the benchmarks,
the full algorithm is able to infer quite a bit.
In our benchmark suite, \ExpandRequired{} infers a median of 13
and a maximum of 75 expansions.

Merely inferring the forced expansions makes almost all the synthesis tasks
solvable.  In many cases, \NoFPEMode{} infers
\emph{all} the expansions.  In \ExpansionsAllForcedNoLC{} of the
\BenchmarksCompletedNoLC{} examples 
solvable by \NoCSMode{}, all expansions
were forced.  However, the remaining \ExpansionsNotAllForcedNoLC{} still require
some enumerative search.
This enumerative search causes the \NoFPEMode{} version of the algorithm to
struggle with some of the more complex benchmarks.
Incorporating \FixProblemElts{} speeds up these slow benchmarks.  When
using full inference (\FixProblemElts{} and \ExpandRequired), the synthesis
algorithm can 
recognize that one of a few expansions must be performed.  Adding in these types of
inferred expansions directs the remaining search even more, both speeding up
existing problems and solving previously unmanageable benchmarks.

\begin{figure}
  \includegraphics{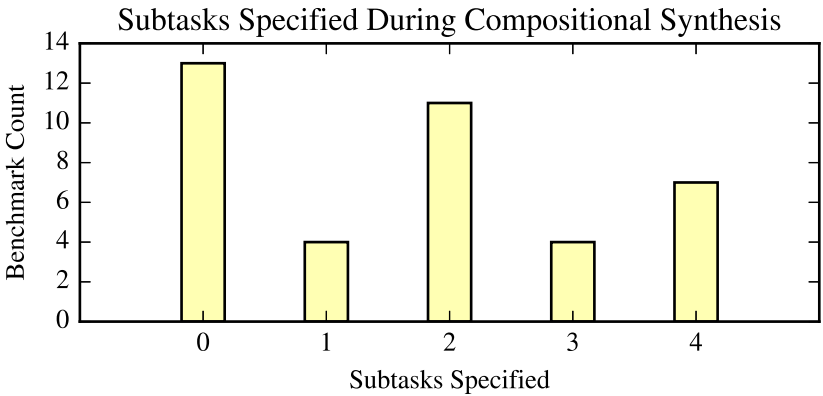}
  \caption{Number of subtasks specified during compositional synthesis.
    Splitting the task into just a few subtasks provides huge performance
    benefits at the cost of a small amount of additional user work.} 
  \label{fig:compositional-graph}
\end{figure}

When combined, these optimizations implement an efficient synthesis algorithm, which can
synthesize lenses between a wide range of data formats.  However, some of the
tasks are still slow, and one remains unsolved.
Using compositional synthesis lets the system scale to the most complex
synthesis tasks, synthesizing all lenses in under 5 seconds.
Additional user interaction is required for compositional synthesis, but the
amount of interaction is minimal, as shown in
Figure~\ref{fig:compositional-graph}.  The number of subtasks used was in no way
the minimal number of subtasks needed for synthesis under 5 seconds, but rather
subtasks were introduced where they naturally arose.

The benchmark that only completes with compositional synthesis is also
the slowest benchmark in \FullMode{},
\CF{aug/xml}\footnote{Since xml syntax is context-free, 
  the source and target regular 
  expressions describe only xml expressions up to depth 3.}.  \Optician{} can
only synthesize a lens for this example when compositional synthesis is used
because it is a 
complex data format, it requires a large
number of expansions, and relatively few expansions are forced.
When not using compositional synthesis, the algorithm must perform a
total of 
398~expansions, of which only 105 are forced.  
The synthesis algorithm is able to force so few expansions because of the highly
repetitive nature of the \CF{aug/xml} specification.  XML tags occur at many
different levels, and they all use the same user-defined data types.
This repetitive nature causes our
expansion inference to find only a few of the large number of required
expansions.  The large fan-out of expansions, combined with the large number
of expansions that must be performed, creates 
a search space too large for our algorithm to effectively search.  However, the
synthesis algorithm is able to succeed on the easier task of finding the desired
transformation when provided with two additional subtasks: synthesis on XML of
depth one, and synthesis of XML of depth up to two.

\paragraph*{Importance of Examples}

\begin{figure}
  \centering
  \includegraphics{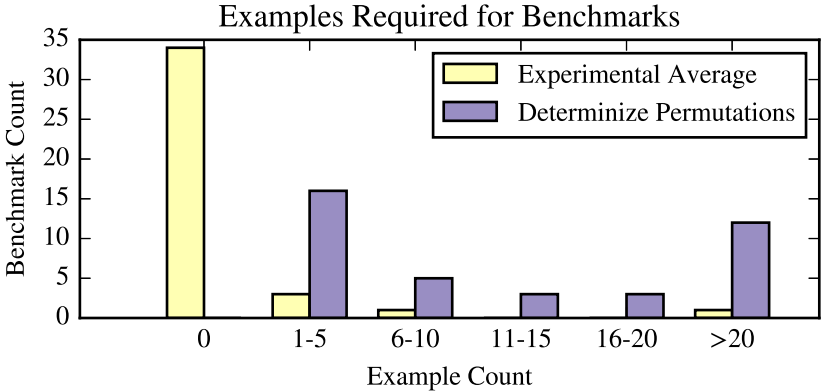}
  \caption{Average number of random examples required to synthesize benchmark
    programs.  {\bf Experimental Average} is the average number of randomly
    generated examples needed to correctly synthesize the lens.  {\bf
      Determinize Permutations} is the theoretical number of examples required
    to determinize the choice all the permutations in \RigidSynth{}.
    In practice, far fewer examples are
    needed to synthesize the correct lens than would be predicted by the number
    required to determinize permutations.}
  \label{fig:exs-reqd}
\end{figure}

To evaluate how many user-supplied examples the algorithm requires in
practice, we \textit{randomly} generated appropriate source/target
pairs, mimicking what a na\"{i}ve user might do.  We did not write the
examples by hand out of concern that our knowledge of the synthesis
algorithm might bias the selection. Figure~\ref{fig:exs-reqd} shows
the number of randomly generated examples it takes to synthesize the
correct lens averaged over ten runs.  The synthesis algorithm almost never needs
any examples: only \ExamplesRequiringNonzeroExamples{} benchmarks need a nonzero number of examples to
synthesize the correct lens and only one, \CF{cust/workitem-probs} required over
10 randomly generated examples.
A clever user may be able to reduce the
number of examples further by selecting examples carefully; we
synthesized \CF{cust/workitem-probs} with only 8 examples.

These numbers are low because there are relatively few well-typed
bijective lenses between any two source and target regular expressions. 
As one would expect, the benchmarks where there are multiple ways to
map source data to the target (and vice versa) require the most examples.
For example, the benchmark \CF{cust/workitem-probs} requires a large number of
examples because it
must differentiate between data in different text fields in both the
source and target and map between them appropriately.  As these text fields are
heavily permuted
(the legacy format ordered fields by a numeric ID, where
the modern format ordered fields alphabetically) and fields can be
omitted, a number of examples are needed to correctly identify the mapping
between fields.

The average number of examples to
infer the correct lens does not tell the whole story.  The system will
stop as soon as it finds a well typed lens that satisfies the supplied examples.
This inferred lens may or may not 
correctly handle unseen examples that correspond to
unexercised portions of the source and target regular expressions.
Figure~\ref{fig:exs-reqd} lists
the number of examples that are required to determinize the generation of
permutations in \RigidSynth{}.
Intuitively, this number represents the maximum number of
examples that a user must supply to guide the synthesis engine if it
always guesses the wrong permutation when multiple permutations can be used to
satisfy the specification. 

The average number of examples is so much lower than the maximum
number of required examples because of correspondences in how we wrote
the regular expressions for the source and target data formats. 
Specifically, when we had corresponding disjunctions in both the
source and the target, we ordered them the same way.  The algorithm
uses the supplied ordering to guide its search, and so the system
requires fewer examples.   We did not write the examples in this style
to facilitate synthesis, but rather because maintaining similar
subparts in similar orderings makes the types much easier to 
read. We expect that most users would do the same.

\paragraph*{Comparison Against Other Tools}
We are the first tool to synthesize bidirectional transformations between data
formats, so there is no tool to which we can make an apple-to-apples comparison.
Instead, we compare against tools for generating unidirectional
transformations instead. 
Figure~\ref{fig:synthesis-times} includes a comparison against two other
well-known tools that synthesize
text transformation and extraction functions from examples -- Flash Fill and FlashExtract.  For this
evaluation, we used the version of these tools distributed through the
PROSE project~\cite{prose}.

To generate specifications for Flash Fill, we generated input/output
specifications by generating random elements of the source language, and
running the lens on those elements to generate elements of the target language.
These were then fed to Flash Fill.

To generate specifications for FlashExtract, we extracted portions of strings
mapped in the generated lens either through an identity
transformation or through a previously synthesized lens, whereas strings that were
mapped through use of \ConstLens{} were considered boilerplate and so not
extracted.

As these tools were designed for a broader audience, they put less of a burden
on the user.  These tools only use input/output examples (for Flash
Fill), or marked text regions (for FlashExtract), as opposed
to \Optician{}'s use of regular expressions to constrain the format of
the input and output.  By using regular expressions,
\Optician{} is able to synthesize significantly more programs
than either existing tool.

Flash Fill and FlashExtract have two tasks: to determine how
the data is transformed, they must also infer the structure of the data, a
difficult job for complex formats.
In particular, neither Flash Fill nor FlashExtract was able to synthesize
transformations or extractions present under two iterations, a type of format
that is notoriously hard to infer.
These types of dual iterations are pervasive in Linux configuration
files, making Flash Fill and FlashExtract ill suited for many of the synthesis
tasks present in our test suite.

Furthermore, as unidirectional transformations, Flash Fill and FlashExtract have
a more expressive calculus.  To guarantee bidirectionality, our syntax must be
highly restrictive, providing a smaller search space to traverse.
%% \paragraph*{Impact of Incompleteness}

%% In synthesizing the lenses for our benchmark suite, we did not miss
%% the ability to use retyping with full regular expression equivalence.
%% The only regular expression equivalences that arose were ones that we
%% had already incorporated in our reduced set of equivalences.
%% For example, the only places where synthesis
%% used \StarRegexType{} equivalences were in varying the number
%% of iterations and in placing separators, both of 
%% which we accounted for in $\SSREquiv$.

%We believe the full generality is not necessary because 
%This lack of impact is likely because, when the languages of two
%regular expressions have 
%bijective lenses between them, the data is naturally grouped in similar ways on each
%side of the regular expression.  If there was a block of text, it would likely
%be written on both sides as \CF{Character*}, instead of \CF{(A*notA)*A*}, where
%\CF{A} is the character \CF{A}, and \CF{notA} is all characters except for
%\CF{A}.  If for some reason it were important that \CF{A} and \CF{notA} were
%separated, we found that it would be important in both data
%representations.  

% end evaluation

% begin related-work
\section{Related Work}
\label{sec:related}

\ifdraft\else \bcp{More that we could include, but perhaps let's save it for the
  conference version if we are luck enough to get to write one... 
  \begin{itemize}
  \item 
    Other
    bidirectional languages include biXid~\cite{bixid} language, which
    expresses bidirectional XML transformations, and XSugar~\cite{xsugar}, which 
    converts bidirectionally between XML and non-XML data and between
    EpiDoc~\cite{epidoc} and Leiden+~\cite{leidenplus} formats for encoding
    ancient documents~\cite{epidocleidenplus}.
  \end{itemize}
}
\fi

In searching for equivalent regular expressions, we focused on
Conway's equational theory rather than
alternative axiomatizations such as Kozen's~\cite{kozen-complete}
and Salomaa's~\cite{salomaa-complete}.  Kozen and Salomaa's
axiomatizations are not equational theories: applying certain
inference rules requires that side conditions must be satisfied.  
Consequently, using these
axiomatizations does not permit a simple search strategy -- our
algorithm could no longer merely apply rewrite rules because it would
need to confirm that the side conditions
are satisfied. To avoid these complications, we focused on 
Conway's equational theory.

The literature on bidirectional programming languages
and on lens-like structures is extensive.  We discussed
highlights in the introduction; readers can also
consult a (slightly dated)
survey~\cite{DBLP:conf/icmt/CzarneckiFHLST09} and
recent theoretical perspectives~\cite{DBLP:journals/chinaf/FischerHP15,
DBLP:conf/birthday/Abou-SalehCGMS16}.
%We focus here on two closely related type-based synthesis tools for string
%transformers described by regular expressions.

While we do not know of any previous efforts to synthesize bidirectional
transformations, there is a good deal of other recent research on
synthesizing unidirectional string
transformations~\cite{singh2012learning,le-pldi-2014,gulwani-popl-2014,perelman2014test,Singh:blinkfill}.
We compared our system to two of these unidirectional string transformers,
Flash Fill~\cite{gulwani-popl-2014} and FlashExtract~\cite{le-pldi-2014}.
We found that these tools were unsuccessful in synthesizing the complex
transformations we are performing -- both these tools synthesized under 5 of our
39 examples.  Furthermore, neither of these tools were able to infer
transformations 
which occurred under two iterations.
Much of this work assumes, like us, that the synthesis engine is provided with a
collection of 
examples.  Our work differs in that we assume the programmer supplies
both examples {\em and} format descriptions in the form of regular expressions.
There is a trade-off here.  On the one hand, a user must have some 
programming expertise to
write regular expression specifications
and it requires some work.
On the other hand, such specifications provide a great deal of information
to the synthesis system, which decreases the number of examples needed
(often to zero), makes the system scale well, and
allows it to handle large, complex formats, as shown
in \S \ref{evaluation}.  By providing these format specifications, the synthesis
engine does not have to both infer the format of the data as well as the
transformations on it, obviating the need to infer tricky formats like those
involving nested iterations.
Furthermore, through focusing on bidirectional transformations we limit the
space of synthesized functions to bijective ones, reducing the search space.

There are many other recent results showing how to synthesize functions from
type-based
specifications~\cite{augustsson-2004,osera+:pldi15,feser-pldi-2015,scherer-icfp-2015,frankle+:popl16,armando+:pldi16}.
These systems enumerate programs of their target language, orienting their
search procedures to process only terms that are well-typed.
Our system is distinctive in that it synthesizes terms in a language with many
type equivalences. 
Perhaps the most similar is \InSynth{}~\cite{gvero-pldi-2013}, a system for
synthesizing terms in the simply-typed lambda calculus that addresses
equivalences on types.  Instead of trying to directly synthesize terms of the
simply-typed lambda calculus, \InSynth{} synthesizes a well-typed term
in the succinct calculus, a language with types
that are equivalent ``modulo isomorphisms of products and
currying''~\cite{gvero-pldi-2013}.
Our type structure is significantly more complex.  In particular, because our 
types do not have full canonical forms, we
use a pseudo-canonical form, which captures part of the equivalence
relation over types.  To preserve completeness, we push some of the remaining
parts of the type equivalence relation into a set of rewriting rules and
other parts into
the \RigidSynth{} algorithm itself.
%% Furthermore, to maintain completeness, we cannot identify all equivalent types
%% into the same syntactic form.
%% Instead \TypeProp{} must traverse
%% the star-semiring equivalences to propose candidate types.

\Morpheus{}~\cite{morpheus} is another synthesis system that uses two
communicating synthesizers to generate programs.  In both \Morpheus{} and
\Optician{}, one synthesizer provides an 
outline for the program, and the other fills in that outline with program
details that satisfy the user's specifications.
This approach works well in large search spaces, which
require some enumerative search.
Our systems differ in that an outline for \Morpheus{} is a sketch---an 
\emph{expression}
containing holes---whereas
an outline for \Optician{} is a pair of DNF regular
expressions, i.e., a 
\emph{type}.  Moreover, in order to implement an efficient
search procedure, we had to create both a new type language and a new
term language for lenses.  Once we did so, we proved our new, more
constrained language
designed for synthesis was just as expressive as the original, more
flexible and compositional language designed for human programmers.
%% To na\"ively search through Boomerang lenses 
%% with one synthesizer processing type equivalences and the other generating
%% lenses, we would lose expressivity; only through changing the language and
%% typing derivations did this approach apply.  In particular, by creating a
%% language closed under composition, all applications of type equivalence can be
%% pushed to the end of the typing derivation.

Many synthesis algorithms work on domain-specific
languages custom built for
synthesis~\cite{flashfill,le-pldi-2014,solar-lezama-thesis-2008,yag+:pldi16}.  We
too built a custom domain-specific 
language for synthesis -- DNF lenses.  We provide the capabilities to
convert specifications in an existing language, Boomerang, to specifications as
DNF regular expressions, and provide the 
capabilities to convert our generated DNF lenses to Boomerang lenses.
But we go further than merely providing a converter to Boomerang, we also
provide completeness results stating exactly which terms of Boomerang we are
able to synthesize. 

\section{Conclusion}
\label{sec:conc}

Data processing systems often need to convert data back-and-forth
between different formats.  Domain-specific languages for generating
bidirectional programs help prevent data corruption in such contexts,
but are unfamiliar and hard to use.  To simplify the development of
bidirectional applications, we have created the first synthesizer of a
bidirectional language, generating
lenses from data format specifications and input/output examples.
To reduce the size of the synthesis search space, our system introduces
a new language of DNF lenses, which are typed by DNF regular expressions.
We have proven our new language sound and complete with respect to
a declarative specification.  We also describe effective optimizations
for efficiently searching through the refined space of lenses.

We evaluated our system on a range of practical examples drawn from
other systems in the literature including Flash Fill and Augeas.  In
general, we found our system to be robust, to require few examples,
and to finish in seconds, even on complex data formats.  We found that our
type-directed synthesis algorithm is able to generate data
transformations too complex for both existing example-directed systems and for a
na\"{i}ve type-directed algorithm, succeeding on \NumMoreThanFlashFill{} more
benchmarks than 
the tested existing alternatives.
We attribute its success to a combination of (1) the information provided
by format specifications, (2) the structure induced by user-specified
names, and (3) the inferences used to guide search.
The approaches we used are generalizable both to other bidirectional languages,
as well as to other domain-specific languages with large numbers of equivalences
on the types.

\ifdraft
\paragraph*{Future Work}
Some of the issues seen in this work have already been noted.  For example, in the
quotient lens paper, the issue of having to hack a place in the source string to
move white-space information from the view to the source was noted.  An extension
we would like to work on would be able to synthesize quotient lenses.

Another approach to having disparate information on each side of the lens is
through the use of symmetric lenses.  Symmetric lenses could be another approach to the
white-space problem.

\bcp{Talk about Dave and Kathleen's work on synthesizing regex types.}
\fi

% end conclusion

\appendix

\ifanon\else
\fi

% We recommend abbrvnat bibliography style.

\bibliographystyle{abbrvnat}

% The bibliography should be embedded for final submission.

\bibliography{local,bcp}

% Appendices.
\ifappendices

\onecolumn
\section{Formal Definitions}
\begin{definition}[Unambiguous Concatenation Language]
  If $\Language_1$ and $\Language_2$ are languages, such that
  for all strings $\String_1,\StringAlt_1 \in \Language_1$, and for all strings
  $\String_2,\StringAlt_2 \in \Language_2$, If $\String_1\Concat\String_2 =
  \StringAlt_1\Concat\StringAlt_2$, then $\Language_1$ is \textit{unambiguously
    concatenable} with $\Language_2$, written
  $\UnambigConcatOf{\Language_1}{\Language_2}$.
\end{definition}

\begin{definition}[Conway's Regular Expression Equivalences]\leavevmode\\
  \begin{tabular}{@{}r@{\hspace{1em}}c@{\hspace{1em}}l@{}r@{}}
    \RegexOr{\Regex}{\emptyset} & $\equiv$ & \Regex{} & \OrIdentityRule{} \\
    $\RegexConcat{\Regex}{\emptyset}$ & $\equiv$ & $\emptyset$ & \EmptyProjectionRightRule{} \\
    $\RegexConcat{\emptyset}{\Regex}$ & $\equiv$ & $\emptyset$ & \EmptyProjectionLeftRule{} \\
    \RegexConcat{(\RegexConcat{\Regex{}}{\Regex'})}{\Regex''} & $\equiv$ & \RegexConcat{\Regex{}}{(\RegexConcat{\Regex'}{\Regex''})} & \ConcatAssocRule{}  \\
    \RegexOr{(\RegexOr{\Regex}{\Regex'})}{\Regex''} & $\equiv$ & \RegexOr{\Regex}{(\RegexOr{\Regex'}{\Regex''})} & \OrAssociativityRule{}  \\
    \RegexOr{\Regex{}}{\RegexAlt{}} & $\equiv$ & \RegexOr{\RegexAlt{}}{\Regex{}} & \OrCommutativityRule{}\\
    \RegexConcat{\Regex{}}{(\RegexOr{\Regex{}'}{\Regex{}''})} & $\equiv$ & \RegexOr{(\RegexConcat{\Regex{}}{\Regex{}'})}{(\RegexConcat{\Regex{}}{\Regex{}''})} & \DistributivityLeftRule{} \\
    \RegexConcat{(\RegexOr{\Regex{}'}{\Regex{}''})}{\Regex{}} & $\equiv$ & \RegexOr{(\RegexConcat{\Regex{}'}{\Regex{}})}{(\RegexConcat{\Regex{}''}{\Regex{}})} & \DistributivityRightRule{} \\
    \RegexConcat{\EmptyString{}}{\Regex{}} & $\equiv$ & \Regex{} & \ConcatIdentityLeftRule{} \\
    \RegexConcat{\Regex{}}{\EmptyString{}} & $\equiv$ & \Regex{} & \ConcatIdentityRightRule{} \\
    \StarOf{(\RegexOr{\Regex{}}{\RegexAlt{}})} & $\equiv$ & \RegexConcat{\StarOf{(\RegexConcat{\StarOf{\Regex{}}}{\RegexAlt{}})}}{\StarOf{\Regex{}}} & \SumstarRule{}\\
    \StarOf{(\RegexConcat{\Regex{}}{\RegexAlt{}})} & $\equiv$ & \RegexOr{\EmptyString{}}{(\RegexConcat{\RegexConcat{\Regex{}}{\StarOf{(\RegexConcat{\RegexAlt{}}{\Regex{}})}}}{\RegexAlt{}})} & \ProductstarRule{} \\
    ${(\Regex{}^*)}^*$ & $\equiv$ & \StarOf{\Regex{}} & \StarstarRule{} \\
    \StarOf{(\RegexOr{\Regex}{\RegexAlt})} & $\equiv$ & $\StarOf{(\RegexConcat{(\RegexOr{\Regex}{\RegexAlt})}{\RegexOr{\RegexAlt}{\RegexConcat{{(\RegexConcat{\Regex}{\StarOf{\RegexAlt}})}^n}{\Regex}}})}\Concat$ & \DicyclicityRule{}\\
             & & $(\EmptyString\Or(\RegexOr{\Regex}{\RegexAlt})\Concat$\\
             & & $({(\RegexConcat{\Regex}{\StarOf{\RegexAlt}})}^0\Or\ldots\Or{(\RegexConcat{\Regex}{\StarOf{\RegexAlt}})}^n))$
  \end{tabular}
%  \caption{Regular Expression Equivalences}
%  \label{fig:regex-equivalence-rules}
\end{definition}

\begin{definition}[Definitional Regular Expression Equivalences]\leavevmode\\
  \begin{tabular}{@{}r@{\hspace{1em}}c@{\hspace{1em}}l@{}r@{}}
    \RegexOr{\Regex}{\emptyset} & $\SSREquiv$ & \Regex{} & \OrIdentityRule{} \\
    $\RegexConcat{\Regex}{\emptyset}$ & $\SSREquiv$ & $\emptyset$ & \EmptyProjectionRightRule{} \\
    $\RegexConcat{\emptyset}{\Regex}$ & $\SSREquiv$ & $\emptyset$ & \EmptyProjectionLeftRule{} \\
    \RegexConcat{(\RegexConcat{\Regex{}}{\Regex'})}{\Regex''} & $\SSREquiv$ & \RegexConcat{\Regex{}}{(\RegexConcat{\Regex'}{\Regex''})} & \ConcatAssocRule{}  \\
    \RegexOr{(\RegexOr{\Regex}{\Regex'})}{\Regex''} & $\SSREquiv$ & \RegexOr{\Regex}{(\RegexOr{\Regex'}{\Regex''})} & \OrAssociativityRule{}  \\
    \RegexOr{\Regex{}}{\RegexAlt{}} & $\SSREquiv$ & \RegexOr{\RegexAlt{}}{\Regex{}} & \OrCommutativityRule{}\\
    \RegexConcat{\Regex{}}{(\RegexOr{\Regex{}'}{\Regex{}''})} & $\SSREquiv$ & \RegexOr{(\RegexConcat{\Regex{}}{\Regex{}'})}{(\RegexConcat{\Regex{}}{\Regex{}''})} & \DistributivityLeftRule{} \\
    \RegexConcat{(\RegexOr{\Regex{}'}{\Regex{}''})}{\Regex{}} & $\SSREquiv$ & \RegexOr{(\RegexConcat{\Regex{}'}{\Regex{}})}{(\RegexConcat{\Regex{}''}{\Regex{}})} & \DistributivityRightRule{} \\
    \RegexConcat{\Regex{}}{\EmptyString{}} & $\SSREquiv$ & \Regex{} & \ConcatIdentityLeftRule{} \\
    \RegexConcat{\Regex{}}{\EmptyString{}} & $\SSREquiv$ & \Regex{} & \ConcatIdentityRightRule{} \\
    \StarOf{\Regex{}} & $\SSREquiv$ & \RegexOr{\EmptyString{}}{(\RegexConcat{\Regex{}}{\StarOf{{\Regex{}}}})} & \UnrollstarLeftRule{} \\
    \StarOf{\Regex{}} & $\SSREquiv$ & \RegexOr{\EmptyString{}}{(\RegexConcat{\StarOf{{\Regex{}}}}{\Regex{}})} & \UnrollstarRightRule{} 
  \end{tabular}
%  \caption{Definitional Regular Expression Equivalences}
%  \label{fig:definitional-equivalence-rules}
\end{definition}
% begin proofs
\section{Proofs}
% proof-dnfrs start
% First we will prove some lemmas.

The proof is split into separate subsections based on what is being done.  The
overall goals are to prove soundness and completeness of DNF regular expressions
with respect to regular expressions, and soundness and completeness of DNF
lenses with respect to lenses.

\begin{itemize}
\item Subsection~\ref{confluence-proofs} defines
  confluence with respect to a property, bisimilarity, and makes some general
  proofs about those properties.  These are used later for the proof of
  confluence of rewriting with respect to semantics, which is used in lens
  completeness.

\item Subsection~\ref{language-proofs} proves some general statements about
  languages, relating to the relationship between nonintersection of pairs of
  languages, and sets of languages, and the relationship between shared prefixes
  and suffixes of pairs of languages.  These are used for proving statements
  about unambiguity of DNF regular expressions from unambiguous regular
  expressions, and vice-versa.

\item Subsection~\ref{basic-property-proofs} proves some intuitive statements
  about lenses and DNF lenses.  These statements are properties like inversion,
  closure under composition for rewriteless DNF lenses, and proves that
  bijective lenses and bijective DNF lenses actually express bijections between
  the languages of their types.

\item Subsection~\ref{dnf-regex} proves soundness and completeness of DNF
  regular expressions to regular expressions.

\item Subsection~\ref{language-rewriting-unambiguity} proves statements relating
  to the retention of unambiguity across languages.  In particular, it proves
  statements about $\ToDNFRegex$ and $\ToRegex$, and also proves statements
  about the retention of unambiguity through rewrites.

\item Subsection~\ref{rewrite-proofs} proves statements about the retention of
  language through proofs, and the equivalences of expressibility of various
  rewrite systems.

\item Subsection~\ref{soundness} proves the soundness of DNF lenses to lenses,
  using the machinery above.

\item Subsection~\ref{dnf-lens-operators} defines operators on DNF lenses, which
  provide combinators similar to the combinators of normal lenses.  This section
  also proves statements about these combinators, like how the combinators act
  similarly to normal lenses.

\item Subsection~\ref{complex-lens-operators} proves more complex properties
  about lens operators.  These more complex statements are needed because DNF
  regular expressions don't have rewrites that order clauses.

\item Subsection~\ref{rewrite-maintenence} proves statements about the ability
  to build up rewrites on DNF regular expressions composed of less complex ones,
  from the rewrites of those less complex DNF regular expressions.  It also
  proves the proof of confluence of rewrites.

\item Subsection~\ref{completeness} proves the completeness of dnf lenses with
  respect to lenses.

\item Subsection~\ref{alg-correctness} proves the algorithm correct.

\item Subsection~\ref{additional-proofs} proves some random statements we make,
  but don't formally express, in the paper.
\end{itemize}

\subsection{Confluence Proofs}
\label{confluence-proofs}

This section begins by defining confluence and bisimilarity.
Next we prove that if a rewrite
system is bisimilar with respect to a property, then the transitive and
reflexive closure of that rewrite
system is too.  Next, a similar statement about transitive and reflexive closure
of rewrite systems for confluence is proven, under the conditions that the
property confluence is defined with respect to is transitive.
Next, propagators are defined, and used in if a rewrite system is confluent with
respect to a property with left and right propagators, then
the transitive and reflexive closures of that rewrite system is confluent with
respect to the same property.

\begin{definition}
  Let $\rightarrow$ and $\Property$ be two binary relations on a set $\Set$.
  We say that $\rightarrow$ is confluent with respect to
  $\Property$, written
  $\IsConfluentWithPropertyOf{\rightarrow}{\Property}$, if, given $x_1,x_2\in\Set$,
  where $\Property(x_1,x_2)$, if $x_1\rightarrow x_1'$ and
  $x_2 \rightarrow x_2'$, then there exists $x_1''$ and $x_2''$ such that
  $x_1'\rightarrow x_1''$, $x_2' \rightarrow x_2''$, and
  $\Property(x_1'',x_2'')$.
\end{definition}

\begin{definition}
  Let $\rightarrow$ and $\Property$ be two binary relations on a set $\Set$.
  We say that $\rightarrow$ is bisimilar through $\Property$, written
  $\IsBisimilarWithPropertyOf{\rightarrow}{\Property}$, if, given
  $x_1,x_2\in\Set$,
  where $\Property(x_1,x_2)$, if $x_1\rightarrow x_1'$ then there
  exists some $x_2$ such that $x_2\rightarrow x_2'$ where
  $\Property(x_1',x_2')$,
  and if $x_2\rightarrow x_2'$, then there exists some $x_1'$ such that
  $x_1\rightarrow x_1'$ where $\Property(x_1',x_2')$.
\end{definition}

\begin{definition}
  Let $\Property$ be a binary relation.  $\StarOf{\Property}$ is the binary
  relation defined via the inference rules
  \begin{mathpar}
    \inferrule[\ReflexivityRule]
    {
    }
    {
      \StarOf{\Property}(x,x)
    }

    \inferrule[\BaseRule]
    {
      \Property(x,y)
    }
    {
      \StarOf{\Property}(x,y)
    }

    \inferrule[\TransitivityRule]
    {
      \StarOf{\Property}(x,y)\\
      \StarOf{\Property}(y,z)
    }
    {
      \StarOf{\Property}(x,z)
    }
  \end{mathpar}
\end{definition}

\begin{lemma}[Bisimilarity Preserved through Star left]
  \label{lem:bisimilarity-star-left}
  Let $\IsBisimilarWithPropertyOf{\rightarrow}{\Property}$.
  If $\Property(x,y)$ and $x\StarOf{\rightarrow} x'$ then there
  exists some $y'$ such that $y\StarOf{\rightarrow} y'$ where
  $\Property(x',y')$
\end{lemma}
\begin{proof}
  By induction on the derivation of $x \StarOf{\rightarrow} x'$.

  \begin{case}[\ReflexivityRule{}]
    \[
      \inferrule*
      {
      }
      {
        x \StarOf{\rightarrow} x
      }
    \]

    Consider the derivation
    \[
      \inferrule*
      {
      }
      {
        y \StarOf{\rightarrow} y
      }
    \]

    and by assumption $\Property(x,y)$.
  \end{case}

  \begin{case}[\BaseRule{}]
    \[
      \inferrule*
      {
        x \rightarrow x'
      }
      {
        x \StarOf{\rightarrow} x'
      }
    \]

    As $\IsBisimilarWithPropertyOf{\rightarrow}{\Property}$, $y \rightarrow y'$
    where $\Property(x',y')$.
    \[
      \inferrule*
      {
        y \rightarrow y'
      }
      {
        y \StarOf{\rightarrow} y'
      }
    \]
  \end{case}

  \begin{case}[\TransitivityRule{}]
    \[
      \inferrule*
      {
        x \StarOf{\rightarrow} x''\\
        x'' \StarOf{\rightarrow} x'
      }
      {
        x \StarOf{\rightarrow} x'
      }
    \]

    By IH, $y \StarOf{\rightarrow} y''$ where $\Property(x'',y'')$.
    By IH, $y'' \StarOf{\rightarrow} y'$ where $\Property(x',y')$.
    \[
      \inferrule*
      {
        y \StarOf{\rightarrow} y''\\
        y'' \StarOf{\rightarrow} y'
      }
      {
        y \StarOf{\rightarrow} y'
      }
    \]
  \end{case}
\end{proof}

\begin{lemma}[Bisimilarity Preserved through Star right]
  \label{lem:bisimilarity-star-right}
  Let $\IsBisimilarWithPropertyOf{\rightarrow}{\Property}$.
  If $\Property(x,y)$ and $x\StarOf{\rightarrow} x'$ then there
  exists some $y'$ such that $y\StarOf{\rightarrow} y'$ where
  $\Property(x',y')$
\end{lemma}
\begin{proof}
  Symmetrically to Lemma~\ref{lem:bisimilarity-star-left}.
\end{proof}

\begin{lemma}[Bisimilarity Preserved through Star]
  \label{lem:bisimilarity-star}
  If $\IsBisimilarWithPropertyOf{\rightarrow}{\Property}$, then
  $\IsBisimilarWithPropertyOf{\StarOf{\rightarrow}}{\Property}$.
\end{lemma}
\begin{proof}
  By application of Lemma~\ref{lem:bisimilarity-star-left} and
  Lemma~\ref{lem:bisimilarity-star-right}.
\end{proof}

\begin{lemma}
  \label{lem:pre-starred-confluence-propagator-like}
  If $\IsConfluentWithPropertyOf{\rightarrow}{\Property}$,
  $\IsBisimilarWithPropertyOf{\rightarrow}{\Property}$,
  $\Property(x,y) \BooleanAnd \Property(y,z) \BooleanImplies
  \Property(x,z)$, and
  $\Property(x,y) \BooleanImplies \Property(x,x) \BooleanAnd \Property(y,y)$
  then if $\Property(x,y)$, $x\StarOf{\rightarrow} x_1$, $y\rightarrow
  x_1$, then there exists some $x_2$, $x_y$ such that
  $x_1\rightarrow x_2$, $y_1\StarOf{\rightarrow} y_2$, and
  $\Property(x_2,y_2)$.
\end{lemma}
\begin{proof}
  By induction on the derivation of $x \StarOf{\rightarrow} x_1$.

  \begin{case}[\ReflexivityRule]
    \[
      \inferrule*
      {
      }
      {
        x \StarOf{\rightarrow} x
      }
    \]

    By $\IsBisimilarWithPropertyOf{\rightarrow}{\Property}$, there exists
    some $x_1$ such that $x \rightarrow x_1$ and $\Property(x_1,y_1)$.
    Furthermore,
    \[
      \inferrule*
      {
      }
      {
        y_1 \StarOf{\Rewrite} y_1
      }
    \]
    so we are done.
  \end{case}

  \begin{case}[\BaseRule{}]
    \[
      \inferrule*
      {
        x \rightarrow x_1
      }
      {
        x \StarOf{\rightarrow} x_1
      }
    \]

    As $\IsConfluentWithPropertyOf{\rightarrow}{\Property}$, there exists
    $x_2$, $y_2$ such that $x_1 \rightarrow x_2$, $y_1 \rightarrow y_2$,
    and $\Property(x_2,y_2)$.
    Furthermore
    \[
      \inferrule*
      {
        y_1 \rightarrow y_2
      }
      {
        y_1 \StarOf{\rightarrow} y_2
      }
    \]
  \end{case}

  \begin{case}[\TransitivityRule{}]
    \[
      \inferrule*
      {
        x \StarOf{\rightarrow} x_1\\
        x_1 \StarOf{\rightarrow} x_2
      }
      {
        x_1 \StarOf{\rightarrow} x_2
      }
    \]

    By IH, there exists $x_3$, $y_2$ such that $x_1 \rightarrow x_3$, and
    $y_1 \StarOf{\rightarrow} y_2$, and $\Property(x_3,y_2)$.

    As $\Property(x,y)$, we have $\Property(x,x)$.  As $\Property(x,x)$, and $x
    \StarOf{\rightarrow} x_1$, then there exists $x'$ such that
    $\Property(x_1,x')$, so $\Property(x_1,x_1)$.
    So, by IH, as $\Property(x_1,x_1)$, $x_1 \StarOf{\rightarrow} x_2$, and
    $x_1 \rightarrow x_3$, there exists $x_4,x_5$ such that $x_2 \rightarrow
    x_4$, $x_3 \StarOf{\rightarrow} x_5$, and $\Property(x_4,x_5)$.

    As $\Property(x_3,y_2)$, and $x_3 \StarOf{\Rewrite} x_5$, then by
    $\IsBisimilarWithPropertyOf{\rightarrow}{\Property}$ and
    Lemma~\ref{lem:bisimilarity-star}, there exists $y_3$ such that
    $y_2 \StarOf{\Rewrite} y_3$, and $\Property(x_5,y_3)$.
    By \TransitivityRule{}, $y_1 \StarOf{\Rewrite} y_3$.
    From before, $x_2 \rightarrow x_4$.
    Because we have $\Property(x_4,x_5)$ and $\Property(x_5,y_3)$, we have
    $\Property(x_4,y_3)$.
  \end{case}
\end{proof}

\begin{lemma}
  \label{lem:starred-confluence-propatator-like}
  If $\IsConfluentWithPropertyOf{\rightarrow}{\Property}$,
  $\IsBisimilarWithPropertyOf{\rightarrow}{\Property}$, and
  $\Property(x,y) \BooleanAnd \Property(y,z) \BooleanImplies
  \Property(x,z)$, and
  $\Property(x,y) \BooleanImplies \Property(x,x) \BooleanAnd \Property(y,y)$
  then if $\Property(x,y)$, $x \StarOf{\rightarrow} x_1$,
  $y \StarOf{\rightarrow} x_1$, then there exists some $x_2$, $x_y$ such that
  $x_1 \StarOf{\rightarrow} x_2$, $y_1\StarOf{\rightarrow} y_2$, and
  $\Property(x_2,y_2)$.
\end{lemma}
\begin{proof}
  By induction on the derivation of $y \StarOf{\rightarrow} y_1$.

  \begin{case}[\ReflexivityRule]
    \[
      \inferrule*
      {
      }
      {
        y \StarOf{\rightarrow} y
      }
    \]

    By $\IsBisimilarWithPropertyOf{\rightarrow}{\Property}$,
    and Lemma~\ref{lem:bisimilarity-star},
    there exists
    some $y_1$ such that $y \StarOf{\rightarrow} y_1$ and $\Property(x_1,y_1)$.
    Furthermore,
    \[
      \inferrule*
      {
      }
      {
        y_1 \StarOf{\Rewrite} y_1
      }
    \]
    so we are done.
  \end{case}

  \begin{case}[\BaseRule{}]
    \[
      \inferrule*
      {
        y \rightarrow y_1
      }
      {
        y \StarOf{\rightarrow} y_1
      }
    \]

    As $\IsConfluentWithPropertyOf{\rightarrow}{\Property}$,
    $\IsBisimilarWithPropertyOf{\rightarrow}{\Property}$,
    $y \rightarrow y_1$, $x \StarOf{\rightarrow} x_1$, and
    $\StarOf{\Property}(x,y)$ if, and only if $\Property(x,y)$,
    by Lemma~\ref{lem:pre-starred-confluence-propagator-like},
    there exists
    $x_2$, $y_2$ such that $x_1 \rightarrow x_2$, $y_1 \StarOf{\rightarrow} y_2$,
    and $\Property(x_2,y_2)$.
    Furthermore
    \[
      \inferrule*
      {
        x_1 \rightarrow x_2
      }
      {
        x_1 \StarOf{\rightarrow} x_2
      }
    \]
  \end{case}

  \begin{case}[\TransitivityRule{}]
    \[
      \inferrule*
      {
        y \StarOf{\rightarrow} y_1\\
        y_1 \StarOf{\rightarrow} y_2
      }
      {
        y \StarOf{\rightarrow} y_2
      }
    \]

    By IH, there exists $x_2$, $y_3$ such that $x_1 \StarOf{\rightarrow} x_2$, and
    $y_1 \StarOf{\Rewrite} y_3$, and $\Property(x_2,y_3)$.

    As $\Property(x,y)$, we have $\Property(y,y)$.  As $\Property(y,y)$, and $y
    \StarOf{\rightarrow} y_1$, then there exists $y'$ such that
    $\Property(y_1,y')$, so $\Property(y_1,y_1)$.
    So, by IH, as $\Property(y_1,y_1)$, $y_1 \StarOf{\rightarrow} y_3$, and
    $y_1 \StarOf{\rightarrow} y_2$, there exists $y_4,y_5$ such that
    $y_3 \StarOf{\rightarrow} y_4$,
    $y_2 \StarOf{\rightarrow} y_5$, and $\Property(y_4,y_5)$.

    As $\Property(x_2,y_3)$, and $y_3 \StarOf{\Rewrite} y_4$, then by
    $\IsBisimilarWithPropertyOf{\rightarrow}{\Property}$ and
    Lemma~\ref{lem:bisimilarity-star}, there exists $x_3$ such that
    $x_2 \StarOf{\Rewrite} x_3$, and $\Property(x_3,y_4)$.
    By \TransitivityRule{}, $x_1 \StarOf{\Rewrite} x_3$.
    From before, $y_2 \rightarrow y_5$.
    As we have $\Property(x_3,y_4)$ and $\Property(y_4,y_5)$, we have
    $\Property(x_3,y_5)$.
  \end{case}
\end{proof}

\begin{definition}
  A property $\Propagator$ is a left propagator for $\Property$ with respect to
  $\rightarrow$ if
  $\IsBisimilarWithPropertyOf{\rightarrow}{\Propagator}$,
  $\IsConfluentWithPropertyOf{\rightarrow}{\Propagator}$,
  $\Propagator(x,y) \BooleanAnd \Propagator(y,z) \BooleanImplies \Propagator(x,z)$,
  $\Propagator(x,y) \BooleanImplies \Propagator(x,x) \BooleanAnd
  \Propagator(y,y))$,
  $\Property(x,y) \BooleanImplies \Propagator(x,x)$, and
  $\Propagator(x,y) \BooleanAnd \Property(y,z) \BooleanImplies \Property(x,z)$.
\end{definition}

\begin{definition}
  A property $\Propagator$ is a right propagator for $\Property$ with respect to
  $\rightarrow$ if
  $\IsBisimilarWithPropertyOf{\rightarrow}{\Property}$,
  $\IsConfluentWithPropertyOf{\rightarrow}{\Property}$,
  $\Propagator(x,y) \BooleanAnd \Propagator(y,z) \BooleanImplies \Propagator(x,z)$,
  $\Propagator(x,y) \BooleanImplies \Propagator(x,x) \BooleanAnd
  \Propagator(y,y))$,
  $\Property(x,y) \BooleanImplies \Propagator(y,y)$, and
  $\Property(x,y) \BooleanAnd \Propagator(y,z) \BooleanImplies \Property(x,z)$.
\end{definition}

\begin{lemma}
  \label{lem:pre-starred-confluence}
  Let $\IsConfluentWithPropertyOf{\rightarrow}{\Property}$.
  Let $\IsBisimilarWithPropertyOf{\rightarrow}{\Property}$.
  Let $\Propagator_L$ be a left propagator for $\Property$ with respect to
  $\rightarrow$.
  If $\Property(x_1,x_2)$, $x_1\StarOf{\rightarrow} x_1'$, $x_2\rightarrow
  x_2'$, then there exists some $x_1''$, $x_2''$ such that
  $x_1'\rightarrow x_1''$, $x_2'\StarOf{\rightarrow} x_2''$, and
  $\Property(x_1'',x_2'')$.
\end{lemma}
\begin{proof}
  By induction on the derivation of $x_1 \StarOf{\rightarrow} x_1'$.

  \begin{case}[\ReflexivityRule]
    \[
      \inferrule*
      {
      }
      {
        x_1 \StarOf{\rightarrow} x_1
      }
    \]

    By $\IsBisimilarWithPropertyOf{\rightarrow}{\Property}$, there exists
    some $x_1'$ such that $x_1 \rightarrow x_1'$ and $\Property(x_1',x_2')$.
    Furthermore,
    \[
      \inferrule*
      {
      }
      {
        x_2' \StarOf{\Rewrite} x_2'
      }
    \]
    so we are done.
  \end{case}

  \begin{case}[\BaseRule]
    \[
      \inferrule*
      {
        x_1 \rightarrow x_1'
      }
      {
        x_1 \StarOf{\rightarrow} x_1'
      }
    \]

    As $\IsConfluentWithPropertyOf{\rightarrow}{\Property}$, there exists
    $x_2$, $y_2$ such that $x_1 \rightarrow x_2$, $y_1 \rightarrow y_2$,
    and $\Property(x_2,y_2)$.
    Furthermore
    \[
      \inferrule*
      {
        y_1 \rightarrow y_2
      }
      {
        y_1 \StarOf{\rightarrow} y_2
      }
    \]
  \end{case}

  \begin{case}[\TransitivityRule{}]
    \[
      \inferrule*
      {
        x \StarOf{\rightarrow} x_1\\
        x_1 \StarOf{\rightarrow} x_2
      }
      {
        x_1 \StarOf{\rightarrow} x_2
      }
    \]

    By IH, there exists $x_3$, $y_2$ such that $x_1 \rightarrow x_3$, and
    $y_1 \StarOf{\Rewrite} y_2$, and $\Property(x_3,y_2)$.

    As $\StarOf{\Propagator_L}(a,b)$ if, and only if $\Propagator_L(a,b)$,
    $\Propagator_L(x_1,x_1)$.

    As $\Property(x,y)$, we have $\Propagator_L(x,x)$.
    As $\Propagator_L(x,x)$ and $x \StarOf{\rightarrow} x_1$, there exists $x'$
    such that $x \StarOf{\rightarrow} x'$, and $\Propagator_L(x_1,x')$, which
    means that $\Propagator_L(x_1,x_1)$.
    As $\Propagator_L$ is a propagator with respect to $\rightarrow$,
    it fills the properties required for
    Lemma~\ref{lem:pre-starred-confluence-propagator-like}, 
    So, as $\Propagator_L(x_1,x_1)$, $x_1 \StarOf{\rightarrow} x_2$, and
    $x_1 \rightarrow x_3$, there exists $x_4,x_5$ such that $x_2 \rightarrow
    x_4$, $x_3 \StarOf{\rightarrow} x_5$, and $\Propagator_L(x_4,x_5)$.

    As $\Property(x_3,y_2)$, and $x_3 \StarOf{\Rewrite} x_5$, then by
    $\IsBisimilarWithPropertyOf{\rightarrow}{\Property}$ and
    Lemma~\ref{lem:bisimilarity-star}, there exists $y_3$ such that
    $y_2 \StarOf{\Rewrite} y_3$, and $\Property(x_5,y_3)$.
    By \TransitivityRule{}, $y_1 \StarOf{\Rewrite} y_3$.
    From before, $x_2 \rightarrow x_4$.
    Because
    $\Propagator_L(x_4,x_5)$ and $\Property(x_5,y_3)$, and
    $\Propagator_L$ is a left propagator, $\Property(x_4,y_3)$.
  \end{case}
\end{proof}

\begin{theorem}[Confluence Preserved Through Star]
  \label{thm:starred-confluence}
  Let $\IsConfluentWithPropertyOf{\rightarrow}{\Property}$.
  Let $\IsBisimilarWithPropertyOf{\rightarrow}{\Property}$.
  Let $\Propagator_L$ be a left propagator for $\Property$ with respect to
  $\rightarrow$.
  Let $\Propagator_R$ be a right propagator for $\Property$ with respect to
  $\rightarrow$.
  If $\Property(x_1,x_2)$, $x_1\StarOf{\rightarrow} x_1'$, $x_2\rightarrow
  x_2'$, then there exists some $x_1''$, $x_2''$ such that
  $x_1'\rightarrow x_1''$, $x_2'\StarOf{\rightarrow} x_2''$, and
  $\Property(x_1'',x_2'')$.
\end{theorem}
\begin{proof}
  By induction on the derivation of $y \StarOf{\rightarrow} y_1$.

  \begin{case}[\ReflexivityRule]
    \[
      \inferrule*
      {
      }
      {
        y \StarOf{\rightarrow} y
      }
    \]

    By $\IsBisimilarWithPropertyOf{\rightarrow}{\Property}$,
    and Lemma~\ref{lem:bisimilarity-star},
    there exists
    some $y_1$ such that $y \StarOf{\rightarrow} y_1$ and $\Property(x_1,y_1)$.
    Furthermore,
    \[
      \inferrule*
      {
      }
      {
        y_1 \StarOf{\Rewrite} y_1
      }
    \]
    so we are done.
  \end{case}

  \begin{case}[\BaseRule{}]
    \[
      \inferrule*
      {
        y \rightarrow y_1
      }
      {
        y \StarOf{\rightarrow} y_1
      }
    \]

    As $\IsConfluentWithPropertyOf{\rightarrow}{\Property}$,
    $\IsBisimilarWithPropertyOf{\rightarrow}{\Property}$,
    $y \rightarrow y_1$, $x \StarOf{\rightarrow} x_1$, and
    $\StarOf{\Property}(x,y)$ if, and only if $\Property(x,y)$,
    by Lemma~\ref{lem:pre-starred-confluence},
    there exists
    $x_2$, $y_2$ such that $x_1 \rightarrow x_2$, $y_1 \StarOf{\rightarrow} y_2$,
    and $\Property(x_2,y_2)$.
    Furthermore
    \[
      \inferrule*
      {
        x_1 \rightarrow x_2
      }
      {
        x_1 \StarOf{\rightarrow} x_2
      }
    \]
  \end{case}

  \begin{case}[\TransitivityRule{}]
    \[
      \inferrule*
      {
        y \StarOf{\rightarrow} y_1\\
        y_1 \StarOf{\rightarrow} y_2
      }
      {
        y \StarOf{\rightarrow} y_2
      }
    \]

    By IH, there exists $x_2$, $y_3$ such that $x_1 \StarOf{\rightarrow} x_2$, and
    $y_1 \StarOf{\Rewrite} y_3$, and $\Property(x_2,y_3)$.

    As $\Property(x,y)$, we have $\Propagator_R(y,y)$.
    As $\Propagator_R(x,x)$ and $y \StarOf{\rightarrow} y_1$, there exists $y'$
    such that $y \StarOf{\rightarrow} y'$, and $\Propagator_R(y_1,y')$, which
    means that $\Propagator_R(y_1,y_1)$.
    As $\Propagator_L$ is a propagator with respect to $\rightarrow$,
    it fills the properties required for
    Lemma~\ref{lem:starred-confluence-propatator-like}, 
    So, as $\Propagator_R(y_1,y_1)$, $y_1 \StarOf{\rightarrow} y_3$, and
    $y_1 \StarOf{\rightarrow} y_2$, there exists $y_4,y_5$ such that
    $y_3 \StarOf{\rightarrow} y_4$,
    $y_2 \StarOf{\rightarrow} y_5$, and $\Propagator_R(y_4,y_5)$.

    As $\Property(x_2,y_3)$, and $y_3 \StarOf{\Rewrite} y_4$, then by
    $\IsBisimilarWithPropertyOf{\rightarrow}{\Property}$ and
    Lemma~\ref{lem:bisimilarity-star}, there exists $x_3$ such that
    $x_2 \StarOf{\Rewrite} x_3$, and $\Property(x_3,y_4)$.
    By \TransitivityRule{}, $x_1 \StarOf{\Rewrite} x_3$.
    From before, $y_2 \rightarrow y_5$.
    Because
    $\Property(x_3,y_4)$ and $\Propagator_R(y_4,y_5)$, and
    $\Propagator_R$ is a right propagator, $\Property(x_3,y_5)$.
  \end{case}
\end{proof}

\begin{definition}
  Let $\Property$ be a binary relation.  $\EquivalenceOf{\Property}$ is the
  binary relation defined via the inference rules
  \begin{mathpar}
    \inferrule[\BaseRule]
    {
      \Property(x,y)
    }
    {
      \EquivalenceOf{\Property}(x,y)
    }
    
    \inferrule[\ReflexivityRule]
    {
    }
    {
      \EquivalenceOf{\Property}(x,x)
    }

    \inferrule[\TransitivityRule]
    {
      \EquivalenceOf{\Property}(x,y)\\
      \EquivalenceOf{\Property}(y,z)
    }
    {
      \EquivalenceOf{\Property}(x,z)
    }

    \inferrule[\SymmetryRule]
    {
      \Property(x,y)
    }
    {
      \EquivalenceOf{\Property}(y,x)
    }
  \end{mathpar}
\end{definition}

\subsection{Language Proofs}
\label{language-proofs}

These proofs prove similar things to unambiguity, but on general languages.

\begin{lemma}
  \label{lem:potentially-empty-concat-implication}
  Let $\Language_1,\ldots,\Language_n,\Language_1',\ldots,\Language_m'$ be
  nonempty languages.
  If $\SequenceUnambigConcatOf{\Language_1;\ldots;\Language_n}$,
  $\SequenceUnambigConcatOf{\Language_1';\ldots;\Language_m'}$, and
  $\SetOf{\String_1\Concat\ldots\Concat\String_n \SuchThat \String_i \in
    \Language_i} \UnambigConcat
  \SetOf{\String_1\Concat\ldots\Concat\String_m \SuchThat \String_i \in
    \Language_i'}$, then
  $\SequenceUnambigConcatOf{\Language_1;\ldots;\Language_n;\Language_1';\ldots;\Language_n'}$
\end{lemma}
\begin{proof}
  Let $\SequenceUnambigConcatOf{\Language_1;\ldots;\Language_n}$,
  $\SequenceUnambigConcatOf{\Language_1';\ldots;\Language_m'}$, and
  $\SetOf{\String_1\Concat\ldots;\String_n \SuchThat \String_i \in
    \Language_i} \UnambigConcat
  \SetOf{\String_1\Concat\ldots\Concat\String_m \SuchThat \String_i \in
    \Language_i'}$
  Let $\String_i,\StringAlt_i\in\Language_i$,
  $\String_i',\StringAlt_i'\in\Language_i'$.
  Let
  $\String_1\Concat\ldots\Concat\String_n\Concat\String_1'\Concat\ldots\Concat\String_m'
  =
  \StringAlt_1\Concat\ldots\Concat\StringAlt_n\Concat\StringAlt_1'\Concat\ldots\Concat\StringAlt_m'$.
  Because $\SetOf{\String_1\Concat\ldots;\String_n \SuchThat \String_i \in
    \Language_i} \UnambigConcat
  \SetOf{\String_1\Concat\ldots\Concat\String_m \SuchThat \String_i \in
    \Language_i'}$, we know
  $\String_1\Concat\ldots\Concat\String_n =
  \StringAlt_1\Concat\ldots\Concat\StringAlt_n$ and
  $\String_1'\Concat\ldots\Concat\String_m' =
  \StringAlt_1'\Concat\ldots\Concat\StringAlt_m'$.
  Because $\SequenceUnambigConcatOf{\Language_1;\ldots;\Language_n}$,
  $\String_i = \StringAlt_i$.
  Because $\SequenceUnambigConcatOf{\Language_1';\ldots;\Language_n'}$,
  $\String_i' = \StringAlt_i'$.
  So
  $\SequenceUnambigConcatOf{\Language_1;\ldots;\Language_n;\Language_1';\ldots;\Language_n'}$
\end{proof}

\begin{lemma}
  \label{lem:unambig-concat-equiv}
  Let $\Language_1,\ldots,\Language_n,\Language_1',\ldots,\Language_m'$ be
  nonempty languages. $\SequenceUnambigConcatOf{\Language_1;\ldots;\Language_n}$,
  $\SequenceUnambigConcatOf{\Language_1';\ldots;\Language_m'}$, and
  $\SetOf{\String_1\Concat\ldots\Concat\String_n \SuchThat \String_i \in
    \Language_i} \UnambigConcat
  \SetOf{\String_1\Concat\ldots\Concat\String_m \SuchThat \String_i \in
    \Language_i'}$ if, and only if
  $\SequenceUnambigConcatOf{\Language_1;\ldots;\Language_n;\Language_1';\ldots;\Language_n'}$
\end{lemma}
\begin{proof}
  \begin{case}[$\Rightarrow$]
    By Lemma~\ref{lem:potentially-empty-concat-implication}.
  \end{case}
  
  \begin{case}[$\Leftarrow$]
    Let $\String,\StringAlt\in
    \SetOf{\String_1\Concat\ldots\Concat\String_n \SuchThat \String_i \in
      \Language_i}$.
    Let $\String',\StringAlt'\in
    \SetOf{\String_1\Concat\ldots\Concat\String_m \SuchThat \String_i \in
      \Language_i'}$.
    Let $\String\Concat\String' = \StringAlt\Concat\StringAlt'$.
    $\String=\String_1\Concat\ldots\Concat\String_n$ where
    $\String_i \in \Language_i$,
    $\StringAlt=\StringAlt_1\Concat\ldots\Concat\StringAlt_n$ where
    $\StringAlt_i \in \Language_i$,
    $\String'=\String_1'\Concat\ldots\Concat\String_m'$ where
    $\String_i' \in \Language_i'$, and
    $\StringAlt'=\StringAlt_1'\Concat\ldots\Concat\StringAlt_m'$ where
    $\StringAlt_i' \in \Language_i'$.
    $\String\Concat\String'=\String_1\Concat\ldots\Concat\String_n
    \Concat\String_1'\Concat\ldots\Concat\String_m'$ and
    $\StringAlt\Concat\StringAlt'=\String_1\Concat\ldots\Concat\String_n
    \Concat\String_1'\Concat\ldots\Concat\String_m'$.
    
    By assumption $\String_i=\StringAlt_i$ and $\String_i'=\StringAlt_i'$.
    This means $\String=\StringAlt$ and $\String'=\StringAlt'$.

    Let $\String_i,\StringAlt_i\in\Language_i$, and let
    $\String_1\Concat\ldots\Concat\String_n =
    \StringAlt_1\Concat\ldots\Concat\StringAlt_n$.
    Consider some strings $\String_i'\in\Language_i'$.
    $\String_1\Concat\ldots\Concat\String_n\Concat
    \String_1'\Concat\ldots\Concat\String_n' =
    \StringAlt_1\Concat\ldots\Concat\StringAlt_n\Concat
    \String_1'\Concat\ldots\Concat\String_n'$.
    
    By assumption, $\String_i=\StringAlt_i$, as desired.
  \end{case}
\end{proof}

\begin{lemma}
  \label{lem:unambig-concat-union-equiv}
  Let $\Language_1,\ldots,\Language_n,\Language_1',\ldots,\Language_m'$ be
  languages.
  Let $\Language_{i,j} =
  \SetOf{\String \Concat \StringAlt \SuchThat \String\in\Language_i \BooleanAnd
    \StringAlt\in\Language_j'}$.
  Let $A = \BigUnion_{i\in\RangeIncInc{1}{n}}\Language_i \neq \SetOf{}$.
  Let $B = \BigUnion_{i\in\RangeIncInc{1}{m}}\Language_i' \neq \SetOf{}$.
  $i \neq j \BooleanImplies \Language_i \Intersect \Language_j = \SetOf{}$
  $i \neq j \BooleanImplies \Language_i' \Intersect \Language_j' = \SetOf{}$
  and
  $\UnambigConcatOf{A}{B}$
  if, and only if
  $(i_1,j_1) \neq (i_2,j_2) \BooleanImplies \Language_{i_1,j_1}'' \Intersect
  \Language_{i_2,j_2}'' = \SetOf{}$ and for all $i\in\RangeIncInc{1}{n}$,
  $j\in\RangeIncInc{1}{m}$, we have
  $\UnambigConcatOf{\Language_i}{\Language_j'}$.
\end{lemma}
\begin{proof}
  \begin{case}[$\Rightarrow$]
    Let $i \neq j \BooleanImplies \Language_i \neq \Language_j$
    $i \neq j \BooleanImplies \Language_i' \neq \Language_j'$
    and
    $\UnambigConcatOf{A}{B}$

    We shall prove $(i_1,j_1) \neq (i_2,j_2) \BooleanImplies
    \Language_{i_1,j_1}''
    \Intersect
    \Language_{i_2,j_2}'' = \SetOf{}$ by contrapositive.
    Let $\String \in \Language_{i_1,j_1}''
    \Intersect
    \Language_{i_2,j_2}''$.  This means that $\String =
    \String_{i_1}\Concat\String_{j_1}$ for some
    $\String_{i_1}\in\Language_{i_1}$ and some
    $\String_{j_1}\in\Language_{j_1}$, and that
    $\String =\String_{i_2}\Concat\String_{j_2}$ for some
    $\String_{i_2}\in\Language_{i_2}$ and some
    $\String_{j_2}\in\Language_{j_2}$.

    Because 
    $\UnambigConcatOf{A}{B}$
    $\String_{i_1} = \String_{i_2}$ and $\String_{j_1} = \String_{j_2}$.
    Because each of $A$ and $B$ are pairwise disjoint, this means
    $i_1 = i_2$ and $j_1 = j_2$.

    Let $\String_i,\StringAlt_i \in \Language_i$.
    Let $\String_j,\StringAlt_j \in \Language_j'$.
    Let $\String_i \Concat \String_j = \StringAlt_i \Concat \StringAlt_j$
    By definition, $\String_i,\StringAlt_i\in A$ and
    $\String_j,\StringAlt_j\in B$.
    By assumption, 
    $\UnambigConcatOf{A}{B}$, so $\String_i = \StringAlt_i$ and $\String_j =
    \StringAlt_j$.
  \end{case}

  \begin{case}[$\Leftarrow$]
    Let
    $(i_1,j_1) \neq (i_2,j_2) \BooleanImplies \Language_{i_1,j_1}'' \Intersect
    \Language_{i_2,j_2}'' = \SetOf{}$ and for all $i\in\RangeIncInc{1}{n}$,
    $j\in\RangeIncInc{1}{m}$, we have
    $\UnambigConcatOf{\Language_i}{\Language_j'}$.

    We prove $i \neq j \BooleanImplies \Language_i \Intersect \Language_j =
    \SetOf{}$ by contrapositive.
    Let $\Language_i \Intersect \Language_j \neq \SetOf{}$.
    Let $\String \in \Language_i \Intersect \Language_j$
    Let $\StringAlt \in B$.
    $\StringAlt \in \Language_k'$ for some $k \in \RangeIncInc{1}{m}$.
    $\String \Concat \StringAlt \in \Language_{i,k}''$ and
    $\String \Concat \StringAlt \in \Language_{j,k}''$.
    By assumption $(i,k) = (j,k)$, so $i = j$.
    
    We prove $i \neq j \BooleanImplies \Language_i' \Intersect \Language_j' =
    \SetOf{}$ in the same way.
    
    Let $\String_1,\String_2 \in A$, $\StringAlt_1,\StringAlt_2 \in B$, and
    $\String_1 \Concat \StringAlt_1 = \String_2 \Concat \StringAlt_2$.
    $\String_1 \in \Language_i$ for some $i$, and $\String_2 \in \Language_j$
    for some $j$, $\StringAlt_1 \in \Language_k'$ for some $k$, and
    $\StringAlt_2 \in \Language_l'$ for some $l$.
    This means $\String_1 \Concat \StringAlt_1 \in \Language_{i,k}''$,
    $\String_2 \Concat \StringAlt_2 \in \Language_{j,l}''$.
    Because $\String_1 \Concat \StringAlt_1 = \String_2 \Concat \StringAlt_2$,
    $(i,k) = (j,l)$.
    So as $\String_1 \in \Language_i, \String_2 \in \Language_i$,
    $\StringAlt_1 \in \Language_k, \StringAlt_2 \in \Language_k'$,
    and $\UnambigConcatOf{\Language_i}{\Language_k'}$, $\String_1 = \String_2$
    and $\StringAlt_2 = \StringAlt_2$.
  \end{case}
\end{proof}

\begin{lemma}
  \label{lem:unambig-union-equiv}
  Let $A = \SetOf{\Language_1,\ldots,\Language_n}$,
  $B = \SetOf{\Language_1',\ldots,\Language_m'}$,
  $C = \SetOf{\Language_1'',\ldots,\Language_{n+m}''}$, be sets of languages
  Such that $A \Union B = C$.
  $(\BigUnion_{i\in\RangeIncInc{1}{n}}\Language_i) \Intersect
  (\BigUnion_{i\in\RangeIncInc{1}{m}}\Language_i') = \emptyset$,
  for all $i,j\in\RangeIncInc{1}{n}$, $i \neq j \BooleanImplies Language_i \Intersect
  \Language_j = \emptyset$, and for all $i,j\in\RangeIncInc{1}{m}$, $i \neq j
  \BooleanImplies \Language_i' \Intersect \Language_j' = \emptyset$ if, and only
  if for all $i,j\in\RangeIncInc{1}{n+m}$, $i \neq j \BooleanImplies \Language_i''
  \Intersect \Language_j'' = \emptyset$.
\end{lemma}
\begin{proof}
  \begin{case}[$\Rightarrow$]
    Let $\Language_i'',\Language_j''\in C$, where $i \neq j$.
    If $\Language_i''\in A$ and $\Language_j''\in A$, then, by pigeonhole
    principle, there exists an $i', j'$ where $i' \neq j'$ such that
    $\Language_i'' = \Language_{i'}$ and $\Language_j''=\Language_{j'}$.
    By assumption, $\Language_{i'} \Intersect \Language_{j'} = \SetOf{}$, so
    $|Language_i'' \Intersect \Language_j'' = \SetOf{}$.
    
    Similarly for if $\Language_i''\in B$ and $\Language_j'' \in B$.
    
    If $\Language_i''\in A$, and $\Language_j'' \in B$.
    $(\BigUnion_{i\in\RangeIncInc{1}{n}}\Language_i) \Intersect
    (\BigUnion_{i\in\RangeIncInc{1}{m}}\Language_i') = \emptyset$.
    By application of distributivity
    $\BigUnion_{(k,l)\in\RangeIncInc{1}{n}\Cross\RangeIncInc{1}{m}}
    \Language_k'' \Intersect \Language_l'' = \SetOf{}$.
    This means that for all
    $(k,l)\in\RangeIncInc{1}{n}\Cross\RangeIncInc{1}{m}$,
    $\Language_k \Intersect \Language_l' = \SetOf{}$.
    In particular, $\Language_i'' \Intersect \Language_j'' = \SetOf{}$.
  \end{case}

  \begin{case}[$\Leftarrow$]
    Let $i,j\in\RangeIncInc{1}{n}$ and $i \neq j$.
    By pigeonhole principle, there exists some $i',j'$ where $i' \neq j'$
    such that
    $\Language_i = \Language_{i'}''$ and $\Language_j = \Language_{j'}''$.
    By assumption, $\Language_{i'}'' \Intersect \Language_{j'}'' = \SetOf{}$,
    so $\Language_i \Intersect \Language_j = \SetOf{}$.

    Similarly for $i,j\in\RangeIncInc{1}{m}$.

    Assume there exists some $\String \in
    (\BigUnion_{i\in\RangeIncInc{1}{n}}\Language_i) \Intersect
    (\BigUnion_{i\in\RangeIncInc{1}{m}}\Language_i')$.
    Then $\String\in\Language_i$ for some $i\in\RangeIncInc{1}{n}$, and
    $\String\in\Language_j'$ for some $j\in\RangeIncInc{1}{m}$.
    There exists some $i',j'$ where $i' \neq j'$ in $\RangeIncInc{1}{n+m}$
    such that $\Language_i = \Language_{i'}''$ and
    $\Language_j = \Language_{j'}''$.  But, by assumption,
    $\Language_{i'}'' \Intersect \Language_{j'}''$, so we have a contraction.
    So there is no $\String \in (\BigUnion_{i\in\RangeIncInc{1}{n}}\Language_i)
    \Intersect (\BigUnion_{i\in\RangeIncInc{1}{m}}\Language_i')$, so
    $(\BigUnion_{i\in\RangeIncInc{1}{n}}\Language_i) \Intersect
    (\BigUnion_{i\in\RangeIncInc{1}{m}}\Language_i') = \SetOf{}$.
  \end{case}
\end{proof}

\subsection{Lens and DNF Basic Property Proofs}
\label{basic-property-proofs}

There are many intuitive facts about DNF lenses.  For example, without rewrites,
they are closed under composition.  Furthermore, we can express the identity
transformation on DNF lenses.  Well-typed DNF lenses and normal lenses express
bijections between their types.  These properties are proven in this section,
and used throughout the paper.

\begin{lemma}[DNF Lens Inversion]
  \label{lem:dnf-lens-inversion}
  \leavevmode
  \begin{enumerate}
  \item If $\DNFLens \OfType \DNFRegex \Leftrightarrow \DNFRegexAlt$, then there
    exists some $\DNFRegex'$, $\DNFRegexAlt'$ such that
    $\DNFLens \OfRewritelessType \DNFRegex' \Leftrightarrow \DNFRegexAlt'$,
    $\DNFRegex \StarOf{\Rewrite} \DNFRegex'$, and $\DNFRegexAlt
    \StarOf{\Rewrite} \DNFRegexAlt'$.
  \item If $\DNFLens \OfRewritelessType \DNFRegex \Leftrightarrow \DNFRegexAlt$,
    then there exists some $n\in\Nats$, $\Sequence_1,\ldots,\Sequence_n$,
    $\SequenceAlt_1,\ldots\SequenceAlt_n$, $\sigma\in\PermutationSetOf{n}$, and
    $\SequenceLens_1;\ldots;\SequenceLens_n$ such that
    for all $i\in\RangeIncInc{1}{n}$, $\SequenceLens_i \OfRewritelessType
    \Sequence_i \Leftrightarrow \SequenceAlt_i$, $\DNFLens =
    (\DNFLensOf{\SequenceLens_0\DNFLSep\ldots\DNFLSep\SequenceLens_n},\sigma)$, $\DNFRegex =
    \DNFOf{\Sequence_1\DNFSep\ldots\DNFSep\Sequence_n}$, and $\DNFRegexAlt =
    \DNFOf{\SequenceAlt_{\sigma(1)} \DNFSep \ldots \DNFSep \SequenceAlt_{\sigma(n)}}$.
  \item If $\SequenceLens \OfRewritelessType \Sequence \Leftrightarrow
    \SequenceAlt$, there exists some $n \in \Nats$,
    $\Atom_1,\ldots,\Atom_n$, $\AtomAlt_1,\ldots,\AtomAlt_n$,
    $String_0,\ldots,\String_n$, $\StringAlt_0,\ldots,\StringAlt_n$,
    $\sigma\in\PermutationSetOf{n}$, and $\AtomLens_1,\ldots,\AtomLens_n$ such that
    for all $i\in\RangeIncInc{1}{n}$, $\AtomLens_i \OfRewritelessType \Atom_i
    \Leftrightarrow \AtomAlt_i$, $\SequenceLens =
    (\SequenceLensOf{(\String_0,\StringAlt_0) \SeqLSep \Atom_1 \SeqLSep \ldots
      \SeqLSep 
      \Atom_n \SeqLSep (\String_n,\StringAlt_n)},\sigma)$, $\Sequence =
    \SequenceOf{\String_0 \SeqSep \Atom_1 \SeqSep \ldots \SeqSep \Atom_n \SeqSep
      \String_n}$, and $\SequenceAlt =
    \SequenceOf{\StringAlt_0 \SeqSep \Atom_{\sigma(1)} \SeqSep \ldots \SeqSep
      \Atom_{\sigma(n)} \SeqSep \StringAlt_n}$.
  \item If $\AtomLens \OfRewritelessType \Atom \Leftrightarrow \AtomAlt$, then
    there exists some $\DNFLens$, $\DNFRegex$, $\DNFRegexAlt$, such that
    $\DNFLens \OfRewritelessType \DNFRegex \Leftrightarrow \DNFRegexAlt$,
    $\AtomLens = \IterateLensOf{\DNFLens}$, $\Atom = \StarOf{\DNFRegex}$, and
    $\AtomAlt = \StarOf{\DNFRegexAlt}$.
  \end{enumerate}
\end{lemma}
\begin{proof}\leavevmode
  \begin{enumerate}
  \item
    Let $\DNFLens \OfType \DNFRegex \Leftrightarrow \DNFRegexAlt$.  The
    only rule that introduces a typing of this form is
    \RewriteDNFRegexLensRule{}.  Because of this, there
    exists some $\DNFRegex'$, $\DNFRegexAlt'$ such that
    $\DNFLens \OfRewritelessType \DNFRegex' \Leftrightarrow \DNFRegexAlt'$,
    $\DNFRegex \StarOf{\Rewrite} \DNFRegex'$, and $\DNFRegexAlt
    \StarOf{\Rewrite} \DNFRegexAlt'$, to build up the typing
    \[
      \inferrule*
      {
        \DNFLens \OfRewritelessType \DNFRegex' \Leftrightarrow \DNFRegexAlt'\\
        \DNFRegex \StarOf{\Rewrite} \DNFRegex'\\
        \DNFRegexAlt \StarOf{\Rewrite} \DNFRegexAlt'
      }
      {
        \DNFLens \OfType \DNFRegex \Leftrightarrow \DNFRegexAlt
      }
    \]
  \item
    Let $\DNFLens \OfRewritelessType \DNFRegex \Leftrightarrow \DNFRegexAlt$.
    The only rule that introduces a typing of this form is \DNFLensRule{}.
    Because of this there exists some $n\in\Nats$, $\Sequence_1,\ldots,\Sequence_n$,
    $\SequenceAlt_1,\ldots\SequenceAlt_n$, $\sigma\in\PermutationSetOf{n}$, and
    $\SequenceLens_1 \DNFLSep \ldots \DNFLSep \SequenceLens_n$ such that
    for all $i\in\RangeIncInc{1}{n}$, $\SequenceLens_i \OfRewritelessType
    \Sequence_i \Leftrightarrow \SequenceAlt_i$, $\DNFLens =
    (\DNFLensOf{\Sequence_0 \DNFLSep \ldots \DNFLSep \Sequence_n},\sigma)$, $\DNFRegex =
    \DNFOf{\Sequence_1 \DNFSep \ldots \DNFSep \Sequence_n}$, $\DNFRegexAlt =
    \DNFOf{\SequenceAlt_{\sigma(1)} \DNFSep \ldots \DNFSep \SequenceAlt_{\sigma(n)}}$,
    $i \neq j \BooleanImplies \Sequence_i \Intersect \Sequence_j = \emptyset$,
    and
    $i \neq j \BooleanImplies \SequenceAlt_i \Intersect \SequenceAlt_j = \emptyset$,
    to build up the typing
    \[
      \inferrule*
      {
        \SequenceLens_i \OfRewritelessType \Sequence_i \Leftrightarrow
        \SequenceAlt_i\\
        i \neq j \BooleanImplies \Sequence_i \Intersect \Sequence_j =
        \emptyset\\
        i \neq j \BooleanImplies \SequenceAlt_i \Intersect \SequenceAlt_j =
        \emptyset
      }
      {
        \DNFLens \OfRewritelessType \DNFRegex \Leftrightarrow \DNFRegexAlt
      }
    \]
  \item
    Let $\SequenceLens \OfRewritelessType \Sequence \Leftrightarrow \SequenceAlt$.
    The only rule that introduces a typing of this form is \SequenceLensRule{}.
    Because of this there exists some $n \in \Nats$,
    $\Atom_1,\ldots,\Atom_n$, $\AtomAlt_1,\ldots,\AtomAlt_n$,
    $String_0,\ldots,\String_n$, $\StringAlt_0,\ldots,\StringAlt_n$,
    $\sigma\in\PermutationSetOf{n}$, and $\AtomLens_1,\ldots,\AtomLens_n$ such that
    for all $i\in\RangeIncInc{1}{n}$, $\AtomLens_i \OfRewritelessType \Atom_i
    \Leftrightarrow \AtomAlt_i$, $\SequenceLens =
    (\SequenceLensOf{(\String_0,\StringAlt_0) \SeqLSep \AtomLens_1 \SeqLSep
      \ldots \SeqLSep 
      \AtomLens_n \SeqLSep (\String_n,\StringAlt_n)},\sigma)$, $\Sequence =
    \SequenceOf{\String_0 \SeqSep \Atom_1 \SeqSep \ldots \SeqSep \Atom_n \SeqSep
      \String_n}$, and $\SequenceAlt =
    \SequenceOf{\StringAlt_0 \SeqSep \Atom_{\sigma(1)} \SeqSep \ldots \SeqSep
      \Atom_{\sigma(n)} \SeqSep \StringAlt_n}$
    to build up the typing
    \[
      \inferrule*
      {
        \AtomLens_i \OfRewritelessType \Atom_i \Leftrightarrow \AtomAlt_i\\
        \SequenceUnambigConcatOf{\String_0;\Atom_1;\ldots;\Atom_n;\String_n}\\
        \SequenceUnambigConcatOf{\StringAlt_0;\AtomAlt_{\sigma(1)};\ldots;\AtomAlt_{\sigma(n)};\StringAlt_n}
      }
      {
        (\SequenceLensOf{(\String_0,\StringAlt_0) \SeqLSep \AtomLens_1 \SeqLSep
          \ldots \SeqLSep
          \AtomLens_n \SeqLSep (\String_n,\StringAlt_n)},\sigma) \OfRewritelessType
        \SequenceOf{\String_0 \SeqSep \Atom_1 \SeqSep \ldots \SeqSep \Atom_n
          \SeqSep \String_n} \Leftrightarrow
        \SequenceOf{\StringAlt_0 \SeqSep \Atom_{\sigma(1)} \SeqSep \ldots
          \SeqSep \Atom_{\sigma(n)} \SeqSep \StringAlt_n}
      }
    \]
  \item
    Let $\AtomLens \OfRewritelessType \Atom \Leftrightarrow \AtomAlt$.
    The only rule that introduces a typing of this form is
    \AtomLensRule{}.
    Because of this, there exists some $\DNFLens$, $\DNFRegex$, $\DNFRegexAlt$, such that
    $\DNFLens \OfRewritelessType \DNFRegex \Leftrightarrow \DNFRegexAlt$,
    $\AtomLens = \IterateLensOf{\DNFLens}$, $\Atom = \StarOf{\DNFRegex}$, and
    $\AtomAlt = \StarOf{\DNFRegexAlt}$ to build up the typing
    \[
      \inferrule*
      {
        \DNFLens \OfRewritelessType \DNFRegex \Leftrightarrow \DNFRegexAlt\\
        \UnambigItOf{DNFRegex}\\
        \UnambigItOf{DNFRegexAlt}
      }
      {
        \IterateLensOf{\DNFLens} \OfRewritelessType
        \StarOf{\DNFRegex} \Leftrightarrow \StarOf{\DNFRegexAlt}
      }
    \]
  \end{enumerate}
\end{proof}

\begin{lemma}
  \label{lem:lens-bij}
  If $\Lens \OfType \Regex \Leftrightarrow \RegexAlt$, then
  $\SemanticsOf{\Lens}$ is a bijection between $\LanguageOf{\Regex}$ and
  $\LanguageOf{\RegexAlt}$.
\end{lemma}
\begin{proof}
  By induction on the typing derivation of the lens
  \begin{case}[\ConstLensType{}]
    \[
      \inferrule*
      {
        \String_1 \in \StarOf{\Sigma}\\
        \String_2 \in \StarOf{\Sigma}
      }
      {
        \ConstLensOf{\String_1}{\String_2} \OfType \String_1 \Leftrightarrow \String_2
      }
    \]

    $\LanguageOf{\String_1} = \SetOf{\String_1}$.
    
    $\LanguageOf{\String_2} = \SetOf{\String_2}$.

    $SemanticsOf{\ConstLensOf{\String_1}{\String_2}} = \SetOf{(\String_1,\String_2)}$.
  \end{case}

  \begin{case}[\IdentityLensType{}]
    \[
      \inferrule*
      {
        \Regex \text{ is strongly unambiguous}
      }
      {
        \IdentityLensOf{\Regex} \OfType \Regex \Leftrightarrow \Regex
      }
    \]

    $\LanguageOf{\Regex} = \LanguageOf{\Regex}$

    The identity relation on $\LanguageOf{\Regex}$ is a bijection.
  \end{case}

  \begin{case}[\IterateLensType{}]
    \[
      \inferrule*
      {
        \Lens \OfType \Regex \Leftrightarrow \RegexAlt\\
        \UnambigItOf{\Regex}\\
        \UnambigItOf{\RegexAlt}
      }
      {
        \IterateLensOf{\Lens} \OfType \StarOf{\Regex} \Leftrightarrow \StarOf{\RegexAlt}
      }
    \]

    Let $\String_1,\String_2$ in $\LanguageOf{\StarOf{\Regex}}$, and
    $(\String_1,\StringAlt) \in \SemanticsOf{\IterateLensOf{\Lens}}$, and
    $(\String_2,\StringAlt) \in \SemanticsOf{\IterateLensOf{\Lens}}$.

    So $\String_1 = \String_{1,1}\Concat\ldots\Concat\String_{1,n}$,
    $\StringAlt = \StringAlt_1\Concat\ldots\StringAlt_n$, and
    $(\String_{1,i},\StringAlt_i) \in \SemanticsOf{\Lens}$.
    So $\String_2 = \String_{2,1}\Concat\ldots\Concat\String_{2,m}$,
    $\StringAlt = \StringAlt_1'\Concat\ldots\StringAlt_m'$, and
    $(\String_{2,i},\StringAlt_i') \in \SemanticsOf{\Lens}$.
    By $\UnambigItOf{\RegexAlt}$, this means that $m=n$, and $\StringAlt_i =
    \StringAlt_i'$.
    So $(\String_{1,i},\StringAlt_i)$, and $(\String_{2,i},\StringAlt_i)$ are
    both in $\SemanticsOf{\Lens}$.
    As $\Lens$ is a bijection, by IH, $\String_{1,i} = \String_{2,i}$, so $\String_1 =
    \String_2$.

    Similarly for $\StringAlt_1,\StringAlt_2$ in
    $\LanguageOf{\StarOf{\RegexAlt}}$.

    Let $\String\in\LanguageOf{\StarOf{\Regex}}$.
    $\String = \String_1\Concat\ldots\Concat\String_n$, where $\String_i \in
    \LanguageOf{\Regex}$.
    By IH, as $\Lens$ is a bijection, there exists $\StringAlt_i\in\LanguageOf{\Regex}$ such that
    $(\String_i,\StringAlt_i) \in \SemanticsOf{\Lens}$.
    So $(\String,\StringAlt_1\Concat\ldots\Concat\StringAlt_n) \in
    \SemanticsOf{\IterateLensOf{\Lens}}$,
    and $\StringAlt_1 \Concat \ldots \Concat \StringAlt_n \in \LanguageOf{\StarOf{\RegexAlt}}$.

    Similarly for $\StringAlt \in \LanguageOf{\StarOf{\RegexAlt}}$.
  \end{case}

  \begin{case}[ConcatLensType{}]
    \[
      \inferrule*
      {
        \Lens_1 \OfType \Regex_1 \Leftrightarrow \RegexAlt_1\\\\
        \Lens_2 \OfType \Regex_2 \Leftrightarrow \RegexAlt_2\\\\
        \UnambigConcatOf{\Regex_1}{\Regex_2}\\
        \UnambigConcatOf{\RegexAlt_1}{\RegexAlt_2}
      }
      {
        \ConcatLensOf{\Lens_1}{\Lens_2} \OfType \Regex_1\Regex_2 \Leftrightarrow \RegexAlt_1\RegexAlt_2
      }
    \]

    Let $\String_1,\String_2$ in $\LanguageOf{\Regex_1\Concat\Regex_2}$, and
    $(\String_1,\StringAlt) \in \SemanticsOf{\ConcatLensOf{\Lens_1}{\Lens_2}}$, and
    $(\String_2,\StringAlt) \in \SemanticsOf{\ConcatLensOf{\Lens_1}{\Lens_2}}$.

    So $\String_1 = \String_{1,1}\Concat\String_{1,2}$,
    $\StringAlt = \StringAlt_1\StringAlt_2$, and
    $(\String_{1,i},\StringAlt_i) \in \SemanticsOf{\Lens_i}$.
    So $\String_2 = \String_{2,1}\Concat\String_{2,2}$,
    $\StringAlt = \StringAlt_1'\Concat\StringAlt_2'$, and
    $(\String_{2,i},\StringAlt_i') \in \SemanticsOf{\Lens_i}$.
    By $\UnambigConcatOf{\RegexAlt_1}{\RegexAlt_2}$,
    $\StringAlt_i = \StringAlt_i'$.
    So $(\String_{1,i},\StringAlt_i)$, and $(\String_{2,i},\StringAlt_i)$ are
    both in $\SemanticsOf{\Lens_i}$.
    As $\Lens_i$ is a bijection, by IH, $\String_{1,i} = \String_{2,i}$, so $\String_1 =
    \String_2$.

    Similarly for $\StringAlt_1,\StringAlt_2$ in
    $\LanguageOf{\RegexAlt_1 \Concat \RegexAlt_2}$.

    Let $\String\in\LanguageOf{\Regex_1\Concat\Regex_2}$.
    $\String = \String_1\Concat\String_2$, where $\String_i \in
    \LanguageOf{\Regex_i}$.
    By IH, as $\Lens_i$ is a bijection, there exists $\StringAlt_i\in\LanguageOf{\DNFRegexAlt}$ such that
    $(\String_i,\StringAlt_i) \in \SemanticsOf{\Lens_i}$.
    So $(\String,\StringAlt_1\Concat\StringAlt_2) \in
    \SemanticsOf{\ConcatLensOf{\Lens_1}{\Lens_2}}$.

    Similarly for $\StringAlt \in \LanguageOf{\RegexAlt_1\Concat\RegexAlt_2}$.
  \end{case}

  \begin{case}[\SwapLensType]
    \[
      \inferrule*
      {
        \Lens_1 \OfType \Regex_1 \Leftrightarrow \RegexAlt_1\\\\
        \Lens_2 \OfType \Regex_2 \Leftrightarrow \RegexAlt_2\\\\
        \UnambigConcatOf{\Regex_1}{\Regex_2}\\
        \UnambigConcatOf{\RegexAlt_2}{\RegexAlt_1}
      }
      {
        \SwapLensOf{\Lens_1}{\Lens_2} \OfType \Regex_1\Regex_2 \Leftrightarrow \RegexAlt_2\RegexAlt_1
      }
    \]

    Let $\String_1,\String_2$ in $\LanguageOf{\Regex_1\Concat\Regex_2}$, and
    $(\String_1,\StringAlt) \in \SemanticsOf{\SwapLensOf{\Lens_1}{\Lens_2}}$, and
    $(\String_2,\StringAlt) \in \SemanticsOf{\SwapLensOf{\Lens_1}{\Lens_2}}$.

    So $\String_1 = \String_{1,1}\Concat\String_{1,2}$,
    $\StringAlt = \StringAlt_2\StringAlt_1$, and
    $(\String_{1,i},\StringAlt_i) \in \SemanticsOf{\Lens_i}$.
    So $\String_2 = \String_{2,1}\Concat\String_{2,2}$,
    $\StringAlt = \StringAlt_2'\Concat\StringAlt_1'$, and
    $(\String_{2,i},\StringAlt_i') \in \SemanticsOf{\Lens_i}$.
    By $\UnambigConcatOf{\RegexAlt_2}{\RegexAlt_1}$,
    $\StringAlt_i = \StringAlt_i'$.
    So $(\String_{1,i},\StringAlt_i)$, and $(\String_{2,i},\StringAlt_i)$ are
    both in $\SemanticsOf{\Lens_i}$.
    As $\Lens_i$ is a bijection, by IH, $\String_{1,i} = \String_{2,i}$,
    so $\String_1 = \String_2$.

    Similarly for $\StringAlt_1,\StringAlt_2$ in
    $\LanguageOf{\RegexAlt_2\Concat\RegexAlt_1}$.

    Let $\String\in\LanguageOf{\Regex_1\Concat\Regex_2}$.
    $\String = \String_1\Concat\String_2$, where $\String_i \in
    \LanguageOf{\Regex_i}$.
    By IH, as $\Lens_i$ is a bijection, there exists $\StringAlt_i\in\LanguageOf{\RegexAlt_i}$ such that
    $(\String_i,\StringAlt_i) \in \SemanticsOf{\Lens_i}$.
    So $(\String,\StringAlt_2\Concat\StringAlt_1) \in
    \SemanticsOf{\SwapLensOf{\Lens_1}{\Lens_2}}$,
    and $\StringAlt_2 \Concat \StringAlt_1 \in \LanguageOf{\Regex_2 \Concat \Regex_1}$.

    Similarly for $\StringAlt \in \LanguageOf{\RegexAlt_2\Concat\RegexAlt_1}$.
  \end{case}

  \begin{case}[\OrLensType{}]
    \[
      \inferrule*
      {
        \Lens_1 \OfType \Regex_1 \Leftrightarrow \RegexAlt_1\\
        \Lens_2 \OfType \Regex_2 \Leftrightarrow \RegexAlt_2\\\\
        \UnambigOrOf{\Regex_1}{\Regex_2}\\
        \UnambigOrOf{\RegexAlt_1}{\RegexAlt_2}
      }
      {
        \OrLensOf{\Lens_1}{\Lens_2} \OfType \Regex_1 \Or \Regex_2
        \Leftrightarrow \RegexAlt_1 \Or \RegexAlt_2
      }
    \]

    Let $\String_1,\String_2$ in $\LanguageOf{\Regex_1\Or\Regex_2}$, and
    $(\String_1,\StringAlt) \in \SemanticsOf{\OrLensOf{\Lens_1}{\Lens_2}}$, and
    $(\String_2,\StringAlt) \in \SemanticsOf{\OrLensOf{\Lens_1}{\Lens_2}}$.

    So $(\String_1,\StringAlt) \in \SemanticsOf{\Lens_1}$ or
    $(\String_1,\StringAlt) \in \SemanticsOf{\Lens_2}$.
    So $(\String_2,\StringAlt) \in \SemanticsOf{\Lens_1}$ or
    $(\String_2,\StringAlt) \in \SemanticsOf{\Lens_2}$.

    As $\UnambigOrOf{\RegexAlt_1}{\RegexAlt_2}$, $\StringAlt_i$ is in only one of
    $\LanguageOf{\RegexAlt_1}$ or $\LanguageOf{\RegexAlt_2}$.

    Let $\StringAlt_i \in \LanguageOf{\RegexAlt_1}$.  This means that
    $(\String_1,\StringAlt) \in \SemanticsOf{\Lens_1}$ and
    $(\String_2,\StringAlt) \in \SemanticsOf{\Lens_1}$.
    As $\Lens_1$ is a bijection, by IH, $\String_1 = \String_2$.
    Similarly if $\StringAlt \in \LanguageOf{\RegexAlt_2}$.

    Similarly for $\StringAlt_1,\StringAlt_2$ in
    $\LanguageOf{\RegexAlt_2\Concat\RegexAlt_1}$.

    Let $\String\in\LanguageOf{\Regex_1\Or\Regex_2}$.
    So $\String$ is either in $\LanguageOf{\Regex_1}$ or $\LanguageOf{\Regex_2}$.
    If $\String \in \LanguageOf{\Regex_1}$, then as $\Lens_1$ is a bijection
    between $\LanguageOf{\Regex_1}$ and $\LanguageOf{\Regex_2}$,
    there exists $\StringAlt \in \LanguageOf{RegexAlt_1}$ such that
    $(\String,\StringAlt) \in \SemanticsOf{\Lens_1}$, so
    $(\String,\StringAlt) \in \SemanticsOf{\OrLensOf{\Lens_1}{\Lens_2}}$,
    and $\StringAlt \in \LanguageOf{\RegexAlt_1 \Or \RegexAlt_2}$.
    Similarly if $\String \in \LanguageOf{\Regex_2}$.

    Similarly for $\StringAlt \in \LanguageOf{\RegexAlt_2\Concat\RegexAlt_1}$.
  \end{case}

  \begin{case}[\ComposeLensType{}]
    \[
      \inferrule*
      {
        \Lens_1 \OfType \Regex_1 \Leftrightarrow \Regex_2\\
        \Lens_2 \OfType \Regex_2 \Leftrightarrow \Regex_3\\
      }
      {
        \ComposeLensOf{\Lens_1}{\Lens_2} \OfType \Regex_1 \Leftrightarrow \Regex_3
      }
    \]

    By IH, $\SemanticsOf{\Lens_1}$ is a bijection between
    $\LanguageOf{\Regex_1}$ and $\LanguageOf{\Regex_2}$.
    By IH, $\SemanticsOf{\Lens_2}$ is a bijection between
    $\LanguageOf{\Regex_2}$ and $\LanguageOf{\Regex_3}$.
    From math, this means that their composition is also a bijection.
  \end{case}
\end{proof}

\begin{lemma}
  \label{lem:rw-dnf-lens-bij}
  \leavevmode
  \begin{itemize}
  \item
    If $\DNFLens \OfRewritelessType \DNFRegex \Leftrightarrow \DNFRegexAlt$, then
    $\SemanticsOf{\DNFLens}$ is a bijection between $\LanguageOf{\DNFRegex}$ and
    $\LanguageOf{\DNFRegexAlt}$.
  \item
    If $\SequenceLens \OfRewritelessType \Sequence \Leftrightarrow \SequenceAlt$, then
    $\SemanticsOf{\SequenceLens}$ is a bijection between $\LanguageOf{\Sequence}$ and
    $\LanguageOf{\SequenceAlt}$.
  \item
    If $\AtomLens \OfRewritelessType \Atom \Leftrightarrow \AtomAlt$, then
    $\SemanticsOf{\AtomLens}$ is a bijection between $\LanguageOf{\Atom}$ and
    $\LanguageOf{\AtomAlt}$.
  \end{itemize}
\end{lemma}
\begin{proof}
  By mutual induction on the typing derivations of DNF lenses, sequence lenses,
  and atom lenses.
  \begin{case}[\IterateAtomType{}]
    \[
      \inferrule*
      {
        \DNFLens \OfRewritelessType \DNFRegex \Leftrightarrow \DNFRegexAlt\\
        \UnambigItOf{\DNFRegex}\\
        \UnambigItOf{\DNFRegexAlt}
      }
      {
        \IterateLensOf{\DNFLens} \OfRewritelessType \StarOf{\DNFRegex}
        \Leftrightarrow \StarOf{\DNFRegexAlt}
      }
    \]
    
    Let $\String_1,\String_2$ in $\LanguageOf{\StarOf{\DNFRegex}}$, and
    $(\String_1,\StringAlt) \in \SemanticsOf{\IterateLensOf{\DNFLens}}$, and
    $(\String_2,\StringAlt) \in \SemanticsOf{\IterateLensOf{\DNFLens}}$.
    
    So $\String_1 = \String_{1,1}\Concat\ldots\Concat\String_{1,n}$,
    $\StringAlt = \StringAlt_1\Concat\ldots\StringAlt_n$, and
    $(\String_{1,i},\StringAlt_i) \in \SemanticsOf{\DNFLens}$.
    So $\String_2 = \String_{2,1}\Concat\ldots\Concat\String_{2,m}$,
    $\StringAlt = \StringAlt_1'\Concat\ldots\StringAlt_m'$, and
    $(\String_{2,i},\StringAlt_i') \in \SemanticsOf{\DNFLens}$.
    By $\UnambigItOf{\DNFRegexAlt}$, this means that $m=n$, and $\StringAlt_i =
    \StringAlt_i'$.
    So $(\String_{1,i},\StringAlt_i)$, and $(\String_{2,i},\StringAlt_i)$ are
    both in $\SemanticsOf{\DNFLens}$.
    As $\DNFLens$ is a bijection, by IH, $\String_{1,i} = \String_{2,i}$, so $\String_1 =
    \String_2$.

    Similarly for $\StringAlt_1,\StringAlt_2$ in
    $\LanguageOf{\StarOf{\DNFRegexAlt}}$.

    Let $\String\in\LanguageOf{\StarOf{\DNFRegex}}$.
    $\String = \String_1\Concat\ldots\Concat\String_n$, where $\String_i \in
    \LanguageOf{\DNFRegex}$.
    By IH, as $\DNFLens$ is a bijection, there exists $\StringAlt_i\in\LanguageOf{\DNFRegexAlt}$ such that
    $(\String_i,\StringAlt_i) \in \SemanticsOf{\DNFLens}$.
    So $(\String,\StringAlt_1\Concat\ldots\Concat\StringAlt_n) \in
    \SemanticsOf{\IterateLensOf{\DNFLens}}$,
    and $\StringAlt_1\Concat\ldots\Concat\StringAlt_n \in \LanguageOf{\DNFRegexAlt}$.

    Similarly for $\StringAlt \in \LanguageOf{\StarOf{\DNFRegexAlt}}$.
  \end{case}

  \begin{case}[\SequenceLensType{}]
    \[
      \inferrule*
      {
        \AtomLens_1 \OfRewritelessType \Atom_1 \Leftrightarrow \AtomAlt_1\\
        \ldots\\
        \AtomLens_n \OfRewritelessType \Atom_n \Leftrightarrow \AtomAlt_n\\
        \sigma \in \PermutationSetOf{n}\\
        \UnambigConcat\SequenceOf{\String_0\SeqSep\Atom_1\SeqSep\ldots\SeqSep\Atom_n\SeqSep\String_n}\\
        \UnambigConcat\SequenceOf{\StringAlt_0\SeqSep\AtomAlt_{\sigma(1)}\SeqSep\ldots\SeqSep\AtomAlt_{\sigma(n)}\SeqSep\StringAlt_n}
      }
      {
        (\SequenceLensOf{(\String_0,\StringAlt_0)\SeqLSep\AtomLens_1\SeqLSep\ldots\SeqLSep\AtomLens_n\SeqLSep(\String_n,\StringAlt_n)},\sigma)
        \OfRewritelessType\\
        \SequenceOf{\String_0\SeqSep\Atom_1\SeqSep\ldots\SeqSep\Atom_n\SeqSep\String_n}
        \Leftrightarrow
        \SequenceOf{\StringAlt_0\SeqSep\AtomAlt_{\sigma(1)}\SeqSep\ldots\SeqSep\AtomAlt_{\sigma(n)}\SeqSep\StringAlt_n}
      }
    \]

    Let $\String_1,\String_2$ in $\LanguageOf{\SequenceOf{\String_0\SeqSep\Atom_1\SeqSep\ldots\SeqSep\Atom_n\SeqSep\String_n}}$, and
    $(\String_1,\StringAlt) \in \SemanticsOf{(\SequenceLensOf{(\String_0,\StringAlt_0)\SeqLSep\AtomLens_1\SeqLSep\ldots\SeqLSep\AtomLens_n\SeqLSep(\String_n,\StringAlt_n)},\sigma)}$, and
    $(\String_2,\StringAlt) \in \SemanticsOf{(\SequenceLensOf{(\String_0,\StringAlt_0)\SeqLSep\AtomLens_1\SeqLSep\ldots\SeqLSep\AtomLens_n\SeqLSep(\String_n,\StringAlt_n)},\sigma)}$.
    
    So $\String_1 = \String_0\Concat\String_{1,1}\Concat\ldots\Concat\String_{1,n}\Concat\String_n$,
    $\StringAlt = \StringAlt_0\Concat\StringAlt_{1,\sigma(i)}\Concat\ldots\StringAlt_{1,\sigma(n)}\Concat\StringAlt_n$, and
    $(\String_{1,i},\StringAlt_{1,i}) \in \SemanticsOf{\AtomLens_i}$.
    So $\String_2 = \String_0\Concat\String_{2,1}\Concat\ldots\Concat\String_{2,n}\Concat\String_n$,
    $\StringAlt = \StringAlt_0\Concat\StringAlt_{2,\sigma(1)}\Concat\ldots\Concat\StringAlt_{2,n}\Concat\StringAlt_{\sigma(n)}$, and
    $(\String_{2,i},\StringAlt_{2,i}) \in \SemanticsOf{\DNFLens}$.
    By
    $\UnambigConcat\SequenceOf{\StringAlt_0\SeqSep\AtomAlt_{\sigma(1)}\SeqSep\ldots\SeqSep\AtomAlt_{\sigma(n)}\SeqSep\StringAlt_n}$,
    this means that $\StringAlt_{1,i} = \StringAlt_{2,i}$.
    So $(\String_{1,i},\StringAlt_{1,i})$, and $(\String_{2,i},\StringAlt_{1,i})$ are
    both in $\SemanticsOf{\AtomLens_i}$.
    As $\AtomLens$ is a bijection, by IH, $\String_{1,i} = \String_{2,i}$, so $\String_1 =
    \String_2$.

    Similarly for $\StringAlt_1,\StringAlt_2$ in
    $\LanguageOf{\StarOf{\DNFRegexAlt}}$.

    Let $\String\in\LanguageOf{\SequenceOf{\String_0\SeqSep\Atom_1\SeqSep\ldots\SeqSep\Atom_n\SeqSep\String_n}}$.
    $\String =
    \String_0\Concat\String_1'\ldots\Concat\String_n'\Concat\String_n$,
    where $\String_i' \in \LanguageOf{\Atom_i}$.
    By IH, as $\Sequence_i$ is a bijection, there exists $\StringAlt_i'\in\LanguageOf{\AtomAlt_i}$ such that
    $(\String_i',\StringAlt_i') \in \SemanticsOf{\AtomLens}$.
    So $(\String,\StringAlt_0\Concat\StringAlt_{\sigma(1)}'\Concat\ldots\Concat\StringAlt_{\sigma(n)}'\Concat\StringAlt_n) \in
    \SemanticsOf{\SequenceLens}$,
    and
    $\StringAlt_0\Concat\StringAlt_{\sigma(1)}'\Concat\ldots\Concat\StringAlt_{\sigma(n)}'\Concat\StringAlt_n
    \in
    \LanguageOf{\SequenceOf{\StringAlt_0\SeqSep\AtomAlt_{\sigma(1)}\SeqSep\ldots\SeqSep\AtomAlt_{\sigma(n)}\SeqSep\StringAlt_n}}$

    Similarly for $\StringAlt \in \LanguageOf{\StarOf{\DNFRegexAlt}}$.
  \end{case}
  
  \begin{case}[\DNFLensType{}]
    \[
      \inferrule*
      {
        \SequenceLens_1 \OfRewritelessType \Sequence_1 \Leftrightarrow \SequenceAlt_1\\
        \ldots\\
        \SequenceLens_n \OfRewritelessType \Sequence_n \Leftrightarrow \SequenceAlt_n\\\\
        \sigma \in \PermutationSetOf{n}\\
        i \neq j \Rightarrow \LanguageOf{\Sequence_{i}} \cap \LanguageOf{\Sequence_{j}}=\emptyset\\
        i \neq j \Rightarrow \LanguageOf{\SequenceAlt_{i}} \cap \LanguageOf{\SequenceAlt_{j}}=\emptyset\\
      }
      {
        (\DNFLensOf{\SequenceLens_1\DNFLSep\ldots\DNFLSep\SequenceLens_n},\sigma)
        \OfRewritelessType\\
        \DNFOf{\Sequence_1\DNFSep\ldots\DNFSep\Sequence_n}
        \Leftrightarrow
        \DNFOf{\SequenceAlt_{\sigma(1)}\DNFSep\ldots\DNFSep\SequenceAlt_{\sigma(n)}}
      }
    \]

    Let $\String_1,\String_2$ in $\LanguageOf{\DNFOf{\Sequence_1\DNFSep\ldots\DNFSep\Sequence_n}}$, and
    $(\String_1,\StringAlt) \in
    \SemanticsOf{(\DNFLensOf{\SequenceLens_1\DNFLSep\ldots\DNFLSep\SequenceLens_n},\sigma)}$,\\
    and $(\String_2,\StringAlt) \in \SemanticsOf{(\DNFLensOf{\SequenceLens_1\DNFLSep\ldots\DNFLSep\SequenceLens_n},\sigma)}$.
    
    So $\exists i.\String_1 \in \LanguageOf{\Sequence_i}$,
    $\StringAlt \in \LanguageOf{\SequenceAlt_{\sigma(i)}}$, and
    $(\String_1,\StringAlt) \in \SemanticsOf{\SequenceLens_i}$.
    So $\exists j.\String_2 \in \LanguageOf{\Sequence_j}$,
    $\StringAlt \in \LanguageOf{\SequenceAlt_{\sigma(j)}}$, and
    $(\String_1,\StringAlt) \in \SemanticsOf{\SequenceLens_j}$.
    As
    $i \neq j \Rightarrow \LanguageOf{\SequenceAlt_{i}} \cap
    \LanguageOf{\SequenceAlt_{j}}=\emptyset$, $i = j$.
    So $\String_1 \in \LanguageOf{\Sequence_i}$, $\String_2 \in
    \LanguageOf{\Sequence_i}$.
    As, by IH, $\SequenceLens_i$ is a bijection,
    $\String_1 = \String_2$.

    Similarly for $\StringAlt_1,\StringAlt_2$ in
    $\LanguageOf{\StarOf{\DNFRegexAlt}}$.

    Let $\String\in\LanguageOf{\DNFOf{\Sequence_1\DNFSep\ldots\DNFSep\Sequence_n}}$.
    So there exists an $i$ such that $\String \in \LanguageOf{\Sequence_i}$.
    By IH, as $\Sequence_i$ is a bijection, there exists $\StringAlt\in\LanguageOf{\SequenceAlt_i}$ such that
    $(\String,\StringAlt) \in \SemanticsOf{\Sequence_i}$.
    So
    $(\String,\StringAlt) \in
    \SemanticsOf{(\DNFLensOf{\SequenceLens_1\DNFLSep\ldots\DNFLSep\SequenceLens_n},\sigma)}$,
    and $\StringAlt \in
    \LanguageOf{\DNFOf{\SequenceAlt_{\sigma(1)}\DNFSep\ldots\DNFSep\SequenceAlt_{\sigma(n)}}}$. 

    Similarly for $\StringAlt \in \LanguageOf{\StarOf{\DNFRegexAlt}}$.
  \end{case}
\end{proof}

\begin{lemma}[Closure of Rewriteless Regular Expressions under Composition]
  \leavevmode
  \label{lem:composition-completeness}
  \begin{enumerate}
  \item If there are two atom lenses
    $\AtomLens_1 \OfRewritelessType \Atom_1 \Leftrightarrow \Atom_2$ and
    $\AtomLens_2 \OfRewritelessType \Atom_2 \Leftrightarrow \Atom_3$,
    then there exists an atom lens
    $\AtomLens \OfRewritelessType \Atom_1 \Leftrightarrow \Atom_3$, such that
    $\SemanticsOf{\AtomLens}=\SetOf{(\String_1,\String_3)\SuchThat
      \exists\String_2
      (\String_1,\String_2)\in\SemanticsOf{\AtomLens_1}\BooleanAnd
      (\String_2,\String_3)\in\SemanticsOf{\AtomLens_2}}$

  \item If there are two sequence lenses
    $\SequenceLens_1 \OfRewritelessType \Sequence_1 \Leftrightarrow \Sequence_2$ and
    $\SequenceLens_2 \OfRewritelessType \Sequence_2 \Leftrightarrow \Sequence_3$,
    then there exists an sequence lens
    $\SequenceLens \OfRewritelessType \Sequence_1 \Leftrightarrow \Sequence_3$, such that
    $\SemanticsOf{\SequenceLens}=\SetOf{(\String_1,\String_3)\SuchThat
      \exists\String_2
      (\String_1,\String_2)\in\SemanticsOf{\SequenceLens_1}\BooleanAnd
      (\String_2,\String_3)\in\SemanticsOf{\SequenceLens_2}}$

  \item If there are two DNF lenses
    $\DNFLens_1 \OfRewritelessType \DNFRegex_1 \Leftrightarrow \DNFRegex_2$ and
    $\DNFLens_2 \OfRewritelessType \DNFRegex_2 \Leftrightarrow \DNFRegex_3$,
    then there exists a DNF lens
    $\DNFLens \OfRewritelessType \DNFRegex_1 \Leftrightarrow \DNFRegex_3$, such that
    $\SemanticsOf{\DNFLens}=\SetOf{(\String_1,\String_3)\SuchThat
      \exists\String_2
      (\String_1,\String_2)\in\SemanticsOf{\DNFLens_1}\BooleanAnd
      (\String_2,\String_3)\in\SemanticsOf{\DNFLens_2}}$
  \end{enumerate}
\end{lemma}

\begin{proof}
  By mutual induction

  \begin{case}[Atom Lenses]
    Let $\StarOf{\DNFRegex_1}$, $\StarOf{\DNFRegex_2}$, $\StarOf{\DNFRegex_3}$
    be three atoms, and $\IterateLensOf{\DNFLens_1} \OfRewritelessType
    \StarOf{\DNFRegex_1} \Leftrightarrow \StarOf{\DNFRegex_2}$ with
    $\IterateLensOf{\DNFLens_2} \OfRewritelessType
    \StarOf{\DNFRegex_2} \Leftrightarrow \StarOf{\DNFRegex_3}$
    lenses between them.
    By induction assumption, there exists the typing of a lens

    \[
      \DNFLens \OfRewritelessType \DNFRegex_1 \Leftrightarrow \DNFRegex_3
    \]

    such that $\SemanticsOf{\DNFLens}=\SetOf{(\String_1,\String_3)\SuchThat
      \exists \String_2
      (\String_1,\String_2)\in\SemanticsOf{\DNFLens_1}\BooleanAnd
      (\String_2,\String_3)\in\SemanticsOf{\DNFLens_2}}$

    $\IterateLensOf{\DNFLens_1}$ and
    $\IterateLensOf{\DNFLens_2}$ came from $\AtomLensRule$, so
    $\UnambigItOf{\DNFRegex_1}$,
    $\UnambigItOf{\DNFRegex_2}$, and
    $\UnambigItOf{\DNFRegex_3}$.

    Consider the lens

    \[
      \inferrule*
      {
        \DNFLens \OfRewritelessType \DNFRegex_1 \Leftrightarrow \DNFRegex_3\\
        \UnambigItOf{\DNFRegex_1}\\
        \UnambigItOf{\DNFRegex_2}
      }
      {
        \IterateLensOf{\DNFLens} \OfRewritelessType
        \StarOf{\DNFRegex_1} \Leftrightarrow \StarOf{\DNFRegex_3}
      }
    \]

    This lens has the semantics

    \begin{tabular}{@{}L@{}L@{}}
      \SemanticsOf{\IterateLensOf{\DNFLens}}
      & = \SetOf{(\String_{1,1}\Concat\ldots\Concat\String_{1,n},
        \String_{3,1}\Concat\ldots\Concat\String_{3,n})
        \SuchThat(\String_{1,i},\String_{3,i})\in\SemanticsOf{\DNFLens}}\\
      & =
        \SetOf{(\String_{1,1}\Concat\ldots\Concat\String_{1,n},
        \String_{3,1}\Concat\ldots\Concat\String_{3,n})
        \SuchThat
        \exists\String_{2,i} (\String_{1,i},\String_{2,i})\in\SemanticsOf{\DNFLens_1}
        \BooleanAnd
        (\String_{2,i},\String_{3,i})\in\SemanticsOf{\DNFLens_2}}\\
      & =
        \SetOf{(\String_{1,1}\Concat\ldots\Concat\String_{1,n},
        \String_{3,1}\Concat\ldots\Concat\String_{3,n}) \\
      & \hspace*{2em}
        \SuchThat
        \exists\String_{2,1}\Concat\ldots\Concat\String_{2,n}\\
      & \hspace*{4em}
        (\String_{1,1}\Concat\ldots\Concat\String_{1,n},
        \String_{2,1}\Concat\ldots\Concat\String_{2,n})
        \in\SemanticsOf{\IterateLensOf{\DNFLens}}\\
      & \hspace*{2.8em}
        \BooleanAnd
        (\String_{2,1}\Concat\ldots\Concat\String_{2,n},
        \String_{3,1}\Concat\ldots\Concat\String_{3,n})
        \in\SemanticsOf{\IterateLensOf{\DNFLens}}}\\
      & =
        \SetOf{(\String_1,\String_3)\SuchThat\exists\String_2
        (\String_1,\String_2)\in\SemanticsOf{\IterateLensOf{\DNFLens_1}} \BooleanAnd
        (\String_2,\String_3)\in\SemanticsOf{\IterateLensOf{\DNFLens_2}}}
    \end{tabular}
  \end{case}

  \begin{case}[Sequence Lenses]
    Let $\SequenceOf{\String_{1,0}\SeqSep\Atom_{1,1}\SeqSep
      \ldots\SeqSep\Atom_{1,n}\SeqSep\String_{1,n}}$ and
    $\SequenceOf{\String_{2,0}\SeqSep\Atom_{2,\Permutation_1(1)}\SeqSep
      \ldots\SeqSep\Atom_{2,\Permutation_1(n)}\SeqSep\String_{2,n}}$
    and $\SequenceOf{\String_{3,0}\SeqSep
      \Atom_{3,\Permutation_2\Compose\Permutation_1(1)}\SeqSep
      \ldots\SeqSep\Atom_{3,\Permutation_2\Compose\Permutation_1(n)}
      \SeqSep\String_{3,n}}$ be sequences,
    with $(\SequenceLensOf{(\String_{1,0},\String_{2,0})\SeqLSep
      \AtomLens_{1,1}\SeqLSep\ldots\SeqLSep\AtomLens_{1,n}
      \SeqLSep(\String_{1,n},\String_{2,n})},\Permutation_1)$ and
    $(\SequenceLensOf{(\String_{2,0},\String_{3,0})\SeqLSep
      \AtomLens_{2,1}\SeqLSep\ldots\SeqLSep\AtomLens_{2,n}
      \SeqLSep(\String_{2,n},\String_{3,n})},\Permutation_2)$ be lenses between them.
    By induction assumption, there is a typing of lenses
    
    \begin{mathpar}
      \AtomLens_i \OfRewritelessType \Atom_{1,i} \Leftrightarrow \Atom_{3,i}
    \end{mathpar}
    
    such that \SemanticsOf{\AtomLens_i} = \SetOf{(\String_1,\String_3)\SuchThat
      \exists \String_2 (\String_1,\String_2)\in\SemanticsOf{\AtomLens_{1,i}}
      \BooleanAnd (\String_2,\String_3)\in\SemanticsOf{\Atom_{2,i}}}
    Define $\Permutation = \Permutation_2\Compose\Permutation_1$.

    $(\SequenceLensOf{(\String_{1,0},\String_{2,0})\SeqLSep
      \AtomLens_{1,1}\SeqLSep\ldots\SeqLSep\AtomLens_{1,n}
      \SeqLSep(\String_{1,n},\String_{2,n})},\Permutation_1)$
    and
    $(\SequenceLensOf{(\String_{2,0},\String_{3,0})\SeqLSep
      \AtomLens_{2,1}\SeqLSep\ldots\SeqLSep\AtomLens_{2,n}
      \SeqLSep(\String_{2,n},\String_{3,n})},\Permutation_2)$
    came from
    \SequenceLensRule{}, so 
    $\SequenceUnambigConcatOf{\SequenceOf{\String_{1,0}\SeqSep\Atom_{1,1}
        \SeqSep\ldots\SeqSep\Atom_{1,n}\SeqSep\String_{1,n}}}$ and
    $\SequenceUnambigConcatOf{\SequenceOf{\String_{3,0}\SeqSep
        \Atom_{\Permutation(3),1}\SeqSep\ldots
        \SeqSep\Atom_{\Permutation(3),n}\SeqSep\String_{3,n}}}$.

    Consider the typing of the lens
    \[
      \inferrule{
        \AtomLens_0 \OfRewritelessType \Atom_{1,0} \Leftrightarrow \Atom_{3,0}\\
        \ldots\\
        \AtomLens_n \OfRewritelessType \Atom_{1,n} \Leftrightarrow \Atom_{3,n}\\
        \sigma\in\PermutationSetOf{n}\\
        \SequenceUnambigConcatOf{\SequenceOf{\String_{1,0}\SeqSep\Atom_{1,1}
            \SeqSep\ldots\SeqSep\Atom_{1,n}\SeqSep\String_{1,n}}}\\
        \SequenceUnambigConcatOf{\SequenceOf{\String_{3,0}\SeqSep
            \Atom_{\Permutation(3),1}\SeqSep\ldots
            \SeqSep\Atom_{\Permutation(3),n}\SeqSep\String_{3,n}}}
      }
      {
        (\SequenceLensOf{(\String_{1,0},\String_{3,0})\SeqLSep\AtomLens_1
          \SeqLSep\ldots\SeqLSep\AtomLens_n
          \SeqLSep(\String_{1,n},\String_{3,n})},\Permutation) \OfRewritelessType\\
        \SequenceOf{\String_{1,0}\SeqSep\Atom_{1,1}\SeqSep\ldots
          \SeqSep\Atom_{1,n}\SeqSep\String_{1,n}} \Leftrightarrow
        \SequenceOf{\String_{3,0}\SeqSep\Atom_{3,\Permutation(1)}\SeqSep\ldots
          \SeqSep\Atom_{3,\Permutation(n)}\SeqSep\String_{3,n}}
      }
    \]

    Furthermore, we can prove the desired property of the semantics.\\\\
    \SemanticsOf{(\SequenceLensOf{(\String_{1,0},\String_{3,0})\SeqLSep
        \AtomLens_1\SeqLSep\ldots\SeqLSep\AtomLens_n
        \SeqLSep(\String_{1,n},\String_{3,n})},\Permutation)}=\\
    \SetOf{(\String_{1,0}\Concat\String_1\Concat\ldots
      \Concat\String_n\Concat\String_{1,n},
      \String_{3,0}\Concat\StringAlt_{\Permutation(1)}\Concat\ldots
      \Concat\StringAlt_{\Permutation(n)}\Concat\String_{1,n})
      \SuchThat(\String_i,\StringAlt_i)\in\SemanticsOf{\AtomLens_i}}=\\
    \SetOf{(\String_{1,0}\Concat\String_1\Concat\ldots
      \Concat\String_n\Concat\String_{1,n},
      \String_{3,0}\Concat\StringAlt_{\Permutation(1)}\Concat\ldots
      \Concat\StringAlt_{\Permutation(n)}\Concat\String_{1,n})
      \SuchThat\exists\String_i'\in\LanguageOf{\Atom_{2,i}}\That
      (\String_i,\String_i')\in\SemanticsOf{\AtomLens_i}\BooleanAnd
      (\String_i',\StringAlt_i)\in\SemanticsOf{\AtomLens_i}}=\\
    \SetOf{(\String_{1,0}\Concat\String_1\Concat\ldots
      \Concat\String_n\Concat\String_{1,n},
      \String_{3,0}\Concat\StringAlt_{\Permutation(1)}\Concat\ldots
      \Concat\StringAlt_{\Permutation(n)}\Concat\String_{1,n})\SuchThat\exists
      \String_i'\in\LanguageOf{\Atom_{2,i}}\\
      \hspace*{2em}
      (\String_{1,0}\Concat\String_1\Concat\ldots
      \Concat\String_n\Concat\String_{1,n},\String_{2,0}\Concat
      \String_{\Permutation_1(1)}'\Concat\ldots\Concat
      \String_{\Permutation_1(n)}'\Concat\String_{2,n})\in\SemanticsOf{(\SequenceLensOf{(\String_{1,0},\String_{2,0})
          \SeqLSep\AtomLens_{1,1}\SeqLSep\ldots\SeqLSep
          \AtomLens_{1,n}\SeqLSep(\String_{1,n},\String_{2,n})},\Permutation_1)}
      \BooleanAnd\\
      \hspace*{2em}
      (\String_{2,0}\Concat\String_{\Permutation_1(1)}'\Concat\ldots\Concat
      \String_{\Permutation_1(n)}'\Concat\String_{2,n},\String_{3,0}\Concat\StringAlt_{\Permutation(1)}\Concat\ldots
      \Concat\StringAlt_{\Permutation(n)}\Concat\String_{1,n})\in\SemanticsOf{(\SequenceLensOf{(\String_{2,0},\String_{3,0})\SeqLSep
          \AtomLens_{2,1}\SeqLSep\ldots\SeqLSep\AtomLens_{2,n}
          \SeqLSep(\String_{2,n},\String_{3,n})},\Permutation_2)}}=\\
    \SetOf{(\String_1,\String_3)\SuchThat\exists\String_2 \in
      \LanguageOf{\SequenceOf{\String_{2,0}\SeqSep\Atom_{2,\Permutation_1(1)}\SeqSep
      \ldots\SeqSep\Atom_{2,\Permutation_1(n)}\SeqSep\String_{2,n}}}\\
      \hspace*{2em}
      (\String_1,\String_2)\in\SemanticsOf{(\SequenceLensOf{(\String_{1,0},\String_{2,0})
          \SeqLSep\AtomLens_{1,1}\SeqLSep\ldots\SeqLSep
          \AtomLens_{1,n}\SeqLSep(\String_{1,n},\String_{2,n})},\Permutation_1)}\BooleanAnd \\
      \hspace*{2em}
      (\String_2,\String_3) \in\SemanticsOf{(\SequenceLensOf{(\String_{2,0},\String_{3,0})
          \SeqLSep\AtomLens_{2,1}\SeqLSep\ldots\SeqLSep
          \AtomLens_{2,n}\SeqLSep(\String_{2,n},\String_{3,n})},\Permutation_2)}}
  \end{case}

  \begin{case}[DNF Lenses]
    Let $\DNFRegex_1=\DNFOf{\Sequence_{1,1}\DNFSep\ldots\DNFSep\Sequence_{1,n}}$ and
    $\DNFRegex_2=\DNFOf{\Sequence_{2,\Permutation_1(1)}\DNFSep\ldots
      \DNFSep\Sequence_{2,\Permutation_1(n)}}$ and\\
    $\DNFRegex_3=\DNFOf{\Sequence_{3,\Permutation_2\Compose\Permutation_1(1)}
      \DNFSep\ldots\DNFSep\Sequence_{3,\Permutation_2\Compose\Permutation_1(n)}}$
    be three DNF regular expressions.\\
    Let $\DNFLens_1=(\DNFLensOf{\SequenceLens_{1,1}\DNFLSep\ldots
      \DNFLSep\SequenceLens_{1,n}},\Permutation_1)\OfRewritelessType
    \DNFRegex_1\Leftrightarrow\DNFRegex_2$ and
    $\DNFLens_2=(\DNFLensOf{\SequenceLens_{2,1}\DNFLSep\ldots
      \DNFLSep\SequenceLens_{2,n}},\Permutation_2)\OfRewritelessType
    \DNFRegex_2\Leftrightarrow\DNFRegex_3$ be lenses between them.
    By induction assumption, there exists a typing of lenses
    \begin{mathpar}
      \SequenceLens_i \OfRewritelessType \Sequence_{1,i} \Leftrightarrow \Sequence_{3,i}
    \end{mathpar}
    Define $\Permutation = \sigma_2\Compose\sigma_1$

    Consider the lens

    \[
      \inferrule*
      {
        \SequenceLens_i \OfRewritelessType \Sequence_{1,i} \Leftrightarrow
        \Sequence_{3,i}\\
        i \neq j \BooleanImplies \Sequence_{1,i} \Intersect \Sequence_{1,j} =
        \emptyset\\
        i \neq j \BooleanImplies \SequenceAlt_{3,i} \Intersect
        \SequenceAlt_{3,j} = \emptyset
      }
      {
        (\DNFLensOf{\SequenceLens_1 \DNFLSep \ldots \DNFLSep \SequenceLens_n},
        \sigma_2\Compose\sigma_1) \OfRewritelessType
        \DNFOf{\Sequence_{1,1}\DNFSep\ldots\DNFSep\Sequence_{1,n}}
        \Leftrightarrow
        \DNFOf{\Sequence_{3,\Permutation_2\Compose\Permutation_1(1)}
          \DNFSep\ldots\DNFSep\Sequence_{3,\Permutation_2\Compose\Permutation_1(n)}}
      }
    \]

    Furthermore, we can prove the desired property of the semantics.\\\\
    \SemanticsOf{(\DNFLensOf{\SequenceLens_1 \DNFLSep \ldots \DNFLSep
        \SequenceLens_n},\Permutation)}=\\
    \SetOf{(\String,\StringAlt)
      \SuchThat \exists i \That
      (\String,\StringAlt)\in\SemanticsOf{\SequenceLens_i}}=\\
    \SetOf{(\String, \StringAlt)
      \SuchThat \exists i \That \exists
      \String'\in\LanguageOf{\Sequence_{2,i}}\That 
      (\String,\String')\in\SemanticsOf{\SequenceLens_{1,i}}\BooleanAnd
      (\String',\StringAlt)\in\SemanticsOf{\SequenceLens_{2,i}}}=\\
    \SetOf{(\String, \StringAlt)
      \SuchThat \exists
      \String'\in\LanguageOf{\DNFOf{\Sequence_{\sigma_1(1)} \DNFSep \ldots
          \DNFSep \Sequence_{\sigma_1(n)}}}\That 
      (\String,\String')\in\SemanticsOf{\SequenceLens_{1,i}}\BooleanAnd
      (\String',\StringAlt)\in\SemanticsOf{\SequenceLens_{2,i}}}
  \end{case}
\end{proof}

\begin{lemma}[Expressibility of Identity on Strongly Unambiguous DNF Regex,
  Clauses, and Atoms]
  \label{lem:strongly-unambiguous-identity-expressible}
  \leavevmode
  \begin{enumerate}
  \item If $\DNFRegex$ is a strongly unambiguous DNF Regular expression, then
    there exists a DNF lens $\DNFLens \OfRewritelessType \DNFRegex \Leftrightarrow
    \DNFRegex$,
    such that $\SemanticsOf{\DNFLens}=
    \SetOf{(\String,\String)\SuchThat\String\in\LanguageOf{\DNFRegex}}$, where
    \DNFLens{} typing includes no rewrite rules.
  \item If $\Sequence$ is a strongly unambiguous sequence, then
    there exists a sequence lens $\SequenceLens \OfRewritelessType \Sequence
    \Leftrightarrow \Sequence$,
    such that $\SemanticsOf{\SequenceLens}=
    \SetOf{(\String,\String)\SuchThat\String\in\LanguageOf{\DNFRegex}}$, where
    \SequenceLens{} typing includes no rewrite rules.
  \item If $\Atom$ is a strongly unambiguous atom, then
    there exists an atom lens $\AtomLens \OfRewritelessType \Atom
    \Leftrightarrow
    \Atom$, such that $\SemanticsOf{\AtomLens}=
    \SetOf{(\String,\String)\SuchThat\String\in\LanguageOf{\DNFRegex}}$, where
    \AtomLens{} typing includes no rewrite rules.
  \end{enumerate}
\end{lemma}
\begin{proof}
  By mutual induction on the structure of the DNF regular expression,
  atom, and clause.
  \begin{case}[\StarAtomType{}]
    Let $\Atom = \StarOf{\DNFRegex}$.
    As $\Atom$ is strongly unambiguous, $\DNFRegex$ is strongly unambiguous,
    and $\UnambigItOf{\DNFRegex}$.
    By IH, there exists $\DNFLens \OfRewritelessType \DNFRegex \Leftrightarrow
    \DNFRegex$ such that $\SemanticsOf{\DNFLens} = \SetOf{(\String,\String)
      \SuchThat \String \in \LanguageOf{\DNFRegex}}$.
    Consider the atom lens
    \[
      \inferrule*
      {
        \DNFLens \OfRewritelessType \DNFRegex \Leftrightarrow \DNFRegex\\
        \UnambigItOf{\DNFRegex}\\
        \UnambigItOf{\DNFRegex}
      }
      {
        \IterateLensOf{\DNFLens} \OfRewritelessType \StarOf{\DNFRegex}
        \Leftrightarrow \StarOf{\DNFRegex}
      }
    \] with typing as desired.

    $\SemanticsOf{\IterateLensOf{\DNFLens}} = \SetOf{(\String_1\Concat\ldots\Concat\String_n,
      \StringAlt_1\Concat\ldots\Concat\StringAlt_n) \SuchThat
      (\String_i,\StringAlt_i) \in \SemanticsOf{\DNFLens}}$.  So through
    semantics of $\DNFLens$,
    $\SemanticsOf{\IterateLensOf{\DNFLens}} = \SetOf{(\String_1\Concat\ldots\Concat\String_n,
      \String_1\Concat\ldots\Concat\String_n) \SuchThat
      \String \in \LanguageOf{\DNFRegex}}$, so through the definition of
    $\StarOf{\DNFRegex}$,
    $\SemanticsOf{\IterateLensOf{\DNFLens}} = \SetOf{(\String,\String)
      \SuchThat \String \in \LanguageOf{\StarOf{\DNFRegex}}}$
  \end{case}

  \begin{case}[\MultiConcatSequenceType{}]
    Let $\Sequence = \SequenceOf{\String_0 \SeqSep \Atom_1 \SeqSep \ldots
      \SeqSep \Atom_n \SeqSep \String_n}$.
    As $\Sequence$ is strongly unambiguous, for all $i$, $\Atom_i$ is strongly
    unambiguous, and
    $\SequenceUnambigConcatOf{\String_0;\Atom_1;\ldots;\Atom_n;\String_n}$.

    By IH, for all $i$, there exists $\AtomLens_i \OfRewritelessType \Atom_i
    \Leftrightarrow \Atom_i$ such that
    $\SemanticsOf{\AtomLens_i} = \SetOf{(\String,\String) \SuchThat \String \in
      \LanguageOf{\Atom_i}}$.

    Consider the typing
    \[
      \inferrule*
      {
        \AtomLens_1 \OfRewritelessType \Atom_1 \Leftrightarrow \Atom_1\\
        \ldots\\
        \AtomLens_n \OfRewritelessType \Atom_n \Leftrightarrow \Atom_n\\
        \Identity \in \PermutationSetOf{n}\\
        \SequenceUnambigConcatOf{\String_0;\Atom_1;\ldots;\Atom_n;\String_n}\\
        \SequenceUnambigConcatOf{\String_0;\Atom_1;\ldots;\Atom_n;\String_n}
      }
      {
        (\SequenceLensOf{(\String_0,\String_0)\SeqLSep\AtomLens_1\SeqLSep\ldots\SeqLSep\AtomLens_n\SeqLSep(\String_n,\String_n)},\Identity) \OfRewritelessType\\
        \SequenceOf{\String_0\SeqSep\Atom_1\SeqSep\ldots\SeqSep\Atom_n\SeqSep\String_n}\Leftrightarrow
        \SequenceOf{\String_0\SeqSep\Atom_1\SeqSep\ldots\SeqSep\Atom_n\SeqSep\String_n}
      }
    \],
    as desired.
    
    $\SemanticsOf{(\SequenceLensOf{(\String_0,\String_0)\SeqLSep\AtomLens_1\SeqLSep\ldots\SeqLSep\AtomLens_n\SeqLSep(\String_n,\String_n)},\Identity)}
    = 
    \SetOf{(\String_0\Concat\StringAlt_1\Concat\ldots\Concat\StringAlt_n\Concat\String_n,\String_0\Concat\StringAlt_1'\Concat\ldots\Concat\StringAlt_n'\Concat\String_n)
      \SuchThat
      (\StringAlt_i,\StringAlt_i') \in \SemanticsOf{\AtomLens_i}}$.  So, through
    the definition of $\AtomLens_i$, 
    $\SemanticsOf{(\SequenceLensOf{(\String_0,\String_0)\SeqLSep\AtomLens_1\SeqLSep\ldots\SeqLSep\AtomLens_n\SeqLSep(\String_n,\String_n)},\Identity)}
    = 
    \SetOf{(\String_0\Concat\StringAlt_1\Concat\ldots\Concat\StringAlt_n\Concat\String_n,\String_0\Concat\StringAlt_1\Concat\ldots\Concat\StringAlt_n\Concat\String_n)
      \SuchThat
      \StringAlt_i \in \LanguageOf{\Atom_i}}$.  So, through the definition of
    $\SequenceOf{\String_0 \SeqSep \Atom_1 \SeqSep \ldots \SeqSep \Atom_n \SeqSep \String_n}$,
    $\SemanticsOf{(\SequenceLensOf{(\String_0,\String_0)\SeqLSep\AtomLens_1\SeqLSep\ldots\SeqLSep\AtomLens_n\SeqLSep(\String_n,\String_n)},\Identity)}
    = 
    \SetOf{(\String,\String)
      \SuchThat
      \String \in
      \LanguageOf{\SequenceOf{\String_0 \SeqSep \Atom_1 \SeqSep \ldots \SeqSep \Atom_n \SeqSep \String_n}}}$,
    as desired.
  \end{case}

  \begin{case}[\MultiOrDNFRegexType{}]
    Let $\DNFRegex = \DNFOf{\Sequence_1\DNFSep\ldots\DNFSep\Sequence_n}$.
    As $\DNFRegex$ is strongly unambiguous, for all $i$, $\Sequence_i$ is strongly
    unambiguous, and
    $i \neq j \BooleanImplies \LanguageOf{\Sequence_i} \Intersect
    \LanguageOf{\Sequence_j} = \SetOf{}$.

    By IH, for all $i$, there exists $\AtomLens_i \OfRewritelessType \Atom_i
    \Leftrightarrow \Atom_i$ such that
    $\SemanticsOf{\AtomLens_i} = \SetOf{(\String,\String) \SuchThat \String \in
      \LanguageOf{\Atom_i}}$.

    Consider the typing
    \[
      \inferrule*
      {
        \SequenceLens_1 \OfRewritelessType \Sequence_1 \Leftrightarrow \Sequence_1\\
        \ldots\\
        \SequenceLens_1 \OfRewritelessType \Sequence_1 \Leftrightarrow \Sequence_1\\
        \Identity \in \PermutationSetOf{n}\\
        i \neq j \Rightarrow \LanguageOf{\Sequence_{i}} \cap \LanguageOf{\Sequence_{j}}=\emptyset\\
        i \neq j \Rightarrow \LanguageOf{\Sequence_{i}} \cap \LanguageOf{\Sequence_{j}}=\emptyset
      }
      {
        (\DNFLensOf{\SequenceLens_1\DNFLSep\ldots\DNFLSep\SequenceLens_n},\Identity) \OfRewritelessType\\
        \DNFOf{\Sequence_1\DNFSep\ldots\DNFSep\Sequence_n}
        \Leftrightarrow \DNFOf{\Sequence_1\DNFSep\ldots\DNFSep\Sequence_n}
      }
    \]
    as desired.
    
    $\SemanticsOf{(\DNFLensOf{\SequenceLens_1\DNFLSep\ldots\DNFLSep\SequenceLens_n},\Identity)}
    = 
    \SetOf{(\String,\StringAlt)
      \SuchThat
      \exists i.
      (\String,\StringAlt) \in \SemanticsOf{\SequenceLens_i}}$.
    So, through the definition of $\SequenceLens_i$,\\
    $\SemanticsOf{(\DNFLensOf{\SequenceLens_1\DNFLSep\ldots \DNFLSep \ldots \SequenceLens_n},\Identity)}
    = 
    \SetOf{(\String,\String)
      \SuchThat \exists i.
      \String \in \LanguageOf{\Sequence_i}}$.
    So, through the definition of
    $\DNFOf{\Sequence_1\DNFSep\ldots\DNFSep\Sequence_n}$,
    $\SemanticsOf{(\DNFLensOf{\SequenceLens_1\DNFLSep\ldots\DNFLSep\SequenceLens_n},\Identity)} 
    = 
    \SetOf{(\String,\String)
      \SuchThat
      \String \in
      \LanguageOf{\DNFOf{\Sequence_1\DNFSep\ldots\DNFSep\Sequence_n}}}$,
    as desired
  \end{case}
\end{proof}

\begin{definition}[Strong Unambiguity on DNF Regular Expressions]
  \leavevmode
  \begin{enumerate}
  \item $\DNFOf{\Sequence_1\DNFSep\ldots\DNFSep\Sequence_n}$ is \emph{strongly umambiguous}
    if $\Sequence_i$ is strongly unambiguous for all $i$, and $i\neq j
    \BooleanImplies
    \LanguageOf{\Sequence_i}\Intersect\LanguageOf{\Sequence_j}=\emptyset$.
  \item $\SequenceOf{\String_0 \SeqSep \Atom_1 \SeqSep \ldots \SeqSep \Atom_n \SeqSep \Sequence_n}$ is
    \emph{strongly unambiguous} if $\Atom_i$ is strongly unambiguous, and
    $\SequenceUnambigConcatOf{\String_0;\Atom_1;\ldots;\Atom_n;\String_n}$.
  \item $\StarOf{\DNFRegex}$ is \emph{strongly unambiguous} if $\DNFRegex$ is
    strongly unambiguous, and $\UnambigItOf{\DNFRegex}$.
  \end{enumerate}
\end{definition}

\begin{lemma}[Strong Unambiguity in Lens Types]
  \label{lem:strong-unambig-lens-types}
  If $\Lens \OfType \Regex \Leftrightarrow \RegexAlt$, then $\Regex$ is strongly
  unambiguous, and $\RegexAlt$ is strongly unambiguous.
\end{lemma}
\begin{proof}
  By induction on the typing derivation of $\Lens$
  \begin{case}[\ConstLensType{}]
    \[
      \inferrule*
      {
        \String_1 \in \StarOf{\Sigma}\\
        \String_2 \in \StarOf{\Sigma}
      }
      {
        \ConstLensOf{\String_1}{\String_2} \OfType \String_1 \Leftrightarrow \String_2
      }
    \]

    Base strings are strongly unambiguous.
  \end{case}

  \begin{case}[\ConcatLensType{}]
    \[
      \inferrule*
      {
        \Lens_1 \OfType \Regex_1 \Leftrightarrow \RegexAlt_1\\\\
        \Lens_2 \OfType \Regex_2 \Leftrightarrow \RegexAlt_2\\\\
        \UnambigConcatOf{\Regex_1}{\Regex_2}\\
        \UnambigConcatOf{\RegexAlt_1}{\RegexAlt_2}
      }
      {
        \ConcatLensOf{\Lens_1}{\Lens_2} \OfType \Regex_1\Regex_2 \Leftrightarrow \RegexAlt_1\RegexAlt_2
      }
    \]

    So by IH, $\Regex_1$, $\Regex_2$, $\RegexAlt_1$, and $\RegexAlt_2$ are all
    strongly unambiguous.

    As $\UnambigConcatOf{\Regex_1}{\Regex_2}$, $\Regex_1 \Concat \Regex_2$ is
    strongly unambiguous.

    As $\UnambigConcatOf{\RegexAlt_1}{\RegexAlt_2}$, $\RegexAlt_1 \Concat \RegexAlt_2$ is
    strongly unambiguous.
  \end{case}

  \begin{case}[\IterateLensType{}]
    \[
      \inferrule*
      {
        \Lens \OfType \Regex \Leftrightarrow \RegexAlt\\
        \UnambigItOf{\Regex}\\
        \UnambigItOf{\RegexAlt}
      }
      {
        \IterateLensOf{\Lens} \OfType \StarOf{\Regex} \Leftrightarrow \StarOf{\RegexAlt}
      }
    \]

    So, by IH, $\Regex$ and $\RegexAlt$ are both strongly unambiguous.

    As $\UnambigItOf{\Regex}$, $\StarOf{\Regex}$ is strongly unambiguous.

    As $\UnambigItOf{\RegexAlt}$, $\StarOf{\RegexAlt}$ is strongly unambiguous.
  \end{case}

  \begin{case}[\SwapLensType{}]
    \[
      \inferrule*
      {
        \Lens_1 \OfType \Regex_1 \Leftrightarrow \RegexAlt_1\\\\
        \Lens_2 \OfType \Regex_2 \Leftrightarrow \RegexAlt_2\\\\
        \UnambigConcatOf{\Regex_1}{\Regex_2}\\
        \UnambigConcatOf{\RegexAlt_2}{\RegexAlt_1}
      }
      {
        \SwapLensOf{\Lens_1}{\Lens_2} \OfType \Regex_1\Regex_2 \Leftrightarrow \RegexAlt_2\RegexAlt_1
      }
    \]

    So by IH, $\Regex_1$, $\Regex_2$, $\RegexAlt_1$, and $\RegexAlt_2$ are all
    strongly unambiguous.

    As $\UnambigConcatOf{\Regex_1}{\Regex_2}$, $\Regex_1 \Concat \Regex_2$ is
    strongly unambiguous.

    As $\UnambigConcatOf{\RegexAlt_2}{\RegexAlt_1}$, $\RegexAlt_2 \Concat \RegexAlt_1$ is
    strongly unambiguous.
  \end{case}

  \begin{case}[\OrLensType{}]
    \[
      \inferrule*
      {
        \Lens_1 \OfType \Regex_1 \Leftrightarrow \RegexAlt_1\\
        \Lens_2 \OfType \Regex_2 \Leftrightarrow \RegexAlt_2\\\\
        \UnambigOrOf{\Regex_1}{\Regex_2}\\
        \UnambigOrOf{\RegexAlt_1}{\RegexAlt_2}
      }
      {
        \OrLensOf{\Lens_1}{\Lens_2} \OfType
        \Regex_1 \Or \Regex_2
        \Leftrightarrow \RegexAlt_1 \Or \RegexAlt_2
      }
    \]

    So by IH, $\Regex_1$, $\Regex_2$, $\RegexAlt_1$, and $\RegexAlt_2$ are all
    strongly unambiguous.

    As $\UnambigOrOf{\Regex_1}{\Regex_2}$, $\Regex_1 \Or \Regex_2$ is
    strongly unambiguous.

    As $\UnambigOrOf{\RegexAlt_2}{\RegexAlt_1}$, $\RegexAlt_2 \Or \RegexAlt_1$ is
    strongly unambiguous.
  \end{case}

  \begin{case}[\ComposeLensType{}]
    \[
      \inferrule*
      {
        \Lens_1 \OfType \Regex_1 \Leftrightarrow \Regex_2\\
        \Lens_2 \OfType \Regex_2 \Leftrightarrow \Regex_3\\
      }
      {
        \ComposeLensOf{\Lens_1}{\Lens_2} \OfType \Regex_1 \Leftrightarrow \Regex_3
      }
    \]

    By IH, $\Regex_1$ is strongly unambiguous, and $\Regex_3$ is strongly
    unambiguous.
  \end{case}

  \begin{case}[\IdentityLensType{}]
    \[
      \inferrule*
      {
        \Regex \text{ is strongly unambiguous}
      }
      {
        \IdentityLensOf{\Regex} \OfType \Regex \Leftrightarrow \Regex
      }
    \]

    By assumption, $\Regex$ is strongly unambiguous.
  \end{case}
\end{proof}

\begin{lemma}[Strong Unambiguity in Rewriteless DNF Lens Types]
  \label{lem:strong-unambig-dnf-lens-types}
  \leavevmode
  \begin{itemize}
  \item If $\DNFLens \OfRewritelessType \DNFRegex \Leftrightarrow \DNFRegexAlt$,
    then $\DNFRegex$ is strongly unambiguous, and $\DNFRegexAlt$ is strongly
    unambiguous.
  \item If $\SequenceLens \OfRewritelessType \Sequence \Leftrightarrow \SequenceAlt$,
    then $\Sequence$ is strongly unambiguous, and $\SequenceAlt$ is strongly
    unambiguous.
  \item If $\AtomLens \OfRewritelessType \Atom \Leftrightarrow \AtomAlt$,
    then $\Atom$ is strongly unambiguous, and $\AtomAlt$ is strongly
    unambiguous.
  \end{itemize}
\end{lemma}
\begin{proof}
  By mutual induction on the typing derivation of $\DNFLens$, $\SequenceLens$,
  and $\AtomLens$.
  \begin{case}[\AtomLensType{}]
    \[
      \inferrule*
      {
        \DNFLens \OfRewritelessType \DNFRegex \Leftrightarrow \DNFRegexAlt\\
        \UnambigItOf{\DNFRegex}\\
        \UnambigItOf{\DNFRegexAlt}
      }
      {
        \IterateLensOf{\DNFLens} \OfRewritelessType \StarOf{\DNFRegex}
        \Leftrightarrow \StarOf{\DNFRegexAlt}
      }
    \]

    By IH, $\DNFRegex$ and $\DNFRegexAlt$ are strongly unambiguous.

    As $\UnambigItOf{\DNFRegex}$, $\StarOf{\DNFRegex}$ is strongly unambiguous.

    As $\UnambigItOf{\DNFRegexAlt}$, $\StarOf{\DNFRegexAlt}$ is strongly
    unambiguous.
  \end{case}

  \begin{case}[\SequenceLensType{}]
    \[
      \inferrule*
      {
        \AtomLens_1 \OfRewritelessType \Atom_1 \Leftrightarrow \AtomAlt_1\\
        \ldots\\
        \AtomLens_n \OfRewritelessType \Atom_n \Leftrightarrow \AtomAlt_n\\
        \sigma \in \PermutationSetOf{n}\\
        \UnambigConcat\SequenceOf{\String_0\SeqSep\Atom_1\SeqSep\ldots\SeqSep\Atom_n\SeqSep\String_n}\\
        \UnambigConcat\SequenceOf{\StringAlt_0\SeqSep\AtomAlt_{\sigma(1)}\SeqSep\ldots\SeqSep\AtomAlt_{\sigma(n)}\SeqSep\StringAlt_n}
      }
      {
        (\SequenceLensOf{(\String_0,\StringAlt_0)\SeqLSep\AtomLens_1\SeqLSep\ldots\SeqLSep\AtomLens_n\SeqLSep(\String_n,\StringAlt_n)},\sigma) \OfRewritelessType\\
        \SequenceOf{\String_0\SeqSep\Atom_1\SeqSep\ldots\SeqSep\Atom_n\SeqSep\String_n}\Leftrightarrow
        \SequenceOf{\StringAlt_0\SeqSep\AtomAlt_{\sigma(1)}\SeqSep\ldots\SeqSep\AtomAlt_{\sigma(n)}\SeqSep\StringAlt_n}
      }
    \]

    By IH, $\Atom_i$ and $\AtomAlt_i$ are strongly unambiguous for all $i$.

    As
    $\UnambigConcat\SequenceOf{\String_0\SeqSep\Atom_1\SeqSep\ldots\SeqSep\Atom_n\SeqSep\String_n}$,
    we have
    $\SequenceOf{\String_0\SeqSep\Atom_1\SeqSep\ldots\SeqSep\Atom_n\SeqSep\String_n}$
    is strongly unambiguous.

    As
    $\UnambigConcat\SequenceOf{\StringAlt_0\SeqSep\AtomAlt_{\sigma(1)}\SeqSep\ldots\SeqSep\AtomAlt_{\sigma(n)}\SeqSep\StringAlt_n}$,
    we have
    $\SequenceOf{\StringAlt_0\SeqSep\AtomAlt_{\sigma(1)}\SeqSep\ldots\SeqSep\AtomAlt_{\sigma(n)}\SeqSep\StringAlt_n}$
    is strongly unambiguous.
  \end{case}

  \begin{case}[\DNFLensType{}]
    \[
      \inferrule*
      {
        \SequenceLens_1 \OfRewritelessType \Sequence_1 \Leftrightarrow \SequenceAlt_1\\
        \ldots\\
        \SequenceLens_n \OfRewritelessType \Sequence_n \Leftrightarrow \SequenceAlt_n\\\\
        \sigma \in \PermutationSetOf{n}\\
        i \neq j \Rightarrow \LanguageOf{\Sequence_{i}} \cap \LanguageOf{\Sequence_{j}}=\emptyset\\
        i \neq j \Rightarrow \LanguageOf{\SequenceAlt_{i}} \cap \LanguageOf{\SequenceAlt_{j}}=\emptyset\\
      }
      {
        (\DNFLensOf{\SequenceLens_1\DNFLSep\ldots\DNFLSep\SequenceLens_n},\sigma) \OfRewritelessType\\
        \DNFOf{\Sequence_1\DNFSep\ldots\DNFSep\Sequence_n}
        \Leftrightarrow \DNFOf{\SequenceAlt_{\sigma(1)}\DNFSep\ldots\DNFSep\SequenceAlt_{\sigma(n)}}
      }
    \]

    By IH, $\Sequence_i$ and $\SequenceAlt_i$ are strongly unambiguous for all
    $i$.

    As $i \neq j \Rightarrow \LanguageOf{\Sequence_{i}} \cap
    \LanguageOf{\Sequence_{j}}=\emptyset$, we have
    $\DNFOf{\Sequence_1\DNFSep\ldots\DNFSep\Sequence_n}$ is strongly
    unambiguous.

    As $i \neq j \Rightarrow \LanguageOf{\SequenceAlt_{i}} \cap
    \LanguageOf{\SequenceAlt_{j}}=\emptyset$, we have
    $\DNFOf{\SequenceAlt_{\sigma(1)}\DNFSep\ldots\DNFSep\SequenceAlt_{\sigma(n)}}$
    is strongly unambiguous.
  \end{case}
\end{proof}

\begin{lemma}[Closure of Rewriteless DNF Lenses Under Inversion]
  \label{lem:closure-inversion}
  \leavevmode
  \begin{enumerate}
  \item If
    $\DNFLens \OfRewritelessType \DNFRegex \Leftrightarrow \DNFRegexAlt$,
    then there exists a dnf lens
    $\InverseOf{\DNFLens} \OfRewritelessType
    \DNFRegexAlt \Leftrightarrow \DNFRegex$
    such that
    $\SemanticsOf{\InverseOf{\DNFRegex}} =
    \SetOf{(\StringAlt,\String) \SuchThat (\String,\StringAlt) \in
      \SemanticsOf{\DNFLens}}$
  \item If
    $\SequenceLens \OfRewritelessType \Sequence \Leftrightarrow \SequenceAlt$,
    then there exists a sequence lens
    $\InverseOf{\SequenceLens} \OfRewritelessType
    \SequenceAlt \Leftrightarrow \Sequence$ such that
    $\SemanticsOf{\InverseOf{\SequenceLens}} =
    \SetOf{(\StringAlt,\String) \SuchThat (\String,\StringAlt) \in
      \SemanticsOf{\SequenceLens}}$
  \item If
    $\AtomLens \OfRewritelessType \Atom \Leftrightarrow \AtomAlt$,
    then there exists an atom lens
    $\InverseOf{\AtomLens} \OfRewritelessType
    \AtomAlt \Leftrightarrow \Atom$ such that
    $\SemanticsOf{\InverseOf{\AtomLens}} =
    \SetOf{(\StringAlt,\String) \SuchThat (\String,\StringAlt) \in
      \SemanticsOf{\AtomLens}}$
  \end{enumerate}
\end{lemma}
\begin{proof}
  By mutual induction on the typing derivation of $\DNFLens$, $\SequenceLens$,
  and $\AtomLens$.

  \begin{case}[\DNFLensRule{}]
    \[
      \inferrule*
      {
        \SequenceLens_1 \OfRewritelessType \Sequence_1 \Leftrightarrow \SequenceAlt_1\\
        \ldots\\
        \SequenceLens_n \OfRewritelessType \Sequence_n \Leftrightarrow \SequenceAlt_n\\\\
        \sigma \in \PermutationSetOf{n}\\
        i \neq j \Rightarrow \LanguageOf{\Sequence_{i}} \cap \LanguageOf{\Sequence_{j}}=\emptyset\\
        i \neq j \Rightarrow \LanguageOf{\SequenceAlt_{i}} \cap \LanguageOf{\SequenceAlt_{j}}=\emptyset\\
      }
      {
        (\DNFLensOf{\SequenceLens_1\DNFLSep\ldots\DNFLSep\SequenceLens_n},\sigma) \OfRewritelessType
        \DNFOf{\Sequence_1\DNFSep\ldots\DNFSep\Sequence_n}
        \Leftrightarrow \DNFOf{\SequenceAlt_{\sigma(1)}\DNFSep\ldots\DNFSep\SequenceAlt_{\sigma(n)}}
      }
    \]

    By IH, there exists
    $\InverseOf{\SequenceLens_i} \OfRewritelessType
    \SequenceAlt_i \Leftrightarrow \Sequence_i$ where
    $\SemanticsOf{\InverseOf{\SequenceLens_i}} =
    \SetOf{(\StringAlt,\String) \SuchThat
      (\String,\StringAlt) \in \SemanticsOf{\SequenceLens_i}}$.

    Consider the typing
    \[
      \inferrule*
      {
        \InverseOf{\SequenceLens_{\sigma(1)}} \OfRewritelessType
        \SequenceAlt_{\sigma(1)} \Leftrightarrow \Sequence_{\sigma(1)}\\
        \ldots\\
        \InverseOf{\SequenceLens_{\InverseOf{\sigma(n)}}} \OfRewritelessType
        \SequenceAlt_{\sigma(n)} \Leftrightarrow \Sequence_{\sigma(n)}\\\\
        \InverseOf{\sigma} \in \PermutationSetOf{n}\\
        i \neq j \Rightarrow \LanguageOf{\SequenceAlt_{i}} \cap \LanguageOf{\SequenceAlt_{j}}=\emptyset\\
        i \neq j \Rightarrow \LanguageOf{\Sequence_{i}} \cap \LanguageOf{\Sequence_{j}}=\emptyset\\
      }
      {
        (\DNFLensOf{\InverseOf{\SequenceLens_{\sigma(1)}}
          \DNFLSep\ldots\DNFLSep
          \InverseOf{\SequenceLens_{\sigma(n)}}},\InverseOf{\sigma}) \OfRewritelessType
        \DNFOf{\SequenceAlt_{\sigma(1)}\DNFSep\ldots\DNFSep\SequenceAlt_{\sigma(n)}}
        \Leftrightarrow
        \DNFOf{\Sequence_{\InverseOf{\sigma}(\sigma(1))}\DNFSep\ldots\DNFSep\Sequence_{\InverseOf{\sigma}(\sigma(n)})}
      }
    \]

    So $(\DNFLensOf{\InverseOf{\SequenceLens_{\sigma(1)}}
      \DNFLSep\ldots\DNFLSep
      \InverseOf{\SequenceLens_{\sigma(n)}}},\InverseOf{\sigma}) \OfRewritelessType
    \DNFOf{\SequenceAlt_{\sigma(1)}\DNFSep\ldots\DNFSep\SequenceAlt_{\sigma(n)}}
    \Leftrightarrow
    \DNFOf{\Sequence_{1}\DNFSep\ldots\DNFSep\Sequence_{n})}$, or in other words
    $(\DNFLensOf{\SequenceLens_{\sigma(1)}
      \DNFLSep\ldots\DNFLSep
      \SequenceLens_{\sigma(n)}},\InverseOf{\sigma}) \OfRewritelessType
    \DNFRegexAlt
    \Leftrightarrow
    \DNFRegex$, as desired.
    
    $\SemanticsOf{(\DNFLensOf{\InverseOf{\SequenceLens_{\sigma(1)}}
        \DNFLSep\ldots\DNFLSep
        \InverseOf{\SequenceLens_{\sigma(n)}}},\InverseOf{\sigma})} =
    \SetOf{(\String,\StringAlt) \SuchThat \exists i.
      (\String,\StringAlt) \in \SemanticsOf{\InverseOf{\SequenceLens_{\sigma(i)}}}} =
    \SetOf{(\StringAlt,\String) \SuchThat \exists i.
      (\String,\StringAlt) \in \SemanticsOf{\SequenceLens_i}} =
    \SetOf{(\StringAlt,\String) \SuchThat
      (\String,\StringAlt) \in \SemanticsOf{\DNFLens}}$, as desired.
  \end{case}

  \begin{case}[\SequenceLensRule{}]
    \[
      \inferrule*
      {
        \AtomLens_1 \OfRewritelessType \Atom_1 \Leftrightarrow \AtomAlt_1\\
        \ldots\\
        \AtomLens_n \OfRewritelessType \Atom_n \Leftrightarrow \AtomAlt_n\\
        \sigma \in \PermutationSetOf{n}\\
        \UnambigConcat\SequenceOf{\String_0\SeqSep\Atom_1\SeqSep\ldots\SeqSep\Atom_n\SeqSep\String_n}\\
        \UnambigConcat\SequenceOf{\StringAlt_0\SeqSep\AtomAlt_1\SeqSep\ldots\SeqSep\AtomAlt_n\SeqSep\StringAlt_n}
      }
      {
        (\SequenceLensOf{(\String_0,\StringAlt_0)\SeqLSep\Atom_1\SeqLSep\ldots\SeqLSep\Atom_n\SeqLSep(\String_n,\StringAlt_n)},\sigma) \OfRewritelessType
        \SequenceOf{\String_0\SeqSep\Atom_1\SeqSep\ldots\SeqSep\Atom_n\SeqSep\String_n}\Leftrightarrow
        \SequenceOf{\StringAlt_0\SeqSep\AtomAlt_{\sigma(1)}\SeqSep\ldots\SeqSep\AtomAlt_{\sigma(n)}\SeqSep\StringAlt_n}
      }
    \]

    By IH, there exists
    $\InverseOf{\AtomLens_i} \OfRewritelessType
    \AtomAlt_i \Leftrightarrow \Atom_i$ where
    $\SemanticsOf{\InverseOf{\AtomLens_i}} =
    \SetOf{(\StringAlt,\String) \SuchThat
      (\String,\StringAlt) \in \SemanticsOf{\AtomLens_i}}$.

    Consider the typing
    \[
      \inferrule*
      {
        \InverseOf{\AtomLens_{\sigma(1)}} \OfRewritelessType
        \AtomAlt_{\sigma(1)} \Leftrightarrow \Atom_{\sigma(n)}\\
        \ldots\\
        \InverseOf{\AtomLens_{\sigma(n)}} \OfRewritelessType
        \AtomAlt_{\sigma(n)} \Leftrightarrow \Atom_{\sigma(n)}\\
        \InverseOf{\sigma} \in \PermutationSetOf{n}\\
        \UnambigConcat\SequenceOf{\StringAlt_0\SeqSep\AtomAlt_1\SeqSep\ldots\SeqSep\AtomAlt_n\SeqSep\StringAlt_n}
        \UnambigConcat\SequenceOf{\String_0\SeqSep\Atom_1\SeqSep\ldots\SeqSep\Atom_n\SeqSep\String_n}\\
      }
      {
        (\SequenceLensOf{(\StringAlt_0,\String_0)\SeqLSep\Atom_1\SeqLSep\ldots\SeqLSep\Atom_n\SeqLSep(\StringAlt_n,\String_n)},\sigma) \OfRewritelessType
        \SequenceOf{\StringAlt_0\SeqSep\AtomAlt_{\sigma(1)}\SeqSep\ldots\SeqSep\AtomAlt_{\sigma(n)}\SeqSep\StringAlt_n}
        \Leftrightarrow
        \SequenceOf{\String_0\SeqSep\Atom_{\InverseOf{\sigma}(\sigma(1))}\SeqSep\ldots\SeqSep\Atom_{\InverseOf{\sigma}(\sigma(n))}\SeqSep\String_n}
      }
    \]

    So $(\SequenceLensOf{(\StringAlt_0,\String_0) \SeqLSep 
      \InverseOf{\AtomLens_{\sigma(1)}}
      \SeqLSep\ldots\SeqLSep
      \InverseOf{\AtomLens_{\sigma(n)}} \SeqLSep 
      (\StringAlt_n,\String_n)},\InverseOf{\sigma}) \OfRewritelessType
    \SequenceOf{\StringAlt_0 \SeqSep \SequenceAlt_{\sigma(1)}
      \SeqSep\ldots\SeqSep
      \SequenceAlt_{\sigma(n)} \SeqSep \StringAlt_n}
    \Leftrightarrow
    \SequenceOf{\String_0 \SeqSep \Sequence_{1}\DNFSep\ldots\DNFSep\Sequence_{n})}$, or in other words
    $(\SequenceLensOf{(\StringAlt_0,\String_0) \SeqLSep 
      \InverseOf{\AtomLens_{\sigma(1)}}
      \SeqLSep\ldots\SeqLSep
      \InverseOf{\AtomLens_{\sigma(n)}} \SeqLSep 
      (\StringAlt_n,\String_n)},\InverseOf{\sigma}) \OfRewritelessType
    \SequenceAlt
    \Leftrightarrow
    \Sequence$, as desired.
    
    $\SemanticsOf{(\SequenceLensOf{(\StringAlt_0,\String_0) \SeqLSep 
      \InverseOf{\AtomLens_{\sigma(1)}}
      \SeqLSep\ldots\SeqLSep
      \InverseOf{\AtomLens_{\sigma(n)}} \SeqLSep 
      (\StringAlt_n,\String_n)},\InverseOf{\sigma})} =
    \SetOf{(\StringAlt_0\String_1'\ldots\String_n'\StringAlt_n,
      \String_0\StringAlt_{\InverseOf{\sigma}(1)}'\ldots\StringAlt_{\InverseOf{\sigma(n)}}'\String_n)
      \SuchThat \forall i.
      (\String_i',\StringAlt_i') \in \SemanticsOf{\InverseOf{\AtomLens_{\sigma(i)}}}} =
    \SetOf{(\StringAlt_0\StringAlt_{\sigma(1)}'\ldots\StringAlt_{\sigma(n)}'\StringAlt_n,
      \String_0\String_1'\ldots\String_n'\String_n) \SuchThat \forall i.
      (\String',\StringAlt') \in \SemanticsOf{\AtomLens_i}} =
    \SetOf{(\StringAlt,\String) \SuchThat
      (\String,\StringAlt) \in \SemanticsOf{\SequenceLens}}$, as desired.
  \end{case}

  \begin{case}[\AtomLensRule{}]
    \[
      \inferrule*
      {
        \DNFLens \OfRewritelessType \DNFRegex \Leftrightarrow \DNFRegexAlt
        \UnambigItOf{\DNFRegex}\\
        \UnambigItOf{\DNFRegexAlt}
      }
      {
        \IterateLensOf{\DNFLens} \OfRewritelessType
        \StarOf{\DNFRegex} \Leftrightarrow \StarOf{\DNFRegexAlt}
      }
    \]

    By IH, there exists
    $\InverseOf{\DNFLens} \OfRewritelessType
    \DNFRegexAlt \Leftrightarrow \DNFRegex$ where
    $\SemanticsOf{\InverseOf{\DNFLens}} =
    \SetOf{(\StringAlt,\String) \SuchThat
      (\String,\StringAlt) \in \SemanticsOf{\DNFLens}}$.

    Consider the typing
    \[
      \inferrule*
      {
        \InverseOf{\DNFLens} \OfRewritelessType \DNFRegexAlt \Leftrightarrow \DNFRegex
        \UnambigItOf{\DNFRegexAlt}\\
        \UnambigItOf{\DNFRegex}
      }
      {
        \IterateLensOf{\InverseOf{\DNFLens}} \OfRewritelessType
        \StarOf{\DNFRegexAlt} \Leftrightarrow \StarOf{\DNFRegex}
      }
    \]

    So
    $\IterateLensOf{\InverseOf{\DNFLens}} \OfRewritelessType
    \StarOf{\DNFRegexAlt}
    \Leftrightarrow
    \StarOf{\DNFRegex}$, or in other words
    $(\IterateLensOf{\InverseOf{\DNFLens}}) \OfRewritelessType
    AtomAlt
    \Leftrightarrow
    Atom$
    as desired.
    
    $\SemanticsOf{\IterateLensOf{\InverseOf{\DNFLens}}} =
    \SetOf{(\String_0\ldots\String_n,
      \StringAlt_0\ldots\StringAlt_n)
      \SuchThat \forall i.
      (\String_i,\StringAlt_i) \in \SemanticsOf{\InverseOf{\DNFLens}}} =
    \SetOf{(\StringAlt_0\ldots\StringAlt_n,\String_0\ldots\String_n)
      \SuchThat \forall i.
      (\String_i,\StringAlt_i) \in \SemanticsOf{\DNFLens}} =
    \SetOf{(\StringAlt,\String) \SuchThat
      (\String,\StringAlt) \in \SemanticsOf{\SequenceLens}}$, as desired.
  \end{case}
\end{proof}

\subsection{DNF Regular Expression and Regular Expression Proofs}
\label{dnf-regex}

This subsection is in the aims of proving that $\ToDNFRegex$ preserves language,
and $\ToRegex$ is its left inverse.  We use this throughout.  These functions
are built on the DNF regular expression operators of $\ConcatDNF$,
$\OrDNF$, and $\ToDNFRegex$.  In this section, we prove that these operators do
as we expect them to, and use these lemmas throughout the paper.

\begin{lemma}[Equivalence of \ConcatSequence{} and \Concat{}]
  If $\LanguageOf{\Regex}=\LanguageOf{\Sequence}$,
  and $\LanguageOf{\RegexAlt}=\LanguageOf{\SequenceAlt}$,
  then $\LanguageOf{\RegexConcat{\Regex}{\RegexAlt}}=\LanguageOf{\ConcatSequenceOf{\Sequence}{\SequenceAlt}}$.
\end{lemma}
\begin{proof}
  Let $\Sequence=\SequenceOf{\String_0\SeqSep\Atom_1\SeqSep\ldots
    \SeqSep\Atom_n\SeqSep\String_n}$, and
  let\\ $\SequenceAlt=[\StringAlt_0\SeqSep\AtomAlt_1\SeqSep\ldots
  \SeqSep\AtomAlt_m\SeqSep\StringAlt_m]$\\
  \begin{tabular}{@{}L@{}L@{}}
    \LanguageOf{\ConcatSequenceOf{\Sequence}{\SequenceAlt}} & = 
                                                              \LanguageOf{\SequenceOf{\String_0\SeqSep\Atom_1\SeqSep\ldots
                                                              \SeqSep\Atom_n\SeqSep\String_n\Concat\StringAlt_0\SeqSep{}
                                                              \AtomAlt_1\SeqSep\ldots\SeqSep\AtomAlt_m\SeqSep\StringAlt_m}} \\
                                                            & = 
                                                              \{\String_0\Concat\String_1'\Concat\ldots\Concat\String_n'\Concat\String_n
                                                              \Concat\StringAlt_0\Concat\StringAlt_1'\Concat\ldots
                                                              \Concat\StringAlt_m'\Concat\StringAlt_m \\
                                                            & \hspace{5em} \SuchThat{} \String_i'\in\LanguageOf{\Atom_i} \BooleanAnd{}
                                                              \StringAlt_i'\in\LanguageOf{\AtomAlt_i}\}\\
                                                            & = 
                                                              \{\String\Concat\StringAlt{} \SuchThat{} \String\in\LanguageOf{\Sequence}
                                                              \BooleanAnd{} \StringAlt\in\LanguageOf{\SequenceAlt}\}\\
                                                            & =
                                                              \{\String\Concat\StringAlt{} \SuchThat{} \String\in\LanguageOf{\Regex}
                                                              \BooleanAnd{} \StringAlt\in\LanguageOf{\RegexAlt}\}\\
                                                            & =
                                                              \LanguageOf{\RegexConcat{\Regex}{\RegexAlt}}
  \end{tabular}
\end{proof}

\begin{lemma}[Equivalence of \ConcatDNF{} and \Concat{}]
  \label{lem:cdnfeq}
  If $\LanguageOf{\Regex}=\LanguageOf{\DNFRegex}$,
  and $\LanguageOf{\RegexAlt}=\LanguageOf{\DNFRegexAlt}$,
  then $\LanguageOf{\RegexConcat{\Regex}{\RegexAlt}}=
  \LanguageOf{\ConcatDNFOf{\DNFRegex}{\DNFRegexAlt}}$.
\end{lemma}
\begin{proof}
  Let $\DNFRegex=\DNFOf{\Sequence_0\DNFSep\ldots\DNFSep\Sequence_n}$, and
  let $\DNFRegexAlt=\DNFOf{\SequenceAlt_0\DNFSep\ldots\DNFSep\SequenceAlt_m}$
  \begin{tabular}{@{}L@{}L@{}}
    \LanguageOf{\ConcatDNFOf{\DNFRegex}{\DNFRegexAlt}} & = 
                                                         \LanguageOf{\DNFOf{\ConcatSequenceOf{\Sequence_i}{\SequenceAlt_j}
                                                         \text{ for $i\in\RangeIncInc{1}{n}$, $j\in\RangeIncInc{1}{m}$}}} \\
                                                       & = 
                                                         \{\String\SuchThat \String\in\ConcatSequenceOf{\Sequence_i}{\SequenceAlt_j}\\
                                                       & \hspace{5em}
                                                         \text{ where $i\in\RangeIncInc{1}{n}$, $j\in\RangeIncInc{1}{m}$}\}\\
                                                       & = 
                                                         \{\String\Concat\StringAlt{} \SuchThat{} \String\in\LanguageOf{\Sequence_i}
                                                         \BooleanAnd{} \StringAlt\in\LanguageOf{\SequenceAlt_j}\}\\
                                                       & \hspace{5em}
                                                         \text{ where $i\in\RangeIncInc{1}{n}$, $j\in\RangeIncInc{1}{m}$}\}\\
                                                       & =
                                                         \{\String\Concat\StringAlt{} \SuchThat{} \String\in\LanguageOf{\DNFRegex}
                                                         \BooleanAnd{} \StringAlt\in\LanguageOf{\DNFRegexAlt}\}\\
                                                       & =
                                                         \{\String\Concat\StringAlt{} \SuchThat{} \String\in\LanguageOf{\Regex}
                                                         \BooleanAnd{} \StringAlt\in\LanguageOf{\RegexAlt}\}\\
                                                       & =
                                                         \LanguageOf{\RegexConcat{\Regex}{\RegexAlt}}
  \end{tabular}
\end{proof}

\begin{lemma}[Equivalence of $\Atom$ and $\AtomToDNFOf{\Atom}$]
  \label{lem:atomtodnfeq}
  $\LanguageOf{\Atom} = \LanguageOf{\AtomToDNFOf{\Atom}}$
\end{lemma}
\begin{proof}
  $\LanguageOf{\AtomToDNFOf{\Atom}} =
  \LanguageOf{\DNFOf{\SequenceOf{\EmptyString \SeqSep \Atom \SeqSep \EmptyString}}}$

  $\LanguageOf{\DNFOf{\SequenceOf{\EmptyString \SeqSep \Atom \SeqSep \EmptyString}}} =
  \SetOf{\String \SuchThat \String \in
    \LanguageOf{\SequenceOf{\EmptyString \SeqSep \Atom \SeqSep \EmptyString}}}$.

  $\LanguageOf{\SequenceOf{\EmptyString \SeqSep \Atom \SeqSep \EmptyString}} =
  \SetOf{\EmptyString\Concat\String\Concat\EmptyString \SuchThat \String \in
    \LanguageOf{\Atom}} = \SetOf{\String \SuchThat \String \in
    \LanguageOf{\Atom}} = \LanguageOf{\Atom}$.

  This means $\LanguageOf{\DNFOf{\SequenceOf{\EmptyString \SeqSep \Atom \SeqSep \EmptyString}}}
  = \SetOf{\String \SuchThat \String \in \LanguageOf{\Atom}} =
  \LanguageOf{\Atom}$.
\end{proof}

\begin{lemma}[Equivalence of \OrDNF{} and \Or{}]
  \label{lem:odnfeq}
  If $\LanguageOf{\Regex}=\LanguageOf{\DNFRegex}$,
  and $\LanguageOf{\RegexAlt}=\LanguageOf{\DNFRegexAlt}$,
  then $\LanguageOf{\RegexOr{\Regex}{\RegexAlt}}=
  \LanguageOf{\OrDNFOf{\DNFRegex}{\DNFRegexAlt}}$.
\end{lemma}
\begin{proof}
  Let $\DNFRegex=\DNFOf{\Sequence_0\DNFSep\ldots\DNFSep\Sequence_n}$, and
  let $\DNFRegexAlt=\DNFOf{\SequenceAlt_0\DNFSep\ldots\DNFSep\SequenceAlt_m}$
  
  \begin{tabular}{@{}L@{}L@{}}
    \LanguageOf{\OrDNFOf{\DNFRegex}{\DNFRegexAlt}} & = 
                                                     \LanguageOf{\DNFOf{\Sequence_0\DNFSep\ldots\DNFSep\Sequence_n\DNFSep
                                                     \SequenceAlt_1\DNFSep\ldots\DNFSep\SequenceAlt_m}}\\
                                                   & = 
                                                     \{\String\SuchThat{} \String\in\Sequence_i\vee\String\in\SequenceAlt_j\\
                                                   & \hspace{5em}
                                                     \text{ where $i\in\RangeIncInc{1}{n}$, $j\in\RangeIncInc{1}{m}$}\}\\
                                                   & = 
                                                     \{\String{} \SuchThat{} \String\in\LanguageOf{\DNFRegex}
                                                     \BooleanOr{} \String\in\LanguageOf{\DNFRegexAlt}\}\\
                                                   & =
                                                     \{\String \SuchThat{} \String\in\LanguageOf{\Regex}
                                                     \BooleanOr{} \String\in\LanguageOf{\RegexAlt}\}\\
                                                   & =
                                                     \LanguageOf{\RegexOr{\Regex}{\RegexAlt}}
  \end{tabular}\\
\end{proof}

\begin{theorem}
  For all regular expressions \Regex{},
  $\LanguageOf{\ToDNFRegexOf{\Regex}}=\LanguageOf{\Regex{}}$.
\end{theorem}
\begin{proof}
  By structural induction.

  Let $\Regex=\String$.
  $\LanguageOf{\ToDNFRegex(\String)}=\LanguageOf{\DNFOf{\SequenceOf{\String}}}=
  \{\String\}=\LanguageOf{\String}$

  Let $\Regex=\emptyset$.
  $\LanguageOf{\ToDNFRegex(\emptyset)}=\LanguageOf{\DNFOf{}} =
  \{\} = \LanguageOf{\emptyset}$.

  Let $\Regex=\StarOf{\Regex'}$.
  By induction assumption, $\LanguageOf{\ToDNFRegex(\Regex')}=
  \LanguageOf{\Regex'}$.\\
  \begin{tabular}{@{}L@{}L@{}}
    \LanguageOf{\ToDNFRegex(\StarOf{\DNFRegex'})} & =
                                                    \LanguageOf{\DNFOf{\SequenceOf{\StarOf{\ToDNFRegex(\Regex')}}}}\\
                                                  & =
                                                    \{\String\SuchThat\String\in
                                                    \LanguageOf{\SequenceOf{\StarOf{\ToDNFRegex(\Regex')}}}\}\\
                                                  & = 
                                                    \{\String\SuchThat{} \String\in\LanguageOf{\StarOf{\ToDNFRegex(\Regex')}}\}\\
                                                  & =
                                                    \{\String_1\Concat\ldots\Concat\String_n\SuchThat{}
                                                    n\in\Nats\\
                                                  & \hspace*{3em}\BooleanAnd\String_i\in\LanguageOf{\ToDNFRegex(\Regex')}\}\\
                                                  & =
                                                    \{\String_1\Concat\ldots\Concat\String_n\SuchThat{}
                                                    n\in\Nats\BooleanAnd\String_i\in\LanguageOf{\Regex'}\}\\
                                                  & = \LanguageOf{\StarOf{\Regex'}}
  \end{tabular}

  Let $\Regex=\RegexConcat{\Regex_1}{\Regex_2}$.
  By induction assumption,
  $\LanguageOf{\ToDNFRegex(\Regex_1)}=\LanguageOf{\Regex_1}$, and
  $\LanguageOf{\ToDNFRegex(\Regex_2)}=\LanguageOf{\Regex_2}$.
  $\ToDNFRegex(\RegexConcat{\Regex_1}{\Regex_2})=
  \ConcatDNFOf{\ToDNFRegex(\Regex_1)}{\ToDNFRegex(\Regex_2)}$.
  By Lemma~\ref{lem:cdnfeq},
  $\RegexConcat{\Regex_1}{\Regex_2}=
  \ConcatDNFOf{\ToDNFRegex(\Regex_1)}{\ToDNFRegex(\Regex_2)}$.

  Let $\Regex=\RegexOr{\Regex_1}{\Regex_2}$.
  By induction assumption,
  $\LanguageOf{\ToDNFRegex(\Regex_1)}=\LanguageOf{\Regex_1}$, and
  $\LanguageOf{\ToDNFRegex(\Regex_2)}=\LanguageOf{\Regex_2}$.
  $\ToDNFRegex(\RegexOr{\Regex_1}{\Regex_2})=
  \OrDNFOf{\ToDNFRegex(\Regex_1)}{\ToDNFRegex(\Regex_2)}$.
  By Lemma~\ref{lem:odnfeq},
  $\RegexOr{\Regex_1}{\Regex_2}=
  \OrDNFOf{\ToDNFRegex(\Regex_1)}{\ToDNFRegex(\Regex_2)}$.
\end{proof}
% proof-dnfrs end

% proof-dnfls start
% First we will prove some lemmas.
\begin{lemma}
  \label{lem:sequence-rx}
  Let $\SequenceOf{\String_0\SeqSep\Atom_1\SeqSep
    \ldots\Atom_n\SeqSep\String_n}$ be a sequence,
  and\\
  $\ToDNFRegex(\ToRegex(\Atom_i))=\DNFOf{\SequenceOf{\Atom_i}}$.
  Then,\\$\ToDNFRegex(\ToRegex(\SequenceOf{\String_0\SeqSep\Atom_1\SeqSep
    \ldots\Atom_n\SeqSep\String_n}))=$\\
  $\DNFOf{\SequenceOf{\String_0\SeqSep\Atom_1\SeqSep
      \ldots\Atom_n\SeqSep\String_n}}$.
\end{lemma}
\begin{proof}
  By induction on $n$.

  Let $n=0$.
  $\Sequence=\SequenceOf{\String_0}$.\\
  $\ToDNFRegex(\ToRegex(\SequenceOf{\String_0}))=
  \ToDNFRegex(\String_0)=\DNFOf{\SequenceOf{\String_0}}$

  Let $n>0$,
  $\Sequence=\SequenceOf{\String_0\SeqSep\Atom_1\SeqSep
    \ldots\Atom_n\SeqSep\String_n}$.\\
  $\ToDNFRegex(\ToRegex(\SequenceOf{\String_0\SeqSep\Atom_1\SeqSep
    \ldots\Atom_n\SeqSep\String_n}))$\\
  $\ToDNFRegex(\ToRegex(\SequenceOf{\String_0\SeqSep\Atom_1\SeqSep
    \ldots\Atom_{n-1}\SeqSep\String_{n-1}})\Concat\ToRegex(\Atom_n)
  \Concat\String_n)$=\\
  $\ToDNFRegex(\ToRegex(\SequenceOf{\String_0\SeqSep\Atom_1\SeqSep
    \ldots\Atom_{n-1}\SeqSep\String_{n-1}}))
  \ConcatDNF
  \ToDNFRegex(\ToRegex(\Atom_n))
  \ConcatDNF\\
  \ToDNFRegex(\String_{n-1})$=
  $\DNFOf{\SequenceOf{\String_0\SeqSep\Atom_1\SeqSep
      \ldots\Atom_{n-1}\SeqSep\String_{n-1}}}
  \ConcatDNF\\
  \DNFOf{\SequenceOf{\Atom_n}}
  \ConcatDNF
  \DNFOf{\SequenceOf{\String_n}}$=
  $\DNFOf{\SequenceOf{\String_0\SeqSep\Atom_1\SeqSep
      \ldots\Atom_n\SeqSep\String_n}}$.
\end{proof}

\begin{lemma}
  \label{lem:dnf-rx}
  Let $\DNFOf{\Sequence_1\DNFSep\ldots\DNFSep\Sequence_n}$ be a sequence,
  and\\ $\ToDNFRegex(\ToRegex(\Sequence_i))=\DNFOf{\Sequence_i}$.
  Then,\\ $\ToDNFRegex(\ToRegex(\DNFOf{\Sequence_1\DNFSep\ldots\DNFSep\Sequence_n}))=
  \DNFOf{\Sequence_1\DNFSep\ldots\DNFSep\Sequence_n}$.
\end{lemma}
\begin{proof}

  By induction on $n$.

  Let $n=0$
  $\ToDNFRegex(\ToRegex(\DNFOf{}))=\ToDNFRegex(\emptyset)=\DNFOf{}$.

  Let $n>0$
  $\ToDNFRegex(\ToRegex(\DNFOf{\Sequence_1\SeqSep\ldots\SeqSep\Sequence_n}))=
  \ToDNFRegex(\ToRegex(\DNFOf{\Sequence_1\SeqSep\ldots\SeqSep\Sequence_{n-1}})
  \Concat\ToRegex(\Sequence_n))$=
  $\ToDNFRegex(\ToRegex(\DNFOf{\Sequence_1\SeqSep\ldots\SeqSep\Sequence_{n-1}}))
  \ConcatDNF\\\ToDNFRegex(\ToRegex(\Sequence_n))$=
  $\DNFOf{\Sequence_1\SeqSep\ldots\SeqSep\Sequence_n}$
\end{proof}

\begin{lemma}[Elimination of $\ToDNFRegex\Compose\ToRegex$]\leavevmode
  \begin{enumerate}
  \item $\ToDNFRegex(\ToRegex(\Atom))=\DNFOf{\SequenceOf{\Atom}}$
  \item $\ToDNFRegex(\ToRegex(\Sequence))=\DNFOf{\Sequence}$
  \item $\ToDNFRegex(\ToRegex(\DNFRegex))=\DNFRegex$
  \end{enumerate}
\end{lemma}
\begin{proof}
  By mutual induction

  Let $\StarOf{\DNFRegex}$ be an atom.
  $\ToDNFRegex(\ToRegex(\StarOf{\DNFRegex}))=
  \ToDNFRegex(\StarOf{\ToRegex(\DNFRegex)})=
  \DNFOf{\SequenceOf{\StarOf{\ToDNFRegex(\ToRegex(\DNFRegex))}}}=
  \DNFOf{\SequenceOf{\StarOf{\DNFRegex}}}$

  Let $\SequenceOf{\String_0\SeqSep\Atom_1\SeqSep\ldots
    \SeqSep\Atom_n\SeqSep\String_n}$ be a sequence.
  $\ToDNFRegex(\ToRegex(\SequenceOf{\String_0\SeqSep\Atom_1\SeqSep\ldots
    \SeqSep\Atom_n\SeqSep\String_n}))$.
  By induction assumption, for each $\Atom_i$,
  $\ToDNFRegex(\ToRegex(\Atom_i))=\DNFOf{\SequenceOf{\Atom_i}}$.
  By Lemma~\ref{lem:sequence-rx},
  $\ToDNFRegex(\ToRegex(\SequenceOf{\String_0\SeqSep\Atom_1\SeqSep\ldots
    \SeqSep\Atom_n\SeqSep\String_n}))=
  \DNFOf{\SequenceOf{\String_0\SeqSep\Atom_1\SeqSep\ldots
      \SeqSep\Atom_n\SeqSep\String_n}}$.

  Let $\DNFOf{\Sequence_1\SeqSep\ldots\SeqSep\Sequence_n}$ be a DNF
  regular expression.
  By induction assumption, for each $\Sequence_i$,
  $\ToDNFRegex(\ToRegex(\Sequence_i))=\DNFOf{\Sequence_i}$.
  By Lemma~\ref{lem:dnf-rx},
  $\ToDNFRegex(\ToRegex(\DNFOf{\Sequence_1\SeqSep\ldots\SeqSep\Sequence_n}))=
  \DNFOf{\Sequence_1\SeqSep\ldots\SeqSep\Sequence_n}$.

\end{proof}

\begin{lemma}
  \label{lem:dnf-or-assoc}
  $(\DNFRegex_1 \OrDNF \DNFRegex_2) \OrDNF \DNFRegex_3 =
  \DNFRegex_1 \OrDNF (\DNFRegex_2 \OrDNF \DNFRegex_3)$
\end{lemma}
\begin{proof}
  Let $\DNFRegex_1 = \DNFOf{\Sequence_{1,1} \DNFSep \ldots \DNFSep
    \Sequence_{1,n_1}}$ and
  $\DNFRegex_2 = \DNFOf{\Sequence_{2,1} \DNFSep \ldots \DNFSep \Sequence_{2,n_2}}$ and\\
  $\DNFRegex_3 = \DNFOf{\Sequence_{3,1} \DNFSep \ldots \DNFSep \Sequence_{3,n_3}}$.
  \[
    \begin{array}{rcl}
      (\DNFRegex_1 \OrDNF \DNFRegex_2) \OrDNF \DNFRegex_3
      & = & \DNFOf{\Sequence_{1,1} \DNFSep \ldots \DNFSep \Sequence_{1,n_1} \DNFSep 
            \Sequence_{2,1} \DNFSep \ldots \DNFSep \Sequence_{2,n_2}} \OrDNF \DNFRegex_3\\
      & = & \DNFOf{\Sequence_{1,1} \DNFSep \ldots \DNFSep \Sequence_{1,n_1} \DNFSep 
            \Sequence_{2,1} \DNFSep \ldots \DNFSep \Sequence_{2,n_2} \DNFSep 
            \Sequence_{3,1} \DNFSep \ldots \DNFSep \Sequence_{3,n_3}}\\
      & = & \DNFRegex_1 \OrDNF
            \DNFOf{\Sequence_{2,1} \DNFSep \ldots \DNFSep \Sequence_{2,n_2} \DNFSep 
            \Sequence_{3,1} \DNFSep \ldots \DNFSep \Sequence_{3,n_3}}\\
      & = & \DNFRegex_1 \OrDNF (\DNFRegex_2 \OrDNF \DNFRegex_3)
    \end{array}
  \]
\end{proof}

\begin{lemma}
  \label{lem:sequence-concat-assoc}
  $(\Sequence_1 \ConcatSequence \Sequence_2) \ConcatSequence \Sequence_3 =
  \Sequence_1 \ConcatSequence (\Sequence_2 \ConcatSequence \Sequence_3)$
\end{lemma}
\begin{proof}
  Let $\Sequence_1 =
  \SequenceOf{\String_{1,0} \SeqSep \Atom_{1,1} \SeqSep \ldots \SeqSep \Atom_{1,n_1} \SeqSep \String_{1,n_1}}$,
  $\Sequence_2 =
  \SequenceOf{\String_{2,1} \SeqSep \Atom_{2,1} \SeqSep \ldots \SeqSep \Atom_{2,n_2} \SeqSep \String_{2,n_2}}$,
  and $\Sequence_3 =
  \SequenceOf{\String_{3,1} \SeqSep \Atom_{3,1} \SeqSep \ldots \SeqSep \Atom_{3,n_3} \SeqSep \String_{3,n_3}}$.
  \[
    \begin{array}{rcl}
      (\Sequence_1 \ConcatSequence \Sequence_2) \ConcatSequence \Sequence_3
      & = & \SequenceOf{\String_{1,1} \SeqSep \Atom_{1,1} \SeqSep \ldots \SeqSep \Atom_{1,n_1} \SeqSep 
            \String_{1,n_1}\Concat\String_{2,1} \SeqSep 
            \Atom_{2,1} \SeqSep \ldots \SeqSep \Atom_{2,n_2} \SeqSep \String_{2,n_2}}
            \ConcatSequence \Sequence_3\\
      & = & \SequenceOf{\String_{1,1} \SeqSep \Atom_{1,1} \SeqSep \ldots \SeqSep \Atom_{1,n_1} \SeqSep 
            \String_{1,n_1}\Concat\String_{2,0} \SeqSep 
            \Atom_{2,1} \SeqSep \ldots \SeqSep \Atom_{2,n_2} \SeqSep \\
      & & \hspace*{1em}
            \String_{2,n_2}\Concat\String_{3,0} \SeqSep 
            \Atom_{3,1} \SeqSep \ldots \SeqSep \Atom_{3,n_3} \SeqSep \String_{3,n_3}}\\
      & = & \Sequence_1 \ConcatSequence
            \SequenceOf{\String_{2,0} \SeqSep 
            \Atom_{2,1} \SeqSep \ldots \SeqSep \Atom_{2,n_2} \SeqSep 
            \String_{2,n_2}\Concat\String_{3,0} \SeqSep 
            \Atom_{3,1} \SeqSep \ldots \SeqSep \Atom_{3,n_3} \SeqSep \String_{3,n_3}}\\
      & = & \Sequence_1 \ConcatSequence (\Sequence_2 \ConcatSequence \Sequence_3)
    \end{array}
  \]
\end{proof}

\begin{lemma}
  \label{lem:dnf-concat-assoc}
  $(\DNFRegex_1 \ConcatDNF \DNFRegex_2) \ConcatDNF \DNFRegex_3 =
  \DNFRegex_1 \ConcatDNF (\DNFRegex_2 \ConcatDNF \DNFRegex_3)$
\end{lemma}
\begin{proof}
  Let $\DNFRegex_1 = \DNFOf{\Sequence_{1,1} \DNFSep \ldots \DNFSep \Sequence_{1,n_1}}$,
  $\DNFRegex_2 = \DNFOf{\Sequence_{2,1} \DNFSep \ldots \DNFSep \Sequence_{2,n_2}}$, and
  $\DNFRegex_3 = \DNFOf{\Sequence_{3,1} \DNFSep \ldots \DNFSep \Sequence_{3,n_3}}$.
  \[
    \begin{array}{rcl}
      (\DNFRegex_1 \ConcatDNF \DNFRegex_2) \ConcatDNF \DNFRegex_3
      & = & \DNFOf{\Sequence_{1,1}\ConcatSequence\Sequence_{2,1} \DNFSep 
            \ldots \DNFSep \ldots \DNFSep \Sequence_{1,n_1}\ConcatSequence\Sequence_{2,n_2}}\\
      & = &
            \DNFOf{(\Sequence_{1,1}\ConcatSequence\Sequence_{2,1})\ConcatSequence\Sequence_{3,1} \DNFSep 
            \ldots \DNFSep \ldots \DNFSep \ldots \DNFSep\\
      & & \hspace*{1em}
            (\Sequence_{1,n_1}\ConcatSequence\Sequence_{2,n_2})\ConcatSequence\Sequence_{3,n_3}}\\
      & = &
            \DNFOf{\Sequence_{1,1}\ConcatSequence(\Sequence_{2,1}\ConcatSequence\Sequence_{3,1}) \DNFSep 
            \ldots \DNFSep \ldots \DNFSep \ldots \DNFSep\\
      & & \hspace*{1em}
            \Sequence_{1,n_1}\ConcatSequence(\Sequence_{2,n_2}\ConcatSequence\Sequence_{3,n_3})}\\
      & = &
            \DNFRegex_1 \ConcatSequence
            \DNFOf{\Sequence_{2,1}\ConcatSequence\Sequence_{3,1} \DNFSep 
            \ldots \DNFSep \ldots \DNFSep \Sequence_{2,n_2}\ConcatSequence\Sequence_{3,n_3}}\\
      & = &
            \DNFRegex_1 \ConcatDNF (\DNFRegex_2 \ConcatDNF \DNFRegex_3)
    \end{array}
  \]
\end{proof}

\begin{lemma}
  \label{lem:dnf-or-identity-left}
  $\DNFOf{} \OrDNF \DNFRegex_1 = \DNFRegex_1$
\end{lemma}
\begin{proof}
  By inspection.
\end{proof}

\begin{lemma}
  \label{lem:dnf-or-identity-right}
  $\DNFRegex_1 \OrDNF \DNFOf{} = \DNFRegex_1$
\end{lemma}
\begin{proof}
  By inspection.
\end{proof}

\begin{lemma}
  \label{lem:dnf-concat-identity-left}
  $\DNFOf{\SequenceOf{\EmptyString}} \ConcatDNF \DNFRegex = \DNFRegex$
\end{lemma}
\begin{proof}
  By inspection,
  $\SequenceOf{\EmptyString} \ConcatSequence \Sequence = \Sequence$.

  Let $\DNFRegex = \DNFOf{\Sequence_1 \DNFSep \ldots \DNFSep \Sequence_n}$.

  \[
    \begin{array}{ccc}
      \DNFOf{\SequenceOf{\EmptyString}} \ConcatDNF \DNFRegex
      & = & \DNFOf{\SequenceOf{\EmptyString}\ConcatSequence\Sequence_1 \DNFSep 
            \ldots \DNFSep \SequenceOf{\EmptyString}\ConcatSequence\Sequence_n}\\
      & = & \DNFOf{\Sequence_1 \DNFSep \ldots \DNFSep \Sequence_n}\\
      & = & \DNFRegex
    \end{array}
  \]
\end{proof}

\begin{lemma}
  \label{lem:dnf-concat-identity-right}
  $\DNFRegex \ConcatDNF \DNFOf{\SequenceOf{\EmptyString}} = \DNFRegex$
\end{lemma}
\begin{proof}
  Done similarly to Lemma~\ref{lem:dnf-concat-identity-left}.
\end{proof}

\begin{lemma}
  \label{lem:dnf-concat-projection-left}
  $\DNFOf{} \ConcatDNF \DNFRegex = \DNFOf{}$
\end{lemma}
\begin{proof}
  By inspection.
\end{proof}

\begin{lemma}
  \label{lem:dnf-concat-projection-right}
  $\DNFRegex \ConcatDNF \DNFOf{} = \DNFOf{}$
\end{lemma}
\begin{proof}
  By inspection.
\end{proof}

\begin{lemma}
  \label{lem:dnf-distribute-right}
  $(\DNFRegex_1 \OrDNF \DNFRegex_2) \ConcatDNF \DNFRegex_3 =
  (\DNFRegex_1 \ConcatDNF \DNFRegex_3) \OrDNF
  (\DNFRegex_2 \ConcatDNF \DNFRegex_3)$
\end{lemma}
\begin{proof}
  Let $\DNFRegex_1 = \DNFOf{\Sequence_{1,1} \DNFSep \ldots \DNFSep
    \Sequence_{1,n_1}}$.
  
  Let $\DNFRegex_2 = \DNFOf{\Sequence_{2,1} \DNFSep \ldots \DNFSep
    \Sequence_{2,n_2}}$.
  
  Let $\DNFRegex_3 = \DNFOf{\Sequence_{3,1} \DNFSep \ldots \DNFSep
    \Sequence_{3,n_3}}$.

  $(\DNFRegex_1 \OrDNF \DNFRegex_2) \ConcatDNF \DNFRegex_3 =
  (\Sequence_{1,1} \DNFSep \ldots \DNFSep \Sequence_{1,n_1} \DNFSep 
  \Sequence_{2,1} \DNFSep \ldots \DNFSep \Sequence_{2,n_2}) \ConcatDNF
  \DNFOf{\Sequence_{3,1} \DNFSep \ldots \DNFSep \Sequence_{3,n_3}}$.
  So, through application of $\ConcatDNF$,
  $\DNFOf{\Sequence_{1,1}\ConcatSequence\Sequence_{3,1} \DNFSep \ldots \DNFSep 
    \Sequence_{1,1}\ConcatSequence\Sequence_{3,n_3} \DNFSep 
    \Sequence_{2,n_2}\ConcatSequence\Sequence_{3,1} \DNFSep \ldots \DNFSep 
    \Sequence_{2,n_2}\ConcatSequence\Sequence_{3,n_3}}$.  This equals
  $\DNFOf{\Sequence_{1,1}\ConcatSequence\Sequence_{3,1} \DNFSep \ldots \DNFSep 
    \Sequence_{1,1}\ConcatSequence\Sequence_{3,n_3} \DNFSep 
    \Sequence_{1,n_1}\ConcatSequence\Sequence_{3,1} \DNFSep \ldots \DNFSep 
    \Sequence_{1,n_1}\ConcatSequence\Sequence_{3,n_3}} \OrDNF
  \DNFOf{\Sequence_{2,1}\ConcatSequence\Sequence_{3,1} \DNFSep \ldots \DNFSep 
    \Sequence_{2,1}\ConcatSequence\Sequence_{3,n_3} \DNFSep 
    \Sequence_{2,n_2}\ConcatSequence\Sequence_{3,1} \DNFSep \ldots \DNFSep 
    \Sequence_{2,n_2}\ConcatSequence\Sequence_{3,n_3}}$, which is
  $(\DNFRegex_1 \ConcatDNF \DNFRegex_2) \OrDNF
  (\DNFRegex_1 \ConcatDNF \DNFRegex_3)$
\end{proof}

\begin{lemma}
  \label{lem:dnf-distribute-singleton-left}
  $\DNFOf{\Sequence} \ConcatDNF (\DNFRegex_1 \OrDNF \DNFRegex_2) =
  (\DNFOf{\Sequence} \ConcatDNF \DNFRegex_1) \OrDNF
  (\DNFOf{\Sequence} \ConcatDNF \DNFRegex_2)$
\end{lemma}
\begin{proof}
  Let $\DNFRegex_1 = \DNFOf{\Sequence_{1,1} \DNFSep \ldots \DNFSep \Sequence_{1,n_1}}$.
  Let $\DNFRegex_2 = \DNFOf{\Sequence_{2,1} \DNFSep \ldots \DNFSep \Sequence_{2,n_2}}$.

  $\DNFOf{\Sequence} \ConcatDNF (\DNFRegex_1 \OrDNF \DNFRegex_2) =
  \DNFOf{\Sequence} \ConcatDNF
  (\DNFOf{\Sequence_{1,1} \DNFSep \ldots \DNFSep \Sequence_{1,n_1}} \OrDNF
  \DNFOf{\Sequence_{2,1} \DNFSep \ldots \DNFSep \Sequence_{2,n_2}})$.
  So, through application of $\ConcatDNF$,
  $\DNFOf{\Sequence\ConcatSequence\Sequence_{1,1} \DNFSep \ldots \DNFSep 
    \Sequence\ConcatSequence\Sequence_{1,n_1} \DNFSep 
    \Sequence\ConcatSequence\Sequence_{2,1} \DNFSep \ldots
    \Sequence\ConcatSequence\Sequence_{2,n_2}}$.
  This equals
  $\DNFOf{\Sequence\ConcatSequence\Sequence_{1,1} \DNFSep \ldots \DNFSep 
    \Sequence\ConcatSequence\Sequence_{1,n_1}} \OrDNF
  \DNFOf{\Sequence\ConcatSequence\Sequence_{2,1} \DNFSep \ldots
    \Sequence\ConcatSequence\Sequence_{2,n_2}}$, which through the definitions,
  equals
  $(\DNFOf{\Sequence} \ConcatDNF \DNFRegex_1) \OrDNF
  (\DNFOf{\Sequence} \ConcatDNF \DNFRegex_2)$.
\end{proof}

\begin{lemma}[$\SSREquiv$ is finer than $\equiv$]
  \label{lem:defequiv-finer-equiv}
  If $\Regex \SSREquiv \RegexAlt$, then $\Regex \equiv \RegexAlt$
\end{lemma}
\begin{proof}
  By induction on the derivation of $\SSREquiv$
  
  \begin{case}[\OrIdentityRule{}]
    Through the use of $\equiv$'s \OrIdentityRule{}.
  \end{case}
  
  \begin{case}[\EmptyProjectionRightRule{}]
    Through the use of $\equiv$'s \EmptyProjectionRightRule{}.
  \end{case}
  
  \begin{case}[\EmptyProjectionLeftRule{}]
    Through the use of $\equiv$'s \EmptyProjectionLeftRule{}.
  \end{case}
  
  \begin{case}[\ConcatAssocRule{}]
    Through the use of $\equiv$'s \ConcatAssocRule{}.
  \end{case}
  
  \begin{case}[\OrAssociativityRule{}]
    Through the use of $\equiv$'s \OrAssociativityRule{}.
  \end{case}

  \begin{case}[\OrCommutativityRule{}]
    Through the use of $\equiv$'s \OrCommutativityRule{}.
  \end{case}

  \begin{case}[\DistributivityLeftRule{}]
    Through the use of $\equiv$'s \DistributivityLeftRule{}.
  \end{case}

  \begin{case}[\DistributivityRightRule{}]
    Through the use of $\equiv$'s \DistributivityRightRule{}.
  \end{case}

  \begin{case}[\ConcatIdentityLeftRule{}]
    Through the use of $\equiv$'s \ConcatIdentityLeftRule{}.
  \end{case}

  \begin{case}[\ConcatIdentityRightRule{}]
    Through the use of $\equiv$'s \ConcatIdentityRightRule{}.
  \end{case}

  \begin{case}[\UnrollstarLeftRule{}]
    Let $\Regex \SSREquiv \RegexAlt$ through an application of
    \UnrollstarLeftRule{}.

    So, without loss of generality, from symmetry, we can say
    $\Regex = \StarOf{\Regex'}$ and
    $\RegexAlt = \EmptyString \Or (\Regex' \Concat \StarOf{\Regex'})$.

    Consider the derivations
    \[
      \inferrule*
      {
        \inferrule*
        {
        }
        {
          \Regex' \equiv \Regex' \Concat \EmptyString
        }
      }
      {
        \StarOf{\Regex'} \equiv \StarOf{(\Regex' \Concat \EmptyString)}
      }
    \]

    \[
      \inferrule*
      {
      }
      {
        \StarOf{\Regex' \Concat \EmptyString}
        \equiv
        \EmptyString \Or
        (\Regex' \Concat
        \StarOf{(\EmptyString \Concat \Regex')} \Concat \EmptyString)
      }
    \]

    \[
      \inferrule*
      {
        \inferrule*
        {
        }
        {
          \Regex' \Concat
          \StarOf{(\EmptyString \Concat \Regex')}
          \Concat \EmptyString
          \equiv
          \Regex' \Concat
          \StarOf{(\EmptyString \Concat \Regex')}
        }
      }
      {
        \EmptyString \Or
        (\Regex' \Concat
        \StarOf{(\EmptyString \Concat \Regex')}
        \Concat \EmptyString)
        \equiv
        \EmptyString \Or
        (\Regex' \Concat
        \StarOf{(\EmptyString \Concat \Regex')})
      }
    \]

    \[
      \inferrule*
      {
        \inferrule*[vdots=1em]
        {
        }
        {
          \Regex' \Concat \EmptyString
          \equiv
          \Regex'
        }
      }
      {
        \EmptyString \Or
        (\Regex' \Concat
        \StarOf{(\EmptyString \Concat \Regex')})
        \equiv
        \EmptyString \Or
        (\Regex' \Concat
        \StarOf{\Regex'})
      }
    \]

    Through repeated application of equational theory transitivity,
    $\Regex \equiv \RegexAlt$.
  \end{case}

  \begin{case}[\UnrollstarRightRule{}]
    Let $\Regex \SSREquiv \RegexAlt$ through an application of
    \UnrollstarLeftRule{}.

    So, without loss of generality, from symmetry, we can say
    $\Regex = \StarOf{\Regex'}$ and
    $\RegexAlt = \EmptyString \Or (\StarOf{\Regex'} \Concat \Regex')$.

    Consider the derivations
    \[
      \inferrule*
      {
        \inferrule*
        {
        }
        {
          \Regex' \equiv \EmptyString \Concat \Regex'
        }
      }
      {
        \StarOf{\Regex'} \equiv \StarOf{(\EmptyString \Concat \Regex')}
      }
    \]

    \[
      \inferrule*
      {
      }
      {
        \StarOf{\EmptyString \Concat \Regex'}
        \equiv
        \EmptyString \Or
        (\EmptyString \Concat
        \StarOf{(\Regex' \Concat \EmptyString)} \Concat \Regex')
      }
    \]

    \[
      \inferrule*
      {
        \inferrule*
        {
        }
        {
          \EmptyString \Concat
          \StarOf{(\Regex' \Concat \EmptyString)}
          \Concat \Regex'
          \equiv
          \StarOf{(\Regex' \Concat \EmptyString)}
          \Concat \Regex'
        }
      }
      {
        \EmptyString \Or
        (\EmptyString \Concat
        \StarOf{(\Regex' \Concat \EmptyString)}
        \Concat \Regex')
        \equiv
        \EmptyString \Or
        (\Regex'
        \StarOf{(\Regex' \Concat \EmptyString)})
      }
    \]

    \[
      \inferrule*
      {
        \inferrule*[vdots=1em]
        {
        }
        {
          \Regex' \Concat \EmptyString
          \equiv
          \Regex'
        }
      }
      {
        \EmptyString \Or
        (\StarOf{(\Regex' \Concat \EmptyString)} \Concat \Regex')
        \equiv
        \EmptyString \Or
        (\Regex' \Concat
        \StarOf{\Regex'})
      }
    \]

    Through repeated application of equational theory transitivity,
    $\Regex \equiv \RegexAlt$.
  \end{case}
\end{proof}

\subsection{Unambiguity Property Proofs}
\label{language-rewriting-unambiguity}

Unambiguity is critical in the typing derivations, so unambiguity proofs are
similarly critical.  In this section, we prove requirements for maintaining
unambiguity.  The bulk of the work for many of these is proven in the language
unambiguity proofs in Subsection~\ref{language-proofs}.  However, this combines
these together to prove things like unambiguity is maintained through
application of the
definitional equivalence rules.

\begin{lemma}
  \label{lem:strong_unambig_or}
  If $\Regex \Or \RegexAlt$ be strongly unambiguous, then
  $\LanguageOf{\Regex} \Intersect \LanguageOf{\RegexAlt} = \SetOf{}$,
  and both $\Regex$ and $\RegexAlt$ are strongly unambiguous.
\end{lemma}
\begin{proof}
  If $\Regex \Or \RegexAlt$ is strongly unambiguous, then either
  $\LanguageOf{\Regex \Or \RegexAlt} = \SetOf{}$, or
  $\LanguageOf{\Regex} \Intersect \LanguageOf{\RegexAlt} = \SetOf{}$, and
  $\Regex$ and $\RegexAlt$ are both strongly unambiguous.

  If the latter, then we are done.

  If the former, then both $\LanguageOf{\Regex} = \SetOf{}$ and
  $\LanguageOf{\RegexAlt} = \SetOf{}$.
  This means they are both strongly unambiguous.  Furthermore,
  $\SetOf{} \Intersect \SetOf{} = \SetOf{}$, so
  $\LanguageOf{\Regex} \Intersect \LanguageOf{\RegexAlt} = \SetOf{}$.
\end{proof}

\begin{lemma}
  Let $\Regex \SSREquiv \RegexAlt$.
  If $\Regex$ is strongly unambiguous, then $\RegexAlt$ is strongly unambiguous.
\end{lemma}
\begin{proof}
  If $\LanguageOf{\Regex} = \SetOf{}$, then $\LanguageOf{\RegexAlt} = \SetOf{}$,
  by Lemma~\ref{lem:defequiv-finer-equiv}.

  For the case where $\LanguageOf{\Regex} \neq \SetOf{}$, we proceed by
  induction on the derivation of equivalence of $\Regex$ and $\RegexAlt$.
  \begin{case}[\OrIdentityRule{} left to right]
    Let the last step of the derivation be \OrIdentityRule{} left to right.
    $\Regex \SSREquiv \Regex \Or \emptyset$.

    $\emptyset$ is strongly unambiguous, as its language is empty.
    $\Regex$ is strongly unambiguous by assumption
    $\LanguageOf{\Regex} \Intersect \LanguageOf{\emptyset} =
    \LanguageOf{\Regex} \Intersect \SetOf{} = \emptyset$, so $\RegexAlt$ is
    strongly unambiguous.
  \end{case}

  \begin{case}[\OrIdentityRule{} right to left]
    Let the last step of the derivation be \OrIdentityRule{} right to left.
    $\RegexAlt \Or \emptyset \SSREquiv \RegexAlt$.

    If $\LanguageOf{\RegexAlt \Or \emptyset} = \SetOf{}$ then
    $\LanguageOf{\RegexAlt} = \SetOf{}$, so $\RegexAlt$ is strongly unambiguous.

    Otherwise $\RegexAlt$ is strongly unambiguous, which is what is desired.
  \end{case}

  \begin{case}[\EmptyProjectionRightRule both directions]
    Let the last step of the derivation be \EmptyProjectionRightRule{}.
    The language of both sides is $\SetOf{}$, by
    Lemma~\ref{lem:defequiv-finer-equiv}.
  \end{case}

  \begin{case}[\EmptyProjectionLeftRule both directions]
    Let the last step of the derivation be \EmptyProjectionLeftRule{}.
    The language of both sides is $\SetOf{}$, by
    Lemma~\ref{lem:defequiv-finer-equiv}.
  \end{case}
  
  \begin{case}[\ConcatAssocRule{} left to right]
    Let the last step of the derivation be \ConcatAssocRule{} left to right.
    $\RegexConcat{(\RegexConcat{\Regex_1}{\Regex_2})}{\Regex_3}
    \SSREquiv
    \RegexConcat{\Regex_1}{(\RegexConcat{\Regex_2}{\Regex_3})}$

    Because \Regex is strongly unambiguous.
    $\Regex_1 \UnambigConcat \Regex_2$ and
    $(\Regex_1 \Concat \Regex_2) \UnambigConcat \Regex_3$

    Let $\String_2,\StringAlt_2 \in \Regex_2$, let $\String_3,\StringAlt_3
    \in \Regex_3$, and let
    $\String_2 \Concat \String_3 = \StringAlt_2 \Concat \StringAlt_3$.
    Consider $\String_1$ in $\Regex_1$ which exists as
    $\LanguageOf{\Regex} \neq \SetOf{}$.
    $(\String_1 \Concat \String_2) \Concat \String_3 =
    (\String_1 \Concat \StringAlt_2) \Concat \StringAlt_3$, so
    $\String_3 = \StringAlt_3$ and
    $\String_1 \Concat \String_2 = \String_1 \Concat \StringAlt_2$,
    so $\String_2 = \StringAlt_2$.

    Let $\String_2 \Concat \String_3 \in \Regex_2 \Concat \Regex_3$,
    $\StringAlt_2 \Concat \StringAlt_3 \in \Regex_2 \Concat \Regex_3$,
    $\String_1,\StringAlt_1 \in \Regex_1$,
    and let $\String_1 \Concat (\String_2 \Concat \String_3) = 
    \StringAlt_1 \Concat (\StringAlt_2 \Concat \StringAlt_3)$.
    This means $(\String_1 \Concat \String_2) \Concat \String_3 =
    (\StringAlt_1 \Concat \StringAlt_2) \Concat \StringAlt_3)$.
    
    So by assumption, $\String_3 = \StringAlt_3$, and
    $\String_1 \Concat \String_2 = \StringAlt_1 \Concat \StringAlt_2$.
    So by assumption, $\String_1 = \StringAlt_1$ and $\String_2 = \StringAlt_2$.
    So, $\String_2 \Concat \String_3 = \StringAlt_2 \Concat \StringAlt_3$,
    and $\String_1 = \StringAlt_1$.
  \end{case}

  \begin{case}[\ConcatAssocRule{} right to left]
    Very similarly to left to right.
  \end{case}

  \begin{case}[\OrAssociativityRule{} left to right]
    $\Regex_1 \Or (\Regex_2 \Or \Regex_3) \SSREquiv
    (\Regex_1 \Or \Regex_2) \Or \Regex_3$.

    $\LanguageOf{\Regex_1} \Intersect \LanguageOf{\Regex_2 \Or \Regex_3} =
    \SetOf{}$.
    This means that $\LanguageOf{\Regex_1} \Intersect
    (\LanguageOf{\Regex_2} \Union \LanguageOf{\Regex_3}) = \SetOf{}$, so through
    distributivity, $\LanguageOf{\Regex_1} \Intersect \LanguageOf{\Regex_2}
    \Union \LanguageOf{\Regex_1} \Intersect \LanguageOf{\Regex_3} = \SetOf{}$.
    This means $\LanguageOf{\Regex_1} \Intersect \LanguageOf{\Regex_2} =
    \SetOf{}$ and
    $\LanguageOf{\Regex_1} \Intersect \LanguageOf{\Regex_3} = \SetOf{}$.

    If $\LanguageOf{\Regex_2 \Or \Regex_3} = \SetOf{}$, then the language of
    each is empty, so they are each strongly unambiguous.
    This means $\Regex_1 \Or \Regex_2$ is strongly unambiguous.

    Furthermore, $\LanguageOf{\Regex_1} \Intersect \LanguageOf{\Regex_3} \Union
    \LanguageOf{\Regex_2} \Intersect \LanguageOf{\Regex_3} = \SetOf{}$ as each
    of the intersections is empty.  So the whole thing is unambiguous.
  \end{case}

  \begin{case}[\OrAssociativityRule{} right to left]
    Done very similarly to the left to right case.
  \end{case}

  \begin{case}[\OrCommutativityRule{}]
    $\Regex_1 \Or \Regex_2 \SSREquiv \Regex_2 \Or \Regex_1$
    So if the languages are empty, then they are both empty.

    Otherwise, $\Regex_1$ is strongly unambiguous, and $\Regex_2$ is strongly
    unambiguous, and $\LanguageOf{\Regex_1} \Intersect \LanguageOf{\Regex_2} =
    \SetOf{}$.

    So $\LanguageOf{\Regex_2} \Intersect \LanguageOf{\Regex_1} = \SetOf{}$, and
    so $\Regex_2 \Or \Regex_1$ is strongly unambiguous.
  \end{case}

  \begin{case}[\DistributivityLeftRule{} left to right]
    $\Regex_1 \Concat (\Regex_2 \Or \Regex_3) \SSREquiv
    (\Regex_1 \Concat \Regex_2) \Or (\Regex_1 \Concat \Regex_3)$.

    If $\LanguageOf{\Regex_1 \Concat (\Regex_2 \Or \Regex_3)} = \SetOf{}$, then
    $(\Regex_1 \Concat \Regex_2) \Or (\Regex_1 \Concat \Regex_3) = \SetOf{}$,
    and we are done.

    If the language is nonempty, so too are the languages of each side, so
    $\Regex_1$ is nonempty, and $\Regex_2 \Or \Regex_3$ is nonempty,
    and $\Regex_1$ is strongly unambiguous, and $\Regex_2 \Or \Regex_3$ is
    strongly unambiguous.

    $\Regex_2 \Or \Regex_3$ being strongly unambiguous implies $\Regex_2$ is
    strongly unambiguous, $\Regex_3$ is strongly unambiguous, and
    $\LanguageOf{\Regex_2} \Intersect \LanguageOf{\Regex_3} = \SetOf{}$, by
    Lemma~\ref{lem:strong_unambig_or}.

    Let $\String_1,\StringAlt_1 \in \LanguageOf{\Regex_1}$,
    $\String_2,\StringAlt_2 \in \LanguageOf{\Regex_2}$,
    $\String_1\Concat\String_2 = \StringAlt_1\Concat\StringAlt_2$.
    $\String$.  Then $\StringAlt_1 \in \LanguageOf{\Regex_1 \Or \Regex_2}$, and
    $\StringAlt_2 \in \LanguageOf{\Regex_1 \Or \Regex_2}$.
    By assumption of strong unambiguity, where the languages are not empty,
    $\String_1  = \String_2$ anc $\StringAlt_1 = \StringAlt_2$.

    Similarly for $\String_1,\StringAlt_1 \in \LanguageOf{\Regex_1}$,
    $\String_3,\StringAlt_3 \in \LanguageOf{\Regex_3}$.

    Assume there exists some $\String \in \LanguageOf{\Regex_1 \Concat \Regex_2}
    \Intersect
    \LanguageOf{\Regex_1 \Concat \Regex_3}$.
    This means $\String = \String_1 \Concat \String_2$, for
    $\String_1\in\LanguageOf{\Regex_1}$ and $\String_2\in\LanguageOf{\Regex_2}$,
    uniquely.
    It means $\String = \StringAlt_1 \Concat \StringAlt_2$, for
    $\StringAlt_1\in\LanguageOf{\Regex_1}$ and
    $\StringAlt_2\in\LanguageOf{\Regex_2}$.
    From assumption, as
    $\String \in \LanguageOf{\Regex_1 \Concat (\Regex_2 \Or \Regex_3)}$,
    $\String_1 = \StringAlt_1$ and $\String_2 = \StringAlt_3$.
    Contradiction, as
    $\LanguageOf{\Regex_2} \Intersect \LanguageOf{\Regex_3} = \SetOf{}$.
    So there is no string in the intersection, or in other words
    $\LanguageOf{\Regex_1 \Concat \Regex_2}
    \Intersect
    \LanguageOf{\Regex_1 \Concat \Regex_3} = \SetOf{}$

    As such, $(\Regex_1 \Concat \Regex_2) \Or (\Regex_1 \Concat \Regex_3)$ is
    strongly unambiguous
  \end{case}

  \begin{case}[\DistributivityLeftRule{} right to left]
    $(\Regex_1 \Concat \Regex_2) \Or (\Regex_1 \Concat \Regex_3)
    \SSREquiv \Regex_1 \Concat (\Regex_2 \Or \Regex_3)$.

    If $\LanguageOf{\Regex_1} = \SetOf{}$, then the language of the entire
    $\Regex$ is empty, and we are done.  Otherwise assume
    $\LanguageOf{\Regex_1} \neq \SetOf{}$.

    From assumption $\Regex_1 \Concat \Regex_2$ is strongly unambiguous,
    $\Regex_1 \Concat \Regex_3$ is strongly unambiguous,
    and $\LanguageOf{\Regex_1 \Concat \Regex_2} \Intersect
    \LanguageOf{\Regex_1 \Concat \Regex_3} = \SetOf{}$.
    
    Assume there exists some
    $\String \in \LanguageOf{\Regex_2} \Intersect \LanguageOf{\Regex_3}$.
    Let $\String_1 \in \LanguageOf{\Regex_1}$.
    This makes
    $\String_1\Concat\String \in \LanguageOf{\Regex_1 \Concat \Regex_2}
    \Intersect \LanguageOf{\Regex_1 \Concat \Regex_3}$.  This is a
    contradiction, so
    $\LanguageOf{\Regex_2} \Intersect \LanguageOf{\Regex_3} = \SetOf{}$.

    Let $\String_1,\StringAlt_1 \in \LanguageOf{\Regex_1}$.
    Let $\String,\StringAlt \in \LanguageOf{\Regex_2 \Or \Regex_3}$.
    Let $\String_1 \Concat \String = \StringAlt_1 \Concat \StringAlt$.
    Assume $\String \in \LanguageOf{\Regex_2}$.  Then $\StringAlt \in
    \LanguageOf{\Regex_2}$, as otherwise $\Regex$ is not strongly unambiguous.
    So as $\String_1 \Concat \String \in \LanguageOf{\Regex_1 \Concat
      \Regex_2}$,
    and $\StringAlt_1 \Concat \StringAlt \in \LanguageOf{\Regex_1 \Concat
      \Regex_2}$, by assumption, $\String_1 = \StringAlt_1$, and $\String =
    \StringAlt$.
    If $\String \not\in \LanguageOf{\Regex_2}$, then $\String \in
    \LanguageOf{\Regex_3}$, and the same argument applies.
  \end{case}

  \begin{case}[\DistributivityRightRule{} both directions]
    Proceeds the same as \DistributivityLeftRule{}.
  \end{case}

  \begin{case}[\ConcatIdentityLeftRule{} left to right]
    $\EmptyString \Concat \Regex' \SSREquiv \Regex'$

    If they have empty languages, we are done.

    If nonempty, then $\Regex'$ is strongly unambiguous, and we are done.
  \end{case}

  \begin{case}[\ConcatIdentityLeftRule{} right to left]
    $\Regex' \SSREquiv \EmptyString \Concat \Regex'$

    Both $\Regex'$ and $\EmptyString$ are strongly unambiguous, by assumption
    and definition, respectively.

    Furthermore, let $\String_1,\StringAlt_1 \in \LanguageOf{\EmptyString}$, and
    $\String_2,\StringAlt_2 \in \LanguageOf{\Regex'}$, and $\String_1 \Concat
    \String_2 = \StringAlt_1 \Concat \StringAlt_2$
    $\String_1 = \StringAlt_1 = \EmptyString$, so $\String_1 = \StringAlt_2$,
    which makes $\String_2 = \StringAlt_2$.
  \end{case}

  \begin{case}[\ConcatIdentityRightRule{} both directions]
    Very similar to \ConcatIdentityLeftRule{}.
  \end{case}

  \begin{case}[\UnrollstarLeftRule{} left to right]
    $\StarOf{\Regex'} \SSREquiv \EmptyString \Or (\Regex' \Concat
    \StarOf{\Regex'})$

    Let $\String \in \LanguageOf{\EmptyString} \Intersect \LanguageOf{\Regex'
      \Concat \StarOf{\Regex'}}$.
    So $\String = \EmptyString$.
    So $\EmptyString \in \LanguageOf{\Regex'}$.
    Contradiction, as if $\EmptyString$ in $\LanguageOf{\Regex'}$, then
    if $\String_1\Concat\String_n = \StringAlt_1\Concat\StringAlt_m$, $n$ no
    longer must equal $m$, as arbitrarily many $\EmptyString$s can be input.

    $\EmptyString$ is strongly unambiguous.

    If $\LanguageOf{\Regex'} = \emptyset$, then
    $\Regex' \Concat \StarOf{\Regex'}$ also has an empty language, and is
    strongly unambiguous.

    If the language is nonempty, $\Regex'$ is strongly unambiguous.

    Let $\String_1,\String_2 \in
    \LanguageOf{\Regex'}$, $\StringAlt_1,\StringAlt_2 \in
    \LanguageOf{\StarOf{\Regex'}}$.  Let $\String_1\Concat\StringAlt_1 =
    \StringAlt_2\Concat\StringAlt_2$.
    $\StringAlt_1 = \StringAlt_{1,1}\Concat\ldots\Concat\StringAlt_{1,n}$ and
    $\StringAlt_2 = \StringAlt_{2,1}\Concat\ldots\Concat\StringAlt_{1,m}$, where
    $\StringAlt_{1,i},\StringAlt_{2,i}\in\LanguageOf{\StarOf{\Regex'}}$.
    Consider
    $\String_1\Concat\StringAlt_{1,1}\Concat\ldots\Concat\StringAlt_{1,n}$
    and
    $\String_2\Concat\StringAlt_{2,1}\Concat\ldots\Concat\StringAlt_{2,m}$.
    As $\Regex'$ is unambiguously iterable, $n+1=m+1$, and
    $\String_1 = \String_2$ and $\StringAlt_{1,i} = \StringAlt_{2,i}$.
    This means that $\StringAlt_1 = \StringAlt_2$.
    So $Regex'$ is unambiguously concatenable with $\StarOf{\Regex'}$.
  \end{case}

  \begin{case}[\UnrollstarLeftRule{} right to left]
    $\EmptyString \Or (\Regex' \Concat \StarOf{\Regex'}) \SSREquiv
    \StarOf{\Regex'}$

    If $\LanguageOf{\Regex'} = \SetOf{}$, then it is vacuously unambiguously
    concatenable, and $\Regex'$ is strongly unambiguous, so $\StarOf{\Regex'}$
    is strongly unambiguous.

    Let $\LanguageOf{\Regex'}$ not be empty.

    Let $\String_1\Concat\ldots\Concat\String_n =
    \StringAlt_1\Concat\ldots\Concat\StringAlt_m$, and
    $\String_i,\StringAlt_i\in\LanguageOf{\Regex'}$.  We want to show that $n=m$
    and $\String_i=\StringAlt_i$.
    This can be done by induction on $n$.

    If $n=0$, then $m=0$, as otherwise $m>0$, which would imply that
    $\EmptyString\in\LanguageOf{\Regex'}$, making $\Regex$ not strongly
    unambiguous.

    If $n\neq 0$, then by the unambiguous concatenability of $\Regex'$ and
    $\StarOf{\Regex'}$, $\String_1 = \StringAlt_1$, and
    $\String_2\Concat\ldots\Concat\String_n =
    \StringAlt_2\Concat\ldots\Concat\StringAlt_n$, and the IH applies.
  \end{case}

  \begin{case}[\UnrollstarRightRule{} both directions]
    Done similarly to \UnrollstarLeftRule{}.
  \end{case}

  \begin{case}[All structural cases]
    As $\SSREquiv$ is finer than $\equiv$, the subparts will have
    the same languages.  If the language of $\Regex$ is empty, then we are done,
    otherwise, each subpart will be strongly unambiguous, by the induction
    hypothesis.  As the top
    level unambiguity condition is based on the language, and the languages of
    the subparts are equal, the top level unambiguity condition will be
    satisfied.
  \end{case}

  \begin{case}[Transitivity of Equational Theory]
    If $\Regex \SSREquiv \Regex'$ and $\Regex \SSREquiv
    \RegexAlt$, then by IH,
    $\Regex'$ is strongly unambiguous, and by IH again, $\RegexAlt$ is strongly
    unambiguous.
  \end{case}
\end{proof}

\begin{lemma}
  \label{lem:distribute-strongly-unambiguous-iff-factor}
  If $\DNFRegex_1 \ConcatDNF (\DNFRegex_2 \OrDNF \DNFRegex_3)$ is strongly
  unambiguous, then 
  $(\DNFRegex_1 \ConcatDNF \DNFRegex_2) \OrDNF
  (\DNFRegex_1 \ConcatDNF \DNFRegex_3)$ is strongly unambiguous.
\end{lemma}
\begin{proof}
  Let $\DNFRegex_1 = \DNFOf{\Sequence_{1,1} \DNFSep \ldots \DNFSep
    \Sequence_{1,n_1}}$.
  
  Let $\DNFRegex_2 = \DNFOf{\Sequence_{2,1} \DNFSep \ldots \DNFSep
    \Sequence_{2,n_2}}$.
  
  Let $\DNFRegex_3 = \DNFOf{\Sequence_{3,1} \DNFSep \ldots \DNFSep
    \Sequence_{3,n_3}}$.
  
  $\DNFRegex_2 \OrDNF \DNFRegex_3 =
  \DNFOf{\Sequence_{2,1} \DNFSep \ldots \DNFSep \Sequence_{2,n_2} \DNFSep 
    \Sequence_{3,1} \DNFSep \ldots \DNFSep \Sequence_{3,n_3}}$.
  
  $\DNFRegex_1 \ConcatDNF (\DNFRegex_2 \OrDNF \DNFRegex_3) =
  \DNFOf{
    \Sequence_{1,1} \ConcatSequence \Sequence_{2,1} \DNFSep  \ldots \DNFSep 
    \Sequence_{1,1}\ConcatSequence\Sequence_{2,n_2} \DNFSep 
    \Sequence_{1,1}\ConcatSequence\Sequence_{3,1} \DNFSep  \ldots \DNFSep 
    \Sequence_{1,1}\ConcatSequence\Sequence_{3,n_3} \DNFSep  \ldots \DNFSep 
    \Sequence_{1,n_1} \ConcatSequence \Sequence_{2,1} \DNFSep  \ldots \DNFSep 
    \Sequence_{1,n_1}\ConcatSequence\Sequence_{2,n_2} \DNFSep 
    \Sequence_{1,n_1}\ConcatSequence\Sequence_{3,1} \DNFSep  \ldots \DNFSep 
    \Sequence_{1,n_1}\ConcatSequence\Sequence_{3,n_3}}$.
  As this is strongly unambiguous, $\Sequence_{1,i} \ConcatSequence
  \Sequence_{j,k}$ is strongly unambiguous for all $i,j,k$.
  Furthermore, by strong unambiguity,
  if $(i_1,j_1,k_1) \neq (i_2,j_2,k_2)$, then
  $\Sequence_{1,i_1} \ConcatSequence \Sequence_{j_1,k_1} \Intersect
  \Sequence_{1,i_2} \ConcatSequence \Sequence_{j_1,k_1}$.

  $(\DNFRegex_1 \ConcatDNF \DNFRegex_2) \OrDNF
  (\DNFRegex_1 \ConcatDNF \DNFRegex_3) =\\
  \DNFOf{
    \Sequence_{1,1}\ConcatSequence\Sequence_{2,1} \DNFSep \ldots
    \Sequence_{1,1}\ConcatSequence\Sequence_{2,n_2} \DNFSep \ldots \DNFSep 
    \Sequence_{1,n_1}\ConcatSequence\Sequence_{2,1} \DNFSep \ldots \DNFSep 
    \Sequence_{1,n_1}\ConcatSequence\Sequence_{2,n_2} \DNFSep 
    \Sequence_{1,1}\ConcatSequence\Sequence_{3,1} \DNFSep \ldots
    \Sequence_{1,1}\ConcatSequence\Sequence_{3,n_3} \DNFSep 
    \Sequence_{1,n_1}\ConcatSequence\Sequence_{3,1} \DNFSep \ldots
    \Sequence_{1,n_1}\ConcatSequence\Sequence_{3,n_3}}$
  
  From before, if $(i_1,j_1,k_1) \neq (i_2,j_2,k_2)$, then
  $\Sequence_{1,i_1} \ConcatSequence \Sequence_{j_1,k_1} \Intersect
  \Sequence_{1,i_2} \ConcatSequence \Sequence_{j_1,k_1} = \SetOf{}$.
  Furthermore, each $\Sequence_{1,i} \ConcatSequence \Sequence_{j,k}$ is still
  strongly unambiguous for all $i,j,k$, so
  $(\DNFRegex_1 \ConcatDNF \DNFRegex_2) \OrDNF
  (\DNFRegex_1 \ConcatDNF \DNFRegex_3)$ is strongly unambiguous.

  The same process can be repeated to show that assumping
  $(\DNFRegex_1 \ConcatDNF \DNFRegex_2) \OrDNF
  (\DNFRegex_1 \ConcatDNF \DNFRegex_3)$ is strongly unambiguous, we can show
  $\DNFRegex \ConcatDNF (\DNFRegex_2 \OrDNF \DNFRegex_3)$ is strongly unambiguous.
\end{proof}

\subsection{Rewrite Equivalence Proofs}
\label{rewrite-proofs}

This subsection goes through equivalence associated with the rewrites.  In this
section, proofs of rewrites not altering the language are proven.  Furthermore,
it is shown that the Definitional Equivalence Rules are finer than the base
axioms.  We prove that both $\ToDNFRegex$ and $\ToRegex$ maintain unambiguity.
Parallel Rewrites with Swap are shown to be equivalent to the definitional
equivalence rules.
We prove that the reflexive and transitive closure of rewrites is equivalent in
expressibility to the reflexive and transitive closure of parallel rewrites.
Lastly, we prove that if the DNF versions of two regular expressions are can be
written to each other, then those two regular expressions are definitionally
equivalent.

\begin{lemma}[Single Rewrites Respecting Language]
  \label{lem:single-rrl}
  \leavevmode
  \begin{itemize}
  \item If $\Atom\RewriteAtom\DNFRegex$, then $\LanguageOf{\Atom} = \LanguageOf{\DNFRegex}$
  \item If $\DNFRegex\Rewrite\DNFRegexAlt$, then $\LanguageOf{\DNFRegex}=\LanguageOf{\DNFRegexAlt}$
  \end{itemize}
\end{lemma}
\begin{proof}
  By mutual induction on the derivation of $\Rewrite$ and $\RewriteAtom$
  \begin{case}[\AtomUnrollstarLeftRule{}]
    \[
      \inferrule*
      {
      }
      {
        \StarOf{\DNFRegex}\RewriteAtom
        \OrDNFOf{\DNFOf{\SequenceOf{\EmptyString}}}{(\ConcatDNFOf{\DNFRegex}{\AtomToDNFOf{\StarOf{\DNFRegex}}})}
      }
    \]

    Let $\ToRegexOf{\DNFRegex} = \Regex$.

    $\StarOf{\Regex} \equiv \EmptyString \Or (\Regex \Concat
    \StarOf{\Regex})$, by Lemma~\ref{lem:defequiv-finer-equiv}.
    By Theorem~\ref{thm:dnfrs},
    $\LanguageOf{\ToDNFRegexOf{\StarOf{\Regex}}} =
    \LanguageOf{\ToDNFRegexOf{(\EmptyString \Or (\Regex \Concat
        \StarOf{\Regex}))}}$.
    So $\LanguageOf{\AtomToDNFOf{\StarOf{(\ToDNFRegexOf{\Regex})}}} =
    \LanguageOf{\ToDNFRegexOf{(\EmptyString \Or (\Regex \Concat
        \StarOf{\Regex}))}}$.
    So by Lemma~\ref{lem:atomtodnfeq}, and application of $\ToDNFRegex$,
    $\LanguageOf{\StarOf{\DNFRegex}} =
    \LanguageOf{\DNFOf{\SequenceOf{\EmptyString}} \OrDNF
      \DNFRegex \ConcatDNF \AtomToDNFOf{(\StarOf{\DNFRegex})}}$, as desired.
  \end{case}
  
  \begin{case}[\AtomUnrollstarRightRule{}]
    \[
      \inferrule*
      {
      }
      {
        \StarOf{\DNFRegex}\RewriteAtom
        \OrDNFOf{\DNFOf{\SequenceOf{\EmptyString}}}{(\ConcatDNFOf{\AtomToDNFOf{\StarOf{\DNFRegex}}}{\DNFRegex})}
      }
    \]

    Let $\ToRegexOf{\DNFRegex} = \Regex$.

    $\StarOf{\Regex} \equiv \EmptyString \Or (\Regex \Concat
    \StarOf{\Regex})$, by Lemma~\ref{lem:defequiv-finer-equiv}.
    By Theorem~\ref{thm:dnfrs},
    $\LanguageOf{\ToDNFRegexOf{\StarOf{\Regex}}} =
    \LanguageOf{\ToDNFRegexOf{(\EmptyString \Or (\StarOf{\Regex} \Concat
        \Regex))}}$.
    So $\LanguageOf{\AtomToDNFOf{\StarOf{(\ToDNFRegexOf{\Regex})}}} =
    \LanguageOf{\ToDNFRegexOf{(\EmptyString \Or (\StarOf{\Regex} \Concat
        \Regex))}}$.
    So by Lemma~\ref{lem:atomtodnfeq}, and application of $\ToDNFRegex$,
    $\LanguageOf{\StarOf{\DNFRegex}} =
    \LanguageOf{\DNFOf{\SequenceOf{\EmptyString}} \OrDNF
      \AtomToDNFOf{(\StarOf{\DNFRegex})} \ConcatDNF \DNFRegex}$, as desired.
  \end{case}

  \begin{case}[\AtomStructuralRewriteRule{}]
    \[
      \inferrule*
      {
        \DNFRegex \Rewrite \DNFRegexAlt
      }
      {
        \StarOf{\DNFRegex} \RewriteAtom \AtomToDNFOf{\StarOf{\DNFRegexAlt}}
      }
    \]

    $\LanguageOf{\DNFRegex} = \LanguageOf{\DNFRegexAlt}$,
    so $\LanguageOf{\StarOf{\DNFRegex}} = \LanguageOf{\StarOf{\DNFRegexAlt}}$.
    Through application of Lemma~\ref{lem:atomtodnfeq},
    $\LanguageOf{\StarOf{\DNFRegex}} =
    \LanguageOf{\AtomToDNFOf{\StarOf{\DNFRegexAlt}}}$.
  \end{case}

  \begin{case}[\DNFStructuralRewriteRule]
    \[
      \inferrule*
      {
        \Atom_j \RewriteAtom \DNFRegex
      }
      {
        \DNFOf{\Sequence_1\DNFSep\ldots\DNFSep\Sequence_{i-1}} \OrDNF
        \DNFOf{\SequenceOf{\String_0\SeqSep\Atom_1\SeqSep\ldots\SeqSep\String_{j-1}}}
        \ConcatDNF \AtomToDNFOf{\Atom_j} \ConcatDNF
        \DNFOf{\SequenceOf{\String_j\SeqSep\ldots\SeqSep\Atom_m\SeqSep\String_m}}
        \OrDNF \DNFOf{\Sequence_{i+1}\DNFSep\ldots\DNFSep\Sequence_n}\Rewrite\\
        \DNFOf{\Sequence_1\DNFSep\ldots\DNFSep\Sequence_{i-1}} \OrDNF
        \DNFOf{\SequenceOf{\String_0\SeqSep\Atom_1\SeqSep\ldots\SeqSep\String_{j-1}}}\ConcatDNF\DNFRegex\ConcatDNF\SequenceOf{\String_j\SeqSep\ldots\SeqSep\Atom_m\SeqSep\String_m} \OrDNF
        \DNFOf{\Sequence_{i+1}\DNFSep\ldots\DNFSep\Sequence_n}
      }
    \]

    As $\LanguageOf{\AtomToDNFOf{\Atom_j}} = \LanguageOf{\DNFRegex}$, by IH and
    Lemma~\ref{lem:atomtodnfeq}, and because the left side is the same as the
    right, except with $\AtomToDNFOf{\Atom_j}$ replacing $\DNFRegex$, the two
  languages are the same.

  \end{case}
\end{proof}

\begin{lemma}[Rewrites Respecting Language]
  \label{lem:rrl}
  If $\DNFRegex \StarOf{\Rewrite} \DNFRegexAlt$, then
  $\LanguageOf{\DNFRegex} = \LanguageOf{\DNFRegexAlt}$
\end{lemma}
\begin{proof}
  By induction on the derivation of $\StarOf{\Rewrite}$

  \begin{case}[\ReflexivityRule{}]
    \[
      \inferrule*
      {
      }
      {
        \DNFRegex \StarOf{\Rewrite} \DNFRegex
      }
    \]

    $\LanguageOf{\DNFRegex} = \LanguageOf{\DNFRegex}$ so we're done.
  \end{case}

  \begin{case}[\BaseRule{}]
    \[
      \inferrule*
      {
        \DNFRegex \Rewrite \DNFRegexAlt
      }
      {
        \DNFRegex \StarOf{\Rewrite} \DNFRegexAlt
      }
    \]
    
    By Lemma~\ref{lem:single-rrl}, as $\DNFRegex \Rewrite \DNFRegexAlt$,
    $\LanguageOf{\DNFRegex} = \LanguageOf{\DNFRegexAlt}$.
  \end{case}

  \begin{case}
    \[
      \inferrule*
      {
        \DNFRegex \StarOf{\Rewrite} \DNFRegex'\\
        \DNFRegex' \StarOf{\Rewrite} \DNFRegexAlt
      }
      {
        \DNFRegex \StarOf{\Rewrite} \DNFRegexAlt
      }
    \]

    By IH, $\LanguageOf{\DNFRegex} = \LanguageOf{\DNFRegex'}$.
    By IH, $\LanguageOf{\DNFRegex'} = \LanguageOf{\DNFRegexAlt}$.
    So $\LanguageOf{\DNFRegex} = \LanguageOf{\DNFRegexAlt}$.
  \end{case}
\end{proof}

\begin{lemma}
  \label{lem:pre-uniqueness-of-empty-in-dnf}
  If $\ToDNFRegexOf{\Regex} = \DNFOf{}$, and $\Regex \equiv \RegexAlt$, then
  $\ToDNFRegexOf{\RegexAlt} = \DNFOf{}$.
\end{lemma}
\begin{proof}
  By induction on the proof of equivalence

  \begin{case}[Structural Equality Rule]
    Then $\RegexAlt = \Regex$, so
    $\ToDNFRegexOf{\RegexAlt} = \ToDNFRegexOf{\Regex} = \DNFOf{}$.
  \end{case}

  \begin{case}[\OrIdentityRule{} left to right]
    $\Regex \equiv \RegexOr{\Regex}{\emptyset}$.
    $\ToDNFRegexOf{(\Regex \Or \emptyset)} =
    \ToDNFRegexOf{\Regex} \OrDNF \ToDNFRegexOf{\emptyset} =
    \DNFOf{} \OrDNF \DNFOf{}$.
  \end{case}

  \begin{case}[\OrIdentityRule{} right to left]
    $\RegexAlt \equiv \RegexOr{\RegexAlt}{\emptyset}$.
    $\ToDNFRegexOf{(\RegexOr{\RegexAlt}{\emptyset})} = \DNFOf{}$. So by
    definition,
    $\ToDNFRegexOf{\RegexAlt} \OrDNF \ToDNFRegexOf{\emptyset} = \DNFOf{}$.
    Again by definition,
    $\ToDNFRegexOf{\RegexAlt} \OrDNF \DNFOf{} = \DNFOf{}$.
    So by Lemma~\ref{lem:dnf-or-identity-right},
    $\ToDNFRegexOf{\RegexAlt} = \DNFOf{}$
  \end{case}

  \begin{case}[\EmptyProjectionRightRule{} left to right]
    $\RegexAlt = \emptyset$ so $\ToDNFRegexOf{\RegexAlt} = \DNFOf{}$
  \end{case}

  \begin{case}[\EmptyProjectionRightRule{} right to left]
    $\RegexAlt = \Regex \Concat \emptyset$, so $\ToDNFRegexOf{\RegexAlt} =
    \ToDNFRegexOf{\Regex} \ConcatDNF \ToDNFRegexOf{\emptyset} =
    \ToDNFRegexOf{\Regex} \ConcatDNF \DNFOf{}$, so by
    Lemma~\ref{lem:dnf-concat-projection-left},
    $\ToDNFRegexOf{\RegexAlt} = \DNFOf{}$.
  \end{case}

  \begin{case}[\EmptyProjectionLeftRule{} both directions]
    Done similarly to \EmptyProjectionRightRule{}.
  \end{case}

  \begin{case}[\ConcatAssocRule{} left to right]
    $(\Regex_1 \Concat \Regex_2) \Concat \Regex_3 \equiv
    \Regex_1 \Concat (\Regex_2 \Concat \Regex_3)$.
    Throguh definitions, and Lemma~\ref{lem:sequence-concat-assoc},
    $\DNFOf{} = \ToDNFRegexOf{((\Regex_1 \Concat \Regex_2) \Concat \Regex_3)} =
    (\ToDNFRegexOf{\Regex_1} \ConcatDNF \ToDNFRegexOf{\Regex_2}) \ConcatDNF
    \ToDNFRegexOf{\Regex_3}) =
    \ToDNFRegexOf{\Regex_1} \ConcatDNF (\ToDNFRegexOf{\Regex_2} \ConcatDNF
    \ToDNFRegexOf{\Regex_3}) =
    \ToDNFRegexOf{(\Regex_1 \Concat (\Regex_2 \Concat \Regex_3))}$
  \end{case}

  \begin{case}[\ConcatAssocRule{} right to left]
    Analogously to left to right
  \end{case}

  \begin{case}[\OrAssociativityRule{} left to right]
    $(\Regex_1 \Or \Regex_2) \Or \Regex_3 \equiv
    \Regex_1 \Or (\Regex_2 \Or \Regex_3)$.
    Through definitions, and Lemma~\ref{lem:dnf-or-assoc},
    $\DNFOf{} = \ToDNFRegexOf{((\Regex_1 \Or \Regex_2) \Or \Regex_3)} =
    (\ToDNFRegexOf{\Regex_1} \OrDNF \ToDNFRegexOf{\Regex_2}) \OrDNF
    \ToDNFRegexOf{\Regex_3}) =
    \ToDNFRegexOf{\Regex_1} \OrDNF (\ToDNFRegexOf{\Regex_2} \OrDNF
    \ToDNFRegexOf{\Regex_3}) =
    \ToDNFRegexOf{(\Regex_1 \Or (\Regex_2 \Or \Regex_3))}$
  \end{case}

  \begin{case}[\OrAssociativityRule{} right to left]
    Analogously to left to right
  \end{case}

  \begin{case}[\OrCommutativityRule{}]
    $\Regex_1 \Or \Regex_2 \equiv \Regex_1 \Or \Regex_2$.
    $\DNFOf{} = \ToDNFRegexOf{(\Regex_1 \Or \Regex_2)} =
    \ToDNFRegexOf{\Regex_1} \OrDNF \ToDNFRegexOf{\Regex_2}$.
    By the definition of $\OrDNF$, $\ToDNFRegexOf{\Regex_1} = \DNFOf{}$, and
    $\ToDNFRegexOf{\Regex_2} = \DNFOf{}$.
    $\ToDNFRegexOf{(\Regex_2 \Or \Regex_1)} =
    \ToDNFRegexOf{\Regex_2} \OrDNF \ToDNFRegexOf{\Regex_1} =
    \DNFOf{} \OrDNF \DNFOf{} = \DNFOf{}$.
  \end{case}

  \begin{case}[\DistributivityLeftRule{} left to right]
    $\RegexConcat{\Regex_1}{(\RegexOr{\Regex_2}{\Regex_3})} \equiv
    \RegexOr{(\RegexConcat{\Regex_1}{\Regex_2})}{(\RegexConcat{\Regex_1}{\Regex_3})}$

    $\ToDNFRegexOf{\Regex_1 \Concat (\Regex_2 \Or \Regex_3)} =
    \ToDNFRegexOf{\Regex_1} \ConcatDNF
    (\ToDNFRegexOf{\Regex_2} \OrDNF \ToDNFRegexOf{\Regex_3}) = \DNFOf{}$.
    By the definition of $\ConcatDNF$, this means $\ToDNFRegexOf{\Regex_1} =
    \DNFOf{}$, or $\ToDNFRegexOf{\Regex_2} \OrDNF \ToDNFRegexOf{\Regex_3} =
    \DNFOf{}$.

    If $\ToDNFRegexOf{\Regex_1} = \DNFOf{}$, then by
    Lemma~\ref{lem:dnf-concat-projection-left},
    $\ToDNFRegexOf{\Regex_1} \ConcatDNF \ToDNFRegexOf{\Regex_2} = \DNFOf{}$ and
    $\ToDNFRegexOf{\Regex_1} \ConcatDNF \ToDNFRegexOf{\Regex_3} = \DNFOf{}$,
    so $\ToDNFRegexOf{((\Regex_1 \Concat \Regex_2) \Or (\Regex_1 \Concat
      \Regex_3)} = 
    (\ToDNFRegexOf{\Regex_1} \ConcatDNF \ToDNFRegexOf{\Regex_2}) \OrDNF
    \ToDNFRegexOf{\Regex_1} \ConcatDNF \ToDNFRegexOf{\Regex_3} = \DNFOf{}$.

    If $\ToDNFRegexOf{\Regex_2} \OrDNF \ToDNFRegexOf{\Regex_3} = \DNFOf{}$,
    then by definition of $\OrDNF$, $\ToDNFRegexOf{\Regex_2} = \DNFOf{}$ and
    $\ToDNFRegexOf{\Regex_3} = \DNFOf{}$.
    By Lemma~\ref{lem:dnf-concat-projection-right},
    $\ToDNFRegexOf{\Regex_1} \ConcatDNF \ToDNFRegexOf{\Regex_2} = \DNFOf{}$ and
    $\ToDNFRegexOf{\Regex_1} \ConcatDNF \ToDNFRegexOf{\Regex_3} = \DNFOf{}$,
    so $\ToDNFRegexOf{((\Regex_1 \Concat \Regex_2) \Or (\Regex_1 \Concat
      \Regex_3)} = 
    (\ToDNFRegexOf{\Regex_1} \ConcatDNF \ToDNFRegexOf{\Regex_2}) \OrDNF
    \ToDNFRegexOf{\Regex_1} \ConcatDNF \ToDNFRegexOf{\Regex_3} = \DNFOf{}$.
  \end{case}

  \begin{case}[\DistributivityLeftRule{} right to left]
    $\RegexOr{(\RegexConcat{\Regex_1}{\Regex_2})}{(\RegexConcat{\Regex_1}{\Regex_3})} \equiv
    \RegexConcat{\Regex_1}{(\RegexOr{\Regex_2}{\Regex_3})}$

    $\ToDNFRegexOf{((\Regex_1 \Concat \Regex_2) \Or (\Regex_1 \Concat \Regex_3))} =
    (\ToDNFRegexOf{\Regex_1} \ConcatDNF \ToDNFRegexOf{\Regex_2}) \OrDNF
    (\ToDNFRegexOf{\Regex_1} \ConcatDNF \ToDNFRegexOf{\Regex_3}) = \DNFOf{}$.
    By the definition of $\OrDNF$, this means
    $\ToDNFRegexOf{\Regex_1} \ConcatDNF \ToDNFRegexOf{\Regex_2} = \DNFOf{}$,
    and
    $\ToDNFRegexOf{\Regex_1} \ConcatDNF \ToDNFRegexOf{\Regex_3} = \DNFOf{}$.

    As $\ToDNFRegexOf{\Regex_1} \ConcatDNF \ToDNFRegexOf{\Regex_2} = \DNFOf{}$.

    If $\ToDNFRegexOf{\Regex_1} = \DNFOf{}$, then by
    Lemma~\ref{lem:dnf-concat-projection-left},
    $\ToDNFRegexOf{\Regex_1} \ConcatDNF (\ToDNFRegexOf{\Regex_2} \OrDNF
    \ToDNFRegexOf{\Regex_3}) = \DNFOf{}$, so
    $\RegexConcat{\Regex_1}{(\RegexOr{\Regex_2}{\Regex_3})} = \DNFOf{}$.

    If $\ToDNFRegexOf{\Regex_1} \neq \DNFOf{}$, then
    $\ToDNFRegexOf{\Regex_2} = \DNFOf{}$ and $\ToDNFRegexOf{\Regex_3} =
    \DNFOf{}$.
    This means $\ToDNFRegexOf{\Regex_2} \OrDNF \ToDNFRegexOf{\Regex_3} =
    \DNFOf{}$.
    So, by Lemma~\ref{lem:dnf-concat-projection-left},
    $\ToDNFRegexOf{\Regex_1} \ConcatDNF (\ToDNFRegexOf{\Regex_2} \OrDNF
    \ToDNFRegexOf{\Regex_3}) = \DNFOf{}$, so
    $\ToDNFRegexOf{(\RegexConcat{\Regex_1}{(\RegexOr{\Regex_2}{\Regex_3})})}$.
  \end{case}

  \begin{case}[\DistributivityRightRule{} both directions]
    Proceeds analogously to \DistributivityLeftRule{}.
  \end{case}

  \begin{case}[\ConcatIdentityLeftRule{} left to right]
    $\RegexConcat{\EmptyString{}}{\RegexAlt} \equiv \RegexAlt$.
    By assumption, $\ToDNFRegexOf{(\RegexConcat{\EmptyString{}}{\RegexAlt})} = \DNFOf{}$
    This means $\ToDNFRegexOf{\EmptyString} \ConcatDNF \ToDNFRegexOf{\RegexAlt}
    = \DNFOf{}$.
    By Lemma~\ref{lem:dnf-concat-identity-left}, $\ToDNFRegexOf{\EmptyString}
    \ConcatDNF \ToDNFRegexOf{\RegexAlt} = \ToDNFRegexOf{\RegexAlt}$, so
    $\ToDNFRegexOf{\RegexAlt} = \DNFOf{}$.
  \end{case}

  \begin{case}[\ConcatIdentityLeftRule{} right to left]
    $\Regex \equiv \RegexConcat{\EmptyString{}}{\Regex}$.
    By assumption, $\ToDNFRegexOf{\Regex} = \DNFOf{}$.
    By Lemma~\ref{lem:dnf-concat-identity-left},
    $\ToDNFRegexOf{(\RegexConcat{\EmptyString{}}{\Regex})} =
    \ToDNFRegexOf{\Regex}$, so $\ToDNFRegexOf{\Regex} = \DNFOf{}$.
  \end{case}

  \begin{case}[\ConcatIdentityRightRule{} both dierections]
    Done analogously to \ConcatIdentityLeftRule{}.
  \end{case}

  \begin{case}[\SumstarRule{}, \ProductstarRule{}, \StarstarRule{},
    \DicyclicityRule{}, Structural \StarRegexType{} Equality]
    In all of these cases, the regular expression on the left is of the form
    $\StarOf{\Regex'}$, for some $\Regex'$.  $\EmptyString \in \StarOf{\Regex'}$
    for all $\Regex'$.  However, $\LanguageOf{\DNFOf{}} = \SetOf{}$, and by
    Theorem~\ref{thm:dnfrs}, $\LanguageOf{\ToDNFRegexOf{\Regex}} =
    \LanguageOf{\Regex}$.  This means that $\ToDNFRegexOf{\StarOf{\Regex'}} \neq
    \DNFOf{}$, for all $\Regex'$, so these rules do not apply.
  \end{case}

  \begin{case}[Structural \OrRegexType{} Equality]
    \[
      \inferrule*
      {
        \Regex_1 \equiv \RegexAlt_1\\
        \Regex_2 \equiv \RegexAlt_2
      }
      {
        \Regex_1 \Or \Regex_2 \equiv \RegexAlt_1 \Or \RegexAlt_2
      }
    \]

    $\ToDNFRegexOf{(\Regex_1 \Or \Regex_2)} =
    \ToDNFRegexOf{\Regex_1} \OrDNF \ToDNFRegexOf{\Regex_2} = \DNFOf{}$.
    By the definition of $\OrDNF$, $\ToDNFRegexOf{\Regex_1} = \DNFOf{}$ and
    $\ToDNFRegexOf{\Regex_2} = \DNFOf{}$.
    So, by induction, $\ToDNFRegexOf{\RegexAlt_1} = \DNFOf{}$ and
    $\ToDNFRegexOf{\RegexAlt_2} = \DNFOf{}$.
    So $\ToDNFRegexOf{\RegexAlt_1} \OrDNF \ToDNFRegexOf{\RegexAlt_2} =
    \ToDNFRegexOf{(\RegexAlt_1 \Or \RegexAlt_2)} = \DNFOf{}$.
  \end{case}

  \begin{case}[Structural \ConcatRegexType{} Equality]
    \[
      \inferrule*
      {
        \Regex_1 \equiv \RegexAlt_1\\
        \Regex_2 \equiv \RegexAlt_2
      }
      {
        \Regex_1 \Concat \Regex_2 \equiv \RegexAlt_1 \Concat \RegexAlt_2
      }
    \]

    $\ToDNFRegexOf{(\Regex_1 \Concat \Regex_2)} =
    \ToDNFRegexOf{\Regex_1} \ConcatDNF \ToDNFRegexOf{\Regex_2} = \DNFOf{}$.
    By the definition of $\ConcatDNF$, $\ToDNFRegexOf{\Regex_1} = \DNFOf{}$ or
    $\ToDNFRegexOf{\Regex_2} = \DNFOf{}$.
    So, by induction, $\ToDNFRegexOf{\RegexAlt_1} = \DNFOf{}$ or
    $\ToDNFRegexOf{\RegexAlt_2} = \DNFOf{}$.
    So $\ToDNFRegexOf{\RegexAlt_1} \ConcatDNF \ToDNFRegexOf{\RegexAlt_2} =
    \ToDNFRegexOf{(\RegexAlt_1 \Concat \RegexAlt_2)} = \DNFOf{}$.
  \end{case}

  \begin{case}[Transitivity of Equational Theories]
    \[
      \inferrule*
      {
        \Regex \equiv \Regex'\\
        \Regex' \equiv \RegexAlt
      }
      {
        \Regex \equiv \RegexAlt
      }
    \]
  \end{case}

  By IH, $\ToDNFRegexOf{\Regex'} = \DNFOf{}$.
  So, by IH, $\ToDNFRegexOf{\RegexAlt} = \DNFOf{}$.
\end{proof}

\begin{lemma}
  \label{lem:uniqueness-of-empty-in-dnf}
  If $\LanguageOf{\Regex} = \SetOf{}$, then $\ToDNFRegexOf{\Regex} = \DNFOf{}$
\end{lemma}
\begin{proof}
  $\LanguageOf{\emptyset} = \SetOf{}$.
  We know $\LanguageOf{\Regex} = \SetOf{}$, so $\Regex \equiv \emptyset$.
  $\ToDNFRegexOf{\emptyset} = \DNFOf{}$.
  So, by Lemma~\ref{lem:pre-uniqueness-of-empty-in-dnf},
  $\ToDNFRegexOf{\Regex} = \DNFOf{}$.
\end{proof}

\begin{lemma}
  \label{lem:retaining-unambiguity-todnf}
  If $\Regex$ is strongly unambiguous as a regular expression, then
  $\ToDNFRegexOf{\Regex}$ is strongly unambiguous as a DNF regular expression.
\end{lemma}
\begin{proof}
  We proceed by induction.
  \begin{case}[\BaseRegexType{}]
    $\ToDNFRegexOf{\String} = \DNFOf{\SequenceOf{\String}}$, which is strongly
    unambiguous.
  \end{case}

  \begin{case}[\EmptyRegexType{}]
    $\ToDNFRegexOf{\emptyset} = \DNFOf{}$, which is strongly unambiguous.
  \end{case}

  \begin{case}[\StarRegexType{}]
    Let $\Regex = \ToDNFRegexOf{(\StarOf{\Regex'})}$ be strongly unambiguous.
    $\ToDNFRegexOf{(\StarOf{\Regex'})} =
    \AtomToDNFOf{\StarOf{(\ToDNFRegexOf{\Regex'})}}$
    By IH, $\ToDNFRegexOf{\Regex'}$ is strongly unambiguous.
    Furthermore, $\LanguageOf{\Regex'} = \LanguageOf{\ToDNFRegexOf{\Regex'}}$ is
    unambiguously iterable, so $\StarOf{(\ToDNFRegexOf{\Regex'})}$ is strongly
    unambiguous.  This means that
    $\AtomToDNFOf{\StarOf{(\ToDNFRegexOf{\Regex'})}}$ is strongly unambiguous.
  \end{case}

  \begin{case}[\ConcatRegexType{}]
    Let $\Regex = \Regex_1 \Concat \Regex_2$ be strongly unambiguous.

    If $\LanguageOf{\Regex} = \SetOf{}$, by
    Lemma~\ref{lem:uniqueness-of-empty-in-dnf}, $\ToDNFRegexOf{\Regex} =
    \DNFOf{}$, which is strongly unambiguous.

    Let $\ToDNFRegexOf{\Regex_1} = \DNFOf{\Sequence_1 \DNFSep \ldots \DNFSep \Sequence_n}$.
    Let $\ToDNFRegexOf{\Regex_2} = \DNFOf{\SequenceAlt_1 \DNFSep \ldots \DNFSep \SequenceAlt_m}$.
    If $\LanguageOf{\Regex} \neq \SetOf{}$, this means that
    $\LanguageOf{\ToDNFRegexOf{\Regex_1}} \neq \SetOf{}$, and
    $\LanguageOf{\ToDNFRegexOf{\Regex_2}} \neq \SetOf{}$.
    This means that $\Sequence_i$ is nonempty, and so is $\SequenceAlt_i$, for
    all $i$.
    Furthermore, as $\Regex$ is strongly unambiguous, and $\LanguageOf{\Regex}
    \neq \SetOf{}$, $\Regex_1$ and $\Regex_2$ are strongly unambiguous,
    which means so too are $\DNFOf{\Sequence_1 \DNFSep \ldots \DNFSep \Sequence_n}$ and
    $\DNFOf{\SequenceAlt_1 \DNFSep \ldots \DNFSep \SequenceAlt_m}$, and so too are
    $\Sequence_i$ and $\SequenceAlt_i$.

    As $\UnambigConcatOf{\ToDNFRegexOf{\Regex_1}}{\ToDNFRegexOf{\Regex_2}}$,
    $i \neq j \BooleanImplies \Sequence_i \Intersect \Sequence_j = \emptyset$, and
    $i \neq j \BooleanImplies \SequenceAlt_i \Intersect \SequenceAlt_j = \emptyset$
    I know from Lemma~\ref{lem:unambig-concat-union-equiv},
    $(i_1,j_1) \neq (i_2,j_2) \BooleanImplies \LanguageOf{\Sequence_{i_1} \ConcatDNF
    \Sequence_{j_1}} \Intersect
    \LanguageOf{\Sequence_{i_2} \ConcatDNF
    \Sequence_{j_2}} = \SetOf{}$.
    and $\LanguageOf{\Sequence_i} \UnambigConcat \LanguageOf{\SequenceAlt_j}$.

    Let $\Sequence_i =
    \SequenceOf{\String_{i,0} \SeqSep \Atom_{i,1} \SeqSep \ldots \SeqSep \Atom_{i,n_i} \SeqSep \String_{i,n_i}}$ and
    $\SequenceAlt_i =
    \SequenceOf{\StringAlt_{i,0} \SeqSep \AtomAlt_{i,1} \SeqSep \ldots \SeqSep \AtomAlt_{i,n_i} \SeqSep \StringAlt_{i,n_i}}$.
    $\SequenceUnambigConcatOf{\String_{i,0},\Atom_{i,1},\ldots,\Atom_{i,n_i},\String_{i,n_i}}$
    $\SequenceUnambigConcatOf{\StringAlt_{i,0},\AtomAlt_{i,1},\ldots,\AtomAlt_{i,n_i},\StringAlt_{i,n_i}}$
    Furthermore, $\Sequence_i$ and $\SequenceAlt_i$ have nonempty languages.
    By Lemma~\ref{lem:unambig-concat-equiv},
    $\SequenceUnambigConcatOf{\String_{i,0},\Atom_{i,1},\ldots,\Atom_{i,n_i},
      \String_{i,n_i}\Concat\StringAlt_{j,0},
      \AtomAlt_{j,1},\ldots,\AtomAlt_{j,n_j},\StringAlt_{j,n_j}}$
    
    As $\Sequence_i$ and $\SequenceAlt_i$ are strongly unambiguous, we know
    $\Atom_{i,j}$ and $\AtomAlt_{i,j}$ are strongly unambiguous.
    So, as $\SequenceUnambigConcatOf{\String_{i,0},\Atom_{i,1},\ldots,\Atom_{i,n_i},
      \String_{i,n_i}\Concat\StringAlt_{j,0},
      \AtomAlt_{j,1},\ldots,\AtomAlt_{j,n_j},\StringAlt_{j,n_j}}$,
    $\Sequence_i \ConcatDNF \SequenceAlt_j$ is strongly unambiguous.

    Furthermore, as 
    $\Sequence_i \ConcatDNF \SequenceAlt_j$ is strongly unambiguous and
    $(i_1,j_1) \neq (i_2,j_2) \BooleanImplies \LanguageOf{\Sequence_{i_1} \ConcatDNF
    \Sequence_{j_1}} \Intersect
    \LanguageOf{\Sequence_{i_2} \ConcatDNF
      \Sequence_{j_2}} = \SetOf{}$, then
    $\DNFOf{\Sequence_1 \ConcatDNF \SequenceAlt_1  \DNFSep \ldots \DNFSep 
    \Sequence_n \ConcatDNF \SequenceAlt_m}$.
  \end{case}

  \begin{case}[\OrRegexType{}]
    Let $\Regex = \Regex_1 \Or \Regex_2$ be strongly unambiguous.

    If $\LanguageOf{\Regex} = \SetOf{}$, by
    Lemma~\ref{lem:uniqueness-of-empty-in-dnf}, $\ToDNFRegexOf{\Regex} =
    \DNFOf{}$, which is strongly unambiguous.

    Otherwise, $\Regex_1$ and $\Regex_2$ are strongly unambiguous,
    and $\LanguageOf{\Regex_1} \Intersect \LanguageOf{\Regex_2} = \SetOf{}$.
    This means $\ToDNFRegexOf{\Regex_1}$ and $\ToDNFRegexOf{\Regex_2}$ are also
    strongly unambiguous, by IH.

    Let $\ToDNFRegexOf{\Regex_1} = \DNFOf{\Sequence_1 \DNFSep \ldots \DNFSep \Sequence_n}$.
    Let $\ToDNFRegexOf{\Regex_2} =
    \DNFOf{\SequenceAlt_1 \DNFSep \ldots \DNFSep \SequenceAlt_n}$.
    Let $\Sequence_i' = \begin{cases*}
      \Sequence_i & if $i \leq n$\\
      \SequenceAlt_{i-n} & otherwise
    \end{cases*}$.
    As $\ToDNFRegexOf{\Regex_1}$ and $\ToDNFRegexOf{\Regex_2}$ are strongly
    unambiguous, $i \neq j \BooleanImplies \LanguageOf{\Sequence_i} \Intersect
    \LanguageOf{\Sequence_j} = \SetOf{}$ and
    $i \neq j \BooleanImplies \LanguageOf{\SequenceAlt_i} \Intersect
    \LanguageOf{\SequenceAlt_j} = \SetOf{}$.
    Furthermore, as $\BigUnion_{i\in\RangeIncInc{1}{n}}\LanguageOf{\Sequence_i}
    \Intersect \BigUnion_{j\in\RangeIncInc{1}{m}}\LanguageOf{\Sequence_j}$,
    from Lemma~\ref{lem:unambig-union-equiv}, $i \neq j \BooleanImplies
    \LanguageOf{\Sequence_i'} \Intersect \LanguageOf{\Sequence_j'} = \SetOf{}$,
    and as each $\Sequence_i$ and $\SequenceAlt_i$ is strongly unambiguous,
    $\ToDNFRegexOf{\Regex_1} \OrDNF \ToDNFRegexOf{\Regex_2}$ is strongly
    unambiguous.
  \end{case}
\end{proof}

\begin{lemma}
  \label{lem:unambig-dnf}
  If $\DNFRegex = \DNFOf{\Sequence_1 \DNFSep \ldots \DNFSep \Sequence_n}$ is strongly
  unambiguous, and for all $i$, $\ToRegexOf{\Sequence_i}$ is strongly
  unambiguous, then $\ToRegexOf{\DNFRegex}$ is strongly unambiguous.
\end{lemma}
\begin{proof}
  By induction on $n$

  \begin{case}[$n=0$]
    $\ToRegexOf{\DNFOf{}} = \emptyset$, which is strongly unambiguous.
  \end{case}

  \begin{case}[$n>0$]
    $\ToRegexOf{\Sequence_1 \DNFSep \ldots \DNFSep \Sequence_n} =
    \ToRegexOf{\Sequence_1 \DNFSep \ldots \DNFSep \Sequence_{n-1}} \Or \ToRegexOf{\Sequence_n}$.
    By IH, $\ToRegexOf{\Sequence_1 \DNFSep \ldots \DNFSep \Sequence_{n-1}}$ is strongly
    unambiguous.
    Furthermore, as $\DNFRegex$ is strongly unambiguous,
    by Lemma~\ref{lem:unambig-union-equiv},
    $\LanguageOf{\ToRegexOf{\Sequence_1 \DNFSep \ldots \DNFSep \Sequence_{n-1}}} \Intersect
    \ToRegexOf{\Sequence_n} = \emptyset$, so
    $\ToRegexOf{\Sequence_n}$ is strongly unambiguous, so the entire thing is
    strongly unambiguous.
  \end{case}
\end{proof}

\begin{lemma}
  \label{lem:unambig-seq}
  If $\Sequence = \SequenceOf{\String_0 \SeqSep \Atom_1 \SeqSep \ldots \SeqSep \Atom_n \SeqSep \String_n}$
  is strongly unambiguous, and for all $i$, $\ToRegexOf{\Atom_i}$ is strongly
  unambiguous, then $\ToRegexOf{\Sequence}$ is strongly unambiguous.
\end{lemma}
\begin{proof}
  By induction on $n$

  \begin{case}[$n=0$]
    $\ToRegexOf{\SequenceOf{\String_0}} = \String_0$, which is strongly unambiguous.
  \end{case}

  \begin{case}[$n>0$]
    $\ToRegexOf{\String_0 \SeqSep \Atom_1 \SeqSep \ldots \SeqSep \Atom_n \SeqSep \String_n} =
    \ToRegexOf{\String_0 \SeqSep \Atom_1 \SeqSep \ldots \SeqSep \Atom_{n-1} \SeqSep \String_{n-1}}
    \Concat \ToRegexOf{\Atom_n} \Concat \ToRegexOf{\String_n}$.
    From $\ToRegexOf{\Atom_n}$ and $\ToRegexOf{\String_n}$,
    we know $\UnambigConcatOf{\ToRegexOf{\Atom_n}}{\String_n}$ because
    the second part will always be $\String_n$, so the first part must be the
    same.
    By IH, $\ToRegexOf{\Sequence_1 \DNFSep \ldots \DNFSep \Sequence_{n-1}}$ is strongly
    unambiguous.
    Furthermore, as $\Sequence$ is strongly unambiguous,
    by Lemma~\ref{lem:unambig-concat-equiv},
    $\LanguageOf{\ToRegexOf{\String_0 \SeqSep \Atom_1 \SeqSep \ldots \SeqSep \Atom_{n-1} \SeqSep \String_{n-1}}}
    \UnambigConcat
    (\ToRegexOf{\Atom_n} \Concat \ToRegexOf{\String_n})$, so
    as each side is also is strongly unambiguous, the entire thing is
    strongly unambiguous.
  \end{case}
\end{proof}

\begin{lemma}\leavevmode
  \label{lem:retaining-unambiguity-toregex}
  \begin{itemize}
  \item 
    If $\DNFRegex$ is strongly unambiguous as a DNF regular expression, then
    $\ToRegexOf{\DNFRegex}$ is strongly unambiguous as a regular expression
  \item
    If $\Sequence$ is strongly unambiguous as a sequence, then
    $\ToRegexOf{\Sequence}$ is strongly unambiguous as a sequence
  \item
    If $\Atom$ is strongly unambiguous as an atom, then
    $\ToRegexOf{\Atom}$ is strongly unambiguous as an atom
  \end{itemize}
\end{lemma}
\begin{proof}
  \begin{case}[\MultiOrDNFRegexType{}]
    Let $\DNFRegex = \DNFOf{\Sequence_1 \DNFSep \ldots \DNFSep \Sequence_n}$.
    By IH, $\ToRegexOf{\Sequence_i}$ is strongly unambiguous.
    By Lemma~\ref{lem:unambig-dnf}, $\ToRegexOf{\DNFRegex}$ is strongly
    unambiguous.
  \end{case}

  \begin{case}[\MultiConcatSequenceType{}]
    Let $\Sequence = \SequenceOf{\String_0 \SeqSep \Atom_1 \SeqSep \ldots \SeqSep \Atom_n \SeqSep \String_n}$.
    By IH, $\ToRegexOf{\Atom_i}$ is strongly unambiguous.
    By Lemma~\ref{lem:unambig-seq}, $\ToRegexOf{\Sequence}$ is strongly
    unambiguous.
  \end{case}

  \begin{case}[StarAtomType]
    Let $\Atom = \StarOf{\DNFRegex}$.
    By IH, $\ToRegexOf{\DNFRegex}$ is strongly unambiguous.
    As $\StarOf{\DNFRegex}$ is strongly unambiguous, $\UnambigItOf{\DNFRegex}$,
    so $\UnambigItOf{\LanguageOf{\DNFRegex}}$, so
    $\UnambigItOf{(\ToRegexOf{\DNFRegex})}$.
    So $\StarOf{\DNFRegex}$ is strongly unambiguous.
  \end{case}
\end{proof}

\begin{definition}[Parallel Rewriting Without Reordering]\leavevmode
  \begin{mathpar}
    \inferrule[\AtomUnrollstarLeftRule{}]
    {
    }
    {
      \StarOf{\DNFRegex}\ParallelRewriteAtom
      \OrDNFOf{\DNFOf{\SequenceOf{\EmptyString}}}
      {(\ConcatDNFOf{\DNFRegex}{\DNFOf{\SequenceOf{\StarOf{\DNFRegex}}}})}
    }
    
    \inferrule[\AtomUnrollstarRightRule{}]
    {
    }
    {
      \StarOf{\DNFRegex}\ParallelRewriteAtom
      \OrDNFOf{\DNFOf{\SequenceOf{\EmptyString}}}
      {(\ConcatDNFOf{\DNFOf{\SequenceOf{\StarOf{\DNFRegex}}}}{\DNFRegex})}
    }

    \inferrule[\ParallelAtomStructuralRewriteRule{}]
    {
      \DNFRegex \ParallelRewrite \DNFRegex'
    }
    {
      \StarOf{\DNFRegex} \ParallelRewriteAtom \DNFOf{\SequenceOf{\StarOf{\DNFRegex'}}}
    }

    \inferrule[\ParallelDNFStructuralRewriteRule{}]
    {
      \DNFRegex = \DNFOf{\Sequence_1 \DNFSep \ldots \DNFSep \Sequence_n}\\
      \forall i. \Sequence_i =
      \SequenceOf{\String_{i,0} \SeqSep \Atom_{i,1} \SeqSep \ldots
        \SeqSep \Atom_{i,n_i} \SeqSep \String_{i,n_i}}\\
      \forall i,j. \Atom_{i,j} \ParallelRewriteAtom \DNFRegex_{i,j}\\
      \forall i. \DNFRegex_i = \DNFOf{\SequenceOf{\String_{i,0}}} \ConcatDNF \DNFRegex_{i,1}
      \ConcatDNF \ldots \ConcatDNF \DNFRegex_{i,n_i} \ConcatDNF
      \DNFOf{\SequenceOf{\String_{i,n_i}}}
    }
    {
      \DNFRegex \ParallelRewrite \DNFRegex_1 \OrDNF \ldots \OrDNF \DNFRegex_n
    }

    \inferrule[\IdentityRewriteRule{}]
    {
    }
    {
      \DNFRegex \ParallelRewrite \DNFRegex
    }
  \end{mathpar}
\end{definition}

\begin{definition}[Parallel Rewriting With Reordering]\leavevmode
  \begin{mathpar}
    \inferrule[\AtomUnrollstarLeftRule{}]
    {
    }
    {
      \StarOf{\DNFRegex}\ParallelRewriteSwapAtom
      \OrDNFOf{\DNFOf{\SequenceOf{\EmptyString}}}
      {(\ConcatDNFOf{\DNFRegex}{\DNFOf{\SequenceOf{\StarOf{\DNFRegex}}}})}
    }

    \inferrule[\AtomUnrollstarRightRule{n}]
    {
    }
    {
      \StarOf{\DNFRegex}\ParallelRewriteSwapAtom
      \OrDNFOf{\DNFOf{\SequenceOf{\EmptyString}}}
      {(\ConcatDNFOf{\DNFOf{\SequenceOf{\StarOf{\DNFRegex}}}}{\DNFRegex})}
    }

    \inferrule[\ParallelSwapAtomStructuralRewriteRule{}]
    {
      \DNFRegex \ParallelRewriteSwap \DNFRegex'
    }
    {
      \StarOf{\DNFRegex} \ParallelRewriteSwapAtom \DNFOf{\SequenceOf{\StarOf{\DNFRegex'}}}
    }

    \inferrule[\DNFReorderRule{}]
    {
      \sigma\in\PermutationSetOf{n}
    }
    {
      \DNFOf{\Sequence_1 \DNFSep \ldots \DNFSep \Sequence_n}
      \ParallelRewriteSwap
      \DNFOf{\Sequence_{\sigma(1)} \DNFSep \ldots \DNFSep \Sequence_{\sigma(n)}}
    }

    \inferrule[\ParallelSwapDNFStructuralRewriteRule{}]
    {
      \DNFRegex = \DNFOf{\Sequence_1 \DNFSep \ldots \DNFSep \Sequence_n}\\
      \forall i. \Sequence_i =
      \SequenceOf{\String_{i,0} \SeqSep \Atom_{i,1} \SeqSep \ldots \SeqSep
        \Atom_{i,n_i} \SeqSep \String_{i,n_i}}\\
      \forall i,j. \Atom_{i,j} \ParallelRewriteSwapAtom \DNFRegex_{i,j}\\
      \forall i. \DNFRegex_i = \DNFOf{\SequenceOf{\String_{i,0}}} \ConcatDNF \DNFRegex_{i,1}
      \ConcatDNF \ldots \ConcatDNF \DNFRegex_{i,n_i} \ConcatDNF
      \DNFOf{\SequenceOf{\String_{i,n_i}}}
    }
    {
      \DNFRegex \ParallelRewriteSwap \DNFRegex_1 \OrDNF \ldots \OrDNF \DNFRegex_n
    }

    \inferrule[\IdentityRewriteRule{}]
    {
    }
    {
      \DNFRegex \ParallelRewriteSwap \DNFRegex
    }
  \end{mathpar}
\end{definition}

\begin{lemma}[$\ParallelRewrite$ Maintained Under Iteration]
  \label{lem:parallel-rewrite-iteration}
  Let $\DNFRegex \ParallelRewrite \DNFRegexAlt$, then
  $\DNFOf{\SequenceOf{\StarOf{\DNFRegex}}} \ParallelRewrite
  \DNFOf{\SequenceOf{\StarOf{\DNFRegexAlt}}}$.
\end{lemma}
\begin{proof}
  Consider the derivation

  \[
    \inferrule*
    {
      \inferrule*
      {
        \DNFRegex \ParallelRewrite \DNFRegexAlt
      }
      {
        \StarOf{\DNFRegex} \ParallelRewriteAtom
        \DNFOf{\SequenceOf{\StarOf{\DNFRegexAlt}}}
      }
    }
    {
      \DNFOf{\SequenceOf{\StarOf{\DNFRegex}}} \ParallelRewrite
      \DNFOf{\SequenceOf{\StarOf{\DNFRegexAlt}}}
    }
  \]
\end{proof}

\begin{lemma}
  \label{lem:final-identity-can-be-ignored}
  If $\DNFRegex \ParallelRewrite \DNFRegex$ through an application of
  \IdentityRewriteRule{}, then $\DNFRegex \ParallelRewrite \DNFRegex$ through an
  application of \ParallelDNFStructuralRewriteRule{}.
\end{lemma}
\begin{proof}
  Let $\DNFRegex \ParallelRewrite \DNFRegex$ through an application of
  \IdentityRewriteRule{}.

  Let $\DNFRegex = \DNFOf{\Sequence_1 \DNFSep \ldots \DNFSep \Sequence_n}$.
  Let $\Sequence_i =
  \SequenceOf{\String_{i,0} \SeqSep \Atom_{i,1} \SeqSep \ldots \SeqSep \Atom_{i,n_i} \SeqSep \String_{i,n_i}}$.
  By Lemma~\ref{lem:identity-atom-in-parallel},
  $\Atom_{i,j} \ParallelRewriteAtom \AtomToDNFOf{\Atom_{i,j}}$.
  Define $\DNFRegex_{i,j}$ as $\AtomToDNFOf{\DNFRegex_{i,j}}$
  
  Define $\DNFRegex_i$ as
  $\DNFOf{\SequenceOf{\String_{i,0}}} \ConcatDNF \DNFRegex_{i,1}
  \ConcatDNF \ldots \ConcatDNF \DNFRegex_{i,n_i} \ConcatDNF
  \DNFOf{\SequenceOf{\String_{i,n_i}}}$,
  so as $\DNFRegex_{i,j} = \AtomToDNFOf{\Atom_{i,j}}$, through the definition
  of $\ConcatDNF$, $DNFRegex_i = \DNFOf{\Sequence_i}$.
  
  By the definition of $\OrDNF$,
  $\DNFOf{\Sequence_1} \OrDNF \ldots \OrDNF \DNFOf{\Sequence_n} =
  \DNFOf{\Sequence_1 \DNFSep \ldots \DNFSep \Sequence_n} = \DNFRegex$.

  \[
    \inferrule*
    {
      \DNFRegex = \DNFOf{\Sequence_1 \DNFSep \ldots \DNFSep \Sequence_n}\\
      \forall i. \Sequence_i =
      \SequenceOf{\String_{i,0} \SeqSep \Atom_{i,1} \SeqSep \ldots \SeqSep \Atom_{i,n_i} \SeqSep \String_{i,n_i}}\\
      \forall i,j. \Atom_{i,j} \ParallelRewriteAtom \DNFRegex_{i,j}\\
      \forall i. \DNFRegex_i = \DNFOf{\SequenceOf{\String_{i,0}}} \ConcatDNF \DNFRegex_{i,1}
      \ConcatDNF \ldots \ConcatDNF \DNFRegex_{i,n_i} \ConcatDNF
      \DNFOf{\SequenceOf{\String_{i,n_i}}}
    }
    {
      \DNFRegex \ParallelRewrite \DNFRegex_1 \OrDNF \ldots \OrDNF \DNFRegex_n
    }
  \]

  So $\DNFRegex \ParallelRewrite \DNFRegex$, with the final rule being an
  application of \IdentityRewriteRule{}.
\end{proof}

\begin{lemma}[$\ParallelRewrite$ Maintained Under $\OrDNF$]
  \label{lem:parallel-rewrite-or}
  Let $\DNFRegex \ParallelRewrite \DNFRegex'$ and $\DNFRegexAlt
  \ParallelRewrite \DNFRegexAlt'$ then
  $\DNFRegex \OrDNF \DNFRegexAlt \ParallelRewrite
  \DNFRegex' \OrDNF \DNFRegexAlt'$.
\end{lemma}
\begin{proof}
  By Lemma~\ref{lem:final-identity-can-be-ignored}, a derivation with the final
  rule being an application of \IdentityRewriteRule{}, can be converted into a
  derivation with the final rule being an application of
  \ParallelDNFStructuralRewriteRule{}.  So we can assume that the final rule of
  each is an application of \ParallelDNFStructuralRewriteRule{}.

  \[
    \inferrule*
    {
      \DNFRegex = \DNFOf{\Sequence_1 \DNFSep \ldots \DNFSep \Sequence_n}\\
      \forall i. \Sequence_i =
      \SequenceOf{\String_{i,0} \SeqSep \Atom_{i,1} \SeqSep \ldots \SeqSep \Atom_{i,n_i} \SeqSep \String_{i,n_i}}\\
      \forall i,j. \Atom_{i,j} \ParallelRewriteAtom \DNFRegex_{i,j}\\
      \forall i. \DNFRegex_i = \DNFOf{\SequenceOf{\String_{i,0}}} \ConcatDNF \DNFRegex_{i,1}
      \ConcatDNF \ldots \ConcatDNF \DNFRegex_{i,n_i} \ConcatDNF
      \DNFOf{\SequenceOf{\String_{i,n_i}}}
    }
    {
      \DNFRegex \ParallelRewrite \DNFRegex_1 \OrDNF \ldots \OrDNF \DNFRegex_n
    }
  \]

  \[
    \inferrule*
    {
      \DNFRegexAlt = \DNFOf{\SequenceAlt_1 \DNFSep \ldots \DNFSep \Sequence_m}\\
      \forall i. \SequenceAlt_i =
      \SequenceOf{\StringAlt_{i,0} \SeqSep \AtomAlt_{i,1} \SeqSep \ldots \SeqSep \AtomAlt_{i,m_i} \SeqSep \StringAlt_{i,m_i}}\\
      \forall i,j. \AtomAlt_{i,j} \ParallelRewriteAtom \DNFRegexAlt_{i,j}\\
      \forall i. \DNFRegexAlt_i = \DNFOf{\SequenceOf{\StringAlt_{i,0}}} \ConcatDNF \DNFRegexAlt_{i,1}
      \ConcatDNF \ldots \ConcatDNF \DNFRegexAlt_{i,n_i} \ConcatDNF
      \DNFOf{\SequenceOf{\StringAlt_{i,n_i}}}
    }
    {
      \DNFRegexAlt \ParallelRewrite \DNFRegexAlt_1 \OrDNF \ldots \OrDNF \DNFRegexAlt_n
    }
  \]

  Define $\Atom_{i,j}'' =
  \begin{cases*}
    \Atom_{i,j} & if $i \leq n$\\
    \AtomAlt_{i-n,j} & if $i > n$
  \end{cases*}$

  Define $\String_{i,j}'' =
  \begin{cases*}
    \String_{i,j} & if $i \leq n$\\
    \StringAlt_{i-n,j} & if $i > n$
  \end{cases*}$

  Define $n_i'' =
  \begin{cases*}
    n_i & if $i \leq n$\\
    m_{i-n} & if $i > n$
  \end{cases*}$

  Define $\Sequence_{i}'' =
  \SequenceOf{\String_{i,0}'' \SeqSep \Atom_{i,1}'' \SeqSep \ldots \SeqSep 
    \Atom_{i,n_i''}'' \SeqSep \String_{i,n_i''}''}$.
  By inspection, $\Sequence_i'' =
  \begin{cases*}
    \Sequence_i & if $i \leq n$\\
    \SequenceAlt_{i-n} & if $i > n$
  \end{cases*}$.
  
  Define $\DNFRegex'' = \DNFOf{\Sequence_1'' \DNFSep \ldots \DNFSep \Sequence_{n+m}''}$.
  By inspection, $\DNFRegex'' = \DNFRegex \OrDNF \DNFRegexAlt$.

  Define $\DNFRegex_{i,j}'' =
  \begin{cases*}
    \DNFRegex_{i,j} & if $i \leq n$\\
    \DNFRegexAlt_{i-n,j} & if $i > n$
  \end{cases*}$.  By inspection $\Atom_{i,j}'' \ParallelRewrite \DNFRegex_{i,j}''$.

  Define $\DNFRegex_i''$ as
  $\DNFOf{\SequenceOf{\String_{i,0}''}} \ConcatDNF \DNFRegex_{i,1}''
  \ConcatDNF \ldots \ConcatDNF \DNFRegex_{i,n_i''}'' \ConcatDNF
  \DNFOf{\SequenceOf{\String_{i,n_i''}''}}$.
  By inspection, $\DNFRegex_i'' =
  \begin{cases*}
    \DNFRegex_i & if $i \leq n$\\
    \DNFRegexAlt_{i-n} & if $i > n$
  \end{cases*}$.
  This means that
  $\DNFRegex_1'' \OrDNF \ldots \OrDNF \DNFRegex_{n+m}'' =
  (\DNFRegex_1 \OrDNF \ldots \OrDNF \DNFRegex_n \OrDNF
  \DNFRegexAlt_1 \OrDNF \ldots \OrDNF \DNFRegexAlt_m) =
  \DNFRegex' \OrDNF \DNFRegexAlt'$.

  Consider the derivation 
  \[
    \inferrule*
    {
      \DNFRegex'' = \DNFOf{\Sequence_1'' \DNFSep \ldots \DNFSep \Sequence_{n+m}''}\\
      \forall i. \Sequence_i'' =
      \SequenceOf{\String_{i,0}'' \SeqSep \Atom_{i,1}'' \SeqSep \ldots \SeqSep \Atom_{i,n_i''}'' \SeqSep \String_{i,n_i''}''}\\
      \forall i,j. \Atom_{i,j}'' \ParallelRewriteAtom \DNFRegex_{i,j}''\\
      \forall i. \DNFRegex_i'' = \DNFOf{\SequenceOf{\String_{i,0}''}} \ConcatDNF \DNFRegex_{i,1}''
      \ConcatDNF \ldots \ConcatDNF \DNFRegex_{i,n_i}'' \ConcatDNF
      \DNFOf{\SequenceOf{\String_{i,n_i''}''}}
    }
    {
      \DNFRegex'' \ParallelRewrite
      \DNFRegex_1'' \OrDNF \ldots \OrDNF \DNFRegex_{n+m}''
    }
  \]
\end{proof}

\begin{lemma}[$\StarOf{\ParallelRewrite}$ Maintained Under Iteration]
  \label{lem:star-parallel-rewrite-iteration}
  Let $\DNFRegex \StarOf{\ParallelRewrite} \DNFRegexAlt$, then
  $\AtomToDNFOf{\StarOf{\DNFRegex}} \StarOf{\ParallelRewrite}
  \AtomToDNFOf{\StarOf{\DNFRegexAlt}}$.
\end{lemma}
\begin{proof}
  By induction on the derivation of $\StarOf{\ParallelRewrite}$.

  \begin{case}[\ReflexivityRule{}]
    \[
      \inferrule*
      {
      }
      {
        \DNFRegex \StarOf{\ParallelRewrite} \DNFRegex
      }
    \]

    By reflexivity rule

    \[
      \inferrule*
      {
      }
      {
        \AtomToDNFOf{\StarOf{\DNFRegex}} \StarOf{\ParallelRewrite}
        \AtomToDNFOf{\StarOf{\DNFRegex}}
      }
    \]
  \end{case}

  \begin{case}[\BaseRule{}]
    \[
      \inferrule*
      {
        \DNFRegex \ParallelRewrite \DNFRegexAlt
      }
      {
        \DNFRegex \StarOf{\ParallelRewrite} \DNFRegexAlt
      }
    \]

    By Lemma~\ref{lem:parallel-rewrite-iteration},
    $\AtomToDNFOf{\StarOf{\DNFRegex}} \ParallelRewrite
    \AtomToDNFOf{\StarOf{\DNFRegexAlt}}$

    Consider the derivation
    \[
      \inferrule*
      {
        \AtomToDNFOf{\StarOf{\DNFRegex}} \ParallelRewrite
        \AtomToDNFOf{\StarOf{\DNFRegexAlt}}
      }
      {
        \AtomToDNFOf{\StarOf{\DNFRegex}} \StarOf{\ParallelRewrite}
        \AtomToDNFOf{\StarOf{\DNFRegexAlt}}
      }
    \]
  \end{case}

  \begin{case}[\TransitivityRule{}]
    \[
      \inferrule*
      {
        \DNFRegex \StarOf{\ParallelRewrite} \DNFRegex'\\
        \DNFRegex' \StarOf{\ParallelRewrite} \DNFRegexAlt
      }
      {
        \DNFRegex \StarOf{\ParallelRewrite} \DNFRegexAlt
      }
    \]

    By IH, $\AtomToDNFOf{\StarOf{\DNFRegex}} \StarOf{\ParallelRewrite}
    \AtomToDNFOf{\StarOf{\DNFRegex'}}$ and
    $\AtomToDNFOf{\StarOf{\DNFRegex'}} \StarOf{\ParallelRewrite}
    \AtomToDNFOf{\StarOf{\DNFRegexAlt}}$.
  \end{case}
\end{proof}

\begin{lemma}[Equivalence of $\ToRegex \Compose \ToDNFRegex$]
  \label{lem:there-and-back-equiv}
  $(\ToRegex \Compose \ToDNFRegex) \Regex \SSREquiv \Regex$
\end{lemma}
\begin{proof}
  By induction on the structure of $\Regex$
  \begin{case}[\BaseRegexType{}]
    $(\ToRegex \Compose \ToDNFRegex)\String =
    \ToRegexOf{\DNFOf{\SequenceOf{\String}}} = \emptyset \Or \String$
    
    $\emptyset \Or \String \SSREquiv \String$.
  \end{case}

  \begin{case}[\EmptyRegexType{}]
    $(\ToRegex \Compose \ToDNFRegex)\emptyset =
    \ToRegexOf{\DNFOf{}} = \emptyset$
    
    $\emptyset \SSREquiv \emptyset$.
  \end{case}

  \begin{case}[\StarRegexType{}]
    $(\ToRegex \Compose \ToDNFRegex)\StarOf{\Regex'} =
    \ToRegexOf{\DNFOf{\SequenceOf{\EmptyString \SeqSep \StarOf{(\ToDNFRegexOf{\Regex'})} \SeqSep \EmptyString}}}
    = \emptyset \Or (\EmptyString \Concat \StarOf{((\ToRegex \Compose \ToDNFRegex)\Regex')}
    \Concat \EmptyString)$
    Then, through application of equational theory transitivity,
    \OrIdentityRule{}, \ConcatIdentityLeftRule{}, and \ConcatIdentityRightRule{},
    We get $(\ToRegex \Compose \ToDNFRegex)\StarOf{\Regex'} \SSREquiv
    \StarOf{((\ToRegex \Compose \ToDNFRegex)\Regex')}$.
    By application of the IH, and transitivity, we get
    $(\ToRegex \Compose \ToDNFRegex)\StarOf{\Regex'} \SSREquiv
    \StarOf{\Regex'}$
  \end{case}

  \begin{case}[\ConcatRegexType{}]
    Let $(\ToRegex \Compose \ToDNFRegex)(\Regex_1 \Concat \Regex_2) =
    \ToRegexOf{(\ToDNFRegexOf{\Regex_1} \ConcatDNF \ToDNFRegexOf{\Regex_2})}$.
    Let $\ToDNFRegexOf{\Regex_1} = \DNFOf{\Sequence_1 \DNFSep \ldots \DNFSep \Sequence_n}$ and
    $\ToDNFRegexOf{\Regex_2} = \DNFOf{\SequenceAlt_1 \DNFSep \ldots \DNFSep \Sequence_m}$.

    $\Regex_1 \SSREquiv (\ToRegex \Compose \ToDNFRegex)(\Regex_1) =
    (\emptyset \Or (\ToRegexOf{\Sequence_1} \Or (\ldots \Or
    (\ToRegexOf{\Sequence_n})\ldots))) \SSREquiv
    \ToRegexOf{\Sequence_1} \Or \ldots \Or \ToRegexOf{\Sequence_n}$
    and
    $\Regex_2 \SSREquiv (\ToRegex \Compose \ToDNFRegex)(\Regex_2) =
    (\emptyset \Or (\ToRegexOf{\SequenceAlt_1} \Or (\ldots \Or
    (\ToRegexOf{\SequenceAlt_m})\ldots))) \SSREquiv
    \ToRegexOf{\SequenceAlt_1} \Or \ldots \Or \ToRegexOf{\SequenceAlt_m}$.

    So by structural \ConcatRegexType{} identity, and transitivity,
    $\Regex_1 \Concat \Regex_2 \SSREquiv
    (\ToRegexOf{\Sequence_1} \Or \ldots \Or \ToRegexOf{\Sequence_n})
    \Concat
    (\ToRegexOf{\SequenceAlt_1} \Or \ldots \Or \ToRegexOf{\SequenceAlt_m})$.

    Through repeated application of \DistributivityLeftRule{} and
    \DistributivityRightRule{},
    $\Regex_1 \Concat \Regex_2 \SSREquiv
    (\ToRegexOf{\Sequence_1}\Concat\ToRegexOf{\SequenceAlt_1} \Or \ldots \Or
    \ToRegexOf{\Sequence_n}\Concat\ToRegexOf{\SequenceAlt_m})$.

    Now, I want to show $\ToRegexOf{\Sequence_i} \Concat \ToRegexOf{\SequenceAlt_j}
    \SSREquiv \ToRegexOf{(\Sequence_i \ConcatDNF \SequenceAlt_j)}$.
    Let $\Sequence_i =
    \SequenceOf{\String_{i,0} \SeqSep \Atom_{i,1} \SeqSep \ldots \SeqSep \Atom_{i,n_i} \SeqSep \String_{i,n_i}}$,
    and $\SequenceAlt_j =
    \SequenceOf{\StringAlt_{j,0} \SeqSep \AtomAlt_{j,1} \SeqSep \ldots \SeqSep \Atom_{j,m_j} \SeqSep \String_{j,n_j}}$.
    $\ToRegexOf{(\Sequence_i \ConcatDNF \Sequence_j)} =
    \ToRegexOf{\SequenceOf{\String_{i,0} \SeqSep \Atom_{i,1} \SeqSep \ldots \SeqSep 
        \Atom_{i,n_i} \SeqSep \String_{i,n_i}\Concat\StringAlt_{j,0} \SeqSep \AtomAlt_{j,1} \SeqSep 
        \ldots \SeqSep \Atom_{j,m_j} \SeqSep \String_{j,n_j}}} =
    \String_{i,0} \Concat (\Atom_{i,1} \Concat (\ldots \Concat \String_{i,n_i}
    \Concat \StringAlt_{j,0} \Concat (\AtomAlt_{j,1} \Concat (\ldots \Concat
    (\Atom_{j,n_j} \Concat \String_{j,n_j})\ldots))\ldots))$
    So through repeated application of \ConcatAssocRule{},
    $\ToRegexOf{(\Sequence_i \ConcatDNF \Sequence_j)} \SSREquiv
    (\String_{i,0} \Concat (\Atom_{i,1} \Concat (\ldots \Concat (\Atom_{i,n_i}
    \Concat \String_{i,n_i})\ldots))) \Concat (\StringAlt_{j,0} \Concat
    (\AtomAlt_{j,1} \Concat (\ldots \Concat (\AtomAlt_{j,n_j} \Concat
    \String_{j,n_j})\ldots)))$.

    Because of this 
    $\Regex_1 \Concat \Regex_2 \SSREquiv
    \ToRegexOf{(\Sequence_1\ConcatDNF\SequenceAlt_1)} \Or \ldots \Or
    \ToRegexOf{(\Sequence_n\ConcatDNF\SequenceAlt_n)}$.
    Through repeated application of \OrAssociativityRule{}, and \OrIdentityRule{},
    $\Regex_1 \Concat \Regex_2 \SSREquiv
    \emptyset \Or (\ToRegexOf{(\Sequence_1\ConcatDNF\SequenceAlt_1)} \Or \ldots
    (\ToRegexOf{(\Sequence_n\ConcatDNF\SequenceAlt_m)})\ldots)$.
    Furthermore, $\emptyset \Or (\ToRegexOf{(\Sequence_1\ConcatDNF\SequenceAlt_1)} \Or \ldots
    (\ToRegexOf{(\Sequence_n\ConcatDNF\SequenceAlt_m)})\ldots) =
    \ToRegexOf{\DNFOf{\Sequence_1\ConcatSequence\SequenceAlt_1 \DNFSep \ldots \DNFSep 
        \Sequence_n\Concat\SequenceAlt_m}} =
    (\ToRegex \Compose \ToDNFRegex)(\Regex_1 \Concat \Regex_2)$ as desired.
  \end{case}

  \begin{case}[\OrRegexType{}]
    Let $(\ToRegex \Compose \ToDNFRegex)(\Regex_1 \Or \Regex_2) =
    \ToRegexOf{(\ToDNFRegexOf{\Regex_1} \OrDNF \ToDNFRegexOf{\Regex_2})}$.
    Let $\ToDNFRegexOf{\Regex_1} = \DNFOf{\Sequence_1 \DNFSep \ldots \DNFSep \Sequence_n}$ and
    $\ToDNFRegexOf{\Regex_2} = \DNFOf{\SequenceAlt_1 \DNFSep \ldots \DNFSep \SequenceAlt_m}$.
    So
    $(\ToRegex \Compose \ToDNFRegex)(\Regex_1 \Or \Regex_2) =
    \ToRegexOf{\DNFOf{\Sequence_1 \DNFSep \ldots \DNFSep \Sequence_n \DNFSep 
        \SequenceAlt_1 \DNFSep \ldots \DNFSep \SequenceAlt_m}} =
    \emptyset \Or (\ToRegexOf{\Sequence_1} \Or (
    \ldots \Or (\ToRegexOf{\SequenceAlt_m})\ldots))$.
    Through applying associativity a lot, and \OrIdentityRule{} once, I get
    $(\ToRegex \Compose \ToDNFRegex)(\Regex_1 \Or \Regex_2) =
    (\emptyset \Or (\ToRegexOf{\Sequence_1} \Or (\ldots \Or
    (\ToRegexOf{\Sequence_n})\ldots)))
    \Or
    (\emptyset \Or (\ToRegexOf{\SequenceAlt_1} \Or (\ldots \Or
    (\ToRegexOf{\SequenceAlt_m})\ldots)))$.

    $(\ToRegex \Compose \ToDNFRegex)(\Regex_1 \Or \Regex_2) =
    (\emptyset \Or (\ToRegexOf{\Sequence_1} \Or (\ldots \Or
    (\ToRegexOf{\Sequence_n})\ldots))) =
    (\ToRegex \Compose \ToDNFRegex)\Regex_1$
    and $(\emptyset \Or (\ToRegexOf{\SequenceAlt_1} \Or (\ldots \Or
    (\ToRegexOf{\SequenceAlt_m})\ldots))) =
    (\ToRegex \Compose \ToDNFRegex)\Regex_2$, so by IH
    $(\ToRegex \Compose \ToDNFRegex)(\Regex_1 \Or \Regex_2) =
    (\emptyset \Or (\ToRegexOf{\Sequence_1} \Or (\ldots \Or
    (\ToRegexOf{\Sequence_n})\ldots))) \SSREquiv
    \Regex_1$
    and $(\emptyset \Or (\ToRegexOf{\SequenceAlt_1} \Or (\ldots \Or
    (\ToRegexOf{\SequenceAlt_m})\ldots))) \SSREquiv
    \Regex_2$.

    Through an application of structural \OrRegexType{} equality,
    $(\emptyset \Or (\ToRegexOf{\Sequence_1} \Or (\ldots \Or
    (\ToRegexOf{\Sequence_n})\ldots)))
    \Or
    (\emptyset \Or (\ToRegexOf{\SequenceAlt_1} \Or (\ldots \Or
    (\ToRegexOf{\SequenceAlt_m})\ldots))) \SSREquiv
    \Regex_1 \Or \Regex_2$, as desired.
  \end{case}
\end{proof}

\begin{lemma}[Equivalence of Preimage of $\ToDNFRegex$]
  \label{lem:preimage-equiv}
  If $\ToDNFRegexOf{\Regex} = \ToDNFRegexOf{\RegexAlt}$, then $\Regex
  \SSREquiv \RegexAlt$.
\end{lemma}
\begin{proof}
  $\ToDNFRegexOf{\Regex} = \ToDNFRegexOf{\RegexAlt}$, so $(\ToRegex \Compose
  \ToDNFRegex) \Regex = (\ToRegex \Compose \ToDNFRegex) \RegexAlt$.
  By Lemma~\ref{lem:there-and-back-equiv},
  $\Regex \SSREquiv (\ToRegex \Compose \ToDNFRegex) \RegexAlt
  \SSREquiv \RegexAlt$
\end{proof}

\begin{lemma}[Equivalence of Adjacent Swapping Permutation of \OrRegexType{}]
  \label{lem:adj-swap-or-in-ssrequiv}
  Let $\Regex_1 \Or \ldots \Or \Regex_n$.  Let $\sigma_i$ be an adjacent
  swapping permutation.  $\Regex_1 \Or \ldots \Or \Regex_n \SSREquiv
  \Regex_{\sigma_i(1)} \Or \ldots \Or \Regex_{\sigma_i(n)}$.
\end{lemma}
\begin{proof}
  $\Regex_1 \Or \ldots \Or \Regex_n \SSREquiv (\Regex_1 \Or \ldots \Or
  \Regex_{i-1}) \Or (\Regex_{i} \Or \Regex_{i+1}) \Or (\Regex_{i+2} \Or \ldots
  \Or \Regex_n)$ by repeated application of associativity.
  
  $\Regex_i \Or \Regex_{i+1} = \Regex_{i+1} \Or \Regex_{i}$ by \OrRegexType{}
  commutativity, so by \OrRegexType{} structural equality,\\
  $(\Regex_1 \Or \ldots \Or
  \Regex_{i-1}) \Or (\Regex_{i} \Or \Regex_{i+1}) \Or (\Regex_{i+2} \Or \ldots
  \Or \Regex_n) \SSREquiv
  (\Regex_1 \Or \ldots \Or
  \Regex_{i-1}) \Or (\Regex_{i+1} \Or \Regex_{i}) \Or (\Regex_{i+2} \Or \ldots
  \Or \Regex_n)$
  
  $(\Regex_1 \Or \ldots \Or
  \Regex_{i-1}) \Or (\Regex_{i+1} \Or \Regex_{i}) \Or (\Regex_{i+2} \Or \ldots
  \Or \Regex_n) \SSREquiv
  \Regex_{\sigma_i(1)} \Or \ldots \Or \Regex_{\sigma_i(n)}$ by repeated
  application of associativity.
  
  So, by the transitivity of equational theories,
  $\Regex_1 \Or \ldots \Or \Regex_n \SSREquiv
  \Regex_{\sigma_i(1)} \Or \ldots \Or \Regex_{\sigma_i(n)}$.
\end{proof}

\begin{lemma}[Expressibility of $\ParallelRewriteSwap$ in $\SSREquiv$
  Up To Preimage]
  \label{lem:express-swap-in-equiv}
  \leavevmode
  \begin{enumerate}
  \item If $\ToDNFRegexOf{\Regex} \ParallelRewriteSwap \ToDNFRegexOf{\RegexAlt}$, then
    $\Regex \SSREquiv \RegexAlt$.
  \item If $\ToDNFRegexOf{\Regex} = \DNFOf{\SequenceOf{\Atom}}$ and
    $\Atom \ParallelRewriteSwapAtom \ToDNFRegexOf{\RegexAlt}$ then
    $\Regex \SSREquiv \RegexAlt$.
  \end{enumerate}
\end{lemma}
\begin{proof}By mutual induction on the derivation of $\ParallelRewriteSwap$ and
  $\ParallelRewriteSwapAtom$.
  \begin{case}[\AtomUnrollstarLeftRule{}]
    Let $\ToDNFRegexOf{\Regex} = \DNFOf{\SequenceOf{\Atom}}$, and
    $\Atom \ParallelRewriteSwapAtom \ToDNFRegexOf{\RegexAlt}$ from
    \AtomUnrollstarLeftRule{}.
    This means that $\Atom = \StarOf{\DNFRegex}$ and
    $\ToDNFRegexOf{\RegexAlt} = \OrDNFOf{\DNFOf{\SequenceOf{\EmptyString}}}
    {(\ConcatDNFOf{\DNFRegex}{\DNFOf{\SequenceOf{\StarOf{\DNFRegex}}}})}$.

    Let $\Regex' = \ToRegexOf{\DNFRegex}$.
    As $\ToDNFRegexOf{\StarOf{\Regex'}} =
    \DNFOf{\SequenceOf{\StarOf{\DNFRegex}}}$, then from
    Lemma~\ref{lem:preimage-equiv}, $\StarOf{\Regex'}
    \SSREquiv \Regex$.
    Similarly, as $\ToDNFRegexOf{(\EmptyString \Or
      (\Regex'\Concat\StarOf{\Regex'}))} =
    \OrDNFOf{\DNFOf{\SequenceOf{\EmptyString}}}
    {(\ConcatDNFOf{\DNFRegex}{\DNFOf{\SequenceOf{\StarOf{\DNFRegex}}}})}$,
    then from
    Lemma~\ref{lem:preimage-equiv}, $\EmptyString \Or
    (\Regex'\Concat\StarOf{\Regex'}) \SSREquiv \RegexAlt$.

    So, through an application of UnrollstarLeftRule{},
    $\Regex \SSREquiv \StarOf{\Regex'} \SSREquiv
    \EmptyString \Or
    (\Regex'\Concat\StarOf{\Regex'}) \SSREquiv \RegexAlt$, as desired.
  \end{case}

  \begin{case}[\AtomUnrollstarRightRule{}]
    Let $\ToDNFRegexOf{\Regex} = \DNFOf{\SequenceOf{\Atom}}$, and
    $\Atom \ParallelRewriteSwapAtom \ToDNFRegexOf{\RegexAlt}$ from
    \AtomUnrollstarRightRule{}.
    This means that $\Atom = \StarOf{\DNFRegex}$ and
    $\ToDNFRegexOf{\RegexAlt} = \OrDNFOf{\DNFOf{\SequenceOf{\EmptyString}}}
    {(\ConcatDNFOf{\DNFOf{\SequenceOf{\StarOf{\DNFRegex}}}}{\DNFRegex})}$.

    Let $\Regex' = \ToRegexOf{\DNFRegex}$.
    As $\ToDNFRegexOf{\StarOf{\Regex'}} =
    \DNFOf{\SequenceOf{\StarOf{\DNFRegex}}}$, then from
    Lemma~\ref{lem:preimage-equiv}, $\StarOf{\Regex'}
    \SSREquiv \Regex$.
    Similarly, as $\ToDNFRegexOf{(\EmptyString \Or
      (\StarOf{\Regex'}\Concat\Regex'))} =
    \OrDNFOf{\DNFOf{\SequenceOf{\EmptyString}}}
    {(\ConcatDNFOf{\DNFOf{\SequenceOf{\StarOf{\DNFRegex}}}}{\DNFRegex})}$,
    then from
    Lemma~\ref{lem:preimage-equiv}, $\EmptyString \Or
    (\StarOf{\Regex'}\Concat\Regex') \SSREquiv \RegexAlt$.
  
    So, through an application of UnrollstarRightRule{},
    $\Regex \SSREquiv \StarOf{\Regex'} \SSREquiv
    \EmptyString \Or
    (\StarOf{\Regex'}\Concat\Regex') \SSREquiv \RegexAlt$, as desired.
  \end{case}
  
  \begin{case}[\ParallelSwapAtomStructuralRewriteRule{}]
    Let $\ToDNFRegexOf{\Regex} = \DNFOf{\SequenceOf{\Atom}}$, and
    $\Atom \ParallelRewriteSwapAtom \ToDNFRegexOf{\RegexAlt}$
    This means that $\Atom = \StarOf{\DNFRegex}$ and
    $\ToDNFRegexOf{\RegexAlt} =
    \DNFOf{\SequenceOf{\DNFRegexAlt})}$ where $\DNFRegex \ParallelRewriteSwap
    \DNFRegexAlt$.

    Let $\ToRegexOf{\DNFRegex} = \Regex'$ and
    $\ToRegexOf{\DNFRegexAlt} = \RegexAlt'$.  By induction assumption, $\Regex'
    \SSREquiv \RegexAlt'$.  By structural equivalence,
    $\StarOf{\Regex'} = \StarOf{\RegexAlt'}$.
    As $\ToDNFRegexOf{\StarOf{\Regex'}} =
    \DNFOf{\SequenceOf{\StarOf{\DNFRegex}}}$, from
    Lemma~\ref{lem:preimage-equiv}, $\StarOf{\Regex'} \SSREquiv
    \Regex$.
    As $\ToDNFRegexOf{\StarOf{\RegexAlt'}} =
    \DNFOf{\SequenceOf{\StarOf{\DNFRegexAlt}}}$, from
    Lemma~\ref{lem:preimage-equiv}, $\StarOf{\RegexAlt'} \SSREquiv
    \RegexAlt$.

    $\Regex \SSREquiv \StarOf{\Regex'} \SSREquiv
    \StarOf{\RegexAlt'} \SSREquiv \RegexAlt$, as desired.
  \end{case}

  \begin{case}[\DNFReorderRule{}]
    Let $\ToDNFRegexOf{\Regex} \ParallelRewriteSwap \ToDNFRegexOf{\RegexAlt}$,
    and the last step of the proof is an application of $\DNFReorderRule$.
    Let $\ToDNFRegexOf{\Regex} = \DNFOf{\Sequence_1 \DNFSep \ldots \DNFSep \Sequence_n}$.  Then,
    for some $\sigma\in\PermutationSetOf{n}$,
    $\ToDNFRegexOf{\RegexAlt} =
    \DNFOf{\Sequence_{\sigma(1)} \DNFSep \ldots \DNFSep \Sequence_{\sigma(n)}}$.

    $\ToRegexOf{\ToDNFRegexOf{\Regex}} =
    \ToRegexOf{\Sequence_1} \Or \ldots \Or \ToRegexOf{\Sequence_n}$ and
    $\ToRegexOf{\ToDNFRegexOf{\RegexAlt}} =
    \ToRegexOf{\Sequence_{\sigma(1)}} \Or \ldots \Or
    \ToRegexOf{\Sequence_{\sigma(n)}}$.

    $\sigma$ can then be broken down into a number of adjacent swapping
    permutations, $\sigma_{i_1} \Compose \ldots \Compose \sigma_{i_m} = \sigma$

    By Lemma~\ref{lem:adj-swap-or-in-ssrequiv}, each $\sigma_{i_j}$ can be applied to a
    sequence of \OrRegexType{}s.

    Consider the derivation
    \[
      \inferrule*
      {
        \inferrule*[vdots=1em]
        {
          \inferrule*[vdots=1em]
          {
            \ToRegexOf{\Sequence_1} \Or \ldots \Or \ToRegexOf{\Sequence_n}
            \SSREquiv
            \ToRegexOf{\Sequence_1} \Or \ldots \Or \ToRegexOf{\Sequence_n}
          }
          {
          }
        }
        {
          \ToRegexOf{\Sequence_{\sigma_{i_m}(1)}} \Or \ldots \Or
          \ToRegexOf{\Sequence_{\sigma_{i_m}(n)}}
          \SSREquiv
          \ToRegexOf{\Sequence_1} \Or \ldots \Or \ToRegexOf{\Sequence_n}
        }
      }
      {
        \ToRegexOf{\Sequence_{(\sigma_{i_1}\Compose\ldots\Compose\sigma_{i_m})(1)}}
        \Or \ldots \Or
        \ToRegexOf{\Sequence_{(\sigma_{i_1}\Compose\ldots\Compose\sigma_{i_m})(n)}}
        \SSREquiv
        \ToRegexOf{\Sequence_1} \Or \ldots \Or \ToRegexOf{\Sequence_n}
      }
    \]

    So, by Lemma~\ref{lem:preimage-equiv},
    $\Regex \SSREquiv \ToRegexOf{\ToDNFRegexOf{\Regex}}$ and
    $\ToRegexOf{\ToDNFRegexOf{\RegexAlt}} \SSREquiv \RegexAlt$.
    Furthermore,
    $\ToRegexOf{\ToDNFRegexOf{\Regex}} \SSREquiv
    \ToRegexOf{\ToDNFRegexOf{\RegexAlt}}$.
    So by the transitivity of an equational theory,
    $\Regex \SSREquiv \RegexAlt$.
  \end{case}

  \begin{case}[\IdentityRewriteRule{}]
    Let $\ToDNFRegexOf{\Regex} \ParallelRewriteSwap \ToDNFRegexOf{\RegexAlt}$
    by an application of \IdentityRewriteRule{}.

    That means $\ToDNFRegexOf{\Regex} = \ToDNFRegexOf{\RegexAlt}$.
    
    So, by Lemma~\ref{lem:preimage-equiv},
    That means that
    $\Regex \SSREquiv \ToRegexOf{\ToDNFRegexOf{\Regex}}
    \SSREquiv \RegexAlt$
  \end{case}

  \begin{case}[\ParallelSwapDNFStructuralRewriteRule{}]
    Let $\ToDNFRegexOf{\Regex} \ParallelRewriteSwap \ToDNFRegexOf{\RegexAlt}$
    by an application of \ParallelSwapDNFStructuralRewriteRule{}.

    \[
      \inferrule*
      {
        \ToDNFRegexOf{\Regex} = \DNFOf{\Sequence_1 \DNFSep \ldots \DNFSep \Sequence_n}\\
        \forall i. \Sequence_i =
        \SequenceOf{\String_{i,0} \SeqSep \Atom_{i,1} \SeqSep \ldots \SeqSep \Atom_{i,n_i} \SeqSep \String_{i,n_i}}\\
        \forall i,j. \Atom_{i,j} \ParallelRewriteSwapAtom \DNFRegex_{i,j}\\
        \forall i. \DNFRegex_i = \DNFOf{\SequenceOf{\String_{i,0}}} \ConcatDNF \DNFRegex_{i,1}
        \ConcatDNF \ldots \ConcatDNF \DNFRegex_{i,n_i} \ConcatDNF
        \DNFOf{\SequenceOf{\String_{i,n_i}}}
      }
      {
        \Regex \ParallelRewriteSwap \DNFRegex_1 \OrDNF \ldots \OrDNF \DNFRegex_n
      }
    \]
    
    $\ToDNFRegexOf{\RegexAlt} = \DNFRegex_1 \OrDNF \ldots \OrDNF \DNFRegex_n$.

    $\Atom_{i,j} \ParallelRewriteSwapAtom \DNFRegex_{i,j}$,
    $\ToDNFRegexOf{\ToRegexOf{\AtomToDNFOf{\Atom_{i,j}}}} =
    \AtomToDNFOf{\Atom_{i,j}}$, and
    $\ToDNFRegexOf{\ToRegexOf{\DNFRegex_{i,j}}} = \DNFRegex_{i,j}$, so by IH,
    $\ToRegexOf{\AtomToDNFOf{\Atom_{i,j}}} \SSREquiv
    \ToRegexOf{\DNFRegex_{i,j}}$.

    Consider the regular expressions
    $\Regex_i =
    \String_{i,0} \Concat \ToRegexOf{\AtomToDNFOf{\Atom_{i,1}}}
    \Concat \ldots \Concat
    \ToRegexOf{\AtomToDNFOf{\Atom_{i,n_i}}} \Concat \String_{i,n_i}$.

    Consider the regular expressions
    $\RegexAlt_i =
    \String_{i,0} \Concat \ToRegexOf{\DNFRegex_{i,j}}
    \Concat \ldots \Concat
    \ToRegexOf{\DNFRegex_{i,n_i}} \Concat \String_{i,n_i}$.

    By structural equality of $\ConcatRegexType$,
    $\Regex_i \SSREquiv \RegexAlt_i$.

    Consider the regular expression
    $\Regex' = \Regex_1 \OrDNF \ldots \OrDNF \Regex_n$ and the regular
    expression
    $\RegexAlt' = \RegexAlt_1 \Or \ldots \Or \RegexAlt_n$.

    By structural equality of $\OrRegexType$,
    $\Regex' \SSREquiv \RegexAlt'$

    $\ToDNFRegexOf{\Regex_i} =
    \DNFOf{\SequenceOf{\String_{i,0}}} \ConcatDNF \AtomToDNFOf{\Atom_{i,1}}
    \ConcatDNF \ldots \ConcatDNF
    \AtomToDNFOf{\Atom_{i,n_i}} \ConcatDNF \DNFOf{\SequenceOf{\String_{i,n_i}}}
    =
    \DNFOf{\SequenceOf{\String_{i,0} \SeqSep \Atom_{i,1}
         \SeqSep \ldots \SeqSep \Atom_{i,n_i} \SeqSep \String_{i,n_i}}}
    = \DNFOf{\Sequence_i}$

    $\ToDNFRegexOf{\Regex'} =
    \ToDNFRegexOf{\Regex_1} \OrDNF \ldots \OrDNF \ToDNFRegexOf{\Regex_n} =
    \DNFOf{\Sequence_1} \OrDNF \ldots \OrDNF \DNFOf{\Sequence_n} =
    \DNFOf{\Sequence_1  \DNFSep  \ldots  \DNFSep  \Sequence_n}$.
    This means, by Lemma~\ref{lem:preimage-equiv}, that
    $\Regex \SSREquiv \Regex'$

    $\ToDNFRegexOf{\RegexAlt_i} =
    \DNFOf{\SequenceOf{\String_{i,0}}} \ConcatDNF \DNFRegex_{i,1}
    \ConcatDNF \ldots \ConcatDNF \DNFRegex_{i,n_i} \ConcatDNF
    \DNFOf{\SequenceOf{\String_{i,n_i}}} =
    \DNFRegex_i$.

    $\ToDNFRegexOf{\RegexAlt'} =
    \ToDNFRegexOf{\RegexAlt_1} \OrDNF \ldots \OrDNF \ToDNFRegexOf{\RegexAlt_n} =
    \DNFRegex_1 \OrDNF \ldots \OrDNF \DNFRegex_n$.
    This means, by Lemma~\ref{lem:preimage-equiv}, that
    $\RegexAlt' \SSREquiv \RegexAlt$
    
    So, $\Regex \SSREquiv \Regex' \SSREquiv \RegexAlt'
    \SSREquiv \RegexAlt$, so by transitivity of equational theories,
    $\Regex \SSREquiv \RegexAlt$.
  \end{case}
\end{proof}

\begin{lemma}[Expressibility of $\EquivalenceOf{\ParallelRewriteSwap}$ in
  $\SSREquiv$]
  \label{lem:express-equiv-swap-in-equiv}
  If $\ToDNFRegexOf{\Regex} \EquivalenceOf{\ParallelRewriteSwap}
  \ToDNFRegexOf{\RegexAlt}$, then
  $\Regex \SSREquiv \RegexAlt$.
\end{lemma}
\begin{proof}
  By induction on the typing derivation of
  $\EquivalenceOf{\ParallelRewriteSwap}$
  \begin{case}[\ReflexivityRule{}]
    Let $\ToDNFRegexOf{\Regex} \EquivalenceOf{\ParallelRewriteSwap}
    \ToDNFRegexOf{\RegexAlt}$, and the last step of the derivation is an
    application of \ReflexivityRule{}.

    This means $\ToDNFRegexOf{\Regex} = \ToDNFRegexOf{\RegexAlt}$.  That means
    $\ToRegexOf{\ToDNFRegexOf{\Regex}} = \ToRegexOf{\ToDNFRegexOf{\RegexAlt}}$.
    By Lemma~\ref{lem:preimage-equiv},
    $\Regex \SSREquiv \ToRegexOf{\ToDNFRegexOf{\Regex}}$.
    By Lemma~\ref{lem:preimage-equiv},
    $\ToRegexOf{\ToDNFRegexOf{\RegexAlt}} \SSREquiv = \RegexAlt$.
    By the transitivity of equational theories,
    $\Regex \SSREquiv \RegexAlt$.
  \end{case}

  \begin{case}[\BaseRule{}]
    Let $\ToDNFRegexOf{\Regex} \EquivalenceOf{\ParallelRewriteSwap}
    \ToDNFRegexOf{\RegexAlt}$, and the last step of the derivation is an
    application of \BaseRule{}.

    This means that $\ToDNFRegexOf{\Regex} \ParallelRewriteSwap
    \ToDNFRegexOf{\RegexAlt}$.

    By Lemma~\ref{lem:express-swap-in-equiv},
    $\Regex \SSREquiv \RegexAlt$.
  \end{case}

  \begin{case}[\SymmetryRule{}]
    Let $\ToDNFRegexOf{\Regex} \EquivalenceOf{\ParallelRewriteSwap}
    \ToDNFRegexOf{\RegexAlt}$, and the last step of the derivation is an
    application of \BaseRule{}.

    This means that $\ToDNFRegexOf{\Regex} \ParallelRewriteSwap
    \ToDNFRegexOf{\RegexAlt}$.

    By Lemma~\ref{lem:express-swap-in-equiv},
    $\Regex \SSREquiv \RegexAlt$.
  \end{case}
\end{proof}

\begin{lemma}[Propagation of $\ParallelRewriteSwap{}$ through $\OrDNF{}$ on
  the left]
  \label{lem:prop_parallel_swap_or_left}
  If $\DNFRegex \ParallelRewriteSwap \DNFRegex'$, then
  $\DNFRegex \OrDNF \DNFRegexAlt \ParallelRewriteSwap \DNFRegex'
  \OrDNF \DNFRegexAlt$
\end{lemma}
\begin{proof}
  This will be done by cases on the last step of the derivation of
  $\ParallelRewriteSwap$
  \begin{case}[\DNFReorderRule{}]
    Let $\DNFRegex \ParallelRewriteSwap \DNFRegex'$ by an application of
    \DNFReorderRule{}.
    This means, for some $\Sequence_1,\ldots,\Sequence_n$, and some $\sigma \in
    \PermutationSetOf{n}$,
    $\DNFRegex = \DNFOf{\Sequence_1 \DNFSep \ldots \DNFSep \Sequence_n}$ and $\DNFRegex' =
    \DNFOf{\Sequence_{\sigma(1)} \DNFSep \ldots \DNFSep \Sequence_{\sigma(n)}}$.
    The DNF regular expression
    $\DNFRegexAlt = \DNFOf{\SequenceAlt_1 \DNFSep \ldots \DNFSep \SequenceAlt_m}$ for some
    $\SequenceAlt_1,\ldots,\SequenceAlt_m$.
    Let $\Identity_m$ be the identity permutation on $m$ elements.
    Define $\sigma' = \ConcatPermutationOf{\sigma}{\Identity_m}$.
    Define $\Sequence_i' =
    \begin{cases*}
      \Sequence_i & if $i \leq n$\\
      \SequenceAlt_{i-n} & otherwise
    \end{cases*}$.
    
    $\DNFOf{\Sequence_1' \DNFSep \ldots \DNFSep \Sequence_{n+m}'} = \DNFRegex \OrDNF
    \DNFRegexAlt$.
    
    $\DNFOf{\Sequence_{\sigma'(1)}' \DNFSep \ldots \DNFSep \Sequence_{\sigma'(n+m)}'} =
    \DNFOf{\Sequence_{\sigma(1)} \DNFSep \ldots \DNFSep \Sequence_{\sigma(n)} \DNFSep 
      \SequenceAlt_1 \DNFSep \ldots \DNFSep \SequenceAlt_m} =
    \DNFRegex' \OrDNF \DNFRegexAlt$.

    Consider the derivation
    \[
      \inferrule*
      {
      }
      {
        \DNFOf{\Sequence_1' \DNFSep \ldots \DNFSep \Sequence_{n+m}'} \ParallelRewriteSwap
        \DNFOf{\Sequence_{\sigma'(1)}' \DNFSep \ldots \DNFSep \Sequence_{\sigma'(n+m)}'}
      }
    \]
    as desired.
  \end{case}
  
  \begin{case}[\ParallelSwapAtomStructuralRewriteRule{}]
    Let $\DNFRegex \ParallelRewriteSwap \DNFRegex'$ by an application of
    \ParallelSwapAtomStructuralRewriteRule{}.
    \[
      \inferrule*
      {
        \DNFRegex = \DNFOf{\Sequence_1 \DNFSep \ldots \DNFSep \Sequence_n}\\
        \forall i. \Sequence_i =
        \SequenceOf{\String_{i,0} \SeqSep \Atom_{i,1} \SeqSep \ldots \SeqSep \Atom_{i,n_i} \SeqSep \String_{i,n_i}}\\
        \forall i,j. \Atom_{i,j} \ParallelRewriteSwapAtom \DNFRegex_{i,j}\\
        \forall i. \DNFRegex_i = \DNFOf{\SequenceOf{\String_{i,0}}} \ConcatDNF \DNFRegex_{i,1}
        \ConcatDNF \ldots \ConcatDNF \DNFRegex_{i,n_i} \ConcatDNF
        \DNFOf{\SequenceOf{\String_{i,n_i}}}
      }
      {
        \DNFRegex \ParallelRewriteSwap \DNFRegex_1 \OrDNF \ldots \OrDNF \DNFRegex_n
      }
    \]

    Let $\DNFRegexAlt = \DNFOf{\SequenceAlt_1 \DNFSep \ldots \DNFSep \SequenceAlt_m}$.
    Let $\SequenceAlt_i =
    \SequenceOf{\StringAlt_{i,0} \SeqSep \AtomAlt_{i,1} \SeqSep \ldots \SeqSep \AtomAlt_{i,m_i} \SeqSep \StringAlt_{i,m_i}}$.
    
    Let $k_i =
    \begin{cases*}
      n_i & if $i \leq n$\\
      m_i & otherwise
    \end{cases*}$
    
    Let $\Sequence_i'' =
    \begin{cases*}
      \Sequence_i & if $i \leq n$\\
      \SequenceAlt_{i-n} & otherwise
    \end{cases*}$.
    Let $\DNFRegex'' = \DNFRegex \OrDNF \DNFRegexAlt =
    \DNFOf{\Sequence_1'' \DNFSep \ldots \DNFSep \Sequence_{n+m}''}$.
    Let $\Atom_{i,j}'' =
    \begin{cases*}
      \Atom_{i,j} & if $i \leq n$\\
      \AtomAlt_{i-n,j} & otherwise
    \end{cases*}$
    Let $\String_{i,j}'' =
    \begin{cases*}
      \String_{i,j} & if $i \leq n$\\
      \StringAlt_{i-n,j} & otherwise
    \end{cases*}$

    Let $\DNFRegex_{i,j}'' =
    \begin{cases*}
      \DNFRegex_{i,j} & if $i \leq n$\\
      \AtomToDNFOf{\AtomAlt_{i-n,j}} & otherwise
    \end{cases*}$
    
    If $i \leq n$, by assumption $\Atom_{i,j}'' = \Atom_{i,j} \ParallelRewriteSwapAtom
    \DNFRegex_{i,j} = \DNFRegex_{i,j}''$.
    If $i > n$, by \ParallelSwapAtomStructuralRewriteRule{},
    $\Atom_{i,j}'' = \AtomAlt_{i-n,j} \ParallelRewriteSwapAtom
    \AtomToDNFOf{\AtomAlt_{i-n,j}} = \DNFRegex_{i,j}''$.
    
    Let $\DNFRegex_i'' =
    \begin{cases*}
      \DNFRegex_i & if $i \leq n$\\
      \DNFOf{\SequenceOf{\StringAlt_{i-n,0}}} \ConcatDNF \AtomAlt_{i-n,1} \SeqSep \ldots \SeqSep 
      \ConcatDNF \AtomAlt_{i-n,m_i} \ConcatDNF \StringAlt_{i-n,m_i} & otherwise
    \end{cases*}$

    For $i > n$, $\DNFRegex_i'' = \DNFOf{\SequenceOf{\StringAlt_{i-n,0}}}
    \ConcatDNF \AtomAlt_{i-n,1} \SeqSep \ldots \SeqSep 
    \ConcatDNF \AtomAlt_{i-n,k_i} \ConcatDNF \StringAlt_{i-n,k_i} =
    \DNFOf{\SequenceAlt_i}$ through application of $\ConcatDNF$ on many
    singleton DNF regular expressions.

    $\DNFRegex_{n+1}'' \OrDNF \ldots \OrDNF \DNFRegex_{n+m}'' =
    \DNFOf{\SequenceAlt_1} \OrDNF \ldots \DNFOf{\SequenceAlt_m} =
    \DNFOf{\SequenceAlt_1 \DNFSep \ldots \DNFSep \SequenceAlt_m}$ through repeated application
    of $\OrDNF$ to singleton DNFs.

    As $\DNFRegex_1'' \OrDNF \ldots \OrDNF \DNFRegex_n''
    = \DNFRegex_1 \OrDNF \ldots \OrDNF \DNFRegex_n
    = \DNFRegex'$,
    and $\DNFRegex_{n+1}'' \OrDNF \ldots \OrDNF \DNFRegex_{n+m}''
    = \DNFRegexAlt$, we get
    $\DNFRegex_1'' \OrDNF \ldots \OrDNF \DNFRegex_{n+m}'' = \DNFRegex' \OrDNF
    \DNFRegexAlt$

    Consider the derivation
    \[
      \inferrule*
      {
        \DNFRegex'' = \DNFOf{\Sequence_1'' \DNFSep \ldots \DNFSep \Sequence_{n+m}''}\\
        \forall i. \Sequence_i'' =
        \SequenceOf{\String_{i,0}'' \SeqSep \Atom_{i,1}'' \SeqSep \ldots \SeqSep \Atom_{i,k_i}'' \SeqSep \String_{i,k_i}''}\\
        \forall i,j. \Atom_{i,j}'' \ParallelRewriteSwapAtom \DNFRegex_{i,j}''\\
        \forall i. \DNFRegex_i'' = \DNFOf{\SequenceOf{\String_{i,0}''}} \ConcatDNF \DNFRegex_{i,1}''
        \ConcatDNF \ldots \ConcatDNF \DNFRegex_{i,n_i}'' \ConcatDNF
        \DNFOf{\SequenceOf{\String_{i,k_i}''}}
      }
      {
        \DNFRegex'' \ParallelRewriteSwap
        \DNFRegex_1'' \OrDNF \ldots \OrDNF \DNFRegex_{n+m}''
      }
    \]
    as desired.
  \end{case}
\end{proof}

\begin{lemma}[Propagation of $\ParallelRewriteSwap{}$ through $\OrDNF{}$ on
  the right]
  If $\DNFRegex \ParallelRewriteSwap \DNFRegex'$, then\\
  $\DNFRegex \ConcatDNF \DNFRegexAlt \ParallelRewriteSwap \DNFRegex'
  \ConcatDNF \DNFRegexAlt$
\end{lemma}
\begin{proof}
  Proceeds as Lemma~\ref{lem:prop_parallel_swap_or_left}, but on the right.
\end{proof}

\begin{lemma}[Propagation of $\EquivalenceOf{\ParallelRewriteSwap{}}$ through
  $\OrDNF{}$ on the left]
  \label{lem:prop-eq-swap-or-left}
  If $\DNFRegex \EquivalenceOf{\ParallelRewriteSwap} \DNFRegex'$, then
  $\DNFRegex \OrDNF \DNFRegexAlt \EquivalenceOf{\ParallelRewriteSwap} \DNFRegex'
  \OrDNF \DNFRegexAlt$
\end{lemma}
\begin{proof}
  By induction on the last step of the derivation of
  $\DNFRegex \EquivalenceOf{\ParallelRewriteSwap} \DNFRegex'$.
  \begin{case}[\ReflexivityRule{}]
    If $\DNFRegex \EquivalenceOf{\ParallelRewriteSwap} \DNFRegex'$ through an
    application of \ReflexivityRule{}, then $\DNFRegex' = \DNFRegex$.
    So, through reflexivity, $\DNFRegex \OrDNF \DNFRegexAlt
    \EquivalenceOf{\ParallelRewriteSwap} \DNFRegex \OrDNF \DNFRegexAlt$
  \end{case}
  \begin{case}[\BaseRule{}]
    If $\DNFRegex \EquivalenceOf{\ParallelRewriteSwap} \DNFRegex'$ through an
    application of \ReflexivityRule{}, then
    $\DNFRegex' \ParallelRewriteSwap \DNFRegex$.
    From Lemma~\ref{lem:prop_parallel_swap_or_left}
    $\DNFRegex \OrDNF \DNFRegexAlt
    \ParallelRewriteSwap \DNFRegex' \OrDNF \DNFRegexAlt$, so
    $\DNFRegex \OrDNF \DNFRegexAlt
    \EquivalenceOf{\ParallelRewriteSwap} \DNFRegex' \OrDNF \DNFRegexAlt$.
  \end{case}
  \begin{case}[\TransitivityRule{}]
    If $\DNFRegex \EquivalenceOf{\ParallelRewriteSwap} \DNFRegex'$ through an
    application of \TransitivityRule{}, then there exists a $\DNFRegex''$ such
    that
    $\DNFRegex \EquivalenceOf{\ParallelRewriteSwap} \DNFRegex''$ and
    $\DNFRegex'' \EquivalenceOf{\ParallelRewriteSwap} \DNFRegex'$.
    By IH, $\DNFRegex \OrDNF \DNFRegexAlt \EquivalenceOf{\ParallelRewriteSwap}
    \DNFRegex'' \OrDNF \DNFRegexAlt$ and
    $\DNFRegex'' \OrDNF \DNFRegexAlt \EquivalenceOf{\ParallelRewriteSwap}
    \DNFRegex' \OrDNF \DNFRegexAlt$.
    
    This gives us the derivation
    \[
      \inferrule*
      {
        \DNFRegex \OrDNF \DNFRegexAlt \EquivalenceOf{\ParallelRewriteSwap}
        \DNFRegex'' \OrDNF \DNFRegexAlt\\
        \DNFRegex'' \OrDNF \DNFRegexAlt \EquivalenceOf{\ParallelRewriteSwap}
        \DNFRegex' \OrDNF \DNFRegexAlt
      }
      {
        \DNFRegex \OrDNF \DNFRegexAlt \EquivalenceOf{\ParallelRewriteSwap}
        \DNFRegex' \OrDNF \DNFRegexAlt
      }
    \]
  \end{case}
\end{proof}

\begin{lemma}[Propagation of $\EquivalenceOf{\ParallelRewriteSwap{}}$ through
  $\OrDNF{}$ on the right]
  \label{lem:prop-eq-swap-or-right}
  If $\DNFRegex \EquivalenceOf{\ParallelRewriteSwap} \DNFRegex'$, then
  $\DNFRegex \OrDNF \DNFRegexAlt \EquivalenceOf{\ParallelRewriteSwap} \DNFRegex'
  \OrDNF \DNFRegexAlt$
\end{lemma}
\begin{proof}
  Proceeds as Lemma~\ref{lem:prop-eq-swap-or-left}, but on the right.
\end{proof}

\begin{lemma}[Propagation of $\EquivalenceOf{\ParallelRewriteSwap{}}$ through
  $\OrDNF{}$]
  \label{lem:prop-eq-swap-or}
  If $\DNFRegex \EquivalenceOf{\ParallelRewriteSwap} \DNFRegex'$ and
  $\DNFRegexAlt \EquivalenceOf{\ParallelRewriteSwap} \DNFRegexAlt'$, then
  $\DNFRegex \OrDNF \DNFRegexAlt \EquivalenceOf{\ParallelRewriteSwap}
  \DNFRegex' \OrDNF \DNFRegexAlt'$.
\end{lemma}
\begin{proof}
  By Lemma~\ref{lem:prop-eq-swap-or-left}, $\DNFRegex \OrDNF \DNFRegexAlt
  \EquivalenceOf{\ParallelRewriteSwap} \DNFRegex' \OrDNF \DNFRegexAlt$.
  By Lemma~\ref{lem:prop-eq-swap-or-right}, $\DNFRegex' \OrDNF \DNFRegexAlt
  \EquivalenceOf{\ParallelRewriteSwap} \DNFRegex' \OrDNF \DNFRegexAlt'$.
  Consider the derivation
  \[
    \inferrule*
    {
      \DNFRegex \OrDNF \DNFRegexAlt
      \EquivalenceOf{\ParallelRewriteSwap} \DNFRegex' \OrDNF \DNFRegexAlt\\
      \DNFRegex' \OrDNF \DNFRegexAlt
      \EquivalenceOf{\ParallelRewriteSwap} \DNFRegex' \OrDNF \DNFRegexAlt'
    }
    {
      \DNFRegex \OrDNF \DNFRegexAlt
      \EquivalenceOf{\ParallelRewriteSwap}
      \DNFRegex' \OrDNF \DNFRegexAlt'
    }
  \]
\end{proof}

\begin{lemma}[Propagation of $\ParallelRewriteSwap{}$ through $\ConcatDNF{}$ on
  the left]
  \label{lem:prop_parallel_swap_concat_left}
  If $\DNFRegex \ParallelRewriteSwap \DNFRegex'$, then
  $\DNFRegex \ConcatDNF \DNFRegexAlt \ParallelRewriteSwap \DNFRegex'
  \ConcatDNF \DNFRegexAlt$
\end{lemma}
\begin{proof}
  By induction on the derivation of $\ParallelRewriteSwap{}$.
  \begin{case}[\DNFReorderRule{}]
    Let $\DNFRegex \ParallelRewriteSwap \DNFRegex'$ by an application of
    \DNFReorderRule{}.
    This means, for some $\Sequence_1,\ldots,\Sequence_n$, and some $\sigma \in
    \PermutationSetOf{n}$,
    $\DNFRegex = \DNFOf{\Sequence_1 \DNFSep \ldots \DNFSep \Sequence_n}$ and $\DNFRegex' =
    \DNFOf{\Sequence_{\sigma(1)} \DNFSep \ldots \DNFSep \Sequence_{\sigma(n)}}$.
    The DNF regular expression
    $\DNFRegexAlt = \DNFOf{\SequenceAlt_1 \DNFSep \ldots \DNFSep \SequenceAlt_m}$ for some
    $\SequenceAlt_1,\ldots,\SequenceAlt_m$.
    Let $\Identity_m$ be the identity permutation on $m$ elements.
    Define $\sigma' = \DistributePermutationOf{\sigma}{\Identity_m}$.
    Define $\Sequence_{i,j} = \Sequence_i \ConcatSequence \SequenceAlt_j$.

    By definition of $\ConcatDNF$,
    $\DNFOf{\Sequence_{1,1} \DNFSep \ldots \DNFSep \Sequence_{n,m}} =
    \DNFOf{\Sequence_1 \ConcatDNF \SequenceAlt_1 \DNFSep  \ldots \DNFSep 
      \Sequence_n \ConcatDNF \SequenceAlt_m} =
    \DNFRegex \ConcatDNF \DNFRegexAlt$.
    
    By the definition of $\ConcatDNF$ and $\DistributePermutation$,
    $\DNFOf{\Sequence_{\sigma'(1,1)} \DNFSep \ldots \DNFSep \Sequence_{\sigma'(n,m)}} =
    \DNFOf{\Sequence_{(\sigma(1),1)} \DNFSep \ldots \DNFSep \Sequence_{(\sigma(n),m)}} =
    \DNFOf{\Sequence_{\sigma(1)} \ConcatDNF \SequenceAlt_{1}  \DNFSep  \ldots \DNFSep 
      \Sequence_{\sigma(n)} \ConcatDNF \SequenceAlt_m} =
    \DNFRegex' \ConcatDNF \DNFRegexAlt$.
  \end{case}

  \begin{case}[\ParallelSwapDNFStructuralRewriteRule{}]
    Let $\DNFRegex \ParallelRewriteSwap \DNFRegex'$ by an application of
    \ParallelSwapDNFStructuralRewriteRule{}.
    \[
      \inferrule*
      {
        \DNFRegex = \DNFOf{\Sequence_1 \DNFSep \ldots \DNFSep \Sequence_n}\\
        \forall i. \Sequence_i =
        \SequenceOf{\String_{i,0} \SeqSep \Atom_{i,1} \SeqSep \ldots \SeqSep \Atom_{i,n_i} \SeqSep \String_{i,n_i}}\\
        \forall i,j. \Atom_{i,j} \ParallelRewriteSwapAtom \DNFRegex_{i,j}\\
        \forall i. \DNFRegex_i = \DNFOf{\SequenceOf{\String_{i,0}}} \ConcatDNF \DNFRegex_{i,1}
        \ConcatDNF \ldots \ConcatDNF \DNFRegex_{i,n_i} \ConcatDNF
        \DNFOf{\SequenceOf{\String_{i,n_i}}}
      }
      {
        \DNFRegex \ParallelRewriteSwap \DNFRegex_1 \OrDNF \ldots \OrDNF \DNFRegex_n
      }
    \]

    Let $\DNFRegexAlt = \DNFOf{\SequenceAlt_1 \DNFSep \ldots \DNFSep \SequenceAlt_m}$.
    Let $\SequenceAlt_i =
    \SequenceOf{\StringAlt_{i,0} \SeqSep \AtomAlt_{i,1} \SeqSep \ldots \SeqSep \AtomAlt_{i,m_i} \SeqSep \StringAlt_{i,m_i}}$.
    Let $\Sequence_{i,j}'' = \Sequence_i \ConcatSequence \Sequence_j$
    Let $\DNFRegex'' = \DNFRegex \ConcatDNF \DNFRegexAlt =
    \DNFOf{\Sequence_{1,1}'' \DNFSep \ldots \DNFSep \Sequence_{n,m}''}$.
    Let $\Atom_{i,j,k}'' =
    \begin{cases*}
      \Atom_{i,j,k} & if $k \leq n_i$\\
      \AtomAlt_{i,j,k-n_i} & otherwise
    \end{cases*}$
    Let $\String_{i,j,k}'' =
    \begin{cases*}
      \String_{i,k} & if $i < n_i$\\
      \String_{i,n_i} \Concat \StringAlt_{j,0} & if $i = n_i$\\
      \StringAlt_{j,k-n_i} & otherwise
    \end{cases*}$

    Let $\DNFRegex_{i,j,k}'' =
    \begin{cases*}
      \DNFRegex_{i,k} & if $i \leq n_i$\\
      \AtomToDNFOf{\AtomAlt_{i,j,k-n_i}} & otherwise
    \end{cases*}$
    
    If $k \leq n_i$, by assumption $\Atom_{i,j,k}'' = \Atom_{i,j} \ParallelRewriteSwapAtom
    \DNFRegex_{i,j,k} = \DNFRegex_{i,j,k}''$.
    If $k > n_i$, by \ParallelSwapAtomStructuralRewriteRule{},
    $\Atom_{i,j,k}'' = \AtomAlt_{j,k-n_i} \ParallelRewriteSwapAtom
    \AtomToDNFOf{\AtomAlt_{j,k-n_i}} = \DNFRegex_{i,j,k}''$.
    
    Let $\DNFRegex_{i,j}'' = \DNFRegex_i \ConcatDNF
    \DNFOf{\SequenceOf{\StringAlt_{i-n,0}}} \ConcatDNF \AtomAlt_{i-n,1} \SeqSep \ldots \SeqSep 
    \ConcatDNF \AtomAlt_{i-n,n_i} \ConcatDNF \StringAlt_{i-n,n_i}$
    
    Through repeated application of $\ConcatDNF$ on singletons,
    $\DNFRegex_{i,j}'' = \DNFRegex_i \ConcatDNF \DNFOf{\SequenceAlt_j}$.

    This means $\DNFRegex_{0,0} \OrDNF \ldots \OrDNF \DNFRegex_{n,m} =
    (\DNFRegex_1 \OrDNF \ldots \OrDNF \DNFRegex_n) \ConcatDNF
    \DNFRegexAlt = \DNFRegex' \ConcatDNF \DNFRegexAlt$.

    Consider the derivation
    \[
      \inferrule*
      {
        \DNFRegex'' = \DNFOf{\Sequence_{1,1}'' \DNFSep \ldots \DNFSep \Sequence_{n,m}''}\\
        \forall i,j. \Sequence_{i,j}'' =
        \SequenceOf{\String_{i,j,0}'' \SeqSep \Atom_{i,j,1}'' \SeqSep \ldots \SeqSep \Atom_{i,j,n_i+m_j}'' \SeqSep \String_{i,j,n_i+m_j}''}\\
        \forall i,j,k. \Atom_{i,j,k}'' \ParallelRewriteSwapAtom \DNFRegex_{i,j,k}''\\
        \forall i,j. \DNFRegex_{i,j}'' = \DNFOf{\SequenceOf{\String_{i,j,0}''}} \ConcatDNF \DNFRegex_{i,j,1}''
        \ConcatDNF \ldots \ConcatDNF \DNFRegex_{i,j,n_i+m_i}'' \ConcatDNF
        \DNFOf{\SequenceOf{\String_{i,j,n_i+m_i}''}}
      }
      {
        \DNFRegex'' \ParallelRewriteSwap
        \DNFRegex_{1,1}'' \OrDNF \ldots \OrDNF \DNFRegex_{n,m}''
      }
    \]
    as desired.
  \end{case}
\end{proof}

\begin{lemma}[Propagation of $\ParallelRewriteSwap{}$ through $\ConcatDNF{}$ on
  the right]
  If $\DNFRegexAlt \ParallelRewriteSwap \DNFRegexAlt'$, then
  $\DNFRegex \ConcatDNF \DNFRegexAlt \ParallelRewriteSwap
  \DNFRegex \ConcatDNF \DNFRegexAlt'$
\end{lemma}
\begin{proof}
  Proceeds as Lemma~\ref{lem:prop_parallel_swap_or_left}, but on the right.
\end{proof}

\begin{lemma}[Propagation of $\EquivalenceOf{\ParallelRewriteSwap{}}$ through
  $\ConcatDNF{}$ on the left]
  \label{lem:prop-eq-swap-concat-left}
  If $\DNFRegex \EquivalenceOf{\ParallelRewriteSwap} \DNFRegex'$, then
  $\DNFRegex \ConcatDNF \DNFRegexAlt \EquivalenceOf{\ParallelRewriteSwap} \DNFRegex'
  \ConcatDNF \DNFRegexAlt$
\end{lemma}
\begin{proof}
  By induction on the last step of the derivation of
  $\DNFRegex \EquivalenceOf{\ParallelRewriteSwap} \DNFRegex'$.
  \begin{case}[\ReflexivityRule{}]
    If $\DNFRegex \EquivalenceOf{\ParallelRewriteSwap} \DNFRegex'$ through an
    application of \ReflexivityRule{}, then $\DNFRegex' = \DNFRegex$.
    So, through reflexivity, $\DNFRegex \ConcatDNF \DNFRegexAlt
    \EquivalenceOf{\ParallelRewriteSwap} \DNFRegex \ConcatDNF \DNFRegexAlt$
  \end{case}
  \begin{case}[\BaseRule{}]
    If $\DNFRegex \EquivalenceOf{\ParallelRewriteSwap} \DNFRegex'$ through an
    application of \ReflexivityRule{}, then
    $\DNFRegex' \ParallelRewriteSwap \DNFRegex$.
    From Lemma~\ref{lem:prop_parallel_swap_concat_left}
    $\DNFRegex \ConcatDNF \DNFRegexAlt
    \ParallelRewriteSwap \DNFRegex' \ConcatDNF \DNFRegexAlt$, so
    $\DNFRegex \ConcatDNF \DNFRegexAlt
    \EquivalenceOf{\ParallelRewriteSwap} \DNFRegex' \ConcatDNF \DNFRegexAlt$.
  \end{case}
  \begin{case}[\TransitivityRule{}]
    If $\DNFRegex \EquivalenceOf{\ParallelRewriteSwap} \DNFRegex'$ through an
    application of \TransitivityRule{}, then there exists a $\DNFRegex''$ such
    that
    $\DNFRegex \EquivalenceOf{\ParallelRewriteSwap} \DNFRegex''$ and
    $\DNFRegex'' \EquivalenceOf{\ParallelRewriteSwap} \DNFRegex'$.
    By IH, $\DNFRegex \ConcatDNF \DNFRegexAlt \EquivalenceOf{\ParallelRewriteSwap}
    \DNFRegex'' \ConcatDNF \DNFRegexAlt$ and
    $\DNFRegex'' \ConcatDNF \DNFRegexAlt \EquivalenceOf{\ParallelRewriteSwap}
    \DNFRegex' \ConcatDNF \DNFRegexAlt$.
    
    This gives us the derivation
    \[
      \inferrule*
      {
        \DNFRegex \ConcatDNF \DNFRegexAlt \EquivalenceOf{\ParallelRewriteSwap}
        \DNFRegex'' \ConcatDNF \DNFRegexAlt\\
        \DNFRegex'' \ConcatDNF \DNFRegexAlt \EquivalenceOf{\ParallelRewriteSwap}
        \DNFRegex' \ConcatDNF \DNFRegexAlt
      }
      {
        \DNFRegex \ConcatDNF \DNFRegexAlt \EquivalenceOf{\ParallelRewriteSwap}
        \DNFRegex' \ConcatDNF \DNFRegexAlt
      }
    \]
  \end{case}
\end{proof}

\begin{lemma}[Propagation of $\EquivalenceOf{\ParallelRewriteSwap{}}$ through
  $\ConcatDNF{}$ on the right]
  \label{lem:prop-eq-swap-concat-right}
  If $\DNFRegex \EquivalenceOf{\ParallelRewriteSwap} \DNFRegex'$, then
  $\DNFRegex \ConcatDNF \DNFRegexAlt \EquivalenceOf{\ParallelRewriteSwap} \DNFRegex'
  \ConcatDNF \DNFRegexAlt$
\end{lemma}
\begin{proof}
  Proceeds as Lemma~\ref{lem:prop-eq-swap-concat-left}, but on the right.
\end{proof}

\begin{lemma}[Propagation of $\EquivalenceOf{\ParallelRewriteSwap{}}$ through
  $\ConcatDNF{}$]
  \label{lem:prop-eq-swap-concat}
  If $\DNFRegex \EquivalenceOf{\ParallelRewriteSwap} \DNFRegex'$ and
  $\DNFRegexAlt \EquivalenceOf{\ParallelRewriteSwap} \DNFRegexAlt'$, then
  $\DNFRegex \ConcatDNF \DNFRegexAlt \EquivalenceOf{\ParallelRewriteSwap}
  \DNFRegex' \ConcatDNF \DNFRegexAlt'$.
\end{lemma}
\begin{proof}
  By Lemma~\ref{lem:prop-eq-swap-concat-left}, $\DNFRegex \ConcatDNF \DNFRegexAlt
  \EquivalenceOf{\ParallelRewriteSwap} \DNFRegex' \ConcatDNF \DNFRegexAlt$.
  By Lemma~\ref{lem:prop-eq-swap-concat-right}, $\DNFRegex' \ConcatDNF \DNFRegexAlt
  \EquivalenceOf{\ParallelRewriteSwap} \DNFRegex' \ConcatDNF \DNFRegexAlt'$.
  Consider the derivation
  \[
    \inferrule*
    {
      \DNFRegex \OrDNF \DNFRegexAlt
      \EquivalenceOf{\ParallelRewriteSwap} \DNFRegex' \OrDNF \DNFRegexAlt\\
      \DNFRegex' \OrDNF \DNFRegexAlt
      \EquivalenceOf{\ParallelRewriteSwap} \DNFRegex' \OrDNF \DNFRegexAlt'
    }
    {
      \DNFRegex \OrDNF \DNFRegexAlt
      \EquivalenceOf{\ParallelRewriteSwap}
      \DNFRegex' \OrDNF \DNFRegexAlt'
    }
  \]
\end{proof}

\begin{lemma}[Propagation of $\EquivalenceOf{\ParallelRewriteSwap{}}$ through
  $\Star{}$]
  \label{lem:prop-eq-swap-star}
  If $\DNFRegex \EquivalenceOf{\ParallelRewriteSwap} \DNFRegexAlt$, then
  $\AtomToDNFOf{\StarOf{\DNFRegex}} \EquivalenceOf{\ParallelRewriteSwap}
  \AtomToDNFOf{\StarOf{\DNFRegexAlt}}$
\end{lemma}
\begin{proof}
  By induction on the derivation of $\EquivalenceOf{\ParallelRewriteSwap{}}$.
  \begin{case}[\ReflexivityRule{}]
    Let $\DNFRegex \EquivalenceOf{\ParallelRewriteSwap} \DNFRegexAlt$, with the
    last step of the derivation being \ReflexivityRule{}.  This means
    $\DNFRegexAlt = \DNFRegex$.  Consider the derivation
    \[
      \inferrule*
      {
      }
      {
        \AtomToDNFOf{\StarOf{\DNFRegex}} \EquivalenceOf{\ParallelRewriteSwap}
        \AtomToDNFOf{\StarOf{\DNFRegex}}
      }
    \]
  \end{case}

  \begin{case}[\BaseRule{}]
    Let $\DNFRegex \EquivalenceOf{\ParallelRewriteSwap} \DNFRegexAlt$, with the
    last step of the derivation being \BaseRule{}.  That means
    $\DNFRegex \ParallelRewriteSwap \DNFRegexAlt$.  Consider the derivation
    \[
      \inferrule*
      {
        \inferrule*
        {
          \inferrule*
          {
            \DNFRegex \ParallelRewriteSwap \DNFRegexAlt
          }
          {
            \StarOf{\DNFRegex} \ParallelRewriteSwapAtom
            \AtomToDNFOf{\StarOf{\DNFRegexAlt}}
          }
        }
        {
          \AtomToDNFOf{\StarOf{\DNFRegex}} \ParallelRewriteSwap
          \AtomToDNFOf{\StarOf{\DNFRegexAlt}}
        }
      }
      {
        \AtomToDNFOf{\StarOf{\DNFRegex}} \EquivalenceOf{\ParallelRewriteSwap}
        \AtomToDNFOf{\StarOf{\DNFRegexAlt}}
      }
    \]

    \begin{case}[\SymmetryRule{}]
      Let $\DNFRegex \EquivalenceOf{\ParallelRewriteSwap} \DNFRegexAlt$, with
      the last step of the derivation being \SymmetryRule{}.  That means
      $\DNFRegexAlt \ParallelRewriteSwap \DNFRegex$.  Consider the derivation
      \[
        \inferrule*
        {
          \inferrule*
          {
            \inferrule*
            {
              \DNFRegexAlt \ParallelRewriteSwap \DNFRegex
            }
            {
              \StarOf{\DNFRegexAlt} \ParallelRewriteSwapAtom
              \AtomToDNFOf{\StarOf{\DNFRegex}}
            }
          }
          {
            \AtomToDNFOf{\StarOf{\DNFRegexAlt}} \ParallelRewriteSwap
            \AtomToDNFOf{\StarOf{\DNFRegex}}
          }
        }
        {
          \AtomToDNFOf{\StarOf{\DNFRegex}} \EquivalenceOf{\ParallelRewriteSwap}
          \AtomToDNFOf{\StarOf{\DNFRegexAlt}}
        }
      \]
    \end{case}
    
    \begin{case}[\TransitivityRule{}]
      Let $\DNFRegex \EquivalenceOf{\ParallelRewriteSwap} \DNFRegexAlt$, with
      the last step of the derivation being \TransitivityRule{}.  That means
      that, for some $\DNFRegex'$, the last step of the derivation is
      \[
        \inferrule*
        {
          \DNFRegex \EquivalenceOf{\ParallelRewriteSwap} \DNFRegex'\\
          \DNFRegex' \EquivalenceOf{\ParallelRewriteSwap} \DNFRegexAlt
        }
        {
          \DNFRegex \EquivalenceOf{\ParallelRewriteSwap} \DNFRegexAlt
        }
      \]

      By induction assumption,
      $\AtomToDNFOf{\StarOf{\DNFRegex}} \EquivalenceOf{\ParallelRewriteSwap}
      \AtomToDNFOf{\StarOf{\DNFRegex'}}$ and
      $\AtomToDNFOf{\StarOf{\DNFRegex'}} \EquivalenceOf{\ParallelRewriteSwap}
      \AtomToDNFOf{\StarOf{\DNFRegexAlt}}$.  Consider the derivation
      \[
        \inferrule*
        {
          \AtomToDNFOf{\StarOf{\DNFRegex}} \EquivalenceOf{\ParallelRewriteSwap}
          \AtomToDNFOf{\StarOf{\DNFRegex'}}\\
          \AtomToDNFOf{\StarOf{DNFRegex'}} \EquivalenceOf{\ParallelRewriteSwap}
          \AtomToDNFOf{\StarOf{\DNFRegexAlt}}
        }
        {
          \AtomToDNFOf{\StarOf{\DNFRegex}} \EquivalenceOf{\ParallelRewriteSwap}
          \AtomToDNFOf{\StarOf{\DNFRegexAlt}}
        }
      \]
    \end{case}
  \end{case}
\end{proof}

\begin{lemma}[Expressibility of $\SSREquiv$ in
  $\EquivalenceOf{\ParallelRewriteSwap}$]
  \label{lem:express-equiv-in-equiv-swap}
  If $\Regex \SSREquiv \RegexAlt$, then
  $\ToDNFRegexOf{\Regex} \EquivalenceOf{\ParallelRewriteSwap}
  \ToDNFRegexOf{\RegexAlt}$.
\end{lemma}
\begin{proof}
  Assume $\Regex \SSREquiv \RegexAlt$.
  Prove by induction on the deduction of $\SSREquiv$
  \begin{case}[Structural Equality Rule]
    Let $\Regex \SSREquiv \RegexAlt$, and the last step of the
    deduction is an application of structural equality rule.
    That means that $\RegexAlt
    = \Regex$.  Through reflexivity, $\ToDNFRegexOf{\Regex}
    \EquivalenceOf{\ParallelRewriteSwap} \ToDNFRegexOf{\Regex} =
    \ToDNFRegexOf{\RegexAlt}$.
  \end{case}

  \begin{case}[\OrIdentityRule{}]
    Let $\Regex \SSREquiv \RegexAlt$, and the last step of the
    deduction is an application of \OrIdentityRule{}.  Without loss of
    generality, from symmetry, $\RegexAlt = \Regex \Or \emptyset$.
  
    $\ToDNFRegexOf{\Regex \Or \emptyset} = \ToDNFRegexOf{\Regex} \OrDNF
    \ToDNFRegexOf{\emptyset} = \ToDNFRegexOf{\Regex} \OrDNF \DNFOf{} =
    \ToDNFRegexOf{\Regex}$.  Through reflexivity, $\ToDNFRegexOf{\Regex}
    \EquivalenceOf{\ParallelRewriteSwap} \ToDNFRegexOf{\Regex} =
    \ToDNFRegexOf{\Regex \Or \emptyset}$
  \end{case}
  
  \begin{case}[\EmptyProjectionRightRule{}]
    Let $\Regex \SSREquiv \RegexAlt$, and the last step of the
    deduction is an application of \EmptyProjectionRightRule{}.  Without
    loss of generality, from symmetry, $\Regex = \Regex' \Concat \emptyset$, and
    $\RegexAlt = \emptyset$.

    $\ToDNFRegexOf{\Regex' \Concat \emptyset} = \ToDNFRegexOf{\Regex} \ConcatDNF
    \ToDNFRegexOf{\emptyset} = \ToDNFRegexOf{\Regex} \ConcatDNF \DNFOf{} =
    \DNFOf{}$.  Through reflexivity, $\ToDNFRegexOf{\Regex} =
    \ToDNFRegexOf{\emptyset} \SSREquiv \ToDNFRegexOf{\emptyset} =
    \ToDNFRegexOf{\RegexAlt}$.
  \end{case}
  
  \begin{case}[\EmptyProjectionLeftRule{}]
    Let $\Regex \SSREquiv \RegexAlt$, and the last step of the
    deduction is an application of \EmptyProjectionRightRule{}.  Without
    loss of generality, from symmetry, $\Regex = \emptyset \Concat \Regex'$, and
    $\RegexAlt = \emptyset$.

    $\ToDNFRegexOf{\emptyset \Concat \Regex'} = \ToDNFRegexOf{\emptyset} \ConcatDNF
    \ToDNFRegexOf{\Regex} = \DNFOf{} \ConcatDNF \ToDNFRegexOf{\Regex} =
    \DNFOf{}$.  Through reflexivity, $\ToDNFRegexOf{\Regex} =
    \ToDNFRegexOf{\emptyset} \EquivalenceOf{\ParallelRewriteSwap}
    \ToDNFRegexOf{\emptyset} = \ToDNFRegexOf{\RegexAlt}$.
  \end{case}

  \begin{case}[\ConcatAssocRule{}]
    Let $\Regex \SSREquiv \RegexAlt$, and the last step of the
    deduction is an application of \ConcatAssocRule{}.  Without loss of
    generality, from symmetry,
    $\Regex = \Regex_1 \Concat (\Regex_2 \Concat \Regex_3)$, and
    $\RegexAlt = (\Regex_1 \Concat \Regex_2) \Concat \Regex_3$.

    $\ToDNFRegexOf{(\Regex_1 \Concat (\Regex_2 \Concat \Regex_3))} =
    \ToDNFRegexOf{\Regex_1} \ConcatDNF (\ToDNFRegexOf{\Regex_2} \ConcatDNF
    \ToDNFRegexOf{\Regex_3}) =
    (\ToDNFRegexOf{\Regex_1} \ConcatDNF \ToDNFRegexOf{\Regex_2}) \ConcatDNF
    \ToDNFRegexOf{\Regex_3} =
    \ToDNFRegexOf{(\Regex_1 \Concat \Regex_2) \Concat \Regex_3}$.  Through
    reflexivity, $\ToDNFRegexOf{\Regex} = \ToDNFRegexOf{\Regex_1} \ConcatDNF
    (\ToDNFRegexOf{\Regex_2} \ConcatDNF \ToDNFRegexOf{\Regex_3})
    \EquivalenceOf{\ParallelRewriteSwap} \ToDNFRegexOf{\Regex_1} \ConcatDNF
    (\ToDNFRegexOf{\Regex_2} \ConcatDNF \ToDNFRegexOf{\Regex_3}) =
    \ToDNFRegexOf{\RegexAlt}$.
  \end{case}
  
  \begin{case}[\OrAssociativityRule{}]
    Let $\Regex \SSREquiv \RegexAlt$, and the last step of the
    deduction is an application of \OrAssociativityRule{}.  Without loss of
    generality, from symmetry,
    $\Regex = \Regex_1 \Or (\Regex_2 \Or \Regex_3)$, and
    $\RegexAlt = (\Regex_1 \Or \Regex_2) \Or \Regex_3$.

    $\ToDNFRegexOf{(\Regex_1 \Or (\Regex_2 \Concat \Regex_3))} =
    \ToDNFRegexOf{\Regex_1} \OrDNF (\ToDNFRegexOf{\Regex_2} \OrDNF
    \ToDNFRegexOf{\Regex_3}) =
    (\ToDNFRegexOf{\Regex_1} \OrDNF \ToDNFRegexOf{\Regex_2}) \OrDNF
    \ToDNFRegexOf{\Regex_3} =
    \ToDNFRegexOf{(\Regex_1 \Or \Regex_2) \Or \Regex_3}$.  Through
    reflexivity, $\ToDNFRegexOf{\Regex} = \ToDNFRegexOf{\Regex_1} \OrDNF
    (\ToDNFRegexOf{\Regex_2} \OrDNF \ToDNFRegexOf{\Regex_3})
    \EquivalenceOf{\ParallelRewriteSwap} \ToDNFRegexOf{\Regex_1} \OrDNF
    (\ToDNFRegexOf{\Regex_2} \OrDNF \ToDNFRegexOf{\Regex_3}) =
    \ToDNFRegexOf{\RegexAlt}$.
  \end{case}

  \begin{case}[\OrCommutativityRule{}]
    Let $\Regex \SSREquiv \RegexAlt$, and the last step of the
    deduction is an application of \OrCommutativityRule{}.
    $\Regex = \Regex_1 \Or \Regex_2$, and
    $\RegexAlt = \Regex_2 \Or \Regex_1$.

    Let $\ToDNFRegexOf{\Regex_1} = \DNFOf{\Sequence_1 \DNFSep \ldots \DNFSep \Sequence_n}$ and
    $\ToDNFRegexOf{\Regex_2} = \DNFOf{\SequenceAlt_1 \DNFSep \ldots \DNFSep \Sequence_m}$.
    $\OrDNFOf
    {\DNFOf{\Sequence_1 \DNFSep \ldots \DNFSep \Sequence_n}}
    {\DNFOf{\SequenceAlt_1 \DNFSep \ldots \DNFSep \Sequence_m}} =
    \DNFOf{\Sequence_1 \DNFSep \ldots \DNFSep \Sequence_n \DNFSep \SequenceAlt_1
      \DNFSep \ldots \DNFSep \SequenceAlt_m}$.
    
    $\OrDNFOf
    {\DNFOf{\SequenceAlt_1 \DNFSep \ldots \DNFSep \Sequence_m}}
    {\DNFOf{\Sequence_1 \DNFSep \ldots \DNFSep \Sequence_n}} =
    \DNFOf{\SequenceAlt_1 \DNFSep \ldots \DNFSep \SequenceAlt_n \DNFSep \Sequence_1 \DNFSep \ldots \DNFSep \Sequence_m}$.

    Let $\Sequence_i' =
    \begin{cases*}
      \Sequence_i & if $i \in \RangeIncInc{1}{n}$\\
      \SequenceAlt_{i-n} & if $i \in \RangeIncInc{n+1}{m}$
    \end{cases*}$
  
    Consider the deduction
    \[
      \inferrule*
      {
        \SwapPermutationOf{\Identity_n}{\Identity_m} \in \PermutationSetOf{n+m}
      }
      {
        \DNFOf{\Sequence_1' \DNFSep \ldots \DNFSep \Sequence_{n+m}'}
        \ParallelRewriteSwap
        \DNFOf{\Sequence_{\SwapPermutationOf{\Identity_n}{\Identity_m}(1)}' \DNFSep \ldots \DNFSep 
        \Sequence_{\SwapPermutationOf{\Identity_n}{\Identity_m}(n+m)}'}
      }
    \]

    \[
      \begin{array}{rcl}
        \DNFOf{\Sequence_1'' \DNFSep \ldots \DNFSep \Sequence_{n+m}''}
        & = & \DNFOf{\Sequence_1 \DNFSep \ldots \DNFSep \Sequence_n \DNFSep 
              \SequenceAlt_1 \DNFSep \ldots \DNFSep \SequenceAlt_n}\\
        & = & \ToDNFRegexOf{\Regex_1} \OrDNF \ToDNFRegexOf{\Regex_2}
      \end{array}
    \]

    \[
      \begin{array}{rcl}
        \DNFOf{\Sequence_{\SwapPermutationOf{\Identity_n}{\Identity_m}(1)}' \DNFSep \ldots \DNFSep 
        \Sequence_{\SwapPermutationOf{\Identity_n}{\Identity_m}(n+m)}'}
        & = & \DNFOf{\Sequence_{\Identity_m(1)+n}' \DNFSep \ldots \DNFSep \Sequence_{\Identity_m(m)+n}' \DNFSep 
              \Sequence_{\Identity_n(1)}' \DNFSep \ldots \DNFSep \Sequence_{\Identity_n(n)}}\\
        & = & \DNFOf{\Sequence_{n+1}' \DNFSep \ldots \DNFSep \Sequence_{n+m} \DNFSep 
              \Sequence_1' \DNFSep \ldots \DNFSep \Sequence_n'}\\
        & = & \DNFOf{\SequenceAlt_1 \DNFSep \ldots \DNFSep \SequenceAlt_m \DNFSep 
              \Sequence_1 \DNFSep \ldots \DNFSep \Sequence_n}\\
        & = & \ToDNFRegexOf{\Regex_2} \OrDNF \ToDNFRegexOf{\Regex_1}
      \end{array}
    \]

    So $\ToDNFRegexOf{\Regex_1} \OrDNF \ToDNFRegexOf{\Regex_2}
    \ParallelRewriteSwap
    \ToDNFRegexOf{\Regex_2} \OrDNF \ToDNFRegexOf{\Regex_1}$, which means
    $\ToDNFRegexOf{\Regex_1} \OrDNF \ToDNFRegexOf{\Regex_2}
    \EquivalenceOf{\ParallelRewriteSwap}
    \ToDNFRegexOf{\Regex_2} \OrDNF \ToDNFRegexOf{\Regex_1}$
  \end{case}

  \begin{case}[\DistributivityLeftRule{}]
    Let $\Regex \SSREquiv \RegexAlt$, and the last step of the
    deduction is an application of \DistributivityLeftRule{}.  Without loss of
    generality, from symmetry,
    $\Regex = \Regex_1 \Concat (\Regex_2 \Or \Regex_3)$, and
    $\RegexAlt = (\Regex_1 \Concat \Regex_2) \Or (\Regex_1 \Concat \Regex_3)$.

    Let $\ToDNFRegexOf{\Regex_1} =
    \DNFOf{\Sequence_{1,1} \DNFSep \ldots \DNFSep \Sequence_{1,n_1}}$.
    Let $\ToDNFRegexOf{\Regex_2} =
    \DNFOf{\Sequence_{2,1} \DNFSep \ldots \DNFSep \Sequence_{2,n_2}}$.
    Let $\ToDNFRegexOf{\Regex_3} =
    \DNFOf{\Sequence_{3,1} \DNFSep \ldots \DNFSep \Sequence_{3,n_3}}$.

    \[
      \begin{array}{rcl}
        \ToDNFRegexOf{(\Regex_1 \Concat (\Regex_2 \Or \Regex_3))}
        & = & \ToDNFRegexOf{\Regex_1} \ConcatDNF (\ToDNFRegexOf{\Regex_2} \OrDNF
              \ToDNFRegexOf{\Regex_3})\\
        & = & \ToDNFRegexOf{\Regex_1} \ConcatDNF
              \DNFOf{\Sequence_{2,1} \DNFSep \ldots \DNFSep \Sequence_{2,n_2} \DNFSep 
              \Sequence_{3,1} \DNFSep \ldots \DNFSep \Sequence_{3,n_3}}\\
        & = & \DNFOf{\Sequence_{1,1}\ConcatSequence\Sequence_{2,1} \DNFSep 
              \ldots \DNFSep \Sequence_{1,1}\ConcatSequence\Sequence_{3,n_3}
              \DNFSep \ldots \\
        &   & \DNFSep \Sequence_{1,n_1}\ConcatSequence\Sequence_{2,n_2} \DNFSep 
              \ldots \DNFSep \Sequence_{1,n_1}\ConcatSequence\Sequence_{3,n_3}}
      \end{array}
    \]

    \[
      \begin{array}{rcl}
        \ToDNFRegexOf{((\Regex_1 \Concat \Regex_2) \Or (\Regex_1 \Concat \Regex_3))}
        & = & (\ToDNFRegexOf{\Regex_1} \ConcatDNF \ToDNFRegexOf{\Regex_2})
              \OrDNF
              (\ToDNFRegexOf{\Regex_1} \ConcatDNF \ToDNFRegexOf{\Regex_3})\\
        & = & \DNFOf{\Sequence_{1,1}\ConcatSequence\Sequence_{2,1} \DNFSep 
              \ldots \DNFSep \ldots \DNFSep 
              \Sequence_{1,n_1}\ConcatSequence\Sequence_{2,n_2}} \OrDNF\\
        &   & \DNFOf{\Sequence_{1,1}\ConcatSequence\Sequence_{3,1} \DNFSep 
              \ldots \DNFSep \ldots \DNFSep 
              \Sequence_{1,n_1}\ConcatSequence\Sequence_{3,n_3}}\\
        & = & \DNFOf{\Sequence_{1,1}\ConcatSequence\Sequence_{2,1} \DNFSep 
              \ldots \DNFSep \ldots \DNFSep 
              \Sequence_{1,n_1}\ConcatSequence\Sequence_{2,n_2} \\
        &   & \DNFSep \Sequence_{1,1}\ConcatSequence\Sequence_{3,1} \DNFSep 
              \ldots \DNFSep \ldots \DNFSep 
              \Sequence_{1,n_1}\ConcatSequence\Sequence_{3,n_3}}
      \end{array}
    \]

    So $\ToDNFRegexOf{(\Regex_1 \Concat (\Regex_2 \Or \Regex_3))}$
    is different from $\ToDNFRegexOf{((\Regex_1 \Concat \Regex_2) \Or (\Regex_1
      \Concat \Regex_3))}$ only by the difference in the ordering of the
    sequences.

    Through using DNFReorderRule{},
    $\ToDNFRegexOf{(\Regex_1 \Concat (\Regex_2 \Or
      \Regex_3))} \ParallelRewriteSwap
    \ToDNFRegexOf{((\Regex_1 \Concat \Regex_2) \Or (\Regex_1
      \Concat \Regex_3))}$, so through the base rule,
    $\ToDNFRegexOf{(\Regex_1 \Concat (\Regex_2 \Or
      \Regex_3))} \EquivalenceOf{\ParallelRewriteSwap}
    \ToDNFRegexOf{((\Regex_1 \Concat \Regex_2) \Or (\Regex_1
      \Concat \Regex_3))}$.
  \end{case}

  \begin{case}[\DistributivityRightRule{}]
    Let $\Regex \SSREquiv \RegexAlt$, and the last step of the
    deduction is an application of \DistributivityRightRule{}.  Without loss of
    generality, from symmetry,
    $\Regex = (\Regex_1 \Or \Regex_2) \Concat \Regex_3$, and
    $\RegexAlt = (\Regex_1 \Concat \Regex_3) \Or (\Regex_2 \Concat \Regex_3)$.

    $\ToDNFRegexOf{(\Regex_1 \Or \Regex_2) \Concat \Regex_3} =
    (\ToDNFRegexOf{\Regex_1} \OrDNF \ToDNFRegexOf{\Regex_2}) \ConcatDNF
    \ToDNFRegexOf{\Regex_3} =
    (\ToDNFRegexOf{\Regex_1} \ConcatDNF \ToDNFRegexOf{\Regex_3}) \OrDNF
    (\ToDNFRegexOf{\Regex_2} \ConcatDNF \ToDNFRegexOf{\Regex_3}) =
    \ToDNFRegexOf{(\Regex_1 \Concat \Regex_3) \Or (\Regex_2 \Concat \Regex_3)}$.
    
    Through reflexivity, $\ToDNFRegexOf{\Regex} = \ToDNFRegexOf{((\Regex_1 \Or
      \Regex_2) \Concat \Regex_3)} \EquivalenceOf{\ParallelRewriteSwap}
    \ToDNFRegexOf{(\Regex_1 \Concat \Regex_3) \Or (\Regex_2 \Concat \Regex_3)}
    = \ToDNFRegexOf{\RegexAlt}$
  \end{case}

  \begin{case}[\ConcatIdentityLeftRule{}]
    Let $\Regex \SSREquiv \RegexAlt$, and the last step of the
    deduction is an application of \DistributivityRightRule{}.  Without loss of
    generality, from symmetry,
    $\Regex = \Regex' \Concat \EmptyString$, and
    $\RegexAlt = \Regex'$.

    $\ToDNFRegexOf{(\Regex' \Concat \EmptyString)} =
    \ToDNFRegexOf{\Regex'} \ConcatDNF \ToDNFRegexOf{\EmptyString} =
    \ToDNFRegexOf{\Regex'} \ConcatDNF \DNFOf{\SequenceOf{\EmptyString}} =
    \ToDNFRegexOf{\Regex'}$.
    
    Through reflexivity,
    $\ToDNFRegexOf{\Regex} = \ToDNFRegexOf{(\Regex' \Concat \EmptyString)}
    \EquivalenceOf{\ParallelRewriteSwap} \ToDNFRegexOf{\Regex'} =
    \ToDNFRegexOf{\RegexAlt}$
  \end{case}

  \begin{case}[\ConcatIdentityRightRule{}]
    Let $\Regex \SSREquiv \RegexAlt$, and the last step of the
    deduction is an application of \DistributivityRightRule{}.  Without loss of
    generality, from symmetry,
    $\Regex = \EmptyString \Concat \Regex'$, and
    $\RegexAlt = \Regex'$.

    $\ToDNFRegexOf{(\EmptyString \Concat \Regex')} =
    \ToDNFRegexOf{\EmptyString} \ConcatDNF \ToDNFRegexOf{\Regex'} =
    \DNFOf{\SequenceOf{\EmptyString}} \ConcatDNF \ToDNFRegexOf{\Regex'}=
    \ToDNFRegexOf{\Regex'}$
    
    Through reflexivity,
    $\ToDNFRegexOf{\Regex} = \ToDNFRegexOf{(\EmptyString \Concat \Regex')}
    \EquivalenceOf{\ParallelRewriteSwap} \ToDNFRegexOf{\Regex'} =
    \ToDNFRegexOf{\RegexAlt}$
  \end{case}

  \begin{case}[\UnrollstarLeftRule{}]
    Let $\Regex \SSREquiv \RegexAlt$, and the last step of the
    deduction is an application of \UnrollstarLeftRule{}.  Without loss of
    generality, from symmetry,
    $\Regex = \StarOf{\Regex'}$, and
    $\RegexAlt = \EmptyString \Or (\Regex' \Concat \StarOf{\Regex'})$.

    $\ToDNFRegexOf{\StarOf{\Regex'}} =
    \AtomToDNFOf{\StarOf{(\ToDNFRegexOf{\Regex'})}}$.
    $\ToDNFRegexOf{(\EmptyString \Or (\Regex' \Concat \StarOf{\Regex'}))} =
    \DNFOf{\SequenceOf{\EmptyString}} \OrDNF (\ToDNFRegexOf{\Regex'} \ConcatDNF
    \AtomToDNFOf{\StarOf{(\ToDNFRegexOf{\Regex'})}})$.
    
    Through \AtomUnrollstarLeftRule{},
    $\ToDNFRegexOf{\Regex} =
    \ToDNFRegexOf{\StarOf{\Regex'}} \ParallelRewriteSwap
    \ToDNFRegexOf{(\EmptyString \Or (\Regex' \Concat \StarOf{\Regex'}))} =
    \ToDNFRegexOf{\RegexAlt}$.
  \end{case}

  \begin{case}[\UnrollstarRightRule{}]
    Let $\Regex \SSREquiv \RegexAlt$, and the last step of the
    deduction is an application of \UnrollstarRightRule{}.  Without loss of
    generality, from symmetry,
    $\Regex = \StarOf{\Regex'}$, and
    $\RegexAlt = \EmptyString \Or (\StarOf{\Regex'} \Concat \Regex')$.

    $\ToDNFRegexOf{\StarOf{\Regex'}} =
    \AtomToDNFOf{\StarOf{(\ToDNFRegexOf{\Regex'})}}$.
    $\ToDNFRegexOf{(\EmptyString \Or (\StarOf{\Regex'} \Concat \Regex'))} =
    \DNFOf{\SequenceOf{\EmptyString}} \OrDNF
    (\AtomToDNFOf{\StarOf{(\ToDNFRegexOf{\Regex'})}} \ConcatDNF
    \ToDNFRegexOf{\Regex'})$.
    
    Through \AtomUnrollstarRightRule{},
    $\ToDNFRegexOf{\Regex} =
    \ToDNFRegexOf{\StarOf{\Regex'}} \ParallelRewriteSwap
    \ToDNFRegexOf{(\EmptyString \Or (\StarOf{\Regex'} \Concat \Regex'))} =
    \ToDNFRegexOf{\RegexAlt}$.
  \end{case}

  \begin{case}[Structural \OrRegexType{} Equality]
    Let $\Regex \SSREquiv \RegexAlt$, through structural equality of
    \OrRegexType{}.
    $\Regex = \Regex_1 \Or \Regex_2$, and
    $\RegexAlt = \RegexAlt_1 \Or \RegexAlt_2$,
    $\Regex_1 \SSREquiv \RegexAlt_1$, and
    $\Regex_2 \SSREquiv \RegexAlt_2$.

    By induction assumption,
    $\ToDNFRegexOf{\Regex_1} \EquivalenceOf{\ParallelRewriteSwap}
    \ToDNFRegexOf{\RegexAlt_1}$ and
    $\ToDNFRegexOf{\Regex_2} \EquivalenceOf{\ParallelRewriteSwap}
    \ToDNFRegexOf{\RegexAlt_2}$.

    By Lemma~\ref{lem:prop-eq-swap-or},
    $\ToDNFRegexOf{\Regex_1} \OrDNF \ToDNFRegexOf{\Regex_2}
    \EquivalenceOf{\ParallelRewriteSwap}
    \ToDNFRegexOf{\RegexAlt_1} \OrDNF \ToDNFRegexOf{\RegexAlt_2}$.
    By the definition of $\ToDNFRegex{}$,
    $\ToDNFRegexOf{(\Regex_1 \Or \Regex_2)}
    \EquivalenceOf{\ParallelRewriteSwap}
    \ToDNFRegexOf{(\RegexAlt_1 \Or \RegexAlt_2)}$, as desired.
  \end{case}

  \begin{case}[Structural \ConcatRegexType{} Equality]
    Let $\Regex \SSREquiv \RegexAlt$, through structural equality of
    \ConcatRegexType{}.
    $\Regex = \Regex_1 \Concat \Regex_2$, and
    $\RegexAlt = \RegexAlt_1 \Concat \RegexAlt_2$,
    $\Regex_1 \SSREquiv \RegexAlt_1$, and
    $\Regex_2 \SSREquiv \RegexAlt_2$.

    By induction assumption,
    $\ToDNFRegexOf{\Regex_1} \EquivalenceOf{\ParallelRewriteSwap}
    \ToDNFRegexOf{\RegexAlt_1}$ and
    $\ToDNFRegexOf{\Regex_2} \EquivalenceOf{\ParallelRewriteSwap}
    \ToDNFRegexOf{\RegexAlt_2}$.

    By Lemma~\ref{lem:prop-eq-swap-concat},
    $\ToDNFRegexOf{\Regex_1} \ConcatDNF \ToDNFRegexOf{\Regex_2}
    \EquivalenceOf{\ParallelRewriteSwap}
    \ToDNFRegexOf{\RegexAlt_1} \ConcatDNF \ToDNFRegexOf{\RegexAlt_2}$.
    By the definition of $\ToDNFRegex{}$,
    $\ToDNFRegexOf{(\Regex_1 \Concat \Regex_2)}
    \EquivalenceOf{\ParallelRewriteSwap}
    \ToDNFRegexOf{(\RegexAlt_1 \Concat \RegexAlt_2)}$, as desired.
  \end{case}

  \begin{case}[Structural \StarRegexType{} Equality]
    Let $\Regex \SSREquiv \RegexAlt$, through structural equality of
    \StarRegexType{}.
    $\Regex = \StarOf{\Regex'}$,
    $\RegexAlt = \StarOf{\RegexAlt'}$, and
    $\Regex' \SSREquiv \RegexAlt'$.

    By induction assumption,
    $\ToDNFRegexOf{\Regex'} \EquivalenceOf{\ParallelRewriteSwap}
    \ToDNFRegexOf{\RegexAlt'}$.

    By Lemma~\ref{lem:prop-eq-swap-star},
    $\AtomToDNFOf{\StarOf{\ToDNFRegexOf{\Regex'}}}
    \EquivalenceOf{\ParallelRewriteSwap}
    \AtomToDNFOf{\StarOf{\ToDNFRegexOf{\Regex'}}}$.
    By the definition of $\ToDNFRegex{}$,
    $\ToDNFRegexOf{\StarOf{\Regex'}}
    \EquivalenceOf{\ParallelRewriteSwap}
    \ToDNFRegexOf{\StarOf{\RegexAlt'}}$, as desired.
  \end{case}

  \begin{case}[Transitivity of Equational Theory]
    Let $\Regex \SSREquiv \RegexAlt$ through the transitivity of an
    equational theory.  This means there exists a $\Regex'$ such that $\Regex
    \SSREquiv \Regex'$ and $\Regex' \SSREquiv \RegexAlt$.
    
    By induction assumption,
    $\ToDNFRegexOf{\Regex} \EquivalenceOf{\ParallelRewriteSwap}
    \ToDNFRegexOf{\Regex'}$ and
    $\ToDNFRegexOf{\Regex'} \EquivalenceOf{\ParallelRewriteSwap}
    \ToDNFRegexOf{\RegexAlt}$.
    
    Consider the derivation
    \[
      \inferrule*
      {
        \ToDNFRegexOf{\Regex} \EquivalenceOf{\ParallelRewriteSwap}
        \ToDNFRegexOf{\Regex'}\\
        \ToDNFRegexOf{\Regex'} \EquivalenceOf{\ParallelRewriteSwap}
        \ToDNFRegexOf{\RegexAlt}
      }
      {
        \ToDNFRegexOf{\Regex} \EquivalenceOf{\ParallelRewriteSwap}
        \ToDNFRegexOf{\RegexAlt}
      }
    \]
  \end{case}
\end{proof}

\begin{theorem}[Equivalence of $\EquivalenceOf{\ParallelRewriteSwap}$ and
  $\SSREquiv$]
  \label{thm:defequiv-equiv-parallelswapequiv}
  $\Regex \SSREquiv \RegexAlt$ if, and only if $\ToDNFRegexOf{\Regex}
  \EquivalenceOf{\ParallelRewriteSwap} \ToDNFRegexOf{\RegexAlt}$
\end{theorem}
\begin{proof}
  The forward direction is proven by Lemma~\ref{lem:express-equiv-in-equiv-swap}.
  The reverse direction is proven by Lemma~\ref{lem:express-equiv-swap-in-equiv}
\end{proof}

\begin{lemma}
  \label{lem:identity-atom-in-parallel}
  $\Atom \ParallelRewriteAtom \AtomToDNFOf{\Atom}$.
\end{lemma}
\begin{proof}
  $\Atom = \StarOf{\DNFRegex}$ for some DNF regular expression.
  Consider the derivation
  \[
    \inferrule*
    {
      \DNFRegex \ParallelRewrite \DNFRegex
    }
    {
      \StarOf{\DNFRegex} \ParallelRewrite \AtomToDNFOf{\StarOf{\DNFRegex}}
    }
  \] as desired.
\end{proof}

\begin{lemma}
  \label{lem:serial-expressible-in-parallel}
  \leavevmode
  \begin{itemize}
  \item If $\Atom \Rewrite \DNFRegexAlt$, then
    $\Atom \ParallelRewrite \DNFRegexAlt$.
  \item If $\DNFRegex \Rewrite \DNFRegexAlt$, then
    $\DNFRegex \ParallelRewrite \DNFRegexAlt$.
  \end{itemize}
\end{lemma}
\begin{proof}
  By mutual induction on the derivation of $\Rewrite$ and $\RewriteAtom$

  \begin{case}[\AtomUnrollstarLeftRule{}]
    \[
      \inferrule*
      {
      }
      {
        \StarOf{\DNFRegex}\RewriteAtom
        \OrDNFOf{\DNFOf{\SequenceOf{\EmptyString}}}{(\ConcatDNFOf{\DNFRegex}{\AtomToDNFOf{\StarOf{\DNFRegex}}})}
      }
    \]

    Consider the derivation
    
    \[
      \inferrule*
      {
      }
      {
        \StarOf{\DNFRegex}\ParallelRewriteAtom
        \OrDNFOf{\DNFOf{\SequenceOf{\EmptyString}}}{(\ConcatDNFOf{\DNFRegex}{\AtomToDNFOf{\StarOf{\DNFRegex}}})}
      }
    \]
  \end{case}

  \begin{case}[\AtomUnrollstarRightRule{}]
    \[
      \inferrule*
      {
      }
      {
        \StarOf{\DNFRegex}\RewriteAtom
        \OrDNFOf{\DNFOf{\SequenceOf{\EmptyString}}}{(\ConcatDNFOf{\AtomToDNFOf{\StarOf{\DNFRegex}}}{\DNFRegex})}
      }
    \]

    Consider the derivation
    
    \[
      \inferrule*
      {
      }
      {
        \StarOf{\DNFRegex}\ParallelRewriteAtom
        \OrDNFOf{\DNFOf{\SequenceOf{\EmptyString}}}{(\ConcatDNFOf{\AtomToDNFOf{\StarOf{\DNFRegex}}}{\DNFRegex})}
      }
    \]
  \end{case}

  \begin{case}[\AtomStructuralRewriteRule{}]
    \[
      \inferrule*
      {
        \DNFRegex \Rewrite \DNFRegexAlt'
      }
      {
        \StarOf{\DNFRegex} \Rewrite \AtomToDNFOf{\StarOf{\DNFRegexAlt'}}
      }
    \]
    
    By IH, $\DNFRegex \ParallelRewrite \DNFRegexAlt'$, so consider the
    derivation
    \[
      \inferrule*
      {
        \DNFRegex \ParallelRewrite \DNFRegexAlt'
      }
      {
        \StarOf{\DNFRegex} \ParallelRewrite \AtomToDNFOf{\StarOf{\DNFRegexAlt'}}
      }
    \]
  \end{case}

  \begin{case}[\DNFStructuralRewriteRule{}]
    \[
      \inferrule*
      {
        \Atom_j \RewriteAtom \DNFRegex
      }
      {
        \DNFOf{\Sequence_1\DNFSep\ldots\DNFSep\Sequence_{i-1}} \OrDNF
        \DNFOf{\SequenceOf{\String_0\SeqSep\Atom_1\SeqSep\ldots\SeqSep\String_{j-1}}}
        \ConcatDNF \AtomToDNFOf{\Atom_j} \ConcatDNF
        \DNFOf{\SequenceOf{\String_j\SeqSep\ldots\SeqSep\Atom_m\SeqSep\String_m}}
        \OrDNF \DNFOf{\Sequence_{i+1}\DNFSep\ldots\DNFSep\Sequence_n}\Rewrite\\
        \DNFOf{\Sequence_1\DNFSep\ldots\DNFSep\Sequence_{i-1}} \OrDNF
        \DNFOf{\SequenceOf{\String_0\SeqSep\Atom_1\SeqSep\ldots\SeqSep\String_{j-1}}}\ConcatDNF\DNFRegex\ConcatDNF\SequenceOf{\String_j\SeqSep\ldots\SeqSep\Atom_m\SeqSep\String_m} \OrDNF
        \DNFOf{\Sequence_{i+1}\DNFSep\ldots\DNFSep\Sequence_n}
      }
    \]

    Define $\Sequence_i$ as
    $\SequenceOf{\String_0\SeqSep\Atom_1\SeqSep\ldots\SeqSep\Atom_m\SeqSep\String_m}$.
    Through the definition of $\OrDNF$ and $\ConcatDNF$,
    $\DNFRegex = \DNFOf{\Sequence_1 \DNFSep \ldots \DNFSep \Sequence_n}$.
    Define $\Sequence_k =
    \SequenceOf{\String_{k,0} \SeqSep \Atom_{k,1} \SeqSep \ldots \SeqSep \Atom_{k,n_k} \SeqSep \String_{k,n_k}}$.
    So, in particular, $\Atom_{i,j} = \Atom_j$, and $n_i = m$.
    Define $\DNFRegex_{k,l} =
    \begin{cases*}
      \DNFRegex & if $(k,l) = (i,j)$\\
      \AtomToDNFOf{\Atom_{k,l}} & otherwise
    \end{cases*}$
    So, for all $k,l$, $\Atom_{k,l} \ParallelRewriteAtom \DNFRegex_{k,l}$, as if
    $(k,l) = (i,j)$, then by assumption
    $\Atom_{i,j} \ParallelRewriteAtom \DNFRegex$, and otherwise,
    from Lemma~\ref{lem:identity-atom-in-parallel},
    $\Atom_{k,l} \ParallelRewriteAtom
    \AtomToDNFOf{\Atom_{k,l}}$.
    
    Define $\DNFRegex_k$ as $\DNFOf{\SequenceOf{\String_{k,0}}} \ConcatDNF \DNFRegex_{k,1}
    \ConcatDNF \ldots \ConcatDNF \DNFRegex_{k,n_k} \ConcatDNF
    \DNFOf{\SequenceOf{\String_{k,n_k}}}$.
    
    \[
      \inferrule*
      {
        \DNFRegex = \DNFOf{\Sequence_1 \DNFSep \ldots \DNFSep \Sequence_n}\\
        \forall i. \Sequence_i =
        \SequenceOf{\String_{i,0} \SeqSep \Atom_{i,1} \SeqSep \ldots \SeqSep \Atom_{i,n_i} \SeqSep \String_{i,n_i}}\\
        \forall i,j. \Atom_{i,j} \ParallelRewriteAtom \DNFRegex_{i,j}\\
        \forall i. \DNFRegex_i = \DNFOf{\SequenceOf{\String_{i,0}}} \ConcatDNF \DNFRegex_{i,1}
        \ConcatDNF \ldots \ConcatDNF \DNFRegex_{i,n_i} \ConcatDNF
        \DNFOf{\SequenceOf{\String_{i,n_i}}}
      }
      {
        \DNFRegex \ParallelRewrite \DNFRegex_1 \OrDNF \ldots \OrDNF \DNFRegex_n
      }
    \]

    So $\DNFRegex_k$, for $k \neq i =
    \DNFOf{\SequenceOf{\String_{k,0}}} \ConcatDNF \AtomToDNFOf{\Atom_{k,1}}
    \ConcatDNF \ldots \ConcatDNF \AtomToDNFOf{\Atom_{k,n_k}} \ConcatDNF
    \DNFOf{\SequenceOf{\String_{k,n_k}}} = \Sequence_k$

    $\DNFRegex_i =
    \DNFOf{\SequenceOf{\String_0\SeqSep\Atom_1\SeqSep\ldots\SeqSep\String_{j-1}}}
    \ConcatDNF \AtomToDNFOf{\Atom_j} \ConcatDNF
    \DNFOf{\SequenceOf{\String_j\SeqSep\ldots\SeqSep\Atom_m\SeqSep\String_m}}$,

    so, through the definition of $\OrDNF$,
    $\DNFRegex_1 \OrDNF \ldots \OrDNF \DNFRegex_n =
    \DNFOf{\Sequence_1\DNFSep\ldots\DNFSep\Sequence_{i-1}} \OrDNF
    \DNFOf{\SequenceOf{\String_0\SeqSep\Atom_1\SeqSep\ldots\SeqSep\String_{j-1}}}\ConcatDNF\DNFRegex\ConcatDNF\SequenceOf{\String_j\SeqSep\ldots\SeqSep\Atom_m\SeqSep\String_m} \OrDNF
    \DNFOf{\Sequence_{i+1}\DNFSep\ldots\DNFSep\Sequence_n}$,
    so we get $\DNFRegex \ParallelRewrite \DNFRegex_1 \OrDNF \ldots \OrDNF \DNFRegex_n =
    \DNFOf{\Sequence_1\DNFSep\ldots\DNFSep\Sequence_{i-1}} \OrDNF
    \DNFOf{\SequenceOf{\String_0\SeqSep\Atom_1\SeqSep\ldots\SeqSep\String_{j-1}}}\ConcatDNF\DNFRegex\ConcatDNF\SequenceOf{\String_j\SeqSep\ldots\SeqSep\Atom_m\SeqSep\String_m} \OrDNF
    \DNFOf{\Sequence_{i+1}\DNFSep\ldots\DNFSep\Sequence_n}$.
  \end{case}
\end{proof}

\begin{lemma}
  \label{lem:star-serial-expressible-in-star-parallel}
  If $\DNFRegex \StarOf{\Rewrite} \DNFRegexAlt$, then
  $\DNFRegex \StarOf{\ParallelRewrite} \DNFRegexAlt$
\end{lemma}
\begin{proof}
  By induction on the derivation of $\StarOf{\Rewrite}$

  \begin{case}[\ReflexivityRule{}]
    \[
      \inferrule*
      {
      }
      {
        \DNFRegex \StarOf{\Rewrite} \DNFRegex
      }
    \]

    Consider the following derivation

    \[
      \inferrule*
      {
      }
      {
        \DNFRegex \StarOf{\ParallelRewrite} \DNFRegex
      }
    \]
  \end{case}

  \begin{case}[\BaseRule{}]
    \[
      \inferrule*
      {
        \DNFRegex \Rewrite \DNFRegexAlt
      }
      {
        \DNFRegex \StarOf{\Rewrite} \DNFRegexAlt
      }
    \]

    By Lemma~\ref{lem:parallel-expressible-in-star-serial},
    $\DNFRegex \StarOf{\ParallelRewrite} \DNFRegexAlt$.
  \end{case}

  \begin{case}[\TransitivityRule{}]
    \[
      \inferrule*
      {
        \DNFRegex \StarOf{\Rewrite} \DNFRegex'\\
        \DNFRegex' \StarOf{\Rewrite} \DNFRegexAlt
      }
      {
        \DNFRegex \StarOf{\Rewrite} \DNFRegexAlt
      }
    \]

    By IH, $\DNFRegex \StarOf{\ParallelRewrite} \DNFRegex'$ and
    $\DNFRegex' \StarOf{\ParallelRewrite} \DNFRegexAlt$.

    Consider the following derivation
    \[
      \inferrule*
      {
        \DNFRegex \StarOf{\ParallelRewrite} \DNFRegex'\\
        \DNFRegex' \StarOf{\ParallelRewrite} \DNFRegexAlt
      }
      {
        \DNFRegex \StarOf{\ParallelRewrite} \DNFRegexAlt
      }
    \]
  \end{case}
\end{proof}

\begin{lemma}
  \label{lem:propagation-of-rewrites-through-atom-structure}
  If $\DNFRegex \StarOf{\Rewrite} \DNFRegexAlt$, then
  $\AtomToDNFOf{\StarOf{\DNFRegex}} \StarOf{\Rewrite}
  \AtomToDNFOf{\StarOf{\DNFRegexAlt}}$.
\end{lemma}
\begin{proof}
  By induction on the derivation of $\StarOf{\Rewrite}$

  \begin{case}[\ReflexivityRule{}]
    \[
      \inferrule*
      {
      }
      {
        \DNFRegex \StarOf{\Rewrite} \DNFRegex
      }
    \]

    Consider the derivation
    
    \[
      \inferrule*
      {
      }
      {
        \AtomToDNFOf{\DNFRegex} \StarOf{\Rewrite} \AtomToDNFOf{\DNFRegex}
      }
    \]
  \end{case}

  \begin{case}[\BaseRule{}]
    \[
      \inferrule*
      {
        \DNFRegex \Rewrite \DNFRegexAlt
      }
      {
        \DNFRegex \StarOf{\Rewrite} \DNFRegexAlt
      }
    \]

    Consider the derivation
    \[
      \inferrule*
      {
        \inferrule*
        {
          \inferrule*
          {
            \DNFRegex \Rewrite \DNFRegexAlt
          }
          {
            \StarOf{\DNFRegex} \RewriteAtom \AtomToDNFOf{\StarOf{\DNFRegexAlt}}
          }
        }
        {
          \AtomToDNFOf{\StarOf{\DNFRegex}} \Rewrite
          \AtomToDNFOf{\StarOf{\DNFRegexAlt}}
        }
      }
      {
        \AtomToDNFOf{\StarOf{\DNFRegex}} \StarOf{\Rewrite}
        \AtomToDNFOf{\StarOf{\DNFRegexAlt}}
      }
    \]
  \end{case}

  \begin{case}[\TransitivityRule{}]
    \[
      \inferrule*
      {
        \DNFRegex \StarOf{\Rewrite} \DNFRegex'\\
        \DNFRegex' \StarOf{\Rewrite} \DNFRegexAlt
      }
      {
        \DNFRegex \StarOf{\Rewrite} \DNFRegexAlt
      }
    \]

    By IH, there exists derivations of
    $\AtomToDNFOf{\StarOf{\DNFRegex}} \StarOf{\Rewrite}
    \AtomToDNFOf{\StarOf{\DNFRegex'}}$ and
    $\AtomToDNFOf{\StarOf{\DNFRegex'}} \StarOf{\Rewrite}
    \AtomToDNFOf{\StarOf{\DNFRegexAlt}}$.

    Consider the derivation
    \[
      \inferrule*
      {
        \AtomToDNFOf{\StarOf{\DNFRegex}} \StarOf{\Rewrite}
        \AtomToDNFOf{\StarOf{\DNFRegex'}}\\
        \AtomToDNFOf{\StarOf{\DNFRegex'}} \StarOf{\Rewrite}
        \AtomToDNFOf{\StarOf{\DNFRegexAlt}}
      }
      {
        \AtomToDNFOf{\StarOf{\DNFRegex}} \StarOf{\Rewrite}
        \AtomToDNFOf{\StarOf{\DNFRegexAlt}}
      }
    \]
  \end{case}
\end{proof}

\begin{lemma}
  \label{lem:propagation-of-star-rewrites-through-or-left}
  If $\DNFRegex_1 \StarOf{\Rewrite} \DNFRegex_2$, then for all $\DNFRegexAlt$,
  $\DNFRegex_1 \OrDNF \DNFRegexAlt \StarOf{\Rewrite}
  \DNFRegex_2 \OrDNF \DNFRegexAlt$
\end{lemma}
\begin{proof}
  By induction on the derivation of \StarOf{\Rewrite}

  \begin{case}[\ReflexivityRule{}]
    \[
      \inferrule*
      {
      }
      {
        \DNFRegex_1 \StarOf{\Rewrite} \DNFRegex_1
      }
    \]

    so, by \ReflexivityRule{}
    
    \[
      \inferrule*
      {
      }
      {
        \DNFRegex_1 \OrDNF \DNFRegexAlt
        \StarOf{\Rewrite}
        \DNFRegex_1 \OrDNF \DNFRegexAlt
      }
    \]
  \end{case}

  \begin{case}[\BaseRule{}]
    \[
      \inferrule*
      {
        \DNFRegex_1 \Rewrite \DNFRegex_2
      }
      {
        \DNFRegex_1 \StarOf{\Rewrite} \DNFRegex_2
      }
    \]

    The only way to get a derivation of \Rewrite{} is with an application of
    \DNFStructuralRewriteRule{}, so by inversion,

    \[
      \inferrule*
      {
        \Atom_j \Rewrite \DNFRegex
      }
      {
        \DNFOf{\Sequence_1\DNFSep\ldots\DNFSep\Sequence_{i-1}} \OrDNF
        \DNFOf{\SequenceOf{\String_0\SeqSep\Atom_1\SeqSep\ldots\SeqSep\String_{j-1}}}
        \ConcatDNF \AtomToDNFOf{\Atom_j} \ConcatDNF
        \DNFOf{\SequenceOf{\String_j\SeqSep\ldots\SeqSep\Atom_m\SeqSep\String_m}}
        \OrDNF \DNFOf{\Sequence_{i+1}\DNFSep\ldots\DNFSep\Sequence_n}\Rewrite\\
        \DNFOf{\Sequence_1\DNFSep\ldots\DNFSep\Sequence_{i-1}} \OrDNF
        \DNFOf{\SequenceOf{\String_0\SeqSep\Atom_1\SeqSep\ldots\SeqSep\String_{j-1}}}\ConcatDNF\DNFRegex\ConcatDNF\SequenceOf{\String_j\SeqSep\ldots\SeqSep\Atom_m\SeqSep\String_m} \OrDNF
        \DNFOf{\Sequence_{i+1}\DNFSep\ldots\DNFSep\Sequence_n}
      }
    \]

    where
    $\DNFRegex_1 = \DNFOf{\Sequence_1\DNFSep\ldots\DNFSep\Sequence_{i-1}} \OrDNF
    \DNFOf{\SequenceOf{\String_0\SeqSep\Atom_1\SeqSep\ldots\SeqSep\String_{j-1}}}
    \ConcatDNF \AtomToDNFOf{\Atom_j} \ConcatDNF
    \DNFOf{\SequenceOf{\String_j\SeqSep\ldots\SeqSep\Atom_m\SeqSep\String_m}}
    \OrDNF \DNFOf{\Sequence_{i+1}\DNFSep\ldots\DNFSep\Sequence_n}$
    and where
    $\DNFRegex_2 = \DNFOf{\Sequence_1\DNFSep\ldots\DNFSep\Sequence_{i-1}} \OrDNF
    \DNFOf{\SequenceOf{\String_0\SeqSep\Atom_1\SeqSep\ldots\SeqSep\String_{j-1}}}\ConcatDNF\DNFRegex\ConcatDNF\SequenceOf{\String_j\SeqSep\ldots\SeqSep\Atom_m\SeqSep\String_m} \OrDNF
    \DNFOf{\Sequence_{i+1}\DNFSep\ldots\DNFSep\Sequence_n}$.

    So, let $\DNFRegexAlt = \DNFOf{\SequenceAlt_1 \DNFSep \ldots \DNFSep \SequenceAlt_{n'}}$.

    Consider the derivation
    \[
      \inferrule*
      {
        \Atom_j \Rewrite \DNFRegex
      }
      {
        \DNFOf{\Sequence_1\DNFSep\ldots\DNFSep\Sequence_{i-1}} \OrDNF
        \DNFOf{\SequenceOf{\String_0\SeqSep\Atom_1\SeqSep\ldots\SeqSep\String_{j-1}}}
        \ConcatDNF \AtomToDNFOf{\Atom_j} \ConcatDNF
        \DNFOf{\SequenceOf{\String_j\SeqSep\ldots\SeqSep\Atom_m\SeqSep\String_m}}
        \OrDNF\\ \DNFOf{\Sequence_{i+1}\DNFSep\ldots\DNFSep\Sequence_n\DNFSep\SequenceAlt_1 \DNFSep \ldots \DNFSep \SequenceAlt_{n'}}\Rewrite\\
        \DNFOf{\Sequence_1\DNFSep\ldots\DNFSep\Sequence_{i-1}} \OrDNF\\
        \DNFOf{\SequenceOf{\String_0\SeqSep\Atom_1\SeqSep\ldots\SeqSep\String_{j-1}}}\ConcatDNF\DNFRegex\ConcatDNF\SequenceOf{\String_j\SeqSep\ldots\SeqSep\Atom_m\SeqSep\String_m} \OrDNF
        \DNFOf{\Sequence_{i+1}\DNFSep\ldots\DNFSep\Sequence_n\DNFSep\SequenceAlt_1 \DNFSep \ldots \DNFSep \SequenceAlt_{n'}}
      }
    \]
    
    $\DNFOf{\Sequence_1\DNFSep\ldots\DNFSep\Sequence_{i-1}} \OrDNF
    \DNFOf{\SequenceOf{\String_0\SeqSep\Atom_1\SeqSep\ldots\SeqSep\String_{j-1}}}
    \ConcatDNF \AtomToDNFOf{\Atom_j} \ConcatDNF\\
    \DNFOf{\SequenceOf{\String_j\SeqSep\ldots\SeqSep\Atom_m\SeqSep\String_m}}
    \OrDNF \DNFOf{\Sequence_{i+1}\DNFSep\ldots\DNFSep\Sequence_n\DNFSep\SequenceAlt_1 \DNFSep \ldots \DNFSep \SequenceAlt_{n'}}
    =\\
    \DNFOf{\Sequence_1\DNFSep\ldots\DNFSep\Sequence_{i-1}} \OrDNF
    \DNFOf{\SequenceOf{\String_0\SeqSep\Atom_1\SeqSep\ldots\SeqSep\String_{j-1}}}
    \ConcatDNF \AtomToDNFOf{\Atom_j} \ConcatDNF
    \DNFOf{\SequenceOf{\String_j\SeqSep\ldots\SeqSep\Atom_m\SeqSep\String_m}}
    \OrDNF (\DNFOf{\Sequence_{i+1}\DNFSep\ldots\DNFSep\Sequence_n} \OrDNF
    \DNFOf{\SequenceAlt_1 \DNFSep \ldots \DNFSep \SequenceAlt_{n'}})$.
    So through associativity of $\OrDNF$, $\DNFOf{\Sequence_1\DNFSep\ldots\DNFSep\Sequence_{i-1}} \OrDNF
    \DNFOf{\SequenceOf{\String_0\SeqSep\Atom_1\SeqSep\ldots\SeqSep\String_{j-1}}}
    \ConcatDNF \AtomToDNFOf{\Atom_j} \ConcatDNF
    \DNFOf{\SequenceOf{\String_j\SeqSep\ldots\SeqSep\Atom_m\SeqSep\String_m}}
    \OrDNF
    \DNFOf{\Sequence_{i+1}\DNFSep\ldots\DNFSep\Sequence_n\DNFSep\SequenceAlt_1 \DNFSep \ldots \DNFSep \SequenceAlt_{n'}}
    = \DNFRegex_1 \OrDNF \DNFRegexAlt$.

    $\DNFOf{\Sequence_1\DNFSep\ldots\DNFSep\Sequence_{i-1}} \OrDNF
    \DNFOf{\SequenceOf{\String_0\SeqSep\Atom_1\SeqSep\ldots\SeqSep\String_{j-1}}}
    \ConcatDNF \DNFRegex \ConcatDNF\\
    \DNFOf{\SequenceOf{\String_j\SeqSep\ldots\SeqSep\Atom_m\SeqSep\String_m}}
    \OrDNF \DNFOf{\Sequence_{i+1}\DNFSep\ldots\DNFSep\Sequence_n\DNFSep\SequenceAlt_1 \DNFSep \ldots \DNFSep \SequenceAlt_{n'}}
    =\\
    \DNFOf{\Sequence_1\DNFSep\ldots\DNFSep\Sequence_{i-1}} \OrDNF
    \DNFOf{\SequenceOf{\String_0\SeqSep\Atom_1\SeqSep\ldots\SeqSep\String_{j-1}}}
    \ConcatDNF \DNFRegex \ConcatDNF
    \DNFOf{\SequenceOf{\String_j\SeqSep\ldots\SeqSep\Atom_m\SeqSep\String_m}}
    \OrDNF (\DNFOf{\Sequence_{i+1}\DNFSep\ldots\DNFSep\Sequence_n} \OrDNF
    \DNFOf{\SequenceAlt_1 \DNFSep \ldots \DNFSep \SequenceAlt_{n'}})$.
    So through associativity of $\OrDNF$, $\DNFOf{\Sequence_1\DNFSep\ldots\DNFSep\Sequence_{i-1}} \OrDNF
    \DNFOf{\SequenceOf{\String_0\SeqSep\Atom_1\SeqSep\ldots\SeqSep\String_{j-1}}}
    \ConcatDNF \DNFRegex \ConcatDNF
    \DNFOf{\SequenceOf{\String_j\SeqSep\ldots\SeqSep\Atom_m\SeqSep\String_m}}
    \OrDNF
    \DNFOf{\Sequence_{i+1}\DNFSep\ldots\DNFSep\Sequence_n\DNFSep\SequenceAlt_1 \DNFSep \ldots \DNFSep \SequenceAlt_{n'}}
    = \DNFRegex_2 \OrDNF \DNFRegexAlt$.
  \end{case}

  \begin{case}[\TransitivityRule{}]
    \[
      \inferrule*
      {
        \DNFRegex_1 \StarOf{\Rewrite} \DNFRegex_3\\
        \DNFRegex_3 \StarOf{\Rewrite} \DNFRegex_2
      }
      {
        \DNFRegex_1 \StarOf{\Rewrite} \DNFRegex_2
      }
    \]

    By IH, $\DNFRegex_1 \OrDNF \DNFRegexAlt \StarOf{\Rewrite}
    \DNFRegex_3 \OrDNF \DNFRegexAlt$.
    By IH, $\DNFRegex_3 \OrDNF \DNFRegexAlt \StarOf{\Rewrite}
    \DNFRegex_2 \OrDNF \DNFRegexAlt$.

    Consider the derivation
    \[
      \inferrule*
      {
        \DNFRegex_1 \OrDNF \DNFRegexAlt \StarOf{\Rewrite}
        \DNFRegex_3 \OrDNF \DNFRegexAlt\\
        \DNFRegex_3 \OrDNF \DNFRegexAlt \StarOf{\Rewrite}
        \DNFRegex_2 \OrDNF \DNFRegexAlt
      }
      {
        \DNFRegex_1 \OrDNF \DNFRegexAlt \StarOf{\Rewrite}
        \DNFRegex_2 \OrDNF \DNFRegexAlt
      }
    \]
  \end{case}
\end{proof}

\begin{lemma}
  \label{lem:propagation-of-star-rewrites-through-or-right}
  If $\DNFRegex_1 \StarOf{\Rewrite} \DNFRegex_2$, then for all $\DNFRegexAlt$,
  $\DNFRegexAlt \OrDNF \DNFRegex_1 \StarOf{\Rewrite}
  \DNFRegexAlt \OrDNF \DNFRegex_2$
\end{lemma}
\begin{proof}
  Proven symmetrically to Lemma~\ref{lem:propagation-of-star-rewrites-through-or-left}.
\end{proof}

\begin{lemma}
  \label{lem:propagation-of-star-rewrites-through-or}
  If $\DNFRegex_1 \StarOf{\Rewrite} \DNFRegex_2$, and
  $\DNFRegexAlt_1 \StarOf{\Rewrite} \DNFRegexAlt_2$, then
  $\DNFRegex_1 \OrDNF \DNFRegexAlt_1 \StarOf{\Rewrite}
  \DNFRegex_2 \OrDNF \DNFRegexAlt_2$
\end{lemma}
\begin{proof}
  By Lemma~\ref{lem:propagation-of-star-rewrites-through-or-left},
  $\DNFRegex_1 \OrDNF \DNFRegex_2 \StarOf{\Rewrite}
  \DNFRegexAlt_1 \OrDNF \DNFRegex_2$.
  By Lemma~\ref{lem:propagation-of-star-rewrites-through-or-right},
  $\DNFRegexAlt_1 \OrDNF \DNFRegex_2 \StarOf{\Rewrite}
  \DNFRegexAlt_1 \OrDNF \DNFRegexAlt_2$.

  Consider the derivation
  \[
    \inferrule*
    {
      \DNFRegex_1 \OrDNF \DNFRegex_2 \StarOf{\Rewrite}
      \DNFRegexAlt_1 \OrDNF \DNFRegex_2\\
      \DNFRegexAlt_1 \OrDNF \DNFRegex_2 \StarOf{\Rewrite}
      \DNFRegexAlt_1 \OrDNF \DNFRegexAlt_2
    }
    {
      \DNFRegex_1 \OrDNF \DNFRegex_2 \StarOf{\Rewrite}
      \DNFRegexAlt_1 \OrDNF \DNFRegexAlt_2
    }
  \]
\end{proof}

\begin{lemma}
  \label{lem:propagation-of-star-rewrites-through-singleton-concat-left}
  If $\DNFRegex_1 \StarOf{\Rewrite} \DNFRegex_2$, then for all $\Sequence$,
  $\DNFRegex_1 \ConcatDNF \DNFOf{\Sequence} \StarOf{\Rewrite}
  \DNFRegex_2 \ConcatDNF \DNFOf{\Sequence}$
\end{lemma}
\begin{proof}
  By induction on the derivation of $\StarOf{\Rewrite}$

  \begin{case}[\ReflexivityRule{}]
    \[
      \inferrule*
      {
      }
      {
        \DNFRegex_1 \StarOf{\Rewrite} \DNFRegex_1
      }
    \]

    so, by \ReflexivityRule{}
    
    \[
      \inferrule*
      {
      }
      {
        \DNFRegex_1 \ConcatDNF \DNFOf{\Sequence}
        \StarOf{\Rewrite}
        \DNFRegex_1 \ConcatDNF \DNFOf{\Sequence}
      }
    \]
  \end{case}

  \begin{case}[\BaseRule{}]
    \[
      \inferrule*
      {
        \DNFRegex_1 \Rewrite \DNFRegex_2
      }
      {
        \DNFRegex_1 \StarOf{\Rewrite} \DNFRegex_2
      }
    \]

    The only way to get a derivation of \Rewrite{} is with an application of
    \DNFStructuralRewriteRule{}, so by inversion,

    \[
      \inferrule*
      {
        \Atom_j \Rewrite \DNFRegex
      }
      {
        \DNFOf{\Sequence_1\DNFSep\ldots\DNFSep\Sequence_{i-1}} \OrDNF
        \DNFOf{\SequenceOf{\String_0\SeqSep\Atom_1\SeqSep\ldots\SeqSep\String_{j-1}}}
        \ConcatDNF \AtomToDNFOf{\Atom_j} \ConcatDNF
        \DNFOf{\SequenceOf{\String_j\SeqSep\ldots\SeqSep\Atom_m\SeqSep\String_m}}
        \OrDNF \DNFOf{\Sequence_{i+1}\DNFSep\ldots\DNFSep\Sequence_n}\Rewrite\\
        \DNFOf{\Sequence_1\DNFSep\ldots\DNFSep\Sequence_{i-1}} \OrDNF
        \DNFOf{\SequenceOf{\String_0\SeqSep\Atom_1\SeqSep\ldots\SeqSep\String_{j-1}}}\ConcatDNF\DNFRegex\ConcatDNF\SequenceOf{\String_j\SeqSep\ldots\SeqSep\Atom_m\SeqSep\String_m} \OrDNF
        \DNFOf{\Sequence_{i+1}\DNFSep\ldots\DNFSep\Sequence_n}
      }
    \]

    Consider the derivation
    \[
      \inferrule*
      {
        \Atom_j \Rewrite \DNFRegex
      }
      {
        \DNFOf{\Sequence_1\ConcatSequence\Sequence\DNFSep\ldots\DNFSep\Sequence_{i-1}\ConcatSequence\Sequence} \OrDNF
        \DNFOf{\SequenceOf{\String_0\SeqSep\Atom_1\SeqSep\ldots\SeqSep\String_{j-1}}}
        \ConcatDNF \AtomToDNFOf{\Atom_j} \ConcatDNF
        \DNFOf{\SequenceOf{\String_j\SeqSep\ldots\SeqSep\Atom_m\SeqSep\String_m}\ConcatSequence\Sequence}
        \OrDNF\\ \DNFOf{\Sequence_{i+1}\ConcatSequence\Sequence\DNFSep\ldots\DNFSep\Sequence_n\ConcatSequence\Sequence}\Rewrite\\
        \DNFOf{\Sequence_1\ConcatSequence\Sequence\DNFSep\ldots\DNFSep\Sequence_{i-1}\ConcatSequence\Sequence} \OrDNF
        \DNFOf{\SequenceOf{\String_0\SeqSep\Atom_1\SeqSep\ldots\SeqSep\String_{j-1}}}\ConcatDNF\DNFRegex\ConcatDNF\DNFOf{\SequenceOf{\String_j\SeqSep\ldots\SeqSep\Atom_m\SeqSep\String_m}\ConcatSequence\Sequence} \OrDNF\\
        \DNFOf{\Sequence_{i+1}\ConcatSequence\Sequence\DNFSep\ldots\DNFSep\Sequence_n\ConcatSequence\Sequence}
      }
    \]

    By the definition of $\ConcatDNF$, using
    Lemma~\ref{lem:dnf-distribute-singleton-left} this is equal to
    $\DNFRegex_1 \ConcatDNF \DNFOf{\Sequence} \Rewrite
    \DNFRegex_2 \ConcatDNF \DNFOf{\Sequence}$, so, consider the derivation

    \[
      \inferrule*
      {
        \inferrule*
        {
          \Atom_j \Rewrite \DNFRegex
        }
        {
          \DNFRegex_1 \ConcatDNF \DNFOf{\Sequence} \Rewrite
          \DNFRegex_2 \ConcatDNF \DNFOf{\Sequence}
        }
      }
      {
        \DNFRegex_1 \ConcatDNF \DNFOf{\Sequence} \StarOf{\Rewrite}
        \DNFRegex_2 \ConcatDNF \DNFOf{\Sequence}
      }
    \]
  \end{case}

  \begin{case}[\TransitivityRule{}]
    \[
      \inferrule*
      {
        \DNFRegex_1 \StarOf{\Rewrite} \DNFRegex_3
        \DNFRegex_3 \StarOf{\Rewrite} \DNFRegex_2
      }
      {
        \DNFRegex_1 \StarOf{\Rewrite} \DNFRegex_2
      }
    \]

    By IH, $\DNFRegex_1 \ConcatDNF \DNFOf{\Sequence} \StarOf{\Rewrite}
    \DNFRegex_3 \ConcatDNF \DNFOf{\Sequence}$.
    By IH, $\DNFOf{\Sequence} \ConcatDNF \DNFRegex_3 \StarOf{\Rewrite}
    \DNFOf{\Sequence} \ConcatDNF \DNFRegex_2$.

    Consider the derivation
    \[
      \inferrule*
      {
        \DNFOf{\Sequence} \ConcatDNF \DNFRegex_1 \StarOf{\Rewrite}
        \DNFOf{\Sequence} \ConcatDNF \DNFRegex_3\\
        \DNFOf{\Sequence} \ConcatDNF \DNFRegex_3 \StarOf{\Rewrite}
        \DNFOf{\Sequence} \ConcatDNF \DNFRegex_2
      }
      {
        \DNFOf{\Sequence} \ConcatDNF \DNFRegex_1 \StarOf{\Rewrite}
        \DNFOf{\Sequence} \ConcatDNF \DNFRegex_2
      }
    \]
  \end{case}
\end{proof}

\begin{lemma}
  \label{lem:propagation-of-star-rewrites-through-concat-right}
  If $\DNFRegex_1 \StarOf{\Rewrite} \DNFRegex_2$, then for all $\DNFRegexAlt$,
  $\DNFRegexAlt \ConcatDNF \DNFRegex_1 \StarOf{\Rewrite}
  \DNFRegexAlt \ConcatDNF \DNFRegex_2$
\end{lemma}
\begin{proof}
  By induction on the derivation of $\StarOf{\Rewrite}$

  \begin{case}[\ReflexivityRule{}]
    \[
      \inferrule*
      {
      }
      {
        \DNFRegex_1 \StarOf{\Rewrite} \DNFRegex_1
      }
    \]

    so, by \ReflexivityRule{}
    
    \[
      \inferrule*
      {
      }
      {
        \DNFRegexAlt \ConcatDNF \DNFRegex_1
        \StarOf{\Rewrite}
        \DNFRegexAlt \ConcatDNF \DNFRegex_1
      }
    \]
  \end{case}

  \begin{case}[\BaseRule{}]
    \[
      \inferrule*
      {
        \DNFRegex_1 \Rewrite \DNFRegex_2
      }
      {
        \DNFRegex_1 \StarOf{\Rewrite} \DNFRegex_2
      }
    \]

    The only way to get a derivation of \Rewrite{} is with an application of
    \DNFStructuralRewriteRule{}, so by inversion,

    \[
      \inferrule*
      {
        \Atom_j \Rewrite \DNFRegex
      }
      {
        \DNFOf{\Sequence_1\DNFSep\ldots\DNFSep\Sequence_{i-1}} \OrDNF
        \DNFOf{\SequenceOf{\String_0\SeqSep\Atom_1\SeqSep\ldots\SeqSep\String_{j-1}}}
        \ConcatDNF \AtomToDNFOf{\Atom_j} \ConcatDNF
        \DNFOf{\SequenceOf{\String_j\SeqSep\ldots\SeqSep\Atom_m\SeqSep\String_m}}
        \OrDNF \DNFOf{\Sequence_{i+1}\DNFSep\ldots\DNFSep\Sequence_n}\Rewrite\\
        \DNFOf{\Sequence_1\DNFSep\ldots\DNFSep\Sequence_{i-1}} \OrDNF
        \DNFOf{\SequenceOf{\String_0\SeqSep\Atom_1\SeqSep\ldots\SeqSep\String_{j-1}}}\ConcatDNF\DNFRegex\ConcatDNF\SequenceOf{\String_j\SeqSep\ldots\SeqSep\Atom_m\SeqSep\String_m} \OrDNF
        \DNFOf{\Sequence_{i+1}\DNFSep\ldots\DNFSep\Sequence_n}
      }
    \]

    Let $\DNFRegexAlt = \DNFOf{\Sequence_1' \DNFSep \ldots \DNFSep \Sequence_{n'}'}$.

    By
    Lemma~\ref{lem:propagation-of-star-rewrites-through-concat-right},
    $\DNFOf{\Sequence_k'} \ConcatDNF \DNFRegex_1 \ParallelRewrite
    \DNFOf{\Sequence_k'} \ConcatDNF \DNFRegex_2$.
    So, through repeated application of
    Lemma~\ref{lem:propagation-of-star-rewrites-through-or},
    $(\DNFOf{\Sequence_1'} \ConcatDNF \DNFRegex_1) \OrDNF \ldots \OrDNF
    (\DNFOf{\Sequence_{n'}'} \ConcatDNF \DNFRegex_1) \ParallelRewrite
    (\DNFOf{\Sequence_1'} \ConcatDNF \DNFRegex_2) \OrDNF \ldots \OrDNF
    (\DNFOf{\Sequence_{n'}'} \ConcatDNF \DNFRegex_2)$

    From Lemma~\ref{lem:dnf-distribute-right},
    $(\DNFOf{\Sequence_1'} \ConcatDNF \DNFRegex_1) \OrDNF \ldots \OrDNF
    (\DNFOf{\Sequence_{n'}'} \ConcatDNF \DNFRegex_1) =
    (\DNFOf{\Sequence_1'} \OrDNF \ldots \OrDNF \DNFOf{\Sequence_{n'}'})
    \ConcatDNF \DNFRegex_1 = \DNFRegexAlt \ConcatDNF \DNFRegex_1$ and
    $(\DNFOf{\Sequence_1'} \ConcatDNF \DNFRegex_2) \OrDNF \ldots \OrDNF
    (\DNFOf{\Sequence_{n'}'} \ConcatDNF \DNFRegex_2) =
    (\DNFOf{\Sequence_1'} \OrDNF
    \ldots \OrDNF \DNFOf{\Sequence_{n'}'}) \ConcatDNF \DNFRegex_2 =
    \DNFRegexAlt \ConcatDNF \DNFRegex_2$.

    So we have
    $\DNFRegexAlt \ConcatDNF \DNFRegex_1 \StarOf{\Rewrite}
    \DNFRegexAlt \ConcatDNF \DNFRegex_2$.
  \end{case}

  \begin{case}[\TransitivityRule{}]
    \[
      \inferrule*
      {
        \DNFRegex_1 \StarOf{\Rewrite} \DNFRegex_3
        \DNFRegex_3 \StarOf{\Rewrite} \DNFRegex_2
      }
      {
        \DNFRegex_1 \StarOf{\Rewrite} \DNFRegex_2
      }
    \]

    By IH, $\DNFRegex_1 \ConcatDNF \DNFRegexAlt \StarOf{\Rewrite}
    \DNFRegex_3 \ConcatDNF \DNFRegexAlt$.
    By IH, $\DNFRegex_3 \ConcatDNF \DNFRegexAlt \StarOf{\Rewrite}
    \DNFRegex_2 \ConcatDNF \DNFRegexAlt$.

    Consider the derivation
    \[
      \inferrule*
      {
        \DNFRegex_1 \ConcatDNF \DNFRegexAlt \StarOf{\Rewrite}
        \DNFRegex_3 \ConcatDNF \DNFRegexAlt\\
        \DNFRegex_3 \ConcatDNF \DNFRegexAlt \StarOf{\Rewrite}
        \DNFRegex_2 \ConcatDNF \DNFRegexAlt
      }
      {
        \DNFRegex_1 \ConcatDNF \DNFRegexAlt \StarOf{\Rewrite}
        \DNFRegex_2 \ConcatDNF \DNFRegexAlt
      }
    \]
  \end{case}
\end{proof}

\begin{lemma}
  \label{lem:propagation-of-star-rewrites-through-sequence}
  Let $\AtomToDNFOf{\Atom_i} \StarOf{\Rewrite} \DNFRegex_i$.
  $\DNFOf{\SequenceOf{\String_0 \SeqSep \Atom_1 \SeqSep \ldots \SeqSep \Atom_n \SeqSep \String_n}}
  \StarOf{\Rewrite}
  \DNFOf{\SequenceOf{\String_0}} \ConcatDNF \DNFRegex_1 \ConcatDNF \ldots
  \ConcatDNF \DNFRegex_n \ConcatDNF \DNFOf{\SequenceOf{\String_n}}$
\end{lemma}
\begin{proof}
  By induction on $n$.
  \begin{case}[$n=0$]
    Through use of \ReflexivityRule{}
    \[
      \inferrule*
      {
      }
      {
        $\DNFOf{\SequenceOf{\String_0}} \StarOf{\Rewrite}
        \DNFOf{\SequenceOf{\String_0}}$
      }
    \]
  \end{case}

  \begin{case}[$n>0$]
    $\DNFOf{\SequenceOf{\String_0 \SeqSep \Atom_1 \SeqSep \ldots \SeqSep \Atom_n \SeqSep \String_n}} =
    \DNFOf{\SequenceOf{\String_0 \SeqSep \Atom_1 \SeqSep \ldots \SeqSep \Atom_{n-1} \SeqSep \String_{n-1}}}
    \ConcatDNF
    \DNFOf{\SequenceOf{\EmptyString \SeqSep \Atom_n \SeqSep \String_n}}$ by the definition of
    $\ConcatDNF$.
    
    From IH,
    $\DNFOf{\SequenceOf{\String_0 \SeqSep \Atom_1 \SeqSep \ldots \SeqSep \Atom_{n-1} \SeqSep \String_{n-1}}}
    \StarOf{\Rewrite}
    \DNFOf{\SequenceOf{\String_0}} \ConcatDNF \DNFRegex_1 \ConcatDNF \ldots
    \ConcatDNF \DNFRegex_{n-1} \ConcatDNF \DNFOf{\SequenceOf{\String_{n-1}}}$
    From
    Lemma~\ref{lem:propagation-of-star-rewrites-through-singleton-concat-left},
    $\DNFOf{\SequenceOf{\String_0 \SeqSep \Atom_1 \SeqSep \ldots \SeqSep \Atom_{n-1} \SeqSep \String_{n-1}}}
    \ConcatDNF
    \DNFOf{\SequenceOf{\EmptyString \SeqSep \Atom_n \SeqSep \String_n}}
    \StarOf{\Rewrite}
    \DNFOf{\SequenceOf{\String_0}} \ConcatDNF \DNFRegex_1 \ConcatDNF \ldots
    \ConcatDNF \DNFRegex_{n-1} \ConcatDNF
    \DNFOf{\SequenceOf{\EmptyString \SeqSep \Atom_n \SeqSep \String_n}}$.

    $\DNFOf{\SequenceOf{\EmptyString \SeqSep \Atom_n \SeqSep \String_n}} =
    \AtomToDNFOf{\Atom_n} \ConcatDNF \DNFOf{\SequenceOf{\String_n}}$
    From
    Lemma~\ref{lem:propagation-of-star-rewrites-through-singleton-concat-left},
    as $\AtomToDNFOf{\Atom_n} \StarOf{\Rewrite} \DNFRegex_n$
    $\DNFOf{\SequenceOf{\EmptyString \SeqSep \Atom_n \SeqSep \String_n}} \StarOf{\Rewrite}
    \DNFRegex_n \ConcatDNF \DNFOf{\SequenceOf{\String_n}}$.

    As $\DNFOf{\SequenceOf{\EmptyString \SeqSep \Atom_n \SeqSep \String_n}} \StarOf{\Rewrite}
    \DNFRegex_n \ConcatDNF \DNFOf{\SequenceOf{\String_n}}$, from
    Lemma~\ref{lem:propagation-of-star-rewrites-through-concat-right},
    $\DNFOf{\SequenceOf{\String_0}} \ConcatDNF \DNFRegex_1 \ConcatDNF \ldots
    \ConcatDNF \DNFRegex_{n-1} \ConcatDNF
    \DNFOf{\SequenceOf{\EmptyString \SeqSep \Atom_n \SeqSep \String_n}} \StarOf{\Rewrite}
    \DNFOf{\SequenceOf{\String_0}} \ConcatDNF \DNFRegex_1 \ConcatDNF \ldots
    \ConcatDNF \DNFRegex_{n-1} \ConcatDNF \DNFRegex_n \ConcatDNF
    \DNFOf{\SequenceOf{\String_n}}$.

    Consider the derivation
    \[
      \inferrule*
      {
        \DNFOf{\SequenceOf{\String_0 \SeqSep \Atom_1 \SeqSep \ldots \SeqSep \Atom_{n-1} \SeqSep \String_{n-1}}}
        \ConcatDNF
        \DNFOf{\SequenceOf{\EmptyString \SeqSep \Atom_n \SeqSep \String_n}}
        \StarOf{\Rewrite}
        \DNFOf{\SequenceOf{\String_0}} \ConcatDNF \DNFRegex_1 \ConcatDNF \ldots
        \ConcatDNF \DNFRegex_{n-1} \ConcatDNF
        \DNFOf{\SequenceOf{\EmptyString \SeqSep \Atom_n \SeqSep \String_n}}\\
        \DNFOf{\SequenceOf{\String_0}} \ConcatDNF \DNFRegex_1 \ConcatDNF \ldots
        \ConcatDNF \DNFRegex_{n-1} \ConcatDNF
        \DNFOf{\SequenceOf{\EmptyString \SeqSep \Atom_n \SeqSep \String_n}} \StarOf{\Rewrite}
        \DNFOf{\SequenceOf{\String_0}} \ConcatDNF \DNFRegex_1 \ConcatDNF \ldots
        \ConcatDNF \DNFRegex_{n-1} \ConcatDNF \DNFRegex_n \ConcatDNF
        \DNFOf{\SequenceOf{\String_n}}
      }
      {
        \DNFOf{\SequenceOf{\String_0 \SeqSep \Atom_1 \SeqSep \ldots \SeqSep \Atom_n \SeqSep \String_n}} =
        \DNFOf{\SequenceOf{\String_0 \SeqSep \Atom_1 \SeqSep \ldots \SeqSep \Atom_{n-1} \SeqSep \String_{n-1}}}
        \ConcatDNF
        \DNFOf{\SequenceOf{\EmptyString \SeqSep \Atom_n \SeqSep \String_n}} \StarOf{\Rewrite}\\
        \DNFOf{\SequenceOf{\String_0}} \ConcatDNF \DNFRegex_1 \ConcatDNF \ldots
        \ConcatDNF \DNFRegex_{n-1} \ConcatDNF \DNFRegex_n \ConcatDNF
        \DNFOf{\SequenceOf{\String_n}}
      }
    \]

    So, $\DNFOf{\SequenceOf{\String_0 \SeqSep \Atom_1 \SeqSep \ldots \SeqSep \Atom_n \SeqSep \String_n}}
    \StarOf{\Rewrite}
    \DNFOf{\SequenceOf{\String_0}} \ConcatDNF \DNFRegex_1 \ConcatDNF \ldots
    \ConcatDNF \DNFRegex_n \ConcatDNF \DNFOf{\SequenceOf{\String_n}}$,
    as desired.
  \end{case}
\end{proof}

\begin{lemma}
  \label{lem:parallel-expressible-in-star-serial}
  \leavevmode
  \begin{itemize}
  \item If $\Atom \ParallelRewriteAtom \DNFRegexAlt$, then
    $\AtomToDNFOf{\DNFRegex} \StarOf{\Rewrite} \DNFRegexAlt$.
  \item If $\DNFRegex \ParallelRewrite \DNFRegexAlt$, then
    $\AtomToDNFOf{\DNFRegex} \StarOf{\Rewrite} \DNFRegexAlt$.
  \end{itemize}
\end{lemma}
\begin{proof}
  By mutual induction on the derivation of $\ParallelRewriteAtom$
  \begin{case}[\AtomUnrollstarLeftRule{}]
    \[
      \inferrule*
      {
      }
      {
        \StarOf{\DNFRegex}\ParallelRewriteAtom
        \OrDNFOf{\DNFOf{\SequenceOf{\EmptyString}}}{(\ConcatDNFOf{\DNFRegex}{\AtomToDNFOf{\StarOf{\DNFRegex}}})}
      }
    \]

    Consider the derivation
    
    \[
      \inferrule*
      {
        \inferrule*
        {
          \inferrule*
          {
          }
          {
            \StarOf{\DNFRegex}\RewriteAtom
            \OrDNFOf{\DNFOf{\SequenceOf{\EmptyString}}}{(\ConcatDNFOf{\DNFRegex}{\AtomToDNFOf{\StarOf{\DNFRegex}}})}
          }
        }
        {
          \AtomToDNFOf{\StarOf{\DNFRegex}} \Rewrite
          \OrDNFOf{\DNFOf{\SequenceOf{\EmptyString}}}{(\ConcatDNFOf{\DNFRegex}{\AtomToDNFOf{\StarOf{\DNFRegex}}})}
        }
      }
      {
        \AtomToDNFOf{\StarOf{\DNFRegex}} \StarOf{\Rewrite}
        \OrDNFOf{\DNFOf{\SequenceOf{\EmptyString}}}{(\ConcatDNFOf{\DNFRegex}{\AtomToDNFOf{\StarOf{\DNFRegex}}})}
      }
    \]
  \end{case}

  \begin{case}[\AtomUnrollstarRightRule{}]
    \[
      \inferrule*
      {
      }
      {
        \StarOf{\DNFRegex}\ParallelRewriteAtom
        \OrDNFOf{\DNFOf{\SequenceOf{\EmptyString}}}{(\ConcatDNFOf{\AtomToDNFOf{\StarOf{\DNFRegex}}}{\DNFRegex})}
      }
    \]

    Consider the derivation
    
    \[
      \inferrule*
      {
        \inferrule*
        {
          \inferrule*
          {
          }
          {
            \StarOf{\DNFRegex}\RewriteAtom
            \OrDNFOf{\DNFOf{\SequenceOf{\EmptyString}}}{(\ConcatDNFOf{\AtomToDNFOf{\StarOf{\DNFRegex}}}{\DNFRegex})}
          }
        }
        {
          \AtomToDNFOf{\StarOf{\DNFRegex}} \Rewrite
          \OrDNFOf{\DNFOf{\SequenceOf{\EmptyString}}}{(\ConcatDNFOf{\DNFRegex}{\AtomToDNFOf{\StarOf{\DNFRegex}}})}
        }
      }
      {
        \AtomToDNFOf{\StarOf{\DNFRegex}} \StarOf{\Rewrite}
        \OrDNFOf{\DNFOf{\SequenceOf{\EmptyString}}}{(\ConcatDNFOf{\DNFRegex}{\AtomToDNFOf{\StarOf{\DNFRegex}}})}
      }
    \]
  \end{case}

  \begin{case}[\ParallelAtomStructuralRewriteRule{}]
    \[
      \inferrule*
      {
        \DNFRegex \ParallelRewrite \DNFRegexAlt'
      }
      {
        \StarOf{\DNFRegex} \ParallelRewrite \AtomToDNFOf{\StarOf{\DNFRegexAlt'}}
      }
    \]
    
    By IH, $\DNFRegex \StarOf{\Rewrite} \DNFRegexAlt'$, so
    by Lemma~\ref{lem:propagation-of-rewrites-through-atom-structure},
    $\AtomToDNFOf{\StarOf{\DNFRegex}} \StarOf{\Rewrite}
    \AtomToDNFOf{\StarOf{\DNFRegexAlt'}}$.
  \end{case}

  \begin{case}[ParallelDNFStructuralRewriteRule{}]
    \[
      \inferrule*
      {
        \DNFRegex = \DNFOf{\Sequence_1 \DNFSep \ldots \DNFSep \Sequence_n}\\
        \forall i. \Sequence_i =
        \SequenceOf{\String_{i,0} \SeqSep \Atom_{i,1} \SeqSep \ldots \SeqSep \Atom_{i,n_i} \SeqSep \String_{i,n_i}}\\
        \forall i,j. \Atom_{i,j} \ParallelRewriteAtom \DNFRegex_{i,j}\\
        \forall i. \DNFRegex_i = \DNFOf{\SequenceOf{\String_{i,0}}} \ConcatDNF \DNFRegex_{i,1}
        \ConcatDNF \ldots \ConcatDNF \DNFRegex_{i,n_i} \ConcatDNF
        \DNFOf{\SequenceOf{\String_{i,n_i}}}
      }
      {
        \DNFRegex \ParallelRewrite \DNFRegex_1 \OrDNF \ldots \OrDNF \DNFRegex_n
      }
    \]

    From the definition of $\OrDNF$,
    $\DNFRegex = \DNFOf{\Sequence_1} \OrDNF \ldots \OrDNF \DNFOf{\Sequence_n}$.
    From IH, $\AtomToDNFOf{\Atom_{i,1}} \StarOf{\Rewrite} \DNFRegex_{i,j}$.
    From Lemma~\ref{lem:propagation-of-star-rewrites-through-sequence},
    $\DNFOf{\Sequence_i} =
    \DNFOf{\SequenceOf{\String_{i,0} \SeqSep \Atom_{i,1} \SeqSep \ldots \SeqSep \Atom_{i,n_i} \SeqSep \String_{i,n_i}}}
    \StarOf{\Rewrite}
    \DNFOf{\SequenceOf{\String_{i,0}}} \ConcatDNF \DNFRegex_{i,1}
    \ConcatDNF \ldots \ConcatDNF \DNFRegex_{i,n_i} \ConcatDNF
    \DNFOf{\SequenceOf{\String_{i,n_i}}} = \DNFRegex_i$,
    so $\DNFOf{\Sequence_i} \StarOf{\Rewrite} \DNFRegex_i$.

    From Lemma~\ref{lem:propagation-of-star-rewrites-through-or},
    $\DNFRegex = \DNFOf{\Sequence_i} \OrDNF \ldots \OrDNF \DNFOf{\Sequence_n}
    \StarOf{\Rewrite}
    \DNFRegex_1 \OrDNF \ldots \OrDNF \DNFRegex_n$.
  \end{case}

  \begin{case}[\IdentityRewriteRule]
    \[
      \inferrule*
      {
      }
      {
        \DNFRegex \ParallelRewrite \DNFRegex
      }
    \]

    Through application of \ReflexivityRule{}
    \[
      \inferrule*
      {
      }
      {
        \DNFRegex \StarOf{\Rewrite} \DNFRegex
      }
    \]
  \end{case}
\end{proof}

\begin{lemma}
  \label{lem:star-parallel-expressible-in-star-serial}
  If $\DNFRegex \StarOf{\ParallelRewrite} \DNFRegexAlt$, then
  $\DNFRegex \StarOf{\Rewrite} \DNFRegexAlt$
\end{lemma}
\begin{proof}
  By induction on the derivation of $\StarOf{\ParallelRewrite}$

  \begin{case}[\ReflexivityRule{}]
    \[
      \inferrule*
      {
      }
      {
        \DNFRegex \StarOf{\ParallelRewrite} \DNFRegex
      }
    \]

    Consider the following derivation

    \[
      \inferrule*
      {
      }
      {
        \DNFRegex \StarOf{\Rewrite} \DNFRegex
      }
    \]
  \end{case}

  \begin{case}[\BaseRule{}]
    \[
      \inferrule*
      {
        \DNFRegex \ParallelRewrite \DNFRegexAlt
      }
      {
        \DNFRegex \StarOf{\ParallelRewrite} \DNFRegexAlt
      }
    \]

    By Lemma~\ref{lem:parallel-expressible-in-star-serial},
    $\DNFRegex \StarOf{\Rewrite} \DNFRegexAlt$.
  \end{case}

  \begin{case}[\TransitivityRule{}]
    \[
      \inferrule*
      {
        \DNFRegex \StarOf{\ParallelRewrite} \DNFRegex'\\
        \DNFRegex' \StarOf{\ParallelRewrite} \DNFRegexAlt
      }
      {
        \DNFRegex \StarOf{\ParallelRewrite} \DNFRegexAlt
      }
    \]

    By IH, $\DNFRegex \StarOf{\Rewrite} \DNFRegex'$ and
    $\DNFRegex' \StarOf{\Rewrite} \DNFRegexAlt$.

    Consider the following derivation
    \[
      \inferrule*
      {
        \DNFRegex \StarOf{\Rewrite} \DNFRegex'\\
        \DNFRegex' \StarOf{\Rewrite} \DNFRegexAlt
      }
      {
        \DNFRegex \StarOf{\Rewrite} \DNFRegexAlt
      }
    \]
  \end{case}
\end{proof}

\begin{theorem}
  \label{thm:parallel-star-equivalence}
  $\DNFRegex \StarOf{\ParallelRewrite} \DNFRegex'$, if, and only if $\DNFRegex
  \StarOf{\Rewrite} \DNFRegex'$
\end{theorem}
\begin{proof}
  Forward direction is proven by
  Lemma~\ref{lem:star-serial-expressible-in-star-parallel}.
  Reverse direction is proven by
  Lemma~\ref{lem:star-parallel-expressible-in-star-serial}.
\end{proof}

\begin{corollary}[$\StarOf{\Rewrite}$ Maintained Under Iteration]
  \label{cor:rewrite-maintained-iteration}
  If $\DNFRegex \StarOf{\Rewrite} \DNFRegexAlt$, then
  $\DNFOf{\SequenceOf{\StarOf{\DNFRegex}}} \StarOf{\Rewrite}
  \DNFOf{\SequenceOf{\StarOf{\DNFRegexAlt}}}$.
\end{corollary}
\begin{proof}
  From Theorem~\ref{thm:parallel-star-equivalence} applied to
  Lemma~\ref{lem:star-parallel-rewrite-iteration}.
\end{proof}

\begin{lemma}[$\ParallelRewrite$ can be expressed in $\ParallelRewriteSwap$]
  \label{parallelrewrite-in-parallelrewriteswap}
  If $\DNFRegex \ParallelRewrite \DNFRegexAlt$ then
  $\DNFRegex \ParallelRewriteSwap \DNFRegexAlt$
\end{lemma}
\begin{proof}
  $\ParallelRewriteSwap$ has all of the inference rules of $\ParallelRewrite$,
  so a straightforward induction using those rules can prove this.
\end{proof}

\begin{lemma}[$\ParallelRewrite$ can be expressed in $\SSREquiv$]
  \label{lem:pr-in-de}
  If $\ToDNFRegexOf{\Regex} \ParallelRewrite \ToDNFRegexOf{\RegexAlt}$,
  then $\Regex \SSREquiv \RegexAlt$.
\end{lemma}
\begin{proof}
  By Lemma~\ref{parallelrewrite-in-parallelrewriteswap},
  $\ToDNFRegexOf{\Regex} \ParallelRewriteSwap \ToDNFRegexOf{\RegexAlt}$, then
  by Lemma~\ref{lem:express-swap-in-equiv},
  $\Regex \SSREquiv \RegexAlt$.
\end{proof}

\begin{lemma}[$\StarOf{\ParallelRewrite}$ can be expressed in
  $\SSREquiv$]
  \label{lem:star-pr-in-de}
  If $\ToDNFRegexOf{\Regex} \StarOf{\ParallelRewrite} \ToDNFRegexOf{\RegexAlt}$, then
  $\Regex \SSREquiv \RegexAlt$.
\end{lemma}
\begin{proof}
  By straightforward induction, using for base rule, Lemma~\ref{lem:pr-in-de},
  for transitivity the transitivity of equational theories, and for reflexivity
  the reflexivity of equational theories.
\end{proof}

\begin{lemma}[$\StarOf{\Rewrite}$ can be expressed in $\SSREquiv$]
  \label{lem:star-rw-in-de}
  If $\ToDNFRegexOf{\Regex} \StarOf{\Rewrite} \ToDNFRegexOf{\RegexAlt}$, then
  $\Regex \SSREquiv \RegexAlt$.
\end{lemma}
\begin{proof}
  By Lemma~\ref{lem:star-pr-in-de} and
  Theorem~\ref{thm:parallel-star-equivalence}.
\end{proof}

\subsection{Lens Soundness}

Using the above machinery, we prove the soundness of DNF lenses.  The unambiguity is
guaranteed through prior unambiguity proofs.  The rewrite portion of DNF lenses
are proven to be correct through the above subsection.  The bulk of this is
showing that the lenses can be built up from their subcomponents, and that
arbitrary permutations can be expressed.

\label{soundness}
\begin{lemma}[Expressibility of Safe Boilerplate Alterations]
  \label{lem:boilerplate-alterations}
  Suppose
  \begin{enumerate}
  \item $\UnambigConcat\SequenceOf{\String_0 \SeqSep \Atom_1 \SeqSep \ldots \SeqSep \Atom_n \SeqSep \String_n}$
  \item $\UnambigConcat\SequenceOf{\StringAlt_0 \SeqSep \Atom_1 \SeqSep \ldots \SeqSep \Atom_n \SeqSep \StringAlt_n}$
  \end{enumerate}
  Then there exists a lens
  $\Lens \OfType \Regex \Leftrightarrow \RegexAlt$ such that
  \begin{enumerate}
  \item $\Regex = \ToRegex(\SequenceOf{\String_0 \SeqSep \Atom_1 \SeqSep \ldots \SeqSep \Atom_n \SeqSep \String_n})$
  \item $\RegexAlt = \ToRegex(\SequenceOf{\StringAlt_0 \SeqSep \Atom_1 \SeqSep \ldots \SeqSep \Atom_n \SeqSep \StringAlt_n})$
  \item $\SemanticsOf{\Lens}=\SetOf{(\String,\StringAlt)\SuchThat
      \String=\String_0\Concat\String_1'\Concat\ldots\Concat\String_n'\Concat\String_n
      \BooleanAnd\\
      \hspace*{6.1em}\StringAlt=\StringAlt_0\Concat\String_1'\Concat\ldots\Concat\String_n'\Concat\StringAlt_n
      \BooleanAnd\\
      \hspace*{6.1em}\String_i\in\LanguageOf{\Atom_i}}$
  \end{enumerate}
\end{lemma}
\begin{proof}
  By induction on $n$.

  Let $n=0$.
  Consider the Lens
  \begin{mathpar}
    \inferrule*
    {
    }
    {
      \ConstLensOf{\String_0}{\StringAlt_0} \OfType \String_0 \Leftrightarrow \StringAlt_0
    }
  \end{mathpar}
  By inspection, this satisfies the desired properties.

  Let $n>0$.
  By induction, there exists a lens $\Lens \OfType \Regex \Leftrightarrow \RegexAlt$
  satisfying the desired properties.
  Consider the lens
  \begin{mathpar}
    \inferrule*[left=\Derivation]
    {
      \Lens \OfType \Regex \Leftrightarrow \RegexAlt\\
      \inferrule*
      {
      }
      {
        \ConstLensOf{\String_n}{\StringAlt_n} \OfType \String_n \Leftrightarrow \StringAlt_n
      }
    }
    {
      \ConcatLensOf{\Lens}{\ConstLensOf{\String_n}{\StringAlt_n}}
      \OfType
      \Regex \Concat \String_n \Leftrightarrow
      \RegexAlt \Concat \StringAlt_n
    }

    \inferrule*
    {
      \Derivation\\
      \IdentityLensOf{\ToRegex(\Atom_n)} \OfType \ToRegex(\Atom_n) \Leftrightarrow \ToRegex(\Atom_n)
    }
    {
      \ConcatLensOf{\ConcatLensOf{\Lens}{\ConstLensOf{\String_n}{\StringAlt_n}}}{\IdentityLensOf{\ToRegex(\Atom_n)}}
      \OfType\\
      \Regex \Concat \String_n \Concat \ToRegex(\Atom_n) \Leftrightarrow
      \Regex \Concat \StringAlt_n \Concat \ToRegex(\Atom_n)
    }
  \end{mathpar}
  By inspection, this satisfies the desired properties.
\end{proof}

\begin{lemma}[Creation of Lens from Identity Perm Sequence Lens]
  \label{lem:id-clause}
  Suppose
  \begin{enumerate}
  \item $\Sequence=\SequenceOf{\String_0  \SeqSep  \Atom_1  \SeqSep  \ldots  \SeqSep  \Atom_n \SeqSep  \String_n}$
  \item $\SequenceAlt=\SequenceOf{\StringAlt_0  \SeqSep  \AtomAlt_1  \SeqSep  \ldots  \SeqSep  \AtomAlt_n  \SeqSep  \StringAlt_n}$
  \item $(\SequenceLensOf{(\String_0,\StringAlt_0) \SeqLSep \AtomLens_1  \SeqLSep  \ldots  \SeqLSep 
      \AtomLens_n \SeqLSep (\String_n,\StringAlt_n)},id) \OfRewritelessType
    \Sequence \Leftrightarrow \SequenceAlt$
  \item For each $\AtomLens_i \OfRewritelessType \Atom_i \Leftrightarrow \AtomAlt_i$,
    there exists a $\Lens_i \OfRewritelessType \ToRegex(\Atom_i) \Leftrightarrow
    \ToRegex(\AtomAlt_i)$ such that $\SemanticsOf{\Lens_i}=\SemanticsOf{\AtomLens_i}$
  \end{enumerate}
  then there exists a $\Lens \OfRewritelessType \ToRegex(\Sequence) \Leftrightarrow \ToRegex(\DNFRegexAlt)$ such that
  $\SemanticsOf{\Lens} =
  \SemanticsOf{(\SequenceLensOf{(\String_0,\StringAlt_0) \SeqLSep \AtomLens_1  \SeqLSep  \ldots  \SeqLSep  \AtomLens_n \SeqLSep (\String_n,\StringAlt_n)},id)}$.
  \begin{proof}
    By induction on $n$.

    Let $n=0$, $(\SequenceLensOf{(\String_0,\StringAlt_0)},id) \OfRewritelessType
    \SequenceOf{\String_0} \Leftrightarrow \SequenceOf{\StringAlt_0}$.
    Then consider
    \begin{mathpar}
      \inferrule[]
      {
      }
      {
        \ConstLensOf{\String_0}{\StringAlt_0}\OfType\String_0\Leftrightarrow\StringAlt_0
      }
    \end{mathpar}

    $\String_0=\ToRegex(\SequenceOf{\String_0})$,
    and
    $\StringAlt_0=\ToRegex(\SequenceOf{\StringAlt_0})$.
    $\SemanticsOf{\ConstLensOf{\String_0}{\StringAlt_0}}=
    \SetOf{\String_0,\StringAlt_0}=
    \SemanticsOf{\SequenceOf{(\String_0,\StringAlt_0)},id)}$.

    Let $n>0$.
    Let $\Sequence'=\SequenceOf{\String_0\SeqSep\Atom_1\SeqSep
      \ldots\SeqSep\Atom_{n-1}\SeqSep\String_{n-1}}$,
    and $\SequenceAlt'=\SequenceOf{\StringAlt_0\SeqSep\AtomAlt_1\SeqSep
      \ldots\SeqSep\AtomAlt_{n-1}\SeqSep\StringAlt_{n-1}}$
    By induction assumption, there exists a typing derivation
    \begin{mathpar}
      \Lens\OfType\ToRegex(\Sequence')\Leftrightarrow\ToRegex(\SequenceAlt')
    \end{mathpar}
    satisfying $\SemanticsOf{\Lens}=\\
    \SemanticsOf{(\SequenceLensOf{(\String_0,\StringAlt_0) \SeqLSep \AtomLens_1  \SeqLSep 
        \ldots  \SeqLSep  \AtomLens_{n-1} \SeqLSep (\String_{n-1},\StringAlt_{n-1})},id)}$

    By problem statement, there exists a typing derivation
    \begin{mathpar}
      \Lens_{\AtomLens_{n}} \OfType
      \ToRegex(\Atom_{n}) \Leftrightarrow \ToRegex(\AtomAlt_{n})
    \end{mathpar}
    satisfying $\SemanticsOf{\Lens_{\Atom_n}}
    =\SemanticsOf{\Atom_n}$.

    Consider the following lens typing
    \begin{mathpar}
      \inferrule*[left=\Derivation{}]
      {
        \Derivation_n\\
        \inferrule*
        {
        }
        {
          \ConstLensOf{\String_n}{\StringAlt_n}
          \OfType
          \String_n \Leftrightarrow \StringAlt_n
        }
      }
      {
        \ConcatLensOf{\Lens_{\AtomLens_n}}{\ConstLensOf{\String_n}{\StringAlt_n}}
        \OfType
        \ToRegex(\Atom_n)\Concat\String_n \Leftrightarrow
        \ToRegex(\AtomAlt_n)\Concat\StringAlt_n
      }

      \inferrule*
      {
        \Lens\OfType\ToRegex(\Sequence) \Leftrightarrow \ToRegex(\SequenceAlt)\\
        \Derivation{}
      }
      {
        \ConcatLensOf
        {\Lens}
        {\ConcatLensOf{\Lens_{\AtomLens_n}}{\ConstLensOf{\String_n}{\StringAlt_n}}}
        \OfType\\
        \ToRegex(\Sequence)\Concat\ToRegex(\Atom_n)\Concat\String_n \Leftrightarrow
        \ToRegex(\SequenceAlt)\Concat\ToRegex(\AtomAlt_n)\Concat\StringAlt_n
      }
    \end{mathpar}

    \SemanticsOf{\ConcatLensOf
      {\Lens}
      {\ConcatLensOf{\Lens_{\AtomLens_n}}{\ConstLensOf{\String_n}{\StringAlt_n}}}}\\
    \hspace*{3em}=\SetOf{(\String,\StringAlt)
      \SuchThat
      \String = \String'\Concat\String''\Concat\String_n\BooleanAnd
      \StringAlt = \StringAlt'\Concat\StringAlt''\Concat\StringAlt_n\BooleanAnd\\
      \hspace*{7em}
      (\String',\StringAlt')\in\SemanticsOf{\Lens}\BooleanAnd
      (\String'',\StringAlt'')\in\SemanticsOf{\Lens_{\AtomLens_n}}}\\
    \hspace*{3em}=\SetOf{
      (\String,\StringAlt)\SuchThat
      \String=
      \String_0\Concat\String_0'\Concat\ldots
      \Concat\String_{n-1}'\Concat\String_{n-1}
      \Concat \String_n \Concat \String_n'\BooleanAnd\\
      \hspace*{7em}\StringAlt=\StringAlt_0\Concat\StringAlt_0'\Concat\ldots
      \Concat\StringAlt_{n-1}'\Concat\StringAlt_{n-1}
      \Concat \StringAlt_n \Concat \StringAlt_n'\BooleanAnd\\
      \hspace*{7em}\String_i'\in\Atom_i\BooleanAnd\StringAlt_i'\in\AtomAlt_i}\\
    \hspace*{3em}=\SemanticsOf{(\SequenceLensOf{(\String_0,\StringAlt_0) \SeqLSep \AtomLens_1  \SeqLSep 
        \ldots  \SeqLSep  \AtomLens_n \SeqLSep (\String_n,\StringAlt_{n-1})},id)}
  \end{proof}
\end{lemma}

\begin{lemma}[Unambiguity of $\Sep$]
  \label{lem:sep_unambiguity}
  Let $\Alphabet$ be an alphabet.  Let $\Alphabet_{\Sep}=\Alphabet\Union\SetOf{\Sep}$,
  where \Sep{} is a character not in \Alphabet{}.
  If $\Language_1, \ldots,
  \Language_n$, are languages in $\StarOf{\Alphabet}$, then
  $\UnambigConcat\SequenceOf{\LanguageOf{\Sep};\Language_1;\LanguageOf{\Sep};
    \ldots;\LanguageOf{\Sep};\Language_n;\LanguageOf{\Sep}}$.
\end{lemma}
\begin{proof}
  We prove this by induction on $n$.

  Let $n=0$.  $\UnambigConcat\SequenceOf{\LanguageOf{\Sep}}$, as
  $\UnambigConcat\SequenceOf{\Language}$, for any language $\Language$.

  Let $n>0$.
  Let $\String_i, \StringAlt_i\in\Language_i$ for all $i\in\RangeIncInc{1}{n}$,
  and let $\Sep\String_1\Sep\ldots\Sep\String_n\Sep=\Sep\StringAlt_1\Sep\ldots\Sep\StringAlt_n\Sep$.
  We want to show that $\String_n\Sep=\StringAlt_n\Sep$.
  If they were not equal, then one string is strictly contained in the other, say without
  loss of generality $\String_n\Sep$ is strictly contained in $\StringAlt_n\Sep$.
  Because of that $\Sep\String_n\Sep$ is contained in $\StringAlt_n\Sep$, so $\Sep$
  is contained in $\StringAlt_n\in\StarOf{\Sigma}$.  This is a contradiction,
  as $\Sep\notin\Sigma$, so we know $\String_n\Sep=\StringAlt_n\Sep$, and so $\String_n=\StringAlt_n$.
  This means that
  $\Sep\String_0\Sep\ldots\Sep\String_{n-1}\Sep=\Sep\StringAlt_0\Sep\ldots\Sep\StringAlt_{n-1}$,
  so by induction, I know $\String_i=\StringAlt_i$ for all $i$.
\end{proof}

\begin{definition}[Adjacent Swapping Permutation]
  Let $\sigma_{i} \in S_n$ be the permutation where
  $\sigma_{i}(i) = i+1$, $\sigma_{i}(i+1) = i$, $\sigma_{i}(k) = k$
  when $k\neq i$, and $k\neq i+1$.
\end{definition}

\begin{lemma}[Expressibility of Adjacent Swapping Permutation Lens]
  \label{lem:adj-perm-exp}
  Suppose
  \begin{enumerate}
  \item $\sigma_i$ is an adjacent element swapping permutation
  \item $\SequenceOf{\Sep \SeqSep \Atom_1 \SeqSep \Sep\ldots\Sep \SeqSep \Atom_n \SeqSep \Sep}$ is a sequence with
    all base strings equal to $\Sep$.
  \end{enumerate}
  Then there exists a typing of a lens $\Lens \OfType \Regex \Leftrightarrow \RegexAlt$ such that
  \begin{enumerate}
  \item $\LanguageOf{\Regex}=\LanguageOf{[\Sep \SeqSep \Atom_1 \SeqSep \ldots \SeqSep \Atom_n \SeqSep \Sep]}$
  \item $\LanguageOf{\RegexAlt}=\LanguageOf{[\Sep \SeqSep \Atom_{\sigma_i(1)} \SeqSep \ldots \SeqSep \Atom_{\sigma_i(n)} \SeqSep \Sep]}$
  \item $\SemanticsOf{\Lens}=
    \SetOf{(\String,\StringAlt)\SuchThat\String=\Sep\Concat\String_1\Concat\Sep\Concat\ldots\Concat\Sep\Concat\String_n\Concat\Sep
      \BooleanAnd\\
      \hspace*{6em}\StringAlt=\Sep\Concat\String_{\sigma_i(1)}\Concat\Sep\Concat\ldots\Concat\Sep\Concat\String_{\sigma_i(n)}\Sep\BooleanAnd\\
      \hspace*{6em}\String_i\in\LanguageOf{\Atom_i}}$
  \end{enumerate}
  \begin{proof}
    By the soundness of regular expressions, define regular expressions
    $\Regex_1, \Regex_2, \Regex_3, \Regex_4$ as
    $\Regex_1=\ToRegex([\Sep \SeqSep \Atom_1 \SeqSep \ldots \SeqSep \Atom_{i-1} \SeqSep \Sep])$,
    $\Regex_2=\ToRegex(\Atom_i)$,
    $\Regex_3=\ToRegex(\Atom_{i+1})$, and
    $\Regex_4=\ToRegex([\Sep \SeqSep \Atom_{i+1} \SeqSep \ldots \SeqSep \Atom_{n} \SeqSep \Sep])$.
    Consider the following deduction
    \begin{mathpar}

      \inferrule*[left=\Derivation{}]
      {
        \inferrule*
        {
        }
        {
          \IdentityLensOf{\Sep} \OfType \Sep \Leftrightarrow \Sep
        }
        \inferrule*
        {
        }
        {
          \IdentityLensOf{\Regex_3} \OfType \Regex_3 \Leftrightarrow \Regex_3
        }
      }
      {
        \SwapLensOf{\IdentityLensOf{\Sep}}{\IdentityLensOf{\Regex_3}} \OfType 
        \Sep\Concat\Regex_3 \Leftrightarrow \Regex_3\Concat\String_i
      }

      \inferrule*[left=\Derivation{}']
      {
        \inferrule*
        {
        }
        {
          \IdentityLensOf{\Regex_2} \OfType \Regex_2 \Leftrightarrow \Regex_2
        }\\
        \Derivation
      }
      {
        \SwapLensOf{\IdentityLensOf{\Regex_2}}{\SwapLensShortOf{\IdentityLensShortOf{\Sep}}{\IdentityLensShortOf{\Regex_3}}} \OfType
        \Regex_2\Concat\Sep\Concat\Regex_3 \Leftrightarrow \Regex_3\Concat\Sep\Concat\Regex_2
      }

      \inferrule*[left=\Derivation{}'']
      {
        \inferrule*
        {
        }
        {
          \IdentityLensOf{\Regex_1} \OfType \Regex_1 \Leftrightarrow \Regex_1
        }\\
        \Derivation{}'
      }
      {
        \ConcatLensOf{\IdentityLensOf{\Regex_1}}{\SwapLensShortOf{\IdentityLensShortOf{\Regex_2}}{\SwapLensShortOf{\IdentityLensShortOf{\Sep}}{\IdentityLensShortOf{\Regex_3}}}} \OfType\\
        \Regex_1\Concat\Regex_2\Concat\Sep\Concat\Regex_3 \Leftrightarrow \Regex_1\Concat\Regex_3\Concat\Sep\Concat\Regex_2
      }

      \inferrule*
      {
        \Derivation{}''\\
        \inferrule*
        {
        }
        {
          \IdentityLensOf{\Regex_4} \OfType \Regex_4 \Leftrightarrow \Regex_4
        }
      }
      {
        \ConcatLensOf{\ConcatLensShortOf{\IdentityLensShortOf{\Regex_1}}{\SwapLensShortOf{\IdentityLensShortOf{\Regex_2}}{\SwapLensShortOf{\IdentityLensShortOf{\Sep}}{\IdentityLensShortOf{\Regex_3}}}}}{\IdentityLensOf{\Regex_4}} \OfType\\
        \Regex_1\Concat\Regex_2\Concat\Sep\Concat\Regex_3\Concat\Regex_4 \Leftrightarrow \Regex_1\Concat\Regex_3\Concat\Sep\Concat\Regex_2\Concat\Regex_4
      }
    \end{mathpar}

    By inspection, the final lens
    $\ConcatLensShortOf{\ConcatLensShortOf{\IdentityLensShortOf{\Regex_1}}{\SwapLensShortOf{\IdentityLensShortOf{\Regex_2}}{\SwapLensShortOf{\IdentityLensShortOf{\Sep}}{\IdentityLensShortOf{\Regex_3}}}}}{\IdentityLensShortOf{\Regex_4}} \OfType
    \Regex_1\Concat\Regex_2\Concat\Sep\Concat\Regex_3\Concat\Regex_4 \Leftrightarrow \Regex_1\Concat\Regex_3\Concat\Sep\Concat\Regex_2\Concat\Regex_4$
    satisfies $\LanguageOf{\Regex_1\Concat\Regex_2\Concat\String_i\Concat\Regex_3\Concat\Regex_4}=\LanguageOf{\SequenceOf{\Sep \SeqSep \Atom_1 \SeqSep \Sep \SeqSep \ldots \SeqSep \Sep \SeqSep \Atom_n \SeqSep \Sep}}$ and
    $\LanguageOf{\Regex_1\Concat\Regex_3\Concat\String_i\Concat\Regex_2\Concat\Regex_4}=\LanguageOf{\SequenceOf{\Sep \SeqSep \Atom_{\sigma_i(1)} \SeqSep \ldots \SeqSep \Atom_{\sigma_i(n)} \SeqSep \Sep}}$
    and has the desired semantics of swapping the strings at spots $i$ and $i+1$.
  \end{proof}
\end{lemma}

\begin{lemma}[Expressibility of Adjacent Swapping Permutation Composition]
  \label{lem:adj-comp-perm-exp}
  Suppose
  \begin{enumerate}
  \item $\sigma=\sigma_{i_1}\Compose\ldots\Compose\sigma_{i_m}$ 
  \item $\SequenceOf{\Sep \SeqSep \Atom_1 \SeqSep \Sep\ldots\Sep \SeqSep \Atom_n \SeqSep \Sep}$ is a sequence with
    all base strings equal to $\Sep$.
  \end{enumerate}
  Then there exists a typing of a lens $\Lens \OfType \Regex \Leftrightarrow \RegexAlt$ such that
  \begin{enumerate}
  \item $\LanguageOf{\Regex}=\LanguageOf{[\Sep \SeqSep \Atom_1 \SeqSep \ldots \SeqSep \Atom_n \SeqSep \Sep]}$
  \item $\LanguageOf{\RegexAlt}=\LanguageOf{[\Sep \SeqSep \Atom_{\sigma(1)} \SeqSep \ldots \SeqSep \Atom_{\sigma(n)} \SeqSep \Sep]}$
  \item $\SemanticsOf{\Lens}=
    \SetOf{(\String,\StringAlt)\SuchThat\String=\Sep\Concat\String_1\Concat\Sep\Concat\ldots\Concat\Sep\Concat\String_n\Concat\Sep
      \BooleanAnd\\
      \hspace*{6em}\StringAlt=\Sep\Concat\String_{\sigma(1)}\Concat\Sep\Concat\ldots\Concat\Sep\Concat\String_{\sigma(n)}\Sep\BooleanAnd\\
      \hspace*{6em}\String_i\in\LanguageOf{\Atom_i}}$
  \end{enumerate}
  \begin{proof}
    By induction on $m$.

    Let $m=0$.  Then $\sigma=\Identity$.  Consider the lens
    $\IdentityLensOf{\ToRegex(\SequenceOf{\Sep \SeqSep \Atom_1 \SeqSep \Sep\ldots\Sep \SeqSep \Atom_n \SeqSep \Sep})} \OfType
    \ToRegex(\SequenceOf{\Sep \SeqSep \Atom_1 \SeqSep \Sep\ldots\Sep \SeqSep \Atom_n \SeqSep \Sep}) \Leftrightarrow
    \ToRegex(\SequenceOf{\Sep \SeqSep \Atom_1 \SeqSep \Sep\ldots\Sep \SeqSep \Atom_n \SeqSep \Sep})$.
    By inspection, this lens satisfies the requirements.

    Let $m>0$.  Let $\sigma'=\sigma_{i_1}\Compose\ldots\Compose\sigma_{i_{m-1}}$.
    Let $\Lens \OfType \Regex \Leftrightarrow \RegexAlt$ be the lens obtained by an
    application of the induction assumption on $\sigma'$.
    Let $\Lens_m \OfType \RegexAlt' \Leftrightarrow \RegexAlt''$ be the lens obtained by
    an application of Lemma~\ref{lem:adj-perm-exp} to the permutation $\sigma_m$ and
    the sequence $\SequenceOf{\Sep \SeqSep \Atom_{\sigma'(1)} \SeqSep \ldots \SeqSep \Atom_{\sigma'(n)} \SeqSep \Sep}$.
    From the induction assumption and the previous lemmas,
    we know $\LanguageOf{\RegexAlt}=
    \LanguageOf{\SequenceOf{\Sep \SeqSep \Atom_{\sigma'(1)} \SeqSep \ldots \SeqSep \Atom_{\sigma'(n)} \SeqSep \Sep}}=
    \LanguageOf{\RegexAlt'}$.
    Consider the following Lens typing

    \begin{mathpar}
      \inferrule*
      {
        \inferrule*
        {
          \Lens \OfType \Regex \Leftrightarrow \RegexAlt\\
          \LanguageOf{\RegexAlt}=\LanguageOf{\RegexAlt'}
        }
        {
          \Lens \OfType \Regex \Leftrightarrow \RegexAlt'
        }\\
        \Lens_m \OfType \RegexAlt' \Leftrightarrow \RegexAlt''
      }
      {
        \ComposeLensOf{\Lens_m}{\Lens} \OfType \Regex \Leftrightarrow \RegexAlt''
      }
    \end{mathpar}

    The language of \Regex{} is already as desired, and
    $\LanguageOf{\RegexAlt''}=
    \LanguageOf{\SequenceOf{\Sep \SeqSep \Atom_{\sigma_m\Compose\sigma'(1)} \SeqSep \ldots \SeqSep \Atom_{\sigma_m\Compose\sigma'(n)}}}=
    \LanguageOf{\SequenceOf{\Sep \SeqSep \Atom_{\sigma(1)} \SeqSep \ldots \SeqSep \Atom_{\sigma(n)}}}$, as desired.
    Furthermore, the composition of the lenses composes the permutations of strings,
    giving the semantics as desired.
  \end{proof}
\end{lemma}

\begin{lemma}[Expressibility of Permutation]
  \label{lem:perm-exp}
  Suppose
  \begin{enumerate}
  \item $\sigma$ is a permutation in $S_n$
  \item $\SequenceOf{\Sep \SeqSep \Atom_1 \SeqSep \Sep\ldots\Sep \SeqSep \Atom_n \SeqSep \Sep}$ is a sequence with
    all base strings equal to $\Sep$.
  \end{enumerate}
  Then there exists a typing of a lens $\Lens \OfType \Regex \Leftrightarrow \RegexAlt$ such that
  \begin{enumerate}
  \item $\LanguageOf{\Regex}=\LanguageOf{[\Sep \SeqSep \Atom_1 \SeqSep \ldots \SeqSep \Atom_n \SeqSep \Sep]}$
  \item $\LanguageOf{\RegexAlt}=\LanguageOf{[\Sep \SeqSep \Atom_{\sigma(1)} \SeqSep \ldots \SeqSep \Atom_{\sigma(n)} \SeqSep \Sep]}$
  \item $\SemanticsOf{\Lens}=
    \SetOf{(\String,\StringAlt)\SuchThat\String=\Sep\Concat\String_1\Concat\Sep\Concat\ldots\Concat\Sep\Concat\String_n\Concat\Sep
      \BooleanAnd\\
      \hspace*{6em}\StringAlt=\Sep\Concat\String_{\sigma(1)}\Concat\Sep\Concat\ldots\Concat\Sep\Concat\String_{\sigma(n)}\Sep\BooleanAnd\\
      \hspace*{6em}\String_i\in\LanguageOf{\Atom_i}}$
  \end{enumerate}
\end{lemma}
\begin{proof}
  By algebra, any permutation can be expressed as the composition of adjacent swapping permutations.
  As such, $\sigma=\sigma_{i_1}\Compose\ldots\Compose\sigma_{i_m}$ for some adjacency swapping
  permutations $\sigma_{i_j}$.
  By Lemma~\ref{lem:adj-comp-perm-exp}, we obtain a lens with the properties desired.
\end{proof}

\begin{lemma}[Creation of Lens from Identity Perm DNF Lens]
  \label{lem:id-dnf}
  Suppose
  \begin{enumerate}
  \item $\DNFRegex = \DNFOf{\Sequence_1  \DNFSep  \ldots  \DNFSep  \Sequence_n}$
  \item $\DNFRegexAlt = \DNFOf{\SequenceAlt_1  \DNFSep  \ldots  \DNFSep  \SequenceAlt_n}$
  \item $(\DNFLensOf{\SequenceLens_1  \DNFLSep  \ldots  \DNFLSep  \SequenceLens_n},id) \OfRewritelessType
    \DNFRegex \Leftrightarrow \DNFRegexAlt$
  \item For each $\SequenceLens_i \OfRewritelessType \Sequence_i \Leftrightarrow \SequenceAlt_i$,
    there exists a $\Lens_i$ such that $\SemanticsOf{\Lens_i}=\SemanticsOf{\SequenceLens_i}$.
  \end{enumerate}
  then there exists a $\Lens \OfRewritelessType \ToRegex(\DNFRegex) \Leftrightarrow \ToRegex(\DNFRegexAlt)$ such that $\SemanticsOf{\Lens} = \SemanticsOf{([\SequenceLens_1  \DNFLSep  \ldots  \DNFLSep  \SequenceLens_n],id)}$.
  \begin{proof}
    By induction on n

    Let $n=0$.
    $\DNFLensOf{} \OfRewritelessType \DNFOf{} \Leftrightarrow \DNFOf{}$.  Then consider
    \begin{mathpar}
      \inferrule*
      {
      }
      {
        \IdentityLensOf{\ToRegex(\DNFOf{})} \OfRewritelessType
        \ToRegex(\DNFOf{}) \Leftrightarrow \ToRegex(\DNFOf{})
      }
    \end{mathpar}
    This has the desired typing, and
    $\SemanticsOf{\IdentityLensOf{\ToRegex(\DNFOf{})}}
    =\SemanticsOf{\IdentityLensOf{\emptyset}}
    =\SetOf{}=\SemanticsOf{\DNFLensOf{}}$.

    Let $n>0$.
    Let $\DNFRegex' = \DNFOf{\Sequence_1 \DNFSep \ldots \DNFSep \Sequence_{n-1}}$, and
    $\DNFRegexAlt' = \DNFOf{\SequenceAlt_1 \DNFSep \ldots \DNFSep \SequenceAlt_{n-1}}$.
    By induction assumption, there exists a derivation of 
    $\Lens \OfType \ToRegex(\DNFRegex') \Leftrightarrow \ToRegex(\DNFRegexAlt')$.
    By problem statement, there exists a typing derivation
    $\Lens_n \OfType \ToRegex(\Sequence_n) \Leftrightarrow \ToRegex(\SequenceAlt_n)$
    Consider the following derivation
    \begin{mathpar}
      \inferrule*
      {
        \Lens \OfType \ToRegex(\DNFRegex') \Leftrightarrow \ToRegex(\DNFRegexAlt')\\
        \Lens_n \OfType \ToRegex(\Sequence_n) \Leftrightarrow \ToRegex(\SequenceAlt_n)
      }
      {
        \OrLensOf{\Lens_n}{\Lens} \OfType \RegexOr{\ToRegex(\DNFRegex')}{\ToRegex(\Sequence_n)} \Leftrightarrow \RegexOr{\ToRegex(\DNFRegexAlt')}{\ToRegex(\Sequence_n)}
      }
    \end{mathpar}
    $\SemanticsOf{\OrLensOf{\Lens}{\Lens_n}}=\SetOf{(\String,\StringAlt)\SuchThat
      (\String,\StringAlt)\in\Lens\BooleanOr(\String,\StringAlt)\in\Lens_n}$\\
    \hspace*{4.6em}$=\SetOf{(\String,\StringAlt)\SuchThat
      (\String,\StringAlt)\in\DNFLensOf{\SequenceLens_1 \DNFLSep \ldots \DNFLSep \SequenceLens_{n-1}}\\
      \hspace*{8em}\BooleanOr(\String,\StringAlt)\in\DNFLensOf{\SequenceLens_n}}$\\
    \hspace*{4.6em}$=\SetOf{(\String,\StringAlt)\SuchThat
      (\String,\StringAlt)\in\SequenceLens_i}$.
  \end{proof}
\end{lemma}

\begin{lemma}[Ineffectiveness of Permutation on DNF Regex Semantics]
  \label{lem:dnfr-perm-sem-ineffective}
  Let $\sigma\in S_n$, and $\DNFOf{\Sequence_1\ldots\Sequence_n}$ be a DNF regex.
  $\LanguageOf{\DNFOf{\Sequence_1 \DNFSep \ldots \DNFSep \Sequence_n}}=
  \LanguageOf{\DNFOf{\Sequence_{\sigma(1)} \DNFSep \ldots \DNFSep \Sequence_{\sigma(n)}}}$.
\end{lemma}
\begin{proof}
  By inspection.
\end{proof}

\begin{lemma}[Ineffectiveness of Permutation on DNF Lens Semantics]
  \label{lem:dnfl-perm-sem-ineffective}
  Let $\sigma\in S_n$, and\\
  $(\DNFLensOf{\SequenceLens_1 \DNFLSep \ldots \DNFLSep \SequenceLens_n},\Identity) \OfRewritelessType
  \DNFOf{\Sequence_1 \DNFSep \ldots \DNFSep \Sequence_n} \Leftrightarrow
  \DNFOf{\SequenceAlt_1 \DNFSep \ldots \DNFSep \SequenceAlt_n}$ be a typing of a DNF lens with
  an identity permutation.
  $\SemanticsOf{(\DNFLensOf{\SequenceLens_1 \DNFLSep \ldots \DNFLSep \SequenceLens_n},\Identity)}
  =\SemanticsOf{(\DNFLensOf{\SequenceLens_1 \DNFLSep \ldots \DNFLSep \SequenceLens_n},\sigma)}$
\end{lemma}
\begin{proof}
  By inspection
\end{proof}

\begin{lemma}[Soundness of DNF, Sequence, and Atom Lenses]\leavevmode
  \label{lem:dnfcal}
  \begin{enumerate}
  \item Let \DNFRegex{} and \DNFRegexAlt{} be two dnf regular expressions, and
    $\DNFLens \OfRewritelessType \DNFRegex \Leftrightarrow \DNFRegexAlt$.  Then
    there exists a \Lens{} such that $\Lens \OfType \ToRegex(\DNFRegex)
    \Leftrightarrow \ToRegex(\DNFRegexAlt)$,
    \SemanticsOf{\Lens}=\SemanticsOf{\DNFLens} 

  \item Let \Sequence{} and \SequenceAlt{} be two clauses, and $\SequenceLens \OfRewritelessType \Sequence \Leftrightarrow \SequenceAlt$.  Then there exists a \Lens{} such that $\Lens \OfType \ToRegex(\Sequence) \Leftrightarrow \ToRegex(\SequenceAlt)$, \SemanticsOf{\Lens}=\SemanticsOf{\SequenceLens}.

  \item Let \Atom{} and \AtomAlt{} be two atoms, and $\AtomLens \OfRewritelessType \Atom \Leftrightarrow \AtomAlt$.  Then there exists a \Lens{}, such that $\Lens \OfRewritelessType \ToRegex(\Atom) \Leftrightarrow \ToRegex(\AtomAlt)$, \SemanticsOf{\Lens}=\SemanticsOf{\AtomLens}.
  \end{enumerate}
  \begin{proof}
    By mutual induction on the structure of the DNF Regex, Sequence, and
    Atom lenses typing.\\ 
    \\
    Let $\DNFLens \OfRewritelessType \DNFRegex \Leftrightarrow \DNFRegexAlt$ be formed from an
    application of\\$\RewriteDNFRegexLensRule{}$.
    \begin{mathpar}
      \inferrule*
      {
        \DNFLens \OfRewritelessType \DNFRegex' \Leftrightarrow \DNFRegexAlt'\\
        \DNFRegex' \Rewrite \DNFRegex\\
        \DNFRegexAlt' \Rewrite \DNFRegexAlt
      }
      {
        \DNFLens \OfRewritelessType \DNFRegex \Leftrightarrow \DNFRegexAlt
      }
    \end{mathpar}
    By induction assumption, there exists a
    $\Lens \OfRewritelessType \ToRegex(\DNFRegex') \Leftrightarrow \ToRegex(\DNFRegexAlt')$,
    and from Lemma~\ref{lem:rrl}, we know
    $\LanguageOf{\DNFRegex}=\LanguageOf{\DNFRegex'}$, and
    $\LanguageOf{\DNFRegexAlt}=\LanguageOf{\DNFRegexAlt'}$.
    Consider the derivation
    \begin{mathpar}
      \inferrule*
      {
        \Lens \OfRewritelessType \ToRegex(\DNFRegex') \Leftrightarrow \ToRegex(\DNFRegexAlt')\\
        \LanguageOf{\ToRegex(\DNFRegex')} = \LanguageOf{\ToRegex(\DNFRegex)}\\
        \LanguageOf{\ToRegex(\DNFRegexAlt')} = \LanguageOf{\ToRegex(\DNFRegexAlt)}
      }
      {
        \Lens \OfRewritelessType \ToRegex(\DNFRegex) \Leftrightarrow \ToRegex(\DNFRegexAlt)
      }
    \end{mathpar}
    This has the desired typing, and by induction assumption, has the desired semantics.\\
    \\
    Let $(\DNFLensOf{\SequenceLens_1 \DNFLSep \ldots \DNFLSep \SequenceLens_n},\sigma) \OfRewritelessType \DNFOf{\Sequence_1 \DNFSep \ldots \DNFSep \Sequence_n} \Leftrightarrow \DNFOf{\SequenceAlt_{\sigma(1)} \DNFSep \ldots \DNFSep \SequenceAlt_{\sigma(n)}}$ be formed from an application of $\DNFLensRule$.
    By Induction assumption, for each $\SequenceLens_i \OfRewritelessType \Sequence_i \Leftrightarrow \SequenceAlt_i$ there exists a $\Lens_i \OfRewritelessType \ToRegex(\Sequence_i) \Leftrightarrow \ToRegex(\SequenceAlt_i)$.\\
    By Lemma~\ref{lem:id-dnf} there exists a $\Lens \OfRewritelessType \ToRegex(\DNFOf{\Sequence_1 \DNFSep \ldots \DNFSep \Sequence_{n}}) \Leftrightarrow \ToRegex(\DNFOf{\SequenceAlt_1 \DNFSep \ldots \DNFSep \SequenceAlt_{n}})$ such that $\SemanticsOf{\Lens}=\SemanticsOf{([\SequenceLens_1 \DNFLSep \ldots\SequenceLens_n],id)}$,
    By Lemma~\ref{lem:dnfl-perm-sem-ineffective},
    $\SemanticsOf{(DNFOf{\SequenceLens_1 \DNFLSep \ldots \DNFLSep \SequenceLens_n},id)}=
    \SemanticsOf{(\DNFOf{\SequenceLens_1 \DNFLSep \ldots \DNFLSep \SequenceLens_n},\sigma)}$.
    By Lemma~\ref{lem:dnfr-perm-sem-ineffective},
    $\LanguageOf{\DNFOf{\SequenceAlt_1 \DNFSep \ldots \DNFSep \SequenceAlt_n}}=
    \LanguageOf{\DNFOf{\SequenceAlt_{\sigma(1)} \DNFSep \ldots \DNFSep \SequenceAlt_{\sigma(n)}}}$.
    Consider the following typing

    \begin{mathpar}
      \inferrule*
      {
        \Lens \OfRewritelessType \ToRegex(\DNFOf{\Sequence_1 \DNFSep \ldots \DNFSep \Sequence_{n}}) \Leftrightarrow \ToRegex(\DNFOf{\SequenceAlt_1 \DNFSep \ldots \DNFSep \SequenceAlt_{n}})\\
        \LanguageOf{\ToRegex(\DNFOf{\SequenceAlt_1 \DNFSep \ldots \DNFSep \SequenceAlt_{n}})} =
        \LanguageOf{\ToRegex(\DNFOf{\SequenceAlt_{\sigma(1)} \DNFSep \ldots \DNFSep \SequenceAlt_{\sigma(n)}})}
      }
      {
        \Lens \OfRewritelessType \ToRegex(\DNFOf{\Sequence_1 \DNFSep \ldots \DNFSep \Sequence_{n}}) \Leftrightarrow \ToRegex(\DNFOf{\SequenceAlt_{\sigma(1)} \DNFSep \ldots \DNFSep \SequenceAlt_{\sigma(n)}})
      }
    \end{mathpar}
    This has the typing and semantics as desired.\\
    \\
    Let $(\SequenceLensOf{(\String_0,\StringAlt_0) \SeqLSep \AtomLens_1 \SeqLSep \ldots \SeqLSep \AtomLens_n \SeqLSep (\String_n,\StringAlt_n)},\sigma \in S_n) \OfRewritelessType \SequenceOf{\String_0  \SeqSep  \Atom_1  \SeqSep  \ldots  \SeqSep  \Atom_n  \SeqSep  \String_n} \Leftrightarrow \SequenceOf{\StringAlt_0 \SeqSep  \AtomAlt_{\sigma(1)}  \SeqSep  \ldots  \SeqSep  \AtomAlt_{\sigma(n)}  \SeqSep  \StringAlt_n}$ be formed from an
    application of\\$\SequenceLensRule{}$.
    By induction assumption, for each
    $\AtomLens_i \OfRewritelessType \Atom_i \Leftrightarrow \AtomAlt_i$ there exists a
    $\Lens_i \OfType \ToRegex(\Regex_i) \Leftrightarrow \ToRegex(\RegexAlt_i)$.
    By Lemma~\ref{lem:id-clause}, there exists a $\Lens \OfType \Regex \Leftrightarrow \RegexAlt$ such that $\SemanticsOf{\Lens}=\SemanticsOf{([(\String_0,\StringAlt_0) \SeqLSep \AtomLens_1 \SeqLSep \ldots \SeqLSep \AtomLens_n \SeqLSep (\String_n,\StringAlt_n)],id)}$,
    $\Regex=\ToRegex(\SequenceOf{\String_0 \SeqSep \Atom_1 \SeqSep \ldots \SeqSep \Atom_n \SeqSep \String_n})$, and
    $\RegexAlt=\ToRegex(\SequenceOf{\StringAlt_0 \SeqSep \AtomAlt_1 \SeqSep \ldots \SeqSep \AtomAlt_n \SeqSep \StringAlt_n})$.
    Define $\RegexAlt_{\Sep}$ as $\ToRegex(\SequenceOf{\Sep \SeqSep \AtomAlt_1 \SeqSep \ldots \SeqSep \AtomAlt_n \SeqSep \Sep})$.
    By Lemma~\ref{lem:boilerplate-alterations}, there exists a
    $\Lens' \OfType \RegexAlt \Leftrightarrow \RegexAlt_{\Sep}$, with semantics of
    merely changing the boilerplate.
    By Lemma~\ref{lem:perm-exp}, there exists a $\Lens'' \OfType \RegexAlt_{\Sep}'
    \Leftrightarrow \RegexAlt_{\Sep}''$ where
    $\SemanticsOf{\RegexAlt_{\Sep}'}=\SemanticsOf{\RegexAlt_{\Sep}}$ and 
    $\SemanticsOf{\RegexAlt_{\Sep}''}=\SemanticsOf{\SequenceOf{\Sep \SeqSep  \AtomAlt_{\sigma(1)}  \SeqSep  \ldots  \SeqSep  \AtomAlt_{\sigma(n)}  \SeqSep  \Sep}}$.
    Lastly, with Lemma~\ref{lem:boilerplate-alterations}, there exists a
    $\Lens''' \OfType \RegexAlt_{\Sep}'' \Leftrightarrow \RegexAlt'$, where
    $\RegexAlt = \ToRegex(\SequenceOf{\StringAlt_0 \SeqSep  \AtomAlt_{\sigma(1)}  \SeqSep  \ldots  \SeqSep  \AtomAlt_{\sigma(n)}  \SeqSep  \StringAlt_n})$.
    Through composition of all these lenses, we finally get a lens with the desired type
    and semantics.\\
    \\
    Let $\IterateLensOf{\DNFLens} \OfRewritelessType \StarOf{\DNFRegex} \Leftrightarrow \StarOf{\DNFRegexAlt}$
    be introduced through an application of \AtomLensRule{}.
    From induction assumption, I know that there exists $\Lens \OfRewritelessType \Regex \Leftrightarrow \RegexAlt$, such that
    $\SemanticsOf{\DNFLens}=\SemanticsOf{\Lens}$,
    \Regex=\ToRegex(\DNFRegex), and
    $\RegexAlt=\ToRegex(\DNFRegexAlt)$.\\
    Consider $\IterateLensOf{\Lens} \OfRewritelessType \StarOf{\Regex} \Leftrightarrow \StarOf{\RegexAlt}$.\\
    By definition, $\StarOf{\Regex}$ and $\StarOf{\RegexAlt}$ are $\ToRegex(\StarOf{\DNFRegex})$
    and $\ToRegex(\StarOf{\Regex})$, respectively.

    \begin{tabular}{RcL}
      \SemanticsOf{\IterateLensOf{\Lens}} & = &
                                                \SetOf{(\String_0\ldots\String_n,\StringAlt_0\ldots\StringAlt_n)\SuchThat
                                                (\String_i,\StringAlt_i)\in\SemanticsOf{\Lens}}\\
                                          & = &
                                                \SetOf{(\String_0\ldots\String_n,\StringAlt_0\ldots\StringAlt_n)\SuchThat
                                                (\String_i,\StringAlt_i)\in\SemanticsOf{\DNFLens}}\\
                                          & = &
                                                \SemanticsOf{\IterateLensOf{\DNFLens}}
    \end{tabular}
  \end{proof}
\end{lemma}

\begin{theorem}
  If there exists a derivation of $\DNFLens \OfType \MapsBetweenTypeOf{\DNFRegex}{\DNFRegexAlt}$,
  then there exist a lens, $\ToLensOf{\DNFLens}$, and regular expressions, $\Regex$ and 
$\RegexAlt$, such that $\ToLensOf{\DNFLens} \OfType \MapsBetweenTypeOf{\Regex}{\RegexAlt}$ and
  $\ToDNFRegexOf{\Regex}=\DNFRegex$ and
  $\ToDNFRegexOf{\RegexAlt}=\DNFRegexAlt$ and
  $\SemanticsOf{\ToLensOf{\DNFLens}}=\SemanticsOf{\DNFLens}$.
\end{theorem}
\begin{proof}
  Let $\DNFLens \OfType \DNFRegex \Leftrightarrow \DNFRegexAlt$.

  By inversion, the last step is
  \[
    \inferrule*
    {
      \DNFLens \OfRewritelessType \DNFRegex' \Leftrightarrow \DNFRegexAlt'\\
      \DNFRegex \StarOf{\Rewrite} \DNFRegex'\\
      \DNFRegexAlt \StarOf{\Rewrite} \DNFRegexAlt'
    }
    {
      \DNFLens \OfType \DNFRegex \Leftrightarrow \DNFRegexAlt
    }
  \]

  From Lemma~\ref{lem:dnfcal}, there exists
  $\Lens \OfType \ToRegexOf{\DNFRegex'} \Leftrightarrow
  \ToRegexOf{\DNFRegexAlt'}$.

  So, as $\ToDNFRegexOf{\ToRegexOf{\DNFRegex}} \StarOf{\Rewrite}
  \ToDNFRegexOf{\ToRegexOf{\DNFRegex'}}$, and
  $\ToDNFRegexOf{\ToRegexOf{\DNFRegexAlt}} \StarOf{\Rewrite}
  \ToDNFRegexOf{\ToRegexOf{\DNFRegexAlt'}}$, from Lemma~\ref{lem:star-rw-in-de},
  $\ToRegexOf{\DNFRegex} \SSREquiv \ToRegexOf{\DNFRegex'}$, and
  $\ToRegexOf{\DNFRegexAlt} \SSREquiv \ToRegexOf{\DNFRegexAlt'}$.
  \[
    \inferrule*
    {
      \Lens \OfType \ToRegexOf{\DNFRegex'} \Leftrightarrow
      \ToRegexOf{\DNFRegexAlt'}\\
      \ToRegexOf{\DNFRegex} \SSREquiv \ToRegexOf{\DNFRegex'}\\
      \ToRegexOf{\DNFRegexAlt} \SSREquiv \ToRegexOf{\DNFRegexAlt'}
    }
    {
      \Lens \OfType \ToRegexOf{\DNFRegex} \SSREquiv
      \ToRegexOf{\DNFRegexAlt}
    }
  \]

  We call this lens $\ToLensOf{\DNFLens} (constructive proof)$.

  Furthermore, we also know that $\ToDNFRegexOf{\ToRegexOf{\DNFRegex}} =
  \DNFRegex$, and similarly for $\DNFRegexAlt$.
\end{proof}
\subsection{DNF Lens Operators}
\label{dnf-lens-operators}

DNF lens operators are defined to give DNF lenses similar capabilities to
lenses.  This allows the proof of many of the cases of completeness to be
trivial, leaving only the complications of proving statements about rewrites,
proving closure under composition, and proving the ability to use rewrites to
express lens retyping.

\begin{definition}[Permutation Functions]\leavevmode\\
  $\ConcatPermutation{} \OfType{}
  \ArrowTypeOf{\PermutationSetOf{n}}
  {\ArrowTypeOf{\PermutationSetOf{m}}{\PermutationSetOf{n+m}}}$\\
  $(\ConcatPermutationOf{\sigma_1}{\sigma_2})(i) =
  \begin{cases*}
    \sigma_1(i) & if $i \leq n$\\
    \sigma_2(i-n)+n & otherwise
  \end{cases*}$\\
  \\\\
  $\SwapPermutation{} \OfType{}
  \ArrowTypeOf{\PermutationSetOf{n}}
  {\ArrowTypeOf{\PermutationSetOf{m}}{\PermutationSetOf{n+m}}}$\\
  $(\SwapPermutationOf{\sigma_1}{\sigma_2})(i) =
  \begin{cases*}
    \sigma_1(i)+n & if $i \leq n$\\
    \sigma_2(i-n) & otherwise
  \end{cases*}$\\
  \\\\
  $\DistributePermutation{} \OfType{}
  \ArrowTypeOf{\PermutationSetOf{n}}
  {\ArrowTypeOf{\PermutationSetOf{m}}{\PermutationSetOf{n\times m}}}$\\
  $(\DistributePermutationOf{\sigma_1}{\sigma_2})(i,j) =
  (\sigma_1(i),\sigma_2(j))$
  \\\\
  $\DistributeSwapPermutation{} \OfType{}
  \ArrowTypeOf{\PermutationSetOf{n}}
  {\ArrowTypeOf{\PermutationSetOf{m}}{\PermutationSetOf{n\times m}}}$\\
  $(\DistributePermutationOf{\sigma_1}{\sigma_2})(i,j) =
  (\sigma_2(j),\sigma_1(i))$
\end{definition}

\begin{definition}[DNF Lens Functions]\leavevmode\\
  $\ConcatSequenceLens{} \OfType{}
  \ArrowTypeOf{\SequenceLensType{}}
  {\ArrowTypeOf{\SequenceLensType{}}{\SequenceLensType{}}}$\\
  $\ConcatSequenceLensOf
  {(\SequenceLensOf{(\String_0,\StringAlt_0) \SeqLSep \AtomLens_1 \SeqLSep \ldots \SeqLSep \AtomLens_n \SeqLSep (\String_n,\StringAlt_n)},\sigma_1)}
  {(\SequenceLensOf{(\String_0',\StringAlt_0') \SeqLSep \AtomLens_1' \SeqLSep \ldots \SeqLSep \AtomLens_m' \SeqLSep (\String_m',\StringAlt_m')},\sigma_2)}=$\\
  \hspace*{2ex}$(\SequenceLensOf{(\String_0,\StringAlt_0) \SeqLSep \AtomLens_1 \SeqLSep \ldots \SeqLSep \AtomLens_n \SeqLSep 
    (\String_n\Concat\String_0',\StringAlt_n\Concat\StringAlt_0') \SeqLSep \AtomLens_1' \SeqLSep 
    \ldots \SeqLSep \AtomLens_m' \SeqLSep (\String_m',\StringAlt_m')},\ConcatPermutationOf{\sigma_1}{\sigma_2})$\\
  \\
  \\$\SwapSequenceLens{} \OfType{}
  \ArrowTypeOf{\SequenceLensType{}}
  {\ArrowTypeOf{\SequenceLensType{}}{\SequenceLensType{}}}$\\
  Let $\String_i'' =
  \begin{cases*}
    \String_i & for $i \in \RangeIncInc{0}{n-1}$ \\
    \String_n \Concat \String_0 & for $i = n$\\
    \String_i' & for $i \in \RangeIncInc{n+1}{n+m}$
  \end{cases*}$
  Let $\StringAlt_i'' =
  \begin{cases*}
    \StringAlt_i' & for $i \in \RangeIncInc{0}{m-1}$ \\
    \StringAlt_m' \Concat \StringAlt_0 & for $i = m$\\
    \StringAlt_i & for $i \in \RangeIncInc{m+1}{m+n}$
  \end{cases*}$\\
  $\SwapSequenceLensOf
  {(\SequenceLensOf{(\String_0,\StringAlt_0) \SeqLSep \AtomLens_1 \SeqLSep \ldots \SeqLSep \AtomLens_n \SeqLSep (\String_n,\StringAlt_n)},\sigma_1)}
  {(\SequenceLensOf{(\String_0',\StringAlt_0') \SeqLSep \AtomLens_1' \SeqLSep \ldots \SeqLSep \AtomLens_m' \SeqLSep (\String_m',\StringAlt_m')},\sigma_2)}=$\\
  \hspace*{2ex}$(\SequenceLensOf{(\String_0'',\StringAlt_0'') \SeqLSep \AtomLens_1 \SeqLSep \ldots \SeqLSep \AtomLens_n \SeqLSep 
    (\String_n'',\StringAlt_n'') \SeqLSep \AtomLens_1' \SeqLSep (\String_{n+1}'',\StringAlt_n'')
    \ldots \SeqLSep \AtomLens_n' \SeqLSep (\String_{n+m}'',\StringAlt_{n+m}'')},\SwapPermutationOf{\sigma_1}{\sigma_2})$\\
  \\
  \\\ConcatDNFLens{} \OfType{}
  \ArrowTypeOf{\DNFLensType{}}{\ArrowTypeOf{\DNFLensType{}}{\DNFLensType{}}}\\
  $\ConcatDNFLensOf{(\DNFLensOf{\SequenceLens_1\DNFSep\ldots\DNFSep\SequenceLens_n},\sigma_1)}
  {(\DNFLensOf{\SequenceLens_1' \DNFLSep \ldots \DNFLSep \SequenceLens_m'},\sigma_2)}=$
  \[
    \begin{array}{rcccl}
      (\DNFLensLeft & \ConcatSequenceLensOf{\SequenceLens_1}{\SequenceLens_1'}\DNFSep & \cdots & \ConcatSequenceLensOf{\SequenceLens_1}{\SequenceLens_m'}\DNFSep \\
          \cdots    & \ConcatSequenceLensOf{\SequenceLens_n}{\SequenceLens_1'}\DNFSep & \cdots & \ConcatSequenceLensOf{\SequenceLens_n}{\SequenceLens_m'} & \DNFLensRight,\DistributePermutationOf{\sigma_1}{\sigma_1})
    \end{array}
  \]
  \\
  \\\SwapDNFLens{} \OfType{}
  \ArrowTypeOf{\DNFLensType{}}{\ArrowTypeOf{\DNFLensType{}}{\DNFLensType{}}}\\
  $\SwapDNFLensOf{(\DNFLensOf{\SequenceLens_1\DNFSep\ldots\DNFSep\SequenceLens_n},\sigma_1)}
  {(\DNFLensOf{\SequenceLens_1' \DNFLSep \ldots \DNFLSep \SequenceLens_m'},\sigma_2)}=$
  \[
    \begin{array}{rcccl}
      (\DNFLensLeft & \SwapSequenceLensOf{\SequenceLens_1}{\SequenceLens_1'}\DNFSep & \cdots & \SwapSequenceLensOf{\SequenceLens_1}{\SequenceLens_m'}\DNFSep \\
          \cdots    & \SwapSequenceLensOf{\SequenceLens_n}{\SequenceLens_1'}\DNFSep & \cdots & \SwapSequenceLensOf{\SequenceLens_n}{\SequenceLens_m'} & \DNFLensRight,\DistributeSwapPermutationOf{\sigma_1}{\sigma_1})
    \end{array}
  \]
  \\
  \\\OrDNFLens{} \OfType{}
  \ArrowTypeOf{\DNFLensType{}}{\ArrowTypeOf{\DNFLensType{}}{\DNFLensType{}}
  }\\
  $\OrDNFLensOf{(\DNFLensOf{\SequenceLens_1 \DNFLSep \ldots \DNFLSep \SequenceLens_n},\sigma_1)}
  ({\DNFLensOf{\SequenceLens_1' \DNFLSep \ldots \DNFLSep \Sequence_m'},\sigma_2)}$=\\
  \hspace*{2ex}$(\DNFLensOf{\SequenceLens_1 \DNFLSep \ldots \DNFLSep \SequenceLens_n \DNFLSep 
    \SequenceLens_1' \DNFLSep \ldots \DNFLSep \Sequence_m'},\ConcatPermutationOf{\sigma_1}{\sigma_2})$\\
  \\
  \\\AtomToDNFLens{} \OfType{}
  \ArrowTypeOf{\AtomLensType{}}{\DNFLensType{}}\\
  $\AtomToDNFLensOf{\AtomLens} = (\DNFLensOf{(\SequenceLensOf{(\EmptyString,\EmptyString) \SeqLSep \AtomLens \SeqLSep (\EmptyString,\EmptyString)},\Identity)},\Identity)$
\end{definition}

\begin{lemma}
  \label{lem:dnf-lens-concat-identity-left}
  $(\DNFLensOf{(\SequenceLensOf{(\EmptyString,\EmptyString)},\Identity_0)},\Identity_1)
  \ConcatDNFLens
  \DNFLens = \DNFLens$,
  where $\Identity_0$ is the identity permutation on $0$ elements, and
  $\Identity_1$ is the identity permutation on $1$ element.
\end{lemma}
\begin{proof}
  Let $\DNFLens = (\DNFLensOf{\SequenceLens_1 \DNFLSep \ldots \DNFLSep \SequenceLens_n},\sigma)$.
  By definition, $(\Identity_1 \DistributePermutation \sigma)(1,i) = (1,\sigma(i))$.
  By definition, $\Identity_0 \ConcatPermutation \sigma = \sigma$.
  Let $\SequenceLens_i =
  (\SequenceLensOf{(\String_{i,0},\StringAlt_{i,0}) \SeqLSep \AtomLens_{i,1} \SeqLSep \ldots
    \AtomLens_{i,n_i} \SeqLSep (\String_{i,n_i},\StringAlt_{i,n_i})},\sigma_i)$.
  So $(\SequenceLensOf{(\EmptyString,\EmptyString)},\Identity_0)
  \ConcatSequenceLens\SequenceLens_i =
  (\SequenceLensOf{(\EmptyString\Concat\String_{i,0},\EmptyString\Concat\StringAlt_{i,0}) \SeqLSep \AtomLens_{i,1} \SeqLSep \ldots
    \AtomLens_{i,n_i} \SeqLSep (\String_{i,n_i},\StringAlt_{i,n_i})},\sigma_i) =
  \SequenceLens_i$.
  So
  $(\DNFLensOf{(\SequenceLensOf{(\EmptyString,\EmptyString)},\Identity_0)},\Identity_1)
  \ConcatDNFLens \DNFLensOf{\SequenceLens_1 \DNFLSep \ldots \DNFLSep \SequenceLens_n} =
  (\DNFLensOf{(\SequenceLensOf{(\EmptyString,\EmptyString)},\Identity_0)
    \ConcatSequenceLens \SequenceLens_1 \DNFLSep  \ldots  \DNFLSep 
    (\SequenceLensOf{(\EmptyString,\EmptyString)},\Identity_0) \ConcatSequenceLens
    \SequenceLens_n},\Identity_1 \DistributePermutation \sigma) =
  (\DNFLensOf{\SequenceLens_1 \DNFLSep \ldots \DNFLSep \SequenceLens_n,\sigma})$.
\end{proof}

\begin{lemma}
  \label{lem:dnf-lens-concat-identity-right}
  $\DNFLens \ConcatDNFLens
  \DNFLensOf{\SequenceLensOf{(\EmptyString,\EmptyString)}} = \DNFLens$
\end{lemma}
\begin{proof}
Done similarly to Lemma~\ref{lem:dnf-lens-concat-identity-left}.
\end{proof}

\begin{lemma}[Typing and Semantics of $\ConcatSequenceLens$]
  \label{lem:typ-sem-concat-seq}
  Let $\SequenceLens_1 \OfType \Sequence_1 \Leftrightarrow \SequenceAlt_1$ and
  $\SequenceLens_2 \OfType \Sequence_2 \Leftrightarrow \SequenceAlt_2$ be the typing of
  two sequence lenses, where
  $\UnambigConcatOf{\LanguageOf{\Sequence_1}}{\LanguageOf{\Sequence_2}}$ and
  $\UnambigConcatOf{\LanguageOf{\SequenceAlt_1}}{\LanguageOf{\SequenceAlt_2}}$.
  Then $\ConcatSequenceLensOf{\SequenceLens_1}{\SequenceLens_2} \OfType
  \ConcatSequenceOf{\Sequence_1}{\Sequence_2} \Leftrightarrow
  \ConcatSequenceOf{\SequenceAlt_1}{\SequenceAlt_2}$ and
  $\SemanticsOf{\ConcatSequenceLensOf{\SequenceLens_1}{\SequenceLens_2}} =
  \SetOf{(\String_1\Concat\String_2,\StringAlt_1\Concat\StringAlt_2) \SuchThat
    (\String_1,\StringAlt_1)\in\SemanticsOf{\SequenceLens_1}
    \BooleanAnd(\String_2,\StringAlt_2)\in\SemanticsOf{\SequenceLens_2}}$
\end{lemma}
\begin{proof}
  By assumption, there exists typing derivations

  \[
    \SequenceLens_1 \OfRewritelessType \Sequence_1 \Leftrightarrow \SequenceAlt_1
  \]
  and
  \[
    \SequenceLens_2 \OfRewritelessType \Sequence_2 \Leftrightarrow \SequenceAlt_2
  \]

  By inversion, we know that the last rule application on each side was
  \DNFLensRule{}, giving

  \[
    \inferrule*
    {
      \AtomLens_i \OfType \Atom_i \Leftrightarrow \AtomAlt_i\\
      \sigma_1 \in \PermutationSetOf{n}\\
      \SequenceUnambigConcatOf{\String_0\SeqSep\Atom_1\SeqSep\ldots\SeqSep\Atom_n\SeqSep\String_n}\\
      \SequenceUnambigConcatOf{\StringAlt_0\SeqSep\AtomAlt_{\sigma_1(1)}\SeqSep\ldots\SeqSep\AtomAlt_{\sigma_1(n)}\SeqSep\StringAlt_n}
    }
    {
      (\SequenceLensOf{(\String_0,\StringAlt_0) \SeqLSep \AtomLens_1 \SeqLSep \ldots \SeqLSep \AtomLens_n},\sigma_1)
      \OfRewritelessType
      \SequenceOf{\String_0 \SeqSep \Atom_1 \SeqSep \ldots \SeqSep \Atom_n \SeqSep \String_n} \Leftrightarrow
      \SequenceOf{\StringAlt_0 \SeqSep \AtomAlt_{\sigma_1(1)} \SeqSep \ldots \SeqSep \AtomAlt_{\sigma_1(n)} \SeqSep \StringAlt_n}
    }
  \]
  and
  \[
    \inferrule*
    {
      \AtomLens_i' \OfType \Atom_i' \Leftrightarrow \AtomAlt_i'\\
      \sigma_2 \in \PermutationSetOf{m}\\
      \SequenceUnambigConcatOf{\String_0'\SeqSep\Atom_1'\SeqSep\ldots\SeqSep\Atom_m'\SeqSep\String_m'}\\
      \SequenceUnambigConcatOf{\StringAlt_0'\SeqSep\AtomAlt_{\sigma_2(1)}'\SeqSep\ldots\SeqSep\AtomAlt_{\sigma_2(n)}'\SeqSep\StringAlt_n'}
    }
    {
      (\SequenceLensOf{(\String_0',\StringAlt_0') \SeqLSep \AtomLens_1' \SeqLSep \ldots \SeqLSep \AtomLens_m' \SeqLSep (\String_m',\StringAlt_m')},\sigma_2)
      \OfRewritelessType
      \SequenceOf{\String_0' \SeqSep \Atom_1' \SeqSep \ldots \SeqSep \Atom_m' \SeqSep \String_m'} \Leftrightarrow
      \SequenceOf{\StringAlt_0' \SeqSep \AtomAlt_{\sigma_2(1)}' \SeqSep \ldots \SeqSep \AtomAlt_{\sigma_2(m)}' \SeqSep \StringAlt_m'}
    }
  \]
  where
  \[
    \begin{array}{rcl}
      \SequenceLens_1 & = &
                            (\SequenceLensOf{(\String_0,\StringAlt_0) \SeqLSep \AtomLens_1 \SeqLSep \ldots \SeqLSep \AtomLens_n},
                            \sigma_1)\\
      \Sequence_1 & = &
                        \SequenceOf{\String_0 \SeqSep \Atom_1 \SeqSep \ldots \SeqSep \Atom_n \SeqSep \String_n}\\
      \SequenceAlt_1 & = &
                           \SequenceOf{\StringAlt_0 \SeqSep \AtomAlt_{\sigma_1(1)} \SeqSep \ldots \SeqSep \AtomAlt_{\sigma_1(n)} \SeqSep \StringAlt_n}\\
      \SequenceLens_2 & = &
                            (\SequenceLensOf{(\String_0',\StringAlt_0') \SeqLSep \AtomLens_1' \SeqLSep \ldots \SeqLSep \AtomLens_m' \SeqLSep (\String_m',\StringAlt_m')},\sigma_2)\\
      \Sequence_2 & = &
                        \SequenceOf{\String_0' \SeqSep \Atom_1' \SeqSep \ldots \SeqSep \Atom_m' \SeqSep \String_m'}\\
      \SequenceAlt_2 & = &
                           \SequenceOf{\StringAlt_0' \SeqSep \AtomAlt_{\sigma_2(1)}' \SeqSep \ldots \SeqSep \AtomAlt_{\sigma_2(m)}' \SeqSep \StringAlt_m'}
    \end{array}
  \]

  Define $\String_i''$ as $\String_i$ for $i \in \RangeIncInc{1}{n-1}$, and as
  $\String_{i-n}'$ for $i \in \RangeIncInc{n+1}{n+m}$, and as
  $\String_n\Concat\String_0'$ for $i=n$.

  Define $\StringAlt_i''$ as $\StringAlt_i$ for $i \in \RangeIncInc{1}{n-1}$,
  and as $\StringAlt_i'$ as $\StringAlt_{i-n}$ for $i \in
  \RangeIncInc{n+1}{n+m}$, and as $\StringAlt_n\Concat\StringAlt_0$ for $i=n$.

  Define $\Atom_i''$ as $\Atom_i$ for $i \in \RangeIncInc{1}{n}$, and as
  $\Atom_{i-n}'$ for $i \in \RangeIncInc{n+1}{n+m}$.

  Define $\AtomAlt_i''$ as $\Atom_i$ for $i \in \RangeIncInc{1}{n}$, and as
  $\AtomAlt_{i-n}'$ for $i \in \RangeIncInc{n+1}{n+m}$.

  Define $\AtomLens_i$ as $\AtomLens_i$ for $i \in \RangeIncInc{1}{n}$, and as
  $\AtomLens_{i-n}'$ for $i \in \RangeIncInc{n+1}{n+m}$.

  From Lemma~\ref{lem:unambig-concat-equiv}, as
  $\SequenceUnambigConcatOf{\String_0 \SeqSep \Atom_1 \SeqSep \ldots \SeqSep \Atom_n \SeqSep \String_n}$,
  $\SequenceUnambigConcatOf{\String_0' \SeqSep \Atom_1' \SeqSep \ldots \SeqSep \Atom_m' \SeqSep \String_m'}$,
  and
  $\UnambigConcatOf
  {\SequenceOf{\String_0 \SeqSep \Atom_1 \SeqSep \ldots \SeqSep \Atom_n \SeqSep \String_n}}
  {\SequenceOf{\String_0' \SeqSep \Atom_1' \SeqSep \ldots \SeqSep \Atom_m' \SeqSep \String_m'}}$,
  then $\SequenceUnambigConcatOf{\String_0;\Atom_1;\ldots;\Atom_n;
    \String_n\Concat\String_0';
    \Atom_1';\ldots;\Atom_m';\String_m'}$, so
  $\SequenceUnambigConcatOf{\String_0'';\Atom_1'';\ldots;\Atom_{n+m}'';\String_{n+m}''}$.

  From Lemma~\ref{lem:unambig-concat-equiv}, as
  $\SequenceUnambigConcatOf{\StringAlt_0;\AtomAlt_{\sigma_1(1)}';\ldots;\AtomAlt_{\sigma_1(n)};\StringAlt_n}$,
  $\SequenceUnambigConcatOf{\StringAlt_0';\AtomAlt_{\sigma_2(1)}';\ldots;\AtomAlt_{\sigma_2(m)}';\StringAlt_m'}$,
  and
  $\UnambigConcatOf
  {\SequenceOf{\StringAlt_0 \SeqSep \AtomAlt_{\sigma_1(1)} \SeqSep \ldots \SeqSep \AtomAlt_{\sigma_1(n)} \SeqSep \StringAlt_n}}
  {\SequenceOf{\StringAlt_0' \SeqSep \AtomAlt_{\sigma_2(1)}' \SeqSep \ldots \SeqSep \AtomAlt_{\sigma_2(m)}' \SeqSep \StringAlt_m'}}$, then
  $\SequenceUnambigConcatOf{\StringAlt_0;\AtomAlt_{\sigma_1(1)};\ldots;\AtomAlt_{\sigma_1(n)};
    \StringAlt_n\Concat\StringAlt_0';
    \AtomAlt_{\sigma_2(1)}';\ldots;\AtomAlt_{\sigma_2(m)}';\StringAlt_m'}$, so
  $\SequenceUnambigConcatOf{\StringAlt_0'';
    \AtomAlt_{\ConcatPermutationOf{\sigma_1}{\sigma_2}(1)}'';\ldots;
    \AtomAlt_{\ConcatPermutationOf{\sigma_1}{\sigma_2}(n+m)}'';\StringAlt_{n+m}''}$.

  Consider the derivation

  \[
    \inferrule*
    {
      \AtomLens_i \OfType \Atom_i \Leftrightarrow \AtomAlt_i\\
      \ConcatPermutationOf{\sigma_1}{\sigma_2} \in \PermutationSetOf{n+m}\\
      \SequenceUnambigConcatOf{\String_0'';\Atom_1'';\ldots;\Atom_{n+m}'';\String_{n+m}''}\\
      \SequenceUnambigConcatOf{\StringAlt_0'';
        \AtomAlt_{\ConcatPermutationOf{\sigma_1}{\sigma_2}(1)}'';\ldots;
        \AtomAlt_{\ConcatPermutationOf{\sigma_1}{\sigma_2}(n+m)}'';\StringAlt_{n+m}''}
    }
    {
      (\SequenceLensOf{(\String_0'',\StringAlt_0'') \SeqLSep \AtomLens_1 \SeqLSep
        \ldots \SeqLSep
        \AtomLens_{n+m} \SeqLSep (\String_{n+m}'',\StringAlt_{n+m}'')},
      \ConcatPermutationOf{\sigma_1}{\sigma_2})
      \OfRewritelessType\\
      \SequenceOf{\String_0'' \SeqSep \Atom_1'' \SeqSep \ldots \SeqSep \Atom_{n+m}'' \SeqSep \String_{n+m}''}
      \Leftrightarrow
      \SequenceOf{\StringAlt_0'' \SeqSep \AtomAlt_1'' \SeqSep \ldots \SeqSep \AtomAlt_{n+m}'' \SeqSep \StringAlt_{n+m}''}
    }
  \]

  We wish to show that this is a derivation of
  $\ConcatSequenceLensOf{\SequenceLens_1}{\SequenceLens_2} \OfType
  \ConcatSequenceOf{\Sequence_1}{\Sequence_2} \Leftrightarrow
  \ConcatSequenceOf{\SequenceAlt_1}{\SequenceAlt_2}$.

  \[
    \begin{array}{rcl}
      (\SequenceLensOf{(\String_0'',\StringAlt_0'') \SeqLSep \AtomLens_1'' \SeqLSep
      \ldots \SeqLSep
      \AtomLens_{n+m}'' \SeqLSep (\String_{n+m}'',\StringAlt_{n+m}'')},
      \ConcatPermutationOf{\sigma_1}{\sigma_2})
      & = & (\SequenceLensOf{(\String_0'',\StringAlt_0'') \SeqLSep \AtomLens_1'' \SeqLSep
            \ldots \SeqLSep \AtomLens_n'' \SeqLSep (\String_n'',\StringAlt_n'') \SeqLSep\\
      & & \AtomLens_{n+1}'' \SeqLSep
          \ldots \SeqLSep \AtomLens_{n+m}'' \SeqLSep (\String_{n+m}'',\StringAlt_{n+m}'')},
          \ConcatPermutationOf{\sigma_1}{\sigma_2})\\
      & = & (\SequenceLensOf{(\String_0,\StringAlt_0) \SeqLSep \AtomLens_1 \SeqLSep
            \ldots \SeqLSep \AtomLens_n \SeqLSep
            (\String_n\Concat\String_0',\StringAlt_n\Concat\StringAlt_0')
            \SeqLSep \\
      & & \AtomLens_0' \SeqLSep
          \ldots \SeqLSep \AtomLens_m' \SeqLSep (\String_m',\StringAlt_m')},
          \ConcatPermutationOf{\sigma_1}{\sigma_2})\\
      & = & \ConcatSequenceLensOf{\SequenceLens_1}{\SequenceLens_2}
    \end{array}
  \]

  \[
    \begin{array}{rcl}
      \SequenceOf{\String_0'' \SeqSep \Atom_1'' \SeqSep \ldots \SeqSep \Atom_{n+m}'' \SeqSep \String_{n+m}''}
      & = & \SequenceOf{\String_0'' \SeqSep \Atom_1'' \SeqSep \ldots \SeqSep \Atom_n'' \SeqSep \String_n'' \SeqSep \Atom_{n+1}'' \SeqSep 
            \ldots\Atom_{n+m}'' \SeqSep \String_{n+m}''}\\
      & = & \SequenceOf{\String_0 \SeqSep \Atom_1 \SeqSep \ldots \SeqSep \Atom_n \SeqSep (\String_n\Concat\String_0') \SeqSep \Atom_1' \SeqSep 
            \ldots\Atom_m' \SeqSep \String_m'}\\
      & = & \ConcatSequenceOf{\Sequence_1}{\Sequence_2}
    \end{array}
  \]

  \[
    \begin{array}{rcl}
      \SequenceOf{\StringAlt_0'' \SeqSep \AtomAlt_{\ConcatPermutationOf{\sigma_1}{\sigma_2}(1)}'' \SeqSep \ldots \SeqSep 
      \AtomAlt_{\ConcatPermutationOf{\sigma_1}{\sigma_2}(n+m)}'' \SeqSep \StringAlt_{n+m}''}
      & = &
            \SequenceOf{\StringAlt_0'' \SeqSep \AtomAlt_{\ConcatPermutationOf{\sigma_1}{\sigma_2}(1)}'' \SeqSep 
            \ldots \SeqSep \AtomAlt_{\ConcatPermutationOf{\sigma_1}{\sigma_2}(n)}'' \SeqSep \\
      &   & \StringAlt_n'' \SeqSep \AtomAlt_{\ConcatPermutationOf{\sigma_1}{\sigma_2}(n+1)}'' \SeqSep 
            \ldots \SeqSep 
            \AtomAlt_{\ConcatPermutationOf{\sigma_1}{\sigma_2}(n+m)}'' \SeqSep \StringAlt_{n+m}''}\\
      & = &
            \SequenceOf{\StringAlt_0'' \SeqSep \AtomAlt_{\sigma_1(1)}'' \SeqSep 
            \ldots \SeqSep \AtomAlt_{\sigma_1(n)}'' \SeqSep 
            \StringAlt_n'' \SeqSep \AtomAlt_{\sigma_2(1)+n}'' \SeqSep 
            \ldots \SeqSep 
            \AtomAlt_{\sigma_2(m)+n}'' \SeqSep \StringAlt_{n+m}''}\\
      & = &
            \SequenceOf{\StringAlt_0 \SeqSep \AtomAlt_{\sigma_1(1)} \SeqSep 
            \ldots \SeqSep \AtomAlt_{\sigma_1(n)} \SeqSep 
            \StringAlt_n\Concat\StringAlt_0'\\
      &   & \SeqSep \AtomAlt_{\sigma_2(1)}' \SeqSep 
            \ldots \SeqSep 
            \AtomAlt_{\sigma_2(m)}' \SeqSep \StringAlt_m'}\\
      & = & \ConcatSequenceOf{\SequenceAlt_1}{\SequenceAlt_2}
    \end{array}
  \]

  So we have a derivation of $\ConcatSequenceLensOf{\SequenceLens_1}{\SequenceLens_2} \OfType
  \ConcatSequenceOf{\Sequence_1}{\Sequence_2} \Leftrightarrow
  \ConcatSequenceOf{\SequenceAlt_1}{\SequenceAlt_2}$

  We also wish to have the desired semantics.

  \[
    \begin{array}{l}
      \SemanticsOf{(\SequenceLensOf{(\String_0'',\StringAlt_0'') \SeqLSep
      \AtomLens_1'' \SeqLSep
      \ldots \SeqLSep
      \AtomLens_{n+m}'' \SeqLSep (\String_{n+m}'',\StringAlt_{n+m}'')},
      \ConcatPermutationOf{\sigma_1}{\sigma_2})}\\
      = \SetOf{(\String_0''\Concat\overline{\String_1}\Concat
      \ldots\Concat\overline{\String_{n+m}}\Concat\String_{n+m}'',
      \StringAlt_0''\Concat\overline{\StringAlt_{\ConcatPermutationOf{\sigma_1}{\sigma_2}(1)}}\Concat
      \ldots\Concat\overline{\StringAlt_{\ConcatPermutationOf{\sigma_1}{\sigma_2}(n+m)}}\Concat\StringAlt_{n+m}''))\\
      \hspace*{3em}\SuchThat
      \forall i\in\RangeIncInc{1}{n+m}.(\overline{\String_i},\overline{\StringAlt_i})\in\SequenceLens_i''}\\
      = \SetOf{(\String_0\Concat\overline{\String_1}\Concat\ldots\Concat
      \overline{\String_n}\Concat\String_n\Concat\String_0'\Concat\overline{\String_{0}'}\Concat
      \ldots\Concat\overline{\String_m'}\Concat\String_m',
      \StringAlt_0\Concat\overline{\StringAlt_{\sigma_1(1)}}\Concat\ldots\Concat
      \overline{\StringAlt_{\sigma_1(n)}}\Concat\StringAlt_n\Concat\StringAlt_0'
      \Concat\overline{\StringAlt_{\sigma_2(0)}'}\Concat
      \ldots\Concat\overline{\StringAlt_{\sigma_2(m)}'}\Concat\StringAlt_m'))\\
      \hspace*{3em}\SuchThat
      (\forall i\in\RangeIncInc{1}{n}.
      (\overline{\String_i},\overline{\StringAlt_i})\in\SequenceLens_i
      \BooleanAnd
      \forall i\in\RangeIncInc{1}{m}.
      (\overline{\String_i'},\overline{\StringAlt_i'})\in\SequenceLens_i'}\\
      = \SetOf{(\String\Concat\String',\StringAlt\Concat\StringAlt')\SuchThat
      (\String,\StringAlt)\in\SemanticsOf{\SequenceLens_1}
      \BooleanAnd
      (\String',\StringAlt')\in\SemanticsOf{\SequenceLens_2}}
    \end{array}
  \]

\end{proof}

\begin{lemma}[Typing and Semantics of $\SwapSequenceLens$]
  \label{lem:typ-sem-swap-seq}
  Let $\SequenceLens_1 \OfType \Sequence_1 \Leftrightarrow \SequenceAlt_1$ and
  $\SequenceLens_2 \OfType \Sequence_2 \Leftrightarrow \SequenceAlt_2$ be the typing of
  two sequence lenses, where
  $\UnambigConcatOf{\LanguageOf{\Sequence_1}}{\LanguageOf{\Sequence_2}}$ and
  $\UnambigConcatOf{\LanguageOf{\SequenceAlt_2}}{\LanguageOf{\SequenceAlt_1}}$
  Then $\ConcatSequenceLensOf{\SequenceLens_1}{\SequenceLens_2} \OfType
  \ConcatSequenceOf{\Sequence_1}{\Sequence_2} \Leftrightarrow
  \ConcatSequenceOf{\SequenceAlt_1}{\SequenceAlt_2}$ and
  $\SemanticsOf{\ConcatSequenceLensOf{\SequenceLens_1}{\SequenceLens_2}} =
  \SetOf{(\String_1\Concat\String_2,\StringAlt_1\Concat\StringAlt_2) \SuchThat
    (\String_1,\StringAlt_1)\in\SemanticsOf{\SequenceLens_1}
    \BooleanAnd(\String_2,\StringAlt_2)\in\SemanticsOf{\SequenceLens_2}}$
\end{lemma}
\begin{proof}
  By assumption, there exists typing derivations

  \[
    \SequenceLens_1 \OfRewritelessType \Sequence_1 \Leftrightarrow \SequenceAlt_1
  \]
  and
  \[
    \SequenceLens_2 \OfRewritelessType \Sequence_2 \Leftrightarrow \SequenceAlt_2
  \]

  By inversion, we know that the last rule application on each side was
  \DNFLensRule{}, giving

  \[
    \inferrule*
    {
      \AtomLens_i \OfType \Atom_i \Leftrightarrow \AtomAlt_i\\
      \sigma_1 \in \PermutationSetOf{n}\\
      \SequenceUnambigConcatOf{\String_0\SeqSep\Atom_1\SeqSep\ldots\SeqSep\Atom_n\SeqSep\String_n}\\
      \SequenceUnambigConcatOf{\StringAlt_0\SeqSep\AtomAlt_{\sigma_1(1)}\SeqSep\ldots\SeqSep\AtomAlt_{\sigma_1(n)}\SeqSep\StringAlt_n}
    }
    {
      (\SequenceLensOf{(\String_0,\StringAlt_0) \SeqLSep \AtomLens_1 \SeqLSep \ldots \SeqLSep \AtomLens_n},\sigma_1)
      \OfRewritelessType
      \SequenceOf{\String_0 \SeqSep \Atom_1 \SeqSep \ldots \SeqSep \Atom_n \SeqSep \String_n} \Leftrightarrow
      \SequenceOf{\StringAlt_0 \SeqSep \AtomAlt_{\sigma_1(1)} \SeqSep \ldots \SeqSep \AtomAlt_{\sigma_1(n)} \SeqSep \StringAlt_n}
    }
  \]
  and
  \[
    \inferrule*
    {
      \AtomLens_i' \OfType \Atom_i' \Leftrightarrow \AtomAlt_i'\\
      \sigma_2 \in \PermutationSetOf{m}\\
      \SequenceUnambigConcatOf{\String_0'\SeqSep\Atom_1'\SeqSep\ldots\SeqSep\Atom_m'\SeqSep\String_m'}\\
      \SequenceUnambigConcatOf{\StringAlt_0'\SeqSep\AtomAlt_{\sigma_2(1)}'\SeqSep\ldots\SeqSep\AtomAlt_{\sigma_2(n)}'\SeqSep\StringAlt_n'}
    }
    {
      (\SequenceLensOf{(\String_0',\StringAlt_0') \SeqLSep \AtomLens_1' \SeqLSep \ldots \SeqLSep \AtomLens_m' \SeqLSep (\String_m',\StringAlt_m')},\sigma_2)
      \OfRewritelessType
      \SequenceOf{\String_0' \SeqSep \Atom_1' \SeqSep \ldots \SeqSep \Atom_m' \SeqSep \String_m'} \Leftrightarrow
      \SequenceOf{\StringAlt_0' \SeqSep \AtomAlt_{\sigma_2(1)}' \SeqSep \ldots \SeqSep \AtomAlt_{\sigma_2(m)}' \SeqSep \StringAlt_m'}
    }
  \]
  where
  \[
    \begin{array}{rcl}
      \SequenceLens_1 & = &
                            (\SequenceLensOf{(\String_0,\StringAlt_0) \SeqLSep \AtomLens_1 \SeqLSep \ldots \SeqLSep \AtomLens_n},
                            \sigma_1)\\
      \Sequence_1 & = &
                        \SequenceOf{\String_0 \SeqSep \Atom_1 \SeqSep \ldots \SeqSep \Atom_n \SeqSep \String_n}\\
      \SequenceAlt_1 & = &
                           \SequenceOf{\StringAlt_0 \SeqSep \AtomAlt_{\sigma_1(1)} \SeqSep \ldots \SeqSep \AtomAlt_{\sigma_1(n)} \SeqSep \StringAlt_n}\\
      \SequenceLens_2 & = &
                            (\SequenceLensOf{(\String_0',\StringAlt_0') \SeqLSep \AtomLens_1' \SeqLSep \ldots \SeqLSep \AtomLens_m' \SeqLSep (\String_m',\StringAlt_m')},\sigma_2)\\
      \Sequence_2 & = &
                        \SequenceOf{\String_0' \SeqSep \Atom_1' \SeqSep \ldots \SeqSep \Atom_m' \SeqSep \String_m'}\\
      \SequenceAlt_2 & = &
                           \SequenceOf{\StringAlt_0' \SeqSep \AtomAlt_{\sigma_2(1)}' \SeqSep \ldots \SeqSep \AtomAlt_{\sigma_2(m)}' \SeqSep \StringAlt_m'}
    \end{array}
  \]

  Define $\String_i''$ as $\String_i$ for $i \in \RangeIncInc{1}{n-1}$, and as
  $\String_{i-n}'$ for $i \in \RangeIncInc{n+1}{n+m}$, and as
  $\String_n\Concat\String_0'$ for $i=n$.

  Define $\StringAlt_i''$ as $\StringAlt_i'$ for $i \in \RangeIncInc{1}{m-1}$,
  and as $\StringAlt_{i-m}$ for $i \in
  \RangeIncInc{m+1}{m+n}$, and as $\StringAlt_m'\Concat\StringAlt_0$ for $i=m$.

  Define $\Atom_i''$ as $\Atom_i$ for $i \in \RangeIncInc{1}{n}$, and as
  $\Atom_{i-n}'$ for $i \in \RangeIncInc{n+1}{n+m}$.

  Define $\AtomAlt_i''$ as $\AtomAlt_i'$ for $i \in \RangeIncInc{1}{m}$, and as
  $\AtomAlt_{i-m}$ for $i \in \RangeIncInc{m+1}{m+n}$.

  Define $\AtomLens_i$ as $\AtomLens_i$ for $i \in \RangeIncInc{1}{n}$, and as
  $\AtomLens_{i-n}'$ for $i \in \RangeIncInc{n+1}{n+m}$.

  From Lemma~\ref{lem:unambig-concat-equiv}, as
  $\SequenceUnambigConcatOf{\String_0;\Atom_1;\ldots;\Atom_n;\String_n}$,
  $\SequenceUnambigConcatOf{\String_0';\Atom_1';\ldots;\Atom_m';\String_m'}$,
  and
  $\UnambigConcatOf
  {\SequenceOf{\String_0 \SeqSep \Atom_1 \SeqSep \ldots \SeqSep \Atom_n \SeqSep \String_n}}
  {\SequenceOf{\String_0' \SeqSep \Atom_1' \SeqSep \ldots \SeqSep \Atom_m' \SeqSep \String_m'}}$,
  then $\SequenceUnambigConcatOf{\String_0;\Atom_1;\ldots;\Atom_n;
    \String_n\Concat\String_0';
    \Atom_1';\ldots;\Atom_m';\String_m'}$, so
  $\SequenceUnambigConcatOf{\String_0'';\Atom_1'';\ldots;\Atom_{n+m}'';\String_{n+m}''}$.

  From Lemma~\ref{lem:unambig-concat-equiv}, as
  $\SequenceUnambigConcatOf{\StringAlt_0';\AtomAlt_{\sigma_2(1)}';\ldots;\AtomAlt_{\sigma_2(m)}';\StringAlt_m'}$,
  $\SequenceUnambigConcatOf{\StringAlt_0;\AtomAlt_{\sigma_1(1)}';\ldots;\AtomAlt_{\sigma_1(n)};\StringAlt_n}$,
  and
  $\UnambigConcatOf
  {\SequenceOf{\StringAlt_0' \SeqSep \AtomAlt_{\sigma_2(1)}' \SeqSep \ldots \SeqSep \AtomAlt_{\sigma_2(m)}' \SeqSep \StringAlt_m'}}
  {\SequenceOf{\StringAlt_0 \SeqSep \AtomAlt_{\sigma_1(1)} \SeqSep \ldots \SeqSep \AtomAlt_{\sigma_1(n)} \SeqSep \StringAlt_n}}$, then
  $\SequenceUnambigConcatOf{\StringAlt_0';\AtomAlt_{\sigma_2(1)}';\ldots;\AtomAlt_{\sigma_2(m)}';
    \StringAlt_m'\Concat\StringAlt_0;
    \AtomAlt_{\sigma_1(1)};\ldots;\AtomAlt_{\sigma_1(n)};\StringAlt_n}$, so
  $\SequenceUnambigConcatOf{\StringAlt_0'';
    \AtomAlt_{\SwapPermutationOf{\sigma_1}{\sigma_2}(1)}'';\ldots;
    \AtomAlt_{\SwapPermutationOf{\sigma_1}{\sigma_2}(n+m)}'';\StringAlt_{n+m}''}$.

  Consider the derivation

  \[
    \inferrule*
    {
      \AtomLens_i \OfType \Atom_i \Leftrightarrow \AtomAlt_i\\
      \SwapPermutationOf{\sigma_1}{\sigma_2} \in \PermutationSetOf{n+m}\\
      \SequenceUnambigConcatOf{\String_0'';\Atom_1'';\ldots;\Atom_{n+m}'';\String_{n+m}''}\\
      \SequenceUnambigConcatOf{\StringAlt_0'';
        \AtomAlt_{\ConcatPermutationOf{\sigma_1}{\sigma_2}(1)}'';\ldots;
        \AtomAlt_{\ConcatPermutationOf{\sigma_1}{\sigma_2}(n+m)}'';\StringAlt_{n+m}''}
    }
    {
      (\SequenceLensOf{(\String_0'',\StringAlt_0'') \SeqLSep \AtomLens_1 \SeqLSep 
        \ldots \SeqLSep 
        \AtomLens_{n+m} \SeqLSep (\String_{n+m}'',\StringAlt_{n+m}'')},
      \SwapPermutationOf{\sigma_1}{\sigma_2})
      \OfRewritelessType\\
      \SequenceOf{\String_0'' \SeqSep \Atom_1'' \SeqSep \ldots \SeqSep \Atom_{n+m}'' \SeqSep \String_{n+m}''}
      \Leftrightarrow
      \SequenceOf{\StringAlt_0'' \SeqSep \AtomAlt_1'' \SeqSep \ldots \SeqSep \AtomAlt_{n+m}'' \SeqSep \StringAlt_{n+m}''}
    }
  \]

  We wish to show that this is a derivation of
  $\SwapSequenceLensOf{\SequenceLens_1}{\SequenceLens_2} \OfType
  \ConcatSequenceOf{\Sequence_1}{\Sequence_2} \Leftrightarrow
  \ConcatSequenceOf{\SequenceAlt_2}{\SequenceAlt_1}$.

  By the definition of $\SwapSequenceLens$, $\String_i''$, and $\StringAlt_i''$,
  $(\SequenceLensOf{(\String_0'',\StringAlt_0'') \SeqLSep \AtomLens_1 \SeqLSep 
        \ldots \SeqLSep 
        \AtomLens_{n+m} \SeqLSep (\String_{n+m}'',\StringAlt_{n+m}'')},
      \SwapPermutationOf{\sigma_1}{\sigma_2}) =
      \SwapSequenceLensOf{\SequenceLens_1}{\SequenceLens_2}$

  \[
    \begin{array}{rcl}
      \SequenceOf{\String_0'' \SeqSep \Atom_1'' \SeqSep \ldots \SeqSep \Atom_{n+m}'' \SeqSep \String_{n+m}''}
      & = & \SequenceOf{\String_0'' \SeqSep \Atom_1'' \SeqSep \ldots \SeqSep \Atom_n'' \SeqSep \String_n'' \SeqSep \Atom_{n+1}'' \SeqSep 
            \ldots\Atom_{n+m}'' \SeqSep \String_{n+m}''}\\
      & = & \SequenceOf{\String_0 \SeqSep \Atom_1 \SeqSep \ldots \SeqSep \Atom_n \SeqSep (\String_n\Concat\String_0') \SeqSep \Atom_1' \SeqSep 
            \ldots\Atom_m' \SeqSep \String_m'}\\
      & = & \ConcatSequenceOf{\Sequence_1}{\Sequence_2}
    \end{array}
  \]

  \[
    \begin{array}{rcl}
      \SequenceOf{\StringAlt_0'' \SeqSep \AtomAlt_{\SwapPermutationOf{\sigma_1}{\sigma_2}(1)}'' \SeqSep \ldots \SeqSep 
      \AtomAlt_{\SwapPermutationOf{\sigma_1}{\sigma_2}(n+m)}'' \SeqSep \StringAlt_{n+m}''}
      & = &
            \SequenceOf{\StringAlt_0'' \SeqSep \AtomAlt_{\SwapPermutationOf{\sigma_1}{\sigma_2}(1)}'' \SeqSep 
            \ldots \SeqSep
            \AtomAlt_{\SwapPermutationOf{\sigma_1}{\sigma_2}(m)}''\\
      &   & \SeqSep 
            \StringAlt_m'' \SeqSep
            \AtomAlt_{\SwapPermutationOf{\sigma_1}{\sigma_2}(n+1)}''\\
      &   & \SeqSep 
            \ldots \SeqSep 
            \AtomAlt_{\SwapPermutationOf{\sigma_1}{\sigma_2}(n+m)}'' \SeqSep \SequenceAlt_{n+m}''}\\
      & = &
            \SequenceOf{\StringAlt_0'' \SeqSep \AtomAlt_{\sigma_2(1)+n}'' \SeqSep 
            \ldots \SeqSep \AtomAlt_{\sigma_2(m)+n}''\\
      &   & \SeqSep 
            \StringAlt_n'' \SeqSep \AtomAlt_{\sigma_1(1)}'' \SeqSep 
            \ldots \SeqSep 
            \AtomAlt_{\sigma(m)}'' \SeqSep \StringAlt_{n+m}''}\\
      & = &
            \SequenceOf{\StringAlt_0' \SeqSep \AtomAlt_{\sigma_2(1)}' \SeqSep 
            \ldots \SeqSep \AtomAlt_{\sigma_2(m)}'\\
      &   & \SeqSep 
            \StringAlt_m'\Concat\StringAlt_0 \SeqSep \AtomAlt_{\sigma_1(1)} \SeqSep 
            \ldots \SeqSep 
            \AtomAlt_{\sigma_1(n)} \SeqSep \StringAlt_n}\\
      & = & \ConcatSequenceOf{\SequenceAlt_1}{\SequenceAlt_2}
    \end{array}
  \]

  So we have a derivation of $\ConcatSequenceLensOf{\SequenceLens_1}{\SequenceLens_2} \OfType
  \ConcatSequenceOf{\Sequence_1}{\Sequence_2} \Leftrightarrow
  \ConcatSequenceOf{\SequenceAlt_1}{\SequenceAlt_2}$

  We also wish to have the desired semantics.

  \[
    \begin{array}{l}
      \SemanticsOf{(\SequenceLensOf{(\String_0'',\StringAlt_0'') \SeqLSep \SequenceLens_1'' \SeqLSep 
      \ldots \SeqLSep 
      \SequenceLens_{n+m}'' \SeqLSep (\String_{n+m}'',\StringAlt_{n+m}'')},
      \SwapPermutationOf{\sigma_1}{\sigma_2})}\\
      = \SetOf{(\String_0''\Concat\overline{\String_1}\Concat
      \ldots\Concat\overline{\String_{n+m}}\Concat\String_{n+m}'',
      \StringAlt_0''\Concat\overline{\StringAlt_{\SwapPermutationOf{\sigma_1}{\sigma_2}(1)}}\Concat
      \ldots\Concat\overline{\StringAlt_{\SwapPermutationOf{\sigma_1}{\sigma_2}(n+m)}}\Concat\StringAlt_{n+m}''))\\
      \hspace*{3em}\SuchThat
      \forall i\in\RangeIncInc{1}{n+m}.(\overline{\String_i},\overline{\StringAlt_i})\in\SequenceLens_i''}\\
      = \SetOf{(\String_0\Concat\overline{\String_1}\Concat\ldots\Concat
      \overline{\String_n}\Concat\String_n\Concat\String_0'\Concat\overline{\String_{0}'}\Concat
      \ldots\Concat\overline{\String_m'}\Concat\String_m',
      \StringAlt_0'\Concat\overline{\StringAlt_{\sigma_2(1)}'}\Concat\ldots\Concat
      \overline{\StringAlt_{\sigma_2(m)}'}\Concat\StringAlt_m'\Concat\StringAlt_0
      \Concat\overline{\StringAlt_{\sigma_1(0)}}\Concat
      \ldots\Concat\overline{\StringAlt_{\sigma_1(n)}}\Concat\StringAlt_n))\\
      \hspace*{3em}\SuchThat
      (\forall i\in\RangeIncInc{1}{n}.
      (\overline{\String_i},\overline{\StringAlt_i})\in\SequenceLens_i
      \BooleanAnd
      \forall i\in\RangeIncInc{1}{m}.
      (\overline{\String_i'},\overline{\StringAlt_i'})\in\SequenceLens_i'}\\
      = \SetOf{(\String\Concat\String',\StringAlt'\Concat\StringAlt)\SuchThat
      (\String,\StringAlt)\in\SemanticsOf{\SequenceLens_1}
      \BooleanAnd
      (\String',\StringAlt')\in\SemanticsOf{\SequenceLens_2}}
    \end{array}
  \]
\end{proof}

\begin{lemma}[Typing and Semantics of $\ConcatDNFLens$]
  \label{lem:typ_sem_concat}
  Let $\DNFLens_1 \OfType \DNFRegex_1 \Leftrightarrow \DNFRegexAlt_1$ and
  $\DNFLens_2 \OfType \DNFRegex_2 \Leftrightarrow \DNFRegexAlt_2$ be the typing of
  two DNF lenses, where
  $\UnambigConcatOf{\LanguageOf{\DNFRegex_1}}{\LanguageOf{\DNFRegex_2}}$ and
  $\UnambigConcatOf{\LanguageOf{\DNFRegexAlt_1}}{\LanguageOf{\DNFRegexAlt_2}}$.
  Then $\ConcatDNFLensOf{\DNFLens_1}{\DNFLens_2} \OfType
  \ConcatDNFOf{\DNFRegex_1}{\DNFRegex_2} \Leftrightarrow
  \ConcatDNFOf{\DNFRegexAlt_1}{\DNFRegexAlt_2}$ and
  $\SemanticsOf{\ConcatDNFLensOf{\DNFLens_1}{\DNFLens_2}} =
  \SetOf{(\String_1\Concat\String_2,\StringAlt_1\Concat\StringAlt_2) \SuchThat
    (\String_1,\StringAlt_1)\in\SemanticsOf{\DNFLens_1}
    \BooleanAnd(\String_2,\StringAlt_2)\in\SemanticsOf{\DNFLens_2}}$
\end{lemma}
\begin{proof}
  By assumption, there exists typing derivations

  \[
    \DNFLens_1 \OfRewritelessType \DNFRegex_1 \Leftrightarrow \DNFRegexAlt_1
  \]
  and
  \[
    \DNFLens_2 \OfRewritelessType \DNFRegex_2 \Leftrightarrow \DNFRegexAlt_2
  \]

  By inversion, we know that the last rule application on each side was
  \DNFLensRule{}, giving

  \[
    \inferrule*
    {
      \SequenceLens_1 \OfRewritelessType \Sequence_1 \Leftrightarrow \SequenceAlt_1\\
      \ldots\\
      \SequenceLens_n \OfRewritelessType \Sequence_n \Leftrightarrow \SequenceAlt_n\\\\
      \sigma_1 \in \PermutationSetOf{n}\\
      i \neq j \Rightarrow \LanguageOf{\Sequence_{i}} \cap \LanguageOf{\Sequence_{j}}=\emptyset\\
      i \neq j \Rightarrow \LanguageOf{\SequenceAlt_{i}} \cap \LanguageOf{\SequenceAlt_{j}}=\emptyset\\
    }
    {
      (\DNFLensOf{\SequenceLens_1\DNFLSep\ldots\DNFLSep\SequenceLens_n},\sigma_1) \OfRewritelessType\\
      \DNFOf{\Sequence_1\DNFSep\ldots\DNFSep\Sequence_n}
      \Leftrightarrow \DNFOf{\SequenceAlt_{\sigma_1(1)}\DNFSep\ldots\DNFSep\SequenceAlt_{\sigma_1(n)}}
    }
  \]
  and
  \[
    \inferrule*
    {
      \SequenceLens_1' \OfRewritelessType \Sequence_1' \Leftrightarrow \SequenceAlt_1\\
      \ldots\\
      \SequenceLens_n' \OfRewritelessType \Sequence_m' \Leftrightarrow \SequenceAlt_m\\\\
      \sigma_2 \in \PermutationSetOf{m}\\
      i \neq j \Rightarrow \LanguageOf{\Sequence_{i}} \cap \LanguageOf{\Sequence_{j}}=\emptyset\\
      i \neq j \Rightarrow \LanguageOf{\SequenceAlt_{i}} \cap \LanguageOf{\SequenceAlt_{j}}=\emptyset\\
    }
    {
      (\DNFLensOf{\SequenceLens_1'\DNFLSep\ldots\DNFLSep\SequenceLens_n'},\sigma_2) \OfRewritelessType\\
      \DNFOf{\Sequence_1'\DNFSep\ldots\DNFSep\Sequence_n'}
      \Leftrightarrow \DNFOf{\SequenceAlt_{\sigma_2(1)}'\DNFSep\ldots\DNFSep\SequenceAlt_{\sigma_2(m)}'}
    }
  \]
  where
  \[
    \begin{array}{rcl}
      \DNFLens_1 & = &
                       (\DNFLensOf{\SequenceLens_1\DNFLSep\ldots\DNFLSep\SequenceLens_n},\sigma_1)\\
      \DNFRegex_1 & = &
                        \DNFOf{\Sequence_1\DNFSep\ldots\DNFSep\Sequence_n}\\
      \DNFRegexAlt_1 & = &
                           \DNFOf{\SequenceAlt_{\sigma_1(1)}\DNFSep\ldots\DNFSep\SequenceAlt_{\sigma_1(n)}}\\
      \DNFLens_2 & = &
                            (\DNFLensOf{\SequenceLens_1'\DNFLSep\ldots\DNFLSep\SequenceLens_n'},\sigma_2)\\
      \DNFRegex_2 & = &
                        \DNFOf{\Sequence_1'\DNFSep\ldots\DNFSep\Sequence_n'}\\
      \DNFRegexAlt_2 & = &
                           \DNFOf{\SequenceAlt_{\sigma_2(1)}'\DNFSep\ldots\DNFSep\SequenceAlt_{\sigma_2(m)}'}
    \end{array}
  \]
  
  Define $\Sequence_{i,j}$ as $\Sequence_i \ConcatSequence \Sequence_j'$.

  Define $\SequenceAlt_{i,j}$ as $\SequenceAlt_i \ConcatSequence
  \SequenceAlt_j'$

  From Lemma~\ref{lem:unambig-concat-union-equiv}, as
  $i \neq j \BooleanImplies \LanguageOf{\Sequence_i} \Intersect
  \LanguageOf{\Sequence_j} = \emptyset$,
  $i \neq j \BooleanImplies \LanguageOf{\Sequence_i'} \Intersect
  \LanguageOf{\Sequence_j'} = \emptyset$,
  and
  $\UnambigConcatOf
  {\DNFOf{\Sequence_1 \DNFSep \ldots \DNFSep \Sequence_n}}
  {\DNFOf{\Sequence_1' \DNFSep \ldots \DNFSep \Sequence_m'}}$,
  then $\UnambigConcatOf{\Sequence_i}{\Sequence_j'}$,
  and $(i_1,j_1) \neq (i_2,j_2) \BooleanImplies \LanguageOf{\Sequence_{i_1,j_1}}
  \Intersect \LanguageOf{\Sequence_{i_2,j_2}} = \emptyset$.

  From Lemma~\ref{lem:unambig-concat-union-equiv}, as
  $i \neq j \BooleanImplies \LanguageOf{\SequenceAlt_i} \Intersect
  \LanguageOf{\SequenceAlt_j} = \emptyset$,
  $i \neq j \BooleanImplies \LanguageOf{\SequenceAlt_i'} \Intersect
  \LanguageOf{\SequenceAlt_j'} = \emptyset$,
  and
  $\UnambigConcatOf
  {\DNFOf{\SequenceAlt_1 \DNFSep \ldots \DNFSep \SequenceAlt_n}}
  {\DNFOf{\SequenceAlt_1' \DNFSep \ldots \DNFSep \SequenceAlt_m'}}$,
  then $\UnambigConcatOf{\SequenceAlt_i}{\SequenceAlt_j'}$,
  and $(i_1,j_1) \neq (i_2,j_2) \BooleanImplies \LanguageOf{\SequenceAlt_{i_1,j_1}}
  \Intersect \LanguageOf{\SequenceAlt_{i_2,j_2}} = \emptyset$.

  So, from Lemma~\ref{lem:typ-sem-concat-seq}, 
  $\ConcatSequenceLensOf{\SequenceLens_i}{\SequenceLens_j} \OfRewritelessType
  \ConcatSequenceOf{\Sequence_i}{\Sequence_j} \Leftrightarrow
  \ConcatSequenceOf{\SequenceAlt_i}{\SequenceAlt_j}$.

  Define $\SequenceLens_{i,j} = \ConcatSequenceLensOf{\SequenceLens_i}{\SequenceLens_j}$

  Consider the derivation

  \[
    \inferrule*
    {
      \SequenceLens_{i,j} \OfType \Sequence_{i,j} \Leftrightarrow \SequenceAlt_{i,j}\\
      \DistributePermutationOf{\sigma_1}{\sigma_2} \in \PermutationSetOf{n\times
        m}\\
      (i_1,j_1) \neq (i_2,j_2) \BooleanImplies \LanguageOf{\Sequence_{i_1,j_1}} \Intersect
      \LanguageOf{\Sequence_{i_2,j_2}} = \emptyset\\
      (i_1,j_1) \neq (i_2,j_2) \BooleanImplies \LanguageOf{\SequenceAlt_{i_1,j_1}} \Intersect
      \LanguageOf{\SequenceAlt_{i_2,j_2}} = \emptyset\\
    }
    {
      (\DNFLensOf{\SequenceLens_{1,1} \DNFLSep \ldots \DNFLSep \SequenceLens_{n,m}},
      \DistributePermutationOf{\sigma_1}{\sigma_2})
      \OfRewritelessType
      \DNFOf{\Sequence_{1,1} \DNFSep \ldots \DNFSep \Sequence_{n,m}}
      \Leftrightarrow
      \DNFOf{\SequenceAlt_{1,1} \DNFSep \ldots \DNFSep \SequenceAlt_{n,m}}
    }
  \]

  We wish to show that this is a derivation of
  $\ConcatDNFLensOf{\DNFLens_1}{\DNFLens_2} \OfType
  \ConcatDNFOf{\DNFRegex_1}{\DNFRegex_2} \Leftrightarrow
  \ConcatDNFOf{\DNFRegexAlt_1}{\DNFRegexAlt_2}$.

  $(\DNFLensOf{\SequenceLens_{1,1} \DNFLSep \ldots \DNFLSep \SequenceLens_{n,m}},
  \DistributePermutationOf{\sigma_1}{\sigma_2}) =
  (\ConcatSequenceLensOf{\SequenceLens_1}{\SequenceLens_1'} \DNFLSep \ldots \DNFLSep 
  \ConcatSequenceLensOf{\SequenceLens_n}{\SequenceLens_m'},\DistributePermutationOf{\sigma_1}{\sigma_2})
  =
  \ConcatDNFLensOf{\DNFLens_1}{\DNFLens_2}$.

  $\DNFOf{\Sequence_{1,1} \DNFSep \ldots \DNFSep \Sequence_{n,m}} = \DNFOf{\Sequence_1
    \ConcatSequence \Sequence_1' \DNFSep  \ldots \DNFSep 
  \Sequence_n \ConcatSequence \Sequence_m'} = \ConcatDNFOf{\DNFRegex_1}{\DNFRegex_2}$

  $\DNFOf{\SequenceAlt_{1,1} \DNFSep \ldots \DNFSep \SequenceAlt_{n,m}} = \DNFOf{\SequenceAlt_1
    \ConcatSequence \SequenceAlt_1' \DNFSep  \ldots \DNFSep 
  \SequenceAlt_n \ConcatSequence \SequenceAlt_m'} = \ConcatDNFOf{\DNFRegexAlt_1}{\DNFRegexAlt_2}$

  So we have a derivation of $\ConcatSequenceLensOf{\SequenceLens_1}{\SequenceLens_2} \OfType
  \ConcatSequenceOf{\Sequence_1}{\Sequence_2} \Leftrightarrow
  \ConcatSequenceOf{\SequenceAlt_1}{\SequenceAlt_2}$

  We also wish to have the desired semantics.

  \[
    \begin{array}{l}
      \SemanticsOf{(\DNFLensOf{\SequenceLens_{1,1} \DNFLSep \ldots \DNFLSep \SequenceLens_{n,m}},
      \DistributePermutationOf{\sigma_1}{\sigma_2})}\\
      = \SetOf{(\String_1\Concat\String_2,\StringAlt_1\Concat\StringAlt_2)
      \SuchThat \exists i,j. (\String_1,\StringAlt_1) \in
      \SemanticsOf{\SequenceLens_i} \BooleanAnd (\String_2,\StringAlt_2) \in
      \SemanticsOf{\SequenceLens_j}}\\
      = \SetOf{(\String_1\Concat\String_2,\StringAlt_1\Concat\StringAlt_2)
      \SuchThat (\String_1,\StringAlt_1) \in
      \SemanticsOf{\DNFLens} \BooleanAnd (\String_2,\StringAlt_2) \in
      \SemanticsOf{\DNFLens'}}
    \end{array}
  \]
\end{proof}

\begin{lemma}[Typing and Semantics of $\SwapDNFLens$]
  \label{lem:typ_sem_swap}
  Let $\DNFLens_1 \OfType \DNFRegex_1 \Leftrightarrow \DNFRegexAlt_1$ and
  $\DNFLens_2 \OfType \DNFRegex_2 \Leftrightarrow \DNFRegexAlt_2$ be the typing of
  two DNF lenses, where
  $\UnambigConcatOf{\LanguageOf{\DNFRegex_1}}{\LanguageOf{\DNFRegex_2}}$ and
  $\UnambigConcatOf{\LanguageOf{\DNFRegexAlt_2}}{\LanguageOf{\DNFRegexAlt_1}}$.
  Then $\ConcatDNFLensOf{\DNFLens_1}{\DNFLens_2} \OfType
  \ConcatDNFOf{\DNFRegex_1}{\DNFRegex_2} \Leftrightarrow
  \ConcatDNFOf{\DNFRegexAlt_1}{\DNFRegexAlt_2}$ and
  $\SemanticsOf{\ConcatDNFLensOf{\DNFLens_1}{\DNFLens_2}} =
  \SetOf{(\String_1\Concat\String_2,\StringAlt_1\Concat\StringAlt_2) \SuchThat
    (\String_1,\StringAlt_1)\in\SemanticsOf{\DNFLens_1}
    \BooleanAnd(\String_2,\StringAlt_2)\in\SemanticsOf{\DNFLens_2}}$
\end{lemma}
\begin{proof}
  By assumption, there exists typing derivations

  \[
    \DNFLens_1 \OfRewritelessType \DNFRegex_1 \Leftrightarrow \DNFRegexAlt_1
  \]
  and
  \[
    \DNFLens_2 \OfRewritelessType \DNFRegex_2 \Leftrightarrow \DNFRegexAlt_2
  \]

  By inversion, we know that the last rule application on each side was
  \DNFLensRule{}, giving

  \[
    \inferrule*
    {
      \SequenceLens_1 \OfRewritelessType \Sequence_1 \Leftrightarrow \SequenceAlt_1\\
      \ldots\\
      \SequenceLens_n \OfRewritelessType \Sequence_n \Leftrightarrow \SequenceAlt_n\\\\
      \sigma_1 \in \PermutationSetOf{n}\\
      i \neq j \Rightarrow \LanguageOf{\Sequence_{i}} \cap \LanguageOf{\Sequence_{j}}=\emptyset\\
      i \neq j \Rightarrow \LanguageOf{\SequenceAlt_{i}} \cap \LanguageOf{\SequenceAlt_{j}}=\emptyset\\
    }
    {
      (\DNFLensOf{\SequenceLens_1\DNFLSep\ldots\DNFLSep\SequenceLens_n},\sigma_1) \OfRewritelessType\\
      \DNFOf{\Sequence_1\DNFSep\ldots\DNFSep\Sequence_n}
      \Leftrightarrow \DNFOf{\SequenceAlt_{\sigma_1(1)}\DNFSep\ldots\DNFSep\SequenceAlt_{\sigma_1(n)}}
    }
  \]
  and
  \[
    \inferrule*
    {
      \SequenceLens_1' \OfRewritelessType \Sequence_1' \Leftrightarrow \SequenceAlt_1\\
      \ldots\\
      \SequenceLens_n' \OfRewritelessType \Sequence_m' \Leftrightarrow \SequenceAlt_m\\\\
      \sigma_2 \in \PermutationSetOf{m}\\
      i \neq j \Rightarrow \LanguageOf{\Sequence_{i}} \cap \LanguageOf{\Sequence_{j}}=\emptyset\\
      i \neq j \Rightarrow \LanguageOf{\SequenceAlt_{i}} \cap \LanguageOf{\SequenceAlt_{j}}=\emptyset\\
    }
    {
      (\DNFLensOf{\SequenceLens_1'\DNFLSep\ldots\DNFLSep\SequenceLens_n'},\sigma_2) \OfRewritelessType\\
      \DNFOf{\Sequence_1'\DNFSep\ldots\DNFSep\Sequence_n'}
      \Leftrightarrow \DNFOf{\SequenceAlt_{\sigma_2(1)}'\DNFSep\ldots\DNFSep\SequenceAlt_{\sigma_2(m)}'}
    }
  \]
  where
  \[
    \begin{array}{rcl}
      \DNFLens_1 & = &
                       (\DNFLensOf{\SequenceLens_1\DNFLSep\ldots\DNFLSep\SequenceLens_n},\sigma_1)\\
      \DNFRegex_1 & = &
                        \DNFOf{\Sequence_1\DNFSep\ldots\DNFSep\Sequence_n}\\
      \DNFRegexAlt_1 & = &
                           \DNFOf{\SequenceAlt_{\sigma_1(1)}\DNFSep\ldots\DNFSep\SequenceAlt_{\sigma_1(n)}}\\
      \DNFLens_2 & = &
                            (\DNFLensOf{\SequenceLens_1'\DNFLSep\ldots\DNFLSep\SequenceLens_n'},\sigma_2)\\
      \DNFRegex_2 & = &
                        \DNFOf{\Sequence_1'\DNFSep\ldots\DNFSep\Sequence_n'}\\
      \DNFRegexAlt_2 & = &
                           \DNFOf{\SequenceAlt_{\sigma_2(1)}'\DNFSep\ldots\DNFSep\SequenceAlt_{\sigma_2(m)}'}
    \end{array}
  \]
  
  Define $\Sequence_{i,j}$ as $\Sequence_i \ConcatSequence \Sequence_j'$.

  Define $\SequenceAlt_{j,i}$ as $\SequenceAlt_j' \ConcatSequence
  \SequenceAlt_i$

  From Lemma~\ref{lem:unambig-concat-union-equiv}, as
  $i \neq j \BooleanImplies \LanguageOf{\Sequence_i} \Intersect
  \LanguageOf{\Sequence_j} = \emptyset$,
  $i \neq j \BooleanImplies \LanguageOf{\Sequence_i'} \Intersect
  \LanguageOf{\Sequence_j'} = \emptyset$,
  and
  $\UnambigConcatOf
  {\DNFOf{\Sequence_1 \DNFSep \ldots \DNFSep \Sequence_n}}
  {\DNFOf{\Sequence_1' \DNFSep \ldots \DNFSep \Sequence_m'}}$,
  then $\UnambigConcatOf{\Sequence_i}{\Sequence_j'}$,
  and $(i_1,j_1) \neq (i_2,j_2) \BooleanImplies \LanguageOf{\Sequence_{i_1,j_1}}
  \Intersect \LanguageOf{\Sequence_{i_2,j_2}} = \emptyset$.

  From Lemma~\ref{lem:unambig-concat-union-equiv}, as
  $i \neq j \BooleanImplies \LanguageOf{\SequenceAlt_i} \Intersect
  \LanguageOf{\SequenceAlt_j} = \emptyset$,
  $i \neq j \BooleanImplies \LanguageOf{\SequenceAlt_i'} \Intersect
  \LanguageOf{\SequenceAlt_j'} = \emptyset$,
  and
  $\UnambigConcatOf
  {\DNFOf{\SequenceAlt_1' \DNFSep \ldots \DNFSep \SequenceAlt_m'}}
  {\DNFOf{\SequenceAlt_1 \DNFSep \ldots \DNFSep \SequenceAlt_n}}$,
  then $\UnambigConcatOf{\SequenceAlt_j'}{\SequenceAlt_i}$,
  and $(j_1,i_1) \neq (j_2,i_2) \BooleanImplies \LanguageOf{\SequenceAlt_{j_1,i_1}}
  \Intersect \LanguageOf{\SequenceAlt_{j_2,i_2}} = \emptyset$.

  So, from Lemma~\ref{lem:typ-sem-swap-seq},
  $\SwapSequenceLensOf{\SequenceLens_i}{\SequenceLens_j} \OfRewritelessType
  \ConcatSequenceOf{\Sequence_i}{\Sequence_j} \Leftrightarrow
  \ConcatSequenceOf{\SequenceAlt_j}{\SequenceAlt_i}$.

  Define $\SequenceLens_{i,j} = \SwapSequenceLensOf{\SequenceLens_i}{\SequenceLens_j}$

  Consider the derivation

  \[
    \inferrule*
    {
      \SequenceLens_{i,j} \OfType \Sequence_{i,j} \Leftrightarrow \SequenceAlt_{j,i}\\
      \DistributeSwapPermutationOf{\sigma_1}{\sigma_2} \in \PermutationSetOf{n\times
        m}\\
      (i_1,j_1) \neq (i_2,j_2) \BooleanImplies \LanguageOf{\Sequence_{i_1,j_1}} \Intersect
      \LanguageOf{\Sequence_{i_2,j_2}} = \emptyset\\
      (j_1,i_1) \neq (j_2,i_2) \BooleanImplies \LanguageOf{\SequenceAlt_{j_1,i_1}} \Intersect
      \LanguageOf{\SequenceAlt_{j_2,i_2}} = \emptyset\\
    }
    {
      (\DNFLensOf{\SequenceLens_{1,1} \DNFLSep \ldots \DNFLSep \SequenceLens_{n,m}},
      \DistributeSwapPermutationOf{\sigma_1}{\sigma_2})
      \OfRewritelessType
      \DNFOf{\Sequence_{1,1} \DNFSep \ldots \DNFSep \Sequence_{n,m}}
      \Leftrightarrow
      \DNFOf{\SequenceAlt_{1,1} \DNFSep \ldots \DNFSep \SequenceAlt_{m,n}}
    }
  \]

  We wish to show that this is a derivation of
  $\SwapDNFLensOf{\DNFLens_1}{\DNFLens_2} \OfType
  \ConcatDNFOf{\DNFRegex_1}{\DNFRegex_2} \Leftrightarrow
  \ConcatDNFOf{\DNFRegexAlt_2}{\DNFRegexAlt_1}$.

  $(\DNFLensOf{\SequenceLens_{1,1} \DNFLSep \ldots \DNFLSep \SequenceLens_{n,m}},
  \DistributeSwapPermutationOf{\sigma_1}{\sigma_2}) =
  (\SwapSequenceLensOf{\SequenceLens_1}{\SequenceLens_1'} \DNFLSep \ldots \DNFLSep 
  \SwapSequenceLensOf{\SequenceLens_n}{\SequenceLens_m'},\DistributeSwapPermutationOf{\sigma_1}{\sigma_2})
  =
  \SwapDNFLensOf{\DNFLens_1}{\DNFLens_2}$.

  $\DNFOf{\Sequence_{1,1} \DNFSep \ldots \DNFSep \Sequence_{n,m}} = \DNFOf{\Sequence_1
    \ConcatSequence \Sequence_1' \DNFSep  \ldots \DNFSep 
  \Sequence_n \ConcatSequence \Sequence_m'} = \ConcatDNFOf{\DNFRegex_1}{\DNFRegex_2}$

  $\DNFOf{\SequenceAlt_{1,1} \DNFSep \ldots \DNFSep \SequenceAlt_{m,n}} = \DNFOf{\SequenceAlt_1'
    \ConcatSequence \SequenceAlt_1 \DNFSep  \ldots \DNFSep 
    \SequenceAlt_m' \ConcatSequence \SequenceAlt_n} = \ConcatDNFOf{\DNFRegexAlt_2}{\DNFRegexAlt_1}$
  
  So we have a derivation of
  $\SwapSequenceLensOf{\SequenceLens_1}{\SequenceLens_2} \OfType
  \ConcatSequenceOf{\Sequence_1}{\Sequence_2} \Leftrightarrow
  \ConcatSequenceOf{\SequenceAlt_2}{\SequenceAlt_1}$

  We also wish to have the desired semantics.

  \[
    \begin{array}{l}
      \SemanticsOf{(\DNFLensOf{\SequenceLens_{1,1} \DNFLSep \ldots \DNFLSep \SequenceLens_{n,m}},
      \DistributeSwapPermutationOf{\sigma_1}{\sigma_2})}\\
      = \SetOf{(\String_1\Concat\String_2,\StringAlt_2\Concat\StringAlt_1)
      \SuchThat \exists i,j. (\String_1,\StringAlt_1) \in
      \SemanticsOf{\SequenceLens_i} \BooleanAnd (\String_2,\StringAlt_2) \in
      \SemanticsOf{\SequenceLens_j}}\\
      = \SetOf{(\String_1\Concat\String_2,\StringAlt_2\Concat\StringAlt_1)
      \SuchThat (\String_1,\StringAlt_1) \in
      \SemanticsOf{\DNFLens} \BooleanAnd (\String_2,\StringAlt_2) \in
      \SemanticsOf{\DNFLens'}}
    \end{array}
  \]
\end{proof}

%\begin{definition}[Adjacent Swapping Permutation]
%  Let $\sigma_{i} \in S_n$ be the permutation where
%  $\sigma_{i}(i) = i+1$, $\sigma_{i}(i+1) = i$, $\sigma_{i}(k) = k$
%  when $k\neq i$, and $k\neq i+1$.
%\end{definition}

\begin{lemma}[Typing and Semantics of $\OrDNFLens$]
  \label{lem:typ_sem_or}
  Let $\DNFLens_1 \OfType \DNFRegex_1 \Leftrightarrow \DNFRegexAlt_1$ and
  $\DNFLens_2 \OfType \DNFRegex_2 \Leftrightarrow \DNFRegexAlt_2$ be the typing of
  two DNF lenses, where
  $\LanguageOf{\DNFRegex_1} \Intersect \LanguageOf{\DNFRegex_2} = \emptyset$
  Then $\OrDNFLensOf{\DNFLens_1}{\DNFLens_2} \OfType
  \OrDNFOf{\DNFRegex_1}{\DNFRegex_2} \Leftrightarrow
  \OrDNFOf{\DNFRegexAlt_1}{\DNFRegexAlt_2}$ and
  $\SemanticsOf{\OrDNFLensOf{\DNFLens_1}{\DNFLens_2}} =
  \SetOf{(\String,\StringAlt) \SuchThat
    (\String,\StringAlt)\in\SemanticsOf{\DNFLens_1}
    \BooleanOr(\String,\StringAlt)\in\SemanticsOf{\DNFLens_2}}$
\end{lemma}
\begin{proof}
  By assumption, there exists typing derivations

  \[
    \DNFLens_1 \OfRewritelessType \DNFRegex_1 \Leftrightarrow \DNFRegexAlt_1
  \]
  and
  \[
    \DNFLens_2 \OfRewritelessType \DNFRegex_2 \Leftrightarrow \DNFRegexAlt_2
  \]

  By inversion, we know that the last rule application on each side was
  \DNFLensRule{}, giving

  \[
    \inferrule*
    {
      \SequenceLens_i \OfType \Sequence_i \Leftrightarrow \SequenceAlt_i\\
      \sigma_1 \in \PermutationSetOf{n}\\
      i \neq j \BooleanImplies \Sequence_{i} \cap \Sequence_{j}=\emptyset\\
      i \neq j \BooleanImplies \SequenceAlt_{i} \cap \SequenceAlt_{j}=\emptyset\\
    }
    {
      (\DNFLensOf{\SequenceLens_1 \DNFLSep \ldots \DNFLSep \SequenceLens_n},\sigma_1)
      \OfRewritelessType
      \DNFOf{\Sequence_1 \DNFSep \ldots \DNFSep \Sequence_n} \Leftrightarrow
      \DNFOf{\SequenceAlt_{\sigma_1(1)} \DNFSep \ldots \DNFSep \SequenceAlt_{\sigma_1(n)}}
    }
  \]
  and
  \[
    \inferrule*
    {
      \SequenceLens_i' \OfType \Sequence_i' \Leftrightarrow \SequenceAlt_i'\\
      \sigma_2 \in \PermutationSetOf{m}\\
      i \neq j \BooleanImplies \Sequence_{i}' \cap \Sequence_{j}'=\emptyset\\
      i \neq j \BooleanImplies \SequenceAlt_{i}' \cap \SequenceAlt_{j}'=\emptyset\\
    }
    {
      (\DNFLensOf{\SequenceLens_1' \DNFLSep \ldots \DNFLSep \SequenceLens_m'},\sigma_2)
      \OfRewritelessType
      \DNFOf{\Sequence_1' \DNFSep \ldots \DNFSep \Sequence_m'} \Leftrightarrow
      \DNFOf{\SequenceAlt_{\sigma_2(1)}' \DNFSep \ldots \DNFSep \SequenceAlt_{\sigma_2(m)}'}
    }
  \]
  where
  $\DNFLens_1 = (\DNFLensOf{\SequenceLens_1 \DNFLSep \ldots \DNFLSep \SequenceLens_n},\sigma_1)$,
  $\DNFRegex_1 = \DNFOf{\Sequence_1 \DNFSep \ldots \DNFSep \Sequence_n}$,
  $\DNFRegexAlt_1 = \DNFOf{\SequenceAlt_{\sigma_1(1)} \DNFSep \ldots \DNFSep \SequenceAlt_{\sigma_1(n)}}$,
  $\DNFLens_2 = (\DNFLensOf{\SequenceLens_1' \DNFLSep \ldots \DNFLSep \SequenceLens_m'},\sigma_2)$,
  $\DNFRegex_2 = \DNFOf{\Sequence_1' \DNFSep \ldots \DNFSep \Sequence_m'}$, and
  $\DNFRegexAlt_2 =
  \DNFOf{\SequenceAlt_{\sigma_2(1)}' \DNFSep \ldots \DNFSep \SequenceAlt_{\sigma_2(m)}'}$.

  Define $\Sequence_i$ as $\Sequence_{i-n}'$ for
  $i \in \RangeIncInc{n+1}{n+m}$.
  Define $\SequenceAlt_i$ as $\SequenceAlt_{i-n}$ for $i \in
  \RangeIncInc{n+1}{n+m}$.
  Define $\SequenceLens_i$ as $\SequenceLens_{i-n}'$ for $i \in
  \RangeIncInc{n+1}{n+m}$.

  If $i \neq j$, and $i,j\in\RangeIncInc{1}{n}$, then $\Sequence_i \cap
  \Sequence_j = \emptyset$ by the derivation of $\DNFLens_1$.
  If $i \neq j$, and $i,j\in\RangeIncInc{n+1}{n+m}$, then $\Sequence_i \cap
  \Sequence_j = \emptyset$ by the derivation of $\DNFLens_2$.
  If $i \neq j$ and $i\in\RangeIncInc{1}{n}$ and $j\in\RangeIncInc{n+1}{n+m}$,
  then $\Sequence_i \cap \Sequence_j = \emptyset$ as $\DNFRegex_1 \cap
  \DNFRegex_2 = \emptyset$, and $\LanguageOf{\Sequence_i} \subset
  \LanguageOf{\DNFRegex_1}$, and $\LanguageOf{\Sequence_j} \subset
  \LanguageOf{\DNFRegex_1}$.
  If $i \neq j$ and $i\in\RangeIncInc{n+1}{n+m}$ and $j\in\RangeIncInc{1}{n}$,
  then $\Sequence_i \cap \Sequence_j = \emptyset$ as $\cap$ is commutative.
  Because of these cases, if $i \neq j$, then $\Sequence_i \cap \Sequence_j = \emptyset$ for all $i,j
  \in \RangeIncInc{1}{n+m}$.

  For a symmetric reason, if $i \neq j$, then $\SequenceAlt_i \cap
  \SequenceAlt_j = \emptyset$, for all $i,j \in \RangeIncInc{1}{n+m}$.

  Consider the derivation

  \[
    \inferrule*
    {
      \SequenceLens_i \OfType \Sequence_i \Leftrightarrow \SequenceAlt_i\\
      \ConcatPermutationOf{\sigma_1}{\sigma_2} \in \PermutationSetOf{n}\\
      i \neq j \BooleanImplies \Sequence_{i} \cap \Sequence_{j}=\emptyset\\
      i \neq j \BooleanImplies \SequenceAlt_{i} \cap \SequenceAlt_{j}=\emptyset\\
    }
    {
      (\DNFLensOf{\SequenceLens_1 \DNFLSep \ldots \DNFLSep \SequenceLens_{n+m}},
      \ConcatPermutationOf{\sigma_1}{\sigma_2})
      \OfRewritelessType
      \DNFOf{\Sequence_1 \DNFSep \ldots \DNFSep \Sequence_{n+m}} \Leftrightarrow
      \DNFOf{\SequenceAlt_{\ConcatPermutationOf{\sigma_1}{\sigma_2}(1)} \DNFSep \ldots
         \DNFSep \SequenceAlt_{\ConcatPermutationOf{\sigma_1}{\sigma_2}(1)(n+m)}}
    }
  \]

  We wish to show that this is a derivation of
  $\OrDNFLensOf{\DNFLens_1}{\DNFLens_2} \OfType
  \OrDNFOf{\DNFRegex_1}{\DNFRegex_2} \Leftrightarrow
  \OrDNFOf{\DNFRegexAlt_1}{\DNFRegexAlt_2}$.

  \[
    \begin{array}{rcl}
      (\DNFLensOf{\SequenceLens_1 \DNFLSep \ldots \DNFLSep \SequenceLens_{n+m}},
      \ConcatPermutationOf{\sigma_1}{\sigma_2})
      & = & (\DNFLensOf{\SequenceLens_1 \DNFLSep \ldots \DNFLSep \SequenceLens_n \DNFLSep 
            \SequenceLens_1' \DNFLSep \ldots \DNFLSep \SequenceLens_m'},
            \ConcatPermutationOf{\sigma_1}{\sigma_2})\\
      & = & \OrDNFLensOf{\DNFLens_1}{\DNFLens_2}
    \end{array}
  \]

  \[
    \begin{array}{rcl}
      \DNFOf{\Sequence_1 \DNFSep \ldots \DNFSep \Sequence_{n+m}}
      & = & \DNFOf{\Sequence_1 \DNFSep \ldots \DNFSep \Sequence_n \DNFSep 
            \Sequence_1' \DNFSep \ldots\Sequence_n'}\\
      & = & \OrDNFOf{\DNFRegex_1}{\DNFRegex_2}
    \end{array}
  \]

  \[
    \begin{array}{rcl}
      \DNFOf{\SequenceAlt_{\ConcatPermutationOf{\sigma_1}{\sigma_2}(1)} \DNFSep \ldots \DNFSep 
      \SequenceAlt_{\ConcatPermutationOf{\sigma_1}{\sigma_2}(n+m)}}
      & = & \DNFOf{\SequenceAlt_{\sigma_1(1)} \DNFSep \ldots \DNFSep \SequenceAlt_{\sigma_1(n)} \DNFSep 
            \SequenceAlt_{\sigma_2(n+1-n)+n} \DNFSep \SequenceAlt_{\sigma_2(n+m-n)+n}}\\
      & = & \DNFOf{\SequenceAlt_{\sigma_1(1)} \DNFSep \ldots \DNFSep \SequenceAlt_{\sigma_1(n)} \DNFSep 
            \SequenceAlt_{\sigma_2(1)+n} \DNFSep \SequenceAlt_{\sigma_2(m)+n}}\\
      & = & \DNFOf{\SequenceAlt_{\sigma_1(1)} \DNFSep \ldots \DNFSep \SequenceAlt_{\sigma_1(n)} \DNFSep 
            \SequenceAlt_{\sigma_2(1)}' \DNFSep \SequenceAlt_{\sigma_2(m)}'}\\
      & = & \OrDNFOf{\DNFRegexAlt_1}{\DNFRegexAlt_2}
    \end{array}
  \]

  So we have a derivation of $\OrDNFLensOf{\DNFLens_1}{\DNFLens_2} \OfType
  \OrDNFOf{\DNFRegex_1}{\DNFRegex_2} \Leftrightarrow
  \OrDNFOf{\DNFRegexAlt_1}{\DNFRegexAlt_2}$

  We also wish to have the desired semantics.

  \[
    \begin{array}{rcl}
      \SemanticsOf{(\DNFLensOf{\SequenceLens_1 \DNFLSep \ldots \DNFLSep \SequenceLens_{n+m}},
      \ConcatPermutationOf{\sigma_1}{\sigma_2})}
      & = & \SetOf{(\String,\StringAlt)\SuchThat
            (\String,\StringAlt)\in\SequenceLens_i\text{ for some $i$}}\\
      & = & \SetOf{(\String,\StringAlt)\SuchThat
            (\String,\StringAlt)\in\SequenceLens_i
            \text{ for some $i\in\RangeIncInc{1}{n}$} \BooleanOr\\
      &   & \hspace*{3em}(\String,\StringAlt)\in\SequenceLens_i
            \text{ for some $i\in\RangeIncInc{n+1}{n+m}$}}\\
      & = & \SetOf{(\String,\StringAlt)\SuchThat
            (\String,\StringAlt)\in\SequenceLens_i
            \text{ for some $i\in\RangeIncInc{1}{n}$} \BooleanOr\\
      &   & \hspace*{3em}(\String,\StringAlt)\in\SequenceLens_i'
            \text{ for some $i\in\RangeIncInc{1}{m}$}}\\
      & = & \SetOf{(\String,\StringAlt) \SuchThat
            (\String,\StringAlt)\in\SemanticsOf{\DNFLens_1}
            \BooleanOr(\String,\StringAlt)\in\SemanticsOf{\DNFLens_2}}
    \end{array}
  \]

\end{proof}

\begin{lemma}[Typing and Semantic of $\AtomToDNFLens$]
  \label{lem:typ_sem_todnflens}
  If $\AtomLens \OfRewritelessType \Atom \Leftrightarrow \AtomAlt$ is the
  typing of a rewriteless Atom lens, then
  $\AtomToDNFLensOf{\AtomLens} \OfRewritelessType \AtomToDNFOf{\Atom}
  \Leftrightarrow \AtomToDNFOf{\AtomAlt}$, and
  $\SemanticsOf{\AtomToDNFLensOf{\AtomLens}} = \SemanticsOf{\AtomLens}$.
\end{lemma}
\begin{proof}
  Let $\AtomLens \OfRewritelessType \Atom \Leftrightarrow \AtomAlt$.
  
  $\SequenceUnambigConcatOf{(\EmptyString;\Atom;\EmptyString)}$ because
  $\LanguageOf{\EmptyString} = \SetOf{\EmptyString}$.
  $\SequenceUnambigConcatOf{(\EmptyString;\AtomAlt;\EmptyString)}$ because
  $\LanguageOf{\EmptyString} = \SetOf{\EmptyString}$.

  As there is only one sequence, the pointwise disjoint condition for DNF lenses
  are true vacuously.
  
  Consider the typing derivation
  \[
    \inferrule*
    {
      \inferrule*
      {
        \AtomLens \OfRewritelessType \Atom \Leftrightarrow \AtomAlt\\
        \SequenceUnambigConcatOf{(\EmptyString;\Atom;\EmptyString)}\\
        \SequenceUnambigConcatOf{(\EmptyString;\AtomAlt;\EmptyString)}
      }
      {
        (\SequenceLensOf{(\EmptyString,\EmptyString) \SeqLSep \AtomLens \SeqLSep (\EmptyString,\EmptyString)},\Identity)
        \OfRewritelessType
        \SequenceOf{\EmptyString \SeqSep \Atom \SeqSep \EmptyString} \Leftrightarrow
        \SequenceOf{\EmptyString \SeqSep \AtomAlt \SeqSep \EmptyString}
      }\\
      i \neq j \BooleanImplies \LanguageOf{\Sequence_i} \Intersect
      \LanguageOf{\Sequence_j} = \emptyset
    }
    {
      (\DNFLensOf{(\SequenceLensOf{(\EmptyString,\EmptyString) \SeqLSep \AtomLens \SeqLSep (\EmptyString,\EmptyString)},\Identity)},\Identity)
      \OfRewritelessType
      \DNFOf{\SequenceOf{\EmptyString \SeqSep \Atom \SeqSep \EmptyString}} \Leftrightarrow
      \DNFOf{\SequenceOf{\EmptyString \SeqSep \AtomAlt \SeqSep \EmptyString}}
    }
  \]
  
  $\AtomToDNFLensOf{\AtomLens} =
  (\DNFLensOf{(\SequenceLensOf{(\EmptyString,\EmptyString) \SeqLSep \AtomLens \SeqLSep (\EmptyString,\EmptyString)},\Identity)},\Identity)$.
  
  $\SemanticsOf{\SequenceLensOf{(\EmptyString,\EmptyString) \SeqLSep \AtomLens \SeqLSep (\EmptyString,\EmptyString)}}
  =
  \SetOf{(\EmptyString,\String,\EmptyString,\EmptyString,\StringAlt,\EmptyString)
    \SuchThat (\String,\StringAlt) \in \LanguageOf{\AtomLens}} =
  \SemanticsOf{\AtomLens}$.

  $\SemanticsOf{(\DNFLensOf{(\SequenceLensOf{(\EmptyString,\EmptyString) \SeqLSep \AtomLens \SeqLSep (\EmptyString,\EmptyString)},\Identity)},\Identity)}
  =
  \SetOf{(\String,\StringAlt) \SuchThat (\String,\StringAlt) \in
    \SemanticsOf{\SequenceLensOf{(\EmptyString,\EmptyString) \SeqLSep \AtomLens \SeqLSep (\EmptyString,\EmptyString)}}}
  = \SetOf{(\String,\StringAlt) \SuchThat (\String,\StringAlt) \in
    \SemanticsOf{\AtomLens}}
  = \SemanticsOf{\AtomLens}$
\end{proof}

\begin{lemma}[Typing and Semantics of $\AtomToDNFLensOf{\IterateLensOf{\cdot}}$]
  \label{lem:typ_sem_it}
  Let $\DNFLens \OfRewritelessType \DNFRegex \Leftrightarrow \DNFRegexAlt$ be
  the typing of a rewriteless DNF lens, where
  $\UnambigItOf{\DNFRegex}$ and $\UnambigItOf{\DNFRegexAlt}$.
  $\DNFLensOf{\SequenceLensOf{\IterateLensOf{\DNFLens}}} \OfRewritelessType
  \DNFOf{\SequenceOf{\StarOf{\DNFRegex}}} \Leftrightarrow
  \DNFOf{\SequenceOf{\StarOf{\DNFRegexAlt}}}$ and
  $\SemanticsOf{\DNFLensOf{\SequenceLensOf{\IterateLensOf{\DNFLens}}}} =
  \SetOf{
    (\String_1\Concat\ldots\Concat\String_n,\StringAlt_1\Concat\ldots\Concat\StringAlt_n)
    \SuchThat
    (\String_i,\StringAlt_i)\in\SemanticsOf{\DNFLens}}$
\end{lemma}
\begin{proof}
  By assumption, there exists a typing derivation

  \[
    \DNFLens \OfRewritelessType \DNFRegex \Leftrightarrow \DNFRegexAlt
  \]

  Consider the typing derivation

  \[
    \inferrule*
    {
      \inferrule*
      {
        \inferrule*
        {
          \DNFLens \OfRewritelessType \DNFRegex \Leftrightarrow \DNFRegexAlt\\
          \UnambigItOf{\DNFRegex}\\
          \UnambigItOf{\DNFRegexAlt}
        }
        {
          \IterateLensOf{\DNFLens} \OfRewritelessType
          \StarOf{\DNFRegex} \Leftrightarrow \StarOf{\DNFRegexAlt}
        }\\
        \SequenceUnambigConcatOf{\EmptyString;\DNFRegex;\EmptyString}\\
        \SequenceUnambigConcatOf{\EmptyString;\DNFRegexAlt;\EmptyString}
      }
      {
        \SequenceLensOf{(\EmptyString,\EmptyString) \SeqLSep \IterateLensOf{\DNFLens} \SeqLSep (\EmptyString,\EmptyString)}
        \OfRewritelessType \SequenceOf{\EmptyString \SeqSep \DNFRegex \SeqSep \EmptyString}
        \Leftrightarrow \SequenceOf{\EmptyString \SeqSep \DNFRegex \SeqSep \EmptyString}
      }
    }
    {
      \DNFLensOf{\SequenceLensOf{\IterateLensOf{\DNFLens}}} \OfRewritelessType
      \DNFOf{\SequenceOf{\StarOf{\DNFRegex}}} \Leftrightarrow
      \DNFOf{\SequenceOf{\StarOf{\DNFRegexAlt}}}
    }
  \]

  And the semantics are shown to be equal to the desired semantics.

  \[
    \begin{array}{rcl}
      \SemanticsOf{\DNFLensOf{\SequenceLensOf{\IterateLensOf{\DNFLens}}}}
      & = &
            \SetOf{
            (\String,\StringAlt)
            \SuchThat
            (\String,\StringAlt)\in\SemanticsOf{\SequenceLensOf{\IterateLensOf{\DNFLens}}}}\\
      & = &
            \SetOf{
            (\EmptyString\Concat\String\Concat\EmptyString,\EmptyString\Concat\StringAlt\Concat\EmptyString)
            \SuchThat
            (\String,\StringAlt)\in\SemanticsOf{\IterateLensOf{\DNFLens}}}\\
      & = &
            \SetOf{
            (\String_1\Concat\ldots\Concat\String_n,\StringAlt_1\Concat\ldots\Concat\StringAlt_n)
            \SuchThat
            (\String_i,\StringAlt_i)\in\SemanticsOf{\DNFLens}}
    \end{array}
  \]
\end{proof}

\subsection{Complex Lens Operator Properties}
\label{complex-lens-operators}

The previous properties of lens operators were merely about the operators, and
how they could be typed.  This portion writes about how lens operators have the
same semantics as lenses with
very complex properties, up to the existence of an identity lens.  Much of this
complication comes from the fact that the DNF regular expression operators don't
have right distributivity.  An analogue to the
commutativity of regular expression $\OrRegexType$ to be expressed using these
properties.

\begin{lemma}[Commutativity of $\OrDNF$]
  \label{lem:or-dnf-commutativity}
  If there exists a lens $\DNFLens \OfRewritelessType \DNFRegex_1 \OrDNF \DNFRegex_2
  \Leftrightarrow \DNFRegexAlt_1 \OrDNF \DNFRegexAlt_2$, then there exists a
  lens
  $\DNFLens \OfRewritelessType \DNFRegex_1 \OrDNF \DNFRegex_2
  \Leftrightarrow \DNFRegexAlt_2 \OrDNF \DNFRegexAlt_1$.
\end{lemma}
\begin{proof}
  Let $\DNFRegex_1 = \DNFOf{\Sequence_{1,1} \DNFSep \ldots \DNFSep
    \Sequence_{1,n}}$.
  
  Let $\DNFRegex_2 = \DNFOf{\Sequence_{2,1} \DNFSep \ldots \DNFSep
    \Sequence_{2,n'}}$.
  
  Let $\DNFRegexAlt_1 = \DNFOf{\SequenceAlt_{1,1} \DNFSep \ldots \DNFSep
    \SequenceAlt_{1,m}}$.
  
  Let $\DNFRegexAlt_2 = \DNFOf{\SequenceAlt_{2,1} \DNFSep \ldots \DNFSep
    \SequenceAlt_{2,m'}}$.
  
  Let $\DNFLens = (\DNFLensOf{\SequenceLens_1 \DNFLSep \ldots \DNFLSep \SequenceLens_{n+m}},\sigma)$

  So $\DNFRegex_1 \OrDNF \DNFRegex_2 =
  \DNFOf{\Sequence_1 \DNFSep \ldots \DNFSep \Sequence_{n+n'}}$, where
  $\Sequence_i =
  \begin{cases*}
    \Sequence_{1,i} & if $i \leq n$\\
    \Sequence_{2,i-n} & if $i > n$
  \end{cases*}$

  So $\DNFRegexAlt_1 \OrDNF \DNFRegexAlt_2 =
  \DNFOf{\SequenceAlt_{\sigma(1)} \DNFSep \ldots \DNFSep \SequenceAlt_{\sigma(m+m')}}$, where
  $\SequenceAlt_{\sigma(i)} =
  \begin{cases*}
    \SequenceAlt_{1,i} & if $i \leq m$\\
    \SequenceAlt_{2,i-m} & if $i > m$
  \end{cases*}$

  By inversion
  \[
    \inferrule*
    {
      \SequenceLens_1 \OfRewritelessType \Sequence_1 \Leftrightarrow \SequenceAlt_1\\
      \ldots\\
      \SequenceLens_n \OfRewritelessType \Sequence_n \Leftrightarrow \SequenceAlt_n\\\\
      \sigma \in \PermutationSetOf{n}\\
      i \neq j \Rightarrow \LanguageOf{\Sequence_{i}} \cap \LanguageOf{\Sequence_{j}}=\emptyset\\
      i \neq j \Rightarrow \LanguageOf{\SequenceAlt_{i}} \cap \LanguageOf{\SequenceAlt_{j}}=\emptyset\\
    }
    {
      (\DNFLensOf{\SequenceLens_1\DNFLSep\ldots\DNFLSep\SequenceLens_n},\sigma) \OfRewritelessType\\
      \DNFOf{\Sequence_1\DNFSep\ldots\DNFSep\Sequence_n}
      \Leftrightarrow \DNFOf{\SequenceAlt_{\sigma(1)}\DNFSep\ldots\DNFSep\SequenceAlt_{\sigma(n)}}
    }
  \]

  Consider the permutation $\sigma'(i) =
  \begin{cases*}
    \sigma(i+m) & if $\sigma(i) \leq m'$\\
    \sigma(i-m) & if $\sigma(i) > m'$
  \end{cases*}$

  Consider the lens
  \[
    \inferrule*
    {
      \SequenceLens_1 \OfRewritelessType \Sequence_1 \Leftrightarrow \SequenceAlt_1\\
      \ldots\\
      \SequenceLens_n \OfRewritelessType \Sequence_n \Leftrightarrow \SequenceAlt_n\\\\
      \sigma' \in \PermutationSetOf{n}\\
      i \neq j \Rightarrow \LanguageOf{\Sequence_{i}} \cap \LanguageOf{\Sequence_{j}}=\emptyset\\
      i \neq j \Rightarrow \LanguageOf{\SequenceAlt_{i}} \cap \LanguageOf{\SequenceAlt_{j}}=\emptyset\\
    }
    {
      (\DNFLensOf{\SequenceLens_1\DNFLSep\ldots\DNFLSep\SequenceLens_n},\sigma')
      \OfRewritelessType\\
      \DNFOf{\Sequence_1\DNFSep\ldots\DNFSep\Sequence_n}
      \Leftrightarrow \DNFOf{\SequenceAlt_{\sigma'(1)}\DNFSep\ldots\DNFSep\SequenceAlt_{\sigma'(n)}}
    }
  \]
  
  So $\SequenceAlt_{\sigma'(i)} =
  \begin{cases*}
    \SequenceAlt_{\sigma(i+m)} = \SequenceAlt_{2,i} & if $\sigma(i) \leq m'$\\
    \SequenceAlt_{\sigma(i-m)} = \SequenceAlt_{1,i} & if $\sigma(i) > m$
  \end{cases*}$

  So $\DNFOf{\SequenceAlt_{\sigma'(1)} \DNFSep \ldots \DNFSep \SequenceAlt_{\sigma'(m+m')}} =
  \DNFRegexAlt_2 \OrDNF \DNFRegexAlt_1$.

  Furthermore, The semantics are the same, as permutation has no impact on the
  semantics of DNF lenses.
\end{proof}

\begin{lemma}[Left Unrolling of $\IterateLens$]
  \label{lem:iterate-lens-unroll-left}
  If $\IterateLensOf{\DNFLens} \OfRewritelessType \StarOf{\DNFRegex}
  \Leftrightarrow \StarOf{\DNFRegexAlt}$ is an atom lens, then
  $\DNFLens' = \DNFLensOf{\SequenceLensOf{(\EmptyString,\EmptyString)}}
  \OrDNFLens
  (\DNFLens \ConcatDNFLens \AtomToDNFLensOf{\IterateLensOf{\DNFLens}})
  \OfRewritelessType
  \DNFOf{\SequenceOf{\EmptyString}}
  \OrDNF
  (\DNFRegex \ConcatDNF \AtomToDNFOf{\StarOf{\DNFRegex}})
  \Leftrightarrow
  \DNFOf{\SequenceOf{\EmptyString}}
  \OrDNF
  (\DNFRegexAlt \ConcatDNF \AtomToDNFOf{\StarOf{\DNFRegexAlt}})$
  is a DNF Lens with
  $\SemanticsOf{\IterateLensOf{\DNFLens}} = \SemanticsOf{\DNFLens'}$
\end{lemma}
\begin{proof}
  So $\StarOf{\DNFRegex}$ and $\StarOf{\DNFRegexAlt}$ are strongly unambiguous
  atoms.

  As such, this means $\UnambigItOf{\DNFRegex}$, $\UnambigItOf{\DNFRegexAlt}$,
  $\DNFRegex$ is strongly unambiguous, and $\DNFRegexAlt$ is strongly
  unambiguous.

  Want to show: because $\UnambigItOf{\DNFRegex}$,
  $\UnambigConcatOf{\DNFRegex}{\AtomToDNFOf{\StarOf{\DNFRegex}}}$.
  Let $\String_1,\String_2 \in \LanguageOf{\DNFRegex}$.
  Let $\StringAlt_1,\StringAlt_2 \in \LanguageOf{\StarOf{\DNFRegex}}$.
  Let $\String_1\Concat\StringAlt_1 = \String_2 \Concat \StringAlt_2$.
  This is $\String_1\Concat\StringAlt_{1,1}\Concat\ldots\Concat\StringAlt_{1,n}$
  and $\String_2\Concat\StringAlt_{2,1}\Concat\ldots\Concat\StringAlt_{2,n'}$.
  where each substring is in $\LanguageOf{\DNFRegex}$.  By unambiguous
  iteration, $n=n'$, and $\String_1 = \String_2$, $\StringAlt_{1,i} =
  \StringAlt_{2,i}$,
  so $\String_1 = \String_2$ and $\StringAlt_1 = \StringAlt_2$.

  As such Lemma~\ref{lem:typ_sem_concat} applies, so
  $(\DNFLens \ConcatDNFLens \AtomToDNFLensOf{\IterateLensOf{\DNFLens}})
  \OfRewritelessType
  (\DNFRegex \ConcatDNF \AtomToDNFOf{\StarOf{\DNFRegex}})
  \Leftrightarrow
  (\DNFRegexAlt \ConcatDNF \AtomToDNFOf{\StarOf{\DNFRegexAlt}})$.

  Want to show: because $\UnambigItOf{\DNFRegex}$,
  $\UnambigOrOf{\DNFOf{\SequenceOf{\EmptyString}}}{\DNFRegex \ConcatDNF
    \AtomToDNFOf{\StarOf{\DNFRegex}}}$.
  $\EmptyString$ is the only element of
  $\LanguageOf{\DNFOf{\SequenceOf{\EmptyString}}}$.
  $\EmptyString \not\in \LanguageOf{\DNFRegex \ConcatDNF
    \AtomToDNFOf{\StarOf{\DNFRegex}}}$, as it cannot be in $\LanguageOf{\DNFRegex}$.
  Otherwise if $\EmptyString \in \LanguageOf{\DNFRegex}$, then for all
  $\String_1\Concat\ldots\Concat\String_n =
  \EmptyString\Concat\String_1\Concat\ldots\Concat\String_n$, betraying
  unambiguous iteration.

  As such Lemma~\ref{lem:typ_sem_or} applies, so
  $\DNFLensOf{\SequenceLensOf{(\EmptyString,\EmptyString)}} \OrDNF
  (\DNFLens \ConcatDNFLens \AtomToDNFLensOf{\IterateLensOf{\DNFLens}})
  \OfRewritelessType
  \DNFOf{\SequenceOf{(\EmptyString,\EmptyString)}} \OrDNF
  (\DNFRegex \ConcatDNF \AtomToDNFOf{\StarOf{\DNFRegex}})
  \Leftrightarrow
  \DNFOf{\SequenceOf{(\EmptyString,\EmptyString)}} \OrDNF
  (\DNFRegexAlt \ConcatDNF \AtomToDNFOf{\StarOf{\DNFRegexAlt}})$.

  $\SemanticsOf{\DNFLensOf{\SequenceLensOf{(\EmptyString,\EmptyString)}} \OrDNF
    (\DNFLens \ConcatDNFLens \AtomToDNFLensOf{\IterateLensOf{\DNFLens}})} =
  \SetOf{(\String,\StringAlt) \SuchThat (\String,\StringAlt) \in
    \SemanticsOf{\DNFOf{\SequenceOf{(\EmptyString,\EmptyString)}}} \BooleanOr
    (\String,\StringAlt) \in
    \SemanticsOf{\DNFLens \ConcatDNFLens
      \AtomToDNFLensOf{\IterateLensOf{\DNFLens}}}} =
  \SetOf{(\String,\StringAlt) \SuchThat (\String,\StringAlt) \in
    \SemanticsOf{\DNFOf{\SequenceOf{(\EmptyString,\EmptyString)}}}} \Union
  \SetOf{(\String,\StringAlt) \SuchThat (\String,\StringAlt) \in
    \SemanticsOf{\DNFLens \ConcatDNFLens
      \AtomToDNFLensOf{\IterateLensOf{\DNFLens}}}}$.
  By the definition of $\ConcatDNFLens$ and $\IterateLens$, this equals
  $\SetOf{(\EmptyString,\EmptyString)} \Union
  \SetOf{(\String_1\Concat\ldots\Concat\String_n,\StringAlt_1\Concat\ldots\Concat\StringAlt_n)
    \SuchThat n \geq 1 \BooleanAnd (\String_i,\StringAlt_i) \in \SemanticsOf{\DNFLens}}$.
  Through combining the zero case and the one case, this is
  $\SetOf{(\String_1\Concat\ldots\Concat\String_n,\StringAlt_1\Concat\ldots\Concat\StringAlt_n)
    \SuchThat (\String_i,\StringAlt_i) \in \SemanticsOf{\DNFLens}} =
  \SemanticsOf{\IterateLensOf{\AtomLens}}$.
\end{proof}

\begin{lemma}[Right Unrolling of $\IterateLens$]
  \label{lem:iterate-lens-unroll-right}
  If $\IterateLensOf{\DNFLens} \OfRewritelessType \StarOf{\DNFRegex}
  \Leftrightarrow \StarOf{\DNFRegexAlt}$ is an atom lens, then
  $\DNFLens' = \DNFLensOf{\SequenceLensOf{(\EmptyString,\EmptyString)}}
  \OrDNFLens
  (\AtomToDNFLensOf{\IterateLensOf{\DNFLens}} \ConcatDNFLens \DNFLens)
  \OfRewritelessType
  \DNFOf{\SequenceOf{\EmptyString}}
  \OrDNF
  (\AtomToDNFOf{\StarOf{\DNFRegex}} \ConcatDNF \DNFRegex)
  \Leftrightarrow
  \DNFOf{\SequenceOf{\EmptyString}}
  \OrDNF
  (\AtomToDNFOf{\StarOf{\DNFRegexAlt}} \ConcatDNF \DNFRegexAlt)$
  is a DNF Lens with
  $\SemanticsOf{\IterateLensOf{\DNFLens}} = \SemanticsOf{\DNFLens'}$
\end{lemma}
\begin{proof}
  This is proven symmetrically to Lemma~\ref{lem:iterate-lens-unroll-left}.
\end{proof}

\begin{lemma}[Left Unrolling of $\IterateLens$ DNF]
  \label{lem:iterate-lens-unroll-left-dnf}
  If $\AtomToDNFLensOf{\IterateLensOf{\DNFLens}}$ is a DNF lens, then
  $\DNFLens' = \DNFLensOf{\SequenceLensOf{(\EmptyString,\EmptyString)}} \OrDNFLens (\DNFLens
  \ConcatDNFLens \AtomToDNFLensOf{\IterateLensOf{\DNFLens}})$ is a DNF Lens with
  $\SemanticsOf{\IterateLensOf{\DNFLens}} = \SemanticsOf{\DNFLens'}$
\end{lemma}
\begin{proof}
  This is through a combination of Lemma~\ref{lem:iterate-lens-unroll-left}, and
  Lemma~\ref{lem:typ_sem_todnflens}.
\end{proof}

\begin{lemma}[Right Unrolling of $\IterateLens$ DNF]
  \label{lem:iterate-lens-unroll-right-dnf}
  If $\AtomToDNFLensOf{\IterateLensOf{\DNFLens}}$ is a DNF lens, then
  $\DNFLens' = \DNFLensOf{\SequenceLensOf{(\EmptyString,\EmptyString)}} \OrDNFLens (\AtomToDNFLensOf{\IterateLensOf{\DNFLens}}
  \ConcatDNFLens \DNFLens)$ is a DNF Lens with
  $\SemanticsOf{\IterateLensOf{\DNFLens}} = \SemanticsOf{\AtomToDNFLensOf{\DNFLens'}}$
\end{lemma}
\begin{proof}
  This is through a combination of Lemma~\ref{lem:iterate-lens-unroll-right}, and
  Lemma~\ref{lem:typ_sem_todnflens}.
\end{proof}

\begin{lemma}[Expressibility of Adjacency Swapping Permutation of Separated
  Concat List]
  \label{lem:adj-swap-perm-concat-express}
  Let for all $i \in \RangeIncInc{1}{n}, \DNFLens_i \OfRewritelessType
  \DNFRegex_i \Leftrightarrow \DNFRegexAlt_i$
  Let $\sigma_i$ be an adjacency swapping permutation, where $1 \leq i < n$.
  There exists a DNF lens $\DNFLens \OfRewritelessType \DNFOf{\SequenceOf{\Sep}}
  \DNFRegex_1 \ConcatDNF \DNFOf{\SequenceOf{\Sep}} \ConcatDNF
  \ldots \ConcatDNF \DNFOf{\SequenceOf{\Sep}}  \ConcatDNF \DNFRegex_n
  \ConcatDNF \DNFOf{\SequenceOf{\Sep}} \Leftrightarrow
  \DNFOf{\SequenceOf{\Sep}} \ConcatDNF \DNFRegexAlt_{\sigma_i(1)} \ConcatDNF \ldots \ConcatDNF
  \DNFRegexAlt_{\sigma_i(n)} \ConcatDNF \DNFOf{\SequenceOf{\Sep}}$, where
  $\SemanticsOf{\DNFRegex} =
  \SetOf{(\Sep\Concat\String_1\Concat\Sep\Concat\ldots\Concat\Sep\String_n\Concat\Sep,
    \Sep\Concat\StringAlt_{\sigma_i(1)}\Concat\Sep\Concat\ldots\Concat\Sep\Concat\StringAlt_{\sigma_i(n)}\Concat\Sep)
    \SuchThat
  (\String_i,\StringAlt_i) \in \SemanticsOf{\DNFLens_i}}$.
\end{lemma}
\begin{proof}
  As $\DNFRegex_i$ and
  $\DNFRegexAlt_i$ are strongly unambiguous, by
  Lemma~\ref{lem:strong-unambig-dnf-lens-types}, and from
  Lemma~\ref{lem:sep_unambiguity},
  we have that\\$\SequenceUnambigConcatOf{\DNFOf{\SequenceOf{\Sep}}
    \DNFRegex_1, \DNFOf{\SequenceOf{\Sep}} , 
    \ldots , \DNFOf{\SequenceOf{\Sep}}, 
    \DNFOf{\SequenceOf{\Sep}} ,\DNFRegex_n ,\DNFOf{\SequenceOf{\Sep}}}$ and
  $\SequenceUnambigConcatOf{\DNFOf{\SequenceOf{\Sep}}
    \DNFRegexAlt_{\sigma_i(1)} , \DNFOf{\SequenceOf{\Sep}} , 
    \ldots , \DNFOf{\SequenceOf{\Sep}}, \DNFRegexAlt_{\sigma_i(n)}
    ,\DNFOf{\SequenceOf{\Sep}}}$

  Consider the lens
  $\DNFOf{\SequenceOf{(\Sep,\Sep)}} \ConcatDNFLens \DNFLens_1 \ConcatDNFLens \ldots
  \ConcatDNFLens ((\DNFOf{\SequenceOf{(\Sep,\Sep)}} \ConcatDNFLens \DNFLens_i)
  \SwapDNFLens (\DNFOf{\SequenceOf{(\Sep,\Sep)}} \ConcatDNFLens \DNFLens_{i+1}))
  \ConcatDNFLens \ldots \ConcatDNFLens \DNFLens_n \ConcatDNFLens \DNFOf{\SequenceOf{(\Sep,\Sep)}}$,
  which by Lemma~\ref{lem:typ_sem_concat} and Lemma~\ref{lem:typ_sem_swap}.
  $\DNFOf{\SequenceOf{(\Sep,\Sep)}} \ConcatDNFLens \DNFLens_1 \ConcatDNFLens \ldots
  \ConcatDNFLens ((\DNFOf{\SequenceOf{(\Sep,\Sep)}} \ConcatDNFLens \DNFLens_i)
  \SwapDNFLens (\DNFOf{\SequenceOf{(\Sep,\Sep)}} \ConcatDNFLens \DNFLens_{i+1}))
  \ConcatDNFLens \ldots \ConcatDNFLens \DNFLens_n \ConcatDNFLens \DNFOf{\SequenceOf{(\Sep,\Sep)}}
  \OfRewritelessType \DNFOf{\SequenceOf{\Sep}} \ConcatDNF
  \DNFRegex_1 \ConcatDNF \DNFOf{\SequenceOf{\Sep}} \ConcatDNF
  \ldots \ConcatDNF \DNFOf{\SequenceOf{\Sep}}  \ConcatDNF \DNFRegex_n
  \ConcatDNF \DNFOf{\SequenceOf{\Sep}} \Leftrightarrow
  \DNFOf{\SequenceOf{\Sep}} \ConcatDNF
  \DNFRegex_1 \ConcatDNF \DNFOf{\SequenceOf{\Sep}} \ConcatDNF
  \ldots \ConcatDNF \DNFOf{\SequenceOf{\Sep}} \ConcatDNF \DNFRegexAlt_{i+1} \ConcatDNF
  \DNFOf{\SequenceOf{\Sep}} \ConcatDNF \DNFRegexAlt_i \DNFOf{\SequenceOf{\Sep}} \ConcatDNF \ldots \ConcatDNF \DNFOf{\SequenceOf{\Sep}}  \ConcatDNF \DNFRegex_n
  \ConcatDNF \DNFOf{\SequenceOf{\Sep}}$ as desired.

  Also by Lemma~\ref{lem:typ_sem_swap}, the semantics are as desired.
\end{proof}

\begin{lemma}[Expressibility of Permutation of Separated Concat List]
  \label{lem:perm-sep-concat}
  Let for all $i \in \RangeIncInc{1}{n}, \DNFLens_i \OfRewritelessType
  \DNFRegex_i \Leftrightarrow \DNFRegexAlt_i$
  Let $\sigma$ be a permutation, where $1 \leq i < n$.
  There exists a DNF lens $\DNFLens \OfRewritelessType \DNFOf{\SequenceOf{\Sep}}
  \DNFRegex_1 \ConcatDNF \DNFOf{\SequenceOf{\Sep}} \ConcatDNF
  \ldots \ConcatDNF \DNFOf{\SequenceOf{\Sep}} \ConcatDNF
  \DNFOf{\SequenceOf{\Sep}} \DNFRegex_n \DNFOf{\SequenceOf{\Sep}} \Leftrightarrow
  \DNFOf{\SequenceOf{\Sep}} \ConcatDNF \DNFRegexAlt_{\sigma(1)} \ConcatDNF \ldots \ConcatDNF
  \DNFRegexAlt_{\sigma(n)} \ConcatDNF \DNFOf{\SequenceOf{\Sep}}$, where $\SemanticsOf{\DNFRegex} =
  \SetOf{(\Sep\Concat\String_1\Concat\Sep\Concat\ldots\Concat\Sep\String_n\Concat\Sep,
    \Sep\Concat\StringAlt_{\sigma(1)}\Concat\Sep\Concat\ldots\Concat\Sep\Concat\StringAlt_{\sigma(n)}\Concat\Sep)
    \SuchThat
  (\String_i,\StringAlt_i) \in \SemanticsOf{\DNFLens_i}}$.
\end{lemma}
\begin{proof}
  As $\DNFRegex_i$ and
  $\DNFRegexAlt_i$ are strongly unambiguous, by
  Lemma~\ref{lem:strong-unambig-dnf-lens-types}, and as
  $\UnambigConcatOf{\DNFOf{\SequenceOf{\Sep}}}{\Language}$, and
  $\UnambigConcatOf{\Language}{\DNFOf{\SequenceOf{\Sep}}}$, for all $\Language$,
  we have that $\DNFOf{\SequenceOf{\Sep}} \ConcatDNF \DNFRegex_1 \ConcatDNF \DNFOf{\SequenceOf{\Sep}} \ConcatDNF
  \ldots \ConcatDNF \DNFOf{\SequenceOf{\Sep}} \ConcatDNF
  \DNFOf{\SequenceOf{\Sep}} \DNFRegex_n \DNFOf{\SequenceOf{\Sep}}$ is strongly
  unambiguous.

  From algebra, $\sigma$ can be decomposed into a series of $\sigma_{i_j}
  \Compose \ldots \Compose \sigma_{i_1}$.

  We proceed by induction

  \begin{case}[$j=0$]
    $\sigma = \Identity$.
    
    Through repeated application of $\ConcatDNFLens$, 
    $\DNFOf{\SequenceOf{(\Sep,\Sep)}} \ConcatDNFLens
    \DNFLens_1 \ConcatDNFLens \ldots \ConcatDNFLens
    \DNFLens_n \ConcatDNFLens \DNFLensOf{\SequenceLensOf{(\Sep,\Sep)}}
    \OfRewritelessType \DNFOf{\SequenceOf{\Sep}}
    \DNFRegex_1 \ConcatDNF \DNFOf{\SequenceOf{\Sep}} \ConcatDNF
    \ldots \ConcatDNF \DNFOf{\SequenceOf{\Sep}} \ConcatDNF
    \DNFOf{\SequenceOf{\Sep}} \DNFRegex_n \DNFOf{\SequenceOf{\Sep}} \Leftrightarrow
    \DNFRegexAlt_{\sigma(1)} \ConcatDNF \ldots \ConcatDNF
    \DNFRegexAlt_{\sigma(n)}$, where
    $\SemanticsOf{\DNFRegex} =
    \SetOf{(\Sep\Concat\String_1\Concat\Sep\Concat\ldots\Concat\Sep\String_n\Concat\Sep,
      \Sep\Concat\StringAlt_{1}\Concat\Sep\Concat\ldots\Concat\Sep\Concat\StringAlt_{n}\Concat\Sep)
      \SuchThat
      (\String_i,\StringAlt_i) \in \SemanticsOf{\DNFLens_i}}$
  \end{case}

  \begin{case}[$j>0$]
    $\sigma = \sigma_{i_j} \Compose \ldots \Compose \sigma_{i_1}$.
    $\sigma' = \sigma_{i_{j-1}} \Compose \ldots \Compose \sigma_{i_1}$.
    $\sigma = \sigma_{i_j} \Compose \sigma'$.
    
    By IH there exists a DNF lens $\DNFLens \OfRewritelessType \DNFOf{\SequenceOf{\Sep}}
    \DNFRegex_1 \ConcatDNF \DNFOf{\SequenceOf{\Sep}} \ConcatDNF
    \ldots \ConcatDNF \DNFOf{\SequenceOf{\Sep}} \ConcatDNF
    \DNFOf{\SequenceOf{\Sep}} \DNFRegex_n \DNFOf{\SequenceOf{\Sep}} \Leftrightarrow
    \DNFOf{\SequenceOf{\Sep}} \ConcatDNF \DNFRegexAlt_{\sigma'(1)} \ConcatDNF \ldots \ConcatDNF
    \DNFRegexAlt_{\sigma'(n)} \ConcatDNF \DNFOf{\SequenceOf{\Sep}}$, where $\SemanticsOf{\DNFRegex} =
    \SetOf{(\Sep\Concat\String_1\Concat\Sep\Concat\ldots\Concat\Sep\String_n\Concat\Sep,
      \Sep\Concat\StringAlt_{\sigma'(1)}\Concat\Sep\Concat\ldots\Concat\Sep\Concat\StringAlt_{\sigma'(n)}\Concat\Sep)
      \SuchThat
      (\String_i,\StringAlt_i) \in \SemanticsOf{\DNFLens_i}}$
    
    As $\DNFRegexAlt_i$ are strongly unambiguous, by
    Lemma~\ref{lem:strongly-unambiguous-identity-expressible}, there exists
    an identity lens $\DNFLens_i' \OfRewritelessType \DNFRegexAlt_i \Leftrightarrow
    \DNFRegexAlt_i$, for each $\DNFRegexAlt_i$.
    
    By Lemma~\ref{lem:adj-swap-perm-concat-express}, there exists a DNF lens
    $\DNFLens' \OfRewritelessType \DNFOf{\SequenceOf{\Sep}}
    \DNFRegexAlt_{\sigma'(1)} \ConcatDNF \DNFOf{\SequenceOf{\Sep}} \ConcatDNF
    \ldots \ConcatDNF \DNFOf{\SequenceOf{\Sep}} \ConcatDNF
    \DNFOf{\SequenceOf{\Sep}} \DNFRegexAlt_{\sigma'(n)} \DNFOf{\SequenceOf{\Sep}} \Leftrightarrow
    \DNFRegexAlt_{\sigma_{i_j}(\sigma'(1))} \ConcatDNF \ldots \ConcatDNF
    \DNFRegexAlt_{\sigma_{i_j}(\sigma'(n))}$, where $\SemanticsOf{\DNFRegex} =
    \SetOf{(\Sep\Concat\StringAlt_{\sigma'(1)}\Concat\Sep\Concat\ldots\Concat\Sep\StringAlt_{\sigma'(n)}\Concat\Sep,
      \Sep\Concat\StringAlt_{\sigma_{i_j}(\sigma'(1))}\Concat\Sep\Concat\ldots\Concat\Sep\Concat\StringAlt_{\sigma_{i_j}(\sigma'(n))}\Concat\Sep)
      \SuchThat
      (\StringAlt_{\sigma'(i)},\StringAlt_{\sigma'(i)}) \in
      \SemanticsOf{\DNFLens_i}}$.

    So, by Lemma~\ref{lem:composition-completeness}, there exists a DNF lens
    $\DNFLens \OfRewritelessType \DNFOf{\SequenceOf{\Sep}}
    \DNFRegex_1 \ConcatDNF \DNFOf{\SequenceOf{\Sep}} \ConcatDNF
    \ldots \ConcatDNF \DNFOf{\SequenceOf{\Sep}} \ConcatDNF
    \DNFOf{\SequenceOf{\Sep}} \DNFRegex_n \DNFOf{\SequenceOf{\Sep}} \Leftrightarrow
    \DNFOf{\SequenceOf{\Sep}} \ConcatDNF \DNFRegexAlt_{\sigma(1)} \ConcatDNF \ldots \ConcatDNF
    \DNFRegexAlt_{\sigma(n)} \ConcatDNF \DNFOf{\SequenceOf{\Sep}}$, where $\SemanticsOf{\DNFRegex} =
    \SetOf{(\Sep\Concat\String_1\Concat\Sep\Concat\ldots\Concat\Sep\String_n\Concat\Sep,
      \Sep\Concat\StringAlt_{\sigma(1)}\Concat\Sep\Concat\ldots\Concat\Sep\Concat\StringAlt_{\sigma(n)}\Concat\Sep)
      \SuchThat
      (\String_i,\StringAlt_i) \in \SemanticsOf{\DNFLens_i}}$
  \end{case}
\end{proof}

\begin{lemma}[Expressibility of Concat Permutation]
  \label{lem:conat-perms}
  Let for all $i \in \RangeIncInc{1}{n}, \DNFLens_i \OfRewritelessType
  \DNFRegex_i \Leftrightarrow \DNFRegexAlt_i$.
  Let $\SequenceUnambigConcatOf{\String_0,\DNFRegex_1,\ldots,\DNFRegexAlt_n,\String_n}$.
  Let $\sigma$ be a permutation.
  Let $\SequenceUnambigConcatOf{\StringAlt_0,\DNFRegexAlt_{\sigma(n)},\ldots,\DNFRegexAlt_{\sigma(n)},\StringAlt_n}$.
  There exists a DNF lens $\DNFLens \OfRewritelessType
  \DNFOf{\SequenceOf{\String_0}} \ConcatDNF
  \DNFRegex_1 \ConcatDNF \DNFOf{\SequenceOf{\String_1}} \ConcatDNF
  \ldots \ConcatDNF \DNFOf{\SequenceOf{\String_{n-1}}} \ConcatDNF
  \DNFRegex_n \ConcatDNF \DNFOf{\SequenceOf{\String_n}} \Leftrightarrow
  \DNFOf{\SequenceOf{\String_0}} \ConcatDNF \DNFRegexAlt_{\sigma(1)} \ConcatDNF \ldots \ConcatDNF
  \DNFRegexAlt_{\sigma(n)} \ConcatDNF \DNFOf{\SequenceOf{\String_n}}$, where $\SemanticsOf{\DNFRegex} =
  \SetOf{(\String_0\Concat\String_1'\Concat\String_1\Concat\ldots\Concat\String_{n-1}\String_n'\Concat\String_n,
    \StringAlt_0\Concat\StringAlt_{\sigma(1)}'\Concat\StringAlt_1\Concat\ldots\Concat\StringAlt_{n-1}\Concat\StringAlt_{\sigma(n)}'\Concat\StringAlt_n)
    \SuchThat
  (\String_i',\StringAlt_i') \in \SemanticsOf{\DNFLens_i}}$.
\end{lemma}
\begin{proof}
  By Lemma~\ref{lem:strong-unambig-dnf-lens-types}, $\DNFRegex_i$ and
  $\DNFRegexAlt_i$ are strongly unambiguous.
  
  By Lemma~\ref{lem:strongly-unambiguous-identity-expressible}, there exists
  $\DNFLens_i' \OfRewritelessType \DNFRegex_i \Leftrightarrow \DNFRegex_i$,
  which are the identity transformations.
  
  By Lemma~\ref{lem:strongly-unambiguous-identity-expressible}, there exists
  $\DNFLens_i'' \OfRewritelessType \DNFRegexAlt_i \Leftrightarrow \DNFRegexAlt_i$,
  which are the identity transformations.

  Consider the lenses
  \[
    \inferrule*
    {
      \inferrule*
      {
      }
      {
        \SequenceLensOf{(\String_i,\Sep)} \OfRewritelessType
        \SequenceOf{\String_i} \Leftrightarrow \SequenceOf{\Sep}
      }
    }
    {
      \DNFLensOf{\SequenceOf{(\String_i,\Sep)}} \OfRewritelessType
      \DNFOf{\SequenceOf{\String_i}} \Leftrightarrow \DNFOf{\SequenceOf{\Sep}}
    }
  \]

  \[
    \inferrule*
    {
      \inferrule*
      {
      }
      {
        \SequenceLensOf{(\Sep,\StringAlt_i)} \OfRewritelessType
        \SequenceOf{\Sep} \Leftrightarrow \SequenceOf{\StringAlt_i}
      }
    }
    {
      \DNFLensOf{\SequenceOf{(\Sep,\StringAlt_i)}} \OfRewritelessType
      \DNFOf{\SequenceOf{\Sep}} \Leftrightarrow
      \DNFOf{\SequenceOf{\StringAlt_i}}
    }
  \]

  Because
  $\SequenceUnambigConcatOf{\Sep,\DNFRegex_1,\ldots,\DNFRegex_n,\Sep}$,
  through repeated application of
  Lemma~\ref{lem:typ_sem_concat},
  $\DNFLensOf{\SequenceOf{(\String_0,\Sep)}} \ConcatDNF \DNFLens_1' \ConcatDNF
  \ldots \ConcatDNF \DNFLens_n' \ConcatDNF
  \DNFLensOf{\SequenceOf{(\String_n,\Sep)}} \OfRewritelessType
  \DNFOf{\SequenceOf{\String_0}} \ConcatDNF \DNFRegex_1 \ConcatDNF \ldots
  \ConcatDNF \DNFRegex_n \ConcatDNF \DNFOf{\SequenceOf{\String_n}}
  \Leftrightarrow
  \DNFOf{\SequenceOf{\Sep}} \ConcatDNF \DNFRegex_1 \ConcatDNF \ldots
  \ConcatDNF \DNFRegex_n \ConcatDNF \DNFOf{\Sep}$, with semantics
  $\SemanticsOf{\DNFLensOf{\SequenceOf{(\String_0,\Sep)}} \ConcatDNF \DNFLens_1' \ConcatDNF
    \ldots \ConcatDNF \DNFLens_n' \ConcatDNF
    \DNFLensOf{\SequenceOf{(\String_n,\Sep)}}} =
  \SetOf{(\String_0\Concat\String_1'\Concat\ldots\Concat\String_n'\Concat\String_n,
    \Sep\Concat\String_1'\Concat\ldots\Concat\String_n'\Concat\Sep) \SuchThat
  \String_i\in\LanguageOf{\DNFRegex_i}}$

  Because
  $\SequenceUnambigConcatOf{\Sep,\DNFRegexAlt_{\sigma(1)},\ldots,\DNFRegexAlt_{\sigma(n)},\Sep}$,
  through repeated application of
  Lemma~\ref{lem:typ_sem_concat},
  $\DNFLensOf{\SequenceOf{(\Sep,\StringAlt_0)}} \ConcatDNF \DNFLens_{\sigma(1)}'' \ConcatDNF
  \ldots \ConcatDNF \DNFLens_{\sigma(n)}'' \ConcatDNF
  \DNFLensOf{\SequenceOf{(\Sep,\StringAlt_n)}} \OfRewritelessType
  \DNFOf{\SequenceOf{\Sep}} \ConcatDNF \DNFRegexAlt_{\sigma(1)} \ConcatDNF \ldots
  \ConcatDNF \DNFRegexAlt_{\sigma(n)} \ConcatDNF \DNFOf{\Sep}
  \Leftrightarrow
  \DNFOf{\SequenceOf{\StringAlt_0}} \ConcatDNF \DNFRegexAlt_{\sigma(1)} \ConcatDNF \ldots
  \ConcatDNF \DNFRegexAlt_{\sigma(n)} \ConcatDNF \DNFOf{\SequenceOf{\StringAlt_n}}$, with semantics
  $\SemanticsOf{\DNFLensOf{\SequenceOf{(\Sep,\StringAlt_0)}} \ConcatDNF \DNFLens_{\sigma(1)}'' \ConcatDNF
    \ldots \ConcatDNF \DNFLens_{\sigma(n)}'' \ConcatDNF
    \DNFLensOf{\SequenceOf{(\Sep,\StringAlt_n)}}} =
  \SetOf{(\Sep\Concat\StringAlt_{\sigma(1)}'\Concat\ldots\Concat\StringAlt_{\sigma(n)}'\Concat\Sep,
    \StringAlt_0\Concat\StringAlt_{\sigma(1)}'\Concat\ldots\Concat\StringAlt_{\sigma(n)}'\Concat\StringAlt_n) \SuchThat
    \String_i\in\LanguageOf{\DNFRegexAlt_i}}$
  
  By Lemma~\ref{lem:perm-sep-concat},
  there exists a lens $\DNFLens \OfRewritelessType \DNFOf{\SequenceOf{\Sep}} \ConcatDNF \DNFRegex_1 \ConcatDNF \ldots
  \ConcatDNF \DNFRegex_n \ConcatDNF \DNFOf{\Sep} \Leftrightarrow
  \DNFOf{\SequenceOf{\Sep}} \ConcatDNF \DNFRegexAlt_{\sigma(1)} \ConcatDNF \ldots
  \ConcatDNF \DNFRegexAlt_{\sigma(n)} \ConcatDNF \DNFOf{\Sep}$, with semantics
  $\SemanticsOf{\DNFLens} = \SetOf{(\Sep\Concat\String_1'\Concat\Sep\Concat\ldots\Concat\Sep\String_n'\Concat\Sep,
    \Sep\Concat\StringAlt_{\sigma(1)}'\Concat\Sep\Concat\ldots\Concat\Sep\Concat\StringAlt_{\sigma(n)}'\Concat\Sep)
    \SuchThat
    (\String_i',\StringAlt_i') \in \SemanticsOf{\DNFLens_i}}$.

  By Lemma~\ref{lem:composition-completeness},
  there exists a lens $\DNFLens' \OfRewritelessType \DNFOf{\SequenceOf{\String_0}} \ConcatDNF \DNFRegex_1 \ConcatDNF \ldots
  \ConcatDNF \DNFRegex_n \ConcatDNF \DNFOf{\SequenceOf{\String_n}}
  \Leftrightarrow
  \DNFOf{\SequenceOf{\StringAlt_0}} \ConcatDNF \DNFRegexAlt_{\sigma(1)} \ConcatDNF \ldots
  \ConcatDNF \DNFRegexAlt_{\sigma(n)} \ConcatDNF
  \DNFOf{\SequenceOf{\StringAlt_n}}$.
  The semantics, through running the strings through, is $\SetOf{(\String_0\Concat\String_1'\Concat\String_1\Concat\ldots\Concat\String_{n-1}\String_n'\Concat\String_n,
    \StringAlt_0\Concat\StringAlt_{\sigma(1)}'\Concat\StringAlt_1\Concat\ldots\Concat\StringAlt_{n-1}\Concat\StringAlt_{\sigma(n)}'\Concat\StringAlt_n)
    \SuchThat
  (\String_i',\StringAlt_i') \in \SemanticsOf{\DNFLens_i}}$.
\end{proof}

\begin{lemma}[Identity Transformation on Adjacent Swapping Or]
  \label{lem:adj-swap-or}
  Let $\DNFRegex_1,\ldots,\DNFRegex_n$ be strongly unambiguous DNF regular
  expressions, where $j \neq k \BooleanImplies
  \LanguageOf{\DNFRegex_i} = \LanguageOf{\DNFRegex_j}$.

  Let $\sigma_i$ be an adjacent swapping permutation.

  There exists a lens $\DNFLens \OfRewritelessType \DNFRegex_1 \OrDNF \ldots
  \OrDNF \DNFRegex_n \Leftrightarrow \DNFRegex_{\sigma_i(1)} \OrDNF \ldots
  \OrDNF \DNFRegex_{\sigma_i(n)}$, such that $\SemanticsOf{\DNFLens} =
  \SetOf{(\String,\String) \SuchThat \String \in \LanguageOf{\DNFRegex_1 \OrDNF
      \ldots \OrDNF \DNFRegex_n}}$.
\end{lemma}
\begin{proof}
  As each DNF regular expression is strongly unambiguous, there exists a DNF
  lens $\DNFLens_j \OfRewritelessType \DNFRegex_j \Leftrightarrow \DNFRegex_j$
  such that $\SemanticsOf{\DNFLens_j} = \SetOf{(\String,\String) \SuchThat
    \String \in \LanguageOf{\DNFRegex_j}}$.
  By assumption, $\UnambigOrOf{\DNFRegex_i}{\DNFRegex_{i+1}}$, so by
  Lemma~\ref{lem:or-dnf-commutativity},
  there exists $\DNFLens' \OfRewritelessType \DNFRegex_i \OrDNF \DNFRegexAlt_i
  \Leftrightarrow \DNFRegexAlt_i \OrDNF \DNFRegex_i$.  By repeated application of
  Lemma~\ref{lem:typ_sem_or},
  $\DNFLens_0 \OrDNF \ldots \OrDNF \DNFLens_{i-1} \OrDNF \DNFLens' \OrDNF
  \DNFLens_{i+2} \OrDNF \ldots \OrDNF \DNFLens_n \OfRewritelessType
  \DNFRegex_1 \OrDNF \ldots \OrDNF \DNFRegex_n
  \Leftrightarrow
  \DNFRegex_1 \OrDNF \ldots \OrDNF \DNFRegex_{i+1} \OrDNF \DNFRegex_i \OrDNF
  \ldots \OrDNF \DNFRegex_n$.

  Because each of the lenses included in this is the identity lens, the overall
  lens is the identity lens.
\end{proof}

\begin{lemma}[Identity Transformation on Or Permutations]
  \label{lem:perm-or}
  Let $\DNFRegex_1,\ldots,\DNFRegex_n$ be strongly unambiguous DNF regular
  expressions, where $j \neq k \BooleanImplies
  \LanguageOf{\DNFRegex_i} = \LanguageOf{\DNFRegex_j}$.

  Let $\sigma$ be a permutation.

  There exists a lens $\DNFLens \OfRewritelessType \DNFRegex_1 \OrDNF \ldots
  \OrDNF \DNFRegex_n \Leftrightarrow \DNFRegex_{\sigma(1)} \OrDNF \ldots
  \OrDNF \DNFRegex_{\sigma(n)}$, such that $\SemanticsOf{\DNFLens} =
  \SetOf{(\String,\String) \SuchThat \String \in \LanguageOf{\DNFRegex_1 \OrDNF
      \ldots \OrDNF \DNFRegex_n}}$.
\end{lemma}
\begin{proof}
  From algebra, there exists a decomposition of $\sigma$ into adjacency
  switching permutations $\sigma = \sigma_{i_j} \Compose \ldots \Compose
  \sigma_{i_1}$.

  We prove this by induction on $n$!

  \begin{case}[$j=0$]
    $\sigma = \Identity$

    As each $\DNFRegex_i$ there exists an identity transformation
    $\DNFLens_i \OfRewritelessType \DNFRegex_i \Leftrightarrow \DNFRegex_i$.

    By repeated application of
    Lemma~\ref{lem:typ_sem_or},
    $\DNFLens_1 \OrDNFLens \ldots \OrDNFLens \DNFLens_n \OfRewritelessType
    \DNFRegex_1 \OrDNF \ldots \OrDNF \DNFRegex_n \Leftrightarrow
    \DNFRegex_1 \OrDNF \ldots \OrDNF \DNFRegex_n$ with semantics of the
    identity, as each of the lenses that built it up have identity semantics.
  \end{case}

  \begin{case}[$j>0$]
    $\sigma = \sigma_{i_j} \Compose \ldots \Compose \sigma_{i_1}$
    Define $\sigma' = \sigma_{i_{j-1}} \Compose \ldots \Compose \sigma_{i_1}$
    By IH, there exists a DNF lens
    $\DNFLens \OfRewritelessType
    \DNFRegex_1 \OrDNF \ldots \OrDNF \DNFRegex_n
    \Leftrightarrow
    \DNFRegex_{\sigma'(1)} \OrDNF \ldots \OrDNF \DNFRegex_{\sigma'(n)}$.

    By Lemma~\ref{lem:adj-swap-or}, there exists a lens
    $\DNFLens' \OfRewritelessType
    \DNFRegex_{\sigma'(1)} \OrDNF \ldots \OrDNF \DNFRegex_{\sigma'(n)}
    \Leftrightarrow
    \DNFRegex_{(\sigma_{i_j} \Compose \sigma')(1)} \OrDNF \ldots \OrDNF
    \DNFRegex_{(\sigma_{i_j} \Compose \sigma')(n)}$,
    so 
    $\DNFLens' \OfRewritelessType
    \DNFRegex_{\sigma'(1)} \OrDNF \ldots \OrDNF \DNFRegex_{\sigma'(n)}
    \Leftrightarrow
    \DNFRegex_{\sigma(1)} \OrDNF \ldots \OrDNF \DNFRegex_{\sigma(n)}$, where
    $\DNFLens$ has the identity semantics.

    By Lemma~\ref{lem:composition-completeness},
    there exists $\DNFLens'' \OfRewritelessType
    \DNFRegex_1 \OrDNF \ldots \OrDNF \DNFRegex_n
    \Leftrightarrow
    \DNFRegex_{\sigma(1)} \OrDNF \ldots \OrDNF \DNFRegex_{\sigma(n)}$.
    As each of its component transformations has identity semantics identity,
    it too has identity semantics.
  \end{case}
\end{proof}

\begin{lemma}[Or Permutating Lenses]
  \label{lem:perm-lens-or}
  Let $n$ a natural number, and for all $i\in\RangeIncInc{1}{n}$,
  $\DNFLens_i \OfRewritelessType \DNFRegex_i \Leftrightarrow \DNFRegexAlt_i$.
  Let $i \neq j \BooleanImplies \DNFRegex_i \Intersect \DNFRegex_j = \emptyset$
  and
  $i \neq j \BooleanImplies \DNFRegexAlt_i \Intersect \DNFRegexAlt_j =
  \emptyset$.
  Let $\sigma$ be a permutation.
  There exists a lens $\DNFLens \OfRewritelessType \DNFRegex_1 \OrDNF \ldots
  \OrDNF \DNFRegex_n \Leftrightarrow \DNFRegexAlt_{\sigma(1)} \OrDNF \ldots
  \OrDNF \DNFRegexAlt_{\sigma(n)}$ such that $\SemanticsOf{\DNFLens} =
  \SetOf{(\String,\StringAlt) \SuchThat \exists i \That (\String,\StringAlt) \in
    \SemanticsOf{\DNFLens_i}}$
\end{lemma}
\begin{proof}
  By Lemma~\ref{lem:typ_sem_or}, there exists $\DNFLens_1 \OrDNF \ldots \OrDNF
  \DNFLens_n \OfRewritelessType \DNFRegex_1 \OrDNF \ldots \OrDNF \DNFRegex_n
  \Leftrightarrow \DNFRegexAlt_1 \OrDNF \ldots \OrDNF \DNFRegexAlt_n$ with
  $\SemanticsOf{\DNFLens_1 \OrDNF \ldots \OrDNF
    \DNFLens_n} = \SetOf{(\String,\StringAlt) \SuchThat \exists i \That (\String,\StringAlt) \in
    \SemanticsOf{\DNFLens_i}}$.
  By Lemma~\ref{lem:perm-or}, there exists a lens $\DNFLens' \OfRewritelessType
  \DNFRegexAlt_1 \OrDNF \ldots \OrDNF \DNFRegexAlt_n \Leftrightarrow
  \DNFRegexAlt_{\sigma(1)} \OrDNF \ldots \OrDNF \DNFRegexAlt_{\sigma(n)}$, with
  $\SemanticsOf{\DNFLens'} = \SetOf{(\String,\String) \SuchThat \String \in
    \LanguageOf{\DNFRegexAlt_1 \OrDNF \ldots \OrDNF \DNFRegexAlt_n}}$.
  By Lemma~\ref{lem:composition-completeness}, there exists $\DNFLens''
  \OfRewritelessType \DNFRegex_1 \OrDNF \ldots \OrDNF \DNFRegex_n
  \Leftrightarrow \DNFRegexAlt_{\sigma(1)} \OrDNF \ldots \OrDNF
  \DNFRegexAlt_{\sigma(n)}$ with semantics $\SetOf{(\String,\StringAlt)
    \SuchThat \exists \String' \That (\String,\String') \in
    \SemanticsOf{\DNFLens_1 \OrDNF \ldots \OrDNF
      \DNFLens_n} \BooleanAnd (\String',\StringAlt) \in \SemanticsOf{\DNFLens'}}$.
  As $\DNFLens'$ is merely the identity, this has the desired semantics.
\end{proof}

\begin{lemma}[Propagation of Unambiguity to Subcomponents $\OrDNF$]
  \label{lem:or-prop-subcomponent}
  If $\DNFRegex \OrDNF \DNFRegexAlt$ is strongly unambiguous, then
  $\DNFRegex$ is strongly unambiguous, $\DNFRegexAlt$ is strongly unambiguous,
  and $\UnambigOrOf{\DNFRegex}{\DNFRegexAlt}$.
\end{lemma}
\begin{proof}
  Let $\DNFRegex = \DNFOf{\Sequence_1 \DNFSep \ldots \DNFSep \Sequence_n}$.

  Let $\DNFRegexAlt = \DNFOf{\SequenceAlt_1 \DNFSep \ldots \DNFSep
    \SequenceAlt_m}$.

  $\DNFRegex \OrDNF \DNFRegexAlt =
  \DNFOf{\Sequence_1 \DNFSep \ldots \DNFSep \Sequence_n \DNFSep \SequenceAlt_1 \DNFSep \ldots \DNFSep \SequenceAlt_m}$.

  This means that, as it is strongly unambiguous, all of the sequences are
  pairwise disjoint, and each sequence is strongly unambiguous.
  By Lemma~\ref{lem:unambig-union-equiv}, this means that
  all of the sequences in $\DNFRegex$ are pairwise disjoint, all the sequences
  in $\DNFRegexAlt$ are pairwise disjoint, and
  $\UnambigOrOf{\DNFRegex}{\DNFRegexAlt}$.
\end{proof}

\begin{lemma}[Reordering of $\OrDNF$ Right]
  \label{lem:or-dnf-reordering-right}
  If there exists a DNF lens $\DNFLens \OfRewritelessType \DNFRegex_1 \OrDNF \ldots
  \OrDNF \DNFRegex_n \Leftrightarrow
  \DNFRegexAlt_1 \OrDNF \ldots \OrDNF \DNFRegexAlt_n$, then for all permutations
  $\sigma \in \PermutationSetOf{n}$, there exists a DNF lens $\DNFLens'
  \OfRewritelessType \DNFRegex_1 \OrDNF \ldots \OrDNF \DNFRegex_n
  \Leftrightarrow
  \DNFRegexAlt_{\sigma(1)} \OrDNF \ldots \OrDNF \DNFRegexAlt_{\sigma(n)}$ where
  $\SemanticsOf{\DNFLens'} = \SemanticsOf{\DNFLens}$.
\end{lemma}
\begin{proof}
  From Lemma~\ref{lem:strong-unambig-dnf-lens-types},
  $\DNFRegexAlt_1 \OrDNF \ldots \OrDNF \DNFRegexAlt_n$ is strongly unambiguous.
  By repeated application of Lemma~\ref{lem:or-prop-subcomponent},
  $i \neq j \BooleanImplies \DNFRegexAlt_i \Intersect \DNFRegexAlt_j =
  \SetOf{}$, and each $\DNFRegexAlt_i$ is strongly unambiguous.

  This means Lemma~\ref{lem:perm-or} applies, so there exists a DNF lens
  $\DNFLens' \OfRewritelessType \DNFRegexAlt_1 \OrDNF \ldots \OrDNF
  \DNFRegexAlt_n \Leftrightarrow
  \DNFRegexAlt_{\sigma(1)} \OrDNF \ldots \OrDNF \DNFRegexAlt_{\sigma(n)}$ such
  that $\SemanticsOf{\DNFLens'}$ is the identity semantics.

  So, by composing $\DNFLens$ with $\DNFLens'$ from
  Lemma~\ref{lem:composition-completeness}, we get\\
  $\DNFLens'' \OfRewritelessType \DNFRegex_1 \OrDNF \ldots \OrDNF \DNFRegex_n
  \Leftrightarrow \DNFRegexAlt_{\sigma(1} \OrDNF \ldots \OrDNF
  \DNFRegexAlt_{\sigma(n)}$, which has $\SemanticsOf{\DNFLens''} =
  \SemanticsOf{\DNFLens}$ as $\DNFLens'$ has identity semantics.
\end{proof}

\begin{lemma}
  \label{lem:id-expressible-on-distribute-left}
  If $\DNFRegex_1 \ConcatDNF (\DNFRegex_2 \OrDNF \DNFRegex_3)$ is strongly
  unambiguous, then there exists a lens
  $\DNFLens \OfRewritelessType
  \DNFRegex_1 \ConcatDNF (\DNFRegex_2 \OrDNF \DNFRegex_3)
  \Leftrightarrow
  (\DNFRegex_1 \ConcatDNF \DNFRegex_2) \OrDNF
  (\DNFRegex_1 \ConcatDNF \DNFRegex_3)$.
\end{lemma}
\begin{proof}
  If $\LanguageOf{\DNFRegex_1 \ConcatDNF (\DNFRegex_2 \OrDNF \DNFRegex_3)} = \SetOf{}$,
  then this is trivial, as
  $\DNFRegex_1 \ConcatDNF (\DNFRegex_2 \OrDNF \DNFRegex_3) =
  \SetOf{}$.
  
  Assume the language is nonempty.
  Let $\DNFRegex_1 = \DNFOf{\Sequence_{1,1} \DNFSep \ldots \DNFSep
    \Sequence_{1,n_1}}$.

  Let $\DNFRegex_2 = \DNFOf{\Sequence_{2,1} \DNFSep \ldots \DNFSep
    \Sequence_{2,n_2}}$.

  Let $\DNFRegex_3 = \DNFOf{\Sequence_{3,1} \DNFSep \ldots \DNFSep \Sequence_{3,n_3}}$.

  $\DNFRegex_2 \OrDNF \DNFRegex_3 =
  \DNFOf{\Sequence_{2,1} \DNFSep \ldots \DNFSep \Sequence_{2,n_2} \DNFSep 
    \Sequence_{3,1} \DNFSep \ldots \DNFSep \Sequence_{3,n_3}}$.

  $\DNFRegex_1 \ConcatDNF (\DNFRegex_2 \OrDNF \DNFRegex_3) =
  \DNFOf{
    \Sequence_{1,1} \ConcatSequence \Sequence_{2,1} \DNFSep  \ldots \DNFSep 
    \Sequence_{1,1}\ConcatSequence\Sequence_{2,n_2} \DNFSep 
    \Sequence_{1,1}\ConcatSequence\Sequence_{3,1} \DNFSep  \ldots \DNFSep 
    \Sequence_{1,1}\ConcatSequence\Sequence_{3,n_3} \DNFSep  \ldots \DNFSep 
    \Sequence_{1,n_1} \ConcatSequence \Sequence_{2,1} \DNFSep  \ldots \DNFSep 
    \Sequence_{1,n_1}\ConcatSequence\Sequence_{2,n_2} \DNFSep 
    \Sequence_{1,n_1}\ConcatSequence\Sequence_{3,1} \DNFSep  \ldots \DNFSep 
    \Sequence_{1,n_1}\ConcatSequence\Sequence_{3,n_3}}$.

  As this is strongly unambiguous, $\Sequence_{1,i} \ConcatSequence
  \Sequence_{j,k}$ is strongly unambiguous for all $i,j,k$.
  Furthermore, by strong unambiguity,
  if $(i_1,j_1,k_1) \neq (i_2,j_2,k_2)$, then
  $\Sequence_{1,i_1} \ConcatSequence \Sequence_{j_1,k_1} \Intersect
  \Sequence_{1,i_2} \ConcatSequence \Sequence_{j_1,k_1}$
  
  $\DNFRegex_1 \ConcatDNF \DNFRegex_2 =
  \DNFOf{\Sequence_{1,1}\ConcatSequence\Sequence_{2,1} \DNFSep \ldots
    \Sequence_{1,1}\ConcatSequence\Sequence_{2,n_2} \DNFSep \ldots \DNFSep 
    \Sequence_{1,n_1}\ConcatSequence\Sequence_{2,1} \DNFSep \ldots \DNFSep 
    \Sequence_{1,n_1}\ConcatSequence\Sequence_{2,n_2}}$.

  $\DNFRegex_1 \ConcatDNF \DNFRegex_3 =
  \DNFOf{\Sequence_{1,1}\ConcatSequence\Sequence_{3,1} \DNFSep \ldots
    \Sequence_{1,1}\ConcatSequence\Sequence_{3,n_3} \DNFSep \\
    \Sequence_{1,n_1}\ConcatSequence\Sequence_{3,1} \DNFSep \ldots \DNFSep
    \Sequence_{1,n_1}\ConcatSequence\Sequence_{3,n_3}}$.
  
  $(\DNFRegex_1 \ConcatDNF \DNFRegex_2) \OrDNF
  (\DNFRegex_1 \ConcatDNF \DNFRegex_3) =\\
  \DNFOf{
    \Sequence_{1,1}\ConcatSequence\Sequence_{2,1} \DNFSep \ldots
    \Sequence_{1,1}\ConcatSequence\Sequence_{2,n_2} \DNFSep \ldots \DNFSep 
    \Sequence_{1,n_1}\ConcatSequence\Sequence_{2,1} \DNFSep \ldots \DNFSep 
    \Sequence_{1,n_1}\ConcatSequence\Sequence_{2,n_2} \DNFSep \\
    \Sequence_{1,1}\ConcatSequence\Sequence_{3,1} \DNFSep \ldots
    \Sequence_{1,1}\ConcatSequence\Sequence_{3,n_3} \DNFSep \ldots
    \Sequence_{1,n_1}\ConcatSequence\Sequence_{3,1} \DNFSep \ldots
    \Sequence_{1,n_1}\ConcatSequence\Sequence_{3,n_3}}$

  From before, if $(i_1,j_1,k_1) \neq (i_2,j_2,k_2)$, then
  $\Sequence_{1,i_1} \ConcatSequence \Sequence_{j_1,k_1} \Intersect
  \Sequence_{1,i_2} \ConcatSequence \Sequence_{j_1,k_1} = \SetOf{}$.

  As $\Sequence_{1,i} \ConcatSequence \Sequence_{j,k}$ is strongly unambiguous,
  there exists
  $\SequenceLens_{i,j,k} \OfRewritelessType
  \Sequence_{1,i} \ConcatSequence \Sequence_{j,k} \Leftrightarrow
  \Sequence_{1,i} \ConcatSequence \Sequence_{j,k}$,
  from Lemma~\ref{lem:strongly-unambiguous-identity-expressible}.

  There exists a unique permutation $\sigma$ that sends
  $\Sequence_{1,i} \ConcatSequence \Sequence_{j,k}$
  in $\DNFRegex_1 \ConcatDNF (\DNFRegex_2 \OrDNF \DNFRegex_3)$
  to $\Sequence_{1,i} \ConcatSequence \Sequence_{j,k}$ in
  $(\DNFRegex_1 \ConcatDNF \DNFRegex_2) \OrDNF
  (\DNFRegex_1 \ConcatDNF \DNFRegex_3)$.  As a permutation is merely a bijective
  between a finite number of elements.
  Note, this permutation is not necessarily the identity permutation.
  In particular, the sequence at position $n_1+1$, if such a sequence exists, in
  $\DNFRegex_1 \ConcatDNF (\DNFRegex_2 \OrDNF \DNFRegex_3)$ is
  $\Sequence_{1,1} \ConcatSequence \Sequence_{3,1}$.
  However, the sequence at position $n_1+1$, if such a sequence exists, in
  $(\DNFRegex_1 \ConcatDNF \DNFRegex_2) \OrDNF
  (\DNFRegex_1 \ConcatDNF \DNFRegex_3)$, is
  $\Sequence_{1,2} \ConcatSequence \Sequence_{2,1}$.

  Consider the derivation
  \[
    \inferrule*
    {
      \SequenceLens \OfRewritelessType
      \Sequence_{1,i} \ConcatSequence \Sequence_{j,k}
      \Leftrightarrow
      \Sequence_{1,i} \ConcatSequence \Sequence_{j,k}\\
      (i_1,j_1,k_1) \neq (i_2,j_2,k_2) \BooleanImplies
      (\Sequence_{1,i_1} \ConcatSequence \Sequence_{j_1,k_1}) \Intersect
      (\Sequence_{1,i_2} \ConcatSequence \Sequence_{j_2,k_2}) = \SetOf{}\\
      (i_1,j_1,k_1) \neq (i_2,j_2,k_2) \BooleanImplies
      (\Sequence_{1,i_1} \ConcatSequence \Sequence_{j_1,k_1}) \Intersect
      (\Sequence_{1,i_2} \ConcatSequence \Sequence_{j_2,k_2}) = \SetOf{}\\
      \sigma \in \PermutationSetOf{n_1\Cross n_2\Cross n_3}
    }
    {
      (\DNFLensLeft \SequenceLens_{1,2,1} \DNFLSep \ldots \DNFLSep \SequenceLens_{1,2,n_2} \DNFLSep 
        \SequenceLens_{1,3,1} \DNFLSep \ldots \DNFLSep \SequenceLens_{1,3,n_3} \DNFLSep \ldots \DNFLSep\\
        \SequenceLens_{n_1,2,1} \DNFLSep \ldots \DNFLSep \SequenceLens_{n_1,2,n_2} \DNFLSep 
        \SequenceLens_{n_1,3,1} \DNFLSep \ldots \DNFLSep
        \SequenceLens_{n_1,3,n_3}\DNFLensRight,\sigma)
      \OfRewritelessType\\
      \DNFRegex_1 \ConcatDNF (\DNFRegex_2 \OrDNF \DNFRegex_3) \Leftrightarrow
      (\DNFRegex_1 \ConcatDNF \DNFRegex_2) \OrDNF (\DNFRegex_1 \ConcatDNF
      \DNFRegex_3) 
    }
  \]

  Furthermore, as each $\SequenceLens$ has the identity transformation, then as
  $\sigma$ has no impact on semantics, the total DNF lens has the identity
  transformation.
\end{proof}

\begin{lemma}
  \label{lem:id-expressible-on-factor-left}
  If $(\DNFRegex_1 \ConcatDNF \DNFRegex_2) \OrDNF
  (\DNFRegex_1 \ConcatDNF \DNFRegex_3)$ is strongly
  unambiguous, then there exists a lens
  $\DNFLens \OfRewritelessType
  (\DNFRegex_1 \ConcatDNF \DNFRegex_2) \OrDNF
  (\DNFRegex_1 \ConcatDNF \DNFRegex_3)
  \Leftrightarrow
  \DNFRegex_1 \ConcatDNF (\DNFRegex_2 \OrDNF \DNFRegex_3)$.
\end{lemma}
\begin{proof}
  By Lemma~\ref{lem:distribute-strongly-unambiguous-iff-factor},
  $\DNFRegex_1 \ConcatDNF (\DNFRegex_2 \OrDNF \DNFRegex_3)$ is strongly
  unambiguous.
  So by Lemma~\ref{lem:id-expressible-on-distribute-left},
  there exists an identity lens
  $\DNFLens \OfRewritelessType
  (\DNFRegex_1 \ConcatDNF \DNFRegex_2) \OrDNF
  (\DNFRegex_1 \ConcatDNF \DNFRegex_3)
  \Leftrightarrow
  \DNFRegex_1 \ConcatDNF (\DNFRegex_2 \OrDNF \DNFRegex_3)$.
  As rewriteless DNF lenses are closed under inversion, there exists a lens
  $\InverseOf{\DNFLens} \OfRewritelessType
  \DNFRegex_1 \ConcatDNF (\DNFRegex_2 \OrDNF \DNFRegex_3)
  \Leftrightarrow
  (\DNFRegex_1 \ConcatDNF \DNFRegex_2) \OrDNF
  (\DNFRegex_1 \ConcatDNF \DNFRegex_3)$.
\end{proof}

\begin{lemma}
  \label{lem:id-expressible-on-distribute-right}
  If $(\DNFRegex_1 \OrDNF \DNFRegex_2) \ConcatDNF \DNFRegex_3$ is strongly
  unambiguous, then there exists a lens
  $\DNFLens \OfRewritelessType
  (\DNFRegex_1 \OrDNF \DNFRegex_2) \ConcatDNF \DNFRegex_3
  \Leftrightarrow
  (\DNFRegex_1 \ConcatDNF \DNFRegex_3) \OrDNF
  (\DNFRegex_2 \ConcatDNF \DNFRegex_3)$.
\end{lemma}
\begin{proof}
  By Lemma~\ref{lem:dnf-distribute-right}, $(\DNFRegex_1 \OrDNF \DNFRegex_2)
  \ConcatDNF \DNFRegex_3 = (\DNFRegex_1 \ConcatDNF \DNFRegex_3) \OrDNF
  (\DNFRegex_2 \ConcatDNF \DNFRegex_3)$, so by
  Lemma~\ref{lem:strongly-unambiguous-identity-expressible}, there is an
  identity lens between them.
\end{proof}

\begin{lemma}
  \label{lem:id-expressible-on-factor-right}
  If $(\DNFRegex_1 \OrDNF \DNFRegex_2) \ConcatDNF \DNFRegex_3$ is strongly
  unambiguous, then there exists a lens
  $\DNFLens \OfRewritelessType
  (\DNFRegex_1 \ConcatDNF \DNFRegex_3) \OrDNF
  (\DNFRegex_2 \ConcatDNF \DNFRegex_3)
  \Leftrightarrow
  (\DNFRegex_1 \OrDNF \DNFRegex_2) \ConcatDNF \DNFRegex_3$.
\end{lemma}
\begin{proof}
  By Lemma~\ref{lem:dnf-distribute-right}, $(\DNFRegex_1 \OrDNF \DNFRegex_2)
  \ConcatDNF \DNFRegex_3 = (\DNFRegex_1 \ConcatDNF \DNFRegex_3) \OrDNF
  (\DNFRegex_2 \ConcatDNF \DNFRegex_3)$, so by
  Lemma~\ref{lem:strongly-unambiguous-identity-expressible}, there is an
  identity lens between them.
\end{proof}

\subsection{Rewrite Property Maintenance}
Here the proof of bisimilarity and confluence on parallel rewrites with respect
to the property of having a lens's semantics is presented.
First a proof must be presented on Parallel Rewrites' ability to be built up
from smaller parts through concatenation.  Because of the lack of a
distributivity rule, this is only maintained up to an identity lens, we cannot
merely concatenate the two rewritten parts.  With this, bisimilarity is proven,
as is confluence.

\label{rewrite-maintenence}
\begin{lemma}[$\ParallelRewrite$ Maintained Under $\ConcatDNF$ up to
  $\IdentityLens$]
  \label{lem:parallel-rewrite-concatenation-to-identity}
  Let $\DNFRegex$ be strongly unambiguous.  Let $\DNFRegexAlt$ be strongly
  unambiguous.
  Let $\UnambigConcatOf{\LanguageOf{\DNFRegex}}{\LanguageOf{\DNFRegexAlt}}$.
  If $\DNFRegex \StarOf{\ParallelRewrite} \DNFRegex'$,
  $\DNFRegexAlt \StarOf{\ParallelRewrite} \DNFRegexAlt'$, and
  $\DNFRegex \ConcatDNF \DNFRegexAlt \StarOf{\ParallelRewrite} \DNFRegex''$
  such that there exists a rewriteless DNF lens
  $\DNFLens \OfRewritelessType
  \DNFRegex' \ConcatDNF \DNFRegexAlt' \Leftrightarrow \DNFRegex''$, and
  $\SemanticsOf{\DNFLens} =
  \SetOf{(\String,\String) \SuchThat \String \in
    \LanguageOf{\DNFRegex \ConcatDNF \DNFRegexAlt}}$.
\end{lemma}
\begin{proof}
  Because $\UnambigConcatOf{\LanguageOf{\DNFRegex}}{\LanguageOf{\DNFRegexAlt}}$,
  $\DNFRegex \ConcatDNF \DNFRegexAlt$ is strongly unambiguous.

  By induction on the derivation of $\StarOf{\ParallelRewrite}$
  \[
    \inferrule*
    {
      \DNFRegex = \DNFOf{\Sequence_1 \DNFSep \ldots \DNFSep \Sequence_n}\\
      \forall i. \Sequence_i =
      \SequenceOf{\String_{i,0} \SeqSep \Atom_{i,1} \SeqSep \ldots \SeqSep \Atom_{i,n_i} \SeqSep \String_{i,n_i}}\\
      \forall i,j. \Atom_{i,j} \ParallelRewriteAtom \DNFRegex_{i,j}\\
      \forall i. \DNFRegex_i = \DNFOf{\SequenceOf{\String_{i,0}}} \ConcatDNF \DNFRegex_{i,1}
      \ConcatDNF \ldots \ConcatDNF \DNFRegex_{i,n_i} \ConcatDNF
      \DNFOf{\SequenceOf{\String_{i,n_i}}}
    }
    {
      \DNFRegex \ParallelRewrite \DNFRegex_1 \OrDNF \ldots \OrDNF \DNFRegex_n
    }
  \]

  \[
    \inferrule*
    {
      \DNFRegexAlt = \DNFOf{\SequenceAlt_1 \DNFSep \ldots \DNFSep \Sequence_m}\\
      \forall i. \SequenceAlt_i =
      \SequenceOf{\StringAlt_{i,0} \SeqSep \AtomAlt_{i,1} \SeqSep \ldots \SeqSep \AtomAlt_{i,m_i} \SeqSep \StringAlt_{i,m_i}}\\
      \forall i,j. \AtomAlt_{i,j} \ParallelRewriteAtom \DNFRegexAlt_{i,j}\\
      \forall i. \DNFRegexAlt_i = \DNFOf{\SequenceOf{\StringAlt_{i,0}}} \ConcatDNF \DNFRegexAlt_{i,1}
      \ConcatDNF \ldots \ConcatDNF \DNFRegexAlt_{i,n_i} \ConcatDNF
      \DNFOf{\SequenceOf{\StringAlt_{i,n_i}}}
    }
    {
      \DNFRegexAlt \ParallelRewrite \DNFRegexAlt_1 \OrDNF \ldots \OrDNF \DNFRegexAlt_n
    }
  \]

  Define $\Atom_{i,j,k}'' =
  \begin{cases*}
    \Atom_{i,k} & if $k \leq n_i$\\
    \AtomAlt_{j,k-n_i} & if $i > n_i$
  \end{cases*}$

  Define $\String_{i,j,k}'' =
  \begin{cases*}
    \String_{i,k} & if $k < n_i$\\
    \String_{i,n_i} \Concat \StringAlt_{j,0} & if $k = n_i$\\
    \StringAlt_{j,k-n_i} & if $i > n$
  \end{cases*}$

  Define $n_{i,j} = n_i + m_j$.

  Define $\Sequence_{i,j}'' =
  \SequenceOf{\String_{i,j,0}'' \SeqSep \Atom_{i,j,1}'' \SeqSep \ldots \SeqSep 
    \Atom_{i,j,n_{i,j}}'' \SeqSep \String_{i,j,n_{i,j}}''}$.
  By inspection,
  $\Sequence_{i,j}'' = \Sequence_i \ConcatSequence \Sequence_j$.
  
  Define $\DNFRegex'' = \DNFOf{\Sequence_{1,1}'' \DNFSep \ldots \DNFSep \Sequence_{n,m}''}$.
  By inspection, $\DNFRegex'' = \DNFRegex \ConcatDNF \DNFRegexAlt$.

  Define $\DNFRegex_{i,j,k}'' =
  \begin{cases*}
    \DNFRegex_{i,k} & if $k \leq n_i$\\
    \DNFRegexAlt_{j,k-n_i} & if $i > n_i$
  \end{cases*}$.
  By inspection $\Atom_{i,j,k}'' \ParallelRewrite \DNFRegex_{i,j,k}''$.

  Define $\DNFRegex_{i,j}''$ as
  $\DNFOf{\SequenceOf{\String_{i,j,0}''}} \ConcatDNF \DNFRegex_{i,j,1}''
  \ConcatDNF \ldots \ConcatDNF \DNFRegex_{i,j,n_{i,j}} \ConcatDNF
  \DNFOf{\SequenceOf{\String_{i,j,n_{i,j}}}}$.
  By inspection, $\DNFRegex_{i,j}'' = \DNFRegex_i \ConcatDNF \DNFRegex_j$
  This means that
  $\DNFRegex_{1,1}'' \OrDNF \ldots \OrDNF \DNFRegex_{1,m}''
  \OrDNF \ldots \OrDNF
  \DNFRegex_{n,1}'' \OrDNF \ldots \OrDNF \DNFRegex_{n,m}'' =
  (\DNFRegex_1 \ConcatDNF \DNFRegexAlt_1) \OrDNF \ldots \OrDNF
  (\DNFRegex_1 \ConcatDNF \DNFRegexAlt_m) \OrDNF \ldots \OrDNF
  (\DNFRegex_n \ConcatDNF \DNFRegexAlt_1) \OrDNF \ldots \OrDNF
  (\DNFRegex_n \ConcatDNF \DNFRegexAlt_m)$
  By repeated application of Lemma~\ref{lem:id-expressible-on-factor-left}
  and Lemma~\ref{lem:id-expressible-on-factor-right}, there
  exists a DNF lens $\DNFLens \OfRewritelessType
  \DNFRegex_{1,1}'' \OrDNF \ldots \OrDNF \DNFRegex_{n,m}'' \Leftrightarrow
  (\DNFRegex_1 \OrDNF \ldots \OrDNF \DNFRegex_n) \ConcatDNF
  (\DNFRegexAlt_1 \OrDNF \ldots \OrDNF \DNFRegexAlt_m)$, so
  $\DNFLens \OfRewritelessType
  \DNFRegex_{1,1}'' \OrDNF \ldots \OrDNF \DNFRegex_{n,m}'' \Leftrightarrow
  \DNFRegex' \ConcatDNF \DNFRegexAlt'$.

  Consider the derivation 
  \[
    \inferrule*
    {
      \DNFRegex'' = \DNFOf{\Sequence_{1,1}'' \DNFSep \ldots \DNFSep \Sequence_{n,m}''}\\
      \forall i,j. \Sequence_{i,j}'' =
      \SequenceOf{\String_{i,j,0}'' \SeqSep \Atom_{i,j,1}'' \SeqSep \ldots \SeqSep \Atom_{i,j,n_{i,j}}'' \SeqSep \String_{i,j,n_{i,j}}''}\\
      \forall i,j. \Atom_{i,j}'' \ParallelRewriteAtom \DNFRegex_{i,j}''\\
      \forall i,j. \DNFRegex_{i,j}'' = \DNFOf{\SequenceOf{\String_{i,j,0}''}} \ConcatDNF \DNFRegex_{i,j,1}''
      \ConcatDNF \ldots \ConcatDNF \DNFRegex_{i,j,n_{i,j}}'' \ConcatDNF
      \DNFOf{\SequenceOf{\String_{i,j,n_{i,j}}''}}
    }
    {
      \DNFRegex'' \ParallelRewrite
      \DNFRegex_{1,1}'' \OrDNF \ldots \OrDNF \DNFRegex_{n,m}''
    }
  \]

  $\DNFRegex \ConcatDNF \DNFRegexAlt \ParallelRewrite
  \DNFRegex_{1,1}'' \OrDNF \ldots \OrDNF \DNFRegex_{n,m}''$,
  $\DNFRegex_{1,1}'' \OrDNF \ldots \OrDNF \DNFRegex_{n,m}''$
  
  If $\DNFRegex \ParallelRewrite \DNFRegex'$ and
  $\DNFRegexAlt \ParallelRewrite \DNFRegexAlt'$, then
  $\DNFRegex \ConcatDNF \DNFRegex \ParallelRewrite \DNFRegex''$ such that there
  exists a rewriteless DNF lens
  $\DNFLens \OfRewritelessType
  \DNFRegex \ConcatDNF \DNFRegexAlt \Leftrightarrow \DNFRegex''$, and
  $\SemanticsOf{\DNFLens} =
  \SetOf{(\String,\String) \SuchThat \String \in
    \LanguageOf{\DNFRegex \ConcatDNF \DNFRegexAlt}}$,
  as desired.
\end{proof}

\begin{lemma}[Swap's Unimportance For Identity]
  \label{lem:swap-unimportance-identity}
  \leavevmode
  \begin{enumerate}
  \item If $\DNFRegex$ is strongly unambiguous and
    $\DNFRegex \ParallelRewriteSwap \DNFRegex_1$ then there exists a
    $\DNFRegex_2$ such that $\DNFRegex \ParallelRewrite \DNFRegex_2$
    and there exists a lens
    $\Lens \OfRewritelessType \DNFRegex_1 \Leftrightarrow \DNFRegex_2$
    such that
    $\SemanticsOf{\Lens} = \SetOf{(\String,\String)
      \SuchThat \String \in \LanguageOf{\DNFRegex}}$
  \item If $\Atom$ is strongly unambiguous and
    $\Atom \ParallelRewriteSwapAtom \DNFRegex_1$ then there exists a
    $\DNFRegex_2$ such that $\Atom \ParallelRewriteAtom \DNFRegex_2$
    and there exists a lens
    $\Lens \OfRewritelessType \DNFRegex_1 \Leftrightarrow \DNFRegex_2$
    such that
    $\SemanticsOf{\Lens} = \SetOf{(\String,\String)
      \SuchThat \String \in \LanguageOf{\Atom}}$
  \end{enumerate}
\end{lemma}
\begin{proof}
  By mutual induction on the derivation of $\ParallelRewriteSwap$.
  \begin{case}[\AtomUnrollstarLeftRule{}]
    Let $\Atom \ParallelRewriteSwapAtom \DNFRegex_1$, and the last step of the
    derivation is an application of \AtomUnrollstarRightRule{}.
    That means $\Atom = \StarOf{\DNFRegex}$ and
    $\DNFRegex_1 = \DNFOf{\SequenceOf{\EmptyString}} \OrDNF
    (\DNFRegex \ConcatDNF \AtomToDNFOf{\StarOf{\DNFRegex}})$.
    
    Consider an application of $\ParallelRewriteAtom$'s
    \AtomUnrollstarRightRule{}.
    $\Atom \ParallelRewrite \DNFRegex_1$.
    By Lemma~\ref{lem:strongly-unambiguous-identity-expressible}, there exists
    a DNF lens
    $\DNFLens \OfRewritelessType \DNFRegex_1 \Leftrightarrow \DNFRegex_1$ and
    $\SemanticsOf{\DNFLens} = \SetOf{(\String,\String) \SuchThat \String \in
      \LanguageOf{\Atom}}$.
  \end{case}
  
  \begin{case}[AtomUnrollstarRightRule{}]
    Let $\Atom \ParallelRewriteSwapAtom \DNFRegex_1$, and the last step of the
    derivation is an application of \AtomUnrollstarLeftRule{}.
    That means $\Atom = \StarOf{\DNFRegex}$ and
    $\DNFRegex_1 = \AtomToDNFOf{\DNFRegex_1'}$, where 
    
    Consider an application of $\ParallelRewriteAtom$'s
    \AtomUnrollstarRightRule{}.
    $\Atom \ParallelRewrite \DNFRegex_1$.
    By Lemma~\ref{lem:strongly-unambiguous-identity-expressible}, there exists
    a DNF lens
    $\DNFLens \OfRewritelessType \DNFRegex_1 \Leftrightarrow \DNFRegex_1$ and
    $\SemanticsOf{\DNFLens} = \SetOf{(\String,\String) \SuchThat \String \in
      \LanguageOf{\Atom}}$.
  \end{case}

  \begin{case}[\ParallelSwapAtomStructuralRewriteRule{}]
    Let $\Atom \ParallelRewriteSwapAtom \DNFRegex_1$, and the last step of the
    derivation is an application of \ParallelSwapAtomStructuralRewriteRule{}.
    That means $\Atom = \StarOf{\DNFRegex}$ and
    $\DNFRegex_1 = \AtomToDNFOf{\StarOf{\DNFRegex_1'}}$, and
    $\DNFRegex \ParallelRewriteSwapAtom \DNFRegex_1'$.

    By IH, there exists $\DNFRegex_2'$ such that
    $\DNFRegex \ParallelRewrite \DNFRegex_2'$, and there exists a
    rewriteless DNF lens
    $\DNFLens \OfRewritelessType \DNFRegex_1' \Leftrightarrow \DNFRegex_2'$.

    By \ParallelSwapAtomStructuralRewriteRule{},
    $\Atom \ParallelRewriteAtom \AtomToDNFOf{\StarOf{\DNFRegex_2'}}$.

    By Lemma~\ref{lem:typ_sem_it},
    $\AtomToDNFLensOf{\IterateLensOf{\DNFLens}} \OfRewritelessType
    \AtomToDNFOf{\StarOf{\DNFRegex_1}} \Leftrightarrow
    \AtomToDNFOf{\StarOf{\DNFRegex_2}}$, with\\
    $\SemanticsOf{\AtomToDNFLensOf{\IterateLensOf{\DNFLens}}}
    = \SetOf{(\String_1\ldots\String_n,\StringAlt_1\ldots\StringAlt_n)
      \SuchThat (\String_i,\StringAlt_i) \in \SemanticsOf{\DNFLens}}\\
    = \SetOf{(\String_1\ldots\String_n,\String_1\ldots\String_n)
      \SuchThat (\String_i,\String_i) \in \LanguageOf{\DNFRegex}}\\
    = \SetOf{(\String,\String) \SuchThat \String \in \LanguageOf{\Atom})}$
  \end{case}

  \begin{case}[\ParallelSwapDNFStructuralRewriteRule{}]
    Let $\DNFRegex \ParallelRewriteSwap \DNFRegex'$, and the last step of the
    derivation is an application of \ParallelSwapDNFStructuralRewriteRule{}.

    \[
      \inferrule*
      {
        \DNFRegex = \DNFOf{\Sequence_1 \DNFSep \ldots \DNFSep \Sequence_n}\\
        \forall i. \Sequence_i =
        \SequenceOf{\String_{i,0} \SeqSep \Atom_{i,1} \SeqSep \ldots \SeqSep \Atom_{i,n_i} \SeqSep \String_{i,n_i}}\\
        \forall i,j. \Atom_{i,j} \ParallelRewriteSwapAtom \DNFRegex_{i,j}\\
        \forall i. \DNFRegex_i = \DNFOf{\SequenceOf{\String_{i,0}}} \ConcatDNF \DNFRegex_{i,1}
        \ConcatDNF \ldots \ConcatDNF \DNFRegex_{i,n_i} \ConcatDNF
        \DNFOf{\SequenceOf{\String_{i,n_i}}}
      }
      {
        \DNFRegex \ParallelRewriteSwap \DNFRegex_1 \OrDNF \ldots \OrDNF \DNFRegex_n
      }
    \]
    and $\DNFRegex' = \DNFRegex_1 \OrDNF \OrDNF \ldots \OrDNF \DNFRegex_n$.

    There exists lenses
    By IH, there exist $\DNFRegexAlt_{i,j}$ and $\DNFLens_{i,j}$, such that
    $\Atom_{i,j} \ParallelRewrite \DNFRegexAlt_{i,j}$,
    $\DNFLens_{i,j} \OfRewritelessType
    \DNFRegex_{i,j} \Leftrightarrow \DNFRegexAlt_{i,j}$,
    and $\SemanticsOf{\DNFLens_{i,j}} =
    \SetOf{(\String,\String) \SuchThat \String \in \LanguageOf{\Atom_{i,j}}}$.

    Define $\DNFRegexAlt_i =
    \DNFOf{\SequenceOf{\String_{i,0}}} \ConcatDNF \DNFRegexAlt_{i,1}
    \ConcatDNF \ldots \ConcatDNF
    \DNFRegexAlt_{i,n_i} \ConcatDNF \DNFOf{\SequenceOf{\String_{i,n_i}}}$.

    Define $\DNFRegexAlt = \DNFRegexAlt_0 \OrDNF \ldots \OrDNF \DNFRegexAlt_n$
    
    By repeated application of Lemma~\ref{lem:typ_sem_concat}, there exists a
    lens
    $\DNFLens_i =
    (\DNFLensOf{(\SequenceLensOf{(\String_{i,0},\String_{i,0})},\Identity)},\Identity)
    \ConcatDNFLens \DNFLens_{i,1}
    \ConcatDNFLens \ldots \ConcatDNFLens
    \DNFLens_{i,n_i} \ConcatDNFLens
    (\DNFLensOf{(\SequenceLensOf{(\String_{i,n_i},\String_{i,n_i})},\Identity)},\Identity)
    \OfRewritelessType
    \DNFOf{\SequenceOf{\String_{i,0}}} \ConcatDNF
    \DNFRegex_{i,1} \ConcatDNF \ldots \ConcatDNF \DNFRegex_{i,n_i}
    \ConcatDNF \DNFOf{\SequenceOf{\String_{i,n_i}}} \Leftrightarrow
    \DNFOf{\SequenceOf{\String_{i,0}}} \ConcatDNF
    \DNFRegexAlt_{i,1} \ConcatDNF \ldots \ConcatDNF \DNFRegexAlt_{i,n_i}
    \ConcatDNF \DNFOf{\SequenceOf{\String_{i,n_i}}}$
    
    By pushing around the definitions of $\ConcatDNF$, this becomes
    $\DNFLens_i \OfRewritelessType
    \DNFRegex_i
    \Leftrightarrow
    \DNFRegexAlt_i$ and
    $\SemanticsOf{\DNFLens_i} = \SetOf{(\String,\String) \SuchThat
      \String \in \LanguageOf{\Sequence_i}}$.

    By repeated applications of Lemma~\ref{lem:typ_sem_or}, there exists a lens
    $\DNFLens =
    \DNFLens_0 \OrDNFLens \ldots \OrDNFLens \DNFLens_n 
    \OfRewritelessType
    \DNFRegex_1 \OrDNF \ldots \OrDNF \DNFRegex_n
    \Leftrightarrow
    \DNFRegexAlt_1 \OrDNF \ldots \OrDNF \DNFRegexAlt_n$.
    
    By pushing around the definitions of $\OrDNF$, this becomes
    $\DNFLens \OfRewritelessType
    \DNFRegex' \Leftrightarrow \DNFRegexAlt$, and
    $\SemanticsOf{\DNFLens} =
    \SetOf{(\String,\String) \SuchThat \String\in\LanguageOf{\DNFRegex}}$

    Furthermore,
    \[
      \inferrule*
      {
        \DNFRegex = \DNFOf{\Sequence_1 \DNFSep \ldots \DNFSep \Sequence_n}\\
        \forall i. \Sequence_i =
        \SequenceOf{\String_{i,0} \SeqSep \Atom_{i,1} \SeqSep \ldots \SeqSep \Atom_{i,n_i} \SeqSep \String_{i,n_i}}\\
        \forall i,j. \Atom_{i,j} \ParallelRewrite \DNFRegexAlt_{i,j}\\
        \forall i. \DNFRegexAlt_i = \DNFOf{\SequenceOf{\String_{i,0}}}
        \ConcatDNF \DNFRegexAlt_{i,1}
        \ConcatDNF \ldots \ConcatDNF \DNFRegexAlt_{i,n_i} \ConcatDNF
        \DNFOf{\SequenceOf{\String_{i,n_i}}}
      }
      {
        \DNFRegex \ParallelRewrite \DNFRegexAlt_1 \OrDNF \ldots \OrDNF \DNFRegexAlt_n
      }
    \]
  \end{case}
  
  \begin{case}[\IdentityRewriteRule{}]
    Let $\DNFRegex \ParallelRewriteSwap \DNFRegex_1$, and the last step of the
    derivation is an application of \IdentityRewriteRule{}.

    This means $\DNFRegex \ParallelRewriteSwap \DNFRegex$

    Consider the application of $\ParallelRewrite{}$'s $\IdentityRewriteRule$,
    $\DNFRegex \ParallelRewrite \DNFRegex$.

    By Lemma~\ref{lem:strongly-unambiguous-identity-expressible}, there exists
    a DNF lens
    $\DNFLens \OfRewritelessType \DNFRegex \Leftrightarrow \DNFRegex$ and
    $\SemanticsOf{\DNFLens} = \SetOf{(\String,\String) \SuchThat \String \in
      \LanguageOf{\DNFRegex}}$.
  \end{case}

  \begin{case}[\DNFReorderRule{}]
    Let $\DNFRegex \ParallelRewriteSwap \DNFRegex_1$, and the last step of the
    derivation is an application of \DNFReorderRule{}.

    Let $\DNFRegex = \DNFOf{\Sequence_1 \DNFSep \ldots \DNFSep \Sequence_n}$.
    This means that there exists a $\sigma$ such that
    $\DNFRegex_1 = \DNFOf{\Sequence_{\sigma(1)} \DNFSep \ldots \DNFSep \Sequence_{\sigma(n)}}$.

    Consider $\DNFRegex \ParallelRewrite \DNFRegex$.
    By Lemma~\ref{lem:strongly-unambiguous-identity-expressible}, there exists
    sequence lenses $\SequenceLens_i$ such that $\SequenceLens_i
    \OfRewritelessType \Sequence_i \Leftrightarrow \Sequence_i$ and
    $\SemanticsOf{\SequenceLens_i} = \SetOf{(\String,\String) \SuchThat \String
      \in \LanguageOf{\Sequence_i}}$

    Consider $(\DNFLensOf{\SequenceLens_1 \DNFLSep \ldots \DNFLSep \SequenceLens_n},\sigma)
    \OfRewritelessType \DNFOf{\Sequence_1 \DNFSep \ldots \DNFSep \Sequence_n}
    \Leftrightarrow \DNFOf{\Sequence_{\sigma(1)} \DNFSep \ldots \DNFSep \Sequence_{\sigma(n)}}$,
    which is typed as desired.
    $\SemanticsOf{(\DNFLensOf{\SequenceLens_1 \DNFLSep \ldots \DNFLSep \SequenceLens_n},\sigma)} =
    \SetOf{(\String,\StringAlt) \SuchThat \exists i. (\String,\StringAlt) \in
      \SemanticsOf{\SequenceLens_i}} =
    \SetOf{(\String,\String) \SuchThat \exists i. \String \in
      \LanguageOf{\Sequence_i}} =
    \SetOf{(\String,\String) \SuchThat \String \in \LanguageOf{\DNFRegex}}$
  \end{case}
\end{proof}

\begin{definition}
  Let $\Lens$ be a lens.
  Define the binary relation, $\DNFLensHasSemanticsOf{\Lens}\subseteq \DNFRegexType
  \Cross\DNFRegexType$, as
  $\SatisfiesDNFLensHasSemanticsOf{\Lens}{\DNFRegex}{\DNFRegexAlt}$ if, and
  only if there exists a DNF Lens $\DNFLens$ such that $\DNFLens
  \OfRewritelessType \DNFRegex \Leftrightarrow \DNFRegexAlt$, and
  $\SemanticsOf{\DNFLens}=\SemanticsOf{\Lens}$.
  $\SatisfiesIdentitySemantics{\DNFRegex}{\DNFRegexAlt}$ is
  shorthand for
  $\SatisfiesDNFLensHasSemanticsOf{\IdentityLensOf{\DNFRegex}}{\DNFRegex}{\DNFRegexAlt}$.
\end{definition}

\begin{lemma}
  \label{lem:copyable_expansions_left_swapless}\leavevmode
  \begin{enumerate}
  \item Let $\DNFLens \OfRewritelessType \DNFRegex \Leftrightarrow
    \DNFRegexAlt$ and $\DNFRegex \ParallelRewrite \DNFRegex'$.  There exists some
    $\DNFRegexAlt'$, $\DNFLens'$ such that $\DNFRegexAlt \ParallelRewrite
    \DNFRegexAlt'$,
    $\DNFLens' \OfRewritelessType
    \DNFRegex' \Leftrightarrow \DNFRegexAlt'$, and $\SemanticsOf{\DNFLens} =
    \SemanticsOf{\DNFLens'}$.
  \item Let $\AtomLens \OfRewritelessType \Atom \Leftrightarrow \AtomAlt$ and
    $\Atom \ParallelRewriteAtom \DNFRegex$.  There exists some $\DNFRegexAlt$,
    $\DNFLens$, such that $\AtomAlt \ParallelRewriteAtom \DNFRegexAlt$,
    $\DNFLens' \OfRewritelessType \DNFRegex
    \Leftrightarrow \DNFRegexAlt$, and $\SemanticsOf{\DNFLens} =
    \SemanticsOf{\AtomLens}$.
  \end{enumerate}
\end{lemma}
\begin{proof}
  By mutual induction on the derivation of $\ParallelRewrite$ and
  $\ParallelRewriteAtom$
  \begin{case}[\AtomUnrollstarLeftRule{}]
    Let $\AtomLens \OfRewritelessType \Atom \Leftrightarrow \AtomAlt$, and
    $\Atom\ParallelRewriteAtom\DNFRegex$ through an application of
    \AtomUnrollstarLeftRule{}.  By inversion, there exists a derivation of
    \[
      \inferrule*
      {
        \DNFLens \OfRewritelessType \DNFRegex' \Leftrightarrow \DNFRegexAlt'\\
        \UnambigItOf{\DNFRegex'}\\
        \UnambigItOf{\DNFRegexAlt'}
      }
      {
        \IterateLensOf{\DNFLens} \OfRewritelessType \StarOf{\DNFRegex'} \Leftrightarrow
        \StarOf{\DNFRegexAlt'}
      }
    \]

    Where $\IterateLensOf{\DNFLens} = \AtomLens$, $\StarOf{\DNFRegex'} = \Atom$,
    and $\StarOf{\DNFRegexAlt'} = \AtomAlt$.

    As \AtomUnrollstarLeftRule{} was applied,
    $\DNFRegex =
    \OrDNFOf{\DNFOf{\SequenceOf{\EmptyString}}}
    {(\ConcatDNFOf{\DNFRegex'}{\DNFOf{\SequenceOf{\StarOf{\DNFRegex'}}}})}$.

    Consider applying \AtomUnrollstarLeftRule{} to $\StarOf{\DNFRegexAlt'}$.
    $\StarOf{\DNFRegexAlt'}\ParallelRewriteAtom
    \OrDNFOf{\DNFOf{\SequenceOf{\EmptyString}}}
    {(\ConcatDNFOf{\DNFRegexAlt'}{\DNFOf{\SequenceOf{\StarOf{\DNFRegexAlt'}}}})}$

    Consider the lenses

    \[
      \inferrule*
      {
        \inferrule*
        {
        }
        {
          \SequenceLensOf{(\EmptyString,\EmptyString)}
        }
      }
      {
        \DNFLensOf{\SequenceLensOf{(\EmptyString,\EmptyString)}}
      }
    \]

    \[
      \inferrule*
      {
        \inferrule*
        {
          \IterateLensOf{\DNFLens} \OfRewritelessType \StarOf{\DNFRegex} \Leftrightarrow
          \StarOf{\DNFRegexAlt}
        }
        {
          \SequenceLensOf{\IterateLensOf{\DNFLens}} \OfRewritelessType
          \SequenceOf{\StarOf{\DNFRegex}} \Leftrightarrow
          \SequenceOf{\StarOf{\DNFRegexAlt}}
        }
      }
      {
        \DNFLensOf{\SequenceLensOf{\IterateLensOf{\DNFLens}}} \OfRewritelessType
        \DNFOf{\SequenceOf{\StarOf{\DNFRegex}}} \Leftrightarrow
        \DNFOf{\SequenceOf{\StarOf{\DNFRegexAlt}}}
      }
    \]

    As $\UnambigItOf{\DNFRegex}$, $\LanguageOf{\DNFRegex} \UnambigConcat
    \LanguageOf{\DNFOf{\SequenceOf{\StarOf{\DNFRegex}}}}$.
    As $\UnambigItOf{\DNFRegex}$, $\EmptyString\not\in\LanguageOf{\DNFRegex}$.
    This means $\EmptyString\not\in\LanguageOf{\DNFRegex \ConcatDNF
      \DNFOf{\SequenceOf{\StarOf{\DNFRegex}}}}$, so
    $\LanguageOf{\DNFOf{\SequenceOf{\EmptyString}}} \Intersect
    \LanguageOf{\DNFRegex \ConcatDNF \DNFOf{\SequenceOf{\StarOf{\DNFRegex}}}} =
    \emptyset$

    Because of this, by Lemma~\ref{lem:typ_sem_concat} and
    Lemma~\ref{lem:typ_sem_or}, there exists the typing
    for the lens
    $\DNFLensOf{\SequenceLensOf{(\EmptyString,\EmptyString)}} \OrDNFLens
    (\DNFLens \ConcatDNFLens \DNFLensOf{\SequenceLensOf{\IterateLensOf{\DNFLens}}})
    \OfRewritelessType
    \DNFOf{\SequenceOf{\EmptyString}} \OrDNF (\DNFRegex \ConcatDNFLens
    \DNFOf{\SequenceOf{\StarOf{\DNFRegex}}}) \Leftrightarrow
    \DNFOf{\SequenceOf{\EmptyString}} \OrDNF (\DNFRegexAlt \ConcatDNFLens
    \DNFOf{\SequenceOf{\StarOf{\DNFRegexAlt}}})$, which is the desired typing.

    \[
      \begin{array}{rcl}
        \SemanticsOf{\DNFLensOf{\SequenceLensOf{(\EmptyString,\EmptyString)}}
        \OrDNFLens
        (\DNFLens \ConcatDNFLens
        \DNFLensOf{\SequenceLensOf{\IterateLensOf{\DNFLens}}})}
        & = & \SetOf{(\String,\StringAlt)\SuchThat
              (\String,\StringAlt)\in\SemanticsOf{\DNFLensOf{\SequenceLensOf{
              (\EmptyString,\EmptyString)}}}\\
        &   & \hspace*{3em}\BooleanOr (\String,\StringAlt)
              \in \SemanticsOf{\DNFLens \ConcatDNFLens
              \DNFLensOf{\SequenceLensOf{\IterateLensOf{\DNFLens}}}}}\\
        & = & \SetOf{(\EmptyString,\EmptyString)} \Union
              \SetOf{(\String,\StringAlt) \SuchThat (\String,\StringAlt) \in
              \SemanticsOf{\DNFLens \ConcatDNFLens
              \DNFLensOf{\SequenceLensOf{\IterateLensOf{\DNFLens}}}}}\\
        & = & \SetOf{(\EmptyString,\EmptyString)} \Union
              \SetOf{(\String_1\Concat\String_2,\StringAlt_1\Concat\StringAlt_2)
              \SuchThat (\String_1,\StringAlt_1) \in
              \SemanticsOf{\DNFLens}\\
        &   & \hspace*{3em}\BooleanAnd
              (\String_2,\StringAlt_2) \in
              \SemanticsOf{\DNFLensOf{\SequenceLensOf{\IterateLensOf{\DNFLens}}}}}\\
        & = & \SetOf{(\EmptyString,\EmptyString)} \Union
              \SetOf{(\String_1\Concat
              (\String_{2,1}\Concat\ldots\Concat\String_{2,n}),
              \StringAlt_1\Concat
              (\StringAlt_{2,1}\Concat\ldots\Concat\String_{2,n}))\\
        & & \hspace*{3em}
            \SuchThat (\String_1,\StringAlt_1) \in
            \SemanticsOf{\DNFLens} \BooleanAnd n\geq 0 \BooleanAnd\\
        & & \hspace*{5em}
            \forall i\in\RangeIncInc{1}{n}.(\String_{2,i},\StringAlt_{2,i}) \in
            \SemanticsOf{\DNFLens}}\\
        & = & \SetOf{(\EmptyString,\EmptyString)} \Union
              \SetOf{(\String_1\Concat\ldots\Concat\String_n,
              \StringAlt_1\Concat\ldots\Concat\StringAlt_n) \SuchThat\\
        & & \hspace*{3em}
              n \geq 1 \BooleanAnd \forall i\in\RangeIncInc{1}{n}.
              (\String_i,\StringAlt_i)\in\SemanticsOf{\DNFLens}}\\
        & = & \SetOf{(\String_1\Concat\ldots\Concat\String_n,
              \StringAlt_1\Concat\ldots\Concat\StringAlt_n) \SuchThat
              n \geq 0\\
        & & \hspace*{3em}\BooleanAnd \forall i\in\RangeIncInc{1}{n}.
              (\String_i,\StringAlt_i)\in\SemanticsOf{\DNFLens}}\\
        & = & \SemanticsOf{\IterateLensOf{\DNFLens}}
      \end{array}
    \]
  \end{case}

  \begin{case}[\AtomUnrollstarRightRule{}]
    Let $\AtomLens \OfRewritelessType \Atom \Leftrightarrow \AtomAlt$, and
    $\Atom\ParallelRewriteAtom\DNFRegex$ through an application of
    \AtomUnrollstarRightRule{}.
  \end{case}

  \begin{case}[\ParallelAtomStructuralRewriteRule{}]
    Let $\AtomLens \OfRewritelessType \Atom \Leftrightarrow \AtomAlt$, and
    $\Atom\ParallelRewriteAtom\DNFRegex$ through an application of
    \ParallelAtomStructuralRewriteRule{}.  By inversion, there exists a derivation of
  
    \[
      \inferrule*
      {
        \DNFLens \OfRewritelessType \DNFRegex' \Leftrightarrow \DNFRegexAlt'\\
        \UnambigItOf{\DNFRegex'}\\
        \UnambigItOf{\DNFRegexAlt'}
      }
      {
        \IterateLensOf{\DNFLens} \OfRewritelessType \StarOf{\DNFRegex'} \Leftrightarrow
        \StarOf{\DNFRegexAlt'}
      }
    \]

    Where $\IterateLensOf{\DNFLens} = \AtomLens$, $\StarOf{\DNFRegex'} = \Atom$,
    and $\StarOf{\DNFRegexAlt'} = \AtomAlt$.

    As \ParallelAtomStructuralRewriteRule{} was applied,
    $\DNFRegex' \ParallelRewrite \DNFRegex''$, and
    $\DNFRegex = \DNFOf{\SequenceOf{\StarOf{\DNFRegex''}}}$.

    By induction hypothesis, there exists some $\DNFLens'$, $\DNFRegexAlt''$,
    such that $\DNFLens' \OfRewritelessType \DNFRegex' \Leftrightarrow
    \DNFRegex''$, and $\SemanticsOf{\DNFLens'} = \SemanticsOf{\DNFLens}$.
    Because $\LanguageOf{\DNFRegex''} = \LanguageOf{\DNFRegex'}$ and
    $\LanguageOf{\DNFRegexAlt''} = \LanguageOf{\DNFRegexAlt'}$,
    $\UnambigItOf{\DNFRegex''}$ and $\UnambigItOf{\DNFRegexAlt''}$.

    Consider the typing
    \[
      \inferrule*
      {
        \inferrule*
        {
          \inferrule*
          {
            \DNFLens' \OfRewritelessType \DNFRegex'' \Leftrightarrow
            \DNFRegexAlt''\\
            \UnambigItOf{\DNFRegex''}\\
            \UnambigItOf{\DNFRegexAlt''}
          }
          {
            \IterateLensOf{\DNFLens'} \OfRewritelessType \StarOf{\DNFRegex''}
            \Leftrightarrow \StarOf{\DNFRegexAlt''}
          }
        }
        {
          \SequenceLensOf{\IterateLensOf{\DNFLens'}} \OfRewritelessType
          \SequenceOf{\StarOf{\DNFRegex''}} \Leftrightarrow
          \SequenceOf{\StarOf{\DNFRegexAlt''}}
        }
      }
      {
        \DNFLensOf{\SequenceLensOf{\IterateLensOf{\DNFLens'}}}
        \OfRewritelessType
        \DNFOf{\SequenceOf{\StarOf{\DNFRegex''}}} \Leftrightarrow
        \DNFOf{\SequenceOf{\StarOf{\DNFRegexAlt''}}}
      }
    \]

    This is the desired typing.  The semantics are as desired as well.

    \[
      \begin{array}{rcl}
        \SemanticsOf{\DNFLensOf{\SequenceLensOf{\IterateLensOf{\DNFLens'}}}}
        & = & \SetOf{(\String_1\Concat\ldots\Concat\String_n,
              \StringAlt_1\Concat\ldots\Concat\StringAlt_n) \SuchThat
              n \geq 0 \BooleanAnd \forall i\in\RangeIncInc{1}{n}
              (\String_i,\StringAlt_i)\in\SemanticsOf{\DNFLens'}}\\
        & = & \SetOf{(\String_1\Concat\ldots\Concat\String_n,
              \StringAlt_1\Concat\ldots\Concat\StringAlt_n) \SuchThat
              n \geq 0 \BooleanAnd \forall i\in\RangeIncInc{1}{n}
              (\String_i,\StringAlt_i)\in\SemanticsOf{\DNFLens}}\\
        & = & \SemanticsOf{\AtomLens}
      \end{array}
    \]
  \end{case}

  \begin{case}[\DNFReorderRule{}]
    Let $\DNFLens \OfRewritelessType \DNFRegex \Leftrightarrow \DNFRegexAlt$, and
    $\DNFRegex\ParallelRewrite\DNFRegex'$ through an application of
    \DNFReorderRule{}.
  \end{case}

  \begin{case}[\ParallelDNFStructuralRewriteRule{}]
    Let $\DNFLens \OfRewritelessType \DNFRegex \Leftrightarrow \DNFRegexAlt$, and
    $\DNFRegex\ParallelRewrite\DNFRegex'$ through an application of
    \ParallelDNFStructuralRewriteRule{}.
  \end{case}

  \begin{case}[\IdentityRewriteRule{}]
    Let $\DNFLens \OfRewritelessType \DNFRegex \Leftrightarrow \DNFRegexAlt$, and
    $\DNFRegex\ParallelRewrite\DNFRegex'$ through an application of
    \IdentityRewriteRule{}.
  \end{case}
\end{proof}

\begin{lemma}
  \label{lem:copyable_expansions_right_swapless}\leavevmode
  \begin{enumerate}
  \item Let $\DNFLens \OfRewritelessType \DNFRegex \Leftrightarrow
    \DNFRegexAlt$ and $\DNFRegexAlt \ParallelRewrite \DNFRegexAlt'$.

    There exists some
    $\DNFRegex'$, $\DNFLens'$ such that $\DNFRegex \ParallelRewrite
    \DNFRegex'$,
    $\DNFLens' \OfRewritelessType
    \DNFRegex' \Leftrightarrow \DNFRegexAlt'$, and $\SemanticsOf{\DNFLens} =
    \SemanticsOf{\DNFLens'}$.
  \item Let $\AtomLens \OfRewritelessType \Atom \Leftrightarrow \AtomAlt$ and
    $\AtomAlt \ParallelRewriteAtom \DNFRegexAlt$.

    There exists some $\DNFRegex$,
    $\DNFLens$, such that $\Atom \ParallelRewriteAtom \DNFRegex$,
    $\DNFLens' \OfRewritelessType \DNFRegex
    \Leftrightarrow \DNFRegexAlt$, and $\SemanticsOf{\DNFLens} =
    \SemanticsOf{\AtomLens}$.
  \end{enumerate}
\end{lemma}
\begin{proof}
  This can be proven symmetrically to Lemma~\ref{lem:copyable_expansions_left_swapless}.
\end{proof}

\begin{lemma}
  \label{lem:bisimilarity-property-parallel-swapless}
  For all lenses $\Lens \OfType \Regex \Leftrightarrow \RegexAlt$,
  $\IsBisimilarWithPropertyOf{\ParallelRewrite}{\DNFLensHasSemanticsOf{\Lens}}$,
  over the set of strongly unambiguous DNF regular expressions.
\end{lemma}
\begin{proof}
  Let $\DNFRegex,\DNFRegexAlt$ be strongly unambiguous DNF regular expressions,
  with $\DNFRegex \DNFLensHasSemanticsOf{\Lens} \DNFRegexAlt$.
  So there exists a rewriteless DNF lens
  $\DNFLens \OfRewritelessType \DNFRegex \Leftrightarrow \DNFRegexAlt$ where
  $\SemanticsOf{\DNFLens} = \SemanticsOf{\Lens}$.

  Let $\DNFRegex \ParallelRewrite \DNFRegex'$.  By
  Lemma~\ref{lem:bisimilarity-property-parallel-swapless}, there exists $\DNFLens'$,
  $\DNFRegexAlt'$ such that $\DNFRegexAlt \ParallelRewrite \DNFRegexAlt'$,
  $\DNFLens' \OfRewritelessType \DNFRegex' \Leftrightarrow \DNFRegexAlt'$,
  and $\SemanticsOf{\DNFLens'} = \SemanticsOf{\DNFLens} = \SemanticsOf{\Lens}$,
  so $\DNFRegex' \DNFLensHasSemanticsOf{\Lens} \DNFRegexAlt'$.

  Let $\DNFRegexAlt \ParallelRewrite \DNFRegexAlt'$.  By
  Lemma~\ref{lem:bisimilarity-property-parallel-swapless}, there exists $\DNFLens'$,
  $\DNFRegex'$ such that $\DNFRegex \ParallelRewrite \DNFRegex'$,
  $\DNFLens' \OfRewritelessType \DNFRegex' \Leftrightarrow \DNFRegexAlt'$,
  and $\SemanticsOf{\DNFLens'} = \SemanticsOf{\DNFLens} = \SemanticsOf{\Lens}$,
  so $\DNFRegex' \DNFLensHasSemanticsOf{\Lens} \DNFRegexAlt'$.
\end{proof}

\begin{lemma}
  \label{lem:bisimilarity-property-star-parallel-swapless}
  For all lenses $\Lens \OfType \Regex \Leftrightarrow \RegexAlt$,
  $\IsBisimilarWithPropertyOf{\StarOf{\ParallelRewrite}}{\DNFLensHasSemanticsOf{\Lens}}$,
  over the set of strongly unambiguous DNF regular expressions.
\end{lemma}
\begin{proof}
  By Lemma~\ref{lem:bisimilarity-star} and
  Lemma~\ref{lem:bisimilarity-property-parallel-swapless}.
\end{proof}

\begin{corollary}[Bisimilarity in Star Sequential]
  \label{cor:bisimilarity-star-sequential}
  By Lemma~\ref{lem:bisimilarity-property-star-parallel-swapless} and
  Theorem~\ref{thm:parallel-star-equivalence}.
\end{corollary}

\begin{lemma}[$\ParallelRewrite$ Maintained Under $\ConcatDNF$ up to
  $\IdentityLens$ on the left]
  \label{lem:star-parallel-rewrite-concatenation-to-identity-left}
  Let $\DNFRegex$ be strongly unambiguous.  Let $\DNFRegexAlt$ be strongly
  unambiguous.
  Let $\UnambigConcatOf{\LanguageOf{\DNFRegex}}{\LanguageOf{\DNFRegexAlt}}$.
  If $\DNFRegex \StarOf{\ParallelRewrite} \DNFRegex'$, then
  $\DNFRegex \ConcatDNF \DNFRegexAlt \StarOf{\ParallelRewrite} \DNFRegex''$ such that there
  exists a rewriteless DNF lens
  $\DNFLens \OfRewritelessType
  \DNFRegex' \ConcatDNF \DNFRegexAlt \Leftrightarrow \DNFRegex''$, and
  $\SemanticsOf{\DNFLens} =
  \SetOf{(\String,\String) \SuchThat \String \in
    \LanguageOf{\DNFRegex \ConcatDNF \DNFRegexAlt}}$.
\end{lemma}
\begin{proof}
  As $\UnambigConcatOf{\LanguageOf{\DNFRegex}}{\LanguageOf{\DNFRegexAlt}}$,
  $\DNFRegex \ConcatDNF \DNFRegexAlt$ is strongly unambiguous.

  We proceed by induction on the derivation of $\StarOf{\ParallelRewrite}$.
  
  \begin{case}[\ReflexivityRule{}]
    \[
      \inferrule*
      {
      }
      {
        $\DNFRegex \StarOf{\ParallelRewrite} \DNFRegex$
      }
    \]

    By reflexivity

    \[
      \inferrule*
      {
      }
      {
        \DNFRegex \ConcatDNF \DNFRegexAlt \StarOf{\ParallelRewrite}
        \DNFRegex \ConcatDNF \DNFRegexAlt
      }
    \]

    Furthermore, as $\DNFRegex \ConcatDNF \DNFRegexAlt$ is strongly unambiguous,
    there exists a lens
    $\DNFLens \OfRewritelessType
    \DNFRegex \ConcatDNF \DNFRegexAlt \Leftrightarrow
    \DNFRegex \ConcatDNF \DNFRegexAlt$.
  \end{case}

  \begin{case}[\BaseRule{}]
    \[
      \inferrule*
      {
        \DNFRegex \ParallelRewrite \DNFRegex'
      }
      {
        \DNFRegex \StarOf{\ParallelRewrite} \DNFRegex'
      }
    \]

    So $\DNFRegex \ParallelRewrite \DNFRegex'$, and by \IdentityRewriteRule{},
    $\DNFRegexAlt \ParallelRewrite \DNFRegexAlt$.

    So Lemma~\ref{lem:parallel-rewrite-concatenation-to-identity} says that
    there exists $\DNFRegex''$ such that
    \[
      \inferrule*
      {
        \DNFRegex \ConcatDNF \DNFRegexAlt \ParallelRewrite \DNFRegex''
      }
      {
        \DNFRegex \ConcatDNF \DNFRegexAlt \StarOf{\ParallelRewrite} \DNFRegex''
      }
    \]
    where there exists a DNF lens
    $\DNFLens \OfRewritelessType \DNFRegex' \ConcatDNF \DNFRegexAlt
    \Leftrightarrow \DNFRegex''$ such that
    $\SemanticsOf{\DNFLens} = \SetOf{(\String,\String) \SuchThat \String \in
      \LanguageOf{\DNFRegex \ConcatDNF \DNFRegexAlt}}$.
  \end{case}

  \begin{case}[\TransitivityRule{}]
    \[
      \inferrule*
      {
        \DNFRegex \StarOf{\ParallelRewrite} \DNFRegex_1\\
        \DNFRegex_1 \StarOf{\ParallelRewrite} \DNFRegex'
      }
      {
        \DNFRegex \StarOf{\ParallelRewrite} \DNFRegex'
      }
    \]

    By IH, there exists $\DNFRegex_1''$ such that
    $\DNFRegex \ConcatDNF \DNFRegexAlt \StarOf{\ParallelRewrite} \DNFRegex_1''$,
    and there exists a DNF lens
    $\DNFLens_1 \OfRewritelessType \DNFRegex_1 \ConcatDNF \DNFRegexAlt
    \Leftrightarrow \DNFRegex_1''$, and
    $\SemanticsOf{\DNFLens_1} =
    \SetOf{(\String,\String) \SuchThat \String \in
      \LanguageOf{\DNFRegex \ConcatDNF \DNFRegexAlt}}$.

    By IH, there exists $\DNFRegex''$ such that
    $\DNFRegex_1 \ConcatDNF \DNFRegexAlt \StarOf{\ParallelRewrite} \DNFRegex''$,
    and there exists a DNF lens
    $\DNFLens \OfRewritelessType \DNFRegex' \ConcatDNF \DNFRegexAlt
    \Leftrightarrow \DNFRegex''$, and
    $\SemanticsOf{\DNFLens} =
    \SetOf{(\String,\String) \SuchThat \String \in
      \LanguageOf{\DNFRegex \ConcatDNF \DNFRegexAlt}}$.

    By Lemma~\ref{lem:bisimilarity-property-star-parallel-swapless}, as
    $\DNFRegex_1 \ConcatDNF \DNFRegexAlt \StarOf{\ParallelRewrite} \DNFRegex''$,
    then there exists $\DNFRegex_1'$, $\DNFLens_1'$ such that
    $\DNFRegex_1'' \StarOf{\ParallelRewrite} \DNFRegex_1'$, and
    $\DNFLens_1' \OfRewritelessType
    \DNFRegex'' \Leftrightarrow \DNFRegex'' \Leftrightarrow \DNFRegex_1'$, with
    the same semantics as $\DNFLens_1$.

    So $\DNFLens \OfRewritelessType \DNFRegex' \ConcatDNF \DNFRegexAlt
    \Leftrightarrow \DNFRegex''$, and
    $\DNFLens_1' \OfRewritelessType
    \DNFRegex'' \Leftrightarrow \DNFRegex'' \Leftrightarrow \DNFRegex_1'$.
    By Lemma~\ref{lem:composition-completeness}, there exists a DNF lens,
    $\DNFLens' \OfRewritelessType \DNFRegex' \ConcatDNF \DNFRegexAlt
    \Leftrightarrow \DNFRegex_1'$.  As both the lenses in the composition are
    the identity lens, this lens is the identity lens, so
    $\SemanticsOf{\DNFLens'} = \SetOf{(\String,\String) \SuchThat
    \String \in \LanguageOf{\DNFRegex \ConcatDNF \DNFRegexAlt}}$

    Furthermore
    \[
      \inferrule*
      {
        \DNFRegex \ConcatDNF \DNFRegexAlt \StarOf{\ParallelRewrite} \DNFRegex_1''\\
        \DNFRegex_1'' \StarOf{\ParallelRewrite} \DNFRegex_1'
      }
      {
        \DNFRegex \ConcatDNF \DNFRegexAlt \StarOf{\ParallelRewrite} \DNFRegex_1'
      }
    \]
  \end{case}
\end{proof}

\begin{lemma}[$\ParallelRewrite$ Maintained Under $\ConcatDNF$ up to
  $\IdentityLens$ on the right]
  \label{lem:star-parallel-rewrite-concatenation-to-identity-right}
  Let $\DNFRegex$ be strongly unambiguous.  Let $\DNFRegexAlt$ be strongly
  unambiguous.
  Let $\UnambigConcatOf{\LanguageOf{\DNFRegexAlt}}{\LanguageOf{\DNFRegex}}$.
  If $\DNFRegex \StarOf{\ParallelRewrite} \DNFRegex'$, then
  $\DNFRegexAlt \ConcatDNF \DNFRegex \StarOf{\ParallelRewrite} \DNFRegex''$ such that there
  exists a rewriteless DNF lens
  $\DNFLens \OfRewritelessType
  \DNFRegexAlt \ConcatDNF \DNFRegex' \Leftrightarrow \DNFRegex''$, and
  $\SemanticsOf{\DNFLens} =
  \SetOf{(\String,\String) \SuchThat \String \in
    \LanguageOf{\DNFRegex \ConcatDNF \DNFRegexAlt}}$.
\end{lemma}
\begin{proof}
  This is done symmetrically to
  Lemma~\ref{lem:star-parallel-rewrite-concatenation-to-identity-left}.
\end{proof}

\begin{lemma}[$\ParallelRewrite$ Maintained Under $\ConcatDNF$ up to
  $\IdentityLens$]
  \label{lem:star-parallel-rewrite-concatenation-to-identity}
  Let $\DNFRegex$ be strongly unambiguous.  Let $\DNFRegexAlt$ be strongly
  unambiguous.
  Let $\UnambigConcatOf{\LanguageOf{\DNFRegexAlt}}{\LanguageOf{\DNFRegex}}$.
  Let $\DNFRegex \StarOf{\ParallelRewrite} \DNFRegex'$.
  Let $\DNFRegexAlt \StarOf{\ParallelRewrite} \DNFRegexAlt'$.
  Then $\DNFRegex \ConcatDNF \DNFRegexAlt \StarOf{\ParallelRewrite} \DNFRegex''$ such
  that
  there exists a rewriteless DNF lens
  $\DNFLens \OfRewritelessType
  \DNFRegex' \ConcatDNF \DNFRegexAlt' \Leftrightarrow \DNFRegex''$, and
  $\SemanticsOf{\DNFLens} =
  \SetOf{(\String,\String) \SuchThat \String \in
    \LanguageOf{\DNFRegex \ConcatDNF \DNFRegexAlt}}$.
\end{lemma}
\begin{proof}
  By Lemma~\ref{lem:star-parallel-rewrite-concatenation-to-identity-right},
  there exists a DNF lens
  $\DNFLens_1 \OfRewritelessType \DNFRegex \ConcatDNF \DNFRegexAlt'
  \Leftrightarrow \DNFRegex_1$, such that
  $\DNFRegex \ConcatDNF \DNFRegexAlt \StarOf{\ParallelRewrite} \DNFRegex_1$ and
  $\SemanticsOf{\DNFLens_1} =
  \SetOf{(\String,\String) \SuchThat \String \in
    \LanguageOf{\DNFRegex \ConcatDNF \DNFRegexAlt}}$.
  
  By Lemma~\ref{lem:star-parallel-rewrite-concatenation-to-identity-left},
  there exists a DNF lens
  $\DNFLens_2 \OfRewritelessType \DNFRegex' \ConcatDNF \DNFRegexAlt'
  \Leftrightarrow \DNFRegex_2$, such that
  $\DNFRegex \ConcatDNF \DNFRegexAlt' \StarOf{\ParallelRewrite} \DNFRegex_2$ and
  $\SemanticsOf{\DNFLens_2} =
  \SetOf{(\String,\String) \SuchThat \String \in
    \LanguageOf{\DNFRegex \ConcatDNF \DNFRegexAlt}}$.
  
  By Lemma~\ref{lem:bisimilarity-property-star-parallel-swapless},
  as $\DNFRegex \ConcatDNF \DNFRegexAlt' \StarOf{\ParallelRewrite} \DNFRegex_2$,
  there exists a DNF lens
  $\DNFLens_1' \OfRewritelessType \DNFRegex_2 \Leftrightarrow \DNFRegex_1'$
  with $\SemanticsOf{\DNFLens_2} = \SetOf{(\String,\String) \SuchThat
    \String \in \LanguageOf{\DNFRegex \ConcatDNF \DNFRegexAlt}}$ and
  $\DNFRegex_1 \StarOf{\ParallelRewrite} \DNFRegex_1'$

  So $\DNFLens_2 \OfRewritelessType \DNFRegex' \ConcatDNF \DNFRegexAlt'
  \Leftrightarrow \DNFRegex_2$, and
  $\DNFLens_1' \OfRewritelessType
  \DNFRegex_2 \Leftrightarrow \DNFRegex_1'$.
  By Lemma~\ref{lem:composition-completeness}, there exists a DNF lens,
  $\DNFLens' \OfRewritelessType \DNFRegex' \ConcatDNF \DNFRegexAlt'
  \Leftrightarrow \DNFRegex_1'$.  As both the lenses in the composition are
  the identity lens, this lens is the identity lens, so
  $\SemanticsOf{\DNFLens'} = \SetOf{(\String,\String) \SuchThat
    \String \in \LanguageOf{\DNFRegex \ConcatDNF \DNFRegexAlt}}$
  
  Furthermore
  \[
    \inferrule*
    {
      \DNFRegex \ConcatDNF \DNFRegexAlt \StarOf{\ParallelRewrite} \DNFRegex_1\\
      \DNFRegex_1 \StarOf{\ParallelRewrite} \DNFRegex_1'
    }
    {
      \DNFRegex \ConcatDNF \DNFRegexAlt \StarOf{\ParallelRewrite} \DNFRegex_1'
    }
  \]
\end{proof}

\begin{corollary}[$\StarOf{\Rewrite}$ Maintained Under $\ConcatDNF$]
  \label{cor:star-rewrite-maintained-concat-to-identity}
  Let $\DNFRegex$ be strongly unambiguous.  Let $\DNFRegexAlt$ be strongly
  unambiguous.
  Let $\UnambigConcatOf{\LanguageOf{\DNFRegexAlt}}{\LanguageOf{\DNFRegex}}$.
  Let $\DNFRegex \StarOf{\Rewrite} \DNFRegex'$.
  Let $\DNFRegexAlt \StarOf{\Rewrite} \DNFRegexAlt'$.
  Then $\DNFRegex \ConcatDNF \DNFRegexAlt \StarOf{\Rewrite} \DNFRegex''$ such
  that
  there exists a rewriteless DNF lens
  $\DNFLens \OfRewritelessType
  \DNFRegex' \ConcatDNF \DNFRegexAlt' \Leftrightarrow \DNFRegex''$, and
  $\SemanticsOf{\DNFLens} =
  \SetOf{(\String,\String) \SuchThat \String \in
    \LanguageOf{\DNFRegex \ConcatDNF \DNFRegexAlt}}$.
\end{corollary}
\begin{proof}
  From Theorem~\ref{thm:parallel-star-equivalence} applied to
  Lemma~\ref{lem:star-parallel-rewrite-concatenation-to-identity}.
\end{proof}

\begin{lemma}[Pre-Confluence of Parallel Rewriting Without Reordering]
  \label{lem:pre-confluence}
  \begin{itemize}\leavevmode
  \item If $\DNFLens \OfRewritelessType \DNFRegex \Leftrightarrow \DNFRegexAlt$,
    $\DNFRegex \ParallelRewrite \DNFRegex'$, and $\DNFRegexAlt \ParallelRewrite
    \DNFRegexAlt'$, then
    \begin{enumerate}
      \item There exists a $\DNFRegex''$ such that $\DNFRegex' \ParallelRewrite
        \DNFRegex''$
      \item There exists a $\DNFRegexAlt''$ such that
        $\DNFRegexAlt' \ParallelRewrite \DNFRegexAlt''$
      \item There exists a $\DNFLens' \OfRewritelessType \DNFRegex''
        \Leftrightarrow \DNFRegex''$ such that $\SemanticsOf{\DNFLens'} =
        \SemanticsOf{\DNFLens}$.
    \end{enumerate}
  \item If $\AtomLens \OfRewritelessType \Atom \Leftrightarrow \AtomAlt$, $\Atom
    \ParallelRewriteAtom \DNFRegex$, and $\AtomAlt \ParallelRewriteAtom
    \DNFRegexAlt$, then
    \begin{enumerate}
    \item There exists a $\DNFRegex'$ such that $\DNFRegex \ParallelRewrite
      \DNFRegex'$
    \item There exists a $\DNFRegexAlt'$ such that
      $\DNFRegexAlt \ParallelRewrite \DNFRegexAlt'$
    \item There exists a $\DNFLens \OfRewritelessType \DNFRegex' \Leftrightarrow
      \DNFRegexAlt'$ such that $\SemanticsOf{\DNFLens} = \SemanticsOf{\AtomLens}$
    \end{enumerate}
  \end{itemize}
\end{lemma}
\begin{proof}
  By mutual induction on the derivation of $\ParallelRewrite$ and
  $\ParallelRewriteAtom$.  We will split into cases by the last step taken in
  each derivation.
  \begin{case}[\AtomUnrollstarLeftRule{},\AtomUnrollstarLeftRule{}]
    Let $\AtomLens \OfType \Atom \Leftrightarrow \AtomAlt$.
    Let $\Atom = \StarOf{\DNFRegex'}$ and $\StarOf{\DNFRegexAlt'}
    \ParallelRewriteAtom
    \DNFOf{\SequenceOf{\EmptyString}} \OrDNF (\DNFRegex'
    \ConcatDNF \DNFOf{\SequenceOf{\StarOf{\DNFRegex'}}})$
    through an application of
    \AtomUnrollstarLeftRule{}.
    Let $\AtomAlt = \StarOf{\DNFRegexAlt}$ and $\StarOf{\DNFRegexAlt}
    \ParallelRewriteAtom \DNFOf{\SequenceOf{\EmptyString}} \OrDNF (\DNFRegexAlt
    \ConcatDNF \DNFOf{\SequenceOf{\StarOf{\DNFRegexAlt}}})$ through an application of
    \AtomUnrollstarLeftRule{}.
    
    \begin{enumerate}
    \item Consider using \IdentityRewriteRule{}
      \[
        \inferrule*
        {
        }
        {
          \DNFOf{\SequenceOf{\EmptyString}} \OrDNF (\DNFRegex
          \ConcatDNF \DNFOf{\SequenceOf{\StarOf{\DNFRegex}}})
          \ParallelRewrite
          \DNFOf{\SequenceOf{\EmptyString}} \OrDNF (\DNFRegex
          \ConcatDNF \DNFOf{\SequenceOf{\StarOf{\DNFRegex}}})
        }
      \]
    \item Consider using \IdentityRewriteRule{}
      \[
        \inferrule*
        {
        }
        {
          \DNFOf{\SequenceOf{\EmptyString}} \OrDNF (\DNFRegexAlt
          \ConcatDNF \DNFOf{\SequenceOf{\StarOf{\DNFRegexAlt}}})
          \ParallelRewrite
          \DNFOf{\SequenceOf{\EmptyString}} \OrDNF (\DNFRegexAlt
          \ConcatDNF \DNFOf{\SequenceOf{\StarOf{\DNFRegexAlt}}})
        }
      \]
    \item 
      By inversion, $\AtomLens = \StarOf{\DNFLens}$, and $\DNFLens
      \OfRewritelessType \DNFRegex \Leftrightarrow \DNFRegexAlt$.
      
      By Lemma~\ref{lem:typ_sem_concat}, Lemma~\ref{lem:typ_sem_or}, and
      Lemma~\ref{lem:typ_sem_it}
      $\DNFLensOf{\SequenceLensOf{(\EmptyString,\EmptyString)}} \OrDNFLens
      (\DNFLens' \ConcatDNFLens \AtomToDNFLensOf{\IterateLensOf{\DNFLens'}})
      \OfRewritelessType \DNFOf{\SequenceOf{\EmptyString}} \OrDNF (\DNFRegex'
      \ConcatDNF \AtomToDNFOf{\StarOf{\DNFRegex'}}) \Leftrightarrow
      \DNFOf{\SequenceOf{\EmptyString}} \OrDNF (\DNFRegexAlt'
      \ConcatDNF \AtomToDNFOf{\StarOf{\DNFRegexAlt'}})$, which is the desired
      typing.

      By Lemma~\ref{lem:iterate-lens-unroll-left},
      $\SemanticsOf{\DNFLensOf{\SequenceLensOf{(\EmptyString,\EmptyString)}}
        \OrDNFLens
        (\DNFLens' \ConcatDNFLens \AtomToDNFLensOf{\IterateLensOf{\DNFLens'}})}
      = \SemanticsOf{\AtomLens}$, which is the desired semantics.
      
    \end{enumerate}
  \end{case}
  \begin{case}[\AtomUnrollstarLeftRule{},\AtomUnrollstarRightRule{}]
    Let $\AtomLens \OfType \Atom \Leftrightarrow \AtomAlt$.
    Let $\Atom = \StarOf{\DNFRegex'}$ and $\StarOf{\DNFRegexAlt'}
    \ParallelRewriteAtom
    \DNFOf{\SequenceOf{\EmptyString}} \OrDNF (\DNFRegex'
    \ConcatDNF \AtomToDNFOf{\StarOf{\DNFRegex'}})$
    through an application of
    \AtomUnrollstarLeftRule{}.
    Let $\AtomAlt = \StarOf{\DNFRegexAlt}$ and $\StarOf{\DNFRegexAlt}
    \ParallelRewriteAtom \DNFOf{\SequenceOf{\EmptyString}} \OrDNF
    (\AtomToDNFOf{\StarOf{\DNFRegexAlt}} \ConcatDNF \DNFRegexAlt)$
    through an application of \AtomUnrollstarRightRule{}.

    \begin{enumerate}
    \item
      \[
        \inferrule*
        {
          \StarOf{\DNFRegex'} \ParallelRewriteAtom
          \DNFOf{\SequenceOf{\EmptyString}} \OrDNF
          (\AtomToDNFOf{\StarOf{\DNFRegex'}}
          \ConcatDNF \DNFRegex')
        }
        {
          \AtomToDNFOf{\StarOf{\DNFRegex'}} \ParallelRewrite
          \DNFOf{\SequenceOf{\EmptyString}} \OrDNF
          (\AtomToDNFOf{\StarOf{\DNFRegex'}}
          \ConcatDNF \DNFRegex')
        }
      \]

      \[
        \inferrule*
        {
        }
        {
          \DNFRegex' \ParallelRewrite \DNFRegex'
        }
      \]

      By Lemma~\ref{lem:parallel-rewrite-concatenation-to-identity} there
      exists a $\DNFRegex_1$ such that
      $\DNFRegex' \ConcatDNF \AtomToDNFOf{\StarOf{\DNFRegex'}} \ParallelRewrite \DNFRegex_1$,
      and there exists $\DNFLens_1 \OfRewritelessType 
      \DNFRegex_1 \Leftrightarrow
      \DNFRegex' \ConcatDNF (\DNFOf{\SequenceOf{\EmptyString}} \OrDNF
      (\AtomToDNFOf{\StarOf{\DNFRegex'}}
      \ConcatDNF \DNFRegex'))$
      where $\DNFLens_1$ has identity semantics.

      \[
        \inferrule*
        {
        }
        {
          \DNFOf{\SequenceOf{\EmptyString}} \ParallelRewrite
          \DNFOf{\SequenceOf{\EmptyString}}
        }
      \]

      So by Lemma~\ref{lem:parallel-rewrite-or},
      $\DNFOf{\SequenceOf{\EmptyString}} \OrDNF
      (\DNFRegex' \ConcatDNF \AtomToDNFOf{\StarOf{\DNFRegex'}}) \ParallelRewrite
      \DNFOf{\SequenceOf{\EmptyString}} \OrDNF \DNFRegex_1$.
      
      Furthermore, as $\DNFLensOf{\SequenceLensOf{(\EmptyString,\EmptyString)}}
      \OfRewritelessType \DNFOf{\SequenceOf{\EmptyString}} \Leftrightarrow
      \DNFOf{\SequenceOf{\EmptyString}}$ has identity semantics, through
      Lemma~\ref{lem:parallel-rewrite-or} we get
      $\DNFLensOf{\SequenceLensOf{(\EmptyString,\EmptyString)}} \OrDNFLens
      \DNFLens_1 \OfRewritelessType \DNFOf{\SequenceOf{\EmptyString}}
      \OrDNF \DNFRegex_1 \Leftrightarrow
      \DNFOf{\SequenceOf{\EmptyString}} \OrDNF
      (\DNFRegex' \ConcatDNF (\DNFOf{\SequenceOf{\EmptyString}} \OrDNF
      (\AtomToDNFOf{\StarOf{\DNFRegex'}}
      \ConcatDNF \DNFRegex')))$, which has the identity semantics.
      
    \item
      \[
        \inferrule*
        {
          \StarOf{\DNFRegexAlt'} \ParallelRewriteAtom
          \DNFOf{\SequenceOf{\EmptyString}} \OrDNF
          (\DNFRegexAlt'
          \ConcatDNF \AtomToDNFOf{\StarOf{\DNFRegexAlt'}})
        }
        {
          \AtomToDNFOf{\StarOf{\DNFRegexAlt'}} \ParallelRewrite
          \DNFOf{\SequenceOf{\EmptyString}} \OrDNF
          (\DNFRegexAlt'
          \ConcatDNF \AtomToDNFOf{\StarOf{\DNFRegexAlt'}})
        }
      \]

      \[
        \inferrule*
        {
        }
        {
          \DNFRegexAlt' \ParallelRewrite \DNFRegexAlt'
        }
      \]

      By Lemma~\ref{lem:parallel-rewrite-concatenation-to-identity} there exists a $\DNFRegexAlt_2$ such that
      $\DNFRegexAlt' \ConcatDNF \AtomToDNFOf{\StarOf{\DNFRegexAlt'}}
      \ParallelRewrite \DNFRegexAlt_2$,
      and there exists $\DNFLens_2 \OfRewritelessType 
      (\DNFOf{\SequenceOf{\EmptyString}} \OrDNF
      (\DNFRegexAlt'
      \ConcatDNF \AtomToDNFOf{\StarOf{\DNFRegexAlt'}})) \ConcatDNF
      \DNFRegexAlt' \Leftrightarrow \DNFRegexAlt_2$
      where $\DNFLens_2$ has identity semantics.

      \[
        \inferrule*
        {
        }
        {
          \DNFOf{\SequenceOf{\EmptyString}} \ParallelRewrite
          \DNFOf{\SequenceOf{\EmptyString}}
        }
      \]

      So by Lemma~\ref{lem:parallel-rewrite-or},
      $\DNFOf{\SequenceOf{\EmptyString}} \OrDNF
      (\AtomToDNFOf{\StarOf{\DNFRegexAlt'}} \ConcatDNF \DNFRegexAlt') \ParallelRewrite
      \DNFOf{\SequenceOf{\EmptyString}} \OrDNF \DNFRegexAlt_2$.

      Furthermore, as $\DNFLensOf{\SequenceLensOf{(\EmptyString,\EmptyString)}}
      \OfRewritelessType \DNFOf{\SequenceOf{\EmptyString}} \Leftrightarrow
      \DNFOf{\SequenceOf{\EmptyString}}$ has identity semantics, through
      Lemma~\ref{lem:parallel-rewrite-or} we get
      $\DNFLensOf{\SequenceLensOf{(\EmptyString,\EmptyString)}} \OrDNFLens
      \DNFLens_2 \OfRewritelessType \DNFOf{\SequenceOf{\EmptyString}} \OrDNF
      ((\DNFOf{\SequenceOf{\EmptyString}} \OrDNF
      (\DNFRegexAlt'
      \ConcatDNF \AtomToDNFOf{\StarOf{\DNFRegexAlt'}})) \ConcatDNF
      \DNFRegexAlt') \Leftrightarrow
      \DNFOf{\SequenceOf{\EmptyString}}
      \OrDNF \DNFRegexAlt_2$, which has the identity semantics.
    \item
      As $\AtomLens \OfRewritelessType \StarOf{\DNFRegex'} \Leftrightarrow
      \StarOf{\DNFRegexAlt'}$, by inversion,
      $\AtomLens = \IterateLensOf{\DNFLens}$, and
      $\DNFLens \OfRewritelessType \DNFRegex' \Leftrightarrow \DNFRegexAlt'$.
      
      Let $\DNFLens' = \DNFLensOf{\SequenceLensOf{(\EmptyString,\EmptyString)}}
      \OrDNFLens (\DNFLens \ConcatDNFLens \AtomToDNFLensOf{\IterateLensOf{\DNFLens}})$
      By Lemma~\ref{lem:iterate-lens-unroll-right},
      $\SemanticsOf{\DNFLens'} = \SemanticsOf{\AtomLens}$, and
      $\DNFLens' \OfRewritelessType \DNFOf{\SequenceOf{\EmptyString}} \OrDNF
      (\DNFRegex' \ConcatDNF \AtomToDNFOf{\StarOf{\DNFRegex}}) \Leftrightarrow
      \DNFOf{\SequenceOf{\EmptyString}} \OrDNF
      (\DNFRegexAlt' \ConcatDNF \AtomToDNFOf{\StarOf{\DNFRegexAlt}})$.
      
      Let $\DNFLens'' = \DNFLensOf{\SequenceLensOf{(\EmptyString,\EmptyString)}}
      \OrDNFLens (\AtomToDNFLensOf{\IterateLensOf{\DNFLens}}
      \ConcatDNFLens \DNFLens)$
      By Lemma~\ref{lem:iterate-lens-unroll-right-dnf},
      $\SemanticsOf{\DNFLens''} =
      \SemanticsOf{\AtomToDNFLensOf{\AtomLens}}$, and
      $\DNFLens'' \OfRewritelessType \DNFOf{\SequenceOf{\EmptyString}} \OrDNF
      (\AtomToDNFOf{\StarOf{\DNFRegex}} \ConcatDNF \DNFRegex') \Leftrightarrow
      \DNFOf{\SequenceOf{\EmptyString}} \OrDNF
      (\AtomToDNFOf{\StarOf{\DNFRegexAlt}} \ConcatDNF \DNFRegexAlt')$.

      Consider the DNF lens
      $\DNFLens''' = \DNFLensOf{\SequenceLensOf{(\EmptyString,\EmptyString)}}
      \OrDNFLens (\DNFLens \ConcatDNFLens \DNFLens'')$.
      $\DNFLens''' \OfRewritelessType
      \DNFOf{\SequenceOf{\EmptyString}} \OrDNF
      (\DNFRegex' \ConcatDNF (\DNFOf{\SequenceOf{\EmptyString}} \OrDNF
      (\AtomToDNFOf{\StarOf{\DNFRegex}} \ConcatDNF \DNFRegex'))) \Leftrightarrow
      \DNFOf{\SequenceOf{\EmptyString}} \OrDNF
      (\DNFRegexAlt' \ConcatDNF (\DNFOf{\SequenceOf{\EmptyString}} \OrDNF
      (\AtomToDNFOf{\StarOf{\DNFRegexAlt}} \ConcatDNF \DNFRegexAlt')))$,
      where $\DNFLens'''$ has the same semantics as $\DNFLens'$, as
      $\DNFLens''$ has the same semantics as $\AtomToDNFLensOf{\AtomLens}$.

      By Lemma~\ref{lem:id-expressible-on-distribute-left}, there exists
      $\DNFLens_3 \OfRewritelessType \DNFOf{\SequenceOf{\EmptyString}} \OrDNF
      (\DNFRegexAlt' \ConcatDNF (\DNFOf{\SequenceOf{\EmptyString}} \OrDNF
      (\AtomToDNFOf{\StarOf{\DNFRegexAlt}} \ConcatDNF \DNFRegexAlt')))
      \Leftrightarrow
      \DNFOf{\SequenceOf{\EmptyString}} \OrDNF
      (\DNFRegexAlt' \ConcatDNF \DNFOf{\SequenceOf{\EmptyString}}) \OrDNF
      (\DNFRegexAlt' \ConcatDNF \AtomToDNFOf{\StarOf{\DNFRegexAlt}}
      \ConcatDNF \DNFRegexAlt')$.

      By Lemma~\ref{lem:id-expressible-on-factor-right}, there exists
      $\DNFLens_4 \OfRewritelessType \DNFOf{\SequenceOf{\EmptyString}} \OrDNF
      (\DNFOf{\SequenceOf{\EmptyString}} \ConcatDNF \DNFRegexAlt') \OrDNF
      (\DNFRegexAlt' \ConcatDNF \AtomToDNFOf{\StarOf{\DNFRegexAlt}}
      \ConcatDNF \DNFRegexAlt')
      \Leftrightarrow
      \DNFOf{\SequenceOf{\EmptyString}} \OrDNF
      ((\DNFOf{\SequenceOf{\EmptyString}} \OrDNF
      \DNFRegexAlt' \ConcatDNF \AtomToDNFOf{\StarOf{\DNFRegexAlt}})
      \ConcatDNF \DNFRegexAlt')$.

      Consider the composition of
      $\DNFLensOf{\SequenceLensOf{(\EmptyString,\EmptyString)}} \OrDNFLens \DNFLens_1$,
      $\DNFLens'''$, $\DNFLens_3$, $\DNFLens_4$, and
      $\DNFLensOf{\SequenceLensOf{(\EmptyString,\EmptyString)}} \OrDNFLens
      \DNFLens_2$

      Because of Lemma~\ref{lem:composition-completeness}, there exists a lens
      $\DNFLens_5 \OfRewritelessType \DNFOf{\SequenceOf{\EmptyString}}
      \OrDNF \DNFRegex_1 \Leftrightarrow \DNFOf{\SequenceOf{\EmptyString}}
      \OrDNF \DNFRegexAlt_2$.  Furthermore, as all lenses except
      $\DNFLens'''$ are the identity lens, $\SemanticsOf{\DNFLens_5} =
      \SemanticsOf{\DNFLens'''} = \SemanticsOf{\AtomLens}$.
    \end{enumerate}
  \end{case}
  \begin{case}[\AtomUnrollstarLeftRule{},\ParallelAtomStructuralRewriteRule{}]
    Let $\AtomLens \OfType \Atom \Leftrightarrow \AtomAlt$.
    Let $\Atom = \StarOf{\DNFRegex}$ and $\StarOf{\DNFRegexAlt}
    \ParallelRewriteAtom
    \DNFOf{\SequenceOf{\EmptyString}} \OrDNF (\DNFRegex'
    \ConcatDNF \AtomToDNFOf{\StarOf{\DNFRegex'}})$
    through an application of
    \AtomUnrollstarLeftRule{}.
    Let $\AtomAlt = \StarOf{\DNFRegexAlt}$ and
    \[
      \inferrule*
      {
        \DNFRegexAlt \ParallelRewrite \DNFRegexAlt'
      }
      {
        \StarOf{\DNFRegexAlt} \ParallelRewriteAtom \AtomToDNFOf{\DNFRegexAlt'}
      }
    \]
    through an application of \ParallelAtomStructuralRewriteRule{}.

    From inversion, $\AtomLens = \IterateLensOf{\DNFLens}$,
    and $\DNFLens \OfRewritelessType \DNFRegex \Leftrightarrow \DNFRegexAlt$.

    From Lemma~\ref{lem:bisimilarity-property-star-parallel-swapless}, there exists
    $DNFRegex'$ such that $\DNFRegex \ParallelRewrite \DNFRegex'$, such that
    there exists a lens $\DNFLens' \OfRewritelessType \DNFRegex'
    \Leftrightarrow \DNFRegexAlt'$, and
    $\SemanticsOf{\DNFLens'} = \SemanticsOf{\DNFLens}$

    \begin{enumerate}
    \item
      \[
        \inferrule*
        {
          \StarOf{\DNFRegex'} \ParallelRewriteAtom
          \DNFOf{\SequenceOf{\EmptyString}} \OrDNF
          (\DNFRegex' \ConcatDNF \AtomToDNFOf{\StarOf{\DNFRegex'}})
        }
        {
          \AtomToDNFOf{\StarOf{\DNFRegex'}} \ParallelRewrite
          \DNFOf{\SequenceOf{\EmptyString}} \OrDNF
          (\DNFRegex' \ConcatDNF \AtomToDNFOf{\StarOf{\DNFRegex'}})
        }
      \]
      
    \item
      \[
        \inferrule*
        {
          \inferrule*
          {
            \DNFRegexAlt \ParallelRewrite \DNFRegexAlt'
          }
          {
            \StarOf{\DNFRegexAlt} \ParallelRewriteAtom
            \AtomToDNFOf{\StarOf{\DNFRegexAlt'}}
          }
        }
        {
          \AtomToDNFOf{\StarOf{\DNFRegex}} \ParallelRewrite
          \AtomToDNFOf{\StarOf{\DNFRegex'}}
        }
      \]

      By Lemma~\ref{lem:parallel-rewrite-concatenation-to-identity} there exists a $\DNFRegexAlt_1$ such that
      $\DNFRegexAlt \ConcatDNF \AtomToDNFOf{\StarOf{\DNFRegexAlt}}
      \ParallelRewrite \DNFRegexAlt_1$,
      and there exists $\DNFLens_1 \OfRewritelessType 
      \DNFRegexAlt' \ConcatDNF \AtomToDNFOf{\StarOf{\DNFRegexAlt'}}
      \Leftrightarrow
      \DNFRegexAlt_1$
      where $\DNFLens_1$ has identity semantics.
      
      \[
        \inferrule*
        {
        }
        {
          \DNFOf{\SequenceOf{\EmptyString}} \ParallelRewrite
          \DNFOf{\SequenceOf{\EmptyString}}
        }
      \]

      So by Lemma~\ref{lem:parallel-rewrite-or},
      $\DNFOf{\SequenceOf{\EmptyString}}
      \OrDNF
      (\DNFRegexAlt \ConcatDNF \AtomToDNFOf{\StarOf{\DNFRegexAlt}})
      \ParallelRewrite
      \DNFOf{\SequenceOf{\EmptyString}} \OrDNF \DNFRegexAlt_1$.
      
      Furthermore, as $\DNFLensOf{\SequenceLensOf{(\EmptyString,\EmptyString)}}
      \OfRewritelessType \DNFOf{\SequenceOf{\EmptyString}} \Leftrightarrow
      \DNFOf{\SequenceOf{\EmptyString}}$ has identity semantics, through
      Lemma~\ref{lem:parallel-rewrite-or} we get
      $\DNFLensOf{\SequenceLensOf{(\EmptyString,\EmptyString)}} \OrDNFLens
      \DNFLens_1 \OfRewritelessType
      \DNFOf{\SequenceOf{\EmptyString}}
      \OrDNF (\DNFRegexAlt' \ConcatDNF \AtomToDNFOf{\StarOf{\DNFRegexAlt'}})
      \Leftrightarrow
      \DNFOf{\SequenceOf{\EmptyString}}
      \OrDNF \DNFRegexAlt_1$, which has the identity semantics.

    \item
      Let $\DNFLens'' = \DNFLensOf{\SequenceLensOf{(\EmptyString,\EmptyString)}}
      \OrDNFLens (\DNFLens' \ConcatDNFLens \AtomToDNFLensOf{\IterateLensOf{\DNFLens'}})$
      By Lemma~\ref{lem:iterate-lens-unroll-right},
      $\SemanticsOf{\DNFLens''} = \SemanticsOf{\IterateLensOf{\DNFLens'}} =
      \SemanticsOf{\AtomLens}$ and
      $\DNFLens'' \OfRewritelessType \DNFOf{\SequenceOf{\EmptyString}} \OrDNF
      (\DNFRegex' \ConcatDNF \AtomToDNFOf{\StarOf{\DNFRegex'}}) \Leftrightarrow
      \DNFOf{\SequenceOf{\EmptyString}} \OrDNF
      (\DNFRegexAlt' \ConcatDNF \AtomToDNFOf{\StarOf{\DNFRegexAlt'}})$.
      
      By Lemma~\ref{lem:composition-completeness}, we can compose lenses, so
      consider $\DNFLens'''$, the composition of the lenses
      $\DNFLens''$ and $\DNFLensOf{\SequenceLensOf{(\EmptyString,\EmptyString)}} \OrDNFLens
      \DNFLens_1$.
      $\DNFLens''' \OfRewritelessType \DNFOf{\SequenceOf{\EmptyString}} \OrDNF
      (\DNFRegex' \ConcatDNF \AtomToDNFOf{\StarOf{\DNFRegex'}}) \Leftrightarrow
      \DNFOf{\SequenceOf{\EmptyString}} \OrDNF \DNFRegexAlt_1$.
      Furthermore, as all lenses in the composition except $\DNFLens'''$ are the
      identity, $\SemanticsOf{\DNFLens'''} = \SemanticsOf{\DNFLens} =
      \SemanticsOf{\AtomLens}$.
    \end{enumerate}
  \end{case}

  \begin{case}[\AtomUnrollstarRightRule{},\AtomUnrollstarLeftRule{}]
    This is easily transformed into the case of
    (\AtomUnrollstarLeftRule{},\AtomUnrollstarRightRule{}), and the solution to
    that case transformed to a solution of this case, through two applications
    of Lemma~\ref{lem:closure-inversion}
  \end{case}

  \begin{case}[\AtomUnrollstarRightRule{},\AtomUnrollstarRightRule{}]
    This proceeds in the same way as
    (\AtomUnrollstarLeftRule{},\AtomUnrollstarLeftRule{}).
  \end{case}

  \begin{case}[\AtomUnrollstarRightRule{},\ParallelAtomStructuralRewriteRule{}]
    This proceeds in the same way as
    (\AtomUnrollstarLeftRule{},\ParallelAtomStructuralRewriteRule{})
  \end{case}

  \begin{case}[\ParallelAtomStructuralRewriteRule{},\AtomUnrollstarLeftRule{}]
    This is easily transformed into the case of
    (\AtomUnrollstarLeftRule{},\ParallelAtomStructuralRewriteRule{}), and the
    solution to
    that case transformed to a solution of this case, through two applications
    of Lemma~\ref{lem:closure-inversion}
  \end{case}

  \begin{case}[\ParallelAtomStructuralRewriteRule{},\AtomUnrollstarRightRule{}]
    This is easily transformed into the case of
    (\AtomUnrollstarRightRule{},\ParallelAtomStructuralRewriteRule{}), and the
    solution to
    that case transformed to a solution of this case, through two applications
    of Lemma~\ref{lem:closure-inversion}
  \end{case}

  \begin{case}[\ParallelAtomStructuralRewriteRule{},\ParallelAtomStructuralRewriteRule{}]
    Let $\AtomLens \OfType \Atom \Leftrightarrow \AtomAlt$.
    Let $\Atom = \StarOf{\DNFRegex}$ and
    \[
      \inferrule*
      {
        \DNFRegex \ParallelRewrite \DNFRegex'
      }
      {
        \StarOf{\DNFRegex} \ParallelRewrite \AtomToDNFOf{\DNFRegex'}
      }
    \]
    through an application of
    \AtomUnrollstarLeftRule{}.
    Let $\AtomAlt = \StarOf{\DNFRegexAlt}$ and
    \[
      \inferrule*
      {
        \DNFRegexAlt \ParallelRewrite \DNFRegexAlt'
      }
      {
        \StarOf{\DNFRegexAlt} \ParallelRewriteAtom \AtomToDNFOf{\DNFRegexAlt'}
      }
    \]
    through an application of \ParallelAtomStructuralRewriteRule{}.

    From inversion, $\AtomLens = \IterateLensOf{\DNFLens}$,
    and $\DNFLens \OfRewritelessType \DNFRegex \Leftrightarrow \DNFRegexAlt$.

    By IH, there exists $\DNFRegex''$, $\DNFRegexAlt''$, and $\DNFLens''$ such that
    $\DNFRegex' \ParallelRewrite \DNFRegex''$, $\DNFRegexAlt' \ParallelRewrite
    \DNFRegexAlt''$, and $\DNFLens'' \OfRewritelessType \DNFRegex''
    \Leftrightarrow \DNFRegexAlt''$ with $\SemanticsOf{\DNFLens''} =
    \SemanticsOf{\DNFLens}$.

    \begin{enumerate}
    \item
      \[
        \inferrule*
        {
          \inferrule*
          {
            \DNFRegex' \ParallelRewrite \DNFRegex''
          }
          {
            \StarOf{\DNFRegex'} \ParallelRewriteAtom
            \AtomToDNFOf{\StarOf{\DNFRegex''}}
          }
        }
        {
          \AtomToDNFOf{\StarOf{\DNFRegex'}} \ParallelRewrite
          \AtomToDNFOf{\StarOf{\DNFRegex''}}
        }
      \]
      
    \item
      \[
        \inferrule*
        {
          \inferrule*
          {
            \DNFRegexAlt' \ParallelRewrite \DNFRegexAlt''
          }
          {
            \StarOf{\DNFRegexAlt'} \ParallelRewriteAtom
            \AtomToDNFOf{\StarOf{\DNFRegexAlt''}}
          }
        }
        {
          \AtomToDNFOf{\StarOf{\DNFRegexAlt'}} \ParallelRewrite
          \AtomToDNFOf{\StarOf{\DNFRegexAlt''}}
        }
      \]

    \item
      As $\SemanticsOf{\DNFLens} = \SemanticsOf{\DNFLens''}$,
      $\SemanticsOf{\AtomLens} = \SemanticsOf{\IterateLensOf{\DNFLens}} =
      \SemanticsOf{\IterateLensOf{DNFLens''}}$.
      Furthermore, as $\LanguageOf{\DNFRegex''} = \LanguageOf{\DNFRegex}$ and
      $\LanguageOf{\DNFRegexAlt''} = \LanguageOf{\DNFRegex''}$,
      $\UnambigItOf{\DNFRegexAlt''}$ and $\UnambigItOf{\DNFRegex''}$.
      This means $\IterateLensOf{\DNFLens''} \OfRewritelessType
      \StarOf{\DNFRegex''} \Leftrightarrow \StarOf{\DNFRegexAlt''}$
      From Lemma~\ref{lem:typ_sem_todnflens},
      $\SemanticsOf{\IterateLensOf{\DNFLens''}} =
      \SemanticsOf{\AtomToDNFLensOf{\IterateLensOf{\DNFLens''}}}$, and\\
      $\AtomToDNFLensOf{\IterateLensOf{\IterateLensOf{\DNFLens''}}}
      \OfRewritelessType \AtomToDNFOf{\StarOf{\DNFRegex''}} \Leftrightarrow
      \AtomToDNFOf{\StarOf{\DNFRegexAlt''}}$.
    \end{enumerate}
  \end{case}

  \begin{case}[\IdentityRewriteRule{},\IdentityRewriteRule{}]
    Let $\DNFLens \OfType \DNFRegex \Leftrightarrow \DNFRegexAlt$.
    Let $\DNFRegex \ParallelRewrite \DNFRegex$ through an application of
    \AtomUnrollstarLeftRule{}.
    Let $\DNFRegexAlt \ParallelRewrite \DNFRegexAlt$ through an application of
    \AtomUnrollstarLeftRule{}.
    \begin{enumerate}
    \item
      \[
        \inferrule*
        {
        }
        {
          \DNFRegex \ParallelRewrite \DNFRegex
        }
      \]
    \item
      \[
        \inferrule*
        {
        }
        {
          \DNFRegexAlt \ParallelRewrite \DNFRegexAlt
        }
      \]
    \item
      $\DNFLens \OfRewritelessType \DNFRegex \Leftrightarrow \DNFRegexAlt$,
      and $\SemanticsOf{\DNFLens} = \SemanticsOf{\DNFLens}$.
    \end{enumerate}
  \end{case}
  
  \begin{case}[\IdentityRewriteRule{},\ParallelDNFStructuralRewriteRule{}]
    Let $\DNFLens \OfType \DNFRegex \Leftrightarrow \DNFRegexAlt$.
    Let $\DNFRegex \ParallelRewrite \DNFRegex$ through an application of
    \AtomUnrollstarLeftRule{}.
    Let $\DNFRegexAlt \ParallelRewrite \DNFRegexAlt'$ through an application of
    \ParallelDNFStructuralRewriteRule{}.

    By Lemma~\ref{lem:bisimilarity-property-parallel-swapless}, there exists
    $\DNFLens'$, $\DNFRegex'$ such that $\DNFRegex \ParallelRewrite \DNFRegex'$,
    $\DNFLens' \OfRewritelessType \DNFRegex' \Leftrightarrow \DNFRegexAlt'$,
    $\SemanticsOf{\DNFLens} = \SemanticsOf{\DNFLens'}$

    \begin{enumerate}
    \item $\DNFRegex \ParallelRewrite \DNFRegex'$
    \item
      \[
        \inferrule*
        {
        }
        {
          \DNFRegex' \ParallelRewrite \DNFRegex'
        }
      \]
    \item
      $\DNFLens' \OfRewritelessType \DNFRegex' \Leftrightarrow \DNFRegexAlt'$
      and
      $\SemanticsOf{\DNFLens} = \SemanticsOf{\DNFLens'}$
    \end{enumerate}
  \end{case}

  \begin{case}[\ParallelDNFStructuralRewriteRule{},\IdentityRewriteRule{}]
    This is easily transformed into the case of
    (\IdentityRewriteRule{},\ParallelDNFStructuralRewriteRule{}), and the
    solution to
    that case transformed to a solution of this case, through two applications
    of Lemma~\ref{lem:closure-inversion}
  \end{case}

  \begin{case}[\ParallelDNFStructuralRewriteRule{},\ParallelDNFStructuralRewriteRule{}]
    Let $\DNFLens \OfType \DNFRegex \Leftrightarrow \DNFRegexAlt$.
    Let $\DNFRegex \ParallelRewrite \DNFRegex'$ through an application of
    \ParallelDNFStructuralRewriteRule{}.
    Let $\DNFRegexAlt \ParallelRewrite \DNFRegexAlt'$ through an application of
    \ParallelDNFStructuralRewriteRule{}.

    By inversion,
    \[
      \inferrule*
      {
        \SequenceLens_1 \OfRewritelessType \Sequence_1 \Leftrightarrow \SequenceAlt_1\\
        \ldots\\
        \SequenceLens_n \OfRewritelessType \Sequence_n \Leftrightarrow \SequenceAlt_n\\\\
        \sigma \in \PermutationSetOf{n}\\
        i \neq j \Rightarrow \LanguageOf{\Sequence_{i}} \cap \LanguageOf{\Sequence_{j}}=\emptyset\\
        i \neq j \Rightarrow \LanguageOf{\SequenceAlt_{i}} \cap \LanguageOf{\SequenceAlt_{j}}=\emptyset\\
      }
      {
        (\DNFLensOf{\SequenceLens_1\DNFLSep\ldots\DNFLSep\SequenceLens_n},\sigma) \OfRewritelessType\\
        \DNFOf{\Sequence_1\DNFSep\ldots\DNFSep\Sequence_n}
        \Leftrightarrow \DNFOf{\SequenceAlt_{\sigma(1)}\DNFSep\ldots\DNFSep\SequenceAlt_{\sigma(n)}}
      }
    \]

    Also by inversion
    \[
      \inferrule*
      {
        \AtomLens_{i,1} \OfRewritelessType \Atom_1 \Leftrightarrow \AtomAlt_{i,1}\\
        \ldots\\
        \AtomLens_{i,n_i} \OfRewritelessType \Atom_n \Leftrightarrow \AtomAlt_{i,n_i}\\
        \sigma_i \in \PermutationSetOf{n_i}\\
        \UnambigConcat\SequenceOf{\String_{i,0}\SeqSep\Atom_{i,1}\SeqSep\ldots\SeqSep\Atom_{i,n_i}\SeqSep\String_{i,n_i}}\\
        \UnambigConcat\SequenceOf{\StringAlt_{i,0}\SeqSep\AtomAlt_{i,\sigma_i(1)}\SeqSep\ldots\SeqSep\AtomAlt_{i,\sigma_i(n_i)}\SeqSep\StringAlt_{i,n_i}}
      }
      {
        (\SequenceLensOf{(\String_{i,0},\StringAlt_{i,0})\SeqLSep\AtomLens_{i,1}\SeqLSep\ldots\SeqLSep\AtomLens_{i,n_i}\SeqLSep(\String_{i,n_i},\StringAlt_{i,n_i})},\sigma_i) \OfRewritelessType\\
        \SequenceOf{\String_{i,0}\SeqSep\Atom_{i,1}\SeqSep\ldots\SeqSep\Atom_{i,n_i}\SeqSep\String_{i,n_i}}\Leftrightarrow
        \SequenceOf{\StringAlt_{i,0}\SeqSep\AtomAlt_{i,\sigma_i(1)}\SeqSep\ldots\SeqSep\AtomAlt_{i,\sigma_i(n_i)}\SeqSep\StringAlt_{i,n_i}}
      }
    \]

    where $\DNFLens =
    \DNFLensOf{\SequenceLens_1\DNFLSep\ldots\DNFLSep\SequenceLens_n},\sigma)$,
    $\SequenceLens_i =
    (\SequenceLensOf{(\String_{i,0},\StringAlt_{i,0})\SeqLSep\AtomLens_{i,1}\SeqLSep\ldots\SeqLSep\AtomLens_{i,n_i}\SeqLSep(\String_{i,n_i},\StringAlt_{i,n_i})},\sigma_i)$,
    $\DNFRegex = \DNFOf{\Sequence_1\DNFSep\ldots\DNFSep\Sequence_n}$,
    $\DNFRegexAlt =
    \DNFOf{\SequenceAlt_{\sigma(1)}\DNFSep\ldots\DNFSep\SequenceAlt_{\sigma(n)}}$,
    $\Sequence_i =
    \SequenceOf{\String_{i,0}\SeqSep\Atom_{i,1}\SeqSep\ldots\SeqSep\Atom_{i,n_i}\SeqSep\String_{i,n_i}}$, and
    $\SequenceAlt_i =
    \SequenceOf{\StringAlt_{i,0}\SeqSep\AtomAlt_{i,\sigma_i(1)}\SeqSep\ldots\SeqSep\AtomAlt_{i,\sigma_i(n_i)}\SeqSep\StringAlt_{i,n_i}}$

    \[
      \inferrule*
      {
        \DNFRegex = \DNFOf{\Sequence_1 \DNFSep \ldots \DNFSep \Sequence_n}\\
        \forall i. \Sequence_i =
        \SequenceOf{\String_{i,0} \SeqSep \Atom_{i,1} \SeqSep \ldots \SeqSep \Atom_{i,n_i} \SeqSep \String_{i,n_i}}\\
        \forall i,j. \Atom_{i,j} \ParallelRewriteAtom \DNFRegex_{i,j}\\
        \forall i. \DNFRegex_i = \DNFOf{\SequenceOf{\String_{i,0}}} \ConcatDNF \DNFRegex_{i,1}
        \ConcatDNF \ldots \ConcatDNF \DNFRegex_{i,n_i} \ConcatDNF
        \DNFOf{\SequenceOf{\String_{i,n_i}}}
      }
      {
        \DNFRegex \ParallelRewrite \DNFRegex_1 \OrDNF \ldots \OrDNF \DNFRegex_n
      }
    \]

    \[
      \inferrule*
      {
        \DNFRegexAlt = \DNFOf{\SequenceAlt_{\sigma(1)} \DNFSep \ldots \DNFSep \SequenceAlt_{\sigma(n)}}\\
        \forall i. \SequenceAlt_{\sigma(i)} =
        \SequenceOf{\StringAlt_{\sigma(i),0} \SeqSep \AtomAlt_{\sigma(i),\sigma_i(1)} \SeqSep \ldots \SeqSep \AtomAlt_{\sigma(i),\sigma_i(n_i)} \SeqSep \StringAlt_{\sigma(i),n_i}}\\
        \forall i,j. \AtomAlt_{\sigma(i),\sigma_i(j)} \ParallelRewriteAtom \DNFRegexAlt_{\sigma(i),\sigma_i(j)}\\
        \forall i. \DNFRegexAlt_{\sigma(i)} = \DNFOf{\SequenceOf{\String_{\sigma(i),0}}} \ConcatDNF \DNFRegex_{\sigma(i),\sigma_i(1)}
        \ConcatDNF \ldots \ConcatDNF \DNFRegex_{\sigma(i),\sigma_i(n_i)} \ConcatDNF
        \DNFOf{\SequenceOf{\String_{\sigma(i),n_i}}}
      }
      {
        \DNFRegexAlt \ParallelRewrite \DNFRegexAlt_{\sigma(1)} \OrDNF \ldots \OrDNF \DNFRegexAlt_{\sigma(n)}
      }
    \]

    By IH, as $\Atom_{i,j} \ParallelRewrite \DNFRegex_{i,j}$,
    $\AtomAlt_{i,j} \ParallelRewrite \DNFRegexAlt_{i,j}$, and
    $\AtomLens_{i,j} \OfRewritelessType
    \Atom_{i,j} \Leftrightarrow \AtomAlt_{i,j}$, then there exists
    $\DNFRegex_{i,j}'$, $\DNFRegexAlt_{i,j}'$, and $\DNFLens_{i,j}$ such that
    $\DNFRegex_{i,j} \ParallelRewrite \DNFRegex_{i,j}'$,
    $\DNFRegexAlt_{i,j} \ParallelRewrite \DNFRegexAlt_{i,j}'$, and
    $\DNFLens_{i,j} \OfRewritelessType \DNFRegex_{i,j} \Leftrightarrow
    \DNFRegexAlt_{i,j}$, where $\SemanticsOf{\DNFLens_{i,j}} =
    \SemanticsOf{\AtomLens_{i,j}}$.
    \begin{enumerate}
    \item
      $\DNFRegex_{i,j} \ParallelRewrite \DNFRegex_{i,j}'$, for all $i,j$.
      \[
        \inferrule*
        {
        }
        {
          \DNFOf{\SequenceOf{\String_{i,j}}} \ParallelRewrite
          \DNFOf{\SequenceOf{\String_{i,j}}}
        }
      \]

      Define $\DNFRegex_i' =
      \DNFOf{\SequenceOf{\String_{i,0}}} \ConcatDNF \DNFRegex_{i,1}'
      \ConcatDNF \ldots \ConcatDNF \DNFRegex_{i,n_i}' \ConcatDNF
      \DNFOf{\SequenceOf{\String_{i,n_i}}}$.

      By repeated application of
      Lemma~\ref{lem:parallel-rewrite-concatenation-to-identity}, there exists $\DNFRegex_i''$
      such that $\DNFRegex_i =
      \DNFOf{\SequenceOf{\String_{i,0}}} \ConcatDNF \DNFRegex_{i,1}
      \ConcatDNF \ldots \ConcatDNF \DNFRegex_{i,n_i} \ConcatDNF
      \DNFOf{\SequenceOf{\String_{i,n_i}}} \ParallelRewrite \DNFRegex_i''$,
      and there exists $\DNFLens_i$ such that
      $\DNFLens_i \OfRewritelessType \DNFRegex_i'' \Leftrightarrow
      \DNFRegex_i'$, and $\DNFLens_i$ has the identity
      semantics on $\LanguageOf{\DNFRegex_i}$.

      By repeated application of Lemma~\ref{lem:parallel-rewrite-concatenation-to-identity},
      $\DNFRegex_1 \OrDNF \ldots \OrDNF \DNFRegex_n \ParallelRewrite
      \DNFRegex_1' \OrDNF \ldots \OrDNF \DNFRegex_n'$.
      Furthermore, through application of Lemma~\ref{lem:typ_sem_or},
      $\DNFLens_1 \OrDNFLens \ldots \OrDNFLens \DNFLens_n \OfRewritelessType
      \DNFRegex_1'' \OrDNF \ldots \OrDNF \DNFRegex_n'' \Leftrightarrow \DNFRegex_1'
      \OrDNF \ldots \OrDNF \DNFRegex_n'$.
      
    \item
      $\DNFRegexAlt_{\sigma(i),\sigma_i(j)} \ParallelRewrite
      \DNFRegexAlt_{\sigma(i),\sigma_i(j)}'$, for all $i,j$.
      \[
        \inferrule*
        {
        }
        {
          \DNFOf{\SequenceOf{\String_{\sigma(i),j}}} \ParallelRewrite
          \DNFOf{\SequenceOf{\String_{\sigma(i),j}}}
        }
      \]

      Define $\DNFRegexAlt_{\sigma(i)}' =
      \DNFOf{\SequenceOf{\String_{\sigma(i),0}}} \ConcatDNF
      \DNFRegex_{\sigma(i),\sigma_i(1)}'
      \ConcatDNF \ldots \ConcatDNF \DNFRegex_{\sigma(i),\sigma_i(n_i)}' \ConcatDNF
      \DNFOf{\SequenceOf{\String_{\sigma(i),\sigma_i(n_i)}}}$.
      
      By repeated application of Lemma~\ref{lem:parallel-rewrite-concatenation-to-identity}, there exists
      $\DNFRegexAlt_{\sigma(i)}''$ such that $\DNFRegexAlt_{\sigma(i)} =
      \DNFOf{\SequenceOf{\StringAlt_{\sigma(i),0}}} \ConcatDNF \DNFRegexAlt_{\sigma(i),\sigma_i(1)}
      \ConcatDNF \ldots \ConcatDNF \DNFRegexAlt_{\sigma(i),\sigma_i(n_i)} \ConcatDNF
      \DNFOf{\SequenceOf{\StringAlt_{\sigma(i),n_i}}} \ParallelRewrite \DNFRegexAlt_i''$,
      and there exists $\DNFLens_{\sigma(i)}'$ such that\\
      $\DNFLens_{\sigma(i)}' \OfRewritelessType
      \DNFRegexAlt_{\sigma(i)}' \Leftrightarrow
      \DNFRegexAlt_{\sigma(i)}''$, and $\DNFLens_{\sigma(i)}'$ has the identity
      semantics on $\LanguageOf{\DNFRegexAlt_{\sigma(i)}}$.
      
      By repeated application of Lemma~\ref{lem:parallel-rewrite-or},
      $\DNFRegexAlt_{\sigma(1)} \OrDNF \ldots \OrDNF \DNFRegexAlt_{\sigma(n)}
      \ParallelRewrite
      \DNFRegexAlt_{\sigma(1)}'' \OrDNF \ldots \OrDNF \DNFRegexAlt_{\sigma(n)}''$.
      Furthermore, through application of Lemma~\ref{lem:typ_sem_or},
      $\DNFLens_{\sigma(1)}' \OrDNFLens \ldots \OrDNFLens \DNFLens_{\sigma(n)}'
      \OfRewritelessType
      \DNFRegexAlt_{\sigma(1)}' \OrDNF \ldots \OrDNF \DNFRegexAlt_{\sigma(n)}'
      \Leftrightarrow
      \DNFRegexAlt_{\sigma(1)}'' \OrDNF \ldots \OrDNF
      \DNFRegexAlt_{\sigma(1)}''$.

    \item
      As $\LanguageOf{\Atom_{i,j}} = \LanguageOf{\DNFRegex_{i,j}'}$, and
      $\LanguageOf{\AtomAlt_{i,j}} = \LanguageOf{\DNFRegexAlt_{i,j}'}$, then
      $\SequenceUnambigConcatOf{\String_{i,0},
        \DNFRegex_{i,1}',\ldots,\DNFRegex_{i,n_i}',\String_{i,n_i}}$.
      and
      
      $\SequenceUnambigConcatOf{\StringAlt_{i,0},
        \DNFRegexAlt_{i,\sigma_i(1)}',\ldots,\DNFRegexAlt_{i,\sigma_i(n_i)}',
        \StringAlt_{i,n_i}}$.
      Then by Lemma~\ref{lem:conat-perms}, with the permutation $\sigma_i$,
      there exists a DNF lens
      $\overline{\DNFLens_i} \OfRewritelessType \DNFRegex_i' \Leftrightarrow
      \DNFRegexAlt_i'$,
      with\\ $\SemanticsOf{\overline{\DNFLens_i}} =
      \SetOf{(\String_{i,0}\String_{i,1}'\ldots\String_{i,n_i}'\String_{i,n_i},
        \StringAlt_{i,0}\StringAlt_{i,\sigma_i(1)}'\ldots\StringAlt_{i,\sigma_i(n_i)}'\StringAlt_{i,n_i})
        \SuchThat
        (\String_{i,j}',\StringAlt_{i,j}') \in \SemanticsOf{\DNFLens_{i,j}}}\\ \hspace*{2em}=
      \SetOf{(\String_{i,0}\String_{i,1}'\ldots\String_{i,n_i}'\String_{i,n_i},
        \StringAlt_{i,0}\StringAlt_{i,\sigma_i(1)}'\ldots\StringAlt_{i,\sigma_i(n_i)}'\StringAlt_{i,n_i})
        \SuchThat
        (\String_{i,j}',\StringAlt_{i,j}') \in \SemanticsOf{\AtomLens_{i,j}}}\\ \hspace*{2em}=
      \SemanticsOf{\SequenceLens_i}$.

      As $\LanguageOf{\DNFRegex_i'} = \LanguageOf{\Sequence_i}$,
      $i \neq j \BooleanImplies \LanguageOf{\DNFRegex_i} \Intersect
      \LanguageOf{\DNFRegex_j} = \SetOf{}$.
      By Lemma~\ref{lem:perm-lens-or}, with the permutation $\sigma$, there exists a
      DNF lens $\overline{\DNFLens} = \OfRewritelessType
      \DNFRegex_1' \OrDNF \ldots \OrDNF \DNFRegex_n'
      \Leftrightarrow
      \DNFRegexAlt_{\sigma(1)}' \OrDNF \ldots \OrDNF \DNFRegexAlt_{\sigma(n)}'$,
      with $\SemanticsOf{\overline{\DNFLens}} =
      \SetOf{(\String,\StringAlt) \SuchThat \exists i. (\String,\StringAlt) \in
        \SemanticsOf{\overline{\DNFLens_i}}} =
      \SetOf{(\String,\StringAlt) \SuchThat \exists i. (\String,\StringAlt) \in
        \SemanticsOf{\SequenceLens_i}} =
      \SemanticsOf{\DNFLens}$.

      Consider the $\DNFLens'$, the composition of
      $\DNFLens_1 \OrDNF \ldots \OrDNF \DNFLens_n$, $\overline{\DNFLens}$, and
      $\DNFLens_1' \OrDNF \ldots \OrDNF \DNFLens_n$,
      $\DNFLens' \OfRewritelessType \DNFRegex_1'' \OrDNF \ldots \OrDNF
      \DNFRegex_n'' \Leftrightarrow
      \DNFRegexAlt_1'' \OrDNF \ldots \OrDNF \DNFRegex_n''$.
      Furthermore, all but $\overline{\DNFLens}$ are identity,
      $\SemanticsOf{\DNFLens'''} = \SemanticsOf{\overline{\DNFLens}} =
      \SemanticsOf{\DNFLens}$.
    \end{enumerate}
  \end{case}
\end{proof}

\begin{theorem}[Confluence of Parallel Rewriting Without Reordering]
  \label{thm:parallel_confluence_noswap}
  For all lenses $\Lens \OfType \Regex \Leftrightarrow \RegexAlt$,
  $\IsConfluentWithPropertyOf{\ParallelRewrite}{\DNFLensHasSemanticsOf{\Lens}}$.
\end{theorem}
\begin{proof}
  Let $\DNFRegex \DNFLensHasSemanticsOf{\Lens} \DNFRegexAlt$.
  This means there exists some $\DNFLens \OfRewritelessType \DNFRegex
  \Leftrightarrow \DNFRegexAlt$ such that $\SemanticsOf{\DNFLens} =
  \SemanticsOf{\Lens}$.
  Let $\DNFRegex \ParallelRewrite \DNFRegex'$ and $\DNFRegexAlt \ParallelRewrite
  \DNFRegexAlt'$.  From Lemma~\ref{lem:pre-confluence}, there exists a
  $\DNFRegex''$, $\DNFRegexAlt''$, $\DNFLens'$ such that $\DNFRegex'
  \ParallelRewrite
  \DNFRegex''$, $\DNFRegexAlt' \ParallelRewrite \DNFRegexAlt''$, $\DNFLens'
  \OfType
  \DNFRegex'' \Leftrightarrow \DNFRegexAlt''$, and $\SemanticsOf{\DNFLens'} =
  \SemanticsOf{\DNFLens}$.  Because $\SemanticsOf{\DNFLens'} =
  \SemanticsOf{\DNFLens} = \SemanticsOf{\Lens}$, $\DNFRegex''
  \DNFLensHasSemanticsOf{\Lens} \DNFRegexAlt''$.
\end{proof}

\begin{lemma}[Identity is a left propagator]
  \label{lem:id-left-prop}
  If $\Lens \OfType \Regex \Leftrightarrow \RegexAlt$ is a lens,
  then $\DNFLensHasSemanticsOf{\IdentityLensOf{\Regex}}$ is a left propagator
  for $\DNFLensHasSemanticsOf{\Lens}$ with respect to $\ParallelRewrite$.
\end{lemma}
\begin{proof}
  If $\Lens \OfType \Regex \Leftrightarrow \RegexAlt$ is a lens,
  by Lemma~\ref{lem:strong-unambig-lens-types},
  $\Regex$ is strongly unambiguous.
  By Lemma~\ref{lem:retaining-unambiguity-todnf}, $\ToDNFRegexOf{\Regex}$ is
  strongly unambiguous.
  As such, $\IdentityLensOf{\Regex} \OfType \Regex \Leftrightarrow \Regex$.
  Consider $\DNFLensHasSemanticsOf{\IdentityLensOf{\Regex}}$.
  
  By Lemma~\ref{lem:bisimilarity-property-parallel-swapless},
  $\IsBisimilarWithPropertyOf{\ParallelRewrite}{\DNFLensHasSemanticsOf{\IdentityLensOf{\Regex}}}$.
  
  By Theorem~\ref{thm:parallel_confluence_noswap},
  $\IsConfluentWithPropertyOf{\ParallelRewrite}{\DNFLensHasSemanticsOf{\IdentityLensOf{\Regex}}}$.
  
  Let $\DNFRegex_1 \DNFLensHasSemanticsOf{\IdentityLensOf{\Regex}} \DNFRegex_2$,
  and $\DNFRegex_2 \DNFLensHasSemanticsOf{\IdentityLensOf{\Regex}} \DNFRegex_3$.
  So there exists $\DNFLens_1$, $\DNFLens_2$ such that
  $\DNFLens_1 \OfRewritelessType \DNFRegex_1 \Leftrightarrow \DNFRegex_2$ and
  $\DNFLens_2 \OfRewritelessType \DNFRegex_2 \Leftrightarrow \DNFRegex_3$,
  where $\SemanticsOf{\DNFLens_1} = \SemanticsOf{\IdentityLensOf{\Regex}} =
  \SemanticsOf{\DNFLens_2}$.
  By Lemma~\ref{lem:composition-completeness}, there exists
  $\DNFLens_3 \OfRewritelessType \DNFRegex_1 \Leftrightarrow \DNFRegex_3$, with
  semantics $\SemanticsOf{\DNFLens_3} = \SemanticsOf{\IdentityLensOf{\Regex}}$,
  as the semantics of each side of the composition was the identity relation
  on $\LanguageOf{\Regex}$.  This means
  $\DNFRegex_1 \DNFLensHasSemanticsOf{\IdentityLensOf{\Regex}} \DNFRegex_3$.

  Let $\DNFRegex \DNFLensHasSemanticsOf{\IdentityLensOf{\Regex}} \DNFRegexAlt$.
  So there exists $\DNFLens \OfRewritelessType \DNFRegex \Leftrightarrow
  \DNFRegexAlt$.  By Lemma~\ref{lem:lens-bij},
  $\SemanticsOf{\Lens}$ is a bijection between $\LanguageOf{\Regex}$ and
  $\LanguageOf{\RegexAlt}$.
  As $\SemanticsOf{\DNFLens} = \SemanticsOf{\Lens}$,
  $\SemanticsOf{\DNFLens}$ is a bijection between $\LanguageOf{\Regex}$ and
  $\LanguageOf{\RegexAlt}$.  $\SemanticsOf{\DNFLens}$ is a bijection between
  $\LanguageOf{\DNFRegex}$ and $\LanguageOf{\DNFRegexAlt}$, by Lemma~\ref{lem:rw-dnf-lens-bij}, so
  $\LanguageOf{\DNFRegex} = \LanguageOf{\Regex}$ and
  $\LanguageOf{\DNFRegexAlt} = \LanguageOf{\RegexAlt}$.
  As $\DNFRegex$ is strongly unambiguous, there exists an identity lens
  by Lemma~\ref{lem:strongly-unambiguous-identity-expressible}
  $\IdentityLens_L \OfRewritelessType \DNFRegex \Leftrightarrow \DNFRegex$,
  such that $\SemanticsOf{\IdentityLens_L} = \SetOf{(\String,\String) \SuchThat
    \String \in \LanguageOf{\DNFRegex}} = \SetOf{(\String,\String) \SuchThat
    \String \in \LanguageOf{\Regex}} = \SemanticsOf{\IdentityLensOf{\Regex}}$.
  This means $\DNFRegex \DNFLensHasSemanticsOf{\IdentityLensOf{\Regex}} \DNFRegex$.
  As $\DNFRegexAlt$ is strongly unambiguous, there exists an identity lens
  $\IdentityLens_R \OfRewritelessType \DNFRegexAlt \Leftrightarrow \DNFRegexAlt$,
  such that $\SemanticsOf{\IdentityLens_R} = \SetOf{(\String,\String) \SuchThat
    \String \in \LanguageOf{\DNFRegexAlt}} = \SetOf{(\String,\String) \SuchThat
    \String \in \LanguageOf{\RegexAlt}} = \SemanticsOf{\IdentityLensOf{\RegexAlt}}$.
  This means $\DNFRegexAlt \DNFLensHasSemanticsOf{\IdentityLensOf{\Regex}} \DNFRegexAlt$.

  Let $\DNFRegex \DNFLensHasSemanticsOf{\Lens} \DNFRegexAlt$.
  This means there exists $\DNFLens \OfRewritelessType \DNFRegex \Leftrightarrow
  \DNFRegexAlt$.
  This means that $\DNFRegex$ is strongly unambiguous, so there exists a DNF
  lens
  $\DNFLens' \OfRewritelessType \DNFRegex \Leftrightarrow \DNFRegex$,
  with $\SemanticsOf{\DNFLens'} = \SetOf{(\String,\String) \SuchThat \String \in
    \LanguageOf{\DNFRegex}}$.  By the same logic as above,
  $\LanguageOf{\DNFRegex} = \LanguageOf{\Regex}$, so $\SemanticsOf{\DNFLens'} =
  \SetOf{(\String,\String) \SuchThat \String \in \LanguageOf{\Regex}} =
  \SemanticsOf{\IdentityLensOf{\Regex}}$.
\end{proof}

\begin{lemma}[Identity is a right propagator]
  \label{lem:id-right-prop}
  If $\Lens \OfType \Regex \Leftrightarrow \RegexAlt$ is a lens,
  then $\DNFLensHasSemanticsOf{\IdentityLensOf{\RegexAlt}}$ is a right propagator
  for $\DNFLensHasSemanticsOf{\Lens}$ with respect to $\ParallelRewrite$.
\end{lemma}
\begin{proof}
  By a symmetric argument to Lemma~\ref{lem:id-left-prop}.
\end{proof}

\begin{lemma}[Confluence of Starred Parallel Rewriting Without Reordering]
  \label{thm:star_parallel_confluence}
  For all lenses $\Lens \OfType \Regex \Leftrightarrow \RegexAlt$,
  $\IsConfluentWithPropertyOf{\StarOf{\ParallelRewrite}}{\DNFLensHasSemanticsOf{\Lens}}$.
\end{lemma}
\begin{proof}
  By Lemma~\ref{lem:bisimilarity-property-parallel-swapless}
  For all lenses $\Lens \OfType \Regex \Leftrightarrow \RegexAlt$,
  $\IsBisimilarWithPropertyOf{\ParallelRewrite}{\DNFLensHasSemanticsOf{\Lens}}$.
  For all lenses $\Lens \OfType \Regex \Leftrightarrow \RegexAlt$,
  $\IsConfluentWithPropertyOf{\ParallelRewrite}{\DNFLensHasSemanticsOf{\Lens}}$.
  By Lemma~\ref{lem:id-left-prop},
  $\DNFLensHasSemanticsOf{\IdentityLensOf{\Regex}}$ is a left propagator for
  $\DNFLensHasSemanticsOf{\Lens}$.
  By Lemma~\ref{lem:id-right-prop},
  $\DNFLensHasSemanticsOf{\IdentityLensOf{\Regex}}$ is a right propagator for
  $\DNFLensHasSemanticsOf{\Lens}$.
  By Theorem~\ref{thm:starred-confluence},
  $\IsConfluentWithPropertyOf{\StarOf{\ParallelRewrite}}{\DNFLensHasSemanticsOf{\Lens}}$.
\end{proof}

\begin{corollary}
  \label{cor:rewrite-confluence}
  For all lenses $\Lens$,
  $\IsConfluentWithPropertyOf{\StarOf{\Rewrite}}{\DNFLensHasSemanticsOf{\Lens}}$
\end{corollary}
\begin{proof}
  By Theorem~\ref{thm:parallel_confluence_noswap}, and
  Theorem~\ref{thm:starred-confluence},
  for all lenses $\Lens$,
  $\IsConfluentWithPropertyOf
  {\StarOf{\ParallelRewrite}}{\DNFLensHasSemanticsOf{\Lens}}$.
  By Theorem~\ref{thm:parallel-star-equivalence},
  for all lenses $\Lens$,
  $\IsConfluentWithPropertyOf
  {\StarOf{\Rewrite}}{\DNFLensHasSemanticsOf{\Lens}}$.
\end{proof}

\subsection{Completeness}
Finally, with all the above machinery, all parts of confluence can be proven.
The final statement is a quick one, with the bulk of the work done by proving a
lemma involving rewrites and lens expressibility.

\label{completeness}

\begin{lemma}
  \label{lem:parallelswapequiv-has-identity-lens}
  Let $\DNFRegex$ be strongly unambiguous, and
  let $\DNFRegex \EquivalenceOf{\ParallelRewriteSwap} \DNFRegexAlt$.
  There exists $\DNFLens$, $\DNFRegex'$, $\DNFRegexAlt'$ such that
  $\DNFRegex \StarOf{\Rewrite} \DNFRegex'$,
  $\DNFRegexAlt' \StarOf{\Rewrite} \DNFRegexAlt'$,
  $\DNFLens \OfRewritelessType \DNFRegex' \Leftrightarrow \DNFRegexAlt'$,
  and $\SemanticsOf{\DNFLens} =
  \SetOf{(\String,\String) \SuchThat \String \in \LanguageOf{\DNFRegex}}$.
\end{lemma}
\begin{proof}
  Proof by induction on the typing of $\EquivalenceOf{\ParallelRewriteSwap}$
  \begin{case}[\ReflexivityRule]
    Let $\DNFRegex \EquivalenceOf{\ParallelRewriteSwap} \DNFRegexAlt$ through an
    application of \ReflexivityRule{}.
    That means $\DNFRegexAlt = \DNFRegex$.

    Consider $\DNFRegex \StarOf{\Rewrite} \DNFRegex$, and
    $\DNFRegexAlt \StarOf{\Rewrite} \DNFRegex$ through applications
    of \ReflexivityRule{}.

    Then, by Lemma~\ref{lem:strongly-unambiguous-identity-expressible}, there
    exists a lens $\DNFLens \OfRewritelessType \DNFRegex \Leftrightarrow
    \DNFRegex$ such that $\SemanticsOf{\DNFLens} = \SetOf{(\String,\String)
      \SuchThat \String \in \LanguageOf{\DNFRegex}}$
  \end{case}

  \begin{case}[\BaseRule]
    Let $\DNFRegex \EquivalenceOf{\ParallelRewriteSwap} \DNFRegexAlt$ through an
    application of \BaseRule{}.
    That means $\DNFRegex \ParallelRewriteSwap \DNFRegexAlt$.

    $\DNFRegexAlt \StarOf{\Rewrite} \DNFRegexAlt$ through an application
    of \ReflexivityRule{}.
    
    By Lemma~\ref{lem:swap-unimportance-identity}, there exists a DNF regular
    expression, $\DNFRegex'$, and a DNF lens $\DNFLens$,
    such that $\DNFRegex \Rewrite \DNFRegex'$,
    $\DNFLens \OfRewritelessType \DNFRegex' \Leftrightarrow \DNFRegexAlt$,
    and $\SemanticsOf{\DNFLens} =
    \SetOf{(\String,\String) \SuchThat \String \in \LanguageOf{\DNFRegex}}$.
    Through an application of \BaseRule{},
    $\DNFRegex \StarOf{\ParallelRewrite} \DNFRegex'$.
    From Theorem~\ref{thm:parallel-star-equivalence},
    $\DNFRegex \StarOf{\Rewrite} \DNFRegex'$, as desired.
  \end{case}

  \begin{case}[\SymmetryRule]
    Let $\DNFRegex \EquivalenceOf{\ParallelRewriteSwap} \DNFRegexAlt$ through an
    application of \SymmetryRule{}.
    That means $\DNFRegexAlt \EquivalenceOf{\ParallelRewriteSwap} \DNFRegex$.

    By IH, there exists DNF regular expressions $\DNFRegexAlt'$, $\DNFRegex'$,
    and a DNF lens $\DNFLens$ such that
    $\DNFRegexAlt \StarOf{\Rewrite} \DNFRegexAlt'$,
    $\DNFRegex \StarOf{\Rewrite} \DNFRegex'$,
    $\DNFLens \OfRewritelessType \DNFRegexAlt' \Leftrightarrow \DNFRegex'$,
    and $\SemanticsOf{\DNFRegexAlt'} = \SetOf{(\String,\String) \SuchThat
      \String \in \LanguageOf{\DNFRegexAlt}}$.

    Because $\EquivalenceOf{\ParallelRewriteSwap}$ is equivalent to
    $\SSREquiv$, $\LanguageOf{\DNFRegex} = \LanguageOf{\DNFRegexAlt}$

    By Lemma~\ref{lem:closure-inversion}, there exists
    $\DNFLens' \OfRewritelessType \DNFRegex' \Leftrightarrow \DNFRegexAlt'$,
    and $\SemanticsOf{\DNFLens'} = \SetOf{(\String,\String) \SuchThat
      \String \in \LanguageOf{\DNFRegex}}$, as desired.
  \end{case}

  \begin{case}[\TransitivityRule{}]
    Let $\DNFRegex \EquivalenceOf{\ParallelRewriteSwap} \DNFRegexAlt$ through an
    application of \TransitivityRule{}.
    That means there exists $\DNFRegex'$ such that
    $\DNFRegex \EquivalenceOf{\ParallelRewriteSwap} \DNFRegex'$ and
    $\DNFRegex' \EquivalenceOf{\ParallelRewriteSwap} \DNFRegexAlt$.

    By IH, there exists DNF regular expressions $\DNFRegex_1$, $\DNFRegex_2$,
    and a DNF lens $\DNFLens_1$ such that
    $\DNFRegex \StarOf{\Rewrite} \DNFRegex_1$,
    $\DNFRegex' \StarOf{\Rewrite} \DNFRegex_2$,
    and $\DNFLens_1 \OfRewritelessType \DNFRegex_1 \Leftrightarrow \DNFRegex_2$.

    By IH, there exists DNF regular expressions $\DNFRegex_3$, $\DNFRegex_4$,
    and a DNF lens $\DNFLens_2$ such that
    $\DNFRegex' \StarOf{\Rewrite} \DNFRegex_3$,
    $\DNFRegexAlt \StarOf{\Rewrite} \DNFRegex_4$,
    and $\DNFLens_2 \OfRewritelessType \DNFRegex_3 \Leftrightarrow \DNFRegex_4$.

    By Lemma~\ref{lem:strongly-unambiguous-identity-expressible}, there exists a
    lens $\DNFLens_{\Identity_1} \OfRewritelessType
    \DNFRegex' \Leftrightarrow \DNFRegex'$, where
    $\SemanticsOf{\DNFLens_{\Identity_1}} = \SetOf{(\String,\String) \SuchThat
      \String \in \LanguageOf{\DNFRegex'}}$.
    
    Because $\SatisfiesIdentitySemantics{\DNFRegex'}{\DNFRegex'}$, and
    by Corollary~\ref{cor:rewrite-confluence}, there exists $\DNFRegex_5$ and
    $\DNFRegex_6$, such that
    $\DNFRegex_2 \StarOf{\Rewrite} \DNFRegex_5$,
    $\DNFRegex_3 \StarOf{\Rewrite} \DNFRegex_6$, and
    $\SatisfiesIdentitySemantics{\DNFRegex_5}{\DNFRegex_6}$.
    That means there exists
    $\DNFLens_{\Identity_2} \OfRewritelessType
    \DNFRegex_5 \Leftrightarrow \DNFRegex_6$, where
    $\SemanticsOf{\DNFLens_{\Identity_2}} = \SetOf{(\String,\String) \SuchThat
      \String \in \LanguageOf{\DNFRegex'}}$.

    By Corollary~\ref{cor:bisimilarity-star-sequential},
    as $\DNFRegex_2 \StarOf{\Rewrite} \DNFRegex_5$, and
    $\DNFLens_1 \OfRewritelessType \DNFRegex_1 \Leftrightarrow \DNFRegex_2$.
    Because $\DNFRegex_2 \StarOf{\Rewrite} \DNFRegex_5$, there exists a
    DNF lens $\DNFLens_3$, and DNF regular expression $\DNFRegex_7$ such that
    $\DNFRegex_1 \StarOf{\Rewrite} \DNFRegex_7$, $\DNFLens_3 \OfRewritelessType
    \DNFRegex_7 \Leftrightarrow \DNFRegex_5$, and
    $\SemanticsOf{\DNFLens_3} =
    \SetOf{(\String,\String) \SuchThat \String \in \LanguageOf{\DNFRegex}}$.

    By Corollary~\ref{cor:bisimilarity-star-sequential},
    as $\DNFRegex_3 \StarOf{\Rewrite} \DNFRegex_6$, and
    $\DNFLens_2 \OfRewritelessType \DNFRegex_3 \Leftrightarrow \DNFRegex_4$.
    Because $\DNFRegex_3 \StarOf{\Rewrite} \DNFRegex_6$, there exists a
    DNF lens $\DNFLens_4$, and DNF regular expression $\DNFRegex_8$ such that
    $\DNFRegex_4 \StarOf{\Rewrite} \DNFRegex_8$, $\DNFLens_4 \OfRewritelessType
    \DNFRegex_6 \Leftrightarrow \DNFRegex_8$, and
    $\SemanticsOf{\DNFLens_3} =
    \SetOf{(\String,\String) \SuchThat \String \in \LanguageOf{\DNFRegex'}}$.

    So there are lenses
    $\DNFLens_3 \OfRewritelessType \DNFRegex_7 \Leftrightarrow \DNFRegex_5$,
    $\DNFLens_{\Identity_2} \OfRewritelessType \DNFRegex_5 \Leftrightarrow \DNFRegex_6$, and
    $\DNFLens_4 \OfRewritelessType \DNFRegex_6 \Leftrightarrow \DNFRegex_8$, so
    by Lemma~\ref{lem:composition-completeness}, there exists a lens
    $\DNFLens_5 \OfRewritelessType \DNFRegex_7 \Leftrightarrow \DNFRegex_8$.
    Because all of these have the semantics of the identity lens on DNF regular
    expressions with the same language,
    $\SemanticsOf{\DNFLens_5} =
    \SetOf{(\String,\String) \SuchThat \String \in \LanguageOf{\DNFRegex}}$.

    Furthermore, $\DNFRegex \StarOf{\Rewrite} \DNFRegex_1$ and
    $\DNFRegex_1 \StarOf{\Rewrite} \DNFRegex_7$, so
    $\DNFRegex \StarOf{\Rewrite} \DNFRegex_7$.
    $\DNFRegexAlt \StarOf{\Rewrite} \DNFRegex_4$ and
    $\DNFRegex_4 \StarOf{\Rewrite} \DNFRegex_8$, so
    $\DNFRegexAlt \StarOf{\Rewrite} \DNFRegex_8$, as desired.
  \end{case}
\end{proof}

\begin{lemma}
  \label{lem:identity-definitional-equivalence}
  Let $\Regex \SSREquiv \RegexAlt$, and let $\Regex$ be strongly
  unambiguous.  There exists $\DNFLens$, $\DNFRegex$, $\DNFRegexAlt$ such that
  $\DNFLens \OfRewritelessType \DNFRegex \Leftrightarrow \DNFRegexAlt$,
  $\ToDNFRegexOf{\Regex} \StarOf{\Rewrite} \DNFRegex$,
  $\ToDNFRegexOf{\RegexAlt} \StarOf{\Rewrite} \DNFRegexAlt$, and
  $\SemanticsOf{\DNFLens} =
  \SetOf{(\String,\String) \SuchThat \String\in\LanguageOf{\Regex}}$
\end{lemma}
\begin{proof}
  From Lemma~\ref{thm:defequiv-equiv-parallelswapequiv}, as 
  $\Regex \SSREquiv \RegexAlt$,
  $\ToDNFRegexOf{\Regex} \EquivalenceOf{\ParallelRewriteSwap}
  \ToDNFRegexOf{\RegexAlt}$.
  Because $\Regex$ is strongly unambiguous, by
  Lemma~\ref{lem:retaining-unambiguity-todnf}, $\ToDNFRegexOf{\Regex}$ is strongly
  unambiguous.
  Because of this, from Lemma~\ref{lem:parallelswapequiv-has-identity-lens},
  there exists $\DNFRegex$, $\DNFRegexAlt$, and $\DNFLens$ such that
  $\ToDNFRegexOf{\Regex} \StarOf{\Rewrite} \DNFRegex$,
  $\ToDNFRegexOf{\RegexAlt} \StarOf{\Rewrite} \DNFRegexAlt$, and
  $\SemanticsOf{\DNFLens} 
  \SetOf{(\String,\String) \SuchThat
    \String\in\LanguageOf{\ToDNFRegexOf{\Regex}}}$.
  From Theorem~\ref{thm:dnfrs}, $\LanguageOf{\ToDNFRegexOf{\Regex}} =
  \LanguageOf{\Regex}$, as desired.
\end{proof}

\begin{lemma}
  \label{lem:pre-completeness}
  If $\Lens \OfType \Regex \Leftrightarrow \RegexAlt$ then there exists
  $\DNFLens$, $\DNFRegex$, $\DNFRegexAlt$ such that $\DNFLens \OfRewritelessType
  \DNFRegex \Leftrightarrow \DNFRegexAlt$, $\ToDNFRegexOf{\Regex}
  \StarOf{\Rewrite} \DNFRegex$, $\ToDNFRegexOf{\RegexAlt} \StarOf{\Rewrite}
  \DNFRegexAlt$, and $\SemanticsOf{\DNFLens} = \SemanticsOf{\Lens}$
\end{lemma}
\begin{proof}
  By induction of the typing derivation of $\Lens \OfType \Regex \Leftrightarrow
  \RegexAlt$.

  Let the last typing rule be an instance of \IterateLensRule{}.
  \[
    \inferrule*
    {
      \Lens \OfType \Regex \Leftrightarrow \RegexAlt
    }
    {
      \IterateLensOf{\Lens} \OfType \StarOf{\Regex} \Leftrightarrow
      \StarOf{\RegexAlt}
    }
  \]

  By IH, there exists $\DNFLens$, $\DNFRegex$, $\DNFRegexAlt$
  such that
  \begin{gather*}
    \DNFLens \OfRewritelessType \DNFRegex \Leftrightarrow \DNFRegexAlt\\
    \ToDNFRegex{\Regex}\StarOf{\Rewrite}\DNFRegex\\
    \ToDNFRegex{\RegexAlt}\StarOf{\Rewrite}\DNFRegexAlt\\
    \SemanticsOf{\DNFLens} = \SemanticsOf{\Lens}
  \end{gather*}

  By Lemma~\ref{lem:typ_sem_it},
  $\DNFLensOf{\SequenceLensOf{\IterateLensOf{\DNFLens}}}
  \OfRewritelessType \DNFOf{\SequenceOf{\StarOf{\DNFRegex}}} \Leftrightarrow
  \DNFOf{\SequenceOf{\StarOf{\DNFRegexAlt}}}$.  By
  Corollary~\ref{cor:rewrite-maintained-iteration},
  $\DNFOf{\SequenceOf{\StarOf{\ToDNFRegexOf{\Regex}}}} \StarOf{\Rewrite}
  \DNFOf{\SequenceOf{\StarOf{\DNFRegex}}}$ and
  $\DNFOf{\SequenceOf{\StarOf{\ToDNFRegexOf{\RegexAlt}}}} \StarOf{\Rewrite}
  \DNFOf{\SequenceOf{\StarOf{\DNFRegexAlt}}}$.  From this, we get

  \[
    \inferrule*
    {
      \DNFLensOf{\SequenceLensOf{\IterateLensOf{\DNFLens}}}
      \OfRewritelessType \DNFOf{\SequenceOf{\StarOf{\DNFRegex}}}
      \Leftrightarrow \DNFOf{\SequenceOf{\StarOf{\DNFRegexAlt}}}\\
      \DNFOf{\SequenceOf{\StarOf{\ToDNFRegexOf{\Regex}}}} \StarOf{\Rewrite}
      \DNFOf{\SequenceOf{\StarOf{\DNFRegex}}}\\
      \DNFOf{\SequenceOf{\StarOf{\ToDNFRegexOf{\RegexAlt}}}} \StarOf{\Rewrite}
      \DNFOf{\SequenceOf{\StarOf{\DNFRegexAlt}}}
    }
    {
      \DNFLensOf{\SequenceLensOf{\IterateLensOf{\DNFLens}}} \OfType
      \ToDNFRegexOf{\StarOf{\Regex}} \Leftrightarrow
      \ToDNFRegexOf{\StarOf{\RegexAlt}}
    }
  \]

  By Lemma~\ref{lem:typ_sem_it},
  $\SemanticsOf{\DNFLensOf{\SequenceLensOf{\IterateLensOf{\DNFLens}}}} =
  \SetOf{(\String_1\Concat\ldots\Concat\String_n,\StringAlt_1\Concat\ldots\Concat\StringAlt_n)
    \SuchThat (\String_i,\StringAlt_i)\in\SemanticsOf{\DNFLens}} =
  \SetOf{(\String_1\Concat\ldots\Concat\String_n,\StringAlt_1\Concat\ldots\Concat\StringAlt_n)
    \SuchThat (\String_i,\StringAlt_i)\in\SemanticsOf{\Lens}} =
  \SemanticsOf{\IterateLensOf{\Lens}}$
  \\
  \\
  
  Let the last typing rule be an instance of \ConstantLensRule{}.

  \[
    \inferrule*
    {
    }
    {
      \ConstLensOf{\String_1}{\String_2}
    }
  \]

  Consider DNF Lens Derivation

  \[
    \inferrule*
    {
      \inferrule*
      {
        \SequenceLensOf{(\String_1,\String_2)} \OfRewritelessType
        \SequenceOf{\String_1} \Leftrightarrow
        \SequenceOf{\String_2}
      }
      {
        \DNFLensOf{\SequenceLensOf{(\String_1,\String_2)}} \OfRewritelessType
        \DNFOf{\SequenceOf{\String_1}} \Leftrightarrow
        \DNFOf{\SequenceOf{\String_2}}
      }\\
      \DNFOf{\SequenceOf{\String_1}} \StarOf{\Rewrite} \DNFOf{\SequenceOf{\String_1}}\\
      \DNFOf{\SequenceOf{\String_2}} \StarOf{\Rewrite} \DNFOf{\SequenceOf{\String_2}}
    }
    {
      \DNFLensOf{\SequenceLensOf{(\String_1,\String_2)}} \OfType
      \DNFOf{\SequenceOf{\String_1}} \Leftrightarrow \DNFOf{\SequenceOf{\String_2}}
    }
  \]

  $\SemanticsOf{\DNFLensOf{\SequenceLensOf{(\String_1,\String_2)}}} =
  \SetOf{(\String_1,\String_2)} =
  \SemanticsOf{\ConstLensOf{\String_1}{\String_2}}$
  \\
  \\

  Let the last typing rule be an instance of \ConcatLensRule{}.
  \[
    \inferrule*
    {
      \Lens_1 \OfType \Regex_1 \Leftrightarrow \RegexAlt_1\\
      \Lens_2 \OfType \Regex_2 \Leftrightarrow \RegexAlt_2
    }
    {
      \ConcatLensOf{\Lens_1}{\Lens_2} \OfType \RegexConcat{\Regex_1}{\Regex_2}
      \Leftrightarrow \RegexConcat{\RegexAlt_1}{\RegexAlt_2}
    }
  \]

  By IH, there exists $\DNFLens_1$, $\DNFRegex_1$,
  $\DNFRegexAlt_1$, $\DNFLens_2$, $\DNFRegex_2$, and $\DNFRegexAlt_2$ such that
  \begin{gather*}
    \DNFLens_1 \OfRewritelessType \DNFRegex_1 \Leftrightarrow \DNFRegexAlt_1 \\
    \ToDNFRegex{\Regex_1 }\StarOf {\Rewrite}\DNFRegex_1 \\
    \ToDNFRegex{\RegexAlt_1} \StarOf {\Rewrite}\DNFRegexAlt_1 \\
    \SemanticsOf{\DNFLens_1} = \SemanticsOf{\Lens_1} \\
    \DNFLens_2 \OfRewritelessType \DNFRegex_2 \Leftrightarrow \DNFRegexAlt_2 \\
    \ToDNFRegex{\Regex_2} \StarOf{\Rewrite} \DNFRegex_2 \\
    \ToDNFRegex{\RegexAlt_2} \StarOf{\Rewrite} \DNFRegexAlt_2 \\
    \SemanticsOf{\DNFLens_2} = \SemanticsOf{\Lens_2}
  \end{gather*}

  From Lemma~\ref{lem:typ_sem_concat}, $\ConcatDNFLensOf{\DNFLens_1}{\DNFLens_2}
  \OfRewritelessType
  \ConcatDNFOf{\DNFRegex_1}{\DNFRegex_2} \Leftrightarrow
  \ConcatDNFOf{\DNFRegexAlt_1}{\DNFRegexAlt_2}$.
  
  By Corollary~\ref{cor:star-rewrite-maintained-concat-to-identity}, and
  Lemma~\ref{lem:dnf-lens-inversion},
  there exists a DNF regular expression, $\DNFRegex_L$, and a DNF lens,
  $\DNFLens_L$, such that $\DNFLens_L \OfRewritelessType \DNFRegex_L
  \Leftrightarrow \ConcatDNFOf{\DNFRegex_1}{\DNFRegex_2}$, where
  $\SemanticsOf{\DNFLens_L} = \SetOf{(\String,\String) \SuchThat
    \String \in
    \LanguageOf{\ToDNFRegexOf{(\ConcatDNFOf{\DNFRegex_1}{\DNFRegex_2})}}}$.
  Furthermore, $\ToDNFRegexOf{(\ConcatDNFOf{\DNFRegex_1}{\DNFRegex_2})}
  \StarOf{\Rewrite} \DNFRegex_L$.
  
  By Corollary~\ref{cor:star-rewrite-maintained-concat-to-identity},
  there exists a DNF regular expression, $\DNFRegex_R$, and a DNF lens,
  $\DNFLens_R$, such that
  $\DNFLens_R \OfRewritelessType \ConcatDNFOf{\DNFRegexAlt_1}{\DNFRegexAlt_2}
  \Leftrightarrow \DNFRegex_R$, where
  $\SemanticsOf{\DNFLens_R} = \SetOf{(\String,\String) \SuchThat
    \String \in
    \LanguageOf{\ToDNFRegexOf{(\ConcatDNFOf{\DNFRegexAlt_1}{\DNFRegexAlt_2})}}}$
  Furthermore, $\ToDNFRegexOf{(\ConcatDNFOf{\DNFRegexAlt_1}{\DNFRegexAlt_2})}
  \StarOf{\Rewrite} \DNFRegex_R$.

  By Lemma~\ref{lem:composition-completeness}, as
  $\DNFLens_L \OfRewritelessType \DNFRegex_L
  \Leftrightarrow \ConcatDNFOf{\DNFRegex_1}{\DNFRegex_2}$,
  $\ConcatDNFLensOf{\DNFLens_1}{\DNFLens_2}
  \OfRewritelessType
  \ConcatDNFOf{\DNFRegex_1}{\DNFRegex_2} \Leftrightarrow
  \ConcatDNFOf{\DNFRegexAlt_1}{\DNFRegexAlt_2}$, and
  $\DNFLens_R \OfRewritelessType \ConcatDNFOf{\DNFRegexAlt_1}{\DNFRegexAlt_2}
  \Leftrightarrow \DNFRegex_R$
  there exists a DNF Lens
  $\DNFLens \OfRewritelessType \DNFRegex_L \Leftrightarrow \DNFRegex_R$, with
  semantics of the composition of the three lenses.  Because the left and right
  lenses are the identity lenses, $\SemanticsOf{\DNFLens} =
  \SemanticsOf{\ConcatDNFLensOf{\DNFLens_1}{\DNFLens_2}}$.

  By Lemma~\ref{lem:typ_sem_concat},
  $\SemanticsOf{\DNFLens} = \SemanticsOf{\ConcatDNFLensOf{\DNFLens_1}{\DNFLens_2}} =
  \SetOf{(\String_1\Concat\String_2,\StringAlt_1\Concat\StringAlt_2) \SuchThat
    (\String_1,\StringAlt_1)\in\SemanticsOf{\DNFLens_1} \BooleanAnd
    (\String_2,\StringAlt_2)\in\SemanticsOf{\DNFLens_2}} =
  \SetOf{(\String_1\Concat\String_2,\StringAlt_1\Concat\StringAlt_2) \SuchThat
    (\String_1,\StringAlt_1)\in\SemanticsOf{\Lens_1} \BooleanAnd
    (\String_2,\StringAlt_2)\in\SemanticsOf{\Lens_2}} =
  \SemanticsOf{\ConcatLensOf{\Lens_1}{\Lens_2}}$.

  \begin{gather*}
    \DNFLens \OfRewritelessType \DNFRegex_L \Leftrightarrow \DNFRegex_R\\
    \ToDNFRegexOf{(\Regex_1 \Concat \Regex_2)} \StarOf{\Rewrite}
    \DNFRegex_L\\
    \ToDNFRegexOf{(\RegexAlt_1 \Concat \RegexAlt_2)} \Leftrightarrow
    \DNFRegex_L\\
    \SemanticsOf{\DNFLens} = \SemanticsOf{\ConcatLensOf{\Lens_1}{\Lens_2}}
  \end{gather*}
  \\
  \\
  Let the last typing rule be an instance of \OrLensRule{}.
  \[
    \inferrule*
    {
      \Lens_1 \OfType \Regex_1 \Leftrightarrow \RegexAlt_1\\
      \Lens_2 \OfType \Regex_2 \Leftrightarrow \RegexAlt_2\\
    }
    {
      \OrLensOf{\Lens_1}{\Lens_2} \OfType \RegexOr{\Regex_1}{\Regex_2}
      \Leftrightarrow \RegexOr{\RegexAlt_1}{\RegexAlt_2}
    }
  \]
  
  By IH, there exists $\DNFLens_1$, $\DNFRegex_1$,
  $\DNFRegexAlt_1$, $\DNFLens_2$, $\DNFRegex_2$, and $\DNFRegexAlt_2$ such that
  \begin{gather*}
    \DNFLens_1 \OfRewritelessType \DNFRegex_1 \Leftrightarrow \DNFRegexAlt_1 \\
    \ToDNFRegex{\Regex_1 }\StarOf {\Rewrite}\DNFRegex_1 \\
    \ToDNFRegex{\RegexAlt_1} \StarOf {\Rewrite}\DNFRegexAlt_1 \\
    \SemanticsOf{\DNFLens_1} = \SemanticsOf{\Lens_1} \\
    \DNFLens_2 \OfRewritelessType \DNFRegex_2 \Leftrightarrow \DNFRegexAlt_2 \\
    \ToDNFRegex{\Regex_2} \StarOf{\Rewrite} \DNFRegex_2 \\
    \ToDNFRegex{\RegexAlt_2} \StarOf{\Rewrite} \DNFRegexAlt_2 \\
    \SemanticsOf{\DNFLens_2} = \SemanticsOf{\Lens_2}
  \end{gather*}

  From Lemma~\ref{lem:typ_sem_or}, $\OrDNFLensOf{\DNFLens_1}{\DNFLens_2}
  \OfRewritelessType
  \OrDNFOf{\DNFRegex_1}{\DNFRegex_2} \Leftrightarrow
  \OrDNFOf{\DNFRegexAlt_1}{\DNFRegexAlt_2}$.
  By Lemma~\ref{lem:propagation-of-star-rewrites-through-or},
  $\ToDNFRegexOf{\RegexOr{\Regex_1}{\Regex_2}} \StarOf{\Rewrite}
  \OrDNFOf{\DNFRegex_1}{\DNFRegex_2}$ and
  $\ToDNFRegexOf{\RegexOr{\RegexAlt_1}{\RegexAlt_2}} \StarOf{\Rewrite}
  \OrDNFOf{\DNFRegexAlt_1}{\DNFRegexAlt_2}$.

  By Lemma~\ref{lem:typ_sem_or},
  $\SemanticsOf{\OrDNFLensOf{\DNFLens_1}{\DNFLens_2}} =
  \SetOf{(\String,\StringAlt) \SuchThat
    (\String,\StringAlt)\in\SemanticsOf{\DNFLens_1} \BooleanOr
    (\String,\StringAlt)\in\SemanticsOf{\DNFLens_2}} =
  \SetOf{(\String,\StringAlt) \SuchThat
    (\String,\StringAlt)\in\SemanticsOf{\Lens_1} \BooleanOr
    (\String,\StringAlt)\in\SemanticsOf{\Lens_2}} =
  \SemanticsOf{\OrLensOf{\Lens_1}{\Lens_2}}$.

  \begin{gather*}
    \OrDNFLensOf{\Lens_1}{\Lens_2} \OfRewritelessType \DNFRegex_1 \OrDNF
    \DNFRegex_2 \Leftrightarrow \DNFRegexAlt_1 \OrDNF \DNFRegexAlt_2\\
    \ToDNFRegexOf{(\Regex_1 \Or \Regex_2)} \StarOf{\Rewrite}
    \DNFRegex_1 \OrDNF \DNFRegex_2\\
    \ToDNFRegexOf{(\RegexAlt_1 \Or \RegexAlt_2)} \StarOf{\Rewrite}
    \DNFRegexAlt_1 \OrDNF \DNFRegexAlt_2\\
    \SemanticsOf{\OrDNFLensOf{\DNFLens_1}{\DNFLens_2}} =
    \SemanticsOf{\OrLensOf{\Lens_1}{\Lens_2}}
  \end{gather*}
  \\
  \\
  Let the last typing rule be an instance of \SwapLensRule{}.
  \[
    \inferrule*
    {
      \Lens_1 \OfType \Regex_1 \Leftrightarrow \RegexAlt_1\\
      \Lens_2 \OfType \Regex_2 \Leftrightarrow \RegexAlt_2\\
    }
    {
      \SwapLensOf{\Lens_1}{\Lens_2} \OfType \RegexConcat{\Regex_1}{\Regex_2}
      \Leftrightarrow \RegexConcat{\RegexAlt_2}{\RegexAlt_1}
    }
  \]

  By IH, there exists $\DNFLens_1$, $\DNFRegex_1$,
  $\DNFRegexAlt_1$, $\DNFLens_2$, $\DNFRegex_2$, and $\DNFRegexAlt_2$ such that.
  \begin{gather*}
    \DNFLens_1 \OfRewritelessType \DNFRegex_1 \Leftrightarrow \DNFRegexAlt_1 \\
    \ToDNFRegex{\Regex_1 }\StarOf {\Rewrite}\DNFRegex_1 \\
    \ToDNFRegex{\RegexAlt_1} \StarOf {\Rewrite}\DNFRegexAlt_1 \\
    \SemanticsOf{\DNFLens_1} = \SemanticsOf{\Lens_1} \\
    \DNFLens_2 \OfRewritelessType \DNFRegex_2 \Leftrightarrow \DNFRegexAlt_2 \\
    \ToDNFRegex{\Regex_2} \StarOf{\Rewrite} \DNFRegex_2 \\
    \ToDNFRegex{\RegexAlt_2} \StarOf{\Rewrite} \DNFRegexAlt_2 \\
    \SemanticsOf{\DNFLens_2} = \SemanticsOf{\Lens_2}
  \end{gather*}

  From Lemma~\ref{lem:typ_sem_swap}, $\SwapDNFLensOf{\DNFLens_1}{\DNFLens_2}
  \OfRewritelessType
  \ConcatDNFOf{\DNFRegex_1}{\DNFRegex_2} \Leftrightarrow
  \ConcatDNFOf{\DNFRegexAlt_2}{\DNFRegexAlt_1}$.
  
  By Corollary~\ref{cor:star-rewrite-maintained-concat-to-identity}, and
  Lemma~\ref{lem:dnf-lens-inversion},
  there exists a DNF regular expression, $\DNFRegex_L$, and a DNF lens,
  $\DNFLens_L$, such that $\DNFLens_L \OfRewritelessType \DNFRegex_L
  \Leftrightarrow \ConcatDNFOf{\DNFRegex_1}{\DNFRegex_2}$, where
  $\SemanticsOf{\DNFLens_L} = \SetOf{(\String,\String) \SuchThat
    \String \in
    \LanguageOf{\ToDNFRegexOf{(\ConcatDNFOf{\DNFRegex_1}{\DNFRegex_2})}}}$.
  Furthermore, $\ToDNFRegexOf{(\ConcatDNFOf{\DNFRegex_1}{\DNFRegex_2})}
  \StarOf{\Rewrite} \DNFRegex_L$.
  
  By Corollary~\ref{cor:star-rewrite-maintained-concat-to-identity},
  there exists a DNF regular expression, $\DNFRegex_R$, and a DNF lens,
  $\DNFLens_R$, such that
  $\DNFLens_R \OfRewritelessType \ConcatDNFOf{\DNFRegexAlt_2}{\DNFRegexAlt_1}
  \Leftrightarrow \DNFRegex_R$, where
  $\SemanticsOf{\DNFLens_R} = \SetOf{(\String,\String) \SuchThat
    \String \in
    \LanguageOf{\ToDNFRegexOf{(\ConcatDNFOf{\DNFRegexAlt_2}{\DNFRegexAlt_1})}}}$
  Furthermore, $\ToDNFRegexOf{(\ConcatDNFOf{\DNFRegexAlt_2}{\DNFRegexAlt_1})}
  \StarOf{\Rewrite} \DNFRegex_R$.

  By Lemma~\ref{lem:composition-completeness}, as
  $\DNFLens_L \OfRewritelessType \DNFRegex_L
  \Leftrightarrow \ConcatDNFOf{\DNFRegex_1}{\DNFRegex_2}$,
  $\SwapDNFLensOf{\DNFLens_1}{\DNFLens_2}
  \OfRewritelessType
  \ConcatDNFOf{\DNFRegex_1}{\DNFRegex_2} \Leftrightarrow
  \ConcatDNFOf{\DNFRegexAlt_2}{\DNFRegexAlt_1}$, and
  $\DNFLens_R \OfRewritelessType \ConcatDNFOf{\DNFRegexAlt_2}{\DNFRegexAlt_1}
  \Leftrightarrow \DNFRegex_R$
  there exists a DNF Lens
  $\DNFLens \OfRewritelessType \DNFRegex_L \Leftrightarrow \DNFRegex_R$, with
  semantics of the composition of the three lenses.  Because the left and right
  lenses are the identity lenses, $\SemanticsOf{\DNFLens} =
  \SemanticsOf{\SwapDNFLensOf{\DNFLens_1}{\DNFLens_2}}$.
  
  By Lemma~\ref{lem:typ_sem_swap},
  $\SemanticsOf{\DNFLens} = \SemanticsOf{\SwapDNFLensOf{\DNFLens_1}{\DNFLens_2}} =
  \SetOf{(\String_1\Concat\String_2,\StringAlt_2\Concat\StringAlt_1) \SuchThat
    (\String_1,\StringAlt_1)\in\SemanticsOf{\DNFLens_1} \BooleanAnd
    (\String_2,\StringAlt_2)\in\SemanticsOf{\DNFLens_2}} =
  \SetOf{(\String_1\Concat\String_2,\StringAlt_2\Concat\StringAlt_1) \SuchThat
    (\String_1,\StringAlt_1)\in\SemanticsOf{\Lens_1} \BooleanAnd
    (\String_2,\StringAlt_2)\in\SemanticsOf{\Lens_2}} =
  \SemanticsOf{\SwapLensOf{\Lens_1}{\Lens_2}}$.

  \begin{gather*}
    \DNFLens \OfRewritelessType \DNFRegex_L \Leftrightarrow \DNFRegex_R\\
    \ToDNFRegexOf{(\RegexConcat{\Regex_1}{\Regex_2})} \StarOf{\Rewrite}
    \DNFRegex_L\\
    \ToDNFRegexOf{(\RegexConcat{\RegexAlt_2}{\RegexAlt_1})} \StarOf{\Rewrite}
    \DNFRegex_R\\
    \SemanticsOf{\DNFLens} = \SemanticsOf{\SwapLensOf{\Lens_1}{\Lens_2}}
  \end{gather*}
  \\
  \\

  Let the last rule be an instance of \ComposeLensRule{}.
  \[
    \inferrule*
    {
      \Lens_1 \OfType \Regex_1 \Leftrightarrow \Regex_2\\
      \Lens_2 \OfType \Regex_2 \Leftrightarrow \Regex_3\\
    }
    {
      \ComposeLensOf{\Lens_2}{\Lens_1} \OfType \Regex_1 \Leftrightarrow \Regex_3
    }
  \]

  By induction assumption, there exists $\DNFLens_1$, $\DNFRegex_1$,
  $\DNFRegex_2$, $\DNFLens_2$, $\DNFRegexAlt_2$, and $\DNFRegex_3$ such that

  \begin{gather*}
    \DNFLens_1 \OfRewritelessType \DNFRegex_1 \Leftrightarrow \DNFRegex_2 \\
    \ToDNFRegexOf{\Regex_1 }\StarOf {\Rewrite}\DNFRegex_1 \\
    \ToDNFRegexOf{\Regex_2} \StarOf {\Rewrite}\DNFRegex_2 \\
    \SemanticsOf{\DNFLens_1} = \SemanticsOf{\Lens_1} \\
    \DNFLens_2 \OfRewritelessType \DNFRegex_2 \Leftrightarrow \DNFRegex_3 \\
    \ToDNFRegexOf{\Regex_2} \StarOf{\Rewrite} \DNFRegex_2' \\
    \ToDNFRegexOf{\Regex_3} \StarOf{\Rewrite} \DNFRegex_3 \\
    \SemanticsOf{\DNFLens_2} = \SemanticsOf{\Lens_2}
  \end{gather*}

  From Lemma~\ref{lem:strongly-unambiguous-identity-expressible}, there exists
  a rewriteless dnf lens
  $\DNFLens_{\IdentityLens} \OfRewritelessType \ToDNFRegex{\Regex_2} \Leftrightarrow
  \ToDNFRegex{\Regex_2}$ where $\SemanticsOf{\DNFLens_{\IdentityLens}} =
  \SetOf{(\String,\String)\SuchThat \String\in\LanguageOf{\Regex_2}}$.
  From Corollary~\ref{cor:rewrite-confluence}, we know that, as
  $\ToDNFRegex{\Regex_2}\StarOf{\Rewrite}\DNFRegex_2$ and as
  $\ToDNFRegex{\Regex_2}\StarOf{\Rewrite}\DNFRegexAlt_2$, there must exist some
  $\DNFRegex_2'$, $\DNFRegexAlt_2'$ such that
  $\DNFRegex_2\StarOf{\Rewrite}\DNFRegex_2'$ and
  $\DNFRegexAlt_2\StarOf{\Rewrite}\DNFRegexAlt_2'$ and there exists a rewriteless dnf
  lens $\DNFLens_{\IdentityLens}' \OfRewritelessType \DNFRegex_2' \Leftrightarrow
  \DNFRegexAlt_2'$ where $\SemanticsOf{\DNFLens_{\IdentityLens}'} =
  \SetOf{(\String,\String)\SuchThat \String\in\LanguageOf{\Regex_2}}$.
  From Corollary~\ref{cor:bisimilarity-star-sequential} and
  Corollary~\ref{cor:bisimilarity-star-sequential}, there exists
  $\DNFRegex_1'$, $\DNFRegex_3'$, $\DNFLens_1'$, and $\DNFLens_2'$ such that
  $\DNFRegex_1\StarOf{\Rewrite}\DNFRegex_1'$, $\DNFRegex_3\StarOf{\Rewrite}\DNFRegex_3'$,
  $\DNFLens_1' \OfRewritelessType \DNFRegex_1' \Leftrightarrow \DNFRegex_2'$,
  $\DNFLens_2' \OfRewritelessType \DNFRegexAlt_2' \Leftrightarrow \DNFRegex_3'$,
  $\SemanticsOf{\DNFLens_1} = \SemanticsOf{\DNFLens_1'}$ and
  $\SemanticsOf{\DNFLens_2} = \SemanticsOf{\DNFLens_2'}$.  From
  Lemma~\ref{lem:composition-completeness}
  rewriteless DNF lenses are closed under composition, so there exists a
  rewriteless DNF lens $\DNFLens' \OfRewritelessType \DNFRegex_1' \Leftrightarrow
  \DNFRegex_2'$ where $\SemanticsOf{\DNFLens'} = \SemanticsOf{\DNFLens_2'}
  \Compose \SemanticsOf{\DNFLens_{\IdentityLens}'} \Compose
  \SemanticsOf{\DNFLens_1'} = \SemanticsOf{\DNFLens_2'} \Compose
  \SemanticsOf{\DNFLens_1'} = \SemanticsOf{\Lens_2} \Compose
  \SemanticsOf{\Lens_1} = \SemanticsOf{\Lens_2\Compose\Lens_1}$.  Furthermore,
  $\ToDNFRegexOf{\Regex_1} \StarOf{\Rewrite} \DNFRegex \StarOf{\Rewrite} \DNFRegex_1'$ so
  $\ToDNFRegex{\Regex_1} \StarOf{\Rewrite} \DNFRegex_1$. $\ToDNFRegex{\Regex_3}
  \StarOf{\Rewrite} \DNFRegex_3 \StarOf{\Rewrite} \DNFRegex_3'$ so $\ToDNFRegex{\Regex_3}
  \StarOf{\Rewrite} \DNFRegex_3'$.

  \begin{gather*}
    \DNFLens' \OfRewritelessType \DNFRegex_1' \Leftrightarrow \DNFRegex_3'\\
    \ToDNFRegexOf{\Regex_1} \StarOf{\Rewrite} \DNFRegex_1'\\
    \ToDNFRegexOf{\Regex_3} \StarOf{\Rewrite} \DNFRegex_3'\\
    \SemanticsOf{\DNFLens'} = \SemanticsOf{\ComposeLensOf{\Lens_1}{\Lens_2}}
  \end{gather*}
  \\
  \\
  Let the last rule be an instance of \RewriteRegexLensRule{}.
  \[
    \inferrule*
    {
      \Lens \OfType \Regex \Leftrightarrow \RegexAlt\\
      \Regex \equiv \Regex'\\
      \RegexAlt \equiv \RegexAlt'
    }
    {
      \Lens \OfType \Regex' \Leftrightarrow \RegexAlt'
    }
  \]
  
  By IH, there exists $\DNFLens$, $\DNFRegex$, $\DNFRegexAlt$ such that
  
  \begin{gather*}
    \DNFLens \OfRewritelessType \DNFRegex \Leftrightarrow \DNFRegexAlt\\
    \ToDNFRegexOf{\Regex} \StarOf{\Rewrite} \DNFRegex\\
    \ToDNFRegexOf{\RegexAlt} \StarOf{\Rewrite} \DNFRegexAlt\\
    \SemanticsOf{\DNFLens} = \SemanticsOf{\Lens}
  \end{gather*}

  As $\Regex' \SSREquiv \Regex$, and $\Regex$ is strongly unambiguous,
  from Lemma~\ref{lem:identity-definitional-equivalence} there exists a
  rewriteless DNF lens $\DNFLens_{\Regex',\Regex}
  \OfRewritelessType \overline{\DNFRegex'} \Leftrightarrow
  \overline{\DNFRegex}$ such that
  $\ToDNFRegexOf{\Regex} \StarOf{\Rewrite} \overline{\DNFRegex}$,
  $\ToDNFRegexOf{\Regex'} \StarOf{\Rewrite} \overline{\DNFRegex'}$, and
  $\SemanticsOf{\DNFLens_{\Regex',\Regex}} = \SetOf{(\String,\String)
    \SuchThat \String \in \LanguageOf{\ToDNFRegexOf{\Regex}}}$.
  
  As $\RegexAlt \SSREquiv \RegexAlt'$, and $\RegexAlt$ is strongly unambiguous,
  from Lemma~\ref{lem:identity-definitional-equivalence} there exists a
  rewriteless DNF lens $\DNFLens_{\RegexAlt,\RegexAlt'}
  \OfRewritelessType \overline{\DNFRegexAlt} \Leftrightarrow
  \overline{\DNFRegexAlt'}$ such that
  $\ToDNFRegexOf{\RegexAlt} \StarOf{\Rewrite} \overline{\DNFRegexAlt}$,
  $\ToDNFRegexOf{\RegexAlt'} \StarOf{\Rewrite} \overline{\DNFRegexAlt'}$, and
  $\SemanticsOf{\DNFLens_{\RegexAlt,\RegexAlt'}} = \SetOf{(\String,\String)
    \SuchThat \String \in \LanguageOf{\ToDNFRegexOf{\RegexAlt}}}$.

  From Lemma~\ref{lem:strongly-unambiguous-identity-expressible}, there exists a
  lens $\DNFLens_{\ToDNFRegexOf{\Regex}} \OfRewritelessType
  \ToDNFRegexOf{\Regex} \Leftrightarrow \ToDNFRegexOf{\Regex}$.  As
  $\ToDNFRegexOf{\Regex} \StarOf{\Rewrite} \overline{\DNFRegex}$ and
  $\ToDNFRegexOf{\Regex} \StarOf{\Rewrite} \DNFRegex$, by
  Corollary~\ref{cor:rewrite-confluence}, there exists some
  $\tilde{\DNFLens_{\ToDNFRegexOf{\Regex}}} \OfRewritelessType
  \tilde{\overline{\DNFRegex}} \Leftrightarrow \tilde{\DNFRegex}$,
  such that
  $\overline{\DNFRegex} \StarOf{\Rewrite} \tilde{\overline{\DNFRegex}}$,
  $\DNFRegex \StarOf{\Rewrite} \tilde{\DNFRegex}$, and
  $\SemanticsOf{\tilde{\DNFLens_{\ToDNFRegexOf{\Regex}}}} =
  \SemanticsOf{\DNFLens_{\ToDNFRegexOf{\Regex}}}$.
  From Corollary~\ref{cor:bisimilarity-star-sequential}, there
  exists a $\tilde{\DNFLens_{\Regex',\Regex}} \OfRewritelessType
  \tilde{\overline{\DNFRegex'}} \Leftrightarrow \tilde{\overline{\DNFRegex}}$
  such that $\overline{\DNFRegex'} \StarOf{\Rewrite}
  \tilde{\overline{\DNFRegex'}}$ and
  $\SemanticsOf{\tilde{\DNFLens_{\Regex',\Regex}}} =
  \SemanticsOf{\DNFLens_{\Regex',\Regex}}$

  From Lemma~\ref{lem:strongly-unambiguous-identity-expressible}, there exists a
  lens $\DNFLens_{\ToDNFRegexOf{\RegexAlt}} \OfRewritelessType
  \ToDNFRegexOf{\RegexAlt} \Leftrightarrow \ToDNFRegexOf{\RegexAlt}$.  As
  $\ToDNFRegexOf{\RegexAlt} \StarOf{\Rewrite} \DNFRegexAlt$ and
  $\ToDNFRegexOf{\RegexAlt} \StarOf{\Rewrite} \overline{\DNFRegexAlt}$, by
  Corollary~\ref{cor:rewrite-confluence}, there exists some
  $\tilde{\DNFLens_{\ToDNFRegexOf{\RegexAlt}}} \OfRewritelessType
  \tilde{\DNFRegexAlt} \Leftrightarrow \tilde{\overline{\DNFRegexAlt}}$,
  such that
  $\DNFRegexAlt \StarOf{\Rewrite} \tilde{\DNFRegexAlt}$,
  $\overline{\DNFRegexAlt} \StarOf{\Rewrite} \tilde{\overline{\DNFRegexAlt}}$, and
  $\SemanticsOf{\tilde{\DNFLens_{\ToDNFRegexOf{\RegexAlt}}}} =
  \SemanticsOf{\DNFLens_{\ToDNFRegexOf{\RegexAlt}}}$.
  From Corollary~\ref{cor:bisimilarity-star-sequential}, there
  exists a $\tilde{\DNFLens_{\RegexAlt,\RegexAlt'}} \OfRewritelessType
  \tilde{\overline{\DNFRegexAlt}} \Leftrightarrow \tilde{\overline{\DNFRegexAlt'}}$
  such that $\overline{\DNFRegexAlt'} \StarOf{\Rewrite}
  \tilde{\overline{\DNFRegexAlt'}}$ and
  $\SemanticsOf{\tilde{\DNFLens_{\RegexAlt',\RegexAlt}}} =
  \SemanticsOf{\DNFLens_{\RegexAlt',\RegexAlt}}$.

  As $\DNFRegex \StarOf{\Rewrite} \tilde{\DNFRegex}$ and
  $\DNFRegexAlt \StarOf{\Rewrite} \tilde{\DNFRegexAlt}$,
  by Corollary~\ref{cor:rewrite-confluence} there exists a lens
  $\underline{\DNFLens} \OfRewritelessType \underline{\DNFRegex} \Leftrightarrow
  \underline{\DNFRegexAlt}$ such that
  $\tilde{\DNFRegex} \StarOf{\Rewrite} \underline{\DNFRegex}$,
  $\tilde{\DNFRegexAlt} \StarOf{\Rewrite} \underline{\DNFRegexAlt}$, and
  $\SemanticsOf{\underline{\DNFLens}} = \SemanticsOf{\DNFLens}$.
  
  From Corollary~\ref{cor:bisimilarity-star-sequential}, there
  exists $\underline{\tilde{\DNFLens_{\ToDNFRegexOf{\Regex}}}}
  \OfRewritelessType \underline{\tilde{\overline{\DNFRegex}}}
  \Leftrightarrow \underline{\DNFRegex}$ such that $\tilde{\overline{\DNFRegex}}
  \StarOf{\Rewrite} \underline{\tilde{\overline{\DNFRegex}}}$ and
  $\SemanticsOf{\underline{\tilde{\DNFLens_{\ToDNFRegexOf{\Regex}}}}} =
  \SemanticsOf{\tilde{\DNFLens_{\ToDNFRegexOf{\Regex}}}}$.
  From Corollary~\ref{cor:bisimilarity-star-sequential}, there
  exists $\underline{\tilde{\DNFLens_{\Regex',\Regex}}}
  \OfRewritelessType \underline{\tilde{\overline{\DNFRegex'}}}
  \Leftrightarrow \underline{\tilde{\overline{\DNFRegex}}}$
  such that $\tilde{\overline{\DNFRegex'}}
  \StarOf{\Rewrite} \underline{\tilde{\overline{\DNFRegex'}}}$ and
  $\SemanticsOf{\underline{\tilde{\DNFLens_{\Regex',\Regex}}}} =
  \SemanticsOf{\tilde{\DNFLens_{\Regex',\Regex}}}$.
  
  From Corollary~\ref{cor:bisimilarity-star-sequential}, there
  exists $\underline{\tilde{\DNFLens_{\ToDNFRegexOf{\RegexAlt}}}}
  \OfRewritelessType \underline{\DNFRegexAlt}
  \Leftrightarrow \underline{\tilde{\overline{\DNFRegexAlt}}}$ such that
  $\tilde{\overline{\DNFRegexAlt}}
  \StarOf{\Rewrite} \underline{\tilde{\overline{\DNFRegexAlt}}}$ and
  $\SemanticsOf{\underline{\tilde{\DNFLens_{\ToDNFRegexOf{\RegexAlt}}}}} =
  \SemanticsOf{\tilde{\DNFLens_{\ToDNFRegexOf{\RegexAlt}}}}$.
  From Corollary~\ref{cor:bisimilarity-star-sequential}, there
  exists $\underline{\tilde{\DNFLens_{\RegexAlt,\RegexAlt'}}}
  \OfRewritelessType \underline{\tilde{\overline{\DNFRegexAlt}}}
  \Leftrightarrow \underline{\tilde{\overline{\DNFRegexAlt'}}}$
  such that $\tilde{\overline{\DNFRegexAlt'}}
  \StarOf{\Rewrite} \underline{\tilde{\overline{\DNFRegexAlt'}}}$ and
  $\SemanticsOf{\underline{\tilde{\DNFLens_{\RegexAlt,\RegexAlt'}}}} =
  \SemanticsOf{\tilde{\DNFLens_{\RegexAlt,\RegexAlt'}}}$.

  From Lemma~\ref{lem:composition-completeness}, there exists a lens
  $\boldsymbol{\DNFLens} \OfRewritelessType
  \underline{\tilde{\overline{\DNFRegex'}}} \Leftrightarrow
  \underline{\tilde{\overline{\DNFRegexAlt'}}}$.  Because the semantics of all
  lenses in the composition for $\boldsymbol{\DNFLens}$ were all the identity
  relation, $\SemanticsOf{\boldsymbol{\DNFLens}} = \SemanticsOf{\DNFLens}$.
  Furthermore,
  $\ToDNFRegexOf{\Regex'} \StarOf{\Rewrite}
  \underline{\tilde{\overline{\DNFRegex'}}}$
  and
  $\ToDNFRegexOf{\RegexAlt'} \StarOf{\Rewrite}
  \underline{\tilde{\overline{\DNFRegexAlt'}}}$, so we have
  \begin{gather*}
    \boldsymbol{\DNFLens} \OfRewritelessType
    \underline{\tilde{\overline{\DNFRegex'}}} \Leftrightarrow
    \underline{\tilde{\overline{\DNFRegexAlt'}}}\\
    \ToDNFRegexOf{\Regex'} \StarOf{\Rewrite}
    \underline{\tilde{\overline{\DNFRegex'}}}\\
    \ToDNFRegexOf{\RegexAlt'} \StarOf{\Rewrite}
    \underline{\tilde{\overline{\DNFRegexAlt'}}}\\
    \SemanticsOf{\boldsymbol{\DNFLens}} = \SemanticsOf{\Lens}
  \end{gather*}
\end{proof}

\begin{theorem}
  If there exists a derivation for $\Lens \OfType \Regex \Leftrightarrow
  \RegexAlt$,
  then there exists a DNF lens $\DNFLens$ such that
  $\DNFLens \OfType (\ToDNFRegexOf{\Regex}) \Leftrightarrow (\ToDNFRegexOf{\RegexAlt})$ and $\SemanticsOf{\Lens}=\SemanticsOf{\DNFLens}$.
\end{theorem}
\begin{proof}
  By Lemma~\ref{lem:pre-completeness}, there exists $\DNFLens$, $\DNFRegex$,
  $\DNFRegexAlt$ such that $\DNFLens \OfRewritelessType \DNFRegex
  \Leftrightarrow \DNFRegexAlt$, $\ToDNFRegexOf{\Regex} \StarOf{\Rewrite}
  \DNFRegex$, $\ToDNFRegexOf{\RegexAlt} \StarOf{\Rewrite} \DNFRegexAlt$, and
  $\SemanticsOf{\DNFLens} = \SemanticsOf{\Lens}$.  Because of that, we have the
  derivation

  \[
    \inferrule*
    {
      \DNFLens \OfRewritelessType \DNFRegex \Leftrightarrow \DNFRegexAlt\\
      \ToDNFRegexOf{\Regex} \StarOf{\Rewrite} \DNFRegex\\
      \ToDNFRegexOf{\RegexAlt} \StarOf{\Rewrite} \DNFRegexAlt
    }
    {
      \DNFLens \OfType \ToDNFRegexOf{\Regex} \Leftrightarrow
      \ToDNFRegexOf{\RegexAlt}
    }
  \]
\end{proof}

\subsection{Algorithm Correctness}
\label{alg-correctness}

We use an auxiliary data structure of a set-of-examples-parse-tree to define the
orderings.

\begin{definition}
  We use \IntList{} to denote list of ints.
\end{definition}

\begin{definition}
  We use \IntListSet{} to denote a set of int lists.
\end{definition}

\begin{definition}
  We use \StringIntListSet{} to denote a set of string and int list pairs.  We
  also require that the int lists are distinct.
\end{definition}

\begin{definition}
  To get the strings out of $\StringIntListSet$, we use $\ProjectStrings$.  In
  particular,\\
  $\ProjectStringsOf{\SetOf{(\String_1,\IntList_1),\ldots,(\String_n,\IntList_n)}}
  = \SetOf{\String_1,\ldots,\String_n}$.
\end{definition}

\begin{definition}
  To get the int lists out of $\StringIntListSet$, we use $\ProjectILS$.  In
  particular,\\
  $\ProjectILSOf{\SetOf{(\String_1,\IntList_1),\ldots,(\String_n,\IntList_n)}}
  = \SetOf{\IntList_1,\ldots,\IntList_n}$.
\end{definition}

\begin{definition}
  Define an exampled atom, exampled sequence, and exampled DNF regular expression as:
  \begin{center}
    \begin{tabular}{l@{\ }c@{\ }l@{\ }>{\itshape\/}r}
      % DNF_REGEX
      \ExampledAtom{},\ExampledAtomAlt{} & \GEq{} & (\StarOf{\ExampledDNFRegex{}},\IntListSet{})
      % & \StarAtomType{}
      \\
      \ExampledSequence{},\ExampledSequenceAlt{} & \GEq{} &
                                                            $(\SequenceOf{\String_0\SeqSep\ExampledAtom_1\SeqSep\ldots\SeqSep\ExampledAtom_n\SeqSep\String_n},\IntListSet)$
                                                            % & \MultiConcatSequenceType{} 
      \\
      \ExampledDNFRegex{},\ExampledDNFRegexAlt{} & \GEq{} &
                                                            $(\DNFOf{\ExampledSequence_1\DNFSep\ldots\DNFSep\ExampledSequence_n},\IntListSet)$ %& \MultiOrDNFRegexType{} 
    \end{tabular}
  \end{center}
\end{definition}

Intuitively, an exampled regular expression is a DNF regular expression, with the
parse trees for a number of strings which match it embedded.

We build the
typing derivation of the form
$\StringIntListSet \in \DNFRegex
\Generates \ExampledDNFRegex$ to express that the strings
$\ProjectStringsOf{\StringIntListSet}$
, labelled by the identifiers
$\ProjectILSOf{\StringIntListSet}$ when they have their parse trees embedded in
$\DNFRegex$, generate $\ExampledDNFRegex$.
Similarly for $\Sequence$ and $\ExampledSequence$, and $\Atom$ and
$\ExampledAtom$.

\begin{definition}
  \begin{mathpar}
    \inferrule[]
    {
      \SetOf{(\String_{1,1},1::\IntList_{1}),\ldots,(\String_{1,n_1},n_1::\IntList_{1}),
        \ldots
        ,(\String_{m,1},1::\IntList_{m}),\ldots,(\String_{m,n_m},n_m::\IntList_{m})}
      \in \DNFRegex \Generates \ExampledDNFRegex
    }
    {
      \SetOf{(\String_{1,1}\Concat\ldots\Concat\String_{1,n_1},\IntList_1),
        \ldots,
        (\String_{m,1}\Concat\ldots\Concat\String_{m,n_m},\IntList_m)} \in
      \StarOf{\DNFRegex} \Generates (\StarOf{\ExampledDNFRegex},\SetOf{\IntList_1,\ldots,\IntList_m})
    }
    
    \inferrule[]
    {
      \SetOf{(\String_{1,1},\IntList_{1}),
        \ldots
        ,(\String_{m,1},\IntList_{m})}
      \in \Atom_1 \Generates \ExampledAtom_1\\
      \ldots\\
      \SetOf{(\String_{1,n},\IntList_{1}),
        \ldots
        ,(\String_{m,n},\IntList_{m})}
      \in \Atom_n \Generates \ExampledAtom_n
    }
    {
      \SetOf{(\String_0'\Concat\String_{1,1}\Concat\ldots\Concat\String_{1,n}\Concat\String_n',\IntList_1),
        \ldots,
        (\String_0'\Concat\String_{m,1}\Concat\ldots\Concat\String_{m,n}\Concat\String_n',\IntList_m)} \in
      \SequenceOf{\String_0' \SeqSep \Atom_1 \SeqSep \ldots \SeqSep \Atom_n
        \SeqSep \String_n'}\\
      \Generates (\SequenceOf{\String_0' \SeqSep \ExampledAtom_1 \SeqSep \ldots \SeqSep \ExampledAtom_n
        \SeqSep \String_n'}, \SetOf{\IntList_1,\ldots,\IntList_m})
    }
    
    \inferrule[]
    {
      \StringIntListSet_1
      \in \Sequence_1 \Generates \ExampledSequence_1\\
      \ldots\\
      \StringIntListSet_n
      \in \Sequence_n \Generates \ExampledSequence_n\\
    }
    {
      \BigUnion_{i\in\RangeIncInc{1}{n}} \StringIntListSet_i
      \in \DNFOf{\Sequence_1 \DNFSep \ldots \DNFSep \Sequence_n}
      \Generates
      (\DNFOf{\ExampledSequence_1 \DNFSep \ldots \DNFSep \ExampledSequence_n},
      \ProjectILSOf{\BigUnion_{i\in\RangeIncInc{1}{n}} \StringIntListSet_i})
    }
  \end{mathpar}
\end{definition}

This is a big typing derivation, and we feel it is clear that, when a DNF
regular expression is strongly unambiguous, this typing derivation is unique for
a given set of strings and DNF regular expression, so it is functional from the
first two arguments of the derivation.  Furthermore, we can perform this
function by doing case analysis on all the possible ways the string is split up
(though it is slow).  We perform this function by performing this embedding the
function into a NFA matching algorithm.  We elide these details.

\begin{definition}
  Define \EmbedExamples{} as the function from DNF Regex $\DNFRegex$ and intlist
  labelled examples $\StringIntListSet$ to
  exampled DNF regex, such that $\StringIntListSet \in \DNFRegex \Generates
  \EmbedExamplesOf{\StringIntListSet}{\DNFRegex}$
\end{definition}

Now, we are going to build up the machinery to define an ordering on exampled
DNF regular expressions, exampled sequences, and exampled atoms.  We need to
define some general orderings first.

\begin{definition}
  Let $\leq$ be an ordering on $A$.  Let $\ListOf{x_1,\ldots,x_n}$ be a list of
  $A$s.  Define $\SortingOf{\ListOf{x_1;\ldots;x_n}}{\leq}$ as a permutation $\sigma
  \in \PermutationSetOf{n}$ such that $\sigma(i) \leq \sigma(j) \BooleanImplies
  x_{\sigma(i)} \leq x_{\sigma(j)}$.
\end{definition}

\begin{definition}
  Let $\leq$ be an ordering on $A$.  Let $\ListOf{x_1;\ldots;x_n}$ be a list of
  $A$s.  Define $\SortOf{\ListOf{x_1;\ldots;x_n}}{\leq} =
  \ListOf{x_{\sigma(1)};\ldots;x_{\sigma(n)}}$  where $\sigma =
  \SortingOf{\ListOf{x_1;\ldots;x_n}}{\leq}$.
\end{definition}

\begin{definition}
  Let $\leq_1$ be an ordering on $A_1$.  Let $\leq_2$ be an ordering on $A_2$.
  Define the product ordering $(\leq_1,\leq_2)$ on $A_1 \times A_2$ as the
  lexicographic ordering on the two elements.
\end{definition}

\begin{definition}
  Let $\leq$ be an ordering on $A$.  We write $\DictionaryOrderOf{\leq}$ for
  the lexicographic ordering on $\ListTypeOf{A}$.
\end{definition}

\begin{property}
  $\ListOf{x_1;\ldots;x_n} \DictionaryOrderOf{\leq} \ListOf{y_1;\ldots;y_m}$
  and $\ListOf{y_1;\ldots;y_m} \DictionaryOrderOf{\leq} \ListOf{x_1;\ldots;x_n}$
  if, and only if
  $n = m$ and for all $i \in \RangeIncInc{1}{n}$
  $x_i \leq y_i$ and $y_i \leq x_i$
\end{property}

\begin{definition}
  Let $\leq$ be an ordering on $A$.  Define $\SetOfListOrderOf{\leq}$ as the
  ordering on $\ListTypeOf{A}$ as:
  $\ListOf{x_1;\ldots;x_n} \SetOfListOrderOf{\leq} \ListOf{y_1;\ldots;y_m}$ if
  $\SortOf{x_1;\ldots;x_n}{\leq} \DictionaryOrderOf{\leq}
  \SortOf{y_1;\ldots;y_m}{\leq}$.
  We also use $\SetOfListOrderOf{\leq}$ to operate on sets, by first converting
  the set to a list, then using that ordering.
\end{definition}

\begin{definition}
  Define an ordering on int list sets, $\ILSLeq$, as
  $\SetOfListOrderOf{\DictionaryOrderOf{\leq}}$, where $\leq$ is the usual on
  integers.
\end{definition}

\begin{lemma}
  \label{lem:spec-set-list-perm}
  If \\
  $\ListOf{x_1;\ldots;x_n} \SetOfListOrderOf{\leq} \ListOf{y_1;\ldots;y_m}$,\\
  $\ListOf{y_1;\ldots;y_m} \SetOfListOrderOf{\leq} \ListOf{x_1;\ldots;x_n}$,\\
  $\sigma_1 = \SortingOf{\ListOf{x_1;\ldots;x_n}}{\leq}$ and\\
  $\sigma_2 = \SortingOf{\ListOf{y_1;\ldots;y_m}}{\leq}$, then\\
  $n = m$,\\
  $x_i \leq y_{(\InverseOf{\sigma_1} \Compose \sigma_2)(i)}$, and\\
  $y_{(\InverseOf{\sigma_1} \Compose \sigma_2)(i)} \leq x_i$
\end{lemma}
\begin{proof}
    Let $\sigma_1 = \SortingOf{\ListOf{x_1;\ldots;x_n}}{\leq}$ and
    $\sigma_2 = \SortingOf{\ListOf{y_1;\ldots;y_m}}{\leq}$
    
    This means that
    $\ListOf{x_{\sigma_1(1)};\ldots;x_{\sigma_1(n)}} \DictionaryOrderOf{\leq}
    \ListOf{y_{\sigma_2(1)};\ldots;y_{\sigma_2(m)}}$ and
    $\ListOf{y_{\sigma_2(1)};\ldots;y_{\sigma_2(m)}} \DictionaryOrderOf{\leq}
    \ListOf{x_{\sigma_1(1)};\ldots;x_{\sigma_1(n)}}$.

    By the above property about dictionary orderings, this means that $n=m$ and
    $x_{\sigma_1(i)} \leq y_{\sigma_2(i)}$ and $y_{\sigma_2(i)} \leq
    x_{\sigma_1(i)}$

    Consider the permutation $\sigma = \InverseOf{\sigma_1} \Compose \sigma_2$.
    We know $x_{\sigma_1(i)} \leq y_{\sigma_2(i)}$ and $y_{\sigma_2(i)} \leq
    x_{\sigma_1(i)}$.  By reordering through the permutation
    $\InverseOf{\sigma_1}$, we get $x_{\InverseOf{\sigma_1} \Compose \sigma_1(i)}
    \leq y_{\InverseOf{\sigma_1} \Compose \sigma_2(i)}$ and
    $y_{\InverseOf{\sigma_1} \Compose \sigma_2(i)} \leq 
    x_{\InverseOf{\sigma_1} \Compose \sigma_1(i)}$.  By simplifying we get
    $x_i
    \leq y_{\sigma(i)}$ and
    $y_{\sigma(i)} \leq x_{i}$
\end{proof}

\begin{lemma}
  \label{lem:set-list-perm}
  $\ListOf{x_1;\ldots;x_n} \SetOfListOrderOf{\leq} \ListOf{y_1;\ldots;y_m}$ and
  $\ListOf{y_1;\ldots;y_m} \SetOfListOrderOf{\leq} \ListOf{x_1;\ldots;x_n}$
  if, and only if, $n = m$ and there exists a permutation $\sigma$ such that
  $x_i \leq y_{\sigma(i)}$ and
  $y_{\sigma(i)} \leq x_i$
\end{lemma}
\begin{proof}
  \begin{case}[$\Rightarrow$]
    By Lemma~\ref{lem:spec-set-list-perm}.
  \end{case}

  \begin{case}[$\Leftarrow$]
    Let $n=m$ and $\sigma$ be a permutation such that
    $x_i \leq y_{\sigma(i)}$ and
    $y_{\sigma(i)} \leq x_i$.

    We know the number of equivalence classes in the two lists is equal, as
    otherwise there would be some equivalence class in one that is not in the
    other, a contradiction with the assumption.

    We proceed by induction on the number of equivalence classes:

    Base Case: no equivalence classes, no elements, trivially true.

    Induction Step:
    Let there be $n+1$ equivalence classes. 
    Let $\sigma_1 = \SortingOf{\ListOf{x_1;\ldots;x_n}}$ and
    $\sigma_2 = \SortingOf{\ListOf{y_1;\ldots;y_n}}$.

    Consider the largest equivalence class.
    We know that all except that equivalence class must map to each other, so
    when we remove that equivalence class, we get that all except the largest
    elements are ordered with $\SetOfListOrderOf{\leq}$, by IH.
    Adding those elements back in, we know they must go at the end.
    Furthermore, they must have the same number of elements on each side $k$, else
    we contradict the assumption.  This means that in
    $\SortOf{\ListOf{x_1;\ldots;x_n}} and \SortOf{\ListOf{y_1;\ldots;y_n}}$
    are ordered such that the $j$th element in the $x$ list is equivalent to the
    $j$th element in the $y$ list, until the end, but the last $k$ elements are
    all equivalent as they are all the largest equivalence class, so we are done.
  \end{case}
\end{proof}

Now we can define what $\ExampledAtomLeq$, $\ExampledSequenceLeq$ and
$\ExampledDNFLeq$ are, mutually.

\begin{definition}\leavevmode
  \begin{itemize}
  \item We say $(\StarOf{\ExampledDNFRegex}, \IntListSet_1) \ExampledAtomLeq
    (\StarOf{\ExampledDNFRegexAlt}, \IntListSet_2)$ if\\
    $(\ExampledDNFRegex,\IntListSet_1)
    (\ExampledDNFLeq,\ILSLeq)
    (\ExampledDNFRegexAlt,\IntListSet_2)$.
  \item We say
    $(\SequenceOf{\String_0 \SeqSep \ExampledAtom_1 \SeqSep \ldots
      \SeqSep \ExampledAtom_n \SeqSep \String_n},\IntListSet_1) \ExampledSequenceLeq
    (\SequenceOf{\StringAlt_0 \SeqSep \ExampledAtomAlt_1 \SeqSep \ldots \SeqSep
      \ExampledAtomAlt_m \SeqSep \StringAlt_n},\IntListSet_2)$ if\\
    $(\ListOf{\ExampledAtom_1 ; \ldots ; \ExampledAtom_n},\IntListSet_1)
    (\DictionaryOrderOf{\ExampledAtomLeq},\ILSLeq)
    (\ListOf{\ExampledAtomAlt_1 ; \ldots ; \ExampledAtomAlt_m},\IntListSet_2)$.
  \item We say
    $(\DNFOf{\ExampledSequence_1 \DNFSep \ldots
      \DNFSep \ExampledSequence_n},\IntListSet_1) \ExampledDNFLeq
    (\DNFOf{\ExampledSequenceAlt_1 \DNFSep \ldots \DNFSep
      \ExampledSequenceAlt_n},\IntListSet_2)$ if\\
    $(\ListOf{\ExampledSequence_1 ; \ldots ; \ExampledSequence_n},\IntListSet_1)
    (\DictionaryOrderOf{\ExampledSequenceLeq},\ILSLeq)
    (\ListOf{\ExampledSequenceAlt_1 ; \ldots ;
      \ExampledSequenceAlt_m},\IntListSet_2)$.
  \end{itemize}
\end{definition}

Now, using this we provide the more formal definition of algorithm, with the
formal use of the examples in Algorithm~\ref{alg:rigidsynthreal}.  We do not
include information about user defined data types.

\begin{algorithm}
  \caption{\RigidSynth}
  \label{alg:rigidsynthreal}
  \begin{algorithmic}[1]
    \Function{\RigidSynthAtom}{($\StarOf{\ExampledDNFRegex},\IntListSet_1),(\StarOf{\ExampledDNFRegexAlt{}},\IntListSet_2)$}
    \If {$\IntListSet_1 \not \ILSLeq \IntListSet_2 \BooleanOr \IntListSet_2 \not \ILSLeq \IntListSet_1$}
    \State $\ReturnVal{\None}$
    \EndIf
    \Switch{$\RigidSynthInternal(\ExampledDNFRegex,\ExampledDNFRegexAlt)$}
    \CaseTwo{\SomeOf{\DNFLens}}{\ReturnVal{\IterateLensOf{\DNFLens}}}
    \EndCaseTwo
    \CaseTwo {\None}{$\ReturnVal{\None}$}
    \EndCaseTwo
    \EndSwitch
    \EndFunction

    \Statex

    \Function{\RigidSynthSequence}{$\ExampledSequence,\ExampledSequenceAlt$}
    \State $(\SequenceOf{\String_0 \SeqSep \ExampledAtom_1 \SeqSep \ldots \SeqSep \ExampledAtom_n
      \SeqSep \String_n},\IntListSet_1) \gets
    \ExampledSequence$
    \State $(\SequenceOf{\StringAlt_0 \SeqSep \ExampledAtomAlt_1 \SeqSep \ldots \SeqSep
      \ExampledAtomAlt_m \SeqSep \StringAlt_m},\IntListSet_2) \gets
    \ExampledSequenceAlt$
    \If {$\IntListSet_1 \not \ILSLeq \IntListSet_2 \BooleanOr \IntListSet_2 \not \ILSLeq \IntListSet_1$}
    \State $\ReturnVal{\None}$
    \EndIf
    \If {$n \neq m$}
    \State $\ReturnVal{\None}$
    \EndIf
    \State $\sigma_1 \gets \SortingOf{\ExampledAtomLeq}{\ListOf{\ExampledAtom_1
        \SeqSep \ldots \SeqSep \ExampledAtom_n}}$
    \State $\sigma_2 \gets \SortingOf{\ExampledAtomLeq}{\ListOf{\ExampledAtomAlt_1
        \SeqSep \ldots \SeqSep \ExampledAtomAlt_n}}$
    \State $\sigma \gets \InverseOf{\sigma_1} \Compose \sigma_2$
    \State $\mathit{EABs} \gets
    \Call{Zip}{\ListOf{\ExampledAtom_1 \SeqSep \ldots \SeqSep
        \ExampledAtom_n},\ListOf{\ExampledAtomAlt_{\sigma(1)} \SeqSep \ldots \SeqSep \ExampledAtomAlt_{\sigma(n)}}}$
    \State $\mathit{alos} \gets
    \Call{\Map}{\RigidSynthAtom,\mathit{EABs}}$
    \Switch{$\Call{AllSome}{\mathit{alos}}$}
    \CaseTwo {\SomeOf{\ListOf{\AtomLens_1 \SeqSep \ldots \SeqSep
          \AtomLens_n}}}{$\ReturnVal{\SomeOf{(\SequenceLensOf{(\String_0,\StringAlt_0)
            \SeqSep \AtomLens_1 \SeqSep \ldots \SeqSep \AtomLens_n \SeqSep (\String_n,\StringAlt_n)},\InverseOf{\sigma})}}$}
    \EndCaseTwo
    \CaseTwo {\None}{$\ReturnVal{\None}$}
    \EndCaseTwo
    \EndSwitch
    \EndFunction

    \Statex
 
    \Function{\RigidSynthInternal}{$\ExampledDNFRegex,\ExampledDNFRegexAlt$}
    \State $(\DNFOf{\ExampledSequence_1 \DNFSep \ldots \DNFSep \ExampledSequence_n},\IntListSet_1) \gets
    \ExampledDNFRegex$
    \State $(\DNFOf{\ExampledSequenceAlt_1 \DNFSep \ldots \DNFSep \ExampledSequenceAlt_m},\IntListSet_2) \gets
    \ExampledDNFRegexAlt$
    \If {$\IntListSet_1 \not \ILSLeq \IntListSet_2 \BooleanOr \IntListSet_2 \not \ILSLeq \IntListSet_1$}
    \State $\ReturnVal{\None}$
    \EndIf
    \If {$n \neq m$}
    \State $\ReturnVal{\None}$
    \EndIf
    \State $\sigma_1 \gets
    \SortingOf{\ExampledSequenceLeq}{\ListOf{\ExampledSequence_1 \DNFSep \ldots
        \DNFSep \ExampledSequence_n}}$
    \State $\sigma_2 \gets
    \SortingOf{\ExampledSequenceLeq}{\ListOf{\ExampledSequenceAlt_1 \DNFSep \ldots
        \DNFSep \ExampledSequenceAlt_n}}$
    \State $\sigma \gets \InverseOf{\sigma_1} \Compose \sigma_2$
    \State $\mathit{ESTQs} \gets
    \Call{Zip}{\ListOf{\ExampledSequence_1 \DNFSep \ldots \DNFSep
        \ExampledSequence_n},\ListOf{\ExampledSequenceAlt_{\sigma(1)} \DNFSep \ldots \DNFSep \ExampledSequenceAlt_{\sigma(n)}}}$
    \State $\mathit{sqlos} \gets
    \Call{\Map}{\RigidSynthSequence,\mathit{ESTQs}}$
    \Switch{$\Call{AllSome}{\mathit{sqlos}}$}
    \CaseTwo {\SomeOf{\ListOf{\SequenceLens_1 \DNFSep \ldots \DNFSep
          \SequenceLens_n}}}{$\ReturnVal{\SomeOf{(\DNFLensOf{\SequenceLens_1
            \DNFSep \ldots \DNFSep \SequenceLens_n},\InverseOf{\sigma})}}$}
    \EndCaseTwo
    \CaseTwo {\None}{$\ReturnVal{\None}$}
    \EndCaseTwo
    \EndSwitch
    \EndFunction
    \Statex

    \Function{\RigidSynth}{$\DNFRegex,\DNFRegexAlt,\Examples$}
    \State $\ListOf{(\String_1,\StringAlt_1);\ldots;(\String_n,\StringAlt_n)}
    \gets \Examples$
    \State $\ExampledDNFRegex \gets
    \EmbedExamplesOf{\ListOf{([1],\String_1);\ldots;([n],\String_n)}}{\DNFRegex}$
    \State $\ExampledDNFRegexAlt \gets
    \EmbedExamplesOf{\ListOf{([1],\StringAlt_1);\ldots;([n],\StringAlt_n)}}{\DNFRegexAlt}$
    \State $\ReturnVal{\RigidSynthInternal(\ExampledDNFRegex,\ExampledDNFRegexAlt)}$
    \EndFunction
  \end{algorithmic}
\end{algorithm}

\begin{lemma}\leavevmode
  \label{lem:rigidsynth-internal-sound}
  \begin{itemize}
  \item
    Let $\Atom$ and $\AtomAlt$ be strongly unambiguous atoms.
    Let $\StringIntListSet_1$ be a string int list set.
    Let $\StringIntListSet_2$ be a string int list set.
    Let $\ExampledAtom$ and $\ExampledAtomAlt$ be exampled atoms.
    Let $\StringIntListSet_1 \in \Atom \Generates \ExampledAtom$.
    Let $\StringIntListSet_2 \in \AtomAlt \Generates \ExampledAtomAlt$.
    If  $\RigidSynthAtom(\ExampledAtom,\ExampledAtomAlt)$ returns
    an atom lens, $\SomeOf{\AtomLens}$, then $\ProjectILSOf{\StringIntListSet_1} =
    \ProjectILSOf{\StringIntListSet_2}$,
    $\AtomLens \OfRewritelessType \Atom
    \Leftrightarrow \AtomAlt$, and for each $(\String,\StringAlt)$ pair with the
    same int list in $\StringIntListSet_1$ and $\StringIntListSet_2$,
    $(\String,\StringAlt) \in \SemanticsOf{\AtomLens}$.
  \item
    Let $\Sequence$ and $\SequenceAlt$ be strongly unambiguous sequences.
    Let $\StringIntListSet_1$ be a string int list set.
    Let $\StringIntListSet_2$ be a string int list set.
    
    Let $\ExampledSequence$ and $\ExampledSequenceAlt$ be exampled sequences.
    Let $\StringIntListSet_1 \in \Sequence \Generates \ExampledSequence$.
    Let $\StringIntListSet_2 \in \SequenceAlt \Generates \ExampledSequenceAlt$.
    If  $\RigidSynthSequence(\ExampledSequence,\ExampledSequenceAlt)$ returns
    a sequence lens, $\SomeOf{\SequenceLens}$, then $\ProjectILSOf{\StringIntListSet_1} =
    \ProjectILSOf{\StringIntListSet_2}$.
    $\SequenceLens \OfRewritelessType \Sequence 
    \Leftrightarrow \SequenceAlt$, and for each $(\String,\StringAlt)$ pair with
    the same int list in $\StringIntListSet_1$ and $\StringIntListSet_2$,
    $(\String,\StringAlt) \in \SemanticsOf{\SequenceLens}$.
  \item
    Let $\DNFRegex$ and $\DNFRegexAlt$ be strongly unambiguous DNF regular expressions.
    Let $\StringIntListSet_1$ be a string int list set.
    Let $\StringIntListSet_2$ be a string int list set.
    Let $\ExampledDNFRegex$ and $\ExampledDNFRegexAlt$ be exampled DNF regular expressions.
    Let $\StringIntListSet_1 \in \ExampledDNFRegex \Generates \ExampledDNFRegex$.
    Let $\StringIntListSet_2 \in \ExampledDNFRegexAlt \Generates \ExampledDNFRegexAlt$.
    If $\RigidSynthInternal(\ExampledDNFRegex,\ExampledDNFRegexAlt)$ returns
    a DNF lens, $\SomeOf{\DNFLens}$, then $\ProjectILSOf{\StringIntListSet_1} =
    \ProjectILSOf{\StringIntListSet_2}$,
    $\DNFLens \OfRewritelessType \DNFRegex 
    \Leftrightarrow \DNFRegexAlt$, and for each $(\String,\StringAlt)$ pair with
    the same int list in $\StringIntListSet_1$ and $\StringIntListSet_2$,
    $(\String,\StringAlt) \in \SemanticsOf{\DNFLens}$.
  \end{itemize}
\end{lemma}
\begin{proof}
  \begin{case}[atom]
    Unfolding definitions.

    Let $\StringIntListSet_1 =
    \SetOf{(\String_1',\IntList_1'),\ldots,(\String_m',\IntList_{m'}')}$.

    Let
    $\StringIntListSet_2 =
    \SetOf{(\StringAlt_1',\IntList_1''),\ldots,(\StringAlt_{m''}',\IntList_{m''}'')}$
    
    By inverison on 
    $\StringIntListSet_1 \in \Atom \Generates \ExampledAtom$ and
    $\StringIntListSet_2 \in \AtomAlt \Generates \ExampledAtomAlt$,
    we know that\\
    $\SetOf{(\String_{1,1}',1::\IntList_1');\ldots;(\String_{1,n_1'}',n_1'::\IntList_1');
      \ldots;(\String_{m',1}',1::\IntList_{m'}');\ldots;(\String_{m',n_{m'}'}',n_{m'}'::\IntList_{m'}')}
    \in \DNFRegex \Generates \ExampledDNFRegex$,\\
    $\SetOf{(\StringAlt_{1,1}',1::\IntList_1'');\ldots;(\StringAlt_{1,n_1''}',n_1''::\IntList_1'');
      \ldots;(\StringAlt_{m'',1}',1::\IntList_{m''}'');\ldots;(\String_{m'',n_{m''}''}',n_{m''}''::\IntList_{m''}'')}
    \in \DNFRegexAlt \Generates \ExampledDNFRegexAlt$,\\
    $\String_{i,1}'\Concat\ldots\Concat\String_{i,n_i'}' = \String_i$,\\
    $\StringAlt_{i,1}'\Concat\ldots\Concat\StringAlt_{i,n_i''}' = \StringAlt_i$,\\
    $\ExampledAtom =
    (\StarOf{\ExampledDNFRegex},\SetOf{\IntList_1',\ldots,\IntList_m'})$, and\\
    $\ExampledAtomAlt =
    (\StarOf{\ExampledDNFRegexAlt},\SetOf{\IntList_1'',\ldots,\IntList_m''})$

    As \RigidSynthAtom{} returns, then it must return
    $\IterateLensOf{\RigidSynthInternal(\ExampledDNFRegex,\ExampledDNFRegexAlt)}$.

    By IH that means that\\
    $\ProjectILSOf{\SetOf{(\String_{1,1}',1::\IntList_1');\ldots;(\String_{1,n_1'}',n_1'::\IntList_1');
      \ldots;(\String_{m',1}',1::\IntList_{m'}');\ldots;(\String_{m',n_{m'}'}',n_{m'}'::\IntList_{m'}')}}\\
    = \ProjectILSOf{\SetOf{(\StringAlt_{1,1}',1::\IntList_1'');\ldots;(\StringAlt_{1,n_1''}',n_1''::\IntList_1'');
      \ldots;(\StringAlt_{m'',1}',1::\IntList_{m''}'');\ldots;(\String_{m'',n_{m''}''}',n_{m''}''::\IntList_{m''}'')}}$,
    
    So, by reindexing,\\
    $\ProjectILSOf{\SetOf{(\String_{1,1},1::\IntList_1);\ldots;(\String_{1,n_1},n_1::\IntList_1);
        \ldots;(\String_{m,1},1::\IntList_m);\ldots;(\String_{m,n_m},n_m::\IntList_m)}}\\
    = \ProjectILSOf{\SetOf{(\StringAlt_{1,1},1::\IntList_1);\ldots;(\StringAlt_{1,n_1},n_1::\IntList_1);
        \ldots;(\StringAlt_{m,1},1::\IntList_m);\ldots;(\String_{m,n_m},n_m::\IntList_m)}}$,\\
    $\StringIntListSet_1 =
    \SetOf{(\String_1,\IntList_1),\ldots,(\String_m,\IntList_{m})}$, and\\
    $\StringIntListSet_2 =
    \SetOf{(\StringAlt_1,\IntList_1),\ldots,(\StringAlt_m,\IntList_{m})}$, so we
    have\\
    $\ProjectILSOf{\SetOf{(\String_1,\IntList_1),\ldots,(\String_m,\IntList_{m})}}
    =
    \ProjectILSOf{\SetOf{(\StringAlt_1,\IntList_1),\ldots,(\StringAlt_m,\IntList_{m})}}$,
    as desired. (also we know by the condition which returns $\None$)

    Furthermore, by IH, for all $i,j$, we have $(\String_{i,j},\StringAlt_{i,j})
    \in \SemanticsOf{\DNFLens}$.  This means that $(\String_{i,1} \Concat \ldots
    \Concat \String_{i,n_i}, \StringAlt_{i,1} \Concat \ldots \Concat
    \StringAlt_{i,n_i}) \in \SemanticsOf{\DNFLens}$ for all $i$, so
    $(\String_i,\StringAlt_i) \in \SemanticsOf{\DNFLens}$ for all $i$.

    Furthermore, by IH, $\DNFLens \OfRewritelessType \DNFRegex \Leftrightarrow
    \DNFRegexAlt$.  Thusly, as $\Atom$ and $\AtomAlt$ are strongly unambiguous,
    $\DNFRegex$ and $\DNFRegexAlt$ are unambiguously iterable, so we have

    \[
      \inferrule*
      {
        \DNFLens \OfRewritelessType \DNFRegex \Leftrightarrow \DNFRegexAlt\\
        \UnambigItOf{\DNFRegex}\\
        \UnambigItOf{\DNFRegexAlt}
      }
      {
        \IterateLensOf{\DNFLens} \OfRewritelessType \StarOf{\DNFRegex}
        \Leftrightarrow \StarOf{\DNFRegexAlt}
      }
    \]
  \end{case}
  
  \begin{case}[sequence]
    Unfolding definitions.

    Let $\StringIntListSet_1 =
    \SetOf{(\String_1'',\IntList_1'),\ldots,(\String_{m'}'',\IntList_{m'}')}$.

    Let
    $\StringIntListSet_2 =
    \SetOf{(\StringAlt_1'',\IntList_1''),\ldots,(\StringAlt_{m''}'',\IntList_{m''}'')}$
    
    By inverison on 
    $\StringIntListSet_1 \in \Sequence \Generates \ExampledSequence$ and
    $\StringIntListSet_2 \in \SequenceAlt \Generates \ExampledSequenceAlt$,
    we know that\\
    $\SetOf{(\String_{1,i}'',\IntList_1');\ldots;(\String_{m',i}'',\IntList_{m'}')}
    \in \Atom_i \Generates \ExampledAtom_i$,\\
    $\SetOf{(\StringAlt_{1,i}'',\IntList_1'');\ldots;(\StringAlt_{m'',i}'',\IntList_{m''}'')}
    \in \AtomAlt_i \Generates \ExampledAtomAlt_i$,\\
    $\String_0'\Concat\String_{i,1}''\Concat\ldots\Concat\String_{i,n}''\Concat\String_n' = \String_i$,\\
    $\StringAlt_0'\Concat\StringAlt_{i,1}''\Concat\ldots\Concat\StringAlt_{i,n'}''\Concat\StringAlt_{n'}' = \StringAlt_i''$,\\
    $\ExampledSequence =
    (\SequenceOf{\String_0' \SeqSep \ExampledAtom_1 \SeqSep \ldots \SeqSep
      \ExampledAtom_n \SeqSep \String_n'},\SetOf{\IntList_1',\ldots,\IntList_{m'}'})$, and\\
    $\ExampledSequenceAlt =
    (\SequenceOf{\String_0' \SeqSep \ExampledAtomAlt_{1} \SeqSep \ldots \SeqSep
      \ExampledAtomAlt_{n'} \SeqSep \String_{n'}'},\SetOf{\IntList_1'',\ldots,\IntList_{m''}''})$

    Let $\sigma_1 =
    \SortingOf{\ListOf{\ExampledAtom_1;\ldots;\ExampledAtom_n}}$.
    Let $\sigma_2 =
    \SortingOf{\ListOf{\ExampledAtom_1;\ldots;\ExampledAtom_n}}$.
    Let $\sigma = \InverseOf{\sigma_1} \Compose \sigma_2$.

    As \RigidSynthSequence{} returns, $\PCF{AllSome}(\mathit{alos})$ must be true, which
    means that for each $i\in\RangeIncInc{1}{n}$,
    $\RigidSynthAtom(\ExampledAtom_i,\ExampledAtomAlt_{\sigma(i)})$ returns a
    lens, $\AtomLens_i$.
    
    By IH, that means that\\
    $\ProjectILSOf{\SetOf{(\String_{1,i}'',\IntList_1');\ldots;(\String_{m',i}'',\IntList_{m'}')}}\\
    =
    \ProjectILSOf{\SetOf{(\StringAlt_{1,\sigma(i)}'',\IntList_1'');\ldots;(\StringAlt_{m'',\sigma(i)}'',\IntList_{m''}'')}}$,
    which immediately implies that $m' = m''$.
    
    So, by reindexing, and aligning the int lists, 
    $\ProjectILSOf{\SetOf{(\String_{1,i},\IntList_1);\ldots;(\String_{m,i},\IntList_{m})}}\\
    =
    \ProjectILSOf{\SetOf{(\StringAlt_{1,\sigma(i)},\IntList_1);\ldots;(\StringAlt_{m,\sigma(i)},\IntList_{m})}}$,\\
    $\StringIntListSet_1 =
    \SetOf{(\String_1,\IntList_1),\ldots,(\String_m,\IntList_{m})}$, and\\
    $\StringIntListSet_2 =
    \SetOf{(\StringAlt_1,\IntList_1),\ldots,(\StringAlt_m,\IntList_{m})}$, so we
    have\\
    $\ProjectILSOf{\SetOf{(\String_1,\IntList_1),\ldots,(\String_m,\IntList_{m})}}\\
    =
    \ProjectILSOf{\SetOf{(\StringAlt_1,\IntList_1),\ldots,(\StringAlt_m,\IntList_{m})}}$,
    as desired. (also we know by the condition which returns $\None$)

    Furthermore, by IH, for all $(i,j)$ we have
    $(\String_{i,j},\StringAlt_{i,\sigma(j)})\in\SemanticsOf{\AtomLens_j}$.
    By the definition of sequence lens semantics,
    $(\String_0'\Concat\String_{i,1}\Concat\ldots\Concat\String_{i,n}\String_n',
    \StringAlt_0'\Concat\StringAlt_{i,\InverseOf{\sigma}(\sigma(1))}\Concat\ldots\Concat\StringAlt_{i,\InverseOf{\sigma}(\sigma(n))}\Concat\StringAlt_n')
    \in \SemanticsOf{(\SequenceLensOf{(\String_0',\StringAlt_0') \SeqLSep \AtomLens_1 \SeqLSep \ldots
        \SeqLSep \AtomLens_n \SeqLSep
        (\String_n',\StringAlt_n')},\InverseOf{\sigma})}$.
    By simplifying, we get 
    $(\String_0'\Concat\String_{i,1}\Concat\ldots\Concat\String_{i,n}\String_n',
    \StringAlt_0'\Concat\StringAlt_{i,1}\Concat\ldots\Concat\StringAlt_{i,n}\Concat\StringAlt_n')
    \in \SemanticsOf{(\SequenceLensOf{(\String_0',\StringAlt_0') \SeqLSep \AtomLens_1 \SeqLSep \ldots
        \SeqLSep \AtomLens_n \SeqLSep
        (\String_n',\StringAlt_n')},\InverseOf{\sigma})}$.
    By simplifying even more, we gets
    $(\String_i,\StringAlt_i)\in\SemanticsOf{(\SequenceLensOf{(\String_0',\StringAlt_0') \SeqLSep \AtomLens_1 \SeqLSep \ldots
        \SeqLSep \AtomLens_n \SeqLSep
        (\String_n',\StringAlt_n')},\InverseOf{\sigma})}$, as desired.

    Furthermore, by IH, $\AtomLens_i \OfRewritelessType \Atom_i \Leftrightarrow
    \AtomAlt_{\sigma(i)}$.  Thusly, as $\Sequence$ and $\SequenceAlt$ are strongly
    unambiguous, 
    they are sequence unambiguously concatenable, so we have:

    \[
      \inferrule*
      {
        \AtomLens_i \OfRewritelessType \Atom_i \Leftrightarrow \AtomAlt_{\sigma(i)}\\
        \SequenceUnambigConcatOf{\String_0';\Atom_1;\ldots;\Atom_n;\String_n'}\\
        \SequenceUnambigConcatOf{\StringAlt_0';\AtomAlt_{\InverseOf{\sigma}(\sigma(1))};\ldots;\AtomAlt_{\InverseOf{\sigma}(\sigma(n))};\StringAlt_n'}
      }
      {
        (\SequenceLensOf{(\String_0',\StringAlt_0') \SeqLSep \AtomLens_1
          \SeqLSep \ldots 
          \SeqLSep \AtomLens_n \SeqLSep
          (\String_n',\StringAlt_n')},\InverseOf{\sigma}) \OfRewritelessType
        \SequenceOf{\String_0' \SeqSep \Atom_1 \SeqSep \ldots \SeqSep \Atom_n
          \SeqSep \String_n'}
        \Leftrightarrow
        \SequenceOf{\String_0' \SeqSep \AtomAlt_{\InverseOf{\sigma}(\sigma(1))}
          \SeqSep \ldots \SeqSep \AtomAlt_{\InverseOf{\sigma}(\sigma(n))}
          \SeqSep \String_n'}
      }
    \]

    Which, by simplifying is:

    \[
      \inferrule*
      {
        \AtomLens_i \OfRewritelessType \Atom_i \Leftrightarrow \AtomAlt_{\sigma(i)}\\
        \SequenceUnambigConcatOf{\String_0';\Atom_1;\ldots;\Atom_n;\String_n'}\\
        \SequenceUnambigConcatOf{\StringAlt_0';\AtomAlt_1;\ldots;\AtomAlt_n;\StringAlt_n'}
      }
      {
        (\SequenceLensOf{(\String_0',\StringAlt_0') \SeqLSep \AtomLens_1
          \SeqLSep \ldots 
          \SeqLSep \AtomLens_n \SeqLSep
          (\String_n',\StringAlt_n')},\InverseOf{\sigma}) \OfRewritelessType
        \SequenceOf{\String_0' \SeqSep \Atom_1 \SeqSep \ldots \SeqSep \Atom_n
          \SeqSep \String_n'}
        \Leftrightarrow
        \SequenceOf{\String_0' \SeqSep \AtomAlt_{1}
          \SeqSep \ldots \SeqSep \AtomAlt_{n}
          \SeqSep \String_n'}
      }
    \]

    As desired.
  \end{case}
  
  \begin{case}[dnfregex]
    Unfolding definitions.

    Let $\StringIntListSet_1 =
    \SetOf{(\String_1'',\IntList_1'),\ldots,(\String_{m'}'',\IntList_{m'}')}$.

    Let
    $\StringIntListSet_2 =
    \SetOf{(\StringAlt_1'',\IntList_1''),\ldots,(\StringAlt_{m''}'',\IntList_{m''}'')}$
    
    By inverison on 
    $\StringIntListSet_1 \in \DNFRegex \Generates \ExampledDNFRegex$ and
    $\StringIntListSet_2 \in \DNFRegexAlt \Generates \ExampledDNFRegexAlt$,
    we know that\\
    $\Set_i = \SetOf{(\String_{i,1}',\IntList_{i,1}');\ldots;(\String_{i,m_i'}',\IntList_{i,m_i'}')}$,\\
    $\Set_i' =
    \SetOf{(\StringAlt_{i,1}',\IntList_{i,1}'');\ldots;(\StringAlt_{i,m_i''}',\IntList_{i,m_i''}'')}$,\\
    $\Set_i \in \Sequence_i \Generates \ExampledSequence_i$,\\
    $\Set_i' \in \SequenceAlt_i \Generates \ExampledSequenceAlt_i$,\\
    $\BigUnion_{i\in\RangeIncInc{1}{n}}\Set_i = \StringIntListSet_1$,\\
    $\BigUnion_{i\in\RangeIncInc{1}{n}}\Set_i' = \StringIntListSet_2$,\\
    $\ExampledDNFRegex = \DNFOf{\ExampledSequence_1 \DNFSep \ldots \DNFSep
      \ExampledSequence_n}$, and\\
    $\ExampledDNFRegexAlt = \DNFOf{\ExampledSequenceAlt_1 \DNFSep \ldots \DNFSep
      \ExampledSequenceAlt_n}$

    Let $\sigma_1 =
    \SortingOf{\ListOf{\ExampledAtom_1;\ldots;\ExampledAtom_n}}$.
    Let $\sigma_2 =
    \SortingOf{\ListOf{\ExampledAtom_1;\ldots;\ExampledAtom_n}}$.
    Let $\sigma = \InverseOf{\sigma_1} \Compose \sigma_2$.

    As \RigidSynthInternal{} returns, $\PCF{AllSome}(\mathit{sqlos})$ must be true, which
    means that for each $i\in\RangeIncInc{1}{n}$,
    $\RigidSynthSequence(\ExampledSequence_i,\ExampledSequenceAlt_{\sigma(i)})$ returns a
    lens, $\SequenceLens_i$.
    
    By IH, that means that
    $\ProjectILSOf{\Set_i}
    =
    \ProjectILSOf{\Set_{\sigma(i)}'}$.
    This means that, $\ProjectILSOf{\StringIntListSet_1} =
    \ProjectILSOf{\StringIntListSet_2}$,as they are each the union of all these
    sets.   We also know this through the fact we
    return a value, as desired.
    This immediately implies that $m' = m''$.
    Furthermore, we can use this to reindex $\Set_i$ as
    $\Set_i =
    \SetOf{(\String_{i,1},\IntList_{i,1});\ldots;(\String_{i,m_i},\IntList_{i,m_i})}$,
    and $\Set_i'$ as
    $\Set_{\sigma(i)}' =
    \SetOf{(\StringAlt_{i,1},\IntList_{i,1});\ldots;(\StringAlt_{i,m_i},\IntList_{i,m_i})}$.

    Furthermore, by IH, for all $(i,j)$ we have
    $(\String_{i,j},\StringAlt_{i,j})\in\SemanticsOf{\SequenceLens_i}$.
    By the definition of dnf lens semantics,
    $\SemanticsOf{(\DNFLensOf{\SequenceLens_1 \DNFLSep \ldots \DNFLSep
        \SequenceLens_n},\InverseOf{\sigma})} =
    \BigUnion_{i\in\RangeIncInc{1}{n}}\SemanticsOf{\SequenceLens_i}$.
    Let $(\String_i,\StringAlt_i)$ arbitrary from $\StringIntListSet_1$ and
    $\StringIntListSet_2$ sharing the same int list $\IntList_i$.
    By them being the union of all $\Set_i$ and $\Set_i'$, there exists some $i',j$
    such that $\String_i = \String_{i,j}$ and $\StringAlt_i = \StringAlt_{i,j}$.
    As such, $(\String_i,\StringAlt_i) \in \SemanticsOf{(\DNFLensOf{\SequenceLens_1 \DNFLSep \ldots \DNFLSep
        \SequenceLens_n},\InverseOf{\sigma})}$, as desired.

    Furthermore, by IH, $\SequenceLens_i \OfRewritelessType \Sequence_i \Leftrightarrow
    \SequenceAlt_{\sigma(i)}$.  Thusly, as $\DNFRegex$ and $\DNFRegexAlt$ are strongly
    unambiguous, 
    they are pairwise disjoint in sequences, so we have:

    \[
      \inferrule*
      {
        \SequenceLens_i \OfRewritelessType \Sequence_i \Leftrightarrow \SequenceAlt_{\sigma(i)}\\
        i \neq j \Rightarrow \LanguageOf{\Sequence_{i}} \cap \LanguageOf{\Sequence_{j}}=\emptyset\\
        i \neq j \Rightarrow \LanguageOf{\SequenceAlt_{i}} \cap \LanguageOf{\SequenceAlt_{j}}=\emptyset
      }
      {
        (\DNFLensOf{\SequenceLens_1
          \DNFLSep \ldots 
          \DNFLSep \SequenceLens_n},\InverseOf{\sigma}) \OfRewritelessType
        \DNFOf{\Sequence_1 \DNFSep \ldots \DNFSep \Sequence_n}
        \Leftrightarrow
        \DNFOf{\SequenceAlt_{\InverseOf{\sigma}(\sigma(1))} \DNFSep \ldots \DNFSep \SequenceAlt_{\InverseOf{\sigma}(\sigma(n))}}
      }
    \]

    Which, by simplifying is:

    \[
      \inferrule*
      {
        \SequenceLens_i \OfRewritelessType \Sequence_i \Leftrightarrow \SequenceAlt_{\sigma(i)}\\
        i \neq j \Rightarrow \LanguageOf{\Sequence_{i}} \cap \LanguageOf{\Sequence_{j}}=\emptyset\\
        i \neq j \Rightarrow \LanguageOf{\SequenceAlt_{i}} \cap \LanguageOf{\SequenceAlt_{j}}=\emptyset
      }
      {
        (\DNFLensOf{\SequenceLens_1
          \DNFLSep \ldots 
          \DNFLSep \SequenceLens_n},\InverseOf{\sigma}) \OfRewritelessType
        \DNFOf{\Sequence_1 \DNFSep \ldots \DNFSep \Sequence_n}
        \Leftrightarrow
        \DNFOf{\SequenceAlt_1 \DNFSep \ldots \DNFSep \SequenceAlt_n}
      }
    \]

    As desired.
  \end{case}
\end{proof}

\begin{lemma}
  \label{lem:rigidsynth-sound}
  Let $\DNFRegex$ and $\DNFRegexAlt$ be strongly unambiguous DNF regular
  expressions.
  Let\\ $\ListOf{(\String_1,\StringAlt_1);\ldots;(\String_m,\StringAlt_m)}$ be a
  list of input-output examples.
  If $\RigidSynth(\ExampledDNFRegex,\ExampledDNFRegexAlt)$ returns
  a DNF lens, $\SomeOf{\DNFLens}$, then  $\DNFLens \OfRewritelessType \DNFRegex 
  \Leftrightarrow \DNFRegexAlt$, and $(\String_i,\StringAlt_i) \in
  \SemanticsOf{\DNFLens}$. 
\end{lemma}
\begin{proof}
  Let $\ListOf{(\String_1,\StringAlt_1);\ldots;(\String_n,\StringAlt_n)} =
  \Examples$.
  
  Let $\ExampledDNFRegex =
  \EmbedExamplesOf{\ListOf{([1],\String_1);\ldots;([n],\String_n)}}{\DNFRegex}$

  Let $\ExampledDNFRegexAlt = \EmbedExamplesOf{\ListOf{([1],\StringAlt_1);\ldots;([n],\StringAlt_n)}}{\DNFRegexAlt}$

  This means that $\ListOf{([1],\String_1);\ldots;([n],\String_n)} \in \DNFRegex
  \Generates \ExampledDNFRegex$ and
  $\ListOf{([1],\StringAlt_1);\ldots;([n],\StringAlt_n)} \in \DNFRegexAlt
  \Generates \ExampledDNFRegexAlt$.

  As \RigidSynth{} returns a lens, then
  $\RigidSynthInternal(\ExampledDNFRegex,\ExampledDNFRegexAlt)$ must return a
  lens.

  By the above statements, Lemma~\ref{lem:rigidsynth-internal-sound} applies, so
  the DNF lens returned by it satisfies $\DNFLens \OfRewritelessType \DNFRegex
  \Leftrightarrow \DNFRegexAlt$, and
  $(\String_i,\StringAlt_i)\in\SemanticsOf{\DNFLens}$.

  As $\DNFLens \OfRewritelessType \DNFRegex \Leftrightarrow \DNFRegexAlt$, we
  know that $\DNFRegex$ and $\DNFRegexAlt$ are strongly unambiguous.

  As such the conditions for Lemma~\ref{lem:exampledordercomplete} apply so
  $\ExampledDNFRegex \ExampledDNFLeq \ExampledDNFRegexAlt$ and
  $\ExampledDNFRegexAlt \ExampledDNFLeq \ExampledDNFRegex$.

  By Lemma~\ref{lem:exampledrigidcompleteorder}, that means that
  $\RigidSynthInternal(\ExampledDNFRegex,\ExampledDNFRegexAlt)$ returns a DNF
  lens, so then so too does $\RigidSynth$.

    Let $\DNFRegex$ and $\DNFRegexAlt$ be strongly unambiguous DNF regular expressions.
    Let $\StringIntListSet_1$ be a string int list set.
    Let $\StringIntListSet_2$ be a string int list set.
    Let $\ExampledDNFRegex$ and $\ExampledDNFRegexAlt$ be exampled DNF regular expressions.
    Let $\StringIntListSet_1 \in \ExampledDNFRegex \Generates \ExampledDNFRegex$.
    Let $\StringIntListSet_2 \in \ExampledDNFRegexAlt \Generates \ExampledDNFRegexAlt$.
    If $\RigidSynthInternal(\ExampledDNFRegex,\ExampledDNFRegexAlt)$ returns
    a sequence lens, $\SomeOf{\DNFLens}$, then $\ProjectILSOf{\StringIntListSet_1} =
    \ProjectILSOf{\StringIntListSet_2}$,
    $\DNFLens \OfRewritelessType \DNFRegex 
    \Leftrightarrow \DNFRegexAlt$, and for each $(\String,\StringAlt)$ pair with
    the same int list in $\StringIntListSet_1$ and $\StringIntListSet_2$,
    $(\String,\StringAlt) \in \SemanticsOf{\DNFLens}$.

\end{proof}

\begin{lemma}
  \label{lem:dnfsynth-sound}
  Let $\DNFRegex$ and $\DNFRegexAlt$ be strongly unambiguous DNF regular
  expressions.
  Let\\ $\ListOf{(\String_1,\StringAlt_1);\ldots;(\String_m,\StringAlt_m)}$ be a
  list of input-output examples.
  If $\SynthDNFLens(\ExampledDNFRegex,\ExampledDNFRegexAlt)$ returns
  a DNF lens, $\SomeOf{\DNFLens}$, then  $\DNFLens \OfType \DNFRegex 
  \Leftrightarrow \DNFRegexAlt$, and $(\String_i,\StringAlt_i) \in
  \SemanticsOf{\DNFLens}$. 
\end{lemma}
\begin{proof}
  If $\SynthDNFLens$ returns, then there must be some $(\Regex',\RegexAlt')$
  popped from the queue which returned a DNF lens $\DNFLens =
  \RigidSynthInternal(\Regex',\RegexAlt')$.
  These regular expressions are 
  such that $\DNFRegex \StarOf{\RewriteDNF} \DNFRegex'$ and $\DNFRegexAlt
  \StarOf{\RewriteDNF} \DNFRegexAlt'$, as the regular expressions are always
  either the originals, or have been added to the queue from an expansion on a
  previously popped element.  Inductively, everything in the queue is an
  expansion on $\DNFRegex$ or $\DNFRegexAlt$.

  As $\DNFRegex$ and $\DNFRegexAlt$ are strongly unambiguous DNF regular
  expressions, $\DNFRegex \StarOf{\RewriteDNF} \DNFRegex'$ means that
  $\DNFRegex'$ is strongly unambiguous, and similarly for $\DNFRegexAlt$ and
  $\DNFRegexAlt'$.

  So, by Lemma~\ref{lem:rigidsynth-sound}, $\DNFLens \OfRewritelessType
  \DNFRegex' \Leftrightarrow \DNFRegexAlt'$.  Furthermore, for each
  $(\String_i,\StringAlt_i)$ pair in the examples
  $(\String_i,\StringAlt_i) \in \SemanticsOf{\DNFLens}$.

  Lastly, we can then build the typing derivation:
  \[
    \inferrule*
    {
      \DNFLens \OfRewritelessType \DNFRegex' \Leftrightarrow \DNFRegexAlt'\\
      \DNFRegex \StarOf{\RewriteDNF} \DNFRegex'\\
      \DNFRegexAlt \StarOf{\RewriteDNF} \DNFRegexAlt'
    }
    {
      \DNFLens \OfRewritelessType \DNFRegex \Leftrightarrow \DNFRegexAlt\\
    }
  \]
\end{proof}

\begin{theorem}
  For all lenses $\Lens$, regular expressions $\Regex$ and $\RegexAlt$, and
  examples $\Examples$, 
  if \\$\Lens = \SynthLens(\Regex,\RegexAlt,\Examples)$, then
  $\Lens \OfType \Regex \Leftrightarrow \RegexAlt$ and for all
  $(\String,\StringAlt)$ in $\Examples$, $(\String,\StringAlt) \in
  \SemanticsOf{\Lens}$.
\end{theorem}
\begin{proof}
  As \SynthLens{} returns, then $\SynthDNFLens(\DNFRegex,\DNFRegexAlt)$ returns,
  where $\DNFRegex = \ToDNFRegexOf{\Regex}$ and $\DNFRegexAlt =
  \ToDNFRegexOf{\RegexAlt}$.  As $\PCF{Validate}(\Regex,\RegexAlt,\Examples)$
  does not error out, then we know $\Regex$ and $\RegexAlt$ are strongly
  unambiguous, and $\Examples$ matches them.  This then means that $\DNFRegex$
  and $\DNFRegexAlt$ are also strongly unambiguous.

  As such, by Lemma~\ref{lem:dnfsynth-sound}, we know that $\DNFLens =
  \SynthDNFLens(\DNFRegex,\DNFRegexAlt)$ returns, that $\DNFLens \OfType
  \DNFRegex \Leftrightarrow \DNFRegex$, and that for each $(\String,\StringAlt)$
  in the examples, $(\String,\StringAlt) \in \SemanticsOf{\DNFLens}$.  By Theorem~\ref{thm:dnfls}, there
  exists $\Regex'$ and $\RegexAlt'$ such that $\ToLensOf{\DNFLens}$ and
  $\ToDNFRegexOf{\Regex'} = \DNFRegex$ and $\ToDNFRegexOf{\RegexAlt'} =
  \DNFRegexAlt$ and $\SemanticsOf{\ToLensOf{\DNFLens}} =
  \SemanticsOf{\DNFLens}$.

  As $\ToDNFRegexOf{\Regex} = \ToDNFRegexOf{\Regex'}$ and
  $\ToDNFRegexOf{\RegexAlt} = \ToDNFRegexOf{\RegexAlt'}$, we have
  $\Regex \SSREquiv \Regex'$ and $\RegexAlt \SSREquiv \RegexAlt'$.

  \[
    \inferrule*
    {
      \ToLensOf{\DNFLens} \OfType \Regex' \Leftrightarrow \RegexAlt'\\
      \Regex \SSREquiv \Regex'\\
      \RegexAlt \SSREquiv \RegexAlt'\\
    }
    {
      \ToLensOf{\DNFLens} \OfType \Regex \Leftrightarrow \RegexAlt
    }
  \]

  Furthermore, as $\SemanticsOf{\DNFLens} = \SemanticsOf{\ToLensOf{\DNFLens}}$,
  we have for each $(\String,\StringAlt)$ in the examples, $(\String,\StringAlt)
  \in \SemanticsOf{\ToLensOf{\DNFLens}}$, as desired.
\end{proof}

\begin{lemma}\leavevmode
  \label{lem:exampledordercomplete}
  \begin{itemize}
  \item
    Let $\Atom$ and $\AtomAlt$ be strongly unambiguous atoms.
    Let $\StringIntListSet_1$ be a string int list set.
    Let $\StringIntListSet_2$ be a string int list set.
    Let $\ProjectILSOf{\StringIntListSet_1} =
    \ProjectILSOf{\StringIntListSet_2}$.
    Let $\ExampledAtom$ and $\ExampledAtomAlt$ be exampled atoms.
    Let $\StringIntListSet_1 \in \Atom \Generates \ExampledAtom$.
    Let $\StringIntListSet_2 \in \AtomAlt \Generates \ExampledAtomAlt$.
    If $\AtomLens \OfRewritelessType \Atom \Leftrightarrow \AtomAlt$,
    and for each $(\String,\StringAlt)$ pair with the same int list in
    $\StringIntListSet_1$ and $\StringIntListSet_2$, $(\String,\StringAlt) \in
    \SemanticsOf{\AtomLens}$, then
    $\ExampledAtom \ExampledAtomLeq \ExampledAtomAlt$ and
    $\ExampledAtomAlt \ExampledAtomLeq \ExampledAtom$.
  \item
    Let $\Sequence$ and $\SequenceAlt$ be strongly unambiguous sequences.
    Let $\StringIntListSet_1$ be a string int list set.
    Let $\StringIntListSet_2$ be a string int list set.
    Let $\ProjectILSOf{\StringIntListSet_1} =
    \ProjectILSOf{\StringIntListSet_2}$.
    Let $\ExampledSequence$ and $\ExampledSequenceAlt$ be exampled sequences.
    Let $\StringIntListSet_1 \in \Sequence \Generates \ExampledSequence$.
    Let $\StringIntListSet_2 \in \SequenceAlt \Generates \ExampledSequenceAlt$.
    If $\SequenceLens \OfRewritelessType \Sequence \Leftrightarrow \SequenceAlt$,
    and for each $(\String,\StringAlt)$ pair with the same int list in
    $\StringIntListSet_1$ and $\StringIntListSet_2$, $(\String,\StringAlt) \in
    \SemanticsOf{\SequenceLens}$, then
    $\ExampledSequence \ExampledSequenceLeq \ExampledSequenceAlt$ and
    $\ExampledSequenceAlt \ExampledSequenceLeq \ExampledSequence$.
  \item
    Let $\DNFRegex$ and $\DNFRegexAlt$ be strongly unambiguous DNF regular expressions.
    Let $\StringIntListSet_1$ be a string int list set.
    Let $\StringIntListSet_2$ be a string int list set.
    Let $\ProjectILSOf{\StringIntListSet_1} =
    \ProjectILSOf{\StringIntListSet_2}$.
    Let $\ExampledDNFRegex$ and $\ExampledDNFRegexAlt$ be exampled DNF regular
    expressions.
    Let $\StringIntListSet_1 \in \ExampledDNFRegex \Generates \ExampledDNFRegex$.
    Let $\StringIntListSet_2 \in \ExampledDNFRegexAlt \Generates \ExampledDNFRegexAlt$.
    If $\DNFLens \OfRewritelessType \DNFRegex \Leftrightarrow \DNFRegexAlt$,
    and for each $(\String,\StringAlt)$ pair with the same int list in
    $\StringIntListSet_1$ and $\StringIntListSet_2$, $(\String,\StringAlt) \in
    \SemanticsOf{\DNFLens}$, then
    $\ExampledDNFRegex \ExampledDNFLeq \ExampledDNFRegexAlt$ and
    $\ExampledDNFRegexAlt \ExampledDNFLeq \ExampledDNFRegex$.
  \end{itemize}
\end{lemma}
\begin{proof}
  We proceed by mutual induction
  \begin{case}[atom case]
    Unfolding definitions.

    Let $\StringIntListSet_1 =
    \SetOf{(\String_1,\IntList_1),\ldots,(\String_m,\IntList_m)}$.

    Let
    $\StringIntListSet_2 =
    \SetOf{(\StringAlt_1,\IntList_1),\ldots,(\StringAlt_m,\IntList_m)}$

    By inversion on $\AtomLens \OfRewritelessType \Atom \Leftrightarrow
    \AtomAlt$, we know $\StarOf{\DNFRegex} = \Atom$, $\StarOf{\DNFRegexAlt} =
    \AtomAlt$, $\IterateLensOf{\DNFLens} = \AtomLens$, and $\DNFLens
    \OfRewritelessType \DNFRegex \Leftrightarrow \DNFRegexAlt$.

    By inversion on $\SemanticsOf{\IterateLensOf{\DNFLens}}$, there exist
    $\String_{1,1}\Concat\ldots\Concat\String_{1,n_1} = \String_1$ through
    $\String_{m,1}\Concat\ldots\Concat\String_{m,n_m} = \String_m$ and
    $\StringAlt_{1,1}\Concat\ldots\Concat\StringAlt_{1,n_1} = \StringAlt_1$ through
    $\StringAlt_{m,1}\Concat\ldots\Concat\String_{m,n_m} = \StringAlt_m$ such that
    $(\String_{i,j},\StringAlt_{i,j}) \in \SemanticsOf{\DNFLens}$.

    Furthermore, this means that $\String_{i,j} \in \LanguageOf{\DNFRegex}$ and
    $\StringAlt_{i,j} \in \LanguageOf{\DNFRegexAlt}$.
    
    By inverison on 
    $\StringIntListSet_1 \in \Atom \Generates \ExampledAtom$ and
    $\StringIntListSet_2 \in \AtomAlt \Generates \ExampledAtomAlt$,
    we know that\\
    $\SetOf{(\String_{1,1}',1::\IntList_1);\ldots;(\String_{1,n_1'}',n_1'::\IntList_1);
      \ldots;(\String_{m,1}',1::\IntList_m);\ldots;(\String_{m,n_m'}',n_m'::\IntList_m)}
    \in \DNFRegex \Generates \ExampledDNFRegex$,\\
    $\SetOf{(\StringAlt_{1,1}',1::\IntList_1);\ldots;(\StringAlt_{1,n_1''}',n_1''::\IntList_1);
      \ldots;(\StringAlt_{m,1}',1::\IntList_m);\ldots;(\String_{m,n_m''}',n_m''::\IntList_m)}
    \in \DNFRegexAlt \Generates \ExampledDNFRegexAlt$,\\
    $\String_{i,1}'\Concat\ldots\Concat\String_{i,n_i'}' = \String_i$,\\
    $\StringAlt_{i,1}'\Concat\ldots\Concat\StringAlt_{i,n_i''}' = \String_i$,\\
    $\ExampledAtom =
    (\StarOf{\ExampledDNFRegex},\SetOf{\IntList_1,\ldots,\IntList_m})$, and\\
    $\ExampledAtomAlt =
    (\StarOf{\ExampledDNFRegexAlt},\SetOf{\IntList_1,\ldots,\IntList_m})$
    
    As $\Atom$ and $\AtomAlt$ are strongly unambiguous, $n_i = n_i' =
    n_i''$, $\String_{i,j} = \String_{i,j}'$, $\StringAlt_{i,j} =
    \StringAlt_{i,j}'$, and $\DNFRegex$ and $\DNFRegexAlt$ are strongly unambiguous.

    By inspection\\$\ProjectILSOf{\SetOf{(\String_{1,1},1::\IntList_1);\ldots;(\String_{1,n_1},n_1::\IntList_1);
      \ldots;(\String_{m,1},1::\IntList_m);\ldots;(\String_{m,n_m},n_m::\IntList_m)}}\\
    = \ProjectILSOf{\SetOf{(\StringAlt_{1,1},1::\IntList_1);\ldots;(\StringAlt_{1,n_1},n_1::\IntList_1);
        \ldots;(\StringAlt_{m,1},1::\IntList_m);\ldots;(\String_{m,n_m},n_m::\IntList_m)}}$

    Using the above facts, we have satisfied the preconditions to use the IH, so
    $\ExampledDNFRegex \ExampledDNFLeq \ExampledDNFRegexAlt$ and
    $\ExampledDNFRegexAlt \ExampledDNFLeq \ExampledDNFRegex$.

    Furthermore, $\SetOf{\IntList_1,\ldots,\IntList_m} \ILSLeq
    \SetOf{\IntList_1,\ldots,\IntList_m}$, so
    $\ExampledAtom \ExampledAtomLeq \ExampledAtomAlt$ and
    $\ExampledAtomAlt \ExampledAtomLeq \ExampledAtom$.
  \end{case}

  \begin{case}[sequence case]
    Unfolding definitions.

    Let $\StringIntListSet_1 =
    \SetOf{(\String_1,\IntList_1),\ldots,(\String_m,\IntList_m)}$.

    Let
    $\StringIntListSet_2 =
    \SetOf{(\StringAlt_1,\IntList_1),\ldots,(\StringAlt_m,\IntList_m)}$

    By inversion on $\SequenceLens \OfRewritelessType \Sequence \Leftrightarrow
    \SequenceAlt$, we know
    $\SequenceOf{\String_0' \SeqSep \Atom_1 \SeqSep \ldots \SeqSep \Atom_n
      \SeqSep \String_n'} = \Sequence$,
    $\SequenceOf{\StringAlt_0' \SeqSep \AtomAlt_{\sigma(1)} \SeqSep \ldots \SeqSep \AtomAlt_{\sigma(n)}
      \SeqSep \StringAlt_n'} = \SequenceAlt$,
    $(\SequenceLensOf{(\String_0',\StringAlt_0') \SeqLSep \AtomLens_1 \SeqLSep
      \ldots \SeqLSep \AtomLens_n \SeqLSep (\String_n',\StringAlt_n')}, \sigma)
    = \SequenceLens$, and $\AtomLens_i
    \OfRewritelessType \Atom_i \Leftrightarrow \AtomAlt_i$.

    By inversion on $\SemanticsOf{(\SequenceLensOf{(\String_0',\StringAlt_0') \SeqLSep \AtomLens_1 \SeqLSep
      \ldots \SeqLSep \AtomLens_n \SeqLSep (\String_n',\StringAlt_n')},
    \sigma)}$,
    there exist
    $\String_0'\Concat\String_{1,1}\Concat\ldots\Concat\String_{1,n}\Concat\String_n' = \String_1$ through
    $\String_0'\Concat\String_{m,1}\Concat\ldots\Concat\String_{m,n}\Concat\String_n' = \String_m$ and
    $\StringAlt_0'\Concat\StringAlt_{1,\sigma(1)}\Concat\ldots\Concat\StringAlt_{1,\sigma(n)}\Concat\StringAlt_n' = \StringAlt_1$ through
    $\StringAlt_0'\Concat\StringAlt_{m,\sigma(1)}\Concat\ldots\Concat\String_{m,\sigma(n)}\Concat\StringAlt_n' = \StringAlt_m$ such that
    $(\String_{i,j},\StringAlt_{i,j}) \in \SemanticsOf{\AtomLens_j}$.

    Furthermore, this means that $\String_{i,j} \in \LanguageOf{\Atom_j}$ and
    $\StringAlt_{i,j} \in \LanguageOf{\AtomAlt_j}$.
    
    By inverison on 
    $\StringIntListSet_1 \in \Sequence \Generates \ExampledSequence$ and
    $\StringIntListSet_2 \in \SequenceAlt \Generates \ExampledSequenceAlt$,
    we know that\\
    $\SetOf{(\String_{1,i}'',\IntList_1);\ldots;(\String_{m,i}'',\IntList_m)}
    \in \Atom_i \Generates \ExampledAtom_i$,\\
    $\SetOf{(\StringAlt_{1,i}'',\IntList_1);\ldots;(\StringAlt_{m,i}'',\IntList_m)}
    \in \AtomAlt_i \Generates \ExampledAtomAlt_i$,\\
    $\String_0'\Concat\String_{i,1}''\Concat\ldots\Concat\String_{i,n}''\Concat\String_n' = \String_i$,\\
    $\StringAlt_0'\Concat\StringAlt_{i,1}''\Concat\ldots\Concat\StringAlt_{i,n}''\Concat\StringAlt_n' = \StringAlt_i$,\\
    $\ExampledSequence =
    (\SequenceOf{\String_0' \SeqSep \ExampledAtom_1 \SeqSep \ldots \SeqSep
      \ExampledAtom_n \SeqSep \String_n'},\SetOf{\IntList_1,\ldots,\IntList_m})$, and\\
    $\ExampledSequenceAlt =
    (\SequenceOf{\String_0' \SeqSep \ExampledAtomAlt_{\sigma(1)} \SeqSep \ldots \SeqSep
      \ExampledAtomAlt_{\sigma(n)} \SeqSep \String_n'},\SetOf{\IntList_1,\ldots,\IntList_m})$
    
    As $\Sequence$ and $\SequenceAlt$ are strongly unambiguous,
    $\String_{i,j} = \String_{i,j}''$, $\StringAlt_{i,j} =
    \StringAlt_{i,j}''$, and $\Atom_i$ and $\AtomAlt_i$ are strongly unambiguous.

    By inspection
    $\ProjectILSOf{\SetOf{(\String_{1,i},\IntList_1);\ldots;(\String_{m,i},\IntList_m)}}
    = \ProjectILSOf{\SetOf{(\StringAlt_{1,i},\IntList_1);\ldots;(\StringAlt_{m,i},\IntList_m)}}$

    Using the above facts, we have satisfied the preconditions to use the IH, so
    $\ExampledAtom_i \ExampledAtomLeq \ExampledAtomAlt_i$ and
    $\ExampledAtomAlt_i \ExampledAtomLeq \ExampledAtom_i$.

    As such, from Lemma~\ref{lem:set-list-perm},
    $\ListOf{\ExampledAtom_1;\ldots;\ExampledAtom_n}
    \SetOfListOrderOf{\ExampledAtomLeq}
    \ListOf{\ExampledAtomAlt_{\sigma(1)};\ldots;\ExampledAtomAlt_{\sigma(n)}}$.

    This means that, as
    $\SetOf{\IntList_1,\ldots,\IntList_m}
    \ILSLeq
    \SetOf{\IntList_1,\ldots,\IntList_m}$,
    we have\\
    $(\ListOf{\ExampledAtom_1;\ldots;\ExampledAtom_n},\SetOf{\IntList_1,\ldots,\IntList_m})
    (\SetOfListOrderOf{\ExampledAtomLeq},\ILSLeq)
    (\ListOf{\ExampledAtomAlt_{\sigma(1)};\ldots;\ExampledAtomAlt_{\sigma(n)}},\SetOf{\IntList_1,\ldots,\IntList_m})$.
    Which means that $\ExampledSequence \ExampledSequenceLeq
    \ExampledSequenceAlt$.
  \end{case}

  \begin{case}[DNF regex case]
    Unfolding definitions.

    Let $\StringIntListSet_1 =
    \SetOf{(\String_1,\IntList_1),\ldots,(\String_m,\IntList_m)}$.

    Let
    $\StringIntListSet_2 =
    \SetOf{(\StringAlt_1,\IntList_1),\ldots,(\StringAlt_m,\IntList_m)}$

    By inversion on $\DNFLens \OfRewritelessType \DNFRegex \Leftrightarrow
    \DNFRegexAlt$, we know
    $\DNFOf{\Sequence_1 \DNFSep \ldots \DNFSep \Sequence_n} = \DNFRegex$,
    $\DNFOf{\SequenceAlt_{\sigma(1)} \DNFSep \ldots \DNFSep
      \SequenceAlt_{\sigma(n)}} = \DNFRegexAlt$,
    $(\DNFLensOf{(\SequenceLens_1 \DNFLSep \ldots \DNFLSep \SequenceLens_n}, \sigma)
    = \SequenceLens$, and $\SequenceLens_i
    \OfRewritelessType \Sequence_i \Leftrightarrow \SequenceAlt_i$.

    By inversion on $\SemanticsOf{(\DNFLensOf{(\SequenceLens_1 \DNFLSep \ldots
        \DNFLSep \SequenceLens_n}, \sigma)}$,
    there exist
    $\SetOf{(\String_{1,1},\StringAlt_{1,1}),\ldots,(\String_{1,m_1},\StringAlt_{1,m_1})} = \Set_1$ through
    $\SetOf{(\String_{n,1},\StringAlt_{n,1}),\ldots,(\String_{n,m_n},\StringAlt_{n,m_n})} = \Set_n$ such that
    $\BigUnion_{i\in\RangeIncInc{1}{n}}\Set_i =
    \SetOf{(\String_1,\StringAlt_1),\ldots,(\String_m,\StringAlt_m)}$,
    such that $\Set_i \subset \SemanticsOf{\SequenceLens_i}$, also with this
    reindexing
    Furthermore, this means that $\String_{i,j} \in \LanguageOf{\Sequence_i}$ and
    $\StringAlt_{i,j} \in \LanguageOf{\SequenceAlt_i}$.
    
    By inverison on 
    $\StringIntListSet_1 \in \DNFRegex \Generates \ExampledDNFRegex$ and
    $\StringIntListSet_2 \in \DNFRegexAlt \Generates \ExampledDNFRegexAlt$,
    we know that\\
    $\SetOf{(\String_{i,1}',\IntList_{i,1}');\ldots;(\String_{i,m_i'}',\IntList_{i,m_i'}')}
    \in \Sequence_i \Generates \ExampledSequence_i$,\\
    $\SetOf{(\StringAlt_{1,i}',\IntList_{i,1}'');\ldots;(\StringAlt_{i,m_i''}',\IntList_{i,m_i''}'')}
    \in \SequenceAlt_i \Generates \ExampledSequenceAlt_i$,\\
    $\BigUnion_{i \in \RangeIncInc{1}{n}}
    \SetOf{\String_{i,1}',\ldots,\String_{i,m_i'}'} =
    \SetOf{\String_1,\ldots,\String_m}$\\
    $\BigUnion_{i \in \RangeIncInc{1}{n}}
    \SetOf{\StringAlt_{i,1}',\ldots,\StringAlt_{i,m_i''}'} =
    \SetOf{\StringAlt_1,\ldots,\StringAlt_m}$\\
    $\ExampledDNFRegex =
    (\DNFOf{\ExampledSequence_1 \DNFSep \ldots \DNFSep
      \ExampledSequence_n},\SetOf{\IntList_1,\ldots,\IntList_m})$, and\\
    $\ExampledDNFRegexAlt =
    (\DNFOf{\ExampledSequenceAlt_{\sigma(1)} \DNFSep \ldots \DNFSep
      \ExampledSequenceAlt_{\sigma(n)}},\SetOf{\IntList_1,\ldots,\IntList_m})$
    
    As $\DNFRegex$ and $\DNFRegexAlt$ are strongly unambiguous,
    $\SetOf{\String_{i,1},\ldots,\String_{i,m_i}} = \SetOf{\String_{i,1}',\ldots,\String_{i,m_i'}'}$,\\
    $\SetOf{\StringAlt_{i,1},\ldots,\StringAlt_{i,m_i}} =
    \SetOf{\StringAlt_{i,1}',\ldots,\StringAlt_{i,m_i'}'}$,\\
    By aligning with their int lists (which are unique), we get
    $\IntListSet_i =
    \SetOf{(\String_{i,1}',\IntList_{i,1}');\ldots;(\String_{i,m_i'}',\IntList_{i,m_i'}')}
    =
    \SetOf{(\String_{i,1},\IntList_{i,1});\ldots;(\String_{i,m_i},\IntList_{i,m_i})}$
    and\\
    $\IntListSet_i' =
    \SetOf{(\StringAlt_{1,i}',\IntList_{i,1}'');\ldots;(\StringAlt_{i,m_i''}',\IntList_{i,m_i''}'')}
    =
    \SetOf{(\StringAlt_{1,i},\IntList_{i,1});\ldots;(\StringAlt_{i,m_i},\IntList_{i,m_i})}$

    By inspection
    $\ProjectILSOf{\SetOf{(\String_{i,1},\IntList_{i,1});\ldots;(\String_{i,m_i},\IntList_{i,m_i})}}
    =
    \ProjectILSOf{\SetOf{(\StringAlt_{1,i},\IntList_{i,1});\ldots;(\StringAlt_{i,m_i},\IntList_{i,m_i})}}$ 

    Using the above facts, we have satisfied the preconditions to use the IH, so
    $\ExampledSequence_i \ExampledSequenceLeq \ExampledSequenceAlt_i$ and
    $\ExampledSequenceAlt_i \ExampledSequenceLeq \ExampledSequence_i$.

    As such, from Lemma~\ref{lem:set-list-perm},
    $\ListOf{\ExampledSequence_1;\ldots;\ExampledSequence_n}
    \SetOfListOrderOf{\ExampledSequenceLeq}
    \ListOf{\ExampledSequenceAlt_{\sigma(1)};\ldots;\ExampledSequenceAlt_{\sigma(n)}}$.

    This means that, as
    $\SetOf{\IntList_1,\ldots,\IntList_m}
    \ILSLeq
    \SetOf{\IntList_1,\ldots,\IntList_m}$,
    we have\\
    $(\ListOf{\ExampledSequence_1;\ldots;\ExampledSequence_n},\SetOf{\IntList_1,\ldots,\IntList_m})
    (\SetOfListOrderOf{\ExampledSequenceLeq},\ILSLeq)
    (\ListOf{\ExampledSequenceAlt_{\sigma(1)};\ldots;\ExampledSequenceAlt_{\sigma(n)}},\SetOf{\IntList_1,\ldots,\IntList_m})$.
    Which means that $\ExampledDNFRegex \ExampledDNFLeq
    \ExampledDNFRegexAlt$.
  \end{case}
\end{proof}

\begin{lemma}\leavevmode
\label{lem:exampledrigidcompleteorder}
  \begin{itemize}
  \item
    Let $\ExampledAtom$ and $\ExampledAtomAlt$ be exampled atoms.
    If $\ExampledAtom \ExampledAtomLeq \ExampledAtomAlt$ and
    $\ExampledAtomAlt \ExampledAtomLeq \ExampledAtom$, then\\
    $\RigidSynthAtom$ returns an atom lens.
  \item
    Let $\ExampledSequence$ and $\ExampledSequenceAlt$ be exampled atoms.
    If $\ExampledSequence \ExampledSequenceLeq \ExampledSequenceAlt$ and
    $\ExampledSequenceAlt \ExampledSequenceLeq \ExampledSequence$, then
    $\RigidSynthSequence$ returns a sequence lens.
  \item
    Let $\ExampledDNFRegex$ and $\ExampledDNFRegexAlt$ be exampled atoms.
    If $\ExampledDNFRegex \ExampledDNFLeq \ExampledDNFRegexAlt$ and
    $\ExampledDNFRegexAlt \ExampledDNFLeq \ExampledDNFRegex$, then
    $\RigidSynthInternal$ returns a DNF lens.
  \end{itemize}
\end{lemma}
\begin{proof}
\begin{case}[atoms]
  Let $(\StarOf{\ExampledDNFRegex},\IntListSet) = \ExampledAtom$
  Let $(\StarOf{\ExampledDNFRegexAlt},\IntListSet') = \ExampledAtomAlt$.
  As $\ExampledAtom \ExampledAtomLeq \ExampledAtomAlt$, we have
  $\ExampledDNFRegex \ExampledDNFLeq \ExampledDNFRegexAlt$.
  By IH, this means that
  $\RigidSynthInternal(\ExampledDNFRegex,\ExampledDNFRegexAlt)$ returns a DNF
  lens, $\DNFLens$, this means that the match goes to the first, returning
  $\IterateLensOf{\DNFLens}$, an atom lens.
\end{case}

\begin{case}[seqs]
  Let $(\SequenceOf{\String_0 \SeqSep \ExampledAtom_1 \SeqSep \ldots \SeqSep \ExampledAtom_n
    \SeqSep \String_n},\IntListSet) = \ExampledSequence$ and
  $(\SequenceOf{\StringAlt_0 \SeqSep \ExampledAtomAlt_1 \SeqSep \ldots \SeqSep
    \ExampledAtomAlt_m \SeqSep \StringAlt_m},\IntListSet') =
  \ExampledSequenceAlt$.

  Let $\sigma_1 = \SortingOf{\ListOf{\ExampledAtom_1;\ldots;\ExampledAtom_n}}$
  Let $\sigma_2 = \SortingOf{\ListOf{\ExampledAtomAlt_1;\ldots;\ExampledAtomAlt_m}}$
  As $\ExampledSequence \ExampledSequenceLeq \ExampledSequenceAlt$ and
  $\ExampledSequenceAlt \ExampledSequenceLeq \ExampledSequence$, we know $n =
  m$.
  By Lemma~\ref{lem:spec-set-list-perm} $\sigma = \InverseOf{\sigma_1} \Compose
  \sigma_2$ is such that 
  $\ExampledAtom_i \leq \ExampledAtomAlt_{\sigma(i)}$ and
  $\ExampledAtomAlt_{\sigma(i)} \leq \ExampledAtom_i$.

  By IH, this means that calling
  $\RigidSynthAtom(\ExampledAtom_i,\ExampledAtomAlt_{\sigma(i)})$ returns an atom lens,
  $\AtomLens_i$. 
  As such, $\PCF{AllSome}$ on all these atom lens options returns a list of atom lenses.
  This then returns the sequence lens $(\SequenceLensOf{(\String_0,\StringAlt_0)
  \SeqLSep \AtomLens_1 \SeqLSep \ldots \SeqLSep \AtomLens_n \SeqLSep
  (\String_n,\StringAlt_n)},\InverseOf{\sigma})$.
\end{case}

\begin{case}[dnf regexps]
  Let $(\DNFOf{\ExampledSequence_1 \DNFSep \ldots \DNFSep
    \ExampledSequence_n},\IntListSet) = \ExampledDNFRegex$ and 
  $(\DNFOf{\ExampledSequenceAlt_1 \DNFSep \ldots \DNFSep
    \ExampledAtomAlt_m},\IntListSet') =
  \ExampledSequenceAlt$.

  Let $\sigma_1 = \SortingOf{\ListOf{\ExampledSequence_1;\ldots;\ExampledSequence_n}}$
  Let $\sigma_2 = \SortingOf{\ListOf{\ExampledSequenceAlt_1;\ldots;\ExampledSequenceAlt_m}}$
  As $\ExampledDNFRegex \ExampledDNFLeq \ExampledDNFRegexAlt$ and
  $\ExampledSequenceAlt \ExampledDNFLeq \ExampledDNFRegex$, we know $n = m$.
  By Lemma~\ref{lem:spec-set-list-perm} $\sigma = \InverseOf{\sigma_1} \Compose
  \sigma_2$ is such that 
  $\ExampledAtom_i \leq \ExampledAtomAlt_{\sigma(i)}$ and
  $\ExampledAtomAlt_{\sigma(i)} \leq \ExampledAtom_i$.

  By IH, this means that calling
  $\RigidSynthSequence(\ExampledSequence_i,\ExampledSequenceAlt_{\sigma(i)})$
  returns a sequence lens, $\SequenceLens_i$. 
  As such, $\PCF{AllSome}$ on all these sequence lens options returns a list of
  sequence lenses. 
  This then returns the DNF lens $(\DNFLensOf{\SequenceLens_1 \DNFLSep \ldots
    \DNFLSep \SequenceLens_n},\InverseOf{\sigma})$.
\end{case}
\end{proof}

\begin{lemma}
\label{lem:rigidsynth-completeness}
  Let $\DNFRegex$ and $\DNFRegexAlt$ be DNF regexes.
  Let $\Examples$ be a set of string pairs.
  If there exists a DNF lens $\DNFLens \OfRewritelessType \DNFRegex
  \Leftrightarrow \DNFRegexAlt$ such that $\Examples \in
  \SemanticsOf{\DNFLens}$, then $\RigidSynth(\DNFRegex,\DNFRegexAlt,\Examples)$
  returns a DNF lens.
\end{lemma}
\begin{proof}
  Let $\ListOf{(\String_1,\StringAlt_1);\ldots;(\String_n,\StringAlt_n)} =
  \Examples$.
  
  Let $\ExampledDNFRegex =
  \EmbedExamplesOf{\ListOf{([1],\String_1);\ldots;([n],\String_n)}}{\DNFRegex}$

  Let $\ExampledDNFRegexAlt = \EmbedExamplesOf{\ListOf{([1],\StringAlt_1);\ldots;([n],\StringAlt_n)}}{\DNFRegexAlt}$

  We know that $\EmbedExamples$ doesn't fail, as each $\String_i \in
  \LanguageOf{\DNFRegex}$ and each $\StringAlt_i \in \LanguageOf{\DNFRegexAlt}$.

  This means that $\ListOf{([1],\String_1);\ldots;([n],\String_n)} \in \DNFRegex
  \Generates \ExampledDNFRegex$ and
  $\ListOf{([1],\StringAlt_1);\ldots;([n],\StringAlt_n)} \in \DNFRegexAlt
  \Generates \ExampledDNFRegexAlt$.

  As $\DNFLens \OfRewritelessType \DNFRegex \Leftrightarrow \DNFRegexAlt$, we
  know that $\DNFRegex$ and $\DNFRegexAlt$ are strongly unambiguous.

  As such the conditions for Lemma~\ref{lem:exampledordercomplete} apply so
  $\ExampledDNFRegex \ExampledDNFLeq \ExampledDNFRegexAlt$ and
  $\ExampledDNFRegexAlt \ExampledDNFLeq \ExampledDNFRegex$.

  By Lemma~\ref{lem:exampledrigidcompleteorder}, that means that
  $\RigidSynthInternal(\ExampledDNFRegex,\ExampledDNFRegexAlt)$ returns a DNF
  lens, so then so too does $\RigidSynth$.
\end{proof}

\begin{lemma}
  \label{lem:synthdnflens-completeness}
  Given DNF regular expressions $\DNFRegex$ and
  $\DNFRegexAlt$, and a set
  of examples $\Examples$, if there exists a DNF lens $\DNFLens$ such that
  $\DNFLens \OfType \DNFRegex \Leftrightarrow \DNFRegexAlt$ and for all
  $(\String,\StringAlt)$ in $\Examples$, $(\String,\StringAlt) \in
  \SemanticsOf{\DNFLens}$, then\\
  $\SynthDNFLens(\DNFRegex,\DNFRegexAlt,\Examples)$ will
  return a DNF lens.
\end{lemma}
\begin{proof}
  If $\DNFLens \OfType \DNFRegex \Leftrightarrow \DNFRegexAlt$, then there exist
  DNF regular expressions $\DNFRegex'$ and $\DNFRegexAlt'$ such that
  $\DNFLens \OfRewritelessType \DNFRegex' \Leftrightarrow \DNFRegexAlt'$, 
  $\DNFRegex \StarOf{\RewriteDNF} \DNFRegex'$, and $\DNFRegexAlt
  \StarOf{\RewriteDNF} \DNFRegexAlt'$.

  However, we may not perform the same single rewrites as $\DNFRegex
  \StarOf{\RewriteDNF} \DNFRegexAlt$, because we infer certain expansions, which
  are then taken earlier.  However, confluence allows us to reorganize the order
  of the expansions (though this may possibly increase the total number of
  expansions, and change the ultimate DNF lens), as was done in the retype case
  of DNF completeness.

  As such, we know that, there exist two DNF regular expressions $\DNFRegex''$
  and $\DNFRegexAlt''$, and a DNF lens $\DNFLens'$ such that $\DNFRegex
  \StarOf{\RewriteDNF} \DNFRegex''$, $\DNFRegexAlt \StarOf{\RewriteDNF}
  \DNFRegexAlt''$, $\DNFLens' \OfRewritelessType \DNFRegex'' \Leftrightarrow
  \DNFRegexAlt''$, $\SemanticsOf{\DNFLens'} = \SemanticsOf{\DNFLens}$, and
  $(\DNFRegex'',\DNFRegexAlt'')$ are enumerated by the queue.

  If there are any $(\DNFRegex''',\DNFRegexAlt''')$ pairs enumerated before
  $(\DNFRegex'',\DNFRegexAlt'')$ such that\\
  $\RigidSynth(\DNFRegex''',\DNFRegexAlt''')$ returns a lens, then that lens is
  returned, and we are done.

  If not, eventually $(\DNFRegex'',\DNFRegexAlt'')$ are enumerated.

  By Lemma~\ref{lem:rigidsynth-completeness}, as $\DNFLens' \OfRewritelessType \DNFRegex'' \Leftrightarrow
  \DNFRegexAlt''$, $\RigidSynth(\DNFRegex'',\DNFRegexAlt'')$ returns. This is
  then immediately returned by \SynthDNFLens{}.
\end{proof} 

\begin{theorem}
  Given regular expressions $\Regex$ and $\RegexAlt$, and a set
  of examples $\Examples$, if there exists a lens $\Lens$ such that
  $\Lens \OfType \Regex \Leftrightarrow \RegexAlt$ and for all
  $(\String,\StringAlt)$ in $\Examples$, $(\String,\StringAlt) \in
  \SemanticsOf{\Lens}$, then $\SynthLens(\Regex,\RegexAlt,\Examples)$ will
  return a lens.
\end{theorem}
\begin{proof}
  Let $\Lens \OfType \Regex \Leftrightarrow \RegexAlt$.  By
  Theorem~\ref{thm:dnflc}, there exists a DNF lens $\DNFLens \OfType \ToDNFRegexOf{\Regex}
  \Leftrightarrow \ToDNFRegexOf{\RegexAlt}$, such that $\SemanticsOf{\DNFLens} =
  \SemanticsOf{\Lens}$.  In particular, this means that $\Examples \subset
  \SemanticsOf{\DNFLens}$.  This means, from
  Lemma~\ref{lem:synthdnflens-completeness},
  $\SynthDNFLens(\ToDNFRegexOf{\Regex},\ToDNFRegexOf{\RegexAlt})$ returns a
  lens, $\DNFLens''$.  From Theorem~\ref{thm:dnfls}, we can then convert using
  $\ToLens$, to get a lens which we return.
\end{proof}

%\begin{lemma}\leavevmode
%  \begin{itemize}
%  \item If $\RigidSynthAtom(\ExampledAtom,\ExampledAtomAlt)$
%    returns a lens, then $\ExampledAtom \ExampledAtomLeq \ExampledAtomAlt$ and
%    $\ExampledAtomAlt \ExampledAtomLeq \ExampledAtom$.
%  \item If $\RigidSynthSequence(\ExampledSequence,\ExampledSequenceAlt)$
%    returns a lens, then $\ExampledSequence \ExampledSequenceLeq \ExampledSequenceAlt$ and
%    $\ExampledSequenceAlt \ExampledSequenceLeq \ExampledSequence$.
%  \item If $\RigidSynthInternal(\ExampledDNFRegex,\ExampledDNFRegexAlt)$
%    returns a lens, then $\ExampledDNFRegex \ExampledDNFLeq \ExampledDNFRegexAlt$ and
%    $\ExampledDNFRegexAlt \ExampledDNFLeq \ExampledDNFRegex$.
%  \end{itemize}
%\end{lemma}
%
%\begin{lemma}\leavevmode
%  \begin{itemize}
%  \item If $\RigidSynthAtom(\ExampledAtom,\ExampledAtomAlt)$
%    returns a lens, then $\ExampledAtom \ExampledAtomLeq \ExampledAtomAlt$ and
%    $\ExampledAtomAlt \ExampledAtomLeq \ExampledAtom$.
%  \item If $\RigidSynthSequence(\ExampledSequence,\ExampledSequenceAlt)$
%    returns a lens, then $\ExampledSequence \ExampledSequenceLeq \ExampledSequenceAlt$ and
%    $\ExampledSequenceAlt \ExampledSequenceLeq \ExampledSequence$.
%  \item If $\RigidSynthInternal(\ExampledDNFRegex,\ExampledDNFRegexAlt)$
%    returns a lens, then $\ExampledDNFRegex \ExampledDNFLeq \ExampledDNFRegexAlt$ and
%    $\ExampledDNFRegexAlt \ExampledDNFLeq \ExampledDNFRegex$.
%  \end{itemize}
%\end{lemma}

\subsection{Additional Proofs}
\label{additional-proofs}

In the paper, some claims were made that aren't necessarily the main theorems.
In this area we prove those claims.

\begin{definition}
  Let $\Rewrite$ be a rewrite rule on regular expressions.
  We define $\Rewrite_{\ToDNFRegex}$ as the rewrite relation on DNF regular
  expressions defined by $\ToDNFRegexOf{\Regex} \Rewrite_{\ToDNFRegex}
  \ToDNFRegexOf{\RegexAlt}$ if $\Regex \Rewrite \RegexAlt$.
\end{definition}

\begin{definition}
  We overload $\DNFLensHasSemanticsOf{\Lens}$ to extend to regular expressions
  with rewriteless DNF lenses after conversion to DNF form.
  In particular, $\Regex \DNFLensHasSemanticsOf{\Lens}
  \RegexAlt$ if there exists $\DNFLens$ such that $\SemanticsOf{\DNFLens} =
  \SemanticsOf{\Lens}$ and $\DNFLens \OfRewritelessType (\ToDNFRegexOf{\Regex})
  \Leftrightarrow (\ToDNFRegexOf{\RegexAlt})$.
\end{definition}

\begin{lemma}
  $\IsConfluentWithPropertyOf{\Rewrite}{\DNFLensHasSemanticsOf{\Lens}}$ implies
  $\IsConfluentWithPropertyOf{\Rewrite_{\ToDNFRegex}}{\DNFLensHasSemanticsOf{\Lens}}$.
\end{lemma}
\begin{proof}
  Let $\DNFLens \OfRewritelessType \DNFRegex \Leftrightarrow \DNFRegexAlt$,
  with $\SemanticsOf{\DNFLens} = \SemanticsOf{\Lens}$.
  Let $\DNFRegex \Rewrite_{\ToDNFRegex} \DNFRegex_1$ and $\DNFRegexAlt
  \Rewrite_{\ToDNFRegex} \DNFRegexAlt_2$.
  This means that there exists $\Regex$, $\RegexAlt$, $\Regex_2$, and $\RegexAlt_2$
  such that $\Regex \Rewrite \Regex_2$, $\RegexAlt \Rewrite \RegexAlt_2$,
  $\ToDNFRegexOf{\Regex} = \DNFRegex$, $\ToDNFRegexOf{\RegexAlt} =
  \DNFRegexAlt$, $\ToDNFRegexOf{\Regex_2} = \DNFRegex_2$, and
  $\ToDNFRegexOf{\RegexAlt_2} = \DNFRegexAlt_2$.
  This means, as $\IsConfluentWithPropertyOf{\Rewrite}{\DNFLensHasSemanticsOf{\Lens}}$,
  there exists $\Regex_3$, $\RegexAlt_3$ such that there exists $\DNFLens_3
  \OfRewritelessType \ToDNFRegexOf{\Regex_3} \Leftrightarrow
  \ToDNFRegexOf{\RegexAlt_3}$ where $\SemanticsOf{\DNFLens_3} =
  \SemanticsOf{\DNFLens}$, $\Regex_2 \Rewrite \Regex_3$ and $\RegexAlt_2
  \Rewrite \RegexAlt_3$.  This means that $\DNFRegex_2 \Rewrite_{\ToDNFRegex}
  \ToDNFRegexOf{\Regex_3}$, and $\DNFRegexAlt_2 \Rewrite_{\ToDNFRegex}
  \ToDNFRegexOf{\Regex_3}$, and as before there exists $\DNFLens_3
  \OfRewritelessType \ToDNFRegexOf{\Regex_3} \Leftrightarrow
  \ToDNFRegexOf{\RegexAlt_3}$ where $\SemanticsOf{\DNFLens_3} =
  \SemanticsOf{\DNFLens}$.  As such,
  $\IsConfluentWithPropertyOf{\Rewrite_{\ToDNFRegex}}{\DNFLensHasSemanticsOf{\Lens}}$.
\end{proof}

\begin{lemma}
  $\IsConfluentWithPropertyOf{\Rewrite_{\ToDNFRegex}}{\DNFLensHasSemanticsOf{\Lens}}$
  implies $\IsConfluentWithPropertyOf{\Rewrite}{\DNFLensHasSemanticsOf{\Lens}}$.
\end{lemma}
\begin{proof}
  Let $\DNFLens \OfRewritelessType \ToDNFRegexOf{\Regex} \Leftrightarrow
  \ToDNFRegexOf{\RegexAlt}$.
  Let $\Regex \Rewrite \Regex_1$ and $\RegexAlt
  \Rewrite \RegexAlt_2$.
  This means that $\ToDNFRegexOf{\Regex} \Rewrite_{\ToDNFRegex}
  \ToDNFRegexOf{\Regex_2}$ and $\ToDNFRegexOf{\RegexAlt} \Rewrite_{\ToDNFRegex}
  \ToDNFRegexOf{\RegexAlt_2}$.  As
  $\IsConfluentWithPropertyOf{\Rewrite_{\ToDNFRegex}}{\DNFLensHasSemanticsOf{\Lens}}$,
  then there exists $\DNFLens_3$, $\DNFRegex_3$ and $\DNFRegexAlt_3$ such that
  $\ToDNFRegexOf{\Regex_2} \Rewrite_{\ToDNFRegex} \DNFRegex_3$,
  $\ToDNFRegexOf{\RegexAlt_2} \Rewrite_{\ToDNFRegex} \DNFRegexAlt_3$,
  $\DNFLens_3 \OfRewritelessType \DNFRegex_3 \Leftrightarrow \DNFRegexAlt_3$,
  and $\SemanticsOf{\DNFLens_3} = \SemanticsOf{\DNFLens}$.
  This means that $\DNFRegex_3 = \ToDNFRegexOf{\Regex_3}$,
  $\DNFRegexAlt_3 = \ToDNFRegexOf{\RegexAlt_3}$.  As such,
  $\DNFLens_3 \OfRewritelessType \ToDNFRegexOf{\Regex_3} \Leftrightarrow
  \ToDNFRegexOf{\RegexAlt_3}$, so
  $\IsConfluentWithPropertyOf{\Rewrite}{\DNFLensHasSemanticsOf{\Lens}}$.
\end{proof}

\begin{lemma}
  If $\IsBisimilarWithPropertyOf{\Rewrite}{\DNFLensHasSemanticsOf{\Lens}}$, then
  $\IsBisimilarWithPropertyOf{\Rewrite_{\ToDNFRegex}}{\DNFLensHasSemanticsOf{\Lens}}$.
\end{lemma}
\begin{proof}
  Let $\DNFLens \OfRewritelessType \DNFRegex \Leftrightarrow \DNFRegexAlt$,
  with $\SemanticsOf{\DNFLens} = \SemanticsOf{\Lens}$.
  Let $\DNFRegex \Rewrite_{\ToDNFRegex} \DNFRegex_2$.
  This means that there exists $\Regex$ and $\Regex_2$
  such that $\Regex \Rewrite \Regex_2$,
  $\ToDNFRegexOf{\Regex} = \DNFRegex$, and $\ToDNFRegexOf{\Regex_2} =
  \DNFRegex_2$.
  This means, as
  $\IsConfluentWithPropertyOf{\Rewrite}{\DNFLensHasSemanticsOf{\Lens}}$,
  there exists $\RegexAlt_2$ such that there exists $\DNFLens_2
  \OfRewritelessType \ToDNFRegexOf{\Regex_2} \Leftrightarrow
  \ToDNFRegexOf{\RegexAlt_2}$ where $\SemanticsOf{\DNFLens_2} =
  \SemanticsOf{\DNFLens}$, and $\ToRegexOf{\DNFRegexAlt} \Rewrite \RegexAlt_2$.
  Symmetrically for if $\DNFRegexAlt \Rewrite_{\ToDNFRegex} \DNFRegexAlt_2$, so
  $\IsBisimilarWithPropertyOf{\Rewrite_{\ToDNFRegex}}{\DNFLensHasSemanticsOf{\Lens}}$.
\end{proof}

\begin{lemma}
  If $\IsBisimilarWithPropertyOf{\Rewrite_{\ToDNFRegex}}{\DNFLensHasSemanticsOf{\Lens}}$, then
  $\IsBisimilarWithPropertyOf{\Rewrite}{\DNFLensHasSemanticsOf{\Lens}}$.
\end{lemma}
\begin{proof}
  Let $\DNFLens \OfRewritelessType \ToDNFRegexOf{\Regex} \Leftrightarrow
  \ToDNFRegexOf{\RegexAlt}$,
  with $\SemanticsOf{\DNFLens} = \SemanticsOf{\Lens}$.
  Let $\Regex \Rewrite \Regex_2$.
  This means that $\ToDNFRegexOf{\Regex} \Rewrite_{\ToDNFRegex}
  \ToDNFRegexOf{\Regex_2}$.  This means that, as
  $\IsBisimilarWithPropertyOf{\Rewrite_{\ToDNFRegex}}{\DNFLensHasSemanticsOf{\Lens}}$
  there exists $\DNFRegexAlt_2$,
  $\DNFLens_2$ such that $\ToDNFRegexOf{\RegexAlt} \Rewrite_{\ToDNFRegex}
  \DNFRegexAlt_2$, and $\DNFLens_2 \OfRewritelessType \ToDNFRegexOf{\Regex_2}
  \Leftrightarrow \DNFRegexAlt_2$, and $\SemanticsOf{\DNFLens_2} =
  \SemanticsOf{\DNFLens}$.  As $\ToDNFRegexOf{\RegexAlt} \Rewrite_{\ToDNFRegex}
  \DNFRegexAlt_2$, we have $\DNFRegexAlt_2 = \ToDNFRegexOf{\RegexAlt_2}$, where
  $\RegexAlt \Rewrite \RegexAlt_2$.  Symmetrically for if $\RegexAlt \Rewrite
  \RegexAlt_2$, so
  $\IsBisimilarWithPropertyOf{\Rewrite}{\DNFLensHasSemanticsOf{\Lens}}$.
\end{proof}

\begin{lemma}
  If $\Rewrite$ doesn't introduce ambiguity, then neither does
  $\Rewrite_{\ToDNFRegex}$.
\end{lemma}
\begin{proof}
  Let $\DNFRegex \Rewrite_{\ToDNFRegex} \DNFRegexAlt$, where $\DNFRegex$ is
  strongly unambiguous.  This means $\DNFRegex =
  \ToDNFRegexOf{\Regex}$ and $\DNFRegexAlt = \ToDNFRegexOf{\RegexAlt}$, and
  $\Regex \Rewrite \RegexAlt$.  As $\ToDNFRegexOf{\Regex}$ is strongly
  unambiguous iff $\Regex$ is, $\Regex$ is strongly unambiguous.  By assumption,
  we now have $\RegexAlt$ is strongly unambiguous, so then
  $\ToDNFRegexOf{\RegexAlt}$ also is.
\end{proof}

\begin{lemma}
  If $\Rewrite_{\ToDNFRegex}$ doesn't introduce ambiguity, then neither does
  $\Rewrite$.
\end{lemma}
\begin{proof}
  Let $\Regex \Rewrite \RegexAlt$, where $\Regex$ is
  strongly unambiguous.  This means 
  $\ToDNFRegexOf{Regex} \Rewrite_{\ToDNFRegex} \ToDNFRegexOf{\RegexAlt}$.
  By assumption, that means $\ToDNFRegexOf{\RegexAlt}$ is strongly unambiguous.
  As $\ToDNFRegexOf{\RegexAlt}$ strongly unambiguous iff $\RegexAlt$ strongly
  unambiguous, by assumption we have $\RegexAlt$ is strongly unambiguous.
\end{proof}

\begin{lemma}
  $\Rewrite_{\ToDNFRegex}$ maintains the language if, and only if $\Rewrite$ does.
\end{lemma}
\begin{proof}
  $\Regex \Rewrite \RegexAlt$.
  $\LanguageOf{\Regex} = \LanguageOf{\ToDNFRegexOf{\Regex}}$.
  $\LanguageOf{\RegexAlt} = \LanguageOf{\ToDNFRegexOf{\RegexAlt}}$. 
  $\ToDNFRegexOf{\Regex} = \LanguageOf{\ToDNFRegexOf{\RegexAlt}}$.
  $\ToDNFRegexOf{\Regex} \Rewrite_{\ToDNFRegex} \ToDNFRegexOf{\RegexAlt}$.
  $\DNFRegex \Rewrite_{\ToDNFRegex} \DNFRegexAlt$.
\end{proof}

Thus, we have confluence, bisimilarity, unambiguity retention, language
retention on rewrites for regular
expressions, iff we also do for the DNF rewrites they induce.  As such, because
confluence and bisimilarity, alongside not introducing ambiguity or changing the
language, is sufficient on DNF rewrites, then it is sufficient on rewrite
rules for regular expressions as a sufficient condition is fulfilled.

\fi

\end{document}

%%% Local Variables:
%%% TeX-master: "main"
%%% End: